\documentclass[11pt,a4paper]{article}
\pdfoutput=1
\usepackage{jheppub}

\usepackage{bbm}
\usepackage{slashed}
\usepackage{mathtools}
\usepackage{amsmath}
\usepackage[matrix,arrow]{xy}
\usepackage{color}
\usepackage{upgreek}
\usepackage{graphicx}
\usepackage{longtable}
\makeatletter
\DeclareFontFamily{OMX}{MnSymbolE}{}
\DeclareSymbolFont{MnLargeSymbols}{OMX}{MnSymbolE}{m}{n}
\SetSymbolFont{MnLargeSymbols}{bold}{OMX}{MnSymbolE}{b}{n}
\DeclareFontShape{OMX}{MnSymbolE}{m}{n}{
    <-6>  MnSymbolE5
   <6-7>  MnSymbolE6
   <7-8>  MnSymbolE7
   <8-9>  MnSymbolE8
   <9-10> MnSymbolE9
  <10-12> MnSymbolE10
  <12->   MnSymbolE12
}{}
\DeclareFontShape{OMX}{MnSymbolE}{b}{n}{
    <-6>  MnSymbolE-Bold5
   <6-7>  MnSymbolE-Bold6
   <7-8>  MnSymbolE-Bold7
   <8-9>  MnSymbolE-Bold8
   <9-10> MnSymbolE-Bold9
  <10-12> MnSymbolE-Bold10
  <12->   MnSymbolE-Bold12
}{}

\let\llangle\@undefined
\let\rrangle\@undefined
\DeclareMathDelimiter{\llangle}{\mathopen}%
                     {MnLargeSymbols}{'164}{MnLargeSymbols}{'164}
\DeclareMathDelimiter{\rrangle}{\mathclose}%
                     {MnLargeSymbols}{'171}{MnLargeSymbols}{'171}
\makeatother
\DeclareFontFamily{U}{mathx}{\hyphenchar\font45}
\DeclareFontShape{U}{mathx}{m}{n}{
      <5> <6> <7> <8> <9> <10>
      <10.95> <12> <14.4> <17.28> <20.74> <24.88>
      mathx10
      }{}
\DeclareSymbolFont{mathx}{U}{mathx}{m}{n}
\DeclareMathSymbol{\bigtimes}{1}{mathx}{"91}

\DeclareFontFamily{OT1}{pzc}{}
\DeclareFontShape{OT1}{pzc}{m}{it}{<-> s * [1.10] pzcmi7t}{}
\DeclareMathAlphabet{\mathpzc}{OT1}{pzc}{m}{it}

\newcommand{\qed}{\nobreak \ifvmode \relax \else
      \ifdim\lastskip<1.5em \hskip-\lastskip
      \hskip1.5em plus0em minus0.5em \fi \nobreak
      \vrule height0.75em width0.5em depth0.25em\fi}

\DeclareMathAlphabet{\mathpzc}{OT1}{pzc}{m}{it}

\DeclareFontFamily{U}{mathx}{\hyphenchar\font45}
\DeclareFontShape{U}{mathx}{m}{n}{
      <5> <6> <7> <8> <9> <10>
      <10.95> <12> <14.4> <17.28> <20.74> <24.88>
      mathx10
      }{}
\DeclareSymbolFont{mathx}{U}{mathx}{m}{n}
\DeclareFontSubstitution{U}{mathx}{m}{n}
\DeclareMathAccent{\widecheck}{0}{mathx}{"71}
\DeclareMathAccent{\wideparen}{0}{mathx}{"75}

\usepackage{enumitem}
\newlist{Description}{description}{1}
\setlist[Description]{labelindent=1.5\parindent,leftmargin=1.5\parindent}

\def\contract{\textrm{\scalebox{1.3}{$\iota$}}}

\def\tHooft{\mbox{'t Hooft }}
\def\gm{\gamma_{\rm m}}

\def\gerel{\gamma_{\mathrm{e},0}}
\def\gi{\mathrm{g}}

\def\NN{\mathcal{N}}
\def\BB{\mathcal{B}}

\def\FF{\mathcal{F}}
\def\EE{\mathcal{E}}
\def\II{\mathcal{I}}
\def\LL{\mathcal{L}}

\def\SS{\mathcal{S}}

\def\DD{\mathcal{D}}
\def\MM{\mathcal{M}}
\def\GG{\mathcal{G}}
\def\OO{\mathcal{O}}
\def\QQ{\mathcal{Q}}
\def\RR{\mathcal{R}}
\def\TT{\mathcal{T}}
\def\HH{\mathcal{H}}

\def\PP{\mathcal{P}}
\def\UU{\mathcal{U}}

\def\XX{\mathcal{X}}
\def\YY{\mathcal{Y}}

\def\cg{\mathpzc{g}}
\def\sc{\mathpzc{c}}
\def\sp{\mathpzc{p}}
\def\sx{{X_{\infty}}}
\def\sy{{\mathcal{Y}_{\infty}}}
\def\ssy{\mathpzc{y}}
\def\sw{\upomega}
\def\fmMM{\overline{\underline{\mathcal{M}}}}
\def\mMM{\mathcal{M}}

\def\fMM{\overline{\underline{\mathcal{M}}}}

\def\fOmega{\overline{\underline{\Omega}}}
\def\fSigma{\overline{\underline{\Sigma}}}
\def\fssy{\overline{\underline{\ssy}}}

\def\fC{\overline{\underline{C}}}

\def\pd{\partial}
\def\del{\partial}
\def\ed{  \textrm{d}}
\def\eD{ \textrm{D}}
\def\im{ \, \textrm{im} }
\def\rnk{ \, \textrm{rnk} \, }
\def\Re{ \, \textrm{Re}  }
\def\Im{ \, \textrm{Im}  }

\def\Hom{ \, \textrm{Hom} \, }

\def\Ad{ \, \textrm{Ad}  }
\def\ad{ \, \textrm{ad} }
\def\Tr{ \, \textrm{Tr}  }
\def\tr{ \, \textrm{tr} }
\def\sgn{ \, \textrm{sgn} }

\def\half{\frac{1}{2}}

\def\rG{\mathrm{G}}

\def\ie{{\it i.e.}}
\def\eg{{\it e.g.}}

\def\etc{{\it etc}}

\def\mfa{\mathfrak{a}}
\def\mfb{\mathfrak{b}}
\def\Span{\mathrm{Span} }

\def\um{\underline{m}}
\def\un{\underline{n}}
\def\up{\underline{p}}
\def\uq{\underline{q}}
\def\ua{\underline{a}}
\def\ub{\underline{b}}
\def\uc{\underline{c}}

\def\bbJ{\mathbb{J}}
\def\tbbJ{\tilde{\mathbb{J}}}

\def\bfm{\mathbf{m}}
\def\bfn{\mathbf{n}}
\def\bfp{\mathbf{p}}

\def\ubfm{\underline{\mathbf{m}}}
\def\ubfn{\underline{\mathbf{n}}}

\newcommand{\overbar}[1]{\mkern 1.5mu\overline{\mkern-1.5mu#1\mkern-1.5mu}\mkern 1.5mu}

\def\bfmbar{\overbar{\mathbf{m}}}
\def\bfnbar{\overbar{\mathbf{n}}}
\def\bfpbar{\overbar{\mathbf{p}}}

\def\ubfmbar{\underline{\overbar{\mathbf{m}}}}
\def\ubfnbar{\underline{\overbar{\mathbf{n}}}}

\def\Zbar{\overbar{Z}}
\def\Pbar{\overbar{P}}
\def\XXbar{\overbar{\mathcal{X}}}
\def\ebar{\overbar{e}}
\def\EEbar{\overbar{\mathcal{E}}}
\def\pdbar{\overbar{\partial}}
\def\Psibar{\overbar{\Psi}}
\def\pibar{\overbar{\pi}}
\def\varphibar{\overbar{\varphi}}
\def\Qbar{\overbar{Q}}
\def\Psibar{\overbar{\Psi}}
\def\psibar{\overbar{\psi}}
\def\xibar{\overbar{\xi}}

\def\tm{\tilde{m}}
\def\tn{\tilde{n}}

\def\Lie{\pounds}

\def\Lsq{{L^2}}

\def\be{\begin{equation}}
\def\ee{\end{equation}}
\def\bea{\begin{eqnarray}}
\def\eea{\end{eqnarray}}

\def\gdot{\dot{\gamma}}
\def\adot{\dot{\alpha}}
\def\bdot{\dot{\beta}}

\def\s{\sigma}
\def\sb{\overbar{\sigma}}

\def\a{\alpha}
\def\b{\beta}

\def\e{\epsilon}

\def\l{\lambda}

\def\k{\kappa}

\def\m{\mu}
\def\n{\nu}

\def\r{\rho}

\def\s{\sigma}

\def\ve{\varepsilon}

\def\um{\underline{m}}
\def\un{\underline{n}}

\def\ge{\tau}

\def\sR{\sigma}

\newcommand{\dpageref}[1]{{page \pageref{#1}}}

\title{Semiclassical framed BPS states}
\author[a]{Gregory W.~Moore,}
\author[b]{Andrew B.~Royston,}
\author[c]{Dieter Van den Bleeken}

\affiliation[a]{NHETC and
Department of Physics and Astronomy, Rutgers University \\
126 Frelinghuysen Rd., Piscataway NJ 08855, USA}
\affiliation[b]{George P.\ \& Cynthia Woods Mitchell Institute for Fundamental Physics and Astronomy, \\
Texas A\&M University, College Station, TX 77843, USA}
\affiliation[c]{Physics Department, Bo\u{g}azi\c{c}i University\\
 34342 Bebek / Istanbul, TURKEY}

\emailAdd{gmoore@physics.rutgers.edu}
\emailAdd{aroyston@physics.tamu.edu}
\emailAdd{dieter.van@boun.edu.tr}

\abstract{We provide a semiclassical description of framed BPS states in four-dimensional $\NN = 2$ super Yang--Mills theories probed by 't Hooft defects, in terms of a supersymmetric quantum mechanics on the moduli space of singular monopoles.  Framed BPS states, like their ordinary counterparts in the theory without defects, are associated with the $\Lsq$ kernel of certain Dirac operators on moduli space, or equivalently with the $\Lsq$ cohomology of related Dolbeault operators.  The Dirac/Dolbeault operators depend on two Cartan-valued Higgs vevs.  We conjecture a map between these vevs and the Seiberg--Witten special coordinates, consistent with a one-loop analysis and checked in examples. The map incorporates all perturbative and nonperturbative corrections that are relevant for the semiclassical construction of BPS states, over a suitably defined weak coupling regime of the Coulomb branch.  We use this map to translate wall crossing formulae and the no-exotics theorem to statements about the Dirac/Dolbeault operators.  The no-exotics theorem, concerning the absence of nontrivial $SU(2)_R$ representations in the BPS spectrum, implies that the kernel of the Dirac operator is chiral, and further translates into a statement that all $\Lsq$ cohomology of the Dolbeault operator is concentrated in the middle degree.  Wall crossing formulae lead to detailed predictions for where the Dirac operators fail to be Fredholm and how their kernels jump.  We explore these predictions in nontrivial examples.  This paper explains the background and arguments behind the results announced in the short note \cite{MRVP3summary}.}


\begin{document}
\begin{flushright} MI-TH-1603 \end{flushright}
\maketitle

%%%%%%%%%%%%%%%%%%%%
%%%%%%%%%%%%%%%%%%%%
\section{Introduction}
%%%%%%%%%%%%%%%%%%%%
%%%%%%%%%%%%%%%%%%%%

The present paper gives a definition of BPS states of supersymmetric $\NN=2$ gauge theories in precise mathematical terms, namely, in terms of the kernels of Dirac-like operators on moduli spaces of magnetic monopoles.  Such a description is only available at weak coupling but, thanks to wall crossing formulae, this is in principle all that is needed to capture the entire BPS spectrum.   An important aspect of this work is the inclusion of certain line defects known as 't Hooft operators.  In general, defects---objects that can be inserted in the path integral, possibly supported on subspaces of spacetime of positive dimension or codimension---have been playing an increasingly important role in work on quantum field theory in the past years. In this paper we consider line defects localized on zero-dimensional submanifolds of space, hence stretching along the time axis. Such defects modify the Hilbert space and Hamiltonian of the theory in an interesting way and lead to the ``framed BPS states'' of the title. 

To put this work in a broader context recall that in  non-Abelian gauge theory Wilson--'t Hooft line operators  were employed in \cite{Wilson:1974sk,'tHooft:1977hy} to study the phases of Yang--Mills theory. Their role in the study of electromagnetic duality was emphasized in \cite{Kapustin:2005py,Kapustin:2006pk}.  In gauge theories with extended supersymmetry these defects can be defined so as to preserve a subset of the supersymmetries of the original theory.  Four-dimensional gauge theories with $\NN =2$ supersymmetry provide excellent laboratories for these studies because they often have two different gauge theory descriptions.  The first is the ultra-violet (UV) microscopic description in terms of the original Yang--Mills gauge group, while the second is an infra-red (IR) low energy effective description in terms of an Abelian gauge theory.  The latter is the quantum-exact long wavelength description of Seiberg and Witten \cite{Seiberg:1994rs,Seiberg:1994aj}.

The relationship between UV Wilson--'t Hooft defects and their IR counterparts in four-dimensional $\NN = 2$ theories was recently explored in \cite{Gaiotto:2010be}, leading to new insights for both the original theories without defects and those in the presence of defects.  A new type of BPS state, dubbed a \emph{framed BPS state} in \cite{Gaiotto:2010be}, has played and continues to play a fundamental role in these developments.\footnote{The adjective ``framed'' refers to the relation of these BPS states to framed quivers.  It is not a terribly good name, but we are stuck with it.}  Framed BPS states are simply BPS states of the theory in the presence of the defect.  They saturate a Bolgomolny bound, the form of which is modified due to the presence of the defect.  In order to minimize confusion, the BPS states in the theory without defects are referred to as \emph{vanilla BPS states}. An important class of framed BPS states, (those which can undergo wall crossing), can be thought of as bound states of the vanilla BPS states with the defect (or defects) at the ``core.'' 

A driving force behind much recent work on supersymmetric gauge and gravity theories has been the effort to give a complete description of wall crossing phenomena for BPS states in those theories that possess a continuous space of vacua.  BPS states are stable at generic points of the vacuum manifold, but as the parameters of the vacuum are dialed through certain co-dimension one walls, they will decay into a number of constituent BPS states, consistent with charge and energy conservation.  Although the ultimate goal is to describe the change in the Hilbert space of BPS states as these \emph{marginal stability walls} are crossed, typically one is limited to describing how certain protected indices that count states with signs---or generalizations of these to spin weighted characters that keep track of spin information---change.  A remarkable wall crossing formula for certain `generalized Donaldson--Thomas invariants' was obtained by Kontsevich and Soibelman \cite{2008arXiv0811.2435K}. Since then, starting with \cite{Gaiotto:2008cd}, a number of physics papers have demonstrated that the Kontsevich--Soibelman formula should apply to BPS states of field theories. 

Framed BPS states undergo wall crossing just as the vanilla states do in $\NN =2$ gauge theories \cite{Gaiotto:2010be,Andriyash:2010qv,Andriyash:2010yf}.  It turns out that the core-halo picture of wall crossing---based on the multi-centered supergravity solutions of \cite{Denef:2000nb,Denef:2000ar}, and employed to obtain primitive and semiprimitive wall crossing formulae for vanilla BPS states in \cite{Denef:2007vg}---is ideally suited to the study of wall crossing for framed BPS states, and a complete description of framed wall crossing was obtained in this way in \cite{Gaiotto:2010be}.  Consistency of the framed wall crossing formula implies the `motivic' Kontsevich--Soibelman wall crossing formula for certain \emph{protected spin characters} of vanilla BPS states, provided there is a sufficiently rich supply of line defects.  This is an excellent example of how one can learn detailed information about a theory by studying it in the presence of defects.

The protected spin characters (PSC's), introduced in \cite{Gaiotto:2010be} for both vanilla and framed BPS states, are \emph{a priori} protected quantities in that they vanish when evaluated on long representations of the supersymmetry algebra.  In order to ensure this property, essential use is made of the internal $SU(2)_R$ symmetry of four-dimensional $\NN = 2$ gauge theories.  While neither the generator $J^3$ of the Cartan of the rotation group, nor the generator $I^3$ of the Cartan of $SU(2)_R$, commute with the supercharges, there is a diagonal combination, $\II^3 = J^3 + I^3$, that does.  The PSC's are characters of the corresponding diagonal $SU(2) \subset SU(2)_{\rm rot} \times SU(2)_R$.  This should be contrasted with the `refined BPS index' of earlier investigations, based on $J^3$.  

However in all known computations of wall crossing on the Coulomb branch of $\NN = 2$ Seiberg--Witten theories for both framed and vanilla BPS states, the two quantities---the PSC and the refined BPS index---behave identically.  Along with other considerations, this led \cite{Gaiotto:2010be} to make the \emph{no-exotics conjecture:}  the Hilbert space of framed BPS states, and the Hilbert space of vanilla BPS states (after factoring out a universal center-of-mass half hypermultiplet factor), over any point on the Coulomb branch of an $\NN = 2$ gauge theory, transform trivially with respect to $SU(2)_R$.  The conjecture has since been proven for pure glue $SU(N)$ theories---\ie\ just vectormultiplets and no matter hypermultiplets---via geometric engineering techniques in \cite{Chuang:2013wt}, a result that has recently been extended to pure glue theories based on simply-laced gauge algebras \cite{DelZotto:2014bga}.  Recently, a more generally applicable argument based on the structure of supersymmetric stress-energy multiplets has been given \cite{CordovaDumitrescu}.

Investigations of framed BPS states have so far been confined to low energy effective descriptions of the physics, \ie\ the Seiberg--Witten description or other related descriptions such as \cite{Lee:2011ph,Cordova:2013bza}.  These have the advantage of being valid in strong coupling regions of the Coulomb branch, but are incapable of probing the core of a BPS particle or line defect on length scales comparable to the inverse of the Higgs vacuum expectation value (vev) determining the Coulomb branch vacuum.  At these length scales the non-Abelian fields can become effectively light, as the Higgs field that gives them mass can deviate significantly from its vev, and thus should not have been integrated out.  

In the case of vanilla BPS states, there is a well known complementary description of them as soliton states of the microscopic UV quantum field theory.\footnote{More precisely, the BPS states carrying nonzero magnetic charge are realized as soliton states, while those carrying zero magnetic charge are ordinary perturbative particle states.}  This description is limited to the weak coupling regime of the Coulomb branch.  It is based on the semiclassical quantization of collective coordinate degrees of freedom associated with the moduli space of classical soliton field configurations (\ie 't Hooft--Polyakov monopoles and Julia--Zee dyons).  Semiclassical quantization leads to a supersymmetric quantum mechanics with the moduli space as target, in which the space of BPS states is identified with the $\Lsq$ kernel of a (twisted) Dirac operator, or equivalently the $\Lsq$ cohomology of a (twisted) Dolbeault operator.  This structure, for $\NN=2$ gauge theories without defects, was uncovered in a series of papers beginning in the mid `90's \cite{Gauntlett:1993sh,Sethi:1995zm,Gauntlett:1995fu,Lee:1996kz,Gauntlett:1999vc,Gauntlett:2000ks}; for a review and complete list of references see \cite{Weinberg:2006rq}.

%%%%%%%%%%%%%%
\subsection{What we do}
%%%%%%%%%%%%%%

In this paper we review,  extend, and streamline the constructions in \cite{Gauntlett:1993sh,Sethi:1995zm,Gauntlett:1995fu,Lee:1996kz,Gauntlett:1999vc,Gauntlett:2000ks}
 in several ways.  First and foremost, we give a semiclassical construction of \emph{framed} BPS states in $\NN = 2$ gauge theories in the presence of line defects, focusing on the class of pure glue theories for arbitrary simple gauge group $G$ in the presence of half BPS 't Hooft line defects.\footnote{The analogous construction for half BPS Wilson line defects has been given in \cite{Tong:2014yla}, and the extension to general Wilson--'t Hooft defects and $\NN = 2$ gauge theories with matter is being considered \cite{BrennanMoore}.}  The construction is based on a collective coordinate approximation of the theory in the presence of defects, where the collective coordinates take values in the moduli space of singular monopoles.  These are moduli spaces of gauge-inequivalent solutions to the Bogomolny equation with prescribed singularities and have been well studied, (though prior to \cite{MRVdimP1} most analyses, with an exception being \cite{Kapustin:2006pk}, were restricted to the case of singular $SU(2)$ monopoles).  These are hyperk\"ahler manifolds, with possible singularities depending on the choice of 't Hooft charges at the defect locations.  In \cite{MRVdimP1} we obtained a formula for their dimension via a generalization of the Callias index theorem \cite{Callias:1977kg}, and gave the result a physical interpretation along the lines of \cite{Weinberg:1979ma,Weinberg:1979zt}, in terms of multiple fundamental mobile monopoles of various types---one type for each simple root---in the presence of the singular monopoles.

The collective coordinate dynamics is obtained from the microscopic field theory in the usual way, by expanding the fields around an (approximate) solution to the equations of motion in which the parameters of the soliton moduli space are promoted to dynamical variables.  The fermionic fields of the $\NN = 2$ field theory possess zero mode excitations around the soliton as well, and we find a supersymmetric quantum mechanics with four supercharges, containing terms of the form derived in the vanilla case \cite{Gauntlett:1999vc,Gauntlett:2000ks}.  We stress, however, that there are additionally new dynamical terms, preserving the same supersymmetries, whose presence is due to the combined effect of having both defects and a nonzero theta angle.\footnote{In the vanilla case these terms collapse to a total time derivative in the collective coordinate Lagrangian, which is nevertheless present and affects the definition of conjugate momenta.  In the case with defects they cannot be written as a total time derivative.}  Boundary terms in the field theory action, localized on the defects and \emph{required} for consistency of the 't Hooft defect boundary conditions with the variational principle, play an important role in this analysis.  After quantization, the new terms of the collective coordinate theory affect the form of the supercharge operator, represented as a Dirac operator, in just the right way so as to make the semiclassical construction of BPS states compatible with the low energy Seiberg--Witten analysis.

Our second main extension relative to previous work concerns the regime of validity of the semiclassical approach.  We emphasize the importance of including the one-loop corrections to the background monopole mass due to integrating out quantum fluctuations around the soliton ansatz.  Indeed, we explain in detail why it is actually an \emph{inconsistent} approximation to the dynamics of the full quantum field theory to consider collective coordinate dynamics while ignoring these corrections.  Doing so 
led some physicists to make mistakes  when comparing semiclassical analyses with predictions from Seiberg--Witten theory.   

Our analysis of these issues in section \ref{ssec:validity} leads us to a   conjecture concerning the form of quantum corrections in the collective coordinate theory.  We suggest that agreement between the two approaches in the weak coupling regime of the Coulomb branch\footnote{Our definition of the weak coupling regime is such that the series expansion around infinity of the Seiberg--Witten dual coordinate $a_{\rm D} = a_{\rm D}(a)$ (or equivalently the prepotential) is convergent.  See item \ref{wcRegime} in section \ref{ssec:fBPSspace} for further discussion.} where both are valid, determines the form of all perturbative and nonperturbative corrections to the Dirac/supercharge operator on moduli space, that are required in order to achieve agreement with the predictions of Seiberg--Witten theory for the BPS spectrum.  The conjecture is corroborated by the known one-loop corrections to the background mass, first computed in \cite{Kaul:1984bp}.  (We give an extension of this one-loop result to the case with defects and general gauge group.)  Further checks, perhaps along the lines of \cite{Rebhan:2004vn,Rebhan:2006fg}, are possible and should be carried out.

Our conjecture takes the form of a map from Seiberg--Witten data $\{ a(u),a_{\rm D}(u),\gamma\}$---the special coordinates, dual special coordinates, and the electromagnetic charge of the BPS state in question---to the data defining the twisted Dirac operator.  The latter consists a real Cartan-valued Higgs vev $X_\infty \in \mathfrak{t}$ and an asymptotic magnetic charge $\gm$ determining the moduli space, together with a second real Higgs vev $\YY_\infty$, determining a triholomorphic Killing field on the moduli space that is used to construct the Dirac operator.  The magnetic charge $\gm$ takes values in either the co-root lattice in the vanilla case, or possibly a shifted copy of the co-root lattice in the case with defects.  The map describes, firstly, a specific choice of weak coupling duality frame, which defines a trivialization of the local system of electromagnetic charges $\gamma \in \Gamma$, and a specific choice for the (Cartan-valued) special coordinates and dual coordinates for the given point $u$ on the Coulomb branch.  With respect to this particular trivialization, $\gm$ is the magnetic component of $\gamma$, while we have
\begin{equation}\label{MPmap}
X_\infty = \Im(\zeta^{-1} a(u))~, \qquad \YY_\infty = \Im(\zeta^{-1} a_{\rm D}(u))~.
\end{equation}
Here $\zeta$ is a phase that is either part of the definition of the line defects in the framed case, or is the phase of minus the central charge in the vanilla case, $Z_{\gamma}(u) = - \zeta |Z_{\gamma}(u)|$.

With the aid of this map we can then make a precise identification of the physical spaces of framed and vanilla BPS states, graded by electromagnetic charge, with subspaces of the kernel of the Dirac operator, corresponding to an appropriately identified eigenvalue of an electric charge operator that commutes with the Dirac operator.  These identifications, \eqref{mainres} and \eqref{mainres2}, are the main results of the paper.  One important issue we do \underline{not} address is the nature of the singularities in the singular monopole moduli spaces. Known singularities are of orbifold type, and will not play an important role in the definition of the $L^2$ kernel. However, if there are more serious singularities they could play an important role in the definition of the $L^2$ kernel. 

In addition to these two extensions, we found that several additional new results were required along the way in order to arrive at the complete picture of semiclassical (framed and vanilla) BPS states summarized by \eqref{mainres} and \eqref{mainres2}.  Let us briefly mention two here.  It is well known that the universal cover of the vanilla monopole moduli space, $\widetilde{\MM}$, factorizes as $\widetilde{\MM} = \mathbb{R}^4 \times \MM_0$, where the flat $\mathbb{R}^4$ factor is associated with the center of mass and $\MM_0$ is an irreducible simply-connected hyperk\"ahler manifold known as the ``strongly centered'' moduli space \cite{Atiyah:1988jp,Hitchin:1995qw}.  It follows from the identification of the moduli space with rational maps to the flag variety (see appendix \ref{app:Dquotient})  that the fundamental group of the moduli space is $\pi_1(\MM) \cong \mathbb{Z}$, for any simple Lie group and magnetic charge, and that this fundamental group acts on the universal cover $\widetilde{\MM}$ by a translation along one of the $\mathbb{R}$ factors associated with the overall phase of the monopole together with with an isometry of $\MM_0$.   However, for a generic magnetic charge, only a subgroup, $L \mathbb{Z} \subset \mathbb{Z}$  of the fundamental group can be associated with the action of asymptotically nontrivial gauge transformations.

What is the value of the positive integer $L$?  This is important from a physical perspective because the action of asymptotically nontrivial gauge transformations is associated with electric charge.  It turns out that a Dirac spinor $\Psi$ on the universal cover, $\widetilde{\MM}$, will descend to a well-defined spinor on the quotient $\widetilde{\MM}/L \mathbb{Z}$, provided that it has properly quantized eigenvalues with respect to the electric charge operator.  However, an additional $\mathbb{Z}_L \equiv \mathbb{Z}/L\mathbb{Z}$ equivariance condition, related to the component of the electric charge parallel to the (dual of the) magnetic charge, will need to be imposed to make it well-defined on $\MM$.  In the case of $SU(2)$ gauge group the answer is known: $L$ is the (integer) value of the magnetic charge along the simple co-root.  As far as we are aware, this question has not been addressed in the case of general simple compact Lie group $G$.  In appendix \ref{app:Dquotient} we prove that in general $L$ is the greatest common divisor of the nonzero components of the \emph{dual} of the magnetic charge along the simple roots, where the dual is defined with respect to the Killing form on the Lie algebra such that long roots have length-squared equal to two.  The associated $\mathbb{Z}_L$ equivariance condition is part of our identification in \eqref{mainres2}.

Another item that appears not to have been spelled out previously, is the manifestation of the internal $SU(2)_R$ symmetry.  Its action in the collective coordinate quantum mechanics is identified with (the lift from the tangent bundle to the Dirac spinor bundle of) the action of the triplet of complex structures.  This action does not commute with the Dirac operator, but it does preserve the kernel.  While it has been previously recognized that there is such an $SU(2)$ action on the kernel \cite{Gauntlett:1999vc}, we connect it directly to the canonical $SU(2)_R$ of the field theory through the collective coordinate ansatz for the field theory fermions.  

Related to to this, it was observed in \cite{deVries:2008ic,deVries:2010vc} that the naive angular momentum operators of the collective coordinate theory, induced from the $SO(3)$ isometry of the moduli space, do not satisfy the proper commutation relations with the supercharges, but that they could be made to do so by adding an additional term to them that involves the action of the complex structures.  We use our identification of $SU(2)_R$ to show that the `naive' angular momentum generators are in fact precisely the generators of the same diagonal subgroup of angular momentum and $SU(2)_R$ that is used in the construction of the protected spin characters.  This result, together with our semiclassical identifications of the spaces of framed and vanilla BPS states, allows us to give a precise identification of the framed and vanilla protected spin characters with index characters of the corresponding Dirac operators.  The construction is spelled out in section \ref{ssec:mathphys}.

With the identification of BPS spaces and symmetry operators in hand, we proceed in the latter half of the paper to applications and examples.  In section \ref{Section:Applications} we give three mathematical conjectures based on our identifications.  The first is a translation of the no-exotics theorem into a statement about the kernel of the Dirac operator, or equivalently about the twisted Dolbeault cohomology---where it takes a particularly striking form, reminiscent of Sen's famous result concerning $\Lsq$ harmonic forms on $\MM$ \cite{Sen:1994yi,Segal:1996eb}.  No-exotics is equivalent to the statement that all nontrivial $\Lsq$ cohomology of the twisted Dolbeault operator is concentrated in the middle degree, and is furthermore primitive with respect to the $\mathfrak{sl}(2)$ Lefschetz action on $(0,\ast)$-forms described in \cite{MR1486984,MR1958088}.\footnote{The raising and lowering operators for the Lefschetz $\mathfrak{sl}(2)$ are constructed from the holomorphic-symplectic form built out of the remaining two K\"ahler forms.}  So in particular, if the dimension of the hyperk\"ahler moduli space is $4N$, then all BPS states are represented by $(0,N)$-forms with respect to any choice of complex structure.  This result applies to both the $\Lsq$ (twisted) Dolbeault cohomology of the singular monopole moduli spaces, $\fMM$, and the strongly centered vanilla moduli spaces $\MM_0$.

The remaining two conjectures concern the translation of wall crossing formulae into statements about the family of Dirac operators, constructed from the data $\{\gm,X_\infty,\YY_\infty\}$.  The identifications of this paper imply that these operators are Fredholm except along certain co-dimension one walls in $X_\infty$-$\YY_\infty$ space.  At these walls the kernels jump in a way determined by the framed or vanilla wall crossing formulae, applied to their index characters.  See section \ref{ssec:mathphys} for the detailed statements.  We hope it will be possible to give an independent proof of the Fredholm property by generalizing the sort of asymptotic analysis made in \cite{Stern:2000ie}.

We also employ our identifications to go in the other direction---using mathematical results to obtain physical insight.  In section \ref{ssec:vanlocus} we show how a simple Lichnerowicz--Weitzenbock  argument involving the square of the Dirac operator implies the existence of a `vanishing locus' in the weak coupling regime of the Coulomb branch, where we can always determine the complete BPS spectrum.  This is very important because wall crossing formulae, for instance, only specify the \emph{change} in the spectrum when a wall is crossed.  One needs other methods, such as the semiclassical ones described here, to determine the actual spectrum in a starting chamber.

In sections \ref{Section:VanillaEx} and \ref{Section:FramedEx} we consider in turn a vanilla and framed example in which explicit $\Lsq$ wavefunctions on the moduli space can be constructed.  The vanilla example concerns the magnetic charge $\{1,1\}$ moduli space for a rank two simple gauge group.  This is an example that has been considered in the past by several groups \cite{Lee:1996kz,Gauntlett:1996cw,Lee:1996if,Gauntlett:1999vc,Jante:2013kha}.  The strongly centered moduli space $\MM_0$ is the single-centered Taub--NUT manifold, and the kernel of the twisted Dirac operator was determined long ago in \cite{Pope:1978zx}.  Our reason for reviewing this example here is to highlight some aspects that were not addressed previously.  In particular we show that the kernel jumps in precisely the way predicted by wall crossing formulae, and we show that the extremum of the radial probability function, constructed from the explicit wavefunctions, agrees precisely with the formula of Denef  \cite{Denef:2000nb,Denef:2000ar}  for the bound state radius.

The example of section \ref{Section:FramedEx} concerns framed BPS states in the presence of a single 't Hooft defect in the $\mathfrak{su}(2)$ gauge theory (of arbitrary 't Hooft charge).  We first review the known framed BPS spectrum in the weak coupling regime, as derived in \cite{Gaiotto:2010be}, extending those results slightly to give a closed-form expression for the generating functional determining all of the framed protected spin characters for this line defect in all weak coupling chambers.  We then consider a special subclass of framed BPS states from the semiclassical point of view, consisting of a single smooth 't Hooft--Polyakov monopole bound to the 't Hooft defect.  We take the opportunity to present the remarkable closed-form expressions of \cite{Cherkis:2007jm, Cherkis:2007qa, Blair:2010vh} for the classical Higgs field and gauge field corresponding to these configurations, and briefly summarize a computation of the moduli space metric directly from these solutions, carried out in \cite{Shah}.  The moduli space is a certain $\mathbb{Z}_{|p|}$ quotient of single centered Taub--NUT, where $p \in\mathbb{Z}$ determines the 't Hooft charge, with $p = \pm 1$ corresponding to the minimal $SO(3)$ 't Hooft defect.\footnote{There are certainly more efficient ways to find this moduli space, which has been known for some time \cite{Cherkis:1997aa,Cherkis:1998xca,Cherkis:1998hi}, but we find the direct approach to be illuminating and of great pedagogical value.}  The Taub--NUT analysis can thus be recycled for this example (with minor modifications to account for the quotient), though the map \eqref{MPmap} between mathematical and physical quantities is completely different in the two examples.  We find perfect agreement between the jumping of the kernel and the predictions from framed wall crossing formulae.  We also explore the implications of our generating functional for Dirac operators on a family of eight-dimensional manifolds describing two smooth monopoles in the presence of a singularity.

%%%%%%%%%%%%%%%%%%
\subsection{Structure of the paper}
%%%%%%%%%%%%%%%%%%

In writing this paper we have attempted to give a complete, self-contained, and pedagogical exposition.  In doing so we necessarily had to include a significant amount of review material, collected from many sources, and the manuscript necessarily became quite lengthy.  We have made every effort to state clearly what is review and give proper references.  For those experts interested primarily in a statement of the new results given without derivation, and especially in the mathematical applications of section \ref{Section:Applications}, we recommend the summary paper \cite{MRVP3summary}.

The structure of the paper is as follows.  After introducing the class of UV theories we consider, section \ref{sec:N2andGMN} reviews essential material on the low energy Seiberg--Witten description of vanilla and framed BPS states including IR line defects, protected spin characters, the core-halo picture, and wall crossing formulae.  Section \ref{sec:clfBPS} concerns the construction of classical BPS field configurations describing dyons in the presence of 't Hooft defects.  We review in detail the essential properties of (singular) monopole moduli spaces, and describe solutions to the ``secondary'' BPS equation that lead to static dyon field configurations.  In section \ref{sec:sc} we carry out the semiclassical quantization of the theory around these soliton field configurations, following the ``moduli space with potential'' approach for the collective coordinate dynamics developed in \cite{Lee:1996kz,Gauntlett:1999vc,Gauntlett:2000ks}.  The inclusion of defects requires a careful analysis of singularities.  Boundary terms in the action, localized on the defects, play an important role.  We motivate the conjectural map \eqref{MPmap} in section \ref{ssec:validity}, while sections \ref{ssec:fBPSspace} and \ref{ssec:vBPSspace} give the precise semiclassical identifications of the spaces of framed and vanilla BPS states.  Section \ref{Section:Applications} details the mathematical applications mentioned above, and two examples are analyzed in sections \ref{Section:VanillaEx} and \ref{Section:FramedEx}.  We describe some future work in section \ref{Section:Further}.

The appendices contain additional material on \ref{N2conventions}: our $\NN = 2$ field theory conventions; \ref{app:bcs}: the analysis of the variational principle and supersymmetry variations in the presence of defects; \ref{app:monmod}: aspects of monopole moduli spaces; \ref{app:cc}: the collective coordinate expansion; \ref{app:holforms}: quantization of collective coordinates via $(0,\ast)$-forms; \ref{app:lemma}: the construction of an inverse to the map \eqref{MPmap} in the asymptotic region of the Coulomb branch; \ref{appendix:TN}: the analysis of the twisted Dirac operator on Taub--NUT; and \ref{wcapp}: framed wall crossing formulae.  Quite a bit of notation has been introduced to meet the goals of completeness and pedagogy, so we have included an index of notation, \ref{app:notation}, listing each item's name or meaning and the place it is first defined.

%%%%%%%%%%%%%%%%%%%%%%
%%%%%%%%%%%%%%%%%%%%%%
\section{$\NN =2$ SYM, line defects, and framed BPS states}\label{sec:N2andGMN}
%%%%%%%%%%%%%%%%%%%%%%
%%%%%%%%%%%%%%%%%%%%%%

In this paper for simplicity we restrict to the case of pure supersymmetric gauge theory with compact, simple gauge group $G$\label{liegroup}.  The UV microscopic field variables are an $\NN = 2$ non-Abelian vector multiplet, consisting of a gauge field $A\label{Adef}$, a complex scalar $\varphi$\label{phidef}, and a pair of Weyl fermions $\psi_A$\label{fermdef}.  The fermions transform as a doublet under the $SU(2)_R$\label{SU2Rdef} internal symmetry group while the gauge and scalar fields are singlets.  All fields are valued in the Lie algebra $\mathfrak{g}$\label{liealg} of the gauge group $G$, which we can equivalently view as the representation space of the adjoint representation.

The bosonic part of the classical action naturally splits into two parts:
\begin{align}\label{Scl}
S =&~ S_{\rm van} + S_{\rm def}~,
\end{align}
with
\begin{align}\label{Svanbos}
S_{\rm van} :=&~  - \frac{1}{g_{0}^2} \int \ed^4 x  \Tr \left( \half F_{\mu\nu} F^{\mu\nu}  + D_\mu \varphi D^\mu \varphibar - \frac{1}{4} ( [\varphi, \varphibar ] )^2  \right) + \frac{\theta_0}{8\pi^2} \int \Tr F \wedge F + \cr
&~ + \textrm{fermi}~,
\end{align}
the standard vanilla action.  The term $S_{\rm def}$ is present when line defects are inserted into the theory, and its form depends on the type of defect under consideration.  We will have more to say about this below, after introducing the defects that we will study in this paper.  Details on the fermionic part of \eqref{Svanbos}, and the supersymmetry of the full action can be found in  appendix \ref{N2conventions}.

Let us settle here some essential notation and conventions.  The bar on $\varphi$ in \eqref{Svanbos} denotes the natural conjugation on the complexified Lie algebra, $\mathfrak{g}_{\mathbb{C}} \simeq \mathfrak{g} \otimes \mathbb{C}$.\label{complexliealg}  We work on $\mathbb{R}^{1,3}$ with signature $(-,+,+,+)$ and orientation $\ed^4 x = \ed x^0 \ed^3 x$, $\epsilon_{0123} = 1$, and use geometric conventions where $F = \half F_{\mu\nu} \ed x^\mu \ed x^\nu$, $F_{\mu\nu} = 2 \pd_{[\mu} A_{\nu]} + [A_\mu ,A_\nu]$\label{Fdef} and $D_\mu \varphi = \pd_\mu \varphi + [A_\mu, \varphi]$.\label{Ddef}  ``$\Tr$'' denotes a symmetric bi-invariant quadratic form on $\mathfrak{g}$.  This choice induces a quadratic form on the vector space dual, $\mathfrak{g}^\ast$\label{liealgdual}.  We follow the convention where the length-squared of a long root is equal to two.  This ensures that Euclidean instantons with integral winding number $k$ will have action $2\pi i k \tau_0$, where $\tau_0 := \frac{4 \pi i}{g_{0}^2} + \frac{\theta_0}{2\pi}$\label{tau0def}.    In terms of the standard Cartan--Killing form we then have for any $T^a\in\mathfrak{g}$ that
\begin{equation}\label{Trdef}
\Tr(T^1 T^2) := -\frac{1}{2h^\vee} \tr(\ad(T^1) \ad(T^2))~.
\end{equation}
We will sometimes also denote this form by $(T^1,T^2) \equiv \Tr(T^1T^2)$.  For $\mathfrak{g} = \mathfrak{su}(N)$, the real Lie algebra of traceless anti-Hermitian matrices, $\Tr = -\tr_{\bf N}$.  The minus sign is inserted for convenience so that the quadratic form is positive-definite on $\mathfrak{g}$.  We always work in conventions where representation matrices for elements of a real Lie algebra are anti-Hermitian.  

The classical theory has a family of vacua labeled by gauge-inequivalent constant values, $\varphi = \varphi_\infty$, satisfying $[\varphi_{\infty}, \varphibar_{\infty}] = 0$.  In this paper we will restrict our considerations to the case of \emph{maximal symmetry breaking}, meaning that $\varphi_{\infty} \in \mathfrak{g}_{\mathbb{C}}$ is regular.  Then $\varphi_{\infty}$ defines a Cartan subalgebra, $\mathfrak{t}_{\mathbb{C}} \subset \mathfrak{g}_{\mathbb{C}}$, which we will henceforth refer to as \emph{the} (complexified) Cartan subalgebra.  The space of gauge-inequivalent $\varphi_\infty$ can be identified with $\mathfrak{t}_{\mathbb{C}}/W$ where the quotient identifies points related by the action of the Weyl group, $W$.  Associated to $\mathfrak{t}_{\mathbb{C}}$ we have a root decomposition of the Lie algebra, 
\begin{equation}\label{rootdecomp}
\mathfrak{g}_{\mathbb{C}} = \mathfrak{t}_{\mathbb{C}} \oplus \bigoplus_{\alpha \in \Delta} (-iE_\alpha) \cdot \mathbb{C}~.
\end{equation}
The $E_\alpha$ are raising/lowering operators and $\Delta$ denotes the set of non-zero roots.  Recall that to each root $\alpha$ we associate a co-root $H_{\alpha}$ defined by the property $\langle \beta, H_\alpha \rangle = 2 (\beta,\alpha)/ (\alpha,\alpha)$, $\forall \beta \in \Delta$.  Here $\langle~,~\rangle$ denotes the canonical pairing\label{canpair} $\mathfrak{t}^\ast \times \mathfrak{t} \to \mathbb{R}$ and $(~,~)$ is the Killing form \eqref{Trdef}---or rather the one induced on $\mathfrak{g}^\ast$ from \eqref{Trdef}; we will use the same notation for both.  (All Killing forms on simple Lie algebras are equivalent up to rescaling so this particular relation does not depend on the choice of Killing form.)  We normalize the $E_{\alpha}$ so that $\{E_{\pm\alpha}, iH_{\alpha} \}$ forms a canonical $\mathfrak{sl}(2)$ subalgebra.  Further details on our Lie algebra conventions can be found in appendix A of \cite{MRVdimP1}.

%%%%%%%%%%%%%%%%%%%%%%
\subsection{Line defects of type $\zeta$}\label{Section:Defects}
%%%%%%%%%%%%%%%%%%%%%%

In the past few years several works have emphasized the importance of one-dimensional (line) and two-dimensional (surface) defects in four-dimensional $\NN = 2$ SYM.  Broadly speaking, defects generalize the Wilson loop in Yang--Mills theory to lines and higher dimensional objects that carry both electric and magnetic charge.  Much as the Wilson loop provides insight into the physics of confinement, these works have shown that defects are excellent probes of low energy strong-coupling phenomena in $\mathcal{N} = 2$ gauge theories.

We follow the approach of \cite{Kapustin:2005py,Kapustin:2006pk} in which defects are defined in the UV theory as operator insertions and/or specified boundary conditions on the field variables in the path integral at the defect locus.  Great progress can be made by studying defects that preserve as much of the original symmetry of the theory as possible, so that one has maximal analytic control.  We work on the Coulomb branch where conformal and superconformal symmetries are broken.  Thus the relevant symmetry algebra in the absence of defects is the super-Poincar\'e algebra, $\mathfrak{s}^0 \oplus \mathfrak{s}^1$, with even (bosonic) subalgebra
\begin{equation}\label{bos}
\mathfrak{s}^0 = \mathfrak{poin}(1,3) \oplus \mathfrak{su}(2)_R \oplus \mathfrak{u}(1)_R~,
\end{equation}
and odd generators $(Q_{\alpha}^{A}, \Qbar_{\dot{\alpha} B})$\label{Qdef} that transform in the representation
\begin{equation}
\left( ({\bf 2}, {\bf 1}; {\bf 2})_{+1} \oplus ({\bf 1}, {\bf 2}; {\bf 2})_{-1} \right)_{\mathbb{R}} ~,
\end{equation}
where the ``$\mathbb{R}$'' indicates that a reality condition is imposed: $(Q_{\alpha}^A)^\dag = \Qbar_{\dot{\alpha}A}$.

We focus on line defects in $\mathbb{R}^{1,3}$ which are located at fixed points $\vec{x}_n \in \mathbb{R}^3$ \label{defposdef} and extend in the time direction.  Generically the unbroken subalgebra of the Poincar\'e algebra is  simply $\mathbb{R}_t$ corresponding to time translations.  Enhanced symmetries occur for special configurations of defects.  For example, a single defect located at $\vec{x} = \vec{x}_0$ will preserve the algebra $\mathfrak{so}(3)$\label{rotationalg} of spatial rotations about $\vec{x}_0$.\footnote{As Lie algebras, $\mathfrak{so}(3) \cong \mathfrak{su}(2)$.  However we reserve the $\mathfrak{so}(3)$ notation specifically for the Lie algebra associated with angular momentum.}  Following \cite{Gaiotto:2010be} we consider defects that also preserve the $SU(2)_R$ symmetry as well as half of the supersymmetry.  In fact there is a $U(1)$ family of possibilities labeled by a phase $\zeta$.\label{zetadef}  They may be characterized as the subalgebra that is fixed under the involution
\begin{align}\label{invol}
& Q_{\alpha}^{A} \to \zeta^{-1} \epsilon^{AB} \sigma_{\alpha\dot{\beta}}^0 \Qbar^{\dot{\beta}}_{B}~, \cr
& \Qbar^{\dot{\alpha}}_{A} \to \zeta \epsilon_{AB} \overbar{\sigma}^{0 \dot{\alpha}\beta} Q_{\beta}^{B}~.
\end{align}
An explicit set of generators invariant under \eqref{invol} is
\begin{equation}\label{Rsusys}
\RR_{\alpha}^A = \zeta^{\half} Q_{\alpha}^A + \zeta^{-\frac{1}{2}} \sigma_{\alpha\dot{\beta}}^0 \Qbar^{\dot{\beta}A}~.
\end{equation}
A \emph{line defect of type} $\zeta$, denoted $L_{\zeta}(\cdots)$\label{Ldef}, is a defect preserving the bosonic symmetry algebra $\mathbb{R}_t \oplus \mathfrak{so}(3) \oplus \mathfrak{su}(2)_R$ together with the supersymmetries \eqref{Rsusys}.

The remaining labels depend on the class of defects under consideration.  Wilson--'t Hooft defects \cite{Kapustin:2005py}, for example, are labeled by equivalence classes of magnetic and electric charges $[(P,Q)] \in \LL$,\label{PQdef} where the set $\LL$ is often taken as
\begin{equation}\label{lineoplattice}
\LL = \left( \Lambda_G  \times \Lambda_{G}^\vee \right) /W~,
\end{equation}
although there are other possibilities \cite{Gaiotto:2010be,Aharony:2013hda}.  Here $W$ is the Weyl group and $\Lambda_G$ is defined by
\begin{equation}\label{cochar}
\Lambda_{G} = \left\{ P \in \mathfrak{t} ~|~ \exp{(2\pi P)} = \mathbbm{1}_G \right\} \cong \Hom{\left(U(1),T\right)}~,
\end{equation}
where $\mathfrak{t}$ is a Cartan subalgebra of $\mathfrak{g}$ and $T \subset G$ is the corresponding Cartan torus.  The dual, $\Lambda_{G}^\vee \cong \Hom{\left(T,U(1)\right)}$, is isomorphic to the character lattice of $G$ and so $\Lambda_G$ is sometimes referred to as the co-character lattice.  The case $(P,Q) = (0,Q) \in \Lambda_G \times \Lambda_{G}^\vee$ corresponds to a supersymmetric Wilson line insertion in an irreducible representation $\rho_{[Q]}$ of $G$, whose dominant weight is given by the representative of the Weyl orbit $[Q] \in \Lambda_{G}^\vee/W$ in the closure of the fundamental Weyl chamber. 

In this paper, for simplicity, we will mostly\footnote{However, as we will explain further at the end of this subsection, allowing for nonzero theta angle implies that our analysis includes a certain subset of Wilson--'t Hooft defects.} restrict ourselves to the case of pure \tHooft defects.  An \tHooft defect of charge $P_n$ and phase $\zeta$ at a point $\vec{x}_n$ is defined by imposing the following boundary conditions on the fields: 
\begin{align}\label{tHooftbc}
& \zeta^{-1} \varphi = \left( \frac{g_{0}^2\theta_0}{8\pi^2} -i \right)\frac{P_n}{2r_n} + O(r_{n}^{-1/2})~, \cr
& F = \left(\sin{\uptheta_n} \ed \uptheta_n \ed \upphi_n + \frac{g_{0}^2 \theta_0}{8\pi^2} \frac{\ed t \ed r_n}{r_{n}^2} \right) \frac{P_n}{2} + O(r_{n}^{-3/2})~, \qquad (r_n \to 0)~,
\end{align}
where $(r_n = |\vec{x} - \vec{x}_n|, \uptheta_n,\upphi_n)$ are spherical coordinates centered on the defect.    Note that the one-forms $\ed \uptheta_n$ and $\ed \upphi_n$ behave as $O(r_{n}^{-1})$ when expanded in a basis of orthonormal one-forms, so both leading terms in $F$ are $O(r_{n}^{-2})$.  $P_n$ is required to be a covariantly constant section of the adjoint bundle restricted to the infinitesimal two-sphere surrounding $\vec{x}_n$.  However, by making patch-wise local gauge transformations in the northern and southern hemisphere, we can conjugate $P_n$ to a constant which, for convenience, we may furthermore assume to be in the Cartan subalgebra defined by $\varphi_{\infty}$.  We will assume that this has been done.  Then single-valuedness of the transition function, $\exp(P_n\upphi_n)$, on the overlap of patches implies $P_n \in \Lambda_G$, \eqref{cochar}.  Local gauge transformations---\ie\ gauge transformations that go to the identity at infinity---can be used to conjugate $P_n$ by a Weyl transformation and thus it is only the Weyl orbit $[P_n]$ that is physically meaningful.

Let us rewrite the complex Higgs field in terms of two real Higgs fields $X,Y$ according to
\begin{equation}
\zeta^{-1} \varphi = Y + i X~.
\end{equation}
Then the boundary conditions \eqref{tHooftbc} can also be expressed in the form
\begin{align}\label{defectbcs}
& B^i = \frac{P_n}{2r_{n}^2} \hat{r}_{n}^i + O(r_{n}^{-3/2})~, \qquad  & X = -\frac{P_n}{2r_n} + O(r_{n}^{-1/2})~, \cr
& E^i = -\tilde{\theta}_0 \cdot \frac{P_n}{2r_{n}^2} \hat{r}_{n}^i + O(r_{n}^{-3/2})~, \qquad & Y =  \tilde{\theta}_0 \cdot \frac{P_n}{2r_n} + O(r_{n}^{-1/2})~,
\end{align}
where $E_i = F_{i0}$ and $B_i = \half \epsilon_{ijk} F^{jk}$ \label{EBdef} are the electric and magnetic field, and we have introduced the notation
\begin{equation}\label{t0tilde}
\tilde{\theta}_0 := \frac{g_{0}^2 \theta_0}{8\pi^2}~,
\end{equation}
as this quantity will appear frequently in the following.  These boundary conditions allow for subleading terms that are still singular.  As we argued in \cite{MRVdimP1}, this type of behavior is observed in explicit solutions and generally can occur for \tHooft defects such that there exists a root $\alpha$ of the Lie algebra with $|\langle\alpha, P_n \rangle| = 1$.  For example, in the solutions of \cite{Cherkis:2007jm, Cherkis:2007qa, Blair:2010vh} the subleading behavior of the Higgs field is regular at the locus of a minimal $SU(2)$ defect, while it is has $1/\sqrt{r_n}$ behavior for the minimal $SO(3)$ defect.  (The leading pole is always of the form \eqref{defectbcs}.)  We will see this explicitly when these solutions are reviewed in subsection \ref{CDmonopole}.

When $\theta_0 = 0$, \eqref{defectbcs} reduce to the boundary conditions considered in \cite{MRVdimP1}, where the indicated subleading behavior of the fields was determined from imposing consistency of the defect boundary conditions with the variational principle for the action.  These arguments can be generalized to the situation under consideration here; some of the details are presented in Appendix \ref{app:bcs}.  In particular, consistency with the variational principle also constrains the form of the defect terms, $S_{\rm def}$, in the action \eqref{Scl}.  Let there be \tHooft defects inserted at points $\vec{x}_n$ and let $S_{\varepsilon_n}^2$ denote the infinitesimal two-sphere of radius $r_n = \varepsilon_n$ surrounding $\vec{x}_n$.  Then we find that a natural choice for these boundary terms is   
\begin{align}\label{Sclbndry}
S_{\rm def} :=&~ \frac{2}{g_{0}^2} \int \ed t \sum_n \Re \bigg\{ \zeta^{-1} \int_{S_{\varepsilon_n}^2} \Tr \left\{ (i F - \star F) \varphi \right\} \bigg\}~, \cr
=&~ \frac{2}{g_{0}^2} \int \ed t \sum_n \Re \bigg\{ \zeta^{-1} \int_{S_{\varepsilon_n}^2} \ed^2 \Omega_n \varepsilon_{n}^2 \hat{r}_{n}^i \Tr \left\{ ( i B_i - E_i) \varphi \right\} \bigg\}~.
\end{align}
In addition to providing a well-defined variational principle, this choice regularizes the energy functional, allowing us to demonstrate the classical analog of a BPS bound for framed BPS states.  This will be reviewed in section \ref{sec:clBPSbnd} when we consider the classical analysis of the full $\NN =2$ theory.

Demonstrating that the variational principle for $S_{\rm van} + S_{\rm def}$ is consistent with the boundary conditions \eqref{defectbcs} is subtle due to the allowed subleading singular behavior.  It requires showing that, on any solution to the second order equations of motion satisfying \eqref{defectbcs}, the following conditions are implied:
\begin{equation}\label{enhancedbcs}
B_i - D_i X = O(r_{n}^{-1/2})~, \quad E_i - D_i Y = O(r_{n}^{-1/2})~, \quad E_i + \tilde{\theta}_0 B_i = O(r_{n}^{-1/2})~,
\end{equation}
as $r_n \to 0$.  This is done in \ref{app:deltahalf}, where it is additionally argued that \eqref{enhancedbcs} hold for a complete basis of fluctuations around any background solution satisfying \eqref{defectbcs}.  In other words, \eqref{enhancedbcs} hold off shell.  When defects are inserted into the theory, the vanilla action is no longer preserved under supersymmetry: there are boundary terms in the supersymmetry variation that are divergent on the infinitesimal two-spheres surrounding the defects.  The addition of the defect action restores half of the supersymmetries---namely the $\RR$ supersymmetries \eqref{Rsusys}.  This is demonstrated in appendix \ref{app:defsusy}, where we find it is essential to make use of \eqref{enhancedbcs}.

When the classical theta angle is non-vanishing, \tHooft defects are sources for the electric field in addition to the magnetic field.  This leads to a manifestation of the Witten effect \cite{Witten:1979ey} for line defects \cite{Kapustin:2005py,Henningson:2006hp}, and the defect action \eqref{Sclbndry} plays a role in making this phenomenon explicit.  The total action contains the terms
\begin{align}\label{SWilson}
S \supset &~ - \int \sum_n \ed t  \int_{S_{\varepsilon_n}^2} \ed^2 \Omega_n \varepsilon_{n}^2 \hat{r}^i \Tr \left\{ \frac{\theta_0}{4\pi^2} A_0 B_i + \frac{2}{g_{0}^2} \Re(\zeta^{-1} \varphi) E_i \right\} \cr
\to &~ - \frac{\theta_0}{2\pi} \sum_n \int \ed t \Tr \left\{ P_n \left( A_0(t,\vec{x}_n) - \Re (\zeta^{-1} \varphi(t,\vec{x}_n)) \right) \right\} ~.
\end{align}
The first term originates from the theta-angle term in the vanilla action while the second originates from \eqref{Sclbndry}.  In the second line we evaluated the magnetic and electric fields on their leading behavior, \eqref{defectbcs}, and integrated over the infinitesimal two-spheres.  Notice that the trace against $P_n$ picks out the Cartan components of the fields $A_0,\varphi$.  Also, while neither $A_0$ nor $\Re(\zeta^{-1} \varphi) = Y$ need be independently well-defined at $\vec{x} = \vec{x}_n$, it can be shown that their difference is; see appendix \ref{app:deltahalf}, and especially \eqref{A0Ysubleading}.

Naively, the way in which $\theta_0$ appears in the boundary conditions \eqref{defectbcs} suggests that it should be thought of as real-valued and not periodic.  However, consider the form of \eqref{SWilson} when $\theta_0 = 2\pi$.  These terms have the same form as a product of (supersymmetric) Wilson line defects in the path integral,
\begin{equation}
\prod_n W_{Q_{n}^{\rm ab}}(\vec{x}_n) = \prod_n \exp \left\{ i \bigg\langle Q_{n}^{\rm ab}, \int \ed t (A_0 - \Re(\zeta^{-1} \varphi)) \bigg\rangle \right\}~,
\end{equation}
with charges $Q_{n}^{\rm ab} = - P_{n}^\ast$.  We remind the reader that $\langle~,~\rangle : \mathfrak{t}^\ast \otimes \mathfrak{t} \to \mathbb{R}$ denotes the canonical pairing between a vector space and its dual, while $P_{n}^\ast \in \mathfrak{t}^\ast$ is the dual element determined by the Killing form \eqref{Trdef} such that $\langle P_{n}^\ast, H \rangle = (P_n, H)$ for all $H \in \mathfrak{t}$.  

Indeed, the charges $Q_{n}^{\rm ab}$ are consistent with \eqref{lineoplattice}, for the electric charges of Wilson--'t Hooft defects.  As explained in \cite{Kapustin:2005py}, an element of \eqref{lineoplattice} is specified by giving a Weyl orbit of the magnetic charge, $[P]$, together with an irreducible representation $\rho_Q$ of $G_P$, the stabilizer of $P$ with respect to the adjoint action.  The Lie algebra $\mathfrak{g}_P$ consists of the Cartan subalgebra $\mathfrak{t} \subset \mathfrak{g}$ together with the root spaces $\mathfrak{g}_{\alpha}$ for those roots $\alpha$ such that $\langle \alpha, P \rangle = 0$.  Hence it is in general the direct sum of an Abelian Lie algebra and a semisimple one.  Correspondingly the highest weight $Q$ decomposes into Abelian and semisimple pieces, $Q = (Q^{\rm ab},Q^{\rm ss})$, and the Wilson line defect is a product of insertions for each factor.  The charges $Q_{n}^{\rm ab} = -P_{n}^\ast$ are examples of the purely Abelian type, so  this special class of Wilson--'t Hooft defects is included in our analysis.  The generic case could be considered by combining the techniques of this paper and the recent paper \cite{Tong:2014yla}.  We will comment briefly on this in section \ref{Section:Further}.

If the charges $(P,Q) =(P, -P^\ast)$ lie in the set $\LL$ of line defect charges we started with, then $\theta_0$ can be taken $2\pi$-periodic provided we allow $\LL$ to undergo monodromy when $\theta_0 \to \theta_0 + 2\pi$.  However it is also possible that $(P,-P^\ast)$ will not be in the set $\LL$ we started with, but in a different set $\LL'$.  The specification of the set of UV line defects should be viewed as part of the defining data of the theory \cite{Aharony:2013hda}.  Denoting $\NN =2$ theories of the type considered here by $G_{\LL}^{\theta_0}$, we would then have $G_{\LL}^{\theta_0+2\pi} \cong G_{\LL'}^{\theta_0}$.  Further shifts of $\theta_0$ by $2\pi$ will eventually bring us back to the set of line defect charges we started with.  For such theories, the periodicity of $\theta_0$ must be assumed larger accordingly.  We will recall an explicit example of this in section \ref{sec:fbps}.

%%%%%%%%%%%%%%%%%%%%%
\subsection{Framed BPS states}
%%%%%%%%%%%%%%%%%%%%% 

The presence of a line defect modifies the spectrum of the theory.  An important subspace of the full Hilbert space, $\HH_{L_\zeta}^{\rm BPS} \subset \HH_{L_\zeta}$\label{hilbdef}, consists of states that are annihilated by a subset of the supersymmetry generators and thus fill out short representations of the supersymmetry algebra.   BPS states of $\NN = 2$ theories in the presence of defects have been dubbed \emph{framed BPS states} in \cite{Gaiotto:2010be}.\footnote{In fact, as we will see shortly, framed BPS states preserve all of the supersymmetries that the line defects preserve.}  The mass of these states saturates a BPS bound, and it is interesting to compare this bound with one obtained for ordinary, ``vanilla'' BPS states in $\NN = 2$ theories.

The $\NN = 2$ supersymmetry algebra takes the form
\begin{align}
& \{ Q_{\alpha}^A~, Q_{\bdot B} \} = 2 {\delta^A}_B (\s^\mu)_{\a\bdot} P_\mu ~, \cr
& \{ Q_{\alpha}^A, Q_{\beta}^B \} = 2 \epsilon_{\alpha\beta} \epsilon^{AB} \bar{Z}~, \qquad \{ \Qbar^{\adot}_{A}, \Qbar^{\bdot}_{ B} \} = 2 \epsilon^{\adot\bdot} \epsilon_{AB} Z~.
\end{align}
See appendix \ref{N2conventions} for more details on our conventions.  Introduce the phase $\zeta$ and the $\RR$-supersymmetries \eqref{Rsusys}, together with
\begin{align}\ \label{Tsusys}
\TT_{\alpha}^A =&~ -i\zeta^{\frac{1}{2}} Q_{\alpha}^A +i \zeta^{-\frac{1}{2}} (\sigma^0)_{\alpha\dot{\beta}} \Qbar^{\dot{\beta} A}~,
\end{align}
such that
\begin{equation}\label{QRT}
Q_{\alpha}^A = \half \zeta^{-\half} ( \RR_{\alpha}^A + i \TT_{\alpha}^A )~, \qquad \Qbar^{\dot{\alpha}A} = \half \zeta^{\half} (\overbar{\sigma}^0)^{\dot{\alpha}\beta} (\RR_{\beta}^A - i \TT_{\beta}^A )~.
\end{equation}
From these relations one finds that $\RR,\TT$ satisfy a reality condition,
\begin{equation}\label{RTconj}
\bar{\RR}^{\dot{\alpha}A} = (\overbar{\sigma}^0)^{\dot{\alpha}\beta} \RR_{\beta}^A~, \qquad \bar{\TT}^{\dot{\alpha} A} =  (\overbar{\sigma}^0)^{\dot{\alpha}\beta} \TT_{\beta}^A~.
\end{equation}
We will refer to any $SU(2)_R$ doublet of Weyl spinors satisfying such a relation as a \emph{symplectic-Majorana--Weyl spinor}.
 
Meanwhile,
\begin{align}\label{RTalgzeta}
& \{ \RR_{\alpha}^A, \RR_{\beta}^B \} = 4 \epsilon_{\a\b} \epsilon^{AB} \left( M + \Re(\zeta^{-1} Z) \right)~,  \cr
& \{ \TT_{\alpha}^A, \TT_{\beta}^B \} = 4 \epsilon_{\a\b} \epsilon^{AB} \left( M - \Re(\zeta^{-1} Z) \right)~,
\end{align}
where $M = - P_0$.  The $\RR$-$\TT$ anticommutator is also nonzero for generic $\zeta$, but we do not write it for reasons to be seen shortly.

In the case \emph{without} defects, both $\RR$ and $\TT$ generate symmetries of the theory.  It follows from \eqref{RTconj}, \eqref{RTalgzeta} that $M \pm \Re(\zeta^{-1} Z)$ are positive semi-definite operators, and thus we derive the bounds $M \pm \Re(\zeta^{-1} Z) \geq 0$.  Furthermore in the case without defects the phase $\zeta$ is arbitrary and thus should be varied to achieve the strongest bound.  Let us define the vanilla phase, $\zeta_{\rm van}$, according to 
\begin{equation}\label{zetavangen}
\zeta_{\rm van} := - \frac{Z}{|Z|}~.
\end{equation}
Then the strongest bound is
\begin{equation}\label{vanillabound}
M \geq |Z|~, \qquad (\textrm{vanilla BPS bound})~,
\end{equation}
and can be achieved in one of two ways.  We can either take $\zeta = \zeta_{\rm van}$ whence the algebra becomes
\begin{align}\label{RTalg}
& \{ \RR_{\alpha}^A, \RR_{\beta}^B \} = 4 \epsilon_{\a\b} \epsilon^{AB} \left( M - |Z|  \right)~,  \cr
& \{ \TT_{\alpha}^A, \TT_{\beta}^B \} = 4 \epsilon_{\a\b} \epsilon^{AB} \left( M + |Z| \right)~, \cr
& \{\RR_{\alpha}^A, \TT_{\beta}^B \} = 0~,
\end{align}
or we can take $\zeta = -\zeta_{\rm van}$ in which case $\RR$ and $\TT$ switch roles in \eqref{RTalg}.  The ordinary BPS states of $\NN = 2$ theories are states that saturate the bound \eqref{vanillabound}: $M = |Z|$.  In the first case with $\zeta = \zeta_{\rm van}$, it is the $\RR$-supersymmetries that are preserved, while in the second case with $\zeta = -\zeta_{\rm van}$, it is the $\TT$-supersymmetries that are preserved.  This distinction corresponds to whether we are considering BPS particle states or anti-particle states.

In the case \emph{with} defects, the presence of the defects already breaks the $\TT$ supersymmetries, and we need only consider the algebra of the $\RR$ supersymmetries.  Furthermore $\zeta$ is fixed by the specification of the defect.  Hence one simply has the bound
\begin{equation}\label{framedbound}
M \geq - \Re(\zeta^{-1} Z)~, \qquad (\textrm{framed BPS bound})~.
\end{equation}
Framed BPS states are states that saturate this bound: $M = -\Re(\zeta^{-1} Z)$.  

Generically this bound allows for states with masses that are lower than masses that would have been allowed by the vanilla bound, \eqref{vanillabound}.  Intuitively, one might expect that framed BPS states correspond to vanilla BPS particles of the original theory that are bound to the defects, with the difference in energy being attributed to the binding energy.  This intuition is confirmed by explicit constructions of framed BPS states, such as in the core-halo picture, originally developed in the supergravity context \cite{Denef:2002ru,Denef:2007vg,Andriyash:2010qv,Andriyash:2010yf} and later adapted to line defects in field theory in \cite{Gaiotto:2010be}.  As similar low energy effective picture had also emerged previously for vanilla BPS states in $\NN = 2$ field theory in \cite{Ritz:2000xa,Argyres:2001pv}.  We will review the core-halo construction in subsection \ref{sec:corehalo} below.  However this picture is an approximation that is good when the bound state radius is large, such that the halo particles are far from the core of the line defect.  When the bound state radius becomes small the usefulness of such a description is not clear.\footnote{Indeed the ``core'' charge that we associate to a given framed BPS state can be different near different boundaries of the same chamber, indicating that the notion of a core can break down in the interior of the chamber.  These ``bound state transformations'' were studied in \eg\ \cite{Andriyash:2010yf}.}  The semiclassical realization of framed BPS states that we will provide in this paper is not limited to large bound state radii; rather it is an approximation scheme in a different parameter---the Yang--Mills coupling.

%%%%%%%%%%%%%%%%%%%%%%
\subsection{Vanilla vacuum structure, BPS spectra, and low energy effective theory}\label{ssec:SWreview}
%%%%%%%%%%%%%%%%%%%%%%

One of the reasons for focusing on the BPS spectrum is that exact results are available thanks to the work initiated by Seiberg and Witten in \cite{Seiberg:1994rs,Seiberg:1994aj}, and reviewed more generally in \eg\ \cite{Lerche:1996xu,AlvarezGaume:1996mv,Freed:1997dp,MR1701615,Tachikawa:2013kta}.  Here we summarize the essential ingredients, first for $\NN = 2$ pure gauge theories without defects.  In section \ref{sec:IRdefects} we discuss how defects are manifested in the low energy effective theory.

The quantum theory has a moduli space of vacua---the Coulomb branch $\BB$---that\label{coulombbdef} can be parameterized by local complex coordinates $\{u^s\}$, $s = 1,\ldots r \equiv \rnk{\mathfrak{g}}$.  These may be understood as parameterizing gauge-inequivalent Higgs vevs; for example when $\mathfrak{g} = \mathfrak{su}(N)$ we may take $u^s = \langle \Tr(\varphi^{1+s}) \rangle$ for $s=1,\ldots,N-1$.  At a generic point in $\BB$ the gauge group $G$ is completely Higgsed to the Cartan torus specified by the regular element $\langle \varphi \rangle \in \mathfrak{g}$.  Classically, there are complex co-dimension one singular loci where some non-Abelian gauge symmetry is restored.  However, quantum mechanically, strong coupling effects prevent this from happening.  Instead these classically singular loci split and separate into multiple quantum loci at which some set of monopoles and/or dyons become massless.  If $\BB^{\rm sing}$ denotes the set of these quantum-singular loci, we define $\BB^* := \BB \setminus \BB^{\rm sing}$\label{coulombast} as the space of vacua with these points removed.

At each $u \in \BB^*$ there is a lattice $\Gamma_{u} \cong \mathbb{Z}^{2r}$ \label{lowecharge} of electric and magnetic charges, equipped with an integral-valued symplectic pairing\footnote{We use the doubled angle bracket notation to avoid conflict with the canonical pairing $\langle~,~\rangle : \mathfrak{t} \times \mathfrak{t}^\ast \to \mathbb{R}$ introduced earlier.}\label{dszpairing} $\llangle~,~\rrangle$.  These lattices are the fibers of a local system over $\BB^\ast$.  The local system has monodromy around the singular loci given by $Sp(2r,\mathbb{Z})$ electromagnetic duality transformations.

Hence for each vacuum $u \in \BB^\ast$ we have a Hilbert space of single particle states, $\HH_{u}^{\textrm{1-part}}$, and this Hilbert space is graded by the conserved electromagnetic charges $\gamma \in \Gamma_u$.  This grading descends to the BPS sector such that
\begin{equation}\label{vanillagrading}
\HH_{u}^{\rm BPS} = \bigoplus_{\gamma \in \Gamma_u} \HH_{u,\gamma}^{\rm BPS}~.
\end{equation}
It is a nontrivial fact that there exists a vector $Z(u) \in \Gamma_{u}^\ast \otimes_{\mathbb{Z}} \mathbb{C}$, in terms of which the central charge operator is a scalar operator in each charge sector, given by $Z_\gamma(u) = Z(u)\cdot \gamma$.  The key point here is that $Z_\gamma(u)$ is linear in $\gamma$.  Together with charge conservation and the BPS bound, this gives BPS states a strong rigidity property.  If a state of charge $\gamma$ is to decay into two constituents (say) of charge $\gamma_1,\gamma_2$, then we must have $\gamma = \gamma_1 + \gamma_2$.  From the BPS bound, the linearity of $Z_\gamma(u)$, and the triangle inequality it then follows that  $M_\gamma(u) \leq M_{\gamma_1}(u) + M_{\gamma_{2}}(u)$.  Thus the decay is kinematically forbidden, unless we have equality.  

This occurs at real co-dimension one walls in $\BB^*$ where the central charges of the constituents, $Z_{\gamma_{1,2}}(u)$, have the same phase.
Hence we define the \emph{vanilla marginal stability walls},  
\begin{equation}\label{vanillawalls}
\widehat{W}(\gamma_1,\gamma_2) := \left\{ u ~ \bigg| ~ \dim{\HH_{u,\gamma_{1,2}}^{\rm BPS}} > 0 ~~ \& ~~  \llangle \gamma_1,\gamma_2\rrangle \neq 0 ~~ \& ~~ Z_{\gamma_1}(u) \Zbar_{\gamma_2}(u) \in \mathbb{R}_+ \right\} \subset \widehat{\BB}~.
\end{equation}
The first condition ensures that the proposed constituents are actually present in the spectrum while the second condition ensures that they can bind.  Recall that the charges can undergo monodromy along paths in $\BB^\ast$.  We have defined the walls as sitting in the universal cover, $\widehat{\BB}$, of $\BB^\ast$, where the fibration of the electromagnetic charge lattice is trivializable.  The image of $\widehat{W}$ under the projection gives co-dimension one walls $W(\gamma_{1},\gamma_{2}) \subset \BB^\ast$.  Thus for every decomposition of a charge $\gamma$ into constituents $\gamma_{1,2}$ with nonzero pairing, and corresponding to BPS particles that are present in the spectrum, there is a marginal stability wall.  Away from all such walls BPS states of charge $\gamma$ are stable, if they exist.  The question of how many BPS states decay across the wall is the subject of wall crossing formulae, which will be briefly reviewed in subsection \ref{sec:wcf} below.

Turning now to the massless degrees of freedom, the leading two derivative form of the low energy effective action is highly constrained by $\NN = 2$ supersymmetry \cite{Grisaru:1979wc,Gates:1983nr,Seiberg:1988ur}.  We work in $\NN = 1$ superspace, where the $\NN = 2$ vector multiplets are composed of $\NN = 1$ vector multiplets with fieldstrengths $W^{\alpha I}$ and $\NN = 1$ chiral multiplets $\Phi^I$, where $I = 1,\ldots,r$.  The two derivative action for the chiral multiplets will be an $\NN = 1$ nonlinear sigma model.  However $\NN =2$ supersymmetry forbids the appearance of a superpotential and furthermore constrains the target geometry to be (rigid) special K\"ahler.  This means that the target manifold can be covered by charts of distinguished coordinate systems, consisting of \emph{special coordinates} $a^I(x)$\label{spcodef} which can be taken as the scalar components of $\Phi^I$, such that the K\"ahler potential takes the form $\mathcal{K} = \Im\left(\frac{\pd \FF}{\pd a^I} \bar{a}^I\right)$, for some locally defined holomorphic function $\FF(a)$ \label{prepotdef} known as the prepotential.  $\NN = 2$ supersymmetry furthermore determines the kinetic terms for the vector multiplets in terms of the second derivatives of $\FF$, such that the Lagrangian density takes the form\footnote{\label{fn:rescale}This agrees with the formula in \cite{Seiberg:1994rs} except for a rescaling of the chiral superfield $\Phi_{\rm here}^I = \sqrt{2} \Phi_{\rm SW}^I$.}
\begin{align}\label{lel}
\LL_{\rm SW} =&~ \frac{1}{4\pi} \Im \left[ \int d^4 \theta \frac{ \pd \FF(\Phi)}{\pd \Phi^I} \bar{\Phi}^I + \int d^2 \theta \frac{ \pd^2 \FF(\Phi) }{\pd \Phi^I \Phi^J}  W^{\alpha I} W_{\alpha}^{J} \right] \cr
=&~ \frac{1}{4\pi} \left\{ - \Im(\tau_{IJ}) \left( \pd_\mu a^I \pd^\mu \bar{a}^J + \half F_{\mu\nu}^I F^{\mu\nu J} \right) + \frac{ \Re(\tau_{IJ})}{4} \epsilon^{\mu\nu\rho\lambda} F_{\mu\nu}^I F_{\rho\lambda}^J + \textrm{fermi} \right\}~, \cr
\end{align}
where $\tau_{IJ}(a) := \frac{\pd^2 \FF(a)}{\pd a^I \pd a^J}$ is required to be positive definite.  The target space of the sigma model is a copy of the Coulomb branch $\BB$.  More precisely, we can identify the Coulomb branch as the manifold  parameterized by the asymptotic values of the scalars, $a^I =\lim_{|\vec{x}| \to \infty} a^I(x)$.  Thus in each patch there will be relations $a^I = a^I(u)$ between these special coordinates and the gauge-invariant coordinates $u^s$.

Associated with the coordinate patch $\{ a^I \}$ is a fixed Lagrangian splitting of the charge lattice $\Gamma_u$ into magnetic and electric components, $\Gamma_{u} = \Gamma_{u}^{\rm m} \oplus \Gamma_{u}^{\rm e}$ \label{emlatticesplit}for all $u$ in the patch.  This splitting is referred to as a choice of duality frame.  The splitting, the vector $Z(u)$, and the special coordinates are related as follows.  First note that the $\mathbb{R}$-linear extension of $\llangle ~,~\rrangle$ to $\Gamma_u \otimes \mathbb{R}$ defines a symplectic vector space\label{sympvecspace} which we denote $V_u := \Gamma_u \otimes \mathbb{R}$.  Choose a basis $\{\mathfrak{a}_1,\ldots,\mathfrak{a}_r\}$ for $V^{\rm m} = \Gamma_{u}^{\rm m} \otimes \mathbb{R}$ and a basis $\{\mathfrak{b}^1,\ldots, \mathfrak{b}^r\}$ for $V^{\rm e} = \Gamma_{u}^{\rm e} \otimes \mathbb{R}$ that are dual with respect to the symplectic pairing,
\begin{equation}\label{Darboux}
\llangle \mathfrak{a}_I, \mathfrak{a}_J \rrangle = \llangle \mathfrak{b}^I, \mathfrak{b}^J \rrangle = 0~, \qquad \llangle \mathfrak{a}_I, \mathfrak{b}^J \rrangle = {\delta_I}^J~,
\end{equation}
and introduce the period vector 
\begin{equation}\label{pvector}
\varpi = a^I \mfa_I + a_{\mathrm{D},I} \mfb^I~, \qquad \textrm{where} \quad a_{\mathrm{D},I} := \frac{\pd \FF}{\pd a^I}~.
\end{equation}
Then the action of the central charge is given by $Z_\gamma(u) = \llangle \gamma, \varpi \rrangle$.  If we introduce the magnetic and electric components of the charge via $\gamma = \gamma_{\rm m}^I \mfa_I - \gamma_{\textrm{e},I} \mfb^I$\label{chargecomps}, this reduces to the standard formula
\begin{equation}\label{ZSW}
Z_{\gamma}(u) = \gamma_{\rm m}^I a_{\mathrm{D},I}(u) + \gamma_{{\rm e}, I} a^I(u)~.
\end{equation}
The mass of vanilla BPS states of charge $\gamma$ is then $M_{\gamma}(u) = |Z_{\gamma}(u)|$.

As we mentioned above the charge lattice is nontrivially fibered over $\BB^\ast$ and on the overlap of patches the two duality frames will be related by an $Sp(2r,\mathbb{Z})$ transformation.  It follows that the two sets of special coordinates $\{a_{\mathrm{D},I},a^J \}$, $\{ a_{\mathrm{D},I}', {a'}^J \}$ will be related by the same transformation and these are precisely the transformations that preserve the special K\"ahler structure.  There is an accompanying duality transformation of the $\NN = 1$ vector multiplet degrees of freedom which is a generalization of the classical electromagnetic duality rotation.

The presence of a massive state of charge $\gamma$---BPS or otherwise---can be inferred in the low energy theory by measuring the flux of the Abelian magnetic and electric fields through the two-sphere at spatial infinity; more precisely we identify,
\begin{equation}\label{IRcharges}
\gamma_{\rm m}^I = \int_{S_{\infty}^2} \frac{F^I}{2\pi} ~, \qquad \gamma_{{\rm e},I} = \int_{S_{\infty}^2} \frac{G_I}{2\pi}~,
\end{equation}
where
\begin{equation}\label{dualfieldstrength}
G_{I} = - (\Im(\tau_{IJ}) \star F^J + \Re(\tau_{IJ}) F^J)~.
\end{equation}
Note that the $\gamma_{{\rm e},I}$ are the components of the Noether charge associated with the $U(1)^r$ global symmetry: An asymptotically nontrivial gauge transformation $A^I \to A^I - \ed \lambda^I$\label{AIdef} with $\lim_{|\vec{x}| \to \infty} \lambda^I \equiv \lambda_{\infty}^I$, generates a global symmetry of \eqref{lel} with a corresponding Noether charge $N = n_{{\rm e},I} \lambda_{\infty}^I$.  In the quantum theory the action of the global gauge transformation on the Hilbert space is represented by the operator $\exp(i \hat{N})$ and $n_{{\rm e},I} \lambda_{\infty}^I$ are the eigenvalues of $\hat{N}$.  Single-valuedness of the transition function for the $U(1)^r$-bundle on the asymptotic two-sphere implies that $\lambda_{\infty}^I = 2\pi n_{\rm m}^I$ should generate a trivial gauge transformation, which in turn should be represented by the identity operator on the Hilbert space.  Thus one infers the Dirac quantization condition $\gamma_{\rm m}^I \gamma_{{\rm e},I} \in \mathbb{Z}$.  

The presence of a massive BPS state must produce a configuration of the IR fields that preserves half of the supersymmetry.  It will be useful to obtain the susy fixed point equations in the IR theory via the Bogomolny trick.  Setting the fermions to zero, the Hamiltonian associated with \eqref{lel} takes the form
\begin{equation}\label{HSW}
H_{\rm SW} = \frac{1}{4\pi} \int_{\mathbb{R}^3} \ed^3 x \Im(\tau_{IJ}) \left\{ \pd_0 a^I \pd_0 a^J + \pd_i a^I \pd^i \bar{a}^J + E_{i}^I E^{i J} + B_{i}^I B^{i J} \right\}~,
\end{equation}
with $E_{i}^I = F_{i0}^I$ and $B_{i}^I = \half \epsilon_{ijk} F^{jk I}$.  This can be written as
\begin{align}\label{IRbogomolny}
H_{\rm SW} =&~ \frac{1}{4\pi} \int_{\mathbb{R}^3} \ed^3 x \bigg\{ || \pd_0 a||^2 + || E_i + i B_i - \pd_i (\zeta^{-1} a^I) ||^2 \bigg\} + \cr
&~ - \frac{1}{2\pi} \Re \left\{ \zeta^{-1} \int_{S_{\infty}^2} (F^I a_{\mathrm{D},I} + G_{I} a^I) \right\}~, 
\end{align}
where $\zeta$ is an arbitrary phase and we have introduced the notation $|| x ||^2 \equiv \Im(\tau_{IJ}) x^I x^J$.  To obtain \eqref{IRbogomolny} from \eqref{HSW} one must make use of Gauss' Law, $\pd^i G_{i I} = 0$; the Bianchi identity, $\pd^i B_{i}^I = 0$; and $\pd^i a_{\mathrm{D},I} = \tau_{IJ} \pd^i a^J$.  Since $\Im(\tau_{IJ})$ is a positive-definite symmetric matrix, the bulk term in \eqref{IRbogomolny} gives a non-negative contribution to the energy functional.  Using \eqref{IRcharges} and \eqref{ZSW}, the boundary term can be expressed in terms of the central charge, so that we arrive at the familiar bound
\begin{equation}
H_{\rm SW} \geq - \Re\left\{ \zeta^{-1} Z_\gamma(u) \right\}~,
\end{equation}
which is saturated on solutions to the first order equations
\begin{equation}\label{IRsusy}
\pd_0 a^I = 0~, \qquad E_{i}^I + i B_{i}^I - \pd_i (\zeta^{-1} a^I) = 0~.
\end{equation}
In the vanilla theory one should choose $\zeta$ such that $- Z_{\gamma}(u) = \zeta |Z_{\gamma}(u)|$, so as to maximize the bound: $H_{\rm SW} \geq |Z_{\gamma}(u)|$.

Solutions to \eqref{IRsusy} with asymptotic flux \eqref{IRcharges}, and which additionally satisfy the second order equations of motion following from \eqref{lel},  provide an IR description of a vanilla BPS state of charge $\gamma$ and mass $|Z_{\gamma}(u)|$, which is valid at large distances where the fields are slowly varying.  In regions where the fields are strongly varying, however, the higher derivative terms that have been neglected in \eqref{lel} will become important.  This is expected since the BPS particle is an excitation of the massive fields that have been integrated out.  Indeed, a nontrivial solution to \eqref{IRsusy} will inevitably become singular; higher derivative corrections should smooth this out so that the only contribution to the energy remains that coming from the asymptotic two-sphere.      

In summary, the quantum-exact two derivative effective action for the massless modes, and the potential spectrum of massive BPS states, is determined from the specification of $\{a_{\mathrm{D}}(u), a(u)\}$ or, equivalently, the specification of the prepotential.  Seiberg and Witten \cite{Seiberg:1994rs,Seiberg:1994aj} expressed this data in terms of a family of auxiliary Riemann surfaces $\Sigma(u;\Lambda)$ equipped with a meromorphic one-form $\lambda_{\rm SW}$.  The family is parameterized by the vacuum parameters $u$, and $\Lambda$ is the dynamical scale of the theory.  (The Riemann surface also depends on the bare masses of $\NN = 2$ matter hypermultiplets when present.)  The choice of duality frame corresponds to a choice of basis of homology cycles $\{\mathfrak{a}_I, \mathfrak{b}^J\}$, and $\{ a_{\mathrm{D},I}(u),a^J(u)\}$ are the integrals of $\lambda_{\rm SW}$ over the corresponding cycles.  

The families $\{\Sigma,\lambda_{\rm SW}\}$ were deduced in \cite{Seiberg:1994rs,Seiberg:1994aj} for all (asymptotically free or conformal) $\NN = 2$ theories with gauge group $SU(2)$, and many results for higher rank gauge groups soon followed.  See \cite{Bhardwaj:2013qia} for an up-to-date compendium with references.  We will not have need of these exact results here.  The form of the weak coupling expansion of the prepotential will be sufficient for our purposes and is reviewed below in section \ref{sec:wce}.

We emphasize that this description determines the possible BPS spectrum: the masses of BPS states are obtained from \eqref{ZSW} when they exist.  Other tools, such as semiclassical methods, quiver techniques, spectral networks, and/or embeddings into string theory, are required in order to determine those $u \in \BB^\ast$ for which the lattice sites $\{n_{\rm m},n_{\rm e}\}$ are populated, and more specifically how many such states there are and what representations of $\mathfrak{so}(3) \oplus \mathfrak{su}(2)_R$ they fall in.

%%%%%%%%%%%%%%%%%%%
\subsection{IR line defects}\label{sec:IRdefects}
%%%%%%%%%%%%%%%%%%%

The IR theory \eqref{lel} is an $\NN = 2$ supersymmetric gauge theory, and thus it admits Wilson--'t Hooft defects that preserve half of the supersymmetry.  IR Wilson--'t Hooft defects of type $\zeta$ are labeled by an IR charge $\gamma_{\rm def} = p^I \mfa_I - q_I \mfb^I$.\label{defcharge}  They are defined through singularity conditions on the IR fields together with the addition of boundary terms to the action that are localized on the defect.  The singularity conditions on the fields should be consistent with the supersymmetry fixed-point equations \eqref{IRsusy} as well as the equations of motion, and should correspond to the insertion of an infinitely heavy dyon of charge $\gamma_{\rm def}$.  The defect charge should satisfy $\llangle \gamma_{\rm def}, \gamma \rrangle \in \mathbb{Z}$ for all $\gamma \in \Gamma_u$, but need not itself be in $\Gamma_u$.  If we have multiple defects with charge $\gamma_{\rm def}^{(n)}$ then these charges should all be mutually local with one another.

In order to determine the form of the fields in the vicinity of the defect, it is useful to formulate the low energy dynamics in a duality-invariant fashion.  First introduce $\mathbb{F} \in \Omega^2(\mathbb{R}^{1,3}) \otimes V_u$.  In a given duality frame corresponding to a basis $\{ \mathfrak{a}_I, \mathfrak{b}^J\}$ of $V_u$, we have $\mathbb{F} = F^I \mathfrak{a}_I - G_I \mathfrak{b}^I$, where $F^I$ are the Abelian fieldstrengths appearing in the component expansion of \eqref{lel}.  There is a complex structure $\II$ on $V_u$ that is compatible with the symplectic structure, $\llangle \II(v), \II(v') \rrangle = \llangle v, v' \rrangle$, $\forall v,v' \in V_u$, and positive such that $(v,v') := \llangle v, \II(v') \rrangle$ is a positive-definite innerproduct.  The components of $\II$ with respect to the Darboux basis are $\tau_{IJ}$.\footnote{More precisely, we introduce the complexified vector space $V_u \otimes_{\mathbb{R}} \mathbb{C}$ which can be decomposed into invariant subspaces $V_{u}^{(0,1)} \oplus V_{u}^{(1,0)}$ where $\II$ acts as $\mp i$.  Then we choose an adapted basis $f_I$ for $V^{(0,1)}$ and $\bar{f}_I$ for $V^{(1,0)}$ such that $f_I = \mfa_I + \tau_{IJ} \mfb^J$.  From this one can infer the action of the complex structure on the Darboux basis:  $\II(\mfa_I) = ({\rm x} {\rm y}^{-1})_{I}^{\phantom{I}J} \mfa_J + ( {\rm y} + {\rm x} {\rm y}^{-1} {\rm x})_{IJ} \mfb^J$, and $\II(\mfb^I) = - {\rm y}^{IJ} \mfa_J - ({\rm y}^{-1} {\rm x})^{I}_{\phantom{I}J} \mfb^J$, where ${\rm x}_{IJ} \equiv \Re(\tau_{IJ})$, ${\rm y}_{IJ} \equiv \Im(\tau_{IJ})$, and ${\rm y}^{IJ} = ({\rm y}^{-1})^{IJ}$.}

The $G_I$ components of $\mathbb{F}$ are not arbitrary: we impose the self-duality constraint, $(\star \otimes \II) \mathbb{F} = \mathbb{F}$, which is solved by \eqref{dualfieldstrength}.  Then $\ed \mathbb{F} = 0$ is equivalent to the Bianchi identity together with the equations of motion for the gauge fields following from \eqref{lel}.  Then it is easy to see that the ansatz
\begin{equation}\label{sdF}
\mathbb{F} \to \half \left( \sin{\uptheta} \ed\uptheta\ed\upphi \otimes \gamma_{\rm def} + \frac{\ed t \ed r}{r^2} \otimes \II(\gamma_{\rm def}) \right)~, \qquad r \to 0~,
\end{equation}
solves the self-duality constraint and describes the singular behavior of the fields produced by a dyon of charge $\gamma_{\rm def}$ at $r =0$.  This ansatz will also be consistent with the equations of motion, $\ed \mathbb{F} = 0$, provided the scalars $a^I$ are functions of $r$ only, to leading order as $r \to 0$.  This is required because the complex structure $\II$ is determined by $\tau_{IJ}(a)$ and thus varies over spacetime if $a$ does.  The behavior of the magnetic and electric fields in the vicinity of the defect can be extracted from \eqref{sdF} and is
\begin{equation}\label{IRdefectbcs}
B_{i}^I \to p^I \frac{\hat{r}_i}{2 r^2}~, \qquad E_{i}^I \to - ((\Im\tau)^{-1})^{IJ} ( q_J +  \Re(\tau_{JK}) p^K ) \frac{\hat{r}_i}{2 r^2}~.
\end{equation}
The BPS equations of the IR theory \eqref{IRsusy} determine the leading behavior of the scalars such that the defect preserves the $\RR$-supersymmetries: $\pd_i (\zeta^{-1} a^I) = E_{i}^I + i B_{i}^I$.

Just as in the UV theory, we are obliged to add boundary terms to the action that are localized on the defects:
\begin{equation}
S^{\rm IR} = \int \ed^4 x  \LL_{\rm SW} + S_{\rm bndry}^{\rm IR}~.
\end{equation}
These terms are responsible for rendering the boundary conditions \eqref{IRdefectbcs} consistent with the variational principle.  The variation of the boundary action with respect to the generator of $\RR$-supersymmetry should also cancel the boundary terms incurred from the corresponding susy variation of the bulk Seiberg--Witten action.  Finally, we expect the boundary action to induce boundary terms in the Hamiltonian that provide a regularized energy functional.  Suppose there are defects at positions $\vec{x}_n$ with charges $\gamma_{\rm def}^{(n)} = p^{(n)I} \mfa_I - q_{I}^{(n)} \mfb^I$.  In appendix \ref{app:IRbndry} we study the variational principle in the presence of line defects and argue that the appropriate boundary action is
\begin{equation}\label{SIRbndry}
S_{\rm def}^{\rm IR} = \frac{1}{2\pi} \sum_n \int \ed t  \int_{S_{\varepsilon_{n}}^2} \left\{ \Re \left[ \zeta^{-1} (F^I a_{\mathrm{D},I} + G_I a^I) \right] - \half q_{I}^{(n)} A_{0}^I \ed\Omega_n \right\}~,
\end{equation}
where $G_I$ is given by \eqref{dualfieldstrength}, $S_{\varepsilon_n}^2$ are infinitesimal two-spheres surrounding the defects at $\vec{x} = \vec{x}_n$, and $\ed\Omega_n = \sin{\uptheta}_n \ed\uptheta_n \ed \upphi_n$.

If we evaluate the flux $G_I$ on its leading behavior, then the last two terms of \eqref{SIRbndry}, (the ones involving $G_I$ and $A_{0}^I$), can be written as $q_{I}^{(n)} \int_{\mathcal{C}_n} \ed t ( \Re(\zeta^{-1} a^I) - A_{0}^I)$, where $\mathcal{C}_n = \mathbb{R}_t \times \{ \vec{x}_n \}$, and therefore can be understood as the insertion of the $\half$-BPS Wilson line defect
\begin{equation}
W(\vec{x}_n, q_{I}^{(n)}) = \exp \left\{ i q_{I}^{(n)} \int_{ \mathcal{C}_n} \ed t \left( \Re(\zeta^{-1} a^I) - A_{0}^I \right) \right\}~,
\end{equation}
in the path integral.  The first term of \eqref{SIRbndry}, (the one involving $F^I$), is analogous to the boundary terms we added in the UV for pure \tHooft defects.  Note that it is incorrect to evaluate $F^I$ on its leading behavior and write this term as an integral of $p^{(n)I} a_{\mathrm{D},I}$ along $\mathcal{C}_n$.   The reason is that the variation of $F^I$ on the infinitesimal two-sphere involves tangential derivatives of $\delta A_{i}^I$ and plays a role in canceling the variation of the bulk action.  In contrast, the variation of $G_I$ involves normal derivatives of $\delta A_{i}^I$ and must be treated independently.  We show in appendix \ref{app:IRbndry} that it is consistent with the defect boundary conditions to impose a Dirichlet-type condition on $G_I$ such that $a^I \delta G_I \to 0$ as $\varepsilon_n \to 0$, and hence it is permissible to evaluate $G_I$ in \eqref{SIRbndry} on its leading behavior.  The asymmetric treatment of these terms is related to the fact that in order to construct a Lagrangian one chooses the electric potentials (say) as the fundamental variables and treats the magnetic potentials as derived quantities.

Let us evaluate the energy of these line defects.  For this purpose we place a single defect of charge $\gamma_{\rm def}$ at the origin, with no other excitations in the system, so that the fields are given everywhere by the supersymmetric dyon solution,
\begin{equation}\label{IRdyon}
\mathbb{F} = \half \left( \sin{\uptheta} \ed\uptheta \ed\upphi \otimes \gamma_{\rm def} + \frac{\ed t \ed r}{r^2} \otimes \II(\gamma_{\rm def}) \right)~.
\end{equation}
The scalar fields are determined from $\pd_i (\zeta^{-1} a^I) = E_{i}^I + i B_{i}^I$ and take on asymptotic values $a^I(u)$ as $r \to \infty$.  

Now consider the energy functional.  In carrying out the Legendre transformation from the Seiberg--Witten Lagrangian to the Hamiltonian\footnote{The Hamiltonian should of course be expressed in terms of $q$'s and $p$'s, not $q$'s and $\dot{q}$'s.  Let $\pi_I,\bar{\pi}_I$ denote the momentum conjugate to $a^I,\bar{a}^I$, and $\pi_{iI}$ the momentum conjugate to $A^{iI}$.  Then $\pd_0 a^I$, $\pd_0 \bar{a}^I$ and $E^{iI}$ should be understood as functionals of the momenta and coordinates, given by $\pd_0 a^I = 4\pi {\rm y}^{IJ} \bar{\pi}_J$, $\pd_0 \bar{a}^I = 4\pi {\rm y}^{IJ} \pi_J$, and $E_{i}^I = -{\rm y}^{IJ} (2\pi \pi_{iJ} + {\rm x}_{JK} B_{i}^K)$, where $\tau_{IJ} = {\rm x}_{IJ} + i {\rm y}_{IJ}$.  We note that $\pi_{iI} = \frac{1}{2\pi} G_{iI}$.}, we pick up additional boundary terms relative to \eqref{HSW} when defects are present:
\begin{align}\label{HSWd1}
H_{\rm SW} =&~ \frac{1}{4\pi} \int_{\UU} \ed^3 x \Im(\tau_{IJ}) \left\{ \pd_0 a^I \pd_0 \bar{a}^J + \pd_i a^I \pd^i \bar{a}^J + E_{i}^I E^{i J} + B_{i}^I B^{i J} \right\} + \cr
&~ +  \frac{1}{2\pi} \int_{S_{\varepsilon}^2} \ed\Omega \varepsilon^2 \hat{r}^i A_{0}^I \left( \Im(\tau_{IJ}) E_{i}^J + \Re(\tau_{IJ}) B_{i}^J \right)~,
\end{align}
where $\UU = \mathbb{R}^3 \setminus \{\vec{0} \}$.  In principle there is an analogous boundary term on the asymptotic two-sphere, which should have also been present in \eqref{HSW}.  However one typically sets that term to zero by working in a gauge where $A_{0}^I \to 0$ as $r \to \infty$.  In contrast gauge transformations cannot be used to remove the leading singularity of the fields in the presence of a defect.  Fortunately the defect action \eqref{SIRbndry} induces a boundary Hamiltonian which, in the case of a single defect, is given by
\begin{equation}
H_{\rm def}^{\rm IR} = -\frac{1}{2\pi} \int_{S_{\varepsilon}^2} \left\{ \Re \left[ \zeta^{-1} (F^I a_{\mathrm{D},I} + G_{I} a^I) \right] - \half q_I A_{0}^I \ed\Omega \right\}~.
\end{equation}
The $A_0$ term here is of precisely the right form to cancel the boundary term in \eqref{HSWd1} upon using the line defect boundary conditions.  Thus
\begin{align}\label{HSWd2}
H^{\rm IR} = H_{\rm SW} + H_{\rm def}^{\rm IR} =&~  \frac{1}{4\pi} \int_{\UU} \ed^3 x \Im(\tau_{IJ}) \left\{ \pd_0 a^I \pd_0 a^J + \pd_i a^I \pd^i \bar{a}^J + E_{i}^I E^{i J} + B_{i}^I B^{i J} \right\} \cr
&~ - \frac{1}{2\pi} \Re \left\{ \zeta^{-1} \int_{S_{\varepsilon}^2} ( F^I a_{\mathrm{D},I} + G_{I} a^I) \right\}~. \raisetag{24pt}
\end{align}
Finally, we apply Bogomolny's trick to the bulk terms as in \eqref{IRbogomolny}.  This time we acquire boundary terms from both the asymptotic and infinitesimal two-sphere.  The latter cancel against the boundary terms already present in \eqref{HSWd2} so that
\begin{align}\label{HIR}
H^{\rm IR} =&~ \frac{1}{4\pi} \int_{\UU} \ed^3 x \bigg\{ || \pd_0 a ||^2 + || E_i + i B_i - \pd_i (\zeta^{-1} a) ||^2 \bigg\} + \cr
&~ - \frac{1}{2\pi} \Re \left\{ \zeta^{-1} \int_{S_{\infty}^2} ( F^I a_{\mathrm{D},I} + G_{I} a^I) \right\}~.
\end{align}
As anticipated, the boundary terms have regularized the energy functional so that one may obtain the BPS bound.  The dyon solution saturates the bound and has energy
\begin{equation}\label{EIRdyon}
E_{\gamma_{\rm def}} = - \Re \left[ \zeta^{-1} Z_{\gamma_{\rm def}}(u) \right]~.
\end{equation}

Note that this Bogomolny bound agrees with the one obtained for framed BPS states in \eqref{framedbound}.  Therefore we interpret the dyon field configuration \eqref{IRdyon}, together with the associated scalar fields satisfying $\pd_i (\zeta^{-1} a^I) = E_{i}^I + i B_{i}^I$, as the manifestation of a framed BPS state of charge $\gamma_{\rm def}$ in the low energy effective theory.\footnote{We mentioned in the discussion below \eqref{framedbound} that framed BPS states can be thought of as vanilla BPS particles  bound to defects; it is important to emphasize that there we were talking about line defects of the UV microscopic theory, as defined in subsection \ref{Section:Defects}.}  

This leads to a natural question which is at the heart of \cite{Gaiotto:2010be}: What is the relation between UV and IR line defects?  The asymptotic weak coupling region of the Coulomb branch provides useful intuition.  At weak coupling there is a natural duality frame where we identify the electric-magnetic splitting, $V_{u} = V_{u}^{\rm m} \oplus V_{u}^{\rm e} \cong \mathfrak{t} \oplus \mathfrak{t}^\ast$, with $\mathfrak{t} \subset \mathfrak{g}$ the Cartan subalgebra determined by the Higgs vev and $\mathfrak{t}^\ast$ its dual space.  There is furthermore a natural polarization vector\footnote{The polarization vector will be defined in section \ref{sec:wce} below; the details are not important for the purposes of this dicussion.} in $\mathfrak{t}$ that can be used to define a set of simple roots\label{defsimpleroots} $\alpha_I$, and hence simple co-roots $H_I \in \mathfrak{t}$, $I = 1,\ldots,r$, and their integral-dual fundamental magnetic weights $\lambda^J \in \mathfrak{t}^\ast$.  Thus a convenient Darboux basis is $\{\mfa_I, \mfb^J\} \cong\{H_I, \lambda^J\}$.  Hence the Abelian fieldstrengths $F^I$ are identified with the Cartan components of the non-Abelian fieldstrength $F$ along the simple co-roots, while the scalars $a^I$ are similarly identified with the Cartan components of $\varphi$.

Now consider the insertion of a UV Wilson line in an irreducible representation $\rho$ of $G$ specified by highest weight $Q$.  We can imagine that the Wilson line will act as an electric source for the Cartan components of the gauge field, but we cannot identify a single set of electric charges.  If $\Delta_\rho$ denotes the set of weights of the irrep $\rho$, then for any $\mu \in \Delta_\rho$ we can construct a set of ``Abelian'' charges, $q_{I}^{(\mu)} = \langle \mu, H_I \rangle$.  Furthermore gauge field configurations with electric sources corresponding to different weights $\mu,\mu' \in \Delta_\rho$ can be related by (non-Abelian) gauge transformations.  Likewise, consider the insertion of a pure UV \tHooft defect specified by a Weyl orbit $[P]$ of $P \in \mathfrak{t}$.  The leading behavior of the Cartan components of the gauge field follows from the boundary conditions \eqref{tHooftbc} and correspond to IR defect boundary conditions with charges $p^I$ where $P = p^I H_I$.  However, we could choose some other representative $P' \in [P]$ and this would give a different set of IR defect charges corresponding to the same UV defect.  Again, in the UV theory, exchanging $P$ for $P'$ can be implemented by a local gauge transformation and that is why it is only the Weyl orbit $[P]$ that is physical.  However there are no such gauge transformations available in the IR theory!

These examples make it clear that a single UV defect can correspond to a number of different IR defects.  More precisely, if we compute a correlator in the UV theory in the presence of a UV defect, then in order to reproduce (the low energy limit of) the correlator in the low energy effective theory we will have to sum over multiple correlators in different ``superselection sectors'' corresponding to the insertion of a number of different IR defects.  

In fact, the situation is even more nontrivial than the discussion above indicates.  Suppose there is a framed BPS state in the spectrum corresponding to a vanilla monopole bound to the UV Wilson line.  In the low energy effective theory the massive non-Abelian fields that smooth out the monopole configuration at short distances have been integrated out.  To the low energy observer measuring the asymptotic Cartan-valued fields where they are sufficiently slowly varying such that the two derivative effective action \eqref{lel} is valid, the bound state system will simply look like a defect carrying both electric and magnetic charge.\footnote{For systems where the bound-state radius is much larger than the effective size of the vanilla particle it is possible to partially resolve the defect in the sense of the core-halo picture described in \ref{sec:corehalo}.}  Hence, even though the UV defect is purely electric, we may have to sum over IR defects that carry magnetic charge as well as electric charge!  See the introduction of \cite{Gaiotto:2010be} for a description of the simplest case in which this phenomenon occurs---namely $SU(2)$ gauge theory probed by a fundamental Wilson line.  Similarly, a purely magnetic UV defect may induce electrically charged framed BPS states represented in the IR by dyonic defects.  

Clearly the spectrum of framed BPS states plays an important role in determining what IR defects must be summed over to reproduce correlation functions in the presence of a given UV defect.  Even in the more mundane cases of two IR defects corresponding to different weights $\mu,\mu' \in \Delta_\rho$, or different representatives $P,P' \in [P]$, one might expect that the proliferation of IR defects is related to the existence of framed BPS states.  All weights of a representation $\rho$ can be related to a given weight by adding or subtracting roots.  Similarly all representatives in the Weyl orbit of an \tHooft charge can be related to a given one by adding or subtracting co-roots.  The roots and co-roots are part of the vanilla charge lattice; they correspond to massive $W$-boson and monopole states.  Does the possibility of these states binding to the Wilson or \tHooft defect likewise encode this degeneracy of IR defects?      

The answer is essentially ``yes'', but in order to explain how this works we must introduce some machinery for counting framed BPS states.  The situation is further complicated (or enriched) by the fact that framed BPS states undergo wall crossing phenomena just as vanilla BPS states do.  We turn to these issues next.

%%%%%%%%%%%%%%%%%%%
\subsection{Framed BPS indices, protected spin characters, and (no) exotics}\label{sec:wcf}
%%%%%%%%%%%%%%%%%%%

For a given (UV) line defect $L_\zeta$, let $\HH_{L_\zeta,u}^{\rm BPS}$ denote the space of framed BPS states above vacuum $u$, saturating the bound \eqref{framedbound}.  It is graded by electromagnetic charges, like the space of vanilla BPS states \eqref{vanillagrading}.  However the charges need not sit in the vanilla lattice $\Gamma_u$.  As the weak coupling discussion above indicates, they will instead sit in a shifted copy, or torsor, of the vanilla charge lattice,
\begin{equation}\label{torsor}
\Gamma_{L,u} = \Gamma_u + \gamma_L~,
\end{equation}
for some $\gamma_L \in V_u$, such that $\llangle \gamma_L, \gamma \rrangle \in \mathbb{Z}$ for all $\gamma \in \Gamma_u$.  One may take $\gamma_L$ to be any one of the IR defect charges $\gamma_{\rm def}$ associated with the UV defect $L$.  For example, consider pure Yang--Mills based on Lie algebra $\mathfrak{g} = \mathfrak{su}(2)$.  At weak coupling we have the natural duality frame $V_{u} \cong \mathfrak{t} \oplus \mathfrak{t}^\ast$.  The vanilla electromagnetic charge lattice is generated by the simple monopole and $W$-boson, whose charges are the co-root $H_\alpha$ and root $\alpha$; $\Gamma_u \cong \Lambda_{\rm cr} \oplus \Lambda_{\rm rt}$.  However, if we consider gauge group $G = SO(3)$ then we can insert an \tHooft defect with charge $P = \half H_\alpha$, which is not in the co-root lattice.  Hence in this case we could take $\gamma_L = \pm \half H_\alpha$.  

Given \eqref{torsor}, the Hilbert space of framed BPS states is graded accordingly:
\begin{equation}\label{Hframed}
\HH_{L_\zeta,u}^{\rm BPS} = \bigoplus_{\gamma \in \Gamma_{L,u}} \HH_{L_\zeta,u,\gamma}^{\rm BPS}~.
\end{equation}
We have the central charge vector $Z(u) \in \Gamma_{L,u}^\ast \otimes_{\mathbb{Z}} \mathbb{C}$, in terms of which the framed BPS bound, \eqref{framedbound}, takes the form
\begin{equation}
M_{\gamma}(u) \geq - \Re(\zeta^{-1} Z_{\gamma}(u))~.
\end{equation}
The masses of framed BPS states saturate the bound; we already saw an example of this in \eqref{EIRdyon}.

The $\HH_{L_\zeta,u,\gamma}^{\rm BPS}$ are special, \emph{short} representation spaces for the supersymmetry algebra preserved by the defect.  Recall that this supersymmetry algebra has an $\mathfrak{so}(3) \oplus \mathfrak{su}(2)_R$ bosonic subalgebra, corresponding to spatial rotations and $SU(2)$ $R$-symmetry, and the four odd generators $\RR_{\alpha}^A$.  A generic, \emph{long} representation has the form $\rho_{\rm hh} \otimes \mathfrak{h}$, where $\rho_{\rm hh}$ is the half-hypermultiplet---the four-dimensional representation $(\half;0) \oplus (0;\half)$ of $\mathfrak{so}(3) \oplus \mathfrak{su}(2)_R$---and $\mathfrak{h}$ is an arbitrary $\mathfrak{so}(3) \oplus \mathfrak{su}(2)_R$ representation.  The presence of the half-hypermultiplet is due to the $\RR_{\alpha}^A$ which form a Clifford algebra with two raising and two lowering operators, while $\mathfrak{h}$ is the representation of the Clifford vacuum.  On the short BPS representations the $\RR_{\alpha}^A$ act trivially, and thus there is no explicit half-hypermultiplet factor: $\HH_{L_\zeta,u,\gamma}^{\rm BPS} \cong \mathfrak{h}$.   

Short representations have a rigidity property thanks to the BPS condition for the mass, together with the linearity property of the central charge.  They are in general stable under deformations of $u$, and can only decay along marginal stability walls.  (These were defined for vanilla BPS states in \eqref{vanillawalls}; there is also a framed analog that will be discussed in the next subsection.)  However, this stability property only holds for ``true'' BPS representations, corresponding to those $\mathfrak{h}$ which do not contain a $\rho_{\rm hh}$ subfactor.  ``Fake'' BPS representations which have a half-hypermultiplet subfactor correspond to the situation where several short representations sum together to produce a long one.  A quantity that counts only the short, protected representations should therefore vanish when $\mathfrak{h}$ contains a $\rho_{\rm hh}$ factor.  The standard quantity that has this property is the \emph{framed BPS index}:
\begin{equation}\label{findex}
\fOmega(L_\zeta,u,\gamma) := \Tr_{\HH_{L_\zeta,u,\gamma}^{\rm BPS}} (-1)^{2 J^3}~,
\end{equation}
where $J^3$ is the $z$-component of angular momentum.  

One can also define a more refined version of the framed BPS indices that keeps track of spin content.  Naively one could generalize the above via $(-1)^{2J^3} \to y^{2J^3}$, where $y$ is a formal variable, but the trace of this does not vanish on the half-hypermultiplet and so it is not, in principle, protected.  One can remedy the situation by making use of $\mathfrak{su}(2)_R$: the trace of $(-1)^{2 J^3} (-y)^{2(J^3 + I^3)}$, where $I^3$ is the third $\mathfrak{su}(2)_R$ generator, does vanish on the half-hypermultiplet.  Hence we define the \emph{framed protected spin character} \cite{Gaiotto:2010be},
\begin{equation}\label{fPSC}
\fOmega(L_\zeta,u,\gamma;y) := \Tr_{\HH_{L_\zeta,u,\gamma}^{\rm BPS}} (-1)^{2J^3} (-y)^{2(J^3+I^3)} = \Tr_{\HH_{L_\zeta,u,\gamma}^{\rm BPS}} y^{2 J^3} (-y)^{2 I^3}~.
\end{equation}
If one evaluates this quantity at $y = -1$ it reduces to the framed BPS index, \eqref{findex}. For a given line defect, the protected spin characters corresponding to different charges can be naturally organized into a generating function by introducing formal variables $X_\gamma$ for the charges:
\begin{equation}
F(L,u,\{X_\gamma\};y):=\sum_{\gamma \in \Gamma_{L,u}}\fOmega(L_\zeta,u,\gamma;y)X_\gamma\qquad \mbox{where}\quad X_{\gamma}X_{\gamma^\prime}=y^{\llangle\gamma,\gamma^\prime\rrangle}X_{\gamma+\gamma^\prime}\,. \label{Fgenfun}
\end{equation}
The particular multiplication rule for the generating function variables $X_\gamma$ is motivated by studying the transformations of the generating function under wall crossing, which we will describe below.  See \cite{Gaiotto:2010be} for a detailed discussion of the properties of $F$.

Let us return to the definition \eqref{fPSC} for a moment.  The appearance of the sum $J^3 + I^3$ in the protected spin character \eqref{fPSC} can also be understood from the observation that, while none of the $\RR_{\alpha}^A$ commute with $\mathfrak{so}(3)$ or $\mathfrak{su}(2)_R$ separately, there is a linear combination, $\epsilon_{\alpha A} \RR^{\alpha A}$, that commutes with the diagonal subalgebra $\mathfrak{su}(2)_{\mathrm{d}} \subset \mathfrak{so}(3) \oplus \mathfrak{su}(2)_R$.  Hence the different weights of a representation of $\mathfrak{su}(2)_{\mathrm{d}}$ are protected.

In fact there is nothing particularly special about the choice $\epsilon_{\alpha A}$.  Let $\kappa_{\alpha}^A$ be an arbitrary symplectic Majorana--Weyl spinor and consider the contraction
\begin{equation}\label{RRkappa}
\RR_{\kappa} := \kappa^{\alpha}_{A} \RR_{\alpha}^A~.
\end{equation}
The symplectic-Majorana--Weyl condition ensures that $\RR_{\kappa}$ is Hermitian.  If we additionally require that $\det{\kappa} :=  -\half \kappa_{\alpha}^A \kappa_{A}^\alpha \neq 0$, then $\RR_\kappa$ will be nondegenerate.  Furthermore, $\RR_{\kappa}$ will be a singlet under a different diagonal subalgebra, $\mathfrak{su}(2)_{\mathrm{d}}^{(\kappa)} \subset  \mathfrak{so}(3) \oplus \mathfrak{su}(2)_R$, related to $\mathfrak{su}(2)_\mathrm{d}$ by an outer automorphism corresponding to an $SO(3)$ conjugation in one of the two factors.  More specifically, if $\mathfrak{su}(2)_\mathrm{d}$ corresponds to the diagonal subalgebra generated by $J^r + I^r$, $r = 1,2,3$, then $\mathfrak{su}(2)_{\mathrm{d}}^{(\kappa)}$ corresponds to the diagonal subalgebra generated by
\begin{equation}\label{twisteddiag}
\mathcal{I}^r := J^r + (R_{\kappa}^{-1})^{r}_{\phantom{r}s}I^s~.
\end{equation}
Here the rotation $R_\kappa \in SO(3)$ is fixed by the choice of $\kappa$ through demanding that $[\II^r, \RR_\kappa] = 0$.  Let the generators of $\mathfrak{su}(2)_R$ and $\mathfrak{so}(3)$ in the doublet representation be $-\frac{i}{2} (\sigma^r)^{A}_{\phantom{A}B}$ and $-\frac{i}{2} (\sigma^r)_{\alpha}^{\phantom{\alpha}\beta}$, where the $\sigma^r$ the Pauli matrices.  Then $[\II^r, \RR_\kappa] = 0$ is equivalent to
\begin{equation}\label{Rotkappa}
(\sigma^r)^{A}_{\phantom{A}B} \kappa_{\alpha}^B = - (R_\kappa)^{r}_{\phantom{r}s} (\sigma^s)_{\a}^{\phantom{\a}\b} \kappa_{\beta}^A~,
\end{equation}
which determines $R_{\kappa}$ uniquely.\footnote{Another way to think about the map $\kappa \mapsto R_{\kappa}$ is as follows.  The symplectic Majorana--Weyl condition implies that $k_{\alpha}^{\phantom{\alpha}A} := \kappa_{\alpha}^{\phantom{\alpha}B} (\sigma^2)_{B}^{\phantom{B}A}$ can be regarded as a quaternion.  Let $q$ be an imaginary quaternion, so in the Lie algebra of $\mathfrak{su}(2)$.  Then \eqref{Rotkappa} is equivalent to the identity $qk = k(k^{-1} q k) = k ({k'}^{-1} q k')$.  Here $k' := k/ (k\bar{k})^{1/2}$ is a unit quaternion and hence in $SU(2)$.  The rotation $R_{\kappa}$ is the image of $k'$ under the standard homomorphism of $SU(2) \to SO(3): k' \mapsto R_{\kappa}$, given by setting ${k'}^{-1} q k' =  R_{\kappa} \cdot q$.}  The special case $\kappa_{\alpha A} = \epsilon_{\alpha A}$ corresponds to $R_{\kappa} = \mathbbm{1}$.  (One can explicitly check that the matrix $R_\kappa$ defined by this equation satisfies $(R_{\kappa})^T R_\kappa = \mathbbm{1}$ and $\det{R_\kappa} = 1$ for any nondegenarate symplectic Majorana--Weyl spinor $\kappa$.)  We mention this freedom in the choice of linear combination \eqref{RRkappa} and corresponding diagonal subgroup $\mathfrak{su}(2)_{\mathrm{d}}^{(\kappa)}$ as it will appear again in the semiclassical analysis to follow.

Indices and protected spin characters can also be defined for vanilla BPS states.  In the vanilla case there is a second set of odd generators in the superalgebra, namely the $\TT_{\alpha}^A$, which commute with the $\RR_{\alpha}^A$.  They generate a second half-hypermultiplet factor, and a generic long representation of the $\NN = 2$ superalgebra takes the form $\rho_{\rm hh} \otimes \rho_{\rm hh} \otimes \mathfrak{h}$.  Here again the Clifford vacuum $\mathfrak{h}$ is a generic $\mathfrak{so}(3) \oplus \mathfrak{su}(2)_R$ representation.  On vanilla BPS representations the $\RR_{\alpha}^A$, say, are realized trivially and we only need to represent the Clifford algebra of the $\TT_{\alpha}^A$.  Thus the vanilla BPS Hilbert spaces $\HH_{u,\gamma}^{\rm BPS}$ appearing in the decomposition \eqref{vanillagrading} are of the form $\rho_{\rm hh} \otimes \mathfrak{h}$.  In order to account for this we introduce the notation
\begin{equation}\label{hhfactor}
\HH_{u,\gamma}^{\rm BPS} = \rho_{\rm hh} \otimes (\HH_0)_{u,\gamma}^{\rm BPS}~.
\end{equation}
A protected quantity should vanish when the representation space being traced over has two half-hypermultiplet factors, but should be nonzero when there is a single factor.  A quantity that has this property is
\begin{equation}
\Tr \left\{ (2J^3) (-1)^{2J^3} (-y)^{2 (J^3 + I^3)} \right\} = x_1 \frac{\pd}{\pd x_1} \left( \Tr \left\{ x_{1}^{2 J^3} x_{2}^{2 I^3} \right\} \right) \bigg|_{x_1 = -x_2 = y} ~.
\end{equation}
On a representation of the form \eqref{hhfactor}, the half-hypermultiplet always contributes a factor of $(y - y^{-1})$ to this trace.  Hence we define the (vanilla) \emph{protected spin character} \cite{Gaiotto:2010be}
\begin{equation}\label{PSC}
(y - y^{-1}) \Omega(u,\gamma;y) := \Tr_{\HH_{u,\gamma}^{\rm BPS}} \left\{ (2J^3) (-1)^{2J^3} (-y)^{2 (J^3 + I^3)} \right\} ~.
\end{equation}
If we specialize to the case $y = -1$ we get the (vanilla) \emph{BPS index}, $\Omega(u,\gamma) = \Omega(u,\gamma;-1)$.  This quantity is more directly defined as
\begin{equation}
\Omega(u,\gamma) := -\half \Tr_{\HH_{u,\gamma}^{\rm BPS}} \left\{ (2J^3)^2 (-1)^{2J^3} \right\} = \Tr_{(\HH_0)_{u,\gamma}^{\rm BPS}} (-1)^{2J^3}~,
\end{equation}
and is also known as the second helicity supertrace.

Notice that if the Clifford vacuum $\mathfrak{h} = (\HH_0)_{u,\gamma}^{\rm BPS}$ is an $\mathfrak{su}(2)_R$ singlet, $\mathfrak{h} = (j;0)$, then the protected spin character \eqref{PSC} reduces to the standard spin character of $\mathfrak{h}$, $\Omega(u,\gamma;y) \to \Tr_{\mathfrak{h}}(y^{2 J^3})$, which gives the dimension of $\mathfrak{h}$ when specialized to $y = 1$.  Similarly, if the space of framed BPS states, $\HH_{L_\zeta,u,\gamma}^{\rm BPS}$, is a trivial $\mathfrak{su}(2)_R$ representation, then the framed protected spin character \eqref{fPSC} evaluated at $y=1$ gives the dimension of $\HH_{L_\zeta,u,\gamma}^{\rm BPS}$.  It is an empirical observation that these conditions hold in all known examples of BPS spectra in $\NN = 2$ theories.  (Framed) BPS states for which they do not hold are referred to as \emph{exotic (framed) BPS states}. The observed absence of exotics motivated\footnote{The original motivation for the conjecture came from observations in \cite{Gaiotto:2010be} of how it was remarkably preserved by explicit wall crossing examples and of how it played a role in allowing for the construction of an elegant algebraic structure on the set of line defects.} the following \emph{no-exotics conjecture} formulated in \cite{Gaiotto:2010be}: there are no exotic (framed) BPS states at smooth points on the Coulomb branch.  This conjecture has since been proven for gauge theories without matter and with simply laced gauge group \cite{Chuang:2013wt,DelZotto:2014bga}.  Recently, a more generally applicable argument has been given \cite{CordovaDumitrescu}.  

We will assume that no-exotics holds for all theories considered in this paper, namely for pure $\NN = 2$ gauge theories with any simple compact Lie algebra $\mathfrak{g}$.  Hence for this class of theories, the (framed) protected spin characters count the number of (framed) BPS states, keeping track of spin information.  In section \ref{ssec:exotics} we will describe what no-exotics implies in the semiclassical regime where BPS states are represented by the kernels of certain Dirac operators on hyperk\"ahler manifolds.

%%%%%%%%%%%%%%%%%%%
\subsection{The core-halo picture and framed wall crossing}\label{sec:corehalo}
%%%%%%%%%%%%%%%%%%%

The framed BPS spin characters and indices \eqref{fPSC}, \eqref{findex}, exhibit wall crossing phenomena, much like their vanilla counterparts.  Somewhat paradoxically, perhaps, it is actually easier in the framed case to give a physically explicit and precise description of this phenomenon: one can determine exactly which subspace of $\HH_{L_\zeta,u}^{\rm BPS}$ enters or leaves the spectrum as a wall is crossed.  The key is the core-halo picture of framed BPS states.

This picture is neither a UV description nor a strictly IR description of the physics, but rather something in between, as we now explain.  We work with the low energy effective variables, and we consider the background of an infinitely heavy dyon of charge $\gamma_{\rm c}$, where ``c'' stands for ``core.''  The field configuration is identical to the IR line defect configuration discussed in subsection \ref{sec:IRdefects} above.  The duality-invariant fieldstrength $\mathbb{F} = F^I \mfa_I - G_I \mfb^I$ takes the same form as \eqref{IRdyon} with $\gamma_{\rm def} \to \gamma_{\rm c}$.  

Here it will be useful to analyze the corresponding scalar field configuration further as follows.  The BPS equation $\pd_i (\zeta^{-1} a^I) = E_{i}^I + i B_{i}^I$ can be written in duality-invariant form by promoting $F^I \to \mathbb{F}$ and $a^I \to \varpi$.  Then, recalling that $Z_{\gamma}(u) = \llangle \gamma, \varpi \rrangle$, we have
\begin{equation}\label{leRfixed}
\llangle \gamma, \mathbb{F}_{0k} \rrangle - \frac{i}{2} \epsilon_{kij} \llangle \gamma, \mathbb{F}^{ij} \rrangle + \pd_k \left( \zeta^{-1} Z_\gamma(u(x)) \right) = 0 ~, \qquad \forall \gamma \in \Gamma_{u}~.
\end{equation}
When the explicit dyon solution is used for $\mathbb{F}$, the imaginary part of this equation is easily integrated and yields a set of implicit attractor-like equations for $u(r)$:
\begin{equation}\label{attractor}
\Im \left[ \zeta^{-1} Z_\gamma(u(r)) \right] =  -\frac{\llangle \gamma, \gamma_{\rm c} \rrangle}{2r} + \Im \left[ \zeta^{-1} Z_\gamma(u) \right]~, \qquad \forall \gamma \in \Gamma_u~,
\end{equation}
where on the right-hand side $u = \lim_{r \to \infty} u(r)$ is constant.  Since the complex structure $\II$ depends on $a^I(r)$ through $\tau_{IJ}(a)$, the real part of \eqref{leRfixed} is a complicated set of nonlinear ODE's.  However, noting that $\mathbb{F}_{0i} = -\pd_i \mathbb{A}_0$, we can derive the relation
\begin{equation}\label{Coulombfield}
\llangle \gamma ,\mathbb{A}_0 \rrangle =  \Re \left[ \zeta^{-1} Z_{\gamma}(u(r)) \right] - \Re \left[ \zeta^{-1} Z_{\gamma}(u) \right]~, \qquad \forall \gamma \in \Gamma_u~,
\end{equation}
which will be useful below.  Here we used that $\mathbb{A}_0 \to 0$ as $r \to \infty$.

Now consider a probe particle propagating in this fixed background.  The probe particle represents a massive vanilla BPS state.  It carries electromagnetic charge $\gamma_{\rm h} \in \Gamma_u$ and has an effective mass $|Z_{\gamma_{\rm h}}(u(r))|$ at a distance $r$ from the core.  The energy of such a particle at rest is the sum of its rest mass and the Coulomb potential due to the background field:
\begin{align}\label{Ehalo}
E_{\rm probe} =&~ |Z_{\gamma_{\rm h}}(u(r))| + \llangle \gamma_{\rm h}, \mathbb{A}_0 \rrangle \cr
=&~  |Z_{\gamma_{\rm h}}(u(r))| + \Re \left[ \zeta^{-1} Z_{\gamma_{\rm h}}(u(r)) \right] - \Re \left[ \zeta^{-1} Z_{\gamma_{\rm h}}(u) \right]~,
\end{align}
where we used \eqref{Coulombfield}.  We see that the energy is minimized when $r = r_{\rm bnd}$ is such that $|Z_{\gamma_{\rm h}}(u(r))| = - \Re[ \zeta^{-1} Z_{\gamma_{\rm h}}(u(r))]$.  As $\zeta$ is a phase, this implies $\Im[ \zeta^{-1} Z_{\gamma_{\rm h}}(u(r))] = 0$, whence, from the attractor equation \eqref{attractor},
\begin{equation}\label{rbound}
r_{\rm bnd} = \frac{\llangle \gamma_{\rm h}, \gamma_{\rm c} \rrangle }{2 \Im[ \zeta^{-1} Z_{\gamma_{\rm h}}(u)] }~.
\end{equation}
If the right-hand side of \eqref{rbound} is negative then there is no stable configuration for probe particles of charge $\gamma_{\rm h}$.  When a stable configuration does exist the energy of the probe is $E_{\rm probe} = - \Re[ \zeta^{-1} Z_{\gamma_{\rm h}}(u)]$.

This formula can also be obtained from a limit of Denef's formula \cite{Denef:2000nb,Gaiotto:2010be}\footnote{The formula we give appears to differ from \cite{Denef:2000nb} by a sign, but this is an illusion.  Our conventions for the expansion of the electromagnetic charge along a Darboux basis are such that $\langle \gamma_1,\gamma_2 \rangle_{\rm us} = - \langle \gamma_1, \gamma_2 \rangle_{\rm Denef}$.} for the bound-state radius of two vanilla constituents of charges $\gamma_{1,2}$,
\begin{equation}\label{rDenef}
r_{\rm Denef} = \half \llangle \gamma_1, \gamma_2 \rrangle \frac{|Z_{\gamma_1 + \gamma_2}(u)|}{\Im[ \overline{Z_{\gamma_1}(u)} Z_{\gamma_2}(u)]} = \frac{\llangle \gamma_1, \gamma_2 \rrangle}{2 \Im[ \zeta_{\rm van}^{-1} Z_{\gamma_1}(u)]}~.
\end{equation}
In the second step we used that $-\zeta_{\rm van}$ is defined as the phase of the central charge of the bound state: $Z_{\gamma_1 + \gamma_2} = - \zeta_{\rm van} |Z_{\gamma_1 + \gamma_2}|$.  Now consider the limit in which particle $2$ is infinitely heavy, $|Z_{\gamma_2}| \to \infty$, so that we identify $\gamma_2 \to \gamma_{\rm c}$ and $\gamma_1 \to \gamma_{\rm h}$.  In this limit, the phase of $Z_{\gamma_1 + \gamma_2}$ agrees with the phase of $Z_{\gamma_2}$, which we identify with $\zeta$ of the defect.  Thus we recover \eqref{rbound}.

Although we considered a single probe particle, the result \eqref{rbound} applies just as well to a collection of probe particles of charge $\gamma_{\rm h}$ since these particles are mutually BPS and do not interact with each other.  In fact any probe of charge $\gamma \in \gamma_{\rm h} \cdot \mathbb{Z}_+$ will have the same bound-state radius since $Z_{\gamma}(u)$ is linear in $\gamma$.  Thus we can have a whole collection of halo particles at a stable distance $r_{\rm bnd}$ from the core of the defect.  If we start in a stable region and dial the Coulomb branch parameters $u$, \emph{and/or} the parameter $\zeta$ characterizing the defect, such that $\Im[ \zeta^{-1} Z_{\gamma_{\rm h}}(u)] \to 0$, then $r_{\rm bnd} \to \infty$.  The halo particles are becoming less and less bound to the defect, until they are completely free.  Note, crucially, that as $r_{\rm bnd}$ becomes much greater than the effective size of the vanilla particle, the probe approximation in which the core-halo picture is derived becomes better and better.

We would like to argue that the core-halo system represents a framed BPS state.  However the total charge of the core-halo system is $\gamma = \gamma_{\rm c} + \gamma_{\rm h}$ and if this configuration truly represents a framed BPS state, its total energy should be $E = - \Re [ \zeta^{-1} Z_{\gamma_{\rm c} + \gamma_{\rm h}}(u) ]$.  So far we have only accounted for the energy of the probe particle.  In the probe approximation the total energy is the sum of this plus the energy of the dyonic background.  Using the regularized energy functional \eqref{HIR}, the energy of the background can be computed and is given by \eqref{EIRdyon} with $\gamma_{\rm def} \to \gamma_{\rm c}$.  Hence, using the linearity of $Z_\gamma$, we indeed saturate the framed BPS bound for the total energy.  

We have so far described all of this in classical language.  One can quantize the non-interacting halo particles and build up the Fock space of halo particle states.  These provide an approximate description of framed BPS states that is good when $r_{\rm bnd}$ is much larger than the length scale set by the Higgs vev, and that becomes exact as $r_{\rm bnd} \to \infty$.  Hence this picture can be used to understand the wall crossing properties of framed BPS states.  We explicitly see that the halo Fock space disappears from the spectrum as we approach the framed marginal stability walls $\widehat{W}(\gamma_{\rm h})$, where
\begin{equation}\label{fmsw}
\widehat{W}(\gamma_{\rm h}) := \bigg\{ (u,\zeta) ~ \bigg| ~ \Omega(u,\gamma_{\rm h}) \neq 0~~ \& ~~  \zeta^{-1} Z_{\gamma_{\rm h}}(u) \in \mathbb{R}_- \bigg\} \subset \widehat{\BB} \times \widehat{\mathbb{C}^\ast}~.
\end{equation}
The first condition ensures that the proposed halo particle actually exists in the vanilla spectrum.\footnote{\label{fn:invisible}One can also define the walls to depend on the core in question, $\widehat{W}(\gamma_{\rm h}, \gamma_{\rm c})$ and impose a binding condition $\llangle \gamma_{\rm h}, \gamma_{\rm c} \rrangle \neq 0$ as part of the definition, analogous to the condition $\llangle \gamma_1,\gamma_2\rrangle \neq 0$ for the vanilla walls \eqref{vanillawalls}.  The definition \eqref{fmsw} works well in conjunction with the framed wall crossing formula, \eqref{genWCtransfo}. It can happen that, for a particular halo charge $\gamma_h$ and line defect, all associated framed BPS states
of charge $\gamma_{\rm c}$  have
$\llangle \gamma_{\rm h}, \gamma_{\rm c} \rrangle =0$. In this case
we say that that halo charge corresponds to an \emph{invisible wall}. In such a case, $S_{\gamma_{\rm h}}$ will simply commute with the generating function $F$, leaving it invariant across the wall.}  The walls are real co-dimension one walls in $(u,\zeta)$ space.  Here we have allowed for an analytic continuation in $\zeta$ so that $\zeta \in \mathbb{C}^\ast = \mathbb{C} \setminus \{0 \}$, as is convenient to do when studying the wall crossing problem.  The electromagnetic charge lattice can undergo monodromy around singular points of the Coulomb branch.  When studying defects it can be useful to lift nontrivial closed paths in $\BB^*$ to paths in $(u,\zeta)$ space which also wrap the $\zeta$ circle.  We have thus defined the walls on the universal cover $\widehat{\BB} \times \widehat{\mathbb{C}^\ast}$ of $\BB^\ast \times \mathbb{C}^\ast$.  One can project these to walls $W(\gamma) \subset \BB^\ast \times \mathbb{C}^\ast$.

The jump in the framed protected spin characters \eqref{fPSC} when $\widehat{W}(\gamma_{\rm h})$ is crossed is most succinctly formulated in terms of the generating function of framed BPS states \eqref{Fgenfun}. Denoting the generating function on the side of the wall where $\pm \Im \zeta^{-1}Z_{\gamma_\mathrm{h}}(u)>0$ with $F^\pm$ respectively, framed wall crossing is equivalent to a conjugation
\begin{equation}\label{genWCtransfo}
F^-(\{X_\gamma\})=S_{\gamma_\mathrm{h}}^{-1} F^+(\{X_\gamma\})S_{\gamma_\mathrm{h}}~.
\end{equation}
For details and the precise form of $S_{\gamma_{\rm h}}$ in general we refer to \cite{Gaiotto:2010be}.  In appendix \ref{wcapp} we work this out explicitly for hypermultiplet halos in the case of $\mathfrak{su}(2)$ SYM.
  
Wall-crossing in the vanilla case is more intricate, but as was shown in \cite{Gaiotto:2010be} can be derived from consistency relations on the framed wall crossing. The simplest case, the so-called primitive wall crossing where both decay products $\gamma_1$ and $\gamma_2$ are primitive charge vectors, was first discussed in \cite{Denef:2007vg}. In that case the vanilla protected spin character \eqref{PSC} jumps in the following simple manner:
\begin{equation}
\Omega^+(\gamma_1+\gamma_2;y)=\Omega^-(\gamma_1+\gamma_2;y)+ \chi_{|\llangle\gamma_1,\gamma_2 \rrangle|}(y) \,\Omega(\gamma_1;y)\Omega(\gamma_2;y)\,,\label{primwc}
\end{equation}
where $\Omega^\pm$ is the PSC on the side of the wall where $\pm \llangle\gamma_1,\gamma_2\rrangle\Im \zeta_\mathrm{van}^{-1}Z_{\gamma_1}(u)>0$, and $\chi_{n}(y)$ is the character \label{ncharacter} of the $n$-dimensional $SU(2)$ representation, $\chi_{n}(y) = (y^n - y^{-n})/(y - y^{-1})$.  The factors $\Omega(\gamma_{1,2};y)$ account for the internal states carried by the constituents, while $\chi_{|\llangle \gamma_1,\gamma_2\rrangle|}(y)$ accounts for the states of the electromagnetic field binding the constituents.

%%%%%%%%%%%%%%%%%%%%%%
\subsection{Weak coupling expansion}\label{sec:wce}
%%%%%%%%%%%%%%%%%%%%%%

In this paper we are interested in comparing the above constructions of (framed) BPS states and wall crossing with the results of a semiclassical analysis.  The semiclassical analysis is in principle valid at any energy scale, provided the effective coupling is weak.  It is therefore worthwhile to recall a few details of the weak coupling expansion of the low energy effective theory.

As we mentioned previously, at weak coupling there is a natural identification of the massless degrees of freedom with the components of the microscopic degrees of freedom along a Cartan subalgebra.  The classical vacua of the theory \eqref{Scl} consist of gauge-inequivalent, constant $\varphi = \varphi_\infty$ such that $[\varphi_\infty, \varphibar_\infty] = 0$.  Although these vacua are parameterized by gauge-invariant data such as $u^s = \langle \Tr(\varphi_{\infty}^{s+1}) \rangle$ for $\mathfrak{su}(N)$, it will be useful to choose a representative $\varphi_\infty$ for each $u \in \BB^\ast$.  Physical quantities of course will not depend on these choices.  First, without loss of generality, we take $\varphi_{\infty}$ to define the Cartan subalgebra $\mathfrak{t}_{\mathbb{C}}$, as described below \eqref{Trdef}.  

This still leaves the gauge redundancy of the Weyl orbit.  We fix it by taking $X_\infty \equiv \Im(\zeta^{-1} \varphi_{\infty})$ to define a notion of positive roots\label{posrootdef}: $\alpha \in \Delta^+ \iff \langle \alpha, X_\infty \rangle > 0$, such that $X_\infty$ is in the \emph{fundamental Weyl chamber}.  This requires that $X_\infty \in \mathfrak{g}$ be regular.  We will assume so for the purposes of this discussion, though later in the paper we will consider families of $X_\infty$ that approach a wall of the fundamental Weyl chamber from within the chamber.  Using this particular real slice of $\varphi_\infty$ to define the fundamental Weyl chamber is motivated by the semiclassical analysis of BPS states, where $X \equiv \Im(\zeta^{-1} \varphi)$ is the real Higgs field that participates in the Bogomolny equation.  We denote by $\alpha_I$ and $H_I \equiv H_{\alpha_I}$, $I = 1,\ldots,r = \rnk{\mathfrak{g}}$, the corresponding simple roots and simple co-roots.  We then identify $\mathfrak{t} \oplus \mathfrak{t}^\ast$ with the symplectic vector space $V_u = V_{u}^{\rm m} \oplus V_{u}^{\rm e}$ and take $\{ H_I, \lambda^J \} = \{ \mfa_I, \mfb^J \}$ as a Darboux basis, where $\lambda^J$ are the fundamental weights.

In order to make contact with the low energy effective Lagrangian \eqref{lel}, one decomposes the fields according to \eqref{rootdecomp}, $A = A^I H_I + \sum_{\alpha \in \Delta} A^\alpha (-i E_\alpha)$, $\varphi = a^I H_I + \sum_{\alpha\in \Delta} \varphi^\alpha (-i E_\alpha)$, \etc,  treating the Cartan components as a fixed background and the root components as quantum fluctuations.  For each root $\alpha$ there is a vector multiplet with charge $\alpha$ and mass $|\langle \alpha, a \rangle|$, where $a = a^I H_I$\label{scalaradef}.  Integrating out these massive vector multiplets leads to \eqref{lel}.  The general form of \eqref{lel} for an arbitrary configuration of Cartan-valued background fields is determined by $\NN = 2$ supersymmetry, so in order to extract the prepotential $\FF$ it is sufficient to consider the special case of constant $a^I, F_{\mu\nu}^I$ and vanishing background fermi fields.  A standard one-loop computation leads to the effective coupling matrix $\tau_{IJ}$ with
\begin{align}\label{1looptau}
\Im(\tau_{IJ}) =&~ \frac{4\pi}{g_{0}^2} \Tr(H_I H_J) + \frac{1}{2\pi} \sum_{\alpha\in\Delta^+} \langle \alpha, H_I \rangle \langle \alpha, H_J \rangle \left\{ \ln \left( \frac{ |\langle \alpha, a \rangle|^2}{2\mu_{0}^2} \right) + 3 \right\}~, \cr
\Re(\tau_{IJ}) =&~ \frac{\theta_0}{2\pi} \Tr(H_I H_J) - \frac{1}{\pi} \sum_{\alpha\in \Delta^+} \langle \alpha, H_I \rangle \langle \alpha, H_J \rangle \theta_\alpha~,
\end{align}
where $\theta_\alpha$ is defined by $\langle \alpha, a \rangle = |\langle \alpha, a \rangle| e^{i \theta_\alpha}$, and the first terms are the classical contributions.  

The one-loop correction to $\Re(\tau_{IJ})$ is finite and originates from the ABJ anomaly of the chiral fermions.  For each root $\alpha \in \Delta^+$ the doublet of Weyl fermions can be packaged into a single Dirac fermion, $\Psi^{(\alpha)}$.  The Yukawa terms generate both a mass and pseudo-mass coupling of the form $\Psibar^{(\alpha)} \left( \langle \alpha, \Re(a)\rangle - i \gamma_5 \langle \alpha, \Im(a)\rangle \right) \Psi^{(\alpha)}$.  By making a unitary chiral rotation of the Dirac fermion---which is a $U(1)_R$ rotation in terms of the original Weyl fermions---one can change field variables to a new Dirac fermion ${\Psi'}^{(\alpha)}$ with mass $|\langle \alpha, a\rangle|$.  The path integral over ${\Psi'}^{(\alpha)}$ can then be handled with standard methods.  The change in integration measure is however anomalous \cite{Fujikawa:1983bg}, resulting in the one-loop correction to the classical theta angle as given in $\Re(\tau_{IJ})$, \eqref{1looptau}.

The one-loop correction to $\Im(\tau_{IJ})$ is divergent and requires renormalization.  Using dimensional regularization in $d = 4-2\epsilon$ dimensions, we introduce the counterterm Lagrangian
\begin{equation}\label{Lct}
\LL_{\rm c.t.} = \frac{h^\vee}{16\pi^2} \left\{ \ln \left( \frac{\mu_{0}^2}{2\pi} \right) - \frac{1}{\epsilon} + \gamma - 3 \right\} \Tr(H_I H_J) F_{\mu\nu}^I F^{\mu\nu J} + \textrm{susy completion} ~. 
\end{equation}
Recall that $h^\vee$ is the dual Coxeter number and from the definition of $\Tr$ in terms of $\tr_{\mathbf{adj}}$, \eqref{Trdef}, it follows that 
\begin{equation}\label{adtr}
\sum_{\alpha\in \Delta} \langle \alpha, H_I \rangle \langle \alpha, H_J \rangle = 2 h^\vee \Tr(H_I H_J)~.
\end{equation}
Together with the one-loop determinant, $\LL_{\rm c.t.}$ produces the one-loop correction to $\Im(\tau_{IJ})$ given in \eqref{1looptau}.

The finite part of $\LL_{\rm c.t.}$ is scheme dependent; we chose it so as to arrive at the standard form of the prepotential, as we now show.  The real and imaginary parts of $\tau_{IJ}$ can be combined into
\begin{align}
\tau_{IJ} =&~ \tau_0 \Tr(H_I H_J) + \frac{i}{2\pi} \sum_{\alpha \in \Delta^+} \langle \alpha, H_I \rangle \langle \alpha, H_J \rangle \left\{ \ln\left( \frac{\langle \alpha, a\rangle^2}{2 \mu_{0}^2} \right) + 3 \right\} \cr
=&~ \frac{i}{2\pi} \sum_{\alpha \in \Delta^+}  \langle \alpha, H_I \rangle \langle \alpha, H_J \rangle \left\{ \ln\left( \frac{\langle \alpha, a\rangle^2}{2 \Lambda^2} \right) + 3 \right\} ~,
\end{align}
where $\tau_0 = \frac{4\pi i}{g_{0}^2} + \frac{\theta_0}{2\pi}$ and we introduced the dynamical scale
\begin{equation}\label{dyscale}
\Lambda := \mu_0 e^{i\pi \tau_0/h^\vee}~.
\end{equation}
If $\mu_0$ is the scale at which the one-loop running coupling takes on the ``bare'' value, $g(\mu_0) = g_0$, then $|\Lambda|$ is the scale where it blows up.  This $\tau_{IJ}$ follows from the perturbative part of the prepotential
\begin{equation}\label{pertprepot}
\FF^{\textrm{1-lp}} = \frac{i}{4\pi} \sum_{\alpha \in \Delta^+} \langle \alpha, a \rangle^2 \ln\left( \frac{\langle \alpha, a\rangle^2}{2 \Lambda^2} \right)~,
\end{equation}
or equivalently the perturbative expression for the dual coordinate,
\begin{equation}\label{pertaD}
a_{\mathrm{D},I}^{\textrm{1-lp}} := \frac{\pd \FF^{\textrm{1-lp}}}{\pd a^I} = \frac{i}{2\pi} \sum_{\alpha \in \Delta_+} \langle \alpha, H_I\rangle \langle \alpha, a \rangle \left\{ \ln\left( \frac{\langle \alpha, a\rangle^2}{2 \Lambda^2} \right) + 1 \right\}~.
\end{equation}
These agree with the usual formulae, provided one remembers the different normalization of the scalar field, $a_{\rm here} = \sqrt{2} a_{\rm SW}$.  (See footnote \ref{fn:rescale}.)

The nontrivial fibration of the electromagnetic charge lattice over the weak coupling regime of the Coulomb branch can be understood from these formulae. To give a simple example, consider the $\mathfrak{su}(2)$ theory with Coulomb branch parameter $u = \half \langle \alpha, a\rangle^2$.  Now let us  increase $\theta_\alpha$ starting from $\theta_{\alpha}^{\rm in}$ while holding $|\langle \alpha, a\rangle|$ fixed.  At each integer $n$ such that $\theta_\alpha = \theta_{\alpha}^{\rm in} + n\pi$ we return to the same point $u$ on the Coulomb branch, but these all give different values of $\Re(\tau)$, \eqref{1looptau}.  At $u$ we can measure the physical magnetic and electric charge as the flux of the low energy magnetic and electric fields through the two-sphere at infinity.  In general these are defined by
\begin{equation}\label{physEflux}
\gamma_{\rm m}^I = \frac{1}{2\pi} \int_{S_{\infty}^2} F^I~, \qquad \gamma_{{\rm e},I}^{\rm phys} := \frac{1}{2\pi} \int_{S_{\infty}^2} \Im(\tau_{IJ}) \star F^J ~.
\end{equation}
Equations \eqref{IRcharges}, \eqref{dualfieldstrength} give a relation between these charges and the quantized electric charge, $\gamma_{{\rm e},I}$:
\begin{equation}\label{chargefibration}
\gamma_{{\rm e},I} = - (\gamma_{{\rm e},I}^{\rm phys} + \Re(\tau_{IJ}) \gamma_{\rm m}^J)~.
\end{equation}
Hence the different values of $\Re(\tau_{IJ})$ that can be associated to the same point $u \in \BB^\ast$, correspond to the different $\gamma_{\rm e}$'s consistent with the given $\gamma_{\rm m}$ and $\gamma_{\rm e}^{\rm phys}$, corresponding to different local trivializations of $\Gamma$ over the patch containing $u$.  Notice that specifying a consistent set $\{a_{\mathrm{D}}(u),a(u)\}$---\ie\ a value of $a(u)$ together with a choice of branch for the logarithm---determines such a local trivialization.

Standard arguments employing holomorphy \cite{Seiberg:1988ur,Seiberg:1994bp} and the $U(1)_R$-symmetry anomaly imply that the perturbative prepotential is one-loop exact.  The exact prepotential, 
\begin{equation}\label{prepot} 
\FF = \FF^{\textrm{1-lp}} + \FF^{\rm np}~,
\end{equation}
includes nonperturbative instanton corrections which are of the general form
\begin{equation}\label{prepotexp}
\FF^{\rm np} =    \frac{1}{2 \pi i} \sum_{k = 1}^{\infty} \FF_{k} \left( \langle \alpha,a \rangle^{-1} \right) \Lambda^{2k h^\vee }~,
\end{equation}
where $\FF_{k}$ is a Weyl-invariant polynomial of degree $2k h^\vee -2$ in $|\Delta_+|$ variables evaluated on $\langle \alpha, a \rangle^{-1}$, $\alpha \in \Delta^+$.  $\FF^{\rm np}$ is exponentially suppressed in the bare coupling $g_0$ relative to $\FF^{\textrm{1-lp}}$, provided none of the $\langle \alpha, a \rangle$ are vanishing.

Thus far we have discussed how one obtains the vanilla low energy effective action \eqref{lel} starting from the microscopic one.  In the presence of line defects it should also be possible to obtain the low energy defect action \eqref{SIRbndry} from the microscopic one \eqref{Sclbndry} in the same way, namely by carrying out a Gaussian path integral for the boundary values of the quantum fluctuation fields.  This would be an interesting computation to do but we leave it for future work.

%%%%%%%%%%%%%%%%%%%%%%
%%%%%%%%%%%%%%%%%%%%%%
\section{Classical framed BPS field configurations}\label{sec:clfBPS}
%%%%%%%%%%%%%%%%%%%%%%
%%%%%%%%%%%%%%%%%%%%%%

In this section we describe the space of classical BPS field configurations for given vev $\varphi_\infty$ and electromagnetic charge $\gamma$.  This provides the starting point for the semiclassical construction of BPS states.  In both this section and the next we are mostly extending well known results for vanilla BPS field configurations and states to framed BPS field configurations and states in the presence of line defects.  However we will obtain some new results even in the vanilla case.  We focus on the case of pure \tHooft defects, where the singular monopole moduli spaces studied in \cite{Kronheimer,MR1624279,Cherkis:1997aa,Cherkis:1998xca,Cherkis:1998hi,Kapustin:2006pk,MRVdimP1,MRVdimP2} play an essential role.

%%%%%%%%%%%%%%%%%%%%%%
\subsection{Hamiltonian and electromagnetic charges}
%%%%%%%%%%%%%%%%%%%%%%

The bosonic Hamiltonian\footnote{In this expression $E_i$, $D_0\varphi$, and $D_0 \varphibar$ should be understood as functionals of the canonical momenta and coordinates.  Defining $\pi,\bar{\pi},\pi_i$ as the momenta canonically conjugate to $\varphi,\varphibar,A^i$ respectively, the relations are $D_0 \varphi = 2 g_{0}^2 \bar{\pi}$, $D_0 \varphibar = 2 g_{0}^2 \pi$, and $E_i = - g_{0}^2 (\pi_i + \tilde{\theta}_0 B_i)$.} associated with the classical action \eqref{Scl} is
\begin{equation}\label{Ham1}
H_{\rm bos} =  \frac{1}{g_{0}^2} \int_{\UU} \ed^3 x \Tr \left( E_i E^i + B_i B^i + D_0 \varphi D_0 \varphibar + D_i \varphi D^i \varphibar - \frac{1}{4} ( [ \varphi, \varphibar ] )^2 \right) + V_{\rm def}~,
\end{equation}
where $\UU = \mathbb{R}^3 \setminus \{\vec{x}_n \}$, $E_i = F_{i0}$, and $B_i = \half \epsilon_{ijk} F^{jk}$. Note that we have used the `Gauss law' constraints associated with $A_0$.  The first of these is local  and can be identified with the $A_0$ equation of motion in the Lagrangian formulation,
\begin{equation}\label{Gauss}
0= D^i E_i - \half \left( [\varphibar, D_0 \varphi] + [\varphi, D_0 \varphibar] \right)~.
\end{equation}
The second is a boundary term,
\begin{align}\label{Gaussbc}
0 = \int_{\pd\UU} \ed^2 S^i \Tr \left\{ A_0 \left(E_i + \tilde{\theta}_0 B_i \right) \right\} = \int_{S_{\infty}^2} \ed^2 \Omega r^2 \hat{r}^i  \Tr  \left\{ A_0 \left(E_i + \tilde{\theta}_0 B_i \right) \right\} ~. 
\end{align}
In the second step we used the boundary conditions \eqref{defectbcs} and \eqref{enhancedbcs} to eliminate the contributions from the infinitesimal two-spheres surrounding the defects.  Since we will be interested in field configurations where the asymptotic flux of $E_i + \tilde{\theta}_0 B_i$ is nontrivial, the simplest way to satisfy \eqref{Gaussbc} is to impose $A_0 \to 0$ as $r \to \infty$.

The defect potential in \eqref{Ham1} follows directly from \eqref{Sclbndry} and is required in the presence of \tHooft defects.  For defects of type $\zeta$ with positions and charges $(\vec{x}_n, P_n)$ it is given by
\begin{equation}
V_{\rm def} = - \frac{2}{g_{0}^2} \sum_n \Re \bigg\{ \zeta^{-1} \int_{S_{\varepsilon_n}^2} \Tr \left\{ (i F - \star F) \varphi \right\} \bigg\} ~.
\end{equation}

The defect potential will serve to regulate the energy functional when evaluated on the singular field configurations required by the defect boundary conditions.  Given this, we then want to impose the standard asymptotic boundary conditions that guarantee finiteness of the energy.  This requires that $\varphi$ and the leading, $O(1/r^2)$, monopole moments of the electric and magnetic field be covariantly constant and mutually commuting sections of the adjoint bundle over the two-sphere at infinity.  Thus by making patchwise gauge transformations as necessary, we may assume them to be constants valued simultaneously in the same Cartan subalgebra.  Having done so, we define the magnetic and electric charges through the asymptotic fluxes as follows:
\begin{align}\label{flux}
 \gm :=&~ \frac{1}{2\pi} \int_{S_{\infty}^2} F  ~, \cr
\gamma_{\rm e}^{\rm phys} :=&~ \frac{2}{g_{0}^2} \int_{S_{\infty}^2} \star F   = - \left( \gamma_{\rm e}^\ast + \frac{\theta_0}{2\pi} \gm \right)~.
\end{align}

We defined $\gamma_{\rm e}^{\rm phys}$ as the actual flux of the electric field.  $\gamma_{\rm e}^\ast \in \mathfrak{t}$ denotes the vector space dual of the electric charge $\gamma_{\rm e}$ with respect to the Killing form $(~,~)$ defined in \eqref{Trdef}, and the electric charge is defined as the conserved Noether charge associated with asymptotically nontrivial gauge transformations that preserve the vaccum $\varphi_\infty$.  The relation $\gamma_{\rm e}^\ast = - (\gamma_{\rm e}^{\rm phys} + \frac{\theta_0}{2\pi} \gm)$ can be derived via the same procedure as outlined for the IR theory under \eqref{dualfieldstrength}.  Recall at weak coupling we have the natural duality frame $V_{u}^{\rm m} \oplus V_{u}^{\rm e} \cong \mathfrak{t} \oplus \mathfrak{t}^\ast$ and Darboux basis $\{ H_I, \lambda^J \}$ of simple co-roots and fundamental weights.  Furthermore, the classical coupling matrix is $\tau_{IJ}^{\rm cl} = \tau_0 (H_I, H_J)$.  Using these, we can see that \eqref{flux} is entirely consistent with the classical limit of the definitions \eqref{IRcharges}, \eqref{dualfieldstrength}, and \eqref{physEflux}.

Furthermore we can identify the vanilla charge lattice $\Gamma_u$ and shifted charge lattice $\Gamma_{L,u}$, \eqref{torsor}, as follows.  The asymptotic gauge field associated with a magnetic charge $\gm$ must be defined patchwise and takes the form
\begin{equation}
A \to \frac{\gm}{2} (\pm 1 - \cos{\uptheta}) \ed\upphi~,
\end{equation}
in spherical coordinates, with the plus sign for the patch containing the north pole, $\uptheta = 0$, and the minus for the patch containing the south pole, $\uptheta = \pi$.  Single-valuedness of the transition function, $\upphi \mapsto \exp(\gm \upphi)$, on the overlap requires that $\exp(2\pi \gm) = \mathbbm{1}_G$, the identity element in the gauge group.  Hence $\gm \in \Lambda_G$, \eqref{cochar}.  However, in the absence of defects, the radial coordinate provides a homotopy of the asymptotic two-sphere to a point at $r = 0$, and thus the closed loop in $G$ defined by the transition function must be homotopically trivial.  This will be the case if and only if it lifts to a closed loop in the simply-connected cover, $\tilde{G}$.  Thus, in the absence of defects, we have the stronger requirement that $\gm \in \Lambda_{\tilde{G}} = \Lambda_{\rm cr}$, the co-root lattice \label{crlatdef}.  Dyons involve exciting fluctuations of the microscopic fields along the root directions of the Lie algebra (as we will see in detail below), and thus as in the perturbative sector, electric charges are confined to the root lattice.  Hence the vanilla electromagnetic charge lattice for pure Yang--Mills theories  is naturally identified with
\begin{equation}\label{wcemid}
\Gamma_u \cong \Lambda_{\rm cr} \oplus \Lambda_{\rm rt}~,
\end{equation}
at weak coupling.  Notice that a priori the Dirac quantization condition, $\langle \gamma_{\rm e}, \gamma_{\rm m} \rangle \in \mathbb{Z}$, allows for a more refined electric lattice, namely the weight lattice $\Lambda_{\rm wt} \cong \oplus_I (\lambda^I \cdot \mathbb{Z})$.  This corresponds to the possibility of coupling matter in any representation\label{repdef} $\rho$ of $\mathfrak{g}$.

When line defects are present the above arguments are modified.  We still have $\gm \in \Lambda_G$, and the same argument concerning single-valuedness of the transition functions on the infinitesimal two-spheres implies that the 't Hooft charges $P_n \in \Lambda_G$.  Now, rather than a homotopy of the asymptotic two-sphere to a point, we have a homotopy of the asymptotic two-sphere to a disjoint union of infinitesimal two-spheres.  Hence it follows that the closed loop $\upphi \mapsto \exp\left\{ (\gm - \sum_n P_n)\upphi \right\}$ should be homotopically trivial in $G$.  Thus in general $\gm$ sits in a torsor for the co-root lattice, $\gm \in  \Lambda_{\rm cr} + (\sum_n P_n)$.  The electric discussion is unmodified for the case of pure \tHooft defects and thus we have $\Gamma_{L,u} = \Gamma_u + \gamma_L$ as in \eqref{torsor} where $\gamma_L$ may be taken as
\begin{equation}
\gamma_L = \sum_{n} P_n~.
\end{equation}
Note that any two $P_n, P_{n}'$ that differ by a Weyl reflection satisfy $P_n - P_{n}' \in \Lambda_{\rm cr}$, hence $\Gamma_{L,u}$ only depends on the Weyl orbits $[P_n]$, consistent with the fact that these are what label physically distinct \tHooft defects.  Note also that the torsor $\Lambda_{\rm cr} + (\sum_n P_n)$ is a subset of $\Lambda_{\rm G}$ which is in turn a sublattice of the magnetic weight lattice, $\Lambda_{\rm mw} = \Lambda_{G_{\rm ad}}$, the co-character lattice of the adjoint form of the gauge group.  $\Lambda_{\rm mw}$\label{mwlatdef} is the integral dual of the root lattice and thus $\gamma_L$ does satisfy the required property $\llangle \gamma_L, \gamma \rrangle \in \mathbb{Z}$ for any $\gamma \in \Gamma_u$.  Here we are using the relation
\begin{equation}
\llangle \gm \oplus \gamma_{\rm e}, \gm' \oplus \gamma_{\rm e}' \rrangle = \langle \gamma_{\rm e}, \gm' \rangle - \langle \gamma_{\rm e}' ,\gm \rangle~,
\end{equation}
between the symplectic pairing on $\Gamma_{L,u}$ and the canonical pairing on $\mathfrak{t}^\ast \times \mathfrak{t}$ that follows from the definition \eqref{Darboux} and the weak coupling identification \eqref{wcemid}.

%%%%%%%%%%%%%%%%%%%
\subsection{Classical BPS bound and BPS field configurations}\label{sec:clBPSbnd}
%%%%%%%%%%%%%%%%%%%

The BPS spectrum is described at the classical level by finite energy field configurations solving first order equations.  The first order equations can be obtained as the fixed point equations of supersymmetry transformations or by finding local minima of the energy functional via Bogomolny's identity \cite{Bogomolny:1975de,Coleman:1976uk}.  We review the former method in appendix \ref{N2conventions}; here we recall the latter.  The canonical Hamiltonian \eqref{Ham1} can be rewritten as
\begin{equation}\label{Ham2}
H_{\rm bos} =  \frac{1}{g_{0}^2} \int_{\UU} \ed^3x \Tr \left\{ \left| -E_i - i B_i + \zeta^{-1} D_i \varphi \right|^2 + \left| \zeta^{-1} D_0 \varphi + \frac{1}{2} [\varphi, \varphibar] \right|^2 \right\} - \Re(\zeta^{-1} Z^{\rm cl})~,
\end{equation}
where
\begin{equation}\label{Zcl}
Z^{\rm cl} =  \frac{2}{g_{0}^2} \int_{S_{\infty}^2} \Tr \left\{ (i F - \star F) \varphi \right\}~.
\end{equation}
This is achieved through a combination of integration by parts, the Bianchi identity, and Gauss' Law, \eqref{Gauss}.  In particular, the integration by parts produces boundary terms on both the infinitesimal and asymptotic two-spheres.  The boundary terms on the infinitesimal ones are exactly canceled by the defect potential, $V_{\rm def}$.  This leaves only the asymptotic boundary term which is directly related to the (classical) central charge.  

The result \eqref{Ham2} implies the classical BPS bound, $M \geq - \Re(\zeta^{-1} Z^{\rm cl})$, which is saturated when the first order equations hold such that the integrand of the bulk term in \eqref{Ham2} vanishes.  In the vanilla case the result is valid for any phase $\zeta$ but the strongest bound is achieved when
\begin{equation}\label{zetavanilla}
\zeta = \zeta_{\rm van}^{\rm cl} \equiv - \frac{Z^{\rm cl}}{|Z^{\rm cl}|}~, \qquad (\textrm{vanilla case})~,
\end{equation}
in which case $M = |Z^{\rm cl}|$ for BPS field configurations.  In the framed case $\zeta$ is instead fixed by the specification of the line defect and will be, in general, different from \eqref{zetavanilla}.  $Z^{\rm cl}$ is the central charge of the field configuration, as can be verified by computing the commutator of supercharges obtained from integrating the corresponding Noether currents.  (See appendix \ref{N2conventions}.)  Using the asymptotic form of $F$ associated with the charges \eqref{flux}, we find
\begin{equation}\label{Zcl2}
Z^{\rm cl} = Z_{\gamma}^{\rm cl}(u) = \tau_0 (\gamma_{\rm m}, a(u)) + \langle \gamma_{\rm e},a(u) \rangle~,
\end{equation}
where $a(u) = \varphi_\infty$.  This is consistent with \eqref{ZSW} where we identify the classical limit of the dual coordinate, $a_{\mathrm{D},I}^{\rm cl} = \tau_0 (H_I, a) = \tau_{IJ}^{\rm cl} a^J$.  The energy functional \eqref{Ham2} has local minima with values $M_{\gamma}^{\rm cl} = -\Re(\zeta^{-1} Z_{\gamma}^{\rm cl}(u))$ at field configurations solving the first order BPS equations
\begin{equation}\label{BPScomplex}
-E_i - i B_i + D_i \left( \zeta^{-1} \varphi \right) = 0~, \qquad D_0 \left( \zeta^{-1} \varphi \right) + \half [ \varphi, \varphibar ] = 0~,
\end{equation}
and the Gauss constraint, \eqref{Gauss}.  

It is convenient to decompose $\varphi$ into two real Higgs fields $X,Y$, \ie\ $\mathfrak{g}$-valued as opposed to $\mathfrak{g}_{\mathbb{C}}$-valued, defined by\footnote{Our $X,Y$ should be compared with the $b,a$ of references \cite{Lee:1998nv,Tong:1999mg,Gauntlett:1999vc}.}
\begin{equation}\label{XYdef}
\zeta^{-1} \varphi = Y + i X~.
\end{equation}
This determines the asymptotic values of $X,Y$, as functions of the vacuum data $(\gamma;u)$ and phase $\zeta$:
\begin{equation}\label{XYasymptotic}
\zeta^{-1} a(u) = Y_\infty + i X_\infty~.
\end{equation}
Note that in the vanilla case, the phase is itself a function of the vacuum data, $\zeta_{\rm van} = \zeta_{\rm van}(\gamma;u)$, determined by \eqref{zetavanilla}, so in this case even the splitting \eqref{XYdef} depends on what vacuum and electromagnetic charge we are considering.  In terms of $X,Y$, the BPS equations take the form
\begin{align}\label{BPSreal}
& B_i = D_i X~, \qquad E_i = D_i Y~, \qquad D_0 X - [Y,X] = 0~, \qquad D_0 Y = 0~,
\end{align}
and using these the Gauss constraint reduces to
\begin{equation}\label{Gauss2}
D^i D_i Y + [X,[X,Y]] = 0~.
\end{equation}

Using \eqref{Zcl2}, \eqref{XYasymptotic} and \eqref{flux} one can determine the real and imaginary parts of $\zeta^{-1} Z_{\gamma}^{\rm cl}(u)$,  
\begin{equation}\label{ZclXY}
\zeta^{-1} Z_{\gamma}^{\rm cl}(u) = - \left[ \frac{4\pi}{g_{0}^2} (\gm, X_\infty) + (\gamma_{\rm e}^{\rm phys}, Y_\infty) \right] + i \left[ \frac{4\pi}{g_{0}^2} (\gm, Y_\infty) - (\gamma_{\rm e}^{\rm phys}, X_\infty) \right]~,
\end{equation}
and thus the classical BPS mass can also be written as
\begin{equation}\label{MBPScl}
M_{\gamma}^{\rm cl} = \frac{4\pi}{g_{0}^2} (\gm, X_\infty) + (\gamma_{\rm e}^{\rm phys}, Y_\infty)~.
\end{equation}
This formula holds in both vanilla and framed cases.  In the vanilla case, $(\zeta_{\rm van}^{\rm cl})^{-1} Z_{\gamma}^{\rm cl} = - |Z_{\gamma}^{\rm cl}|$ so we get a constraint that the imaginary part of \eqref{ZclXY} should vanish.  However one can show that this constraint automatically holds provided the BPS equations \eqref{BPSreal} are satisfied \cite{Tong:1999mg}.  These equations, together with some integration by parts, yield
\begin{align}\label{XYasrel}
0 =&~ - \int_{\UU} \ed^3x \Tr \left\{ X [X,[X,Y]] \right\} = \int_{\UU} \ed^3 x \Tr \{ X D^i E_i \} \cr
=&~ \int_{\UU} d^3 x \pd^i \Tr \left\{ X E_i  \right\} -  \int_{\UU} \ed^3 x \Tr \{ D^i X E_i \} \cr
=&~ \int_{\UU} \ed^3 x \pd^i \Tr \left\{ X E_i - Y B_i \right\}~.
\end{align}
Recall that $\UU = \mathbb{R}^3 \setminus \{\vec{x}_n \}$ and the boundary $\pd \UU$ consists of the asymptotic two-sphere as well as infinitesimal ones surrounding the $\vec{x}_n$.  However, in the vanilla case, the only boundary is $S_{\infty}^2$.  The flux conditions \eqref{flux} imply
\begin{equation}\label{BEasymptotic}
B_i = \frac{\gm}{2 r^2} \hat{r}_i + o(r^{-2})~, \qquad E_i = \frac{ g_{0}^2 \gamma_{\rm e}^{\rm phys} }{8 \pi r^2} \hat{r}_i + o(r^{-2})~.
\end{equation}
Plugging \eqref{BEasymptotic} into \eqref{XYasrel} leads to the linear constraint, 
\begin{equation}\label{vanconstraint}
0 = \frac{4\pi}{g_{0}^2} (\gm, Y_\infty) - (\gamma_{\rm e}^{\rm phys}, X_\infty)~, \qquad \textrm{(vanilla case)}.
\end{equation}

This constraint nevertheless does imply that the map $a(u) \mapsto (X_\infty,Y_\infty)$ given by \eqref{XYasymptotic} is not one-to-one in the vanilla case.  Indeed, under a $U(1)_R$ rotation $a \to e^{i\vartheta} a$ we have that $\zeta_{\rm v}^{\rm cl} \to e^{i \vartheta} \zeta_{\rm v}^{\rm cl}$, and therefore $Y_\infty, X_\infty$ are invariant.  The pre-image of the point $(X_\infty,Y_\infty)$ is a closed loop in the Coulomb branch, at fixed values of $|a^I|$.  Hence, classically, each solution to the BPS system \eqref{BPSreal}, \eqref{Gauss2} with boundary data $(\gamma;X_\infty,Y_\infty)$ provides a family of solutions for all $a(u)$ in the pre-image.

It is interesting to ask what happens to $\Im(\zeta^{-1} Z^{\rm cl})$ when defects are present.  It is still proportional to the boundary term of \eqref{XYasrel} coming from the asymptotic two-sphere.  However instead of vanishing this contribution should now match the boundary terms from the infinitesimal two-spheres:
\begin{align}\label{ImZdef}
(\gamma_{\rm e}^{\rm phys}, X_\infty) - \frac{4\pi}{g_{0}^2} (\gm, Y_\infty) = \sum_n \lim_{\varepsilon_n \to 0} \frac{2}{g_{0}^2} \int_{S_{\varepsilon_n}^2} \ed^2 \Omega_n \varepsilon_{n}^2 \hat{r}_{n}^i \Tr \left\{ X E_i - Y B_i \right\} ~,
\end{align}
where $\varepsilon_n = |\vec{x} - \vec{x}_n|$.  This is quite a nontrivial statement since the right-hand side must evidently be finite and generically nonzero, as $\Im(\zeta^{-1} Z^{\rm cl})$ need not vanish.  

Indeed, using the defect boundary conditions \eqref{defectbcs} one finds that the leading divergence of the quantity in the trace cancels.  Naively, though, there could still be a problem when the leading divergence of one field, say $X \sim O(\varepsilon_{n}^{-1})$, multiplies the first subleading divergence of another field, say $\delta E_i \sim O(\varepsilon_{n}^{-3/2})$.  In fact the analysis of appendix \ref{app:deltahalf}, which showed the consistency of \eqref{defectbcs} with the variational principle, comes in handy here as well.  One result we prove there, (see \eqref{XYsubleading}), is that on any solution to the equations of motion, in particular a BPS solution, the boundary conditions \eqref{defectbcs} imply that
\begin{equation}\label{XYC}
\YY := \frac{4\pi}{g_{0}^2} Y + \frac{\theta_0}{2\pi} X = \YY(\vec{x}_n) + O(\varepsilon_{n}^{1/2})~, \qquad \textrm{as } \varepsilon_n \to 0~,
\end{equation}
where $\YY(\vec{x}_n)$ is finite and Cartan-valued.  The right-hand side of \eqref{XYC} is more restrictive than the $O(\varepsilon_{n}^{-1/2})$ behavior one might expect from a naive application of \eqref{defectbcs}.  Taking a covariant derivative of \eqref{XYC} and using the BPS equations, we also get
\begin{equation}
 E_i + \tilde{\theta}_0 B_i = O(\varepsilon_{n}^{-1/2})~,
\end{equation}
where, recall, $\tilde{\theta}_0 = \frac{g_{0}^2 \theta_0}{8\pi^2}$.

These two results can be used to evaluate the right-hand side of \eqref{ImZdef}:
\begin{align}
\lim_{\varepsilon_n \to 0} \frac{2}{g_{0}^2} \int_{S_{\varepsilon_n}^2} \ed^2 \Omega_n \varepsilon_{n}^2 \hat{r}_{n}^i \Tr \left\{ X E_i - Y B_i \right\} =&~ - \lim_{\varepsilon_n \to 0} \frac{2}{g_{0}^2} \int_{S_{\varepsilon_n}^2} \ed^2 \Omega_n \varepsilon_{n}^2 \hat{r}_{n}^i \Tr \left\{ \left( \tilde{\theta}_0 X + Y \right) B_i \right\} \cr
=&~ - (P_n, \YY(\vec{x}_n))~. \raisetag{24pt}
\end{align}
It is useful to express the left-hand side of \eqref{ImZdef} in terms of the asymptotic value of $\YY$ as well.  Setting $\YY_{\infty}^{\rm cl} := \frac{4\pi}{g_{0}^2} Y_{\infty} + \frac{\theta_0}{2\pi} X_\infty$, we use \eqref{flux} to write
\begin{equation}
\frac{4\pi}{g_{0}^2} (\gm, Y_\infty) - (\gamma_{\rm e}^{\rm phys}, X_\infty) = (\gm, \YY_{\infty}^{\rm cl}) + \langle \gamma_{\rm e}, X_\infty \rangle~.
\end{equation}
Combining this with \eqref{ImZdef} and \eqref{ZclXY}, we find, in the presence of multiple \tHooft defects of type $\zeta$ with positions and charges $(\vec{x}_n,P_n)$,
\begin{equation}\label{strangeid}
\Im( \zeta^{-1} Z_{\gamma}^{\rm cl}(u) ) = (\gm, \YY_{\infty}^{\rm cl}) + \langle \gamma_{\rm e}, X_\infty \rangle =  \sum_n (P_n, \YY(\vec{x}_n))~.
\end{equation}
If there are no defects then we recover the vanilla result \eqref{vanconstraint}.  This result will be useful later when we discuss the collective coordinate approximation to the dynamics in the presence of defects.  

The quantity $\YY_{\infty}^{\rm cl}$ is closely related to the classical limit of the dual special coordinate $a_{\mathrm{D}}$.  Recall that $a_{\mathrm{D},I}^{\rm cl} = \tau_0 (H_I, a)$.  These are naturally contracted with the fundamental weights $\lambda^I \in \mathfrak{t}^\ast$, which comprise the electric part of our weak coupling Darboux basis.  However we can define a $\mathfrak{t}$-valued dual coordinate by taking the (vector space) dual with respect to the Killing form \eqref{Trdef}:
\begin{equation}\label{tvaluedaD}
a_{\mathrm{D}} := \sum_I a_{\mathrm{D},I} (\lambda^I)^\ast~.
\end{equation}
Since $\lambda^I$ and $H_J$ are integral-dual bases, we see that $a_{\mathrm{D}}^{\rm cl} = a_{\mathrm{D},I}^{\rm cl} (\lambda^I)^\ast = \tau_0 a$.  Then define
\begin{equation}\label{YYcl}
\YY_{\infty}^{\rm cl} := \Im(\zeta^{-1} a_{\mathrm{D}}^{\rm cl}) = \frac{4\pi}{g_{0}^2} \Re(\zeta^{-1} a) + \frac{\theta_0}{2\pi} \Im(\zeta^{-1} a)  = \frac{4\pi}{g_{0}^2} Y_\infty + \frac{\theta_0}{2\pi} X_\infty~,
\end{equation}
and observe that it agrees with our previous definition of $\YY_{\infty}^{\rm cl}$.  This will serve as an important hint for how quantum corrections are to be accounted for in the semiclassical analysis to follow.

Returning to the analysis of the BPS equations, we are interested in solutions to \eqref{BPSreal}, \eqref{Gauss2} modulo the group of gauge transformations.  For the purpose of constructing solutions to these equations it is convenient to initially fix part of the gauge freedom by choosing a ``generalized temporal'' gauge where
\begin{equation}\label{temporalgauge}
A_0 = Y~.
\end{equation}
Note that this condition is consistent with the defect boundary conditions \eqref{defectbcs}, in that $Y$ and $A_0$ are required to have the same singular behavior in the vicinity of defects.  It violates the condition $\lim_{|\vec{x}| \to \infty} A_0 = 0$ arising from the global Gauss constraint, \eqref{Gaussbc}, but that can be remedied by transforming the final solution by a gauge transformation $\cg$ that asymptotes to the time-dependent Cartan-valued phase $\cg_\infty = \exp{(-Y_\infty t)}$.  This transformation will preserve all defect and asymptotic boundary conditions, while changing $A_0$ such that it vanishes asymptotically.  It is the generalization of the Gibbons--Manton gauge \cite{Gibbons:1986df} for the Julia--Zee dyon \cite{Julia:1975ff}.  

In the gauge \eqref{temporalgauge} the last three of \eqref{BPSreal} reduce to the statement that field configurations are time-independent,
\begin{equation}\label{timeindependent}
\pd_0 A_i = \pd_0 X = \pd_0 Y = 0~,
\end{equation}
while the first of \eqref{BPSreal} and \eqref{Gauss2} become the ``primary'' and ``secondary'' BPS equations:
\begin{equation}\label{BPS}
 \textrm{primary:} ~B_i = D_i X~, \qquad \qquad \textrm{secondary:} ~D^i D_i Y + [X,[X,Y]] = 0~.
\end{equation}
Notice that the primary equation is the standard BPS monopole equation \cite{Prasad:1975kr,Bogomolny:1975de} for a pair $\{A,X\} \in \Omega^1(\PP) \times \Omega^0(\ad_{\PP})$, where $\pi : \PP \to \UU$ denotes the principal $G$-bundle over $\UU = \mathbb{R}^3 \setminus \{\vec{x}_n \}$ and $\ad_{\PP}$ is the associated adjoint bundle.  It can be solved independently of the secondary equation.  We still have to identify solutions that are related by gauge transformations that preserve \eqref{temporalgauge}.  These are the time-independent transformations.  This construction leads to the singular monopole moduli spaces that were studied in \cite{Kronheimer,MR1624279,Cherkis:1997aa,Cherkis:1998xca,Cherkis:1998hi,Kapustin:2006pk,MRVdimP1} and will be reviewed in the next subsection.

The secondary equation is a linear equation for $Y$ in the (singular) monopole background; the boundary value problem has a unique solution as we will describe later. (See section \ref{sec:cldyon}.)  With the solution for $Y$ in hand, the electric field, and in particular the electric flux through the two-sphere at infinity, is determined.  Thus it should be clear that for a fixed collection of \tHooft defects $\{\zeta;P_n, \vec{x}_n\}$, solutions might not exist for a given set of asymptotic data, $\{\gm,\gamma_{\rm e} ; X_\infty,Y_\infty\}$, or equivalently---using \eqref{XYasymptotic}---for a given set of $\{\gamma;u\}$.  The logical flow is as follows.  First, $\{\gm; X_\infty\}$ determine a moduli space of solutions to the primary BPS equation, which might be empty.  We will discuss conditions for this in the following subsection.  Assuming this space is nonempty, for a given point, \ie\ gauge-equivalence class $[\{A,X\}]$, and a given $Y_\infty$, we obtain a unique solution to the secondary equation and hence through the relation $E_i = D_i Y$ a unique electric charge.  While this would be fine classically, Dirac quantization imposes that the electric charge sit in a discrete lattice.  One may attempt to accommodate such a $\gamma_{\rm e}$ by moving around to different points in moduli space; in other words the solution $Y$ to the secondary BPS equation will determine the electric charge \emph{as a function on moduli space} \cite{Lee:1998nv,Bak:1999hp,Tong:1999mg}.  Then there might or might not exist a locus where this function takes on the given value $\gamma_{\rm e}$.  This question will be analyzed in detail in section \ref{sec:cldyon}.

In what follows it will sometimes be useful to cast the BPS equations in a different form.  One introduces a  fourth Euclidian direction, $x^4$, endow $\mathbb{R}^4$ with a flat Euclidian metric, and orient it so that $ \ed^3 x\wedge \ed x^4$ is positive.  We define a gauge field on $\mathbb{R}^4$,
\begin{equation}\label{hatA}
\hat{A} = \hat{A}_a \ed x^{a} = A_i \ed x^i + X \ed x^4~,
\end{equation}
with $a = 1,\ldots, 4$.  Using this gauge field we extend the covariant derivative, $\eD_{(3)} \to \hat{\eD}$, and fieldstrength, $F_{(3)} \to \hat{F}$.  All fields are independent of $x^4$ so $\hat{F} = F_{(3)} + (D_i X) \ed x^i\wedge \ed x^4$.  Then the primary BPS equation is equivalent to the self-duality equation for $\hat{F}$,
\begin{equation}\label{sd}
\hat{\star} \hat{F} = \hat{F}~,
\end{equation}
while the secondary equation is simply the gauge-covariant Laplace equation:
\begin{equation}\label{4dLaplace}
\hat{D}^a \hat{D}_a Y = 0~.
\end{equation}
%

%%%%%%%%%%%%%%%%%%%%%
\subsection[Singular and vanilla monopole moduli spaces]{Singular and vanilla monopole moduli spaces\footnote{We avoid using the terminology ``framed'' to refer to the moduli spaces of monopoles in the presence of 't Hooft defects because that adjective already has another meaning in the context of monopole moduli spaces.  Unfortunately, ``singular'' isn't an ideal choice either because the actual moduli space of singular monopole configurations might or might not have singularities.}}
%%%%%%%%%%%%%%%%%%%%%

Let the data $\{\vec{x}_n,P_n\}$ for a set of \tHooft defects be given, such that the behavior of $X$ and $B_i = \half \epsilon_{ijk} F^{jk}$ is specified by \eqref{defectbcs} in the vicinity of the defects.  Suppose we are also given a Higgs vev $X_\infty \in \mathfrak{t}$ and a magnetic charge $\gm \in \Lambda_{\rm cr} + (\sum_n P_n) \subset \mathfrak{t}$.  These specify the asymptotic behavior of the fields:
\begin{align}\label{XBasymptotics}
& X = X_\infty - \frac{\gm}{2r} + o(1/r)~, \qquad B_i = \frac{\gm}{2 r^2} \hat{r}_i + o(1/r^2)~, \qquad \textrm{as } r \to \infty~.
\end{align}
Note that for solutions to the Bogomolny equation it is sufficient to give only the Higgs field asymptotics in \eqref{defectbcs} and \eqref{XBasymptotics}, as the asymptotics for the magnetic field follow.

In order to define the moduli space, we first define the group of local gauge transformations.  Consider the action of gauge transformations in the vicinity of an \tHooft defect.  Although two charges $P,P' \in \Lambda_G$ related by a Weyl transformation are physically equivalent, it will be convenient to define the moduli space for a given set of $P_n \in \Lambda_G$, rather than for a given set of Weyl orbits of \tHooft charges.  Thus we require elements in the group of local gauge transformations to leave the $P_n$ invariant.  If $\cg$ is a gauge transformation, let $\cg_n := \cg |_{S_{\varepsilon_n}^2}$ be the restriction to the infinitesimal two-sphere surrounding $\vec{x}_n$.  We define
\begin{equation}\label{localgts}
\GG_{\{P_n\}}^0 := \left\{ \cg : \UU \to G ~|~ \Ad_{\cg_{n}}(P_n) = P_n~, \forall n~, \textrm{ and } \lim_{r \to \infty} \cg = 1_G \right\}~.
\end{equation}
Since the principal $G$-bundle over $\UU$ can be nontrivial,\footnote{It will be nontrivial iff any of the $P_n \in \Lambda_G$ satisfy $P_n \notin \Lambda_{\rm cr}$ --- \ie\ if there is nontrivial \tHooft flux.} we should really speak of a collection of smooth patch-wise transformations $\cg_\alpha : \UU_\alpha \to G$ with $\{ \UU_\alpha \}$ an open cover for $\UU$ and the $\cg_\alpha$ patched together appropriately via the transition functions $\cg_{\alpha\beta}$ of the bundle.  Similar remarks of course apply to the Higgs field and gauge field.  We understand ``$\cg,X,A$'' to denote such collections.  Also, in order to be more precise about \eqref{localgts}, if $\GG_{\{P_n \}}^0 \ni \cg = \exp(\epsilon)$, then we require $\epsilon = \epsilon_{\infty}^{(1)}/r + o(1/r)$ as $r \to \infty$ where $\epsilon_{\infty}^{(1)} \in \mathfrak{t}$, and $\epsilon = \epsilon_n + O(|\vec{x} - \vec{x}_n|^{1/2})$ as $\vec{x} \to \vec{x}_n$, where $\epsilon_n : S_{\varepsilon_n}^2  \to \mathfrak{g}$ satisfies $\cg_n = \exp(\epsilon_n)$ and $[\epsilon_n ,P_n] = 0$.  These conditions are such that gauge transformations in $\GG_{\{P_n\}}^0$ preserve the boundary conditions on the fields.

A gauge transformation $\cg \in \GG_{\{P_n\}}^0$ acts on the fields sending $\{A,X\} \to \{A',X'\}$ with
\begin{equation}\label{finitegt}
A = \Ad(\cg^{-1})(A') + \cg^\ast \theta~, \qquad \Phi = \Ad(\cg^{-1})(\Phi')~,
\end{equation}
where $\theta$ is the Maurer--Cartan form on $G$; for matrix groups, $\cg^\ast \theta = \cg^{-1} \ed\cg$ and $\Ad(\cg)(H) = \cg H \cg^{-1}$.  If $\cg = \exp(\epsilon)$ then these transformations correspond to the infinitesimal action $A \to A' = A - \eD \epsilon$, $\Phi \to \Phi' = \Phi + \ad(\epsilon)(\Phi) = \Phi + [\epsilon,\Phi]$.  The Bogomolny equation transforms covariantly while the defect and asymptotic boundary conditions are invariant.

The moduli space of singular monopoles is the space of gauge equivalence classes of solutions satisfying the required boundary conditions.  We denote it by
\begin{align}\label{Mdef}
& \fMM\left( \{\vec{x}_{n}, P_{n}\}_{n = 1}^{N_t} ; \gamma_{\rm m} ; X_{\infty} \right) := \cr
& ~  \left\{ \{A,X\} ~ \bigg| ~ B_i = D_i X~,\begin{array}{l} X = - \frac{1}{2 |\vec{x} - \vec{x}_n|} P_n + O(|\vec{x}-\vec{x}_n|^{-1/2})~,~ \vec{x} \to \vec{x}_n~, \\  X = X_\infty - \frac{1}{2|\vec{x}|} \gm + o(|\vec{x}|^{-1})~, ~ |\vec{x}| \to \infty \end{array}  \right\} \bigg/ \GG_{\{ P_n \}}^0~. \qquad ~ ~
\end{align}
We will sometimes abbreviate the \tHooft defect data to $L$ and write $\fMM(L;\gm;X_\infty)$.  If there are no \tHooft defects then there are no special points $\vec{x}_n$ and we have only the asymptotic boundary conditions, while $\GG$ simply becomes the space of smooth maps $\cg : \mathbb{R}^3 \to G$ that go to the identity at infinity.  These ``vanilla moduli spaces'' $\MM(\gm;X_\infty)$ have been intensively studied since their introduction over thirty years ago \cite{Weinberg:1979zt,Manton:1981mp,Nahm:1981nb,Hitchin:1982gh,Taubes:1983bm,Donaldson:1985id,Atiyah:1985dv}.  Classic texts are \cite{Jaffe:1980mj,Atiyah:1988jp,Coleman50}; modern reviews with extensive references include \cite{Harvey:1996ur,MR2068924,Tong:2005un,Weinberg:2006rq}.  Singular monopoles and their moduli spaces were first considered by Kronheimer \cite{Kronheimer}, with further important results obtained in \cite{MR1624279,Cherkis:1997aa,Cherkis:1998xca,Cherkis:1998hi,Kapustin:2006pk}.  See the introductory section of \cite{MRVdimP1} for a more detailed account of the previous literature on singular monopoles and their moduli spaces.

% % % % % % % % % % % 
\subsubsection{Dimension and hyperk\"ahler structure}
% % % % % % % % % % % 

The space $\fMM$ is finite dimensional, and furthermore has a natural and compatible Riemannian metric and quaternionic structure, making it a hyperk\"ahler manifold possibly with singular loci.  The derivation of these facts starts with a study of the tangent space $T_{[\hat{A}]} \fMM$ at a point $[\hat{A}] = [\{A,X\}]$.  Tangent vectors $\delta \in T_{[\hat{A}]} \fMM$ are in one-to-one correspondence with \emph{bosonic zero modes}.  These are deformations $\delta \hat{A}$ that 
\begin{itemize}
\item solve the linearized self-duality equation,
\begin{equation}\label{linearsd}
\hat{D}_{[a} \delta \hat{A}_{b]} = \half \epsilon_{ab}^{\phantom{ab}cd} \hat{D}_c \delta \hat{A}_d~,
\end{equation}
where $\hat{D}$ is the covariant derivative with respect to the background solution $\hat{A}$, and
\item are not pure gauge.  To quantify this latter condition we make use of the metric on field configuration space that is naturally defined by the kinetic terms of the energy functional \eqref{Ham1}:\footnote{A more standard normalization of this metric in the physics literature is $g_{\rm phys}(\delta_1,\delta_2) := \frac{4\pi}{g_{0}^2} g(\delta_1,\delta_2)$, in terms of which the collective coordinate Lagrangian to be constructed in \ref{ssec:ccL} has canonically normalized kinetic terms.  The normalization we have given, however, will turn out to be more natural for comparing the semiclassical analysis with the Seiberg--Witten low energy effective one.  Note that the metric $g$ is independent of the Yang--Mills coupling $g_0$ since the equations defining the zero modes are. \label{fn:physmet}}
\begin{equation}\label{metC}
g(\delta_1,\delta_2) := \frac{1}{2\pi} \int_{\UU} \ed^3 x \Tr \left\{ \delta_1 \hat{A}_a \delta_2 \hat{A}^a \right\}~.
\end{equation}
Choosing $\delta_2 = \delta_\epsilon$ to be the tangent vector corresponding to a local gauge transformation generated by $\epsilon(\vec{x}) \in \mathfrak{g}$, $\delta_{\epsilon} \hat{A} = - \hat{\eD} \epsilon$, we find that $g(\delta,\delta_\epsilon) = 0$ if and only if
\begin{equation}\label{gaugeorth}
\hat{D}^a \delta \hat{A}_a = 0~.
\end{equation}
\end{itemize}
An asymptotic analysis of \eqref{linearsd} and \eqref{gaugeorth} using the leading asymptotics of the background fields shows that solutions $\delta \hat{A}_a$ are at worst $O(\varepsilon_{n}^{-1/2})$ as $\varepsilon_n = |\vec{x} - \vec{x}_n| \to 0$ and $O(r^{-2})$ as $r = |\vec{x}| \to \infty$.  Hence the metric \eqref{metC} is well defined and there are no boundary terms in the argument leading to \eqref{gaugeorth}.

Equations \eqref{linearsd} and \eqref{gaugeorth} comprise four constraints on the zero modes that can be usefully packaged into a single Dirac equation.  In \cite{MRVdimP1} we generalized the Callias index theorem \cite{Callias:1977kg} for such Dirac operators on open Euclidean space to the case of $\UU = \mathbb{R}^3 \setminus \{ \vec{x}_n \}$ with background fields satisfying \tHooft defect boundary conditions at the $\vec{x}_n$.  This, together with a vanishing theorem for the adjoint operator, led to a formula for the dimension of $T_{[\hat{A}]}\fMM$ and hence the dimension of $\fMM$ at smooth points:
\begin{equation}\label{dim1}
\dim_{\mathbb{R}} \fMM = \sum_{\alpha \in \Delta} \left( \frac{ \langle \alpha, X_{\infty} \rangle \langle \alpha, \gamma_{\rm m} \rangle }{ | \langle \alpha, X_\infty \rangle |} + \sum_{n}  | \langle \alpha, P_{n} \rangle | \right) ~.
\end{equation}
In this form the dimension formula exhibits manifest invariance under Weyl conjugation of the asymptotic data $\{\gm,X_\infty\}$ and independent Weyl conjugation of each \tHooft charge $P_n$.  Such conjugations amount to a relabeling of the roots for the appropriate term, but all roots are being summed over in each term.  Thus the dimension only depends on the Weyl orbits $[\{\gm,X_\infty\}]$, $[P_n]$.  This is natural since any of these Weyl conjugations can be achieved by a gauge transformation (which is not in $\GG_{\{ P_n\} }$ but acts isometrically on $(\fMM,g)$).

The dimension formula can be cast into a physically intuitive form as follows \cite{MRVdimP1}.  In accordance with our discussion following \eqref{rootdecomp}, let $X_\infty$ define a splitting of the roots into positive and negative, $\Delta = \Delta^+ \cup \Delta^-$, with $\alpha \in \Delta^+$ iff $\langle \alpha, X_\infty \rangle > 0$.  The sum over roots becomes twice the sum over positive roots, after which the first term simplifies to $\langle \alpha,\gm\rangle$.  For each $P_n$ let $P_{n}^- \in [P_n]$ denote the representative in the closure of the anti-fundamental Weyl chamber: $\langle \alpha, P_{n}^- \rangle \leq 0, \forall \alpha \in \Delta^+$.  Now define the \emph{relative magnetic charge}
\begin{equation}\label{relcharge}
\tilde{\gamma}_{\rm m} := \gm - \sum_n P_{n}^- ~.
\end{equation}
This is a generalization of Kronheimer's ``non-Abelian'' charge \cite{Kronheimer} for $\mathfrak{su}(2)$ to arbitrary simple $\mathfrak{g}$.  Then, using the freedom to replace $P_n \to P_{n}^-$ in \eqref{dim1}, one can show
\begin{equation}\label{dim2}
\dim_{\mathbb{R}} \fMM = 4 \langle \varrho, \tilde{\gamma}_{\rm m} \rangle~,
\end{equation}
where $\varrho := \half \sum_{\alpha \in \Delta^+} \alpha$ is the Weyl element.  Notice that $\tilde{\gamma}_{\rm m}$ sits in the co-root lattice and thus we may write $\tilde{\gamma}_{\rm m} = \sum_I \tilde{n}_{\rm m}^I H_I$, where $H_I$ are the simple co-roots and the $\tilde{n}_{\rm m}^I$ are integers.  This finally leads to
\begin{equation}\label{dim3}
\dim_{\mathbb{R}} \fMM = 4 \sum_{I=1}^{r} \tilde{n}_{\rm m}^I ~.
\end{equation}
This gives the formal dimension of $\fMM$ in the sense that we assumed the existence of the background solution $\hat{A}$ that we expanded around.  One needs an independent argument to determine when $\fMM$ is nonempty.  

Following Weinberg's interpretation in the vanilla case \cite{Weinberg:1979zt}, the result \eqref{dim3} suggests we have $\tilde{n}_{\rm m}^I$ smooth (\ie 't Hooft--Polyakov) fundamental monopoles of type $I$ for each $I = 1,\ldots,r$, in the presence of the defects.  Each fundamental monopole has four degrees of freedom associated with it---three for its position and one phase whose conjugate momentum corresponds to electric charge.  A natural conjecture, then, is that $\fMM$ is nonempty iff all $\tilde{n}_{\rm }^I \geq 0$.  This has been proven in the vanilla case \cite{Taubes:1981gw}, (where one should additionally assume at least one of the $n_{\rm m}^I$ is strictly positive).  In \cite{MRVdimP2} we found support for it using realizations of singular monopole configurations via intersecting brane systems for $\mathfrak{g} = \mathfrak{su}(N)$ theories.  

A second assumption we make is that $\fMM$ is connected.  This is also motivated by the physical picture of constituent fundamental monopoles.  In the vanilla case it is proven through the relation to moduli spaces of rational maps \cite{Donaldson:1985id}.  We believe an analogous result will follow from a rational map construction of singular monopoles, as appears for example in \cite{MR2755487}.  See also appendix A of \cite{Bullimore:2015lsa}.

If $\delta \hat{A}_a$ is a solution to \eqref{linearsd}, \eqref{gaugeorth}, then so is $\bar\eta^{r}_{ab} \delta \hat{A}^b$, with 
\begin{equation}\label{asdtHooft}
(\bar{\eta}^{r})_{ab} := \epsilon^{r}_{\phantom{r}ab4}-(\delta_{a}^r\delta_{b4}-\delta_{b}^r\delta_{a4})~, \qquad r = 1,2,3~,
\end{equation}
the anti-self-dual 't Hooft symbols.  This follows from the fact that the background $\hat{F}_{ab}$ is self-dual.  The $\bar{\eta}$ are a basis for the anti-self-dual matrices in four dimensions, but there is nothing particularly special about this choice.  With the advantage of hindsight, it turns out to be useful to rotate this basis by the same $SO(3)$ transformation, $R_\kappa$, that appeared in \eqref{Rotkappa}.\footnote{The reader might object to this as, at the moment, there is no connection between the discussion that led to the introduction of $R_\kappa$ in section \ref{sec:wcf} and the discussion here.  However we will make the connection later when we discuss the role of $SU(2)_R$ symmetry in the collective coordinate theory.  Here we attempt to reduce the proliferation of notation by anticipating that result.}  We define
\begin{equation}\label{R4cs}
(\mathbbm{j}^r)_{ab} := (R_{\kappa})^{r}_{\phantom{r}s} (\bar{\eta}^s)_{ab}~.
\end{equation}
Then the tangent space $T_{[\hat{A}]} \fMM$ at any point $[ \hat{A}]$ is equipped with a natural quaternionic structure; the triplet of endomorphisms
\begin{equation}\label{quatstructure}
(\bbJ^r\delta \hat A)_a := (\mathbbm{j}^r)_{ab} \delta\hat A^b ~,
\end{equation}
satisfy the quaternionic algebra
\begin{equation}\label{quatalgebra}
\bbJ^r \bbJ^s = - \delta^{rs} \mathbbm{1} + \epsilon^{rs}_{\phantom{rs}t} \bbJ^t~,
\end{equation}
where $\mathbbm{1}$ is the identity map on $T_{[\hat{A}]} \fMM$.  This follows from the algebra satisfied by the $\bar{\eta}$: $\bar{\eta}^r \bar{\eta}^s = -\delta^{rs} - \epsilon^{rs}_{\phantom{rs}t} \bar{\eta}^t$, together with the fact that $R_\kappa$, being an element of $SO(3)$, preserves this algebra.  

Away from singular loci one may show that these almost complex structures on $\fMM$ are integrable. Furthermore the restriction of \eqref{metC} to $T_{[\hat{A}]} \fMM$ at each point $[\hat{A}] \in \fMM$ defines a Riemannian metric on $\fMM$ (which we also denote by $g$), and the three complex structures \eqref{quatstructure} are covariantly constant with respect to this metric \cite{Atiyah:1988jp}.  Thus, away from possible singular loci, $(\fMM,g, \bbJ^r)$ is a hyperk\"ahler manifold.  We denote the corresponding triplet of K\"ahler forms by $\sw^r$; they are related to the metric and complex structures via
\begin{equation}\label{comtripple}
\sw^r(U,V) = g(U,\bbJ^r(V))~,
\end{equation}
for all vector fields $U,V$.

One can introduce a local \emph{real} coordinate system $\{ z^m \}$, $m = 1,\ldots, 4\sum_I \tilde{n}_{\rm m}^I$, on $\fMM$.  The coordinates parameterize a smooth family of gauge-inequivalent solutions to \eqref{sd}: $\hat{A} = \hat{A}(\vec{x};z^m)$.  Noticing that $\pd_m \hat{A} \equiv \frac{\pd}{\pd z^m} \hat{A}$ solves the linearized equations, \eqref{linearsd}, \eqref{gaugeorth}, we naturally obtain a set of zero modes,
\begin{equation}\label{coordflows}
\delta_m \hat{A}_a := \pd_m \hat{A}_a - \hat{D}_a \varepsilon_m~,
\end{equation}
associated with the coordinate tangent vectors $\delta_m$.  For each $m$, $\varepsilon_{m}(\vec{x};z)$ is the generator of a local gauge transformation that is uniquely determined by requiring $\delta_m$ to be orthogonal to local gauge transformations: $\hat{D}^a (\pd_m \hat{A}_a - \hat{D}_a \varepsilon_m) = 0$.  We denote the components of the metric with respect to this coordinate basis in the usual way: $g_{mn} := g(\delta_m,\delta_n)$.  Meanwhile \eqref{comtripple} is equivalent to $(\sw^r)_{mn} = g_{mp} (\bbJ^r)_{n}^{\phantom{n}p}$.

The collection $\{\hat{A}_a, \varepsilon_m\}$ are in fact the horizontal components of the connection one-form on the universal bundle $G \to \PP_{\rm uni} \to \mathbb{R}^4 \times \fMM$ of Atiyah and Singer \cite{MR742394}.  Extending the definition of the covariant derivative to the moduli space,
\begin{equation}\label{Dmdef}
D_m := \pd_m + \ad(\varepsilon_m)~,
\end{equation}
the bosonic zero modes may be understood as the mixed components of the curvature, $\ad(\delta_m \hat{A}_a) = [D_m, \hat{D}_a]$.  We denote the moduli space components of the curvature
\begin{equation}\label{modunicurve}
\phi_{mn} := \pd_m \varepsilon_n - \pd_n \varepsilon_m + [\varepsilon_m, \varepsilon_n]~,
\end{equation}
and note the remarkable identity
\begin{equation}\label{unibundle}
\hat{D}^a \hat{D}_a \phi_{mn} = 2[ \delta_m \hat{A}^a , \delta_n \hat{A}_a]~.
\end{equation}
This identity follows from manipulating the Jacobi identify for $(\hat{D}_a,D_m)$.  For example, one has $\hat{D}_a \phi_{mn} = -2 D_{[m} \delta_{n]} \hat{A}_a$.  

We also note the compact expression for the Christoffel symbols with all indices down, $\Gamma_{mnp} = g_{mq} \Gamma^{q}_{\phantom{q}np}$:
\begin{equation}\label{ccChristoffel}
\Gamma_{mnp} = \frac{1}{2\pi} \int_{\UU} \Tr \left\{ \delta_m \hat{A}_a D_n \delta_{p} \hat{A}^a \right\}~,
\end{equation}
which follows directly from \eqref{metC}.  One can check that this formula is consistent with $\Gamma_{mnp} = \Gamma_{m(np)}$.  These results will be useful below.

The vanilla moduli spaces $\MM(\gm;X_\infty)$ are smooth and complete hyperk\"ahler manifolds.  However this is generically not the case for singular monopole moduli spaces, which can have co-dimension four or higher singularities corresponding to the phenomenon of monopole bubbling \cite{Kapustin:2006pk}.  In this process an \tHooft defect can emit or absorb smooth monopoles, changing its charge accordingly, such that the asymptotic magnetic charge $\gm$ is preserved.  In the situation where a defect absorbs some smooth monopoles such that its charge changes $P \to P'$, with $P' \notin [P]$, it is expected that the singular locus of $\fMM(P;\gm;X_\infty)$ can be naturally identified with $\fMM(P';\gm;X_\infty)$.  Note that the smaller space $\fMM(P';\gm;X_\infty)$ may itself possess singular loci, so that in general we can have a nested sequence of singularities.  As for the nature of the singularities, in examples they are always found to be of orbifold type, but it does not appear to be known whether this is true in general.  

We do not consider an absorption or emission process in which $P \to P'$, where $P' \in [P]$ as monopole bubbling.  The two \tHooft defects are physically equivalent and the moduli spaces are isometric.  By making a gauge transformation on the latter configuration we can conjugate $P'$ back to $P$, so that the entire process can be described by motion on a single moduli space.  Such processes were dubbed ``monopole extraction'' in \cite{MRVdimP2}, where we studied them in an intersecting D-brane picture.  The analog of this process from the point of view of the low energy effective $\NN = 2$ theory will be described in section \ref{ssec:tropical}.

In fact gauge transformations that leave the asymptotic data fixed preserve both the metric and quaternionic structure of $\fMM$.  Thus the moduli spaces $\fMM(P;\gm,X_\infty)$ and $\fMM(P' ;\gm,X_\infty)$ for $P' \in [P]$ are related by a hyperk\"ahler isometry.  We use this to identify such $\fMM$'s, so that in this way our semiclassical constructions to follow only depend on the Weyl orbit $[P]$.

% % % % % % % % % % %
\subsubsection{Isometries from symmetries}\label{sec:modisom}
% % % % % % % % % % %

Let us begin by recalling some facts about the vanilla moduli spaces $\MM(\gm;X_\infty)$.  At each point in $\MM$ there is a distinguished four-dimensional subspace of the tangent space.  If $\hat{A}(x^i)$ solves \eqref{sd} then so does $\hat{A}(\vec{x} - \vec{x}_{\rm cm})$ for any fixed $\vec{x}_{\rm cm}$.  These are the center-of-mass position moduli.  The corresponding zero modes are $\delta_i \hat{A}_a = \pd_i \hat{A} - \hat{D}_a \varepsilon_i$, where $\varepsilon_i$ must solve $\hat{D}^a \left( \pd_i \hat{A}_a - \hat{D}_a \varepsilon_i \right) = 0$.  However, upon using \eqref{temporalgauge} and \eqref{timeindependent}, one finds that this equation is identical to the equation of motion for $A_i$ following from \eqref{Scl} if we identify $\varepsilon_i = A_i$.  Therefore $\varepsilon_i = A_i$ is a solution and furthermore it is the unique solution since $\hat{D}^2$ has trivial kernel.  It follows that $\delta_i \hat{A}_a = \hat{F}_{ia}$.  These zero modes give a three-dimensional subspace of the tangent space.  

A fourth zero mode can be obtained from these using the quaternionic structure: $\delta_{4} \hat{A}_a = \hat{F}_{4a}$.  This zero mode, $\delta_4 \hat{A}_a = (-D_i X, 0)$, corresponds to the infinitesimal action of an \emph{asymptotically nontrivial} gauge transformation---in particular one that is not in the group of local gauge transformations $\GG_{\{P_n\}}^0$ that we mod out by to construct the moduli space.  This zero mode together with the $\delta_i \hat{A}_a$ form an invariant subspace under \eqref{quatstructure}.  Together they can be summarized as $\delta_a \hat{A}_b = \hat{F}_{ab}$.

In fact one can go further and show that the vector fields on $\MM$ associated with these zero modes are covariantly constant.  Hence the simply-connected cover, $\widetilde{\MM}$, of $\MM$ factorizes into a direct product of a flat $\mathbb{R}^4$ factor generated by these vector fields, and a ``relative'' or ``centered'' moduli space $\MM_0$ \cite{Atiyah:1988jp}.  We will demonstrate some of these facts below in subsection \ref{sssec:Mfactor}, after introducing a little more technology.

Defects break translational symmetry and hence $\fMM$ cannot have the corresponding isometries.  Global gauge transformations, however, do still generate symmetries of $\fMM$.  

The remaining spacetime symmetry inherited from the Poincar\'e group is the $SO(3)$ of spatial rotations.  Rotations map a solution of the Bogomolny equation to a new, physically inequivalent solution.  The asymptotic boundary conditions are preserved by the rotation.  The line defect boundary conditions however will only be preserved if there is a single line defect and it is located at the fixed point of the rotation.  In this case the action of the spatial rotation on $\hat{A}_a(\vec{x},z)$ can be equivalently represented by a diffeomorphism of $\fMM$.

It will be useful in the following to introduce a bit of notation.  Let $\{ \delta^{E} \}$ represent a collection of infinitesimal symmetry transformations that act on $\hat{A}$.  They comprise spatial rotations, effectively-acting global gauge transformations and, in the vanilla case, spatial translations.  The symmetry transformations satisfy $(\delta^E \delta^F - \delta^F \delta^E) = f^{EF}_{\phantom{EF}G} \delta^G$, where the $f^{EF}_{\phantom{EF}G}$ are the structure constants of the symmetry algebra.  If the action of $\delta^E$ on $\hat{A}$ can equivalently be represented by an infinitesimal diffeomorphism of $\fMM$, then there exist vector fields $\{K^E(z) \}$ such that
\begin{equation}\label{symEonA}
\delta^E \hat{A}_a = (\delta_E z^m) \pd_m \hat{A}_a \equiv -(K^E)^m \pd_m \hat{A}_a~.
\end{equation}
Furthermore the symmetry algebra of transformations on $\hat{A}$ implies that the $K^E$ must satisfy
\begin{equation}\label{Kvecalg}
[K^E, K^F] = f^{EF}_{\phantom{EF}G} K^G~,
\end{equation}
where $[K^E, K^F] := (K^E)^m \pd_m K^F - (K^F)^m \pd_m K^E$ is the usual commutator of vector fields.  The sign in the definition \eqref{symEonA} is necessary in order that the $K^E$ satisfy the same algebra as the transformations $\delta_E$.\footnote{It could be absorbed into the definition of $K^E$ resulting in a sign change on the right-hand side of \eqref{Kvecalg}, but when we consider the symmetries of the collective coordinate theory induced from these it will be better to have this sign exposed from the beginning.}  Note that $\delta^E \hat{A}_a$ is not orthogonal to local gauge transformations but can be made so by performing a compensating gauge transformation,
\begin{align}\label{gaugeorthsymE}
\tilde{\delta}^E \hat{A}_a :=&~ \delta^E \hat{A}_a + \hat{D}_a \left( (K^E)^m \varepsilon_m \right) = -(K^E)^m \delta_m \hat{A}_a~,
\end{align}
with gauge parameter $\epsilon^E := -(K^E)^m \varepsilon_m \in \GG_{ \{P_n\} }^0$.  

We would like the determine how the transformation $\delta^E$ acts on the zero modes $\delta_m \hat{A}_a$ themselves, as defined through \eqref{coordflows}.  Starting from the definition \eqref{coordflows} and using \eqref{symEonA}, \eqref{gaugeorthsymE} one finds
\begin{equation}
\tilde{\delta}^E (\delta_m \hat{A}_a) =  -\left( \pd_m(K^E)^n \right) \delta_n \hat{A}_a - (K^E)^n D_m \delta_n \hat{A}_a - \hat{D}_a (\tilde{\delta}^E \varepsilon_{m})~,
\end{equation}
for the gauge-orthogonal variation.  In order to determine the variation of the local gauge parameter $\varepsilon_m$ one must use its defining property, $\hat{D}^2 \varepsilon_m = \hat{D}^a \pd_m \hat{A}_a$.  Taking the $\delta^E$ variation of both sides of this equation and making use of \eqref{unibundle} we find
\begin{equation}
\tilde{\delta}^E \varepsilon_m = -(K^E)^n \phi_{nm}~.
\end{equation}
Here we use that $\hat{D}^2$ is invertible acting on elements of $\GG_{ \{ P_n \} }^0$.  Then, with $\hat{D}_a \phi_{mn} = -2 D_{[m} \delta_{n]} \hat{A}_a$ we obtain
\begin{equation}
\tilde{\delta}^E (\delta_m \hat{A}_a) =  -\left( \pd_m(K^E)^n \right) \delta_n \hat{A}_a - (K^E)^n D_n \delta_m \hat{A}_a~,
\end{equation}
whence
\begin{equation}\label{Liezm}
\delta^E (\delta_m \hat{A}_a) = -\left( \pd_m(K^E)^n \right) - (K^E)^n \pd_n \delta_{m} \hat{A}_a \equiv  \Lie_{-K^E} (\delta_m \hat{A}_a)~.
\end{equation}
In the last step we noted that this quantity is precisely the Lie derivative along $-K^E$ of $\delta_m \hat{A}_a$, viewed as a co-vector on $\fMM$.

It follows that the variation $\delta^E$ of any moduli space quantity constructed from the bosonic zero modes will be by Lie derivative with respect to $-K^E$.  In particular this holds for both the metric and K\"ahler forms, \eqref{metC} and \eqref{comtripple}, which we collect together as
\begin{equation}\label{metJcomp}
(g_{mn}, (\sw^r)_{mn}) = (\delta_{ab}, (\mathbbm{j}^r)_{ab}) \frac{1}{2\pi} \int_{\UU} \ed^3 x \Tr \left\{ \delta_m \hat{A}^a \delta_n \hat{A}^b \right\}~,
\end{equation}
where, recall, $(\mathbbm{j}^r)_{ab}$ was defined in \eqref{R4cs}.  Both the metric and K\"ahler forms are covariantly constant with respect to the Levi--Civita connection, and so it is convenient to write the Lie derivative in terms of the corresponding covariant derivative operator, $\nabla_m$.  Hence on the one hand we have
\begin{align}
\delta^E g_{mn} =&~ - \Lie_{K^E} g_{mn} = -\nabla_m (K^E)_n - \nabla_n (K^E)_m~, \cr
\delta^E (\sw^r)_{mn} =&~ -\Lie_{K^E} (\sw^r)_{mn} = - \nabla_m (K^{E})^p (\sw^r)_{pn} + \nabla_n (K^E)^p (\sw^r)_{pm} ~.
\end{align}

On the other hand we can compute the symmetry variation of these quantities directly for a given symmetry transformation of $\delta_m \hat{A}_a$, viewing this quantity now as a function of $\vec{x}$.  For example, in the vanilla case, the generators of spatial translations, $\delta^i$, act on the zero modes via $\delta^i(\delta \hat{A}_a) = \delta^{ij} \pd_j (\delta \hat{A}_a)$.  Since the metric and K\"ahler forms are defined in terms of integrals over all of $\mathbb{R}^3$ (again, in the vanilla case), it is clear that they are invariant.  The metric is also invariant with respect to the generators of angular momentum, since it is the spatial integral of a scalar quantity.  Meanwhile both the metric and complex structures are invariant under global gauge transformations since $\Tr$ is a bi-invariant form under the adjoint action.  Let us denote the angular momentum and gauge transformations by $\delta^{r}$, $r = 1,2,3$, and $\delta^A$ respectively, where $A$ runs over a basis of independent, effectively-acting global gauge transformations.  Then we have $K^E = (K^i,K^r, K^A)$, where the $K^i$ are only present in the vanilla case, and we have just argued that
\begin{align}\label{Killvecs}
& \Lie_{K^i} g_{mn} = 0~ \textrm{(vanilla case)}~, \qquad  \Lie_{K^r} g_{mn} = 0~, \qquad \Lie_{K^A} g_{mn} = 0~, \quad \textrm{and} \cr
& \Lie_{K^i} (\sw^r)_{mn} = 0 ~ \textrm{(vanilla case)}~, \qquad  \Lie_{K^A} (\sw^r)_{mn} = 0~.
\end{align}
The first line implies that all of the $K^E$ are Killing vector fields.  The last is the definition of a \emph{triholomorphic} vector field on a hyperk\"ahler manifold.  Hence the $(K^i,K^A)$ are triholomorphic Killing vector fields; they generate isometries which additionally preserve the quaternionic structure.  They will be discussed further in the next section.

The Killing vectors $K^r$ associated with angular momentum are not triholomorphic, but they do act on the $\sw^s$ nicely.  To derive this action we must spell out the spacetime action $\delta^r$:
\begin{equation}\label{angAvar}
\delta^r \hat{A}^a = -\epsilon^{rjk} x_j \pd_k \hat{A}^a + (\ell^r)^{a}_{\phantom{a}b} \hat{A}^b~, \qquad r = 1,2,3~.
\end{equation}
Here $\ell^r$ are $\mathfrak{so}(3)$ representation matrices for the direct sum of the vector and trivial representation; explicitly
\begin{equation}\label{elldef}
(\ell^r)^{j}_{\phantom{j}k} = - \epsilon^{rj}_{\phantom{rj}k}~, \qquad (\ell^r)^{j}_{\phantom{j}4} = (\ell^r)^{4}_{\phantom{4}k} = (\ell^r)^{4}_{\phantom{4}4} = 0~.
\end{equation}
Then, integration by parts together with the identity
\begin{equation}
(\bar{\eta}^s)_{ac} (\ell^r)^{c}_{\phantom{c}b} - (\bar{\eta}^s)_{bc} (\ell^r)^{c}_{\phantom{c}a} =  \epsilon^{sr}_{\phantom{sr}t} (\bar{\eta}^t)_{ab}~,
\end{equation}
leads to the action
\begin{equation}\label{angJaction}
\delta^r (\sw^s)_{mn} = -\Lie_{K^r} (\sw^s)_{mn} = - (R_{\kappa}^{-1})^{r}_{\phantom{r}u} \epsilon^{us}_{\phantom{us}t} (\sw^t)_{mn}~.
\end{equation}
Since $\Lie_{K^r} g_{mn} = 0$ we can also express this result in terms of the complex structures: $\Lie_{K^r} \bbJ^s = (R_{\kappa}^{-1})^{r}_{\phantom{r}u} \epsilon^{us}_{\phantom{us}t} \bbJ^t$.  This will be needed later when we discuss the action of symmetries on the collective coordinate dynamics.

% % % % % % % % % % % %
\subsubsection{Triholomorphic $U(1)$'s and the ${\rm G}$-map}\label{sssec:isoG}
% % % % % % % % % % % % 

As we just saw, the moduli spaces $\fMM$ have a number of triholomorphic isometries, induced from asymptotically nontrivial gauge transformations.  These are related to turning on electric charges and will be important for us in the following.  Asymptotically nontrivial gauge transformations that leave the boundary conditions invariant sit in $\TT_{\{ P_n \} }$, which we define via its Lie algebra,
\begin{align}\label{globalgts}
{\rm Lie}(\TT_{ \{ P_n \} }) :=&~ \left\{ \epsilon : \UU \to \mathfrak{g} ~\bigg|~ \begin{array}{l} \lim_{\vec{x} \to \vec{x}_n} \epsilon = \epsilon_n + O(|\vec{x} - \vec{x}_n|^{1/2})~,~ \textrm{with } [\epsilon_n, P_n] = 0~, \\\lim_{|\vec{x}| \to \infty} \epsilon = \epsilon_{\infty}^{(0)} + \epsilon_{\infty}^{(1)}/|\vec{x}| + o(1/|\vec{x}|)~,~ \textrm{with } \epsilon_{\infty}^{(0,1)} \in \mathfrak{t} \end{array} \right\} ~. \quad
\end{align}
The BPS equation is gauge-covariant, so the gauge transformation of a solution will be another solution.  However, we do not necessarily get a new solution for all $\mathpzc{g} \in \TT_{\{ P_n \} }$.  First, two elements of $\TT_{\{ P_n \} }$ that approach the same asymptotic value differ by an element of $\GG_{\{ P_n \} }^0$, the group of local gauge transformations.  The Lie algebra of $\GG_{\{P_n\}}^0$ is defined as in \eqref{globalgts} but with $\epsilon_{\infty}^{(0)} = 0$.  Since the moduli space $\fMM$ is defined through a quotient by $\GG_{\{ P_n \} }^0$, we get an action of $\TT_{\{ P_n \} }/\GG_{\{ P_n \} }^0 \cong T$, the Cartan torus, on $\fMM$.

This action leaves both the metric and quaternionic structure invariant.  Hence, to each element of $T$ for which the action is nontrivial, we get a nontrivial triholomorphic isometry.  We denote the derivative of this map by
\begin{equation}\label{Gdef}
{\rm G}: \mathfrak{t} \rightarrow \mathfrak{isom}_{\mathbb{H}}(\fMM)~,
\end{equation}  
which is a Lie algebra homomorphism from the Cartan subalgebra of $\mathfrak{g}$ into the Lie algebra of triholomorphic Killing vectors.  Concretely, ${\rm G}$ is constructed as follows.  Given an element $H \in \mathfrak{t}$, we find the unique solution\footnote{Suppose there were two different solutions $\epsilon_{H}^{1,2}$.  Then the difference $\epsilon_{H}^{12} \equiv \epsilon_{H}^1 - \epsilon_{H}^2 \in {\rm Lie}(\GG_{\{P_n\}}^0) \cap \ker{\hat{D}^2}$.  However it is easy to see that this space is trivial and thus $\epsilon_{H}^1 = \epsilon_{H}^2$: integrating by parts $0 = \int \Tr \{ \epsilon_{H}^{12} \hat{D}^2 \epsilon_{H}^{12} \}$ one finds that $\epsilon_{H}^{12}$ must be covariantly constant.  The asymptotic boundary condition then implies it must vanish.} to the boundary value problem $\hat{D}^2 \epsilon_H = 0$, $\lim_{|\vec{x}| \to \infty} \epsilon_H(\vec{x}) = H$, for $\epsilon_H \in {\rm Lie}(\TT_{ \{ P_n \} })$.  Then the tangent vector field $\delta_H = {\rm G}(H)$ corresponds to the zero mode generated by the gauge transformation with respect to $\epsilon_H$: $\delta_H \hat{A} = -\hat{D} \epsilon_H$.  For those $H$ such that this zero mode is nonvanishing, $\delta_H$ is a nontrivial triholomorphic Killing vector.

${\rm G}$ will in general have a nontrivial kernel.  For example, consider a smooth $\mathfrak{g}$-monopole obtained by embedding a single $\mathfrak{su}(2)$ monopole along a simple root $\alpha_I$, such that the only nonzero components of $\{A,X\}$ are those along $E_{\pm \alpha_I}$ and $H_I$.  Thus in order for $\exp(H) \in T$ to act nontrivially it is necessary that $\langle \alpha_I, H \rangle \neq 0$.  In general we are interested in $\mathfrak{t} / \ker{\rm G}$; it is nonzero elements in this space that map to nonzero triholomorphic Killing vectors on $\fMM$.

One could restrict consideration to generic charges such that all components of the relative magnetic charge are nonzero.  In this case the kernel of ${\rm G}$ vanishes.  However we feel it is worthwhile to give a complete description that includes non-generic cases.  In particular, such a construction shows one how to embed the results of detailed analyses performed in cases of low rank gauge groups into higher rank gauge groups.  In contrast to the vanilla case, the embedding procedure is not entirely obvious when defects are present.  The reason is that, although we consider a non-generic relative magnetic charge from the perspective of the higher rank gauge group, the asymptotic magnetic charge $\gm$ can be generic.  (This is achieved by adjusting the sum of 't Hooft charges accordingly so that the relative magnetic charge remains non-generic.)  We have relegated the details of the construction to appendix \ref{app:embed} since they are somewhat tangential to the main development of the paper, and in the following we just summarize the essential results.

Let $\tilde{\gamma}_{\rm m} = \sum_I \tilde{n}_{\rm m}^I H_I$ denote the expansion of the relative magnetic charge along the basis of simple co-roots.  We partition the labels of the simple co-roots according to whether the corresponding $\tilde{n}_{\rm m}^I$ is zero or not:
\begin{equation}\label{corootpart}
\{ I_A ~|~ \tilde{n}_{\rm m}^{I_A} > 0~,~ A = 1,\ldots,d \} ~\cup~ \{ I_M ~|~ \tilde{n}_{\rm m}^{I_M} = 0~,~ M = 1,\ldots,r-d \}~,
\end{equation}
Recall that $r = \rnk{\mathfrak{g}}$ and we assume $0 < d \leq r$, as $d=0$ implies $\dim{\fMM} = 0$.  Let $\mathfrak{D}$ denote the Dynkin diagram of our simple $\mathfrak{g}$, and define $\mathfrak{D}^{\rm ef}$ to be the diagram obtained by deleting those nodes corresponding to the $I_M$ and any lines attached to them.  $\mathfrak{D}^{\rm ef}$ will be the Dynkin diagram of a semisimple Lie algebra that we denote $\mathfrak{g}^{\rm ef}$.  Let $\{H_A\}$ be a basis of simple co-roots of the Cartan subalgebra $\mathfrak{t}^{\rm ef}$, and let $i_\ast : \mathfrak{g}^{\rm ef} \hookrightarrow \mathfrak{g}$ be the natural embedding such that $i_{\ast}(H_A) = H_{I_A}$. Note that this implies $d=\rnk{\mathfrak{g}^{\rm ef}}$\label{defd}.  In the appendix we use this embedding to construct an embedding of singular monopole moduli spaces which is dimension preserving.  Thus it is likely a (hyperk\"ahler) isomorphism of moduli spaces, but we cannot rule out the possibility of a discrete cover.  Using this embedding we then argue that the map
\begin{equation}\label{iastdef}
{\rm G} \circ i_{\ast} : \mathfrak{t}^{\rm ef} \to \mathfrak{isom}_{\mathbb{H}}(\fMM)
\end{equation}
is an \emph{injective} Lie algebra homomorphism.

We would like to exponentiate this to a Lie group homomorphism that gives an effective torus action of triholomorphic isometries on $\fMM$.  ${\rm G}$ acts by gauge transformations which act through the adjoint representation, and this representation is only faithful for the adjoint form of the group.  We should therefore use the exponential map associated with the adjoint form of the effective gauge group, $G^{\rm ef}_{\rm ad}$.  This is the semisimple Lie group with Lie algebra $\mathfrak{g}^{\rm ef}$ and trivial center.  The corresponding exponential of $\mathfrak{t}^{\rm ef}$ gives a Cartan torus $T_{\rm ad}^{\rm ef}$\label{Tefad} that acts effectively on $\fMM$ via triholomorphic isometries.

The fundamental magnetic weights $h^{A} \in \mathfrak{t}^{\rm ef}$ form an integral basis for the co-character lattice of $G_{\rm ad}^{\rm ef}$; so that $\exp(2\pi h^A)$ is the identity.  The difference $h^{I_A} - i_\ast(h^A)$ is nonzero but in the kernel of ${\rm G}$.  Hence the vector fields
\begin{equation}\label{KAdef}
K^A := ({\rm G} \circ i_\ast)(h^{A}) = {\rm G}(h^{I_A})~, \qquad A = 1,\ldots, d~,
\end{equation}
are triholomorphic Killing fields that generate $2\pi$-periodic triholomorphic isometries of $\fMM$.  If $h^{I_M}$ are the remaining fundamental magnetic weights of $\mathfrak{g}$ then ${\rm G}(h^{I_M}) = 0$.  Hence the action of ${\rm G}$ on a generic element of $\mathfrak{t}$ can be expressed in terms of the $K^A$ via linearity.  Since the fundamental magnetic weights are integral-dual to the simple roots we have, for example, $X_\infty = \sum_A \langle \alpha_{I_A}, X_\infty\rangle h^{I_A} + \sum_M \langle \alpha_{I_M}, X_\infty \rangle h^{I_M}$, and this implies
\begin{equation}\label{GXKexpand}
{\rm G}(X_\infty) = \sum_{A=1}^d \langle \alpha_{I_A}, X_\infty \rangle K^A~,
\end{equation}
The construction in \ref{app:embed} also makes it clear that the metric and quaternionic structure of $\fMM$ only depend on the $\langle \alpha_{I_A}, X_\infty \rangle$ and not the $\langle \alpha_{I_M}, X_\infty \rangle$.

Note that, as long as we are not in the case of the pure 't Hooft defect where $\dim{\fMM} = 0$ and there are no $\alpha_{I_A}$, ${\rm G}(X_\infty)$ is nontrivial since $\langle \alpha,X_\infty\rangle \neq 0$ for all nonzero roots $\alpha$.  In the vanilla case the global gauge transformation corresponding to ${\rm G}(X_\infty)$ is the one generated by $X$ itself: $\epsilon_{X_\infty} = X$.  This zero mode and the triplet corresponding to translations generate the flat $\mathbb{R}^4$ factor of the simply-connected cover, $\widetilde{\MM}$, that we mentioned above.  We are now in a better position to establish this factorization property of the vanilla moduli spaces.

% % % % % % % % % % % % % % %
\subsubsection{Periodic isometries and factorization of the vanilla moduli space}\label{sssec:Mfactor}
% % % % % % % % % % % % % % %

We wish to show that ${\rm G}(X_\infty)$ is covariantly constant in the vanilla case.  We proceed by first establishing two useful results that hold generally in the singular case as well.  (See \cite{Atiyah:1988jp} for a different argument.)    Let $\epsilon_H \in {\rm Lie}(\TT_{\{P_n\}})$ be a global gauge transformation with ${\rm G}(H)$ the corresponding Killing vector; \ie\ $\hat{D}_a \epsilon_H = - {\rm G}(H)^m \delta_m \hat{A}_a$.  Then with $\UU = \mathbb{R}^3 \setminus \{ \vec{x}_n \}$ we have
\begin{align}\label{covKilling}
& \frac{1}{2\pi} \int_{\UU} \ed^3 x \Tr \left\{ [ \delta_m \hat{A}^a, \delta_n \hat{A}_a] \epsilon_H \right\} =  \frac{1}{2\pi} \int_{\UU} \ed^3 x \Tr \left\{ \delta_{m} \hat{A}^a \, [ \delta_{n} \hat{A}_a, \epsilon_H ]\right\} \cr
& \qquad \qquad = \frac{1}{2\pi}  \int_{\UU} \ed^3 x \Tr \left\{ \delta_{m} \hat{A}^a \left( D_{n} \hat{D}_a \epsilon_H - \hat{D}_{a} D_{n} \epsilon_H \right) \right\} \cr
& \qquad \qquad = - \frac{1}{2\pi}  \int_{\UU} \ed^3 x \Tr \left\{ \delta_m \hat{A}^a \left[ (\pd_n {\rm G}(H)^p) \delta_p \hat{A}_a + {\rm G}(H)^p D_n \delta_p \hat{A}_a \right] \right\} \cr
& \qquad \qquad = - g_{mq} \left( \pd_n {\rm G}(H)^q + \Gamma^{q}_{\phantom{q}np} {\rm G}(H)^p \right) = - \nabla_n {\rm G}(H)_m ~.
\end{align}
In going from the first to second line we used the interpretation of the zero modes as the mixed components of the curvature of the connection on the universal bundle: $\ad(\delta_n \hat{A}_a) = [D_n, \hat{D}_a]$.  In going from the second to third line we noted that the $\hat{D}_a D_n \epsilon_H$ term vanishes.  This follows via integration by parts, \eqref{gaugeorth}, and noting that the asymptotics of the zero modes, given below \eqref{gaugeorth}, and of $\epsilon_H$ as $|\vec{x}| \to \infty$ and $\vec{x} \to \vec{x}_n$ are such that the boundary terms vanish.  Observe that antisymmetry of the initial expression implies that ${\rm G}(H)$ is Killing.

The second result is
\begin{align}\label{intid1}
& \int_{\UU} \ed^3 x \Tr \left\{ X [\delta_m \hat{A}_a, \delta_n \hat{A}^a] \right\}  = - \int_{\UU} \ed^3 x \Tr \left\{ [ \delta_{[m} \hat{A}^{a}, X] \delta_{n]} \hat{A}_a \right\} \cr
& \qquad = - \int_{\UU} \ed^3 x \Tr \left\{ (D_{[m} \hat{D}^{a} X) \delta_{n]} \hat{A}_a \right\} + \int_{\UU} \ed^3 x \Tr \left\{ (\hat{D}_a D_{[m} X) \delta_{n]} \hat{A}^a \right\} \cr
& \qquad = -\half \epsilon^{ijk} \int_{\UU} \ed^3 x \Tr \left\{ (D_{[m} F_{|jk|} ) \delta_{n]} \hat{A}_i \right\}  = \epsilon^{ijk} \int_{\UU} \ed^3 x \Tr \left\{ (D_j \delta_{[m} \hat{A}_{|k|} ) \delta_{n]} \hat{A}_i \right\} \cr
& \qquad = \half \epsilon^{ijk} \int_{\UU} \ed^3 x \, \pd_i \Tr \left\{ \delta_{m} \hat{A}_j \delta_n \hat{A}_k \right\}  =  0~.
\end{align}
In going from the first to second line we used $\ad(\delta_m \hat{A}_a) = [D_m ,\hat{D}_a]$.  The second term of the second line vanishes upon integrating by parts, using \eqref{gaugeorth} and the zero mode asymptotics.  Next we used the Bogomolny equation to replace $D_i X$ with the fieldstrength, and then again used a Jacobi identity for the curvature of the connection on the universal bundle: $D_m F_{ij} = -2 D_{[i} \delta_{|m|} A_{j]}$.  Due to the antisymmetry on $mn$ the resulting term is a total derivative, and the zero mode asymptotics ensure that the boundary terms vanish.

Equations \eqref{covKilling} and \eqref{intid1} hold in both the singular and vanilla cases.  In the vanilla case---and only in that case---we have $X = \epsilon_{X_\infty}$.  (In the singular case $X \notin {\rm Lie}(\TT_{ \{P_n\} })$ due to the 't Hooft poles.)  Therefore by combining these two results we deduce
\begin{equation}\label{GXcovcon}
\nabla_m {\rm G}(X_\infty)_n = 0~, \qquad (\textrm{vanilla case only})~.
\end{equation}
Henceforth in this subsection we focus on the vanilla case exclusively.

Recall that ${\rm G}(X_\infty)$ spans one direction of a four-dimensional subspace of the tangent space that is closed with respect to the action of the quaternionic structure.  This is the subspace corresponding to the four bosonic zero modes $\delta_a \hat{A}_b = \hat{F}_{ab}$.  Indeed one can compute from the definitions, \eqref{quatstructure}, \eqref{metJcomp}, that
\begin{equation}
(\bbJ^r)_{a}^{\phantom{a}m} = \left\{ \begin{array}{l l} - (\mathbbm{j}^r)_{a}^{\phantom{a}b} ~,~~ & m = b~, \cr
0~, & \textrm{otherwise}~. \end{array} \right.
\end{equation}
Then if $K^i$ are the Killing vectors generating translations, $(K^i)^m \delta_{m} \hat{A}_a = (K^i)^j \delta_j \hat{A}_a = \delta^{ij} \hat{F}_{ja}$, we find 
\begin{equation}
\bbJ^r \left( {\rm G}(X_\infty) \right) = - (R_\kappa)^{r}_{\phantom{r}i} K^i~.
\end{equation}
Hence \eqref{GXcovcon} implies that $\nabla_m (K^i)_n =0$ as well.  (Recall that $R_\kappa$ is just a given fixed element of $SO(3)$.)

It then follows from the de Rham decomposition theorem that the universal cover of the moduli space decomposes into a direct product,
\begin{equation}\label{ucoverfactor}
\widetilde{\MM}(\gm;X_\infty) = \mathbb{R}_{\rm cm}^3 \times \mathbb{R}_{X_\infty} \times \MM_0(\gm;X_\infty)~.
\end{equation}
Here the simply-connected hyperk\"ahler manifold $\MM_0(\gm;X_\infty)$ is (the analog for general simple Lie group of) the ``strongly centered'' monopole moduli space identified in \cite{Hitchin:1995qw}.\footnote{Two warnings concerning notation:  First, references \cite{Atiyah:1988jp,Hitchin:1995qw} use the notation $\widetilde{\MM}_0$ for the strongly centered moduli space because a different space---the ``centered'' monopole moduli space---was defined and denoted $\MM_0$ in \cite{Atiyah:1988jp}.  The relation for $SU(2)$ monopoles of charge $k$ is that the strongly centered moduli space is a $k$-fold covering of the centered one.  For higher rank gauge groups and generic Higgs vevs, there is no analogous relationship and it is the generalization of the strongly centered moduli space that is more useful.  Following most of the physics literature, \eg\ the review \cite{Weinberg:2006rq} , we denote the strongly centered moduli space $\MM_0$.  Second, in the same $SU(2)$ context, \cite{Atiyah:1988jp} uses the notation $\widetilde{\MM}$ to denote a $k$-fold covering of the full moduli space $\MM$ that is not simply-connected.  Our $\widetilde{\MM}$ in \eqref{ucoverfactor} is the universal cover of $\MM$.}  The vanilla monopole moduli space is a quotient of $\widetilde{\MM}$ by a discrete normal subgroup, $\mathbb{D}$, of the isometry group of $\widetilde{\MM}$ and and generally takes the form
\begin{equation}\label{Mfactor}
\MM(\gm;X_\infty) = \mathbb{R}_{\rm cm}^3 \times \frac{\mathbb{R}_{X_\infty} \times \MM_0(\gm;X_\infty)}{\mathbb{D}}~.
\end{equation}
Here $\mathbb{D} \cong \pi_1(\MM)$ is the group of deck transformations of the universal cover.  On the one hand, it is known from the rational map construction \cite{Donaldson:1985id,MR804459,MR1625475} that $\pi_1(\MM) \cong \mathbb{Z}$.  On the other hand, we have an effective $T_{\rm ad}^{\rm ef} \cong U(1)^d$ action on $\MM$ by triholomorphic isometries, induced from asymptotically nontrivial gauge transformations.  This group action induces a homomorphism $\mu : \pi_1(T_{\rm ad}^{\rm ef}) \cong \Lambda_{\rm mw}^{\rm ef} \to \pi_1(\MM) \cong \mathbb{Z}$\label{mudef} from the magnetic weight lattice of the effective Lie algebra to the fundamental group of $\MM$.  Is this map onto?  If not, what is the image, $\im(\mu)$, as a subgroup of $\mathbb{Z}$?  The rational map construction provides a convenient language for answering these questions, and the answers will be important for us later.

The following decomposition of $\mathbb{D}$, into transformations that can be generated from the action of  gauge transformations and those that cannot, seems not to have been spelled out in the literature for generic Higgs vev and magnetic charge.  We claim that the group homomorphism $\mu : \Lambda_{\rm mw}^{\rm ef} \to \mathbb{Z}$ is given by pairing a magnetic weight $h \in \Lambda_{\rm mw}^{\rm ef}$ with the dual of the effective magnetic charge $\gamma_{\rm m}^{\rm ef}:=\sum_{A = 1}^d n_{\rm m}^{I_A} H_{A}$. This is equivalent to the contraction of $i_\ast(h)$ and the magnetic charge with respect to the Killing form $(~,~)$:
\begin{equation}\label{muclaim}
\mu(h) = \langle (\gamma_{\rm m}^{\rm ef})^\ast , h \rangle = (\gamma_{\rm m}^{\rm ef}, h) = (\gm, i_\ast(h))~.
\end{equation}
A proof of this result using the rational map construction is given in appendix \ref{app:Dquotient}.  We will see momentarily that it is consistent with known results in the case of $\mathfrak{g}^{\rm ef} = \mathfrak{su}(2)$.  

In order to clarify what $\im(\mu)$ is, let us expand $(\gamma_{\rm m}^{\rm ef})^\ast$ in the basis of simple roots of $\mathfrak{g}^{\rm ef}$:
\begin{equation}\label{dualgm}
(\gamma_{\rm m}^{\rm ef})^\ast = \sum_{A = 1}^d n_{\rm m}^{I_A} H_{A}^\ast = \sum_{A=1}^d n_{\rm m}^{I_A} \sp^{A} \alpha_{A} \equiv \sum_{A = 1}^d \ell^A \alpha_{A}~,
\end{equation}
where $\sp^A := 2/\alpha_{A}^2 = 2/\alpha_{I_A}^2  \in \{1,2,3\}$ and for convenience we have defined the positive integers $\ell^A := \sp^A n_{\rm m}^{I_A}$.  The $\ell^A$ are the components of $(\gamma_{\rm m}^{\rm ef})^\ast$ along $\alpha_A$ or equivalently the components of $\gm^\ast$ along $\alpha_{I_A}$.  Now, $\im(\mu)$ will be generated by $\mu(h^B)$, where $h^B$ are the fundamental magnetic weights of $\mathfrak{g}^{\rm ef}$, and we have from \eqref{dualgm} that $\mu(h^B) = \ell^B$.  It follows that
\begin{equation}
\im(\mu) = \gcd(\ell^1,\ldots,\ell^d)  \mathbb{Z} \equiv L \mathbb{Z}~,
\end{equation}
where we've denoted by $L$ the greatest common divisor of the $\ell^A$.  Let the corresponding subgroup of the group of deck transformations be 
\begin{equation}
\mathbb{D}_{\gi} \subseteq \mathbb{D}~, \qquad  \textrm{such that} \quad \mathbb{D}/\mathbb{D}_{\gi} \cong \mathbb{Z}/L\mathbb{Z} \equiv \mathbb{Z}_L~,  
\end{equation}
the cyclic group of order $L$.  Here the subscript ``g'' stands for ``gauge-induced deck transformations.''  Then we can write \eqref{Mfactor} as
\begin{equation}\label{Dhkfactor}
\MM(\gm;X_\infty) = \mathbb{R}_{\rm cm}^3 \times \left(\frac{\mathbb{R}_{X_\infty} \times \MM_0(\gm;X_\infty)}{\mathbb{D}_{\gi}}\right) \bigg/ \mathbb{Z}_L~.
\end{equation}
This refinement is useful because we can give a fairly explicit description of gauge-induced isometries and their action on the $\mathbb{R}_{X_\infty} \times \MM_0$ factor of the universal cover via the $\rG$-map, which will furthermore be important later for understanding the decomposition of electric charge into ``center of mass'' and ``relative'' pieces.

For any $h \in \Lambda_{\rm mw}^{\rm ef}$, the isometry $\exp(2\pi \rG(i_\ast(h))) : \MM \to \MM$ is the trivial isometry of $\MM$ since it corresponds to a gauge transformation that asymptotes to the identity and hence leaves all points of $\MM$ fixed.  This isometry lifts to a trivial isometry of the universal cover $\widetilde{\MM}$ iff $h \in \ker(\mu)$.  Since $\ker(\mu)$ is a rank $d-1$ sublattice of $\Lambda_{\rm mw}^{\rm ef}$, this leads to a $(d-1)$-dimensional torus of effectively acting gauge-induced isometries of $\MM_0$, as we now explain.  First, recall that translations along $\mathbb{R}_{X_\infty}$ are generated by $\rG(X_\infty)$.  Second, observe that for any $H \in \mathfrak{t}$,
\begin{align}\label{gGXGH}
g({\rm G}(X_\infty),{\rm G}(H)) =&~ \frac{1}{2\pi} \int_{\mathbb{R}^3} \ed^3 x \Tr \left\{ \hat{D}_a X \hat{D}^a \epsilon_H \right\} =  \frac{1}{2\pi} \int_{S_{\infty}^2} \ed^2 S^i \Tr \left\{ (D_i X) H \right\} \cr
=&~ (\gm, H)~,
\end{align}
Hence the vector field $\rG(H)$ is \emph{metric-orthogonal} to $\rG(X_\infty)$ iff $H$ is \emph{Killing-orthogonal} to $\gm$.\footnote{In particular the G-map is \emph{not} metric preserving.  This explains how it is possible to reconcile the following two statements that naively sound contradictory:  (1) For $d > 1$, a  generic vev $X_\infty$ generates an irrational direction in the Cartan torus and hence $\rG(X_\infty)$ generates an irrational direction in the torus of triholomorphic isometries of $\MM$; nevertheless, (2) there exists a subtorus of triholomorphic isometries generated by Killing vectors that are metric-orthogonal to $\rG(X_\infty)$.}  Any $h \in \ker{\mu}$ satisfies $g(\rG(X_\infty), \rG(i_\ast(h))) = 0$ and therefore maps to a triholomorphic Killing field $\rG(i_\ast(h))$ with legs along $\MM_0$ only; in other words it generates triholomorphic gauge-induced isometries of $\MM_0$.  Furthermore $\exp(2\pi \rG(i_\ast(h)))$ is the trivial isometry.

Let us denote a basis of generators for the sublattice $\ker(\mu) \subset \Lambda_{\rm mw}^{\rm ef}$ by $\{h_{0}^A\}_{A=1}^{d-1}$, such that
\begin{equation}
\ker(\mu) = \bigoplus_{A=1}^{d-1} h_{0}^A \cdot \mathbb{Z}~.
\end{equation}
Then the $d-1$ triholomorphic Killing vectors
\begin{equation}\label{K0Adef}
K_{0}^A := \rG(i_\ast(h_{0}^A))
\end{equation}
restrict trivially to triholomorphic Killing vectors on $\MM_0$, where they generate $2\pi$-periodic  isometries.  Since $\MM_0$ is simply-connected, the closed curves generated by these Killing vectors must be homotopically trivial.  Via the Smith normal form procedure, for example, one can exhibit an explicit $GL(d,\mathbb{Z})$ change of basis transformation,
\begin{equation}
\{h^A\}_{A=1}^{d} \mapsto \{h_{0}^A\}_{A=1}^{d-1} \cup \{ h_\gi \}~,
\end{equation}
mapping the fundamental magnetic weights to the generators $\{ h_{0}^{I_A} \}$ of the kernel together with an element we denote $h_\gi \in \Lambda_{\rm mw}^{\rm ef}$ such that $\mu(h_\gi)$ generates the image of $\mu$.  In particular,
\begin{equation}\label{phiiso}
\phi_\gi := \exp(2\pi \rG(i_\ast(h_\gi))) : \widetilde{\MM} \to \widetilde{\MM}~,
\end{equation}
can be taken as the generator of the subgroup of gauge-induced deck transformations, $\mathbb{D}_\gi$.  Of course $\phi_\gi = \exp(2\pi \rG(i_\ast(h_\gi + h')))$ for any $h' \in \ker(\mu)$.  The closed curves generated by $\rG(i_\ast(h_\gi))$ and $\rG(i_\ast(h_\gi + h'))$ are homotopically equivalent and correspond to the same element of $\mathbb{D}_{\gi}$.

It is of interest to determine precisely how $\phi_\gi$ acts on the $\mathbb{R}_{X_\infty} \times \MM_0$ factor of $\widetilde{\MM}$, which carries the same metric as $\MM$, \eqref{metC}.  To that end we introduce a slightly different basis for $\mathfrak{t}^{\rm ef}$ that is adapted to this factorization.  Namely, we take the $\{h_{0}^A\}_{A=1}^{d-1}$ as before, but we choose the remaining element to be proportional to $X_{\infty}^{\rm ef}$, so that under the G-map it corresponds to a generator of translations  along $\mathbb{R}_{X_\infty}$.  There is a natural choice for the normalization of this element suggested by the asymptotic analysis of the moduli space metric, namely
\begin{equation}\label{hcmdef}
h_{\rm cm} := \frac{1}{(\gm,X_\infty)} X_{\infty}^{\rm ef} = \frac{1}{(\gm,X_\infty)} \sum_{A=1}^d \langle \alpha_{I_A}, X_\infty \rangle h^{A}~.
\end{equation}
If we define a global coordinate $\chi$ on $\mathbb{R}_{X_\infty}$ by identifying the corresponding coordinate vector field with
\begin{equation}\label{pdchi}
\pd_\chi := \rG(i_\ast(h_{\rm cm}))~,
\end{equation}
then $\chi$ can be identified with the sum total of the phases of the constituent monopoles in the asymptotic region of moduli space \cite{Lee:1996kz}.  If we also introduce global coordinates $\vec{x}_{\rm cm}$ on the $\mathbb{R}_{\rm cm}^3$ by identifying $K^i = \delta^{ij} \pd_{x_{\rm cm}^j}$, then the metric on $\MM$ can be written
\begin{equation}\label{productmetric}
\ed s_{\MM}^2 = (\gm, X_\infty) \left( \ed \vec{x}_{\rm cm} \cdot \ed \vec{x}_{\rm cm} + \frac{\ed \chi^2}{(\gm,X_\infty)^2} \right) + \ed s_{\MM_0}^2~.
\end{equation}

In order to determine the action of $\phi_\gi$ on $\widetilde{\MM}$, we would like to decompose $h_\gi$ with respect to the basis $\{h_{0}^{A}; h_{\rm cm} \}$ of $\mathfrak{t}^{\rm ef}$.  By linearity of the G-map, this equivalently tells us how $\rG(i_\ast(h_\gi))$ decomposes along $\pd_\chi$ and the $K_{0}^A$.  In particular, the former determines the translation of $\chi$ while the latter describes a triholomorphic isometry of $\MM_0$ in terms of the basic generators \eqref{K0Adef}.  Now, in order to decompose $h_\gi$ with respect to the basis $\{h_{0}^A; h_{\rm cm}\}$ we require the integral dual basis of $(\mathfrak{t}^{\rm ef})^\ast$.  However the dual of $h_{\rm cm}$ is none other than the (dual of the) effective magnetic charge, since
\begin{equation}
\langle (\gamma_{\rm m}^{\rm ef})^\ast, h_{\rm cm} \rangle = 1~, \qquad \langle (\gamma_{\rm m}^{\rm ef})^\ast, h_{0}^A \rangle = 0~, ~\forall A~.
\end{equation}
Let $\beta_A$ be the remaining components of the dual basis, defined by the properties
\begin{equation}
\langle \beta_A, h_{0}^{B} \rangle = \delta_{A}^{\phantom{A}B}~, \qquad \langle \beta_A, h_{\rm cm} \rangle = 0~.
\end{equation}
An explicit expression for these in terms of the simple roots $\alpha_A$ can be obtained by making use of the inverse of the $GL(d,\mathbb{Z})$ transformation used to obtain the $h_{0}^A$.  Then we have that
\begin{equation}
h_\gi = \langle (\gamma_{\rm m}^{\rm ef})^\ast, h_\gi \rangle h_{\rm cm} + \sum_{A=1}^{d-1} \langle \beta_A, h_\gi \rangle h_{0}^A~,
\end{equation}
and observe that $\langle (\gamma_{\rm m}^{\rm ef})^\ast, h_\gi \rangle = \mu(h_\gi) = L$.  

If follows that the action of $\phi_\gi$ on $\widetilde{\MM}$, is
\begin{equation}\label{hkdeckgen}
\phi_\gi(\widetilde{\MM}) = \mathbb{R}_{\rm cm}^3 \times \left( \exp(2\pi L \pd_\chi) \cdot \mathbb{R}_{X_\infty} \right) \times \phi_{\gi,0}(\MM_0)~,
\end{equation}
where
\begin{equation}\label{phihk0}
\phi_{\gi,0} := \exp\left(2\pi \sum_{A=1}^{d-1} \langle \beta_A, h_\gi \rangle K_{0}^A \right) : \MM_0 \to \MM_0
\end{equation}
is a triholomorphic gauge-induced isometry of $\MM_0$ (assuming $d>1$).  This action defines the quotient of $\mathbb{R}_{X_\infty} \times \MM_0$ by $\mathbb{D}_\gi$ appearing in \eqref{Dhkfactor}.  In particular $\phi_\gi$ translates $\chi$ by an amount $2\pi L$.  

In appendix \ref{app:Dquotient2} we describe one particular  construction of a basis $\{h_{0}^A\}$ and dual basis $\{\beta_A\}$ for $\ker(\mu)$, and we compute the coefficients $\langle \beta_A, h_\gi \rangle$ appearing in \eqref{phihk0} for this choice.  These are only meaningful up to $GL(d-1,\mathbb{Z})$ transformations that change the basis of the $K_{0}^A$ but preserve the essential property that they generate $2\pi$-periodic isometries.  The main observation from the particular expressions we obtain in the appendix, which is unaffected by such transformations, is that the coefficients depend on ratios of the quantities $\langle \alpha_{I_A},X_\infty \rangle$ appearing in \eqref{hcmdef} and, for generic Higgs vev, they are irrational numbers.  Hence generically, as pointed out in \cite{Lee:1996kz}, \emph{no power of $\phi_{\gi,0}$ will give the trivial isometry of $\MM_0$.}  This means that generically, for $d> 1$, the quotient \eqref{Dhkfactor}, and hence \eqref{Mfactor}, is not presentable as a quotient of $\mathbbm{R}_{\rm cm}^3 \times S^1 \times \MM_0$ by a finite cyclic group.

We note however in the case $d =1$, corresponding to $\mathfrak{g}^{\rm ef} = \mathfrak{su}(2)$, that $\rG(X_\infty)$ is the only triholomorphic Killing vector on $\MM$ up to rescaling.  $h_{\rm cm}$ does generate a closed circle and there are no $h_{0}^A$.  The same formulae above apply, but with $\phi_{\gi,0}$ the trivial isometry.  Hence in this case $\mathbb{D}_\gi$ does act only on $\mathbb{R}_{X_\infty}$, where it acts by translating $\chi \to \chi + 2\pi L$ with $L = \sp n_{\rm m}$.  Here $n_{\rm m}$ is the usual $\mathfrak{su}(2)$ magnetic monopole charge while $\sp \in \{1,2,3\}$ accounts for the possibility of embedding the $\mathfrak{su}(2)$ monopole along a short root of $\mathfrak{g}$ with length-squared one-half or one-third that of the long root.  The quotient by $\mathbb{D}_\gi$ thus produces a circle parameterized by $\chi \sim \chi + 2\pi L$, and from \eqref{Dhkfactor} we have
\begin{equation}\label{su2Mfactor}
\MM = \mathbb{R}_{\rm cm}^3 \times \frac{S^1 \times \MM_0}{\mathbb{Z}_L}~, \qquad (\mathfrak{g}^{\rm ef} = \mathfrak{su}(2) ~ \textrm{only}).
\end{equation}
This is in agreement with known results \cite{Atiyah:1988jp}.  Furthermore in the $\mathfrak{su}(2)$ case our definition of the coordinate $\chi$ agrees with the phase of the resultant of the rational map corresponding to the monopole \cite{Atiyah:1988jp,Hitchin:1995qw}.

Finally, there is still the question of how the remaining quotient group $\mathbb{D}/\mathbb{D}_\gi \cong \mathbb{Z}_L$ of (non-gauge-induced) isometries acts on $\widetilde{\MM}$.  Let $\phi : \widetilde{\MM} \to \widetilde{\MM}$ denote the generator of $\mathbb{D}$.  Then we must have that
\begin{equation}\label{phiL}
\phi^L = \phi_\gi~.
\end{equation}
(One might have thought that $\phi^L \sim \phi_{\gi}$ up to homotopy only.  However, both $\phi^L$ and $\phi_{\gi}$ have to be actual isometries---$\phi_{\gi}$ by construction and $\phi^L$ because $\phi$ is.  Therefore they would differ by an isometry that generates a homotopically trivial loop.  Such an isometry would be generated by $2\pi \rG(i_\ast(h))$ for some $h \in \ker{\mu}$, but we know that such isometries are the trivial isometry on $\widetilde{\MM}$.)  Furthermore, since the action of $\phi_\gi$, \eqref{hkdeckgen}, factorizes into a uniform translation of $\mathbb{R}_{X_\infty}$ and an isometry of $\MM_0$, it is clear that $\phi$ must do the same in order for   \eqref{phiL} to hold.  Hence there exists an isometry $\phi_0$ of $\MM_0$ such that
\begin{equation}\label{phi0L}
\phi_{0}^L = \phi_{\gi,0}~,
\end{equation}
and $\phi$ acts by
\begin{equation}\label{deckgen}
\phi(\widetilde{\MM}) = \mathbbm{R}_{\rm cm}^3 \times \left( \exp(2\pi \pd_\chi) \cdot \mathbbm{R}_{X_\infty} \right) \times \phi_0(\MM_0)~.
\end{equation}
This action, together with \eqref{phi0L}, \eqref{phihk0}, and \eqref{productmetric}, defines the quotient by $\mathbb{D}$ in \eqref{Mfactor} in a sufficiently explicit manner for our purposes.  

In the $\mathfrak{su}(2)$ context of \eqref{su2Mfactor} where $\phi_{\gi,0}$ is the trivial isometry, we have that $\phi_{0}$ is an isometry of $\MM_0$ such that $\phi_{0}^L$ is trivial.  In particular, in this context, the strongly centered moduli space $\MM_0$ is an $L$-fold covering of another space, $\MM_0/\sim_{\phi_0}$, which was called the centered moduli space in \cite{Atiyah:1988jp}.

%%%%%%%%%%%%%%%%%%%
\subsection{Classical dyons, bound-state radii, and wall crossing}\label{sec:cldyon}
%%%%%%%%%%%%%%%%%%%

Having dealt with the primary BPS equation and its space of solutions, one is then instructed to solve the secondary equation \eqref{4dLaplace} in the monopole background.  Once the boundary value $Y_\infty$ is specified the solution for $Y$ will be unique.  Since $E_i = D_i Y$, this implies that the full configuration describes a dyon, with an electric charge that is determined in terms of $Y_\infty$ and the monopole data.  In particular the electric charge will be a function on the moduli space $\fMM$ that can be determined explicitly in terms of the metric and the Killing vector ${\rm G}(Y_\infty)$.  Fixing the electric charge constrains some of the moduli to fixed values that can be interpreted as bound-state radii between constituents.  This leads to a classical understanding of wall crossing.  These points were first uncovered for the vanilla case in \cite{Lee:1998nv,Bak:1999hp}.  We follow and expand on the discussion in \cite{Tong:1999mg}.

% % % % % % % % % % % % % % %
\subsubsection{The space of framed BPS field configurations}
% % % % % %  %% % % % % % % %

Above we indicated that the secondary BPS equation, $\hat{D}^2 Y = 0$, subject to the asymptotic and defect boundary conditions, has a unique solution.  To see this we argue as follows.  First the relation
\begin{equation}\label{curlyY}
Y = \frac{g_{0}^2}{4\pi} \YY - \tilde{\theta}_0 X~,
\end{equation}
from \eqref{XYC} implies that $Y$ satisfies the requisite defect boundary conditions.  The $X$ term properly accounts for the pole while the $\YY$ term does not contribute to it at all.  Then since the secondary equation is linear, and is satisfied by $X$ as a consequence of the Bogomolny equation, it follows that $\hat{D}^2 Y = 0$ iff $\hat{D}^2 \YY = 0$.  Now we also have that $\YY = \YY_{\infty}^{\rm cl} + y_1/|\vec{x}| + o(1/|\vec{x}|)$ as $|\vec{x}|\to \infty$, with $\YY_{\infty}^{\rm cl}$ given in \eqref{YYcl}, so $\YY \in {\rm Lie}(\TT_{\{P_n\}})$.  Hence there is a unique solution to $\hat{D}^2 \YY = 0$, namely $\YY = \epsilon_{\YY_{\infty}^{\rm cl}}$, and thus a unique solution for $Y$.

Given $Y$, we can construct the electric field $E_i = D_i Y$ and determine the electric charge.  We begin by showing that electric charge can only be excited along the root directions $\alpha_{I_A}$ associated with the effective Lie algebra $\mathfrak{g}^{\rm ef}$.  Working in the gauge \eqref{nicegauge}, where the background takes the form $\hat{A} = \Ad(\cg)( i_\ast(\hat{A}^{\rm ef}) + \hat{A}^\perp)$, we see that $\YY$ must be of the form
\begin{equation}
\YY = \Ad(\cg) \left( i_\ast(\YY^{\rm ef}) + \YY_{\infty}^\perp\right) = \Ad(\cg)(i_\ast(\YY^{\rm ef})) + \YY_{\infty}^\perp~,
\end{equation}
where $\YY^{\rm ef}$ satisfies $(\hat{D}^{\rm ef})^2 \YY^{\rm ef} = 0$.  Here the $\perp$ superscript on an element $H \in \mathfrak{t}$ indicates the projection of that element to the orthogonal complement of the subspace $i_{\ast}(\mathfrak{t}^{\rm ef}) \subset \mathfrak{t}$ with respect to the Killing form.  See appendix \ref{app:embed} for further details.  Then we have the electric field
\begin{equation}\label{EYef}
E_i = \hat{D}_i Y = \frac{g_{0}^2}{4\pi} \hat{D}_i \left(\YY - \frac{\theta_0}{2\pi} X \right)  = \frac{g_{0}^2}{4\pi}  \Ad(\cg) \left( i_\ast( \hat{D}_{i}^{\rm ef} \YY^{\rm ef}) \right) - \tilde{\theta}_0 B_i~.
\end{equation}
In particular, using \eqref{BEasymptotic}, the (dual of the) electric charge, $\gamma_{\rm e}^\ast$, comes entirely from the $\YY^{\rm ef}$ term of \eqref{EYef}, and hence $\gamma_{\rm e}^\ast \in i_\ast(\mathfrak{t}^{\rm ef})$.  The gauge transformation $\cg$ goes to the identity at infinity and therefore cannot affect the leading asymptotics.  Therefore $\gamma_{\rm e}^\ast$ can be expanded in the $H_{I_A}$, or equivalently $\gamma_{\rm e}$ can be expanded in the $\alpha_{I_A}$:
\begin{equation}
\gamma_{\rm e} = \sum_{A=1}^d \langle \gamma_{\rm e}, h^{I_A} \rangle \alpha_{I_A} = - \sum_{A=1}^d \left( \gamma_{\rm e}^{\rm phys} + \frac{\theta_0}{2\pi} \gm , h^{I_A} \right) \alpha_{I_A} ~.
\end{equation}

We can obtain a useful expression for the coefficient of $\gamma_{\rm e}$ along $\alpha_{I_A}$ via the following.  As we mentioned, $\YY = \epsilon_{\YY_{\infty}^{\rm cl}}$ generates a global gauge transformation.  Consider the innerproduct of the triholomorphic Killing vectors ${\rm G}(\YY_{\infty}^{\rm cl})$ and $K^A = {\rm G}(h^{I_A})$:
\begin{align}\label{gtoKilling}
g(\rG(\YY_{\infty}^{\rm cl}),K^A) =&~  \frac{1}{2\pi} \int_{\UU} \ed^3 x \Tr \left\{ \hat{D}^a \YY \hat{D}_a \epsilon_{h^{I_A}} \right\} \cr
=&~ \frac{2}{g_{0}^2} \int_{S_{\infty}^2} \lim_{r \to \infty} \ed^2 \Omega r^2 \hat{r}^i \Tr \left\{ \left( E_i + \tilde{\theta}_0 B_i\right) h^{I_A} \right\} \cr
=&~ \left( \gamma_{\rm e}^{\rm phys} + \frac{\theta_0}{2\pi} \gm, h^{I_A} \right)~,
\end{align}
where in the second equality we integrated by parts and used $\hat{D}^2 \YY = 0$.  The boundary conditions on $\YY, \epsilon_{h^{I_A}} \in {\rm Lie}( \TT_{\{ P_n\}})$ exclude the appearance of boundary terms associated with the defects.  Hence we have the electric charge
\begin{equation}\label{gecl}
\gamma_{\rm e} = - \sum_{A = 1}^d g(\rG(\YY_{\infty}^{\rm cl}),K^A) \alpha_{I_A} = -\sum_{A,B=1}^d \langle \alpha_{I_A} , \YY_{\infty}^{\rm cl} \rangle g(K^A,K^B) \alpha_{I_B}~.
\end{equation}
This exhibits the charge as a function of the asymptotic data $X_\infty,Y_\infty$ as well as the metric on $\fMM$, which will generally depend on the moduli $z^m$ as well as the asymptotic data $\gm, X_\infty$.  Thus for given IR data $(X_\infty,Y_\infty,\gm)$, $\gamma_{\rm e}$ is a function on $\fMM$, as has been found in \cite{Lee:1998nv,Bak:1999hp,Tong:1999mg} in the vanilla case.    

To summarize, solutions to the BPS equations always have $\gamma_{\rm e}^\ast \in i_\ast(\mathfrak{t}^{\rm ef})$.  Furthermore, requiring that the charge be properly quantized, $\gamma_{\rm e} \in \Lambda_{\rm rt}$, $\gamma_{\rm e} = n_{\rm e}^I \alpha_I$, imposes $d$ constraints among the asymptotic data and moduli.  Altogether we have
\begin{equation}\label{econstraints}
n_{\rm e}^{I_M} = 0~, \qquad n_{\rm e}^{I_A} = - g_{mn} {\rm G}(\YY_\infty^{\rm cl})^m (K^A)^n ~,
\end{equation}
in order for the BPS field configuration to exist.  This is on top of the additional requirement that the coefficients of the relative magnetic charge be non-negative in order that $\fMM$ is nonempty.  As a special case, the pure \tHooft defects have $\tilde{n}_{\rm m}^I = 0, \forall I$, so there are no $I_A$.  Hence these must have $\gamma_{\rm e} = 0$, and their physical electric charge is due entirely to the Witten effect: $\gamma_{\rm e}^{\rm phys} = -\frac{\theta_0}{2\pi} \gm$.  We will denote  the submanifold cut out of $\fMM$ by the $d$ ``$A$-type'' equations of \eqref{econstraints}---\ie\ the latter set of equations---as follows:
\begin{equation}\label{fBPSfc}
\fSigma(L; \gamma_{\rm m}, \gamma_{\rm e}; X_\infty,Y_\infty) := \left\{[\hat{A}]\,\vert\,
 n_{\rm e}^{I_B} = - g(\rG(\YY_{\infty}^{\rm cl}),K^B) ~, \forall B \right\} \subseteq  \fMM(L;\gamma_{\rm m};X_\infty)~,
\end{equation} 
where $L$ denotes the collection of line defect data, $(\zeta,P_n, \vec{x}_n)$, and $\gamma_{\rm e} = \sum_A n_{\rm e}^{I_A} \alpha_{I_A}$.  This is the \emph{moduli space of classical framed BPS field configurations}.  We will also write $\fSigma(L;\gamma; a)$ where we recall the relation $a = \zeta(Y_\infty + i X_\infty)$.

It is easy to see that the triholomorphic isometries of $\fMM$ descend to isometries of $\fSigma$.  Recall that these isometries originate from asymptotically nontrivial gauge transformations that preserve the asymptotic data.  They preserve not just the magnetic data defining $\fMM$, but the electric data as well.  Hence they map points in $\fSigma$ to points in $\fSigma$.  Since they preserve the metric of $\fMM$, they will preserve the induced metric on $\fSigma$.  The same is true concerning the isometries of $\fMM$ originating from spatial rotations.  The full set of BPS equations are covariant under spatial rotations, and these rotations preserve all asymptotic data and line defect data (provided there is a single line defect and that it is located at the fixed point of the rotation).  Hence $\fSigma$, when nonempty, inherits all of the isometries of $\fMM$. 

We will provide some physical intuition for the equations defining $\fSigma$ in subsection \ref{sec:dyonmod}.  Here we simply note that these equations might not have solutions for a given $(\YY_\infty,\gamma_{\rm e})$---especially if the $g(K^A,K^B)$ appearing in \eqref{gecl} are bounded on $\fMM$.    Asymptotic analysis of the metric, \eg\ \cite{Gibbons:1995yw,Lee:1996kz} for the vanilla case, suggests this is the case.  It would be interesting to see how far one can push this line of thought in the general setting, but we will not pursue it further here.  We also note that the second expression of \eqref{gecl} makes it clear that $\langle \gamma_{\rm e}, \YY_{\rm \infty}^{\rm cl}\rangle = - || \rG(\YY_{\infty}^{\rm cl}) ||_{g}^2$, minus the norm squared of the Killing field $\rG(\YY_{\infty}^{\rm cl})$.   Therefore $\fSigma$ is a subspace of a level set of the function $|| \rG(\YY_{\infty}^{\rm cl})||_{g}^2$ on $\MM$, and when there is only one $A$-type electric charge it corresponds to a full level set.  This function will play the role of a potential on the monopole moduli space in the collective coordinate dynamics described below.

% % % % % % % % % % % % % %
\subsubsection{The space of vanilla BPS field configurations}\label{sec:vandyon}
% % % % % % % % % % % % % % 

In the vanilla case a more refined analysis is required due to the factorization \eqref{Mfactor} of the moduli space.  Let us analyze how the expression for the electric charge, \eqref{gecl}, decomposes with respect to the direct product structure of the moduli space metric, \eqref{productmetric}.  We will assume that the only nonzero components of $\gamma_{\rm e}$ are those along the $\alpha_{I_A}$; if this is not the case then the space of BPS field configurations for the corresponding pair of $(\gm,\gamma_{\rm e})$ is empty.

In subsection \ref{sssec:Mfactor} we discussed a change of basis of $\mathfrak{t}^{\rm ef}$ from the fundamental magnetic weights, $\{h^{A}\}_{A=1}^d$, to the set $\{ h_{0}^A\}_{A=1}^{d-1} \cup \{h_{\rm cm}\}$, where $\pd_\chi \equiv \rG(i_\ast(h_{\rm cm}))$ generates translations along $\mathbb{R}_{X_\infty}$ and $K_{0}^A \equiv \rG(i_\ast(h_{0}^A))$ generates $2\pi$-periodic isometries of $\MM_0$.  Correspondingly, there is a change of integral dual basis of $(\mathfrak{t}^{\rm ef})^\ast$, from the simple roots, $\{\alpha_A\}_{A=1}^d$, to the set $\{\beta_A\}_{A=1}^{d-1} \cup \{ (\gamma_{\rm m}^{\rm ef})^\ast\}$.  The expressions for $h_{\rm cm}$ as a linear combination of fundamental magnetic weights and $(\gamma_{\rm m}^{\rm ef})^\ast$ as a linear combination of simple roots are given in \eqref{hcmdef} and \eqref{dualgm} respectively.  

The $\mathbb{Z}$-linear span of the $h_{0}^A$ gives the kernel of the homomorphism $\mu : \Lambda_{\rm mw}^{\rm ef} \to \mathbb{Z}$, defined by $\mu(h) = \langle (\gamma_{\rm m}^{\rm ef})^\ast ,h \rangle$.  Since the $\{h_{0}^A\}$ only span a $(d-1)$-dimensional sublattice of $\Lambda_{\rm mw}$, the set of elements in $(\mathfrak{t}^{\rm ef})^\ast$ that are integral-dual to the $h_{0}^A$ is of the form $(\ker{\mu})^\ast \cong \mathbb{Z}^{d-1} \times \mathbb{R}$.  The $\{ \beta_A \}_{A =1}^{d-1}$ generate the particular slice $\Span_{\mathbb{Z}} \{\beta_{A} \} \subset (\ker{\m})^\ast$ determined by the condition that $\langle i_\ast(\beta_A) , X_\infty \rangle = 0, \forall A$.  For a given $\gamma_{\rm e} \in \Lambda_{\rm rt}$ we define
\begin{align}\label{echarge0def}
q_{\rm cm} :=&~ \langle \gamma_{\rm e}, i_\ast(h_{\rm cm})\rangle = \frac{\langle \gamma_{\rm e}, X_\infty\rangle}{(\gm,X_\infty)} \in \mathbb{R}~, \cr
N_{{\rm e},0}^A :=&~ \langle \gamma_{\rm e}, i_\ast(h_{0}^A) \rangle \in \mathbb{Z}~, \quad A = 1,\ldots. d-1~.
\end{align}
The $N_{{\rm e},0}^A$ are integers.  We also define the \emph{relative electric charge}
\begin{equation}\label{geequiv}
\gamma_{{\rm e},0} := \gamma_{\rm e} - \frac{\langle \gamma_{\rm e}, X_\infty \rangle}{(\gm, X_\infty)} \gm^\ast = \sum_{A=1}^{d-1} N_{{\rm e},0}^A \, i_\ast(\beta_A) \in \{ \beta \in \mathfrak{t}^\ast ~|~ \langle \beta, X_\infty \rangle = 0 \} \cap (\ker{\mu})^\ast ~,
\end{equation}
as the part of $\gamma_{\rm e}$ that has zero pairing with $X_\infty$.  Then we have the decomposition
\begin{align}\label{gebetaexp}
\gamma_{\rm e} =&~ \langle \gamma_{\rm e}, i_\ast(h_{\rm cm}) \rangle \gamma_{\rm m}^\ast + \sum_{A=1}^{d-1} \langle \gamma_{\rm e}, i_\ast(h_{0}^A) \rangle i_\ast(\beta_A)  = q_{\rm cm} \gm^\ast + \gamma_{{\rm e},0}~.
\end{align}

We can obtain an alternative expression for $\gamma_{\rm e}$ by applying the same expansions to the right-hand side of the first equality in \eqref{gecl}, which expresses $\gamma_{\rm e}$ as a $\Lambda_{\rm rt}$-valued function on moduli space.  Making use of
\begin{equation}
\sum_{A=1}^d \langle \alpha_{I_A}, h_{\rm cm} \rangle K^A = \frac{1}{(\gm,X_\infty)} \sum_{A=1}^d \langle \alpha_{I_A}, X_\infty \rangle \rG(h^{I_A}) = \frac{1}{(\gm,X_\infty)} \rG(X_\infty)~,
\end{equation}
and
\begin{equation}
\sum_{A=1}^d \langle \alpha_{I_A}, i_\ast(h_{0}^{I_B}) \rangle K^A = \rG(i_\ast(h_{0}^B)) \equiv K_{0}^B~,
\end{equation}
we find that
\begin{align}\label{gecmrel}
\gamma_{\rm e} =&~ - \frac{g(\rG(\YY_{\infty}^{\rm cl}), \rG(X_\infty))}{(\gm,X_\infty)} \gm^\ast - \sum_{B=1}^{d-1} g(\rG(\YY_{\infty}^{\rm cl}),K_{0}^B) \, i_\ast(\beta_B) \cr
=&~ - \frac{(\gm, \YY_{\infty}^{\rm cl})}{(\gm,X_\infty)} \gm^\ast -  \sum_{A=1}^{d-1} g(\rG(\YY_{\infty}^{\rm cl}),K_{0}^A)\, i_\ast(\beta_A) ~.
\end{align}

Equating \eqref{gecmrel} and \eqref{gebetaexp} we find the ``constraint''
\begin{equation}\label{vconstraint2}
\langle \gamma_{\rm e}, X_\infty \rangle + (\gm, \YY_{\rm \infty}^{\rm cl}) = 0~,
\end{equation}
from the $\gm^\ast$ terms.  However using \eqref{YYcl}, one sees that this is just a rewriting of \eqref{vanconstraint}, and therefore is satisfied for any BPS field configuration.  In particular this relation does not impose any conditions on the moduli.  We refer to the $q_{\rm cm} \gm^\ast$ component of the electric charge as the  ``Julia--Zee component'' since it generalizes the electric charge of the $\mathfrak{su}(2)$ dyon solution of \cite{Julia:1975ff}.  

Meanwhile, equating the $\beta_A$ terms leads to
\begin{equation}\label{M0slice1}
N_{{\rm e},0}^A = - g\left(\rG(\YY_{\infty}^{\rm cl}), K_{0}^A\right)~, \quad A = 1,\ldots, d-1~.
\end{equation}
Since the Killing vectors $K_{0}^A$ have legs along $\MM_0$ only and the metric is a product metric, the right-hand side of this expression only depends on the projection of $\rG(\YY_{\infty}^{\rm cl})$ onto $T\MM_0 \subset T\MM$.  This can be made explicit by expanding $\YY_{\infty}^{\rm cl}$ in the basis $\{ h_{\rm cm}, h_{0}^{I_B}\}$ as well:
\begin{equation}\label{Cartandecomp}
\YY_{\infty}^{\rm cl} = \left\langle \gm^\ast, \YY_{\infty}^{\rm cl} \right\rangle i_\ast(h_{\rm cm}) + \sum_{B=1}^{d-1} \left\langle i_\ast(\beta_B), \YY_{\infty}^{\rm cl} \right\rangle i_\ast(h_{0}^B) + (\YY_{\infty}^{\rm cl})^\perp~,
\end{equation}
which then leads to
\begin{equation}
\rG(\YY_{\infty}^{\rm cl}) = \frac{(\gm, \YY_{\infty}^{\rm cl})}{(\gm, X_\infty)} \rG(X_\infty) + \sum_{A=1}^{d-1} \left\langle i_\ast(\beta_B), \YY_{\infty}^{\rm cl} \right\rangle K_{0}^A  = \frac{(\gm, \YY_{\infty}^{\rm cl})}{(\gm, X_\infty)} \rG(X_\infty) + \rG_0(\YY_{\infty}^{\rm cl})~.
\end{equation}
In the last step we introduced the map,
\begin{equation}
\rG_0 : \mathfrak{t} \to \mathfrak{isom}_{\mathbb{H}}(\MM_0)~,
\end{equation}
which can be defined by projecting $\rG(H) \in \mathfrak{isom}_{\mathbb{H}}(\MM)$ onto the subbundle of $T\MM$ that is (metric-) orthogonal to $\rG(X_\infty)$,
\begin{equation}\label{vGdecomp}
\rG_0(H) := \rG(H) - \frac{g\left(\rG(H), \rG(X_\infty)\right)}{||\rG(X_\infty)||_{g}^2} \rG(X_\infty) = \rG(H) - \frac{(\gm,H)}{(\gm, X_\infty)} \rG(X_\infty)~.
\end{equation}
The intersection of this subbundle with $\mathfrak{isom}_{\mathbb{H}}(\MM)$ is $\mathfrak{isom}_{\mathbb{H}}(\MM_0)$.  Hence we have
\begin{equation}\label{M0slice2}
N_{{\rm e},0}^B = - g\left(\rG_0(\YY_{\infty}^{\rm cl}), K_{0}^B\right) = - \sum_{A=1}^{d-1} \left\langle i_\ast(\beta_A), \YY_{\infty}^{\rm cl} \right\rangle g\left( K_{0}^A, K_{0}^B\right)~.
\end{equation}
Observe that neither the component $\left( \gm, \YY_{\infty}^{\rm cl}\right)$ of $\YY_{\infty}^{\rm cl}$, which is in any event fixed by \eqref{vconstraint2}, nor the components $(\YY_{\infty}^{\rm cl})^\perp$ of $\YY_{\infty}^{\rm cl}$ in the orthogonal complement of $i_\ast(\mathfrak{t}^{\rm ef})$ in $\mathfrak{t}$, participate in the conditions \eqref{M0slice2}.

The equations \eqref{M0slice2} cut out a hypersurface of co-dimension $d-1$ in $\MM_0$ that we call the \emph{strongly centered moduli space of classical BPS field configurations}:
\begin{align}\label{relBPSfc}
&\Sigma_0(\gm,\gamma_{\rm e}; X_\infty,Y_\infty) := \bigg\{ N_{{\rm e},0}^A = - g\left(\rG_0(\YY_{\infty}^{\rm cl}), K_{0}^A\right)~, \forall A \bigg\} \subseteq \MM_0(\gm;X_\infty)~.
\end{align}
We again emphasize that $N_{{\rm e},0}^A \in \mathbb{Z}$ and the triholomorphic Killing vectors $K_{0}^A$ generate triholomorphic isometries of $\MM_0$ with $2\pi$-periodicity.  In the case $\mathfrak{g}^{\rm ef} = \mathfrak{su}(2)$, the set of equations \eqref{M0slice1} is empty and $\Sigma_0 = \MM_0$.

As we mentioned in the framed case, the asymptotically nontrivial gauge transformations that are responsible for the triholomorphic isometries of the magnetic monopole moduli space preserve both the magnetic and electric boundary conditions.  Hence those isometries of $\MM_0$ will descend to isometries of $\Sigma_0$, and the identifications \eqref{hkdeckgen} will define the discrete quotient of $\mathbb{R}_{X_\infty} \times \Sigma_0$ by $\mathbb{D}_\gi$.  We do not have such a simple argument that the identifications imposed by $\phi$, the generator of $\mathbb{D}$, \eqref{deckgen}, should restrict to an action on $\mathbb{R}_{X_\infty} \times \Sigma_0$.  However in this case we can observe that the equations defining $\Sigma_0$ depend only on the local data of the metric and therefore take the same form on all leaves of the $\mathbb{Z}_L$ cover, $(\mathbb{R}_{X_\infty} \times \MM_0)/\mathbb{D}_\gi$ of $(\mathbb{R}_{X_\infty} \times \MM_0)/\mathbb{D}$.  Hence we expect that the action of $\phi$ on $\mathbb{R}_{X_\infty} \times \MM_0$ does induce a well-defined action on $\mathbb{R}_{X_\infty} \times \Sigma_0$.  Therefore, using this action, we define the \emph{moduli space of classical BPS field configurations}
\begin{equation}\label{BPSfc}
\Sigma(\gm,\gamma_{\rm e}; X_\infty,Y_\infty) := \mathbb{R}_{\rm cm}^3 \times \frac{\mathbb{R}_{X_\infty} \times \Sigma_0(\gm,\gamma_{\rm e}; X_\infty,Y_\infty)}{\mathbb{D}}~.
\end{equation}
We will also denote these spaces via the shorthand $\Sigma(\gamma;a)$ and $\Sigma_0(\gamma; a)$, and we remind the reader of the rather nontrivial relationship between $X_\infty,Y_\infty$ and $a$, which involves the electromagnetic charge as well:
\begin{align}
Y_\infty + i X_\infty = (\zeta_{\rm van}^{\rm cl})^{-1} a = - \frac{|Z^{\rm cl}|}{Z^{\rm cl}} a ~, \quad \textrm{with}  \quad Z^{\rm cl} =&~ \tau_0 (\gm,a) + \langle \gamma_{\rm e},a\rangle~.
\end{align}
As we discussed following \eqref{vanconstraint}, this relationship between $X_\infty,Y_\infty$ and $a$ ensures the constraint \eqref{vconstraint2} is satisfied.  As a consequence, any $a$ related by an overall phase rotation, $a \to e^{i\vartheta} a$ will give the same $X_\infty, Y_\infty$, and for given $\gamma$, $\Sigma(\gamma;a)$ only depends on the equivalence classes $[a]$ defined by $a \sim e^{i\vartheta} a$.

% % % % % % % % % % % % % % % 
\subsubsection{Classical bound-state radii and classical wall crossing}\label{sec:dyonmod}
% % % % % % % % % % % % % % % 

Let us provide some physical intuition for the constraints \eqref{econstraints}, and their vanilla version in \eqref{relBPSfc}, that determine respectively the spaces of framed and vanilla BPS field configurations, $\fSigma$ and $\Sigma$.  The mechanism at work was understood in \cite{Lee:1998nv,Tong:1999mg,Bak:1999hp}; see also the discussion in \cite{Weinberg:2006rq}.

First let us suppose that $\YY_{\infty}^{\rm cl} = 0$, implying $\gamma_{\rm e} = 0$, so that the moduli space is the magnetic monopole moduli space, $\fSigma = \fMM$ or $\Sigma = \MM$.   This space has dimension $4 |\tilde{\gamma}_{\rm m}|$ or $4|\gm|$ in the framed or vanilla cases respectively.  Restricting consideration to the framed case for the moment, recall the physical interpretation of this dimension: there are $|\tilde{\gamma}_{\rm m}| = \sum_{A} \tilde{n}_{\rm m}^{I_A}$ fundamental 't Hooft--Polyakov monopoles in the system, consisting of $\tilde{n}_{\rm m}^{I_A} > 0$ monopoles of type $I_A$, for each $A = 1,\ldots,d$.  Choose any one line defect $L$ at position $\vec{x}_0$.  Then, at least when all monopoles are well separated from each other and the defects, a good set of coordinates on $\fMM$ should be the set of position vectors for each fundamental monopole relative to $\vec{x}_0$, together with a phase coordinate for each fundamental monopole.  In particular, there are no net forces exerted on the fundamental monopoles by the defects, nor on the fundamental monopoles by each other.  

Now suppose we turn on the scalar vev $\YY_{\infty}^{\rm cl}$.  In order to maintain a BPS field configuration, the scalar field $\YY$ gets a nontrivial profile determined by $\hat{D}^2 \YY = 0$ and the asymptotic boundary condition.  This in turn sources the combination $E_i + \tilde{\theta}_0 B_i$, whose leading asymptotics determine $\gamma_{\rm e}$.  If $\gamma_{\rm e} = \sum_{A} n_{\rm e}^{I_A} \alpha_{I_A}$\label{neIA}, then the monopoles of type $I_A$ become dyons carrying net\footnote{There is no invariant way to distribute this charge among the constituents as they can exchange charge with each other in dynamical processes.  The prototypical example where such processes have been studied in detail is \cite{Atiyah:1988jp}. It is only the net charge of each type that is conserved.} electric charge $n_{\rm e}^{I_A}$, with respect to the $U(1) \subset T_{\rm ad}^{\rm ef}$ generated by $h^{I_A}$.  Note that it only makes sense to ascribe the component of electric charge $n_{\rm e}^{I_A}$ to the dyons of type $I_A$ when the different types are well-separated, such that gauge transformations that asymptote to $h^{I_A}$ effectively only act on the fields in the vicinity of these monopoles.  With the secondary scalar field and the electric charge turned on, the various position-dependent forces on the constituents no longer cancel pointwise.  For a generic configuration of positions, a defect will exert a net force on each type of dyon, and dyons of different types will exert net forces on each other.  Note however that dyons of the same type do not exert forces on one another---the moduli space of $N$ identical dyons is still $4N$-dimensional.  

Hence, if we associate a position vector $\vec{x}^A$ to the \emph{center of mass of the dyons of each type}, relative to $\vec{x}_0$ say, then it is this position that is being constrained.  The $d$ equations $g_{mn} {\rm G}(\YY_{\infty}^{\rm cl})^m (K_A)^n + n_{\rm e}^{I_A} = 0$ that define $\fSigma$, \eqref{fBPSfc}, can be thought of as fixing the $d$ relative distances between the $d+1$ vectors $\{\vec{x}_0, \vec{x}^A\}$.  It is at these values of the relative distances that the net forces on each type of fundamental dyon vanish.  Again, this picture should be reasonable when all of these points are well-separated (relative to the scales set by the components of the Higgs vev, $\langle \alpha_{I_A}, X_\infty \rangle$).

In the vanilla case the picture is essentially the same, except that there is one class of choices of $\{\YY_{\infty}^{\rm cl}, \gamma_{\rm e}\}$ such that the equations defining $\Sigma_0$, \eqref{relBPSfc}, are trivially satisfied: namely $\{\YY_{\infty}^{\rm cl}, \gamma_{\rm e}\} = \{C_1 X_\infty, C_2 \gamma_{\rm m}^\ast\}$, where $C_1,C_2$ are proportionality constants.  (The ratio $C_1/C_2$ will be fixed by the constraint \eqref{vconstraint2}.)  In this case both sides of the equation in \eqref{relBPSfc} are zero.  The right-hand side is zero because ${\rm G}(X_\infty)$ is metric-orthogonal to all of the $K_{A}^0$.  The left-hand side is zero because $\gamma_{\rm e} = C \gamma_{\rm m}^\ast$ means $n_{\rm e}^{I_A} = C \sp^A n_{\rm m}^{I_A}$ for all $A$, whence $N_{{\rm e},0}^A = 0$.  Hence for this class of $\{\YY_{\infty}^{\rm cl}, \gamma_{\rm e}\}$ the relative positions $\vec{x}^A$ are unconstrained.  This type of electric charge, which we referred to as the ``Julia--Zee'' charge, is associated with the $\mathbb{R} \subset T_{\rm ad}^{\rm ef}$ generated by $X_\infty$, and corresponds to the ``center of mass phase'' direction, $\mathbb{R}_{X_\infty}$, in $\Sigma$, \eqref{BPSfc}.  The unconstrained combination of the $\vec{x}^A$ is the overall center-of-mass coordinate, $\vec{x}_{\rm cm}$, parameterizing the $\mathbb{R}_{\rm cm}^3$ factor of $\Sigma$.  The $(d-1)$ equations defining $\Sigma_0$, and hence $\Sigma$, can then be thought of as constraining the relative distances of the $d$ vectors $\{ \vec{x}^{A}\}$.  The relative distances that get fixed in either the framed or vanilla case provide a classical notion of bound-state radii.

As we have mentioned previously, for given $\{\YY_{\infty}^{\rm cl}, \gamma_{\rm e}\}$, the equations defining $\fSigma$ and $\Sigma_0$, might not have solutions.  This is especially clear if the metric innerproducts of the triholomorphic Killing vectors, $g(K^A,K^B)$ in the framed case and $g(K_{0}^A,K_{0}^B)$ in the vanilla case, are bounded functions.  This is expected in the framed case and known in the vanilla case.  Let us denote these functions by $g(K^A,K^B) \equiv g^{AB}$.  In the framed case, the equations determining $\fSigma$ are $\sum_B g^{AB} \langle \alpha_{I_B}, \YY_{\infty}^{\rm cl} \rangle + n_{\rm e}^{I_A} = 0$.  Suppose, for the sake of argument, that that the $g^{AB}$ have upper and lower bounds $g^{AB}_{\rm min} \leq g^{AB}(z^m) \leq g^{AB}_{\rm max}$.  Then for fixed $\YY_{\infty}^{\rm cl}$, if we take the $n_{\rm e}$ too large in magnitude, the equations will not have a solution for any $\{ z^m \}$; $\fSigma$ will be empty.  

Suppose instead we fix a $\gamma_{\rm e}$ and an initial $\YY_{\infty}^{\rm cl}$ such that there is a solution.  Then by dialing any of the components $\langle \alpha_{I_B}, \YY_{\infty}^{\rm cl} \rangle$, (by changing the components $\langle \alpha_{I_B} , Y_\infty \rangle$ while holding $\langle \alpha_{I_B}, X_\infty \rangle$ fixed such that the metric remains fixed), we might eventually cross a value for which the system of equations no longer has a solution.  This point represents a co-dimension one wall in $Y_\infty$ space across which $\fSigma$ ceases to exist.  Walls can also occur in $X_\infty$ space: since the metric depends on $X_\infty$, $g^{AB}_{\rm min}$ and/or $g^{AB}_{\rm max}$ can depend the components $\langle \alpha_{I_A}, X_\infty \rangle$.  By changing any one of these components, the bounds can be modified such that a wall is crossed, where $\fSigma$ is nonempty on one side and empty on the other.  The walls in $\{Y_\infty,X_\infty\}$ space are translated to walls in $\{a,\zeta\}$ space via the relation $\zeta^{-1} a = Y_\infty + i X_\infty$.  Analogous statements can be made in the vanilla case.  

The walls just described are in some sense a classical analog of the marginal stability walls defined in \eqref{vanillawalls}, \eqref{fmsw}, and the corresponding jumping phenomena for the vanilla spaces $\Sigma_0$ has been noted previously \cite{Lee:1998nv,Tong:1999mg}.  The analogy is not precise, as the discussion here concerns the existence of BPS field configurations in the classical field theory, whereas marginal stability walls are defined for BPS states in the quantum theory.  One expects the existence of a classical BPS field configuration carrying the appropriate charges to be a necessary prerequisite for the existence of the corresponding BPS state.  It does not shed light, however, on the degeneracy of such states, nor is it a sufficient condition for their existence.  For example, there is a moduli space of classical solutions describing two identical $\mathfrak{su}(2)$ dyons, but there are no $\Lsq$ normalizable wavefunctions on the strongly centered moduli space and hence no (one-particle) BPS state carrying this charge, at least in the weak coupling regime of the Coulomb branch where the semiclassical analysis is reliable.

%%%%%%%%%%%%%%%%%%%%%%
%%%%%%%%%%%%%%%%%%%%%%
\section{Semiclassical framed BPS states}\label{sec:sc}
%%%%%%%%%%%%%%%%%%%%%%
%%%%%%%%%%%%%%%%%%%%%%

In the last section we discussed families of classical field configurations labeled by charges $\gm \oplus\gamma_{\rm e}$ and saturating a (classical) BPS bound.  In this section we define corresponding states in the quantum theory using semiclassical techniques.  They will be eigenstates of the magnetic and electric charge operators and of the Hamiltonian, with eigenvalues $\gamma = \gm \oplus \gamma_{\rm e}$ and $M_{\gamma}$, which continue to preserve half of the supersymmetry and saturate the BPS bound involving the (quantum-corrected) central charge and mass.

%%%%%%%%%%%%%%%%%%%%%
\subsection{Semiclassical expansion and the moduli space approximation}\label{sec:scphilo}
%%%%%%%%%%%%%%%%%%%%%

Classical solutions with nonzero $\tilde{\gamma}_{\rm m}$ of \eqref{relcharge} are solitons: local minima of the energy functional in topologically nontrivial sectors of field configuration space.\footnote{The presence of \tHooft defects changes the notion of ``trivial'' topological sector of field configuration space.  We take the trivial sector to be the one where there is no further magnetic charge in the system beyond the one due to the defects, $\gm = \sum_n P_{n}^-$, such that $\tilde{\gamma}_{\rm m} = 0$.}  Indeed, the allowed values of $\tilde{\gamma}_{\rm m}$, or the allowed asymptotic magnetic charges if we prefer, label the different connected components of configuration space.  The semiclassical expansion of a quantum theory around a static classical soliton is an expansion in the coupling $g_0$.  The charges of BPS states are determined by topological and symmetry considerations; they are quantized and therefore cannot receive corrections.  The masses however will receive quantum corrections to their classical value, \eqref{MBPScl}.  

The formalism for studying a quantum theory in a soliton sector and systematically computing corrections to classical quantities such as the soliton mass was developed in several works in the mid `70's \cite{Dashen:1974ci,Dashen:1974cj, Goldstone:1974gf,Gervais:1974dc, Tomboulis:1975gf,Callan:1975yy, Christ:1975wt, Gervais:1975pa,Tomboulis:1975qt}.  See \cite{Jackiw:1977yn,Rajaraman:1982is,Dorey:2002ik,Tong:2005un} for pedagogical reviews.  The basic idea is to quantize the fluctuations of the fields around the classical soliton solution, making a mode expansion in terms of eigenfunctions of the appropriate differential operator, $\Delta$, which is obtained from the linearization of the time-independent equations of motion around the background configuration.  However there will be zero modes in this expansion, corresponding to tangent vectors of the moduli space $\fMM$.  In order to avoid the usual problems associated with zero-frequency modes, one exchanges those degrees of freedom for an equal number of alternative degrees of freedom, obtained by promoting the moduli $z^m$ to dynamical variables $z^m(t)$.  The moduli thus become \emph{collective coordinates}.  In the full quantum field theory, one is changing coordinates on configuration space from the original fields $\hat{A}$ to collective coordinates and fluctuations $(z^m(t), \hat{a})$ via
\begin{equation}\label{solitoncov}
\hat{A}(t,\vec{x}) = \hat{A}^{\rm cl}(\vec{x}; z(t)) + g_0 \, \hat{a}(t,\vec{x};z(t))~,
\end{equation}
where the zero-frequency modes are excluded from $\hat{a}$ either by hand or by imposing constraints.\footnote{In more detail: \eqref{solitoncov} can be viewed as a partial definition of the fluctuation field $\hat{a}$.  However this definition must be supplemented with the information that $\hat{a}$ is not a completely generic $\mathbb{R}^4 \otimes \mathfrak{g}$-valued function on spacetime.  The eigenmodes of $\Delta$ corresponding to zero eigenvalue are to be excluded.  This can be implemented by either explicitly replacing the $\hat{a}$ term of \eqref{solitoncov} with a mode expansion where these modes are absent, or by giving a set of constraints that $\hat{a}$ must satisfy.  In the latter case, this set of constraints together with \eqref{solitoncov} define $\hat{a}$.}  Note that the linear differential operator $\Delta$, controlling the spectrum of fluctuations, will depend on the moduli, and hence so will its eigenmodes.  A factor of $g_0$ has been extracted from the fluctuation field so that the kinetic term for the fluctuation will have canonical normalization.

One must supplement the change of variables \eqref{solitoncov} with a transformation for the remaining degrees of freedom which, in a covariant formulation, will include $(A_0,Y,\psi_A)$ as well as ghosts.  Let us denote the collection of original fields by $\Phi$ and the collection of fluctuation fields around the soliton as $\phi$ so that the change of coordinates in the path integral is $\Phi \mapsto (z^m,\phi)$.  One then needs to determine the corresponding change of momentum variables, $\Pi \mapsto (p_m, \pi)$, where $(\Pi,p_m,\pi)$ are conjugate to $(\Phi,z^m,\phi)$ respectively, such that the total transformation on phase space, $(\Phi;\Pi) \mapsto (z^m,\phi;p_m,\pi)$ is canonical.  This ensures that the quantum commutator is preserved under the phase space transformation.  With the canonical transformation in hand, one can finally write the Hamiltonian in the new variables that governs the dynamics in the soliton sector.

In fact, carrying out this transformation explicitly and exactly is rather difficult, and for the most part it has only been done in scalar theories with kink solitons.\footnote{Tomboulis and Woo \cite{Tomboulis:1975qt} quantized the one-monopole sector Yang--Mills--Higgs theory with $SU(2)$ gauge group, where the moduli space is $\MM = \mathbb{R}^3 \times S^1$.  See \cite{Papageorgakis:2014jia} for a recent analysis of a class of multi-component scalar theories with generic moduli spaces.}  Nevertheless some general lessons can be extracted from these examples, arguably the most important of which is the following.  The classical configuration with time-dependent moduli, $\hat{A}^{\rm cl}(\vec{x};z(t))$, is generally not an exact solution to the time-dependent equations of motion.  This manifests itself in the form of a tadpole term in the Hamiltonian for the fluctuation field $\hat{a}$.  This term has the form $\sim p^2\hat{a}$, which is consistent with the fact that $\hat{A}^{\rm cl}$ is a solution when $z^m$ is constant (and hence the momentum $p_m$ vanishes).  In order to have a well-defined perturbation theory, one wants the tadpole to be subleading to the quadratic order terms for $\hat{a}$, so that the tadpole can be viewed as part of the interaction Hamiltonian.  This is what necessitates\footnote{In the rare circumstance where one has access to exact time-\emph{dependent} solutions, this approximation is no longer necessary and the more powerful WKB methods of \cite{Dashen:1974ci,Dashen:1974cj} become available.} the \emph{small velocity assumption},
\begin{equation}\label{manton}
\dot{z}^m \sim O(g_{0})~.
\end{equation}
Thus we see that the Manton approximation \cite{Manton:1981mp}, originally envisioned as an approximation scheme for finding \emph{classical} time-dependent solutions to the equations of motion,  emerges as part of the \emph{semiclassical} analysis, where the small parameter controlling the time variation of the classical field is identified with the semiclassical expansion parameter $g_0$.

Under this assumption, the generic soliton-sector Hamiltonian is expected to reduce to the following form:
\begin{equation}\label{solitonHam}
H = M_{\gamma}^{\rm cl} + \half p_m (g_{\rm phys})^{mn} p_n + \half  \int \ed^3 x \Tr \left\{ \pi \cdot \pi + \phi \cdot \Delta \phi \right\} + O(g_{0})~,
\end{equation}
where $(g_{\rm phys})_{mn}$ is the canonically normalized moduli space metric; see footnote \ref{fn:physmet}.  Again, $\phi$ denotes the collection of fluctuation fields around the soliton and $\pi$ the collection of conjugate momentum densities, while $\Delta$ is the appropriate differential operator obtained from linearizing the time-independent equations of motion around the classical soliton.  The classical mass $M_{\gamma}^{\rm cl} \sim O(g_{0}^{-2})$, as we can see from \eg\ \eqref{MBPScl}.

Now on the one hand, it follows from \eqref{manton} that $p_m \sim O(g_{0}^{-1})$.  On the other hand the metric $(g_{\rm phys})_{mn} \sim O(g_{0}^{-2})$, implying $(g_{\rm phys})^{mn} \sim O(g_{0}^2)$.  Thus the collective coordinate kinetic term is $O(1)$, the \emph{same order} as the quadratic Hamiltonian for the fluctuation fields.  Meanwhile the interaction Hamiltonian starts at $O(g_0)$.  The reason we emphasize this is the following.  The $\pi,\phi$ fields can be expanded in eigenmodes of the differential operator $\Delta$, with coefficients that create/annihilate perturbative particle states in the presence of the soliton.  If we restrict ourselves to the `vacuum' of the soliton sector, so that there are no perturbative excitations in the incoming and outgoing states, then this term reduces to a sum over the zero-point energies of the modes, resulting in the one-loop correction to the vacuum energy, $M_{\gamma}$.  In \cite{Kaul:1984bp} it is shown that the leading divergence cancels between bose and fermi contributions, however there is a remaining logarithmic divergence due to a mismatch in the density of states.  With the counterterms already fixed in the perturbative sector, \eg\ \eqref{Lct}, it is a nontrivial test of renormalizability that they cancel the divergences in soliton sectors as well.  The finite remainder is the physical one-loop correction to the soliton mass.  It follows from \eqref{manton} and \eqref{solitonHam} that it is \emph{inconsistent}, from the point of view of the full quantum field theory, to consider collective coordinate dynamics while ignoring the one-loop correction to the soliton mass.

This point is absolutely crucial if one wants to compare consistently results of a semiclassical analysis of the \emph{field theory} with other field theory approaches, such as the low energy effective theory of Seiberg and Witten.  We feel that it has been rather under-appreciated in the literature, so we repeat, step by step, the reasoning that leads to it:
\begin{enumerate}
\item  Taking the static classical soliton field configuration and allowing the moduli to become time-dependent does not generally provide solutions to the time dependent equations of motion.  When there is either acceleration and/or we are considering motion on a moduli space with curvature,  the classical field profile with time-dependent moduli will source the equation of motion for the part of the field orthogonal to the tangent space to moduli space in field configuration space---\ie\ the `quantum fluctuation field' $\hat{a}$ in \eqref{solitoncov}.
\item In order apply standard QFT perturbation theory in the soliton sector, where one introduces a set of particle creation and annihilation operators that diagonalize the quadratic Hamiltonian for the fluctuation field, it is necessary that these tadpole terms in the action, corresponding to the above mentioned source term in the equation of motion, be small---that is of the same order as the interaction Hamiltonian, so that they can be treated as interactions in perturbation theory.
\item Since the strength of the tadpole terms is controlled by either velocity-squared or acceleration of the collective coordinates, in other words expressions that involve two time derivatives acting on the collective coordinates, we take the time derivatives to be of the same order as the small parameter controlling the quantum perturbation theory.  This is quite literally the QFT analog of Manton's classical field theory approximation scheme.  We assume $\dot{z} = O(g_0)$, $\ddot{z} = O(g_{0}^2)$, \etc.
\item Under this assumption, we then consider the expansion of the Hamiltonian around the soliton field configuration.  The leading piece is the classical mass of the static soliton and, as is generally the case for solitons, this term scales parametrically as $O(g_{0}^{-2})$.  The next terms in the Hamiltonian appear at $O(1)$ and they are of two types:  the collective coordinate Hamiltonian and the quadratic Hamiltonian for the fluctuation fields.  Since these two sets of terms are parametrically the same order in $g_0$, it is invalid to ignore one in favor of the other.
\item  If we restrict our external states to those which do not have any perturbative particle states excited above the soliton, then the only contribution at $O(1)$ from the quadratic order fluctuation Hamiltonian to expectation values of observables between soliton states comes from the sum over zero-point energies---\ie\ the vacuum bubble.
\item  In the perturbative sector of the theory the energy of the vacuum is infinite and we renormalize it to zero.  However, having done that, via the introduction of appropriate counterterms---see \eqref{Lct}, the vacuum energy of the soliton sector relative to the perturbative sector is finite and physically meaningful.  This is what gives rise to the one-loop correction to the monopole mass.
\item  In the quantum theory, under the Manton scaling assumption, this $O(1)$ correction to the leading $O(1/g_{0}^2)$ mass is the same order as corrections to the mass coming from collective coordinate kinetic and/or potential energy.  Hence it is inconsistent to ignore it.
\end{enumerate}

Before discussing the one-loop correction to the soliton mass in detail, we must point out that the form of the Hamiltonian \eqref{solitonHam} is actually insufficient for our purposes on two counts.  First, we have so far neglected the possibility that other fields in addition to $\hat{A}$ might contain zero modes.  In the vanilla case it is well known, and in the framed case it was demonstrated in \cite{MRVdimP1}, that fermions coupled to Yang--Mills--Higgs theory generically have zero modes.  If the fermions transform in a representation $\rho$ of the gauge group, the zero modes form a basis of sections for a certain vector bundle over $\fMM$, namely the index bundle of the Dirac operator controlling the fermionic spectrum \cite{Taubes:1984je,Manton:1993aa}.  These zero modes also play an important role in the semiclassical analysis \cite{Sethi:1995zm,Cederwall:1995bc,Gauntlett:1995fu,Henningson:1995hj} and will be discussed further in the next sections.  Note that for the class of pure $\NN =2$ gauge theories we are considering, the relevant representation is the adjoint, and in this case the index bundle is isomorphic to the tangent bundle.  We will show below how one can construct a natural quaternion-linear isomorphism between the spaces of bosonic and fermionic zero modes.

Second, consider applying what has been discussed so far to the case of classical dyons with charge $\gamma = \gm \oplus \gamma_{\rm e}$, described by solutions to the primary and secondary BPS equations, \eqref{BPS}.  Then the moduli space in question is not the full monopole moduli space but rather the hypersurface $\fSigma \subset \fMM$ in the framed case, or $\Sigma \subset \MM$ in the vanilla case, defined in \eqref{fBPSfc}, \eqref{BPSfc} respectively.  Correspondingly the metric $(g_{\rm phys})_{mn}$ in \eqref{solitonHam} would be the induced metric on this surface.  It would be interesting to consider the dynamics of the collective coordinates parameterizing this surface, but this is not the right approach for the regime we will be focusing on.  Rather we should formulate the collective coordinate dynamics on the full monopole moduli, treating the effects of turning on the secondary Higgs vev $Y_\infty$ as a perturbation.  There is a well-established formalism for doing this---namely, we will be considering dynamics on a ``moduli space with potential'' \cite{Bak:1999da,Bak:1999ip,Bak:1999sv,Gauntlett:1999vc}.  

This is the natural approach to take for a semiclassical analysis, provided one restricts attention to electric charges $\gamma_{\rm e} \ll O(g_{0}^{-2})$.  The equations defining the hypersurfaces $\fSigma,\Sigma$, balance the components of the electric charge, $\{n_{\rm e}^{I} \}$, against components of $\YY_{\infty}^{\rm cl}$ times metric innerproducts of canonically normalized triholomorphic Killing vectors; see \eqref{econstraints} and \eqref{M0slice1}.  The metric is independent of $g_{0}$ while $\YY_{\infty}^{\rm cl}$ scales as $O(g_{0}^{-2})$ times $Y_\infty$.  Hence for electric charges $\gamma_{\rm e} \ll O(g_{0}^{-2})$ the components $\langle \alpha_I, Y_\infty \rangle$ should be taken much smaller than the natural length scales appearing in the metric, which are the $\langle \alpha_I, X_\infty \rangle$.  Therefore it is natural to treat the effects of turning on $Y_\infty$ as a perturbation, on the same footing as allowing time dependence via \eqref{manton}.  This observation has also been nicely explained from the point of view of moduli space dynamics in \cite{Peeters:2001np,Lee:2011ph}.

These two modifications to \eqref{solitonHam}---fermionic zero modes and a moduli space potential---depend on the collective coordinate degrees of freedom only.  Thus they can be obtained by the standard truncation to collective coordinates which we will review below.

Now let us return to the one-loop correction to the vacuum energy, or soliton mass, originating from the terms in \eqref{solitonHam} that are quadratic in the fluctuation fields $\phi,\pi$.  Reference \cite{Kaul:1984bp} showed how the mismatch in the density of states can be extracted from the index density
\begin{equation}\label{Idensity}
I(m^2) := \Tr_{\Lsq[\mathbb{R}^3, \mathbb{C}^2\otimes \mathfrak{g}]} \left\{ \frac{m^2}{L^\dag L + m^2} - \frac{m^2}{L L^\dag + m^2} \right\}~,
\end{equation}
where the trace is over the Hilbert space of adjoint-valued $\Lsq$ spinors on $\mathbb{R}^3$, and $L$ is the same Dirac operator whose zero modes determine the dimension of the moduli space.  (See \eqref{Dirac2} below.)  The density, in turn, was computed in \cite{Weinberg:1979ma} by making use of the Callias index theorem \cite{Callias:1977kg}.  This leads to an explicit result for the one-loop correction to the mass, obtained in \cite{Kaul:1984bp} for the case of gauge algebra $\mathfrak{g} = \mathfrak{su}(2)$.  See also \cite{Rebhan:2004vn,Rebhan:2006fg} where comparison is made to the Seiberg--Witten formula for the monopole mass.

The generalization of \eqref{Idensity} to the case of $\NN = 2$ SYM with simple compact gauge group $G$ and an arbitrary number of 't Hooft defect insertions is a byproduct of the analysis in \cite{MRVdimP1}, and it can be used to determine the one-loop correction to the monopole mass $M_{\gm}$ for this class of theories.  The details of this computation will be given elsewhere, while here we simply state the result.  The one-loop correction, computed in the same renormalization scheme as \eqref{Lct}, is
\begin{align}\label{QMcorrection}
\Delta M_{\gm} =&~ \frac{1}{2\pi} \sum_{\alpha \in \Delta^+} \langle \alpha, \gm \rangle \langle \alpha, X_\infty \rangle \left\{ \ln \left( \frac{ \langle \alpha, X_\infty \rangle^2}{2\mu_{0}^2} \right) + 1 + O\left( \frac{\langle\alpha, Y_\infty\rangle^2}{\langle \alpha, X_\infty \rangle^2}\right) \right\} + \cr
&~ + \frac{1}{\pi} \sum_{\alpha \in \Delta^+} \langle \alpha, \gm \rangle \langle \alpha, Y_\infty \rangle \theta_\alpha~.
\end{align}

A couple of comments are in order.  Notice that the first line is an approximation that is valid when
\begin{equation}
|\langle \alpha, Y_\infty \rangle| \ll \langle \alpha, X_\infty \rangle~, \quad \forall \alpha \in \Delta^+~.
\end{equation}
This assumption is necessary in order for the index density \eqref{Idensity} to be the relevant quantity for determining the spectral asymmetry of fluctuations.  (We also just argued it is natural if one wishes to consider electric charges that are not parametrically large.)  If it is violated then the monopole background is not the appropriate background to expand around.  As emphasized in \cite{Peeters:2001np,Lee:2011ph}, it is quite natural from the point of view of the collective coordinate dynamics to tie this small parameter to the small parameter controlling the collective coordinate velocities.  Thus we will assume
\begin{equation}\label{scassumption}
 \max_{\alpha \in \Delta^+} \left\{ \frac{|\langle \alpha, Y_\infty \rangle|}{\langle \alpha, X_\infty\rangle} \right\} \lesssim O(g_0) \sim \dot{z}^m~,
\end{equation}
for most of this section.  However in \ref{ssec:validity} we will conjecture how to extend results for certain protected quantities to all orders in $|\langle \alpha,Y_\infty \rangle|/\langle \alpha, X_\infty \rangle$.  We will refer to the approximation in these quantities as the \emph{weak potential energy approximation}.

Second, $\theta_\alpha$ is the argument of $\langle \alpha,a\rangle$ as in \eqref{1looptau}.  The origin of this term is the same as there: the natural fermionic variables to use in studying the fluctuation spectrum are related by a chiral $U(1)_R$ rotation to the canonical field theory ones, $\psi^A$.  Hence there is an ABJ anomaly contribution to the sum over the spectrum, that appears to have been previously overlooked in this context.  Note that with \eqref{scassumption} this term is $O(g_0)$ relative to the terms in the first line of \eqref{QMcorrection} and therefore it is consistent to keep it while neglecting the $O(g_{0}^2)$ corrections.      

Recall from \eqref{flux} that if the electric charge $\gamma_{\rm e} = 0$ then $\gamma_{\rm e}^{\rm phys} = - \frac{\theta_0}{2\pi} \gm$.  Thus the classical contribution to the monopole mass, \eqref{MBPScl}, is $M_{\gm}^{\rm cl} = \frac{4\pi}{g_{0}^2} (\gm,X_\infty) - \frac{\theta_0}{2\pi} (\gm, Y_\infty)$.  Combining the classical piece and the one-loop correction using \eqref{adtr}, one finds
\begin{align}\label{Qmonomass}
M_{\gm}^{\textrm{1-lp}} =&~ M_{\gm}^{\rm cl} + \Delta M_{\gm} \cr
=&~ \frac{1}{2\pi} \sum_{\alpha \in \Delta^+} \langle \alpha, \gm \rangle \langle \alpha, X_\infty \rangle \left\{ \ln \left( \frac{ \langle \alpha, X_\infty \rangle^2}{2 |\Lambda|^2} \right) + 1 + O\left( \frac{\langle\alpha, Y_\infty\rangle^2}{\langle \alpha, X_\infty \rangle^2}\right) \right\} + \cr
&~ + \frac{1}{\pi} \sum_{\alpha \in \Delta^+} \langle \alpha, \gm \rangle \langle \alpha, Y_\infty \rangle \left( \theta_\alpha - \frac{\theta_0}{2h^\vee} \right)~,
\end{align}
where $\Lambda$ is the dynamical scale, \eqref{dyscale}.

This result is consistent with the framed BPS mass $M_{\gm} = - \Re(\zeta^{-1} Z_{\gm})$, as computed using the one-loop corrected Seiberg--Witten central charge.\footnote{It is a separate matter to show that the one-loop corrections to the central charge, $Z_{\gm}$, \emph{as computed in the microscopic (UV) theory}, are consistent with the Seiberg--Witten central charge \cite{Rebhan:2004vn,Rebhan:2006fg}; we will assume that these results can also be generalized to the case at hand. \label{fn:Z1loop}}  From the one-loop dual coordinate, \eqref{pertaD}, we determine the central charge associated with a purely magnetic charge:
\begin{equation}
Z_{\gm}^{\textrm{1-lp}} = \frac{i}{2\pi} \sum_{\alpha \in \Delta^+} \langle \alpha, \gm \rangle \langle \alpha, a \rangle \left\{ \ln \left( \frac{ \langle \alpha , a\rangle^2}{2\Lambda^2} \right) + 1 \right\}~.
\end{equation}
Then, using \eqref{XYasymptotic}, we find
\begin{align}\label{Zgm1loop}
- \Re\left(\zeta^{-1} Z_{\gm}^{\textrm{1-lp}}\right) =&~ \frac{1}{2\pi} \sum_{\alpha \in \Delta^+} \langle \alpha, X_\infty \rangle \langle \alpha, \gm \rangle \left\{ \ln \left( \frac{ \langle \alpha, X_\infty \rangle^2 + \langle \alpha, Y_\infty \rangle^2}{2 |\Lambda|^2} \right) + 1 \right\} \cr
&~ +  \frac{1}{\pi} \sum_{\alpha \in \Delta^+} \langle \alpha, \gm \rangle \langle \alpha, Y_\infty \rangle \left( \theta_\alpha - \frac{\theta_0}{2h^\vee} \right)~,
\end{align}
which agrees with \eqref{Qmonomass} and predicts the form of the corrections in $\langle \alpha, Y_\infty \rangle^2/ \langle \alpha, X_\infty \rangle^2$.

These results are valid in both framed and vanilla cases.  In the framed case it is interesting to observe that even when there are no non-Abelian monopoles present, $\tilde{\gamma}_{\rm m} = 0$, the energy of the ``vacuum'' still receives quantum corrections.  However, for configurations where the non-Abelian monopoles screen the defects such that $\gm  =0$, the correction vanishes.

%%%%%%%%%%%%%%%%%%%%%
\subsection{Fermionic zero modes}
%%%%%%%%%%%%%%%%%%%%%

Although we put the fermion fields to zero in the classical BPS solutions, these fields can have zero-mass excitations around the solution that play an important role in the moduli space approximation.  Here we review the space of fermionic zero modes, emphasizing its quaternionic structure and how this is related to transformations under $SU(2)_R$.

The $\NN=2$ theory comes with an $SU(2)_R$ doublet of Weyl fermions $\psi^A$. As discussed in appendix \ref{N2conventions}, it is natural to split $\psi^A$ into two symplectic-Majorana--Weyl components, $\psi^A=\zeta^{\frac{1}{2}}(\rho^A+i\l^A)$, with
\begin{equation}\label{smwdef}
\rho(x),\lambda(x) \in S^{+}_{\mathrm{smw}} \otimes \mathfrak{g}~, \qquad S_{\rm smw}^+ :=\left\{\varepsilon\in S^+\otimes \mathbb{C}^2\,|\, \bar\varepsilon^A=\overbar{\sigma}^0\varepsilon^A\right\}~,
\end{equation}
where $S^+$ denotes the space of positive chirality Weyl spinors, and the $SU(2)_R$ index $A$ corresponds to the $\mathbb{C}^2$ factor.  Spinor indices will usually be suppressed; canonical placement and contraction for spinor indices is understood.  The change of basis $\psi^A \mapsto \{ \lambda^A, \rho^A \}$ mirrors that of the supersymmetires, $Q^A \mapsto \{ \RR^A, \TT^A \}$.  We will find that $\rho^A$ possesses zero modes while $\lambda^A$ does not.  

Note that the map $\mathfrak{c}$ defined by $\psi_{\alpha}^A \mapsto \mathfrak{c}(\psi)_{\alpha}^A = (\s^0)_{\a\adot} \psibar^{\adot A}$ defines a real structure on the complex four-dimensional vector space $S^+ \otimes \mathbb{C}^2$.  $S_{\rm smw}^+$ is the fixed-point locus of $\mathfrak{c}$ and is thus a real four-dimensional vector space.

$S_{\rm smw}^+$ can be endowed with a quaternionic structure, making it a quaternionic one-dimensional vector space.  As the symplectic-Majorana--Weyl condition, (\ie\ $\mathfrak{c}$-fixed point condition), is invariant under the $SU(2)_{R}$ action, the following are three well-defined complex structures  on $S^{+}_{\mathrm{smw}}$, satisfying the quaternion algebra:
\begin{equation}\label{quat2}
(\mathcal{J}^r\{\rho,\lambda\})^A \equiv (\mathcal{J}^r)^{A}_{\phantom{A}B} \{\rho,\lambda\}^B :=  (-i\sR^r)^{A}_{\phantom{A}B} \{\rho,\lambda\}^B~,\qquad r=1,2,3\,,
\end{equation}
with $(\sigma^r)^{A}_{\phantom{A}B}$ the standard Pauli-matrices.

After this purely algebraic aside we can go back to the dynamics of the theory. The equations of motion for $\lambda^A$ and $\rho^A$ can be derived from the action in the form (\ref{action2ndform}) and are equivalent to the following two Dirac equations
\bea
0 &=& -i \left(D_0 \rho^A + [Y,\rho^A]\right) +\sigma^0\overbar{\sigma}^i D_i\l^A-i[X,\l^A]\,,\label{firstfermeq}\\
0 &=& i \left(D_0 \lambda^A - [Y,\lambda^A] \right) + \sigma^0\overbar{\sigma}^i D_i \r^A+i [X,\r^A] \,.\label{secfermeq0}
\eea
Let us introduce the Euclidean sigma matrices 
\begin{equation}\label{taumat}
(\tau^a)_{\a}^{\phantom{\a}\b} = ( \s^0 \sb^i, -i \mathbbm{1})_{\a}^{\phantom{\a}\b}~, \qquad (\bar{\tau}^a)_{\a}^{\phantom{\a}\b} = (\s^0 \sb^i, i\mathbbm{1})_{\a}^{\phantom{\a}\b}~.
\end{equation}
Assuming fields to be time-independent and working in the generalized temporal gauge, $A_0 = Y$, these equations reduce to
\begin{align}
L \rho^A \equiv i \bar{\tau}^a \hat{D}_a \rho^A =&~ 0\,, \label{Dirac2}  \\ 
L^\dag \lambda^A \equiv -i \tau^a \hat{D}_a \lambda^A  =&~  2[Y,\rho^A]\, .  \label{secfermeq}
\end{align}
The operators $L,L^\dag$ are Hilbert-space adjoints of each other acting on $\Lsq[\UU,S_{\rm smw}^+ \otimes \mathfrak{g}]$.  Indeed, one can show\footnote{Take the conjugate of $L \varepsilon^A = 0$ and use the identity $(\bar{\tau}^{a \ast})_{\adot}^{\phantom{\adot}\bdot} = - \epsilon_{\adot\dot{\delta}} (\sb^0)^{\dot{\delta}\alpha} (\bar{\tau}^a)_{\a}^{\phantom{\alpha}\beta} (\s^0)_{\beta\gdot} \epsilon^{\gdot\bdot}$.} that $L \varepsilon^A = 0 \iff L (\sigma^0 \bar{\varepsilon}) = 0$.  Thus $\mathfrak{c}$ maps $\ker{L}$ to $\ker{L}$ and the symplectic-Majorana--Weyl condition can be consistently imposed.  Using the fact that $\bar{\tau}^{[a} \tau^{b]}$ is anti-self-dual together with the fact that the background fieldstrength $\hat{F}_{ab}$ is self-dual, one finds that
\begin{eqnarray}
L L^\dag &=& - \hat{D}^a \hat{D}_a\, ,
\end{eqnarray}
a positive-definite operator.  It follows that $L^\dag$ has no zero modes and so \eqref{secfermeq} has a unique solution for $\l_A$, given a $\r_A$. 

In contrast the operator $L$ does have zero modes.  We define
\begin{equation}\label{fzmspace}
\mathcal{S}_{[\hat{A}]} := \left\{ \rho^A \in \Lsq[\mathbb{R}^3, S^+_{\mathrm{smw}} \otimes \mathfrak{g}] ~ |~ L\rho^A = 0 \right\}~,
\end{equation}
where the notation is meant to indicate that there is a vector space of zero modes for each point $[\hat{A}] \in \fMM$.  In fact $L$ is precisely the Dirac operator that appears in the analysis of bosonic zero modes and determines the dimension of the moduli space.  The real dimension of the moduli space is twice the dimension of $\ker{L}$, when viewed as an operator acting on the space of complex-valued spinors, $\psi \in \Lsq[\mathbb{R}^3, S^+ \otimes \mathfrak{g}]$.  Here $L$ is acting on spinors valued in the real vector space $S_{\rm smw}^+ \otimes \mathfrak{g}$, and for each (complex) solution in $S^+$ we obtain two real independent solutions in $S_{\rm smw}^+$.  Hence the dimensions of $\SS_{[\hat{A}]}$ and $T_{[\hat{A}]} \fMM$ as real vector spaces agree.

In fact $\SS_{[\hat{A}]}$ can be given a quaternionic structure: the $SU(2)_R$ action \eqref{quat2} commutes with the Dirac operator and descends to a quaternionic structure on the space of fermionic zero modes.  

% % % % % % % % % % % % %  % %  
\subsubsection{The quaternionic map between bosonic and fermionic zero modes}
% % % % % % % % % % % % % % %

In this subsection we demonstrate that the space of fermionic zero modes $\mathcal{S}$ is naturally quaternion-isomorphic to the space of bosonic zero modes $T\fMM$, when each of them is equipped  respectively with the quaternionic structures (\ref{quat2}) and (\ref{quatstructure}). Before we continue let us stress that in this subsection we assume all spinorial variables are classically \emph{commuting} fields. The physical fermions will be related to the the objects defined here by Grassmann-valued coefficients, as we discuss in the next section.

We can construct a family of quaternion-linear isomorphisms explicitly as follows. Given a solution to the Dirac equation \eqref{Dirac2}, $\rho^{A}$, one obtains a bosonic zero mode through the map $T_{\kappa} : \mathcal{S} \to T \fMM$, given by
\begin{equation}\label{Tmap}
\rho^A \xmapsto{T_\kappa} \delta \hat A_a = 2\kappa_A \bar\ge_a \rho^A~.
\end{equation}
Note that the family $T_\kappa$ is parameterized by a constant symplectic-Majorana-Weyl spinor $\kappa^A \in S^+_{\mathrm{smw}}$. The origin of this map lies in the action of the broken supersymmetries (\ref{etavar1}, \ref{etavar2}), where $\eta^A=\kappa^A$ is the susy parameter.  We use the same symbol, $\kappa \in S_{\rm smw}^+$, as for the symplectic Majorana--Weyl spinor appearing in \eqref{RRkappa}, which was used to construct a supercharge $\RR_{\kappa} := \kappa^{\alpha}_A \RR_{\alpha}^A$ that commutes with the diagonal of spatial rotations and $\mathfrak{su}(2)_R$.  Indeed, we will see later when we study the supersymmetry of the collective coordinate theory that these $\kappa$'s are one and the same.  One can explicitly check that indeed $\delta \hat A_a$ satisfies the linearized self-duality equation \eqref{linearsd} and gauge orthogonality condition \eqref{gaugeorth}, provided $\rho^{A}$ satisfies the Dirac equation \eqref{Dirac2}.  To do this one notes first that \eqref{linearsd} and \eqref{gaugeorth} are equivalent to the Dirac equation $L (\tau^a \delta \hat{A}_a) = 0$ for the bispinor ${(\tau^a)_\a}^\b \delta \hat{A}_a$, and second that ${(\tau^a)_\a}^\b \delta \hat{A}_a = 4 \kappa_{A}^\beta \rho_{\alpha}^A$ with \eqref{Tmap}.

For those $\kappa^A$ for which $\det \kappa := -\frac{1}{2}\kappa^B_\beta \kappa^\beta_B\neq 0$, the map $T_\kappa$ has a well-defined inverse
\begin{equation}\label{inverseTmap}
\delta \hat A_a \xmapsto{T^{-1}_\kappa} \rho^{A} =  -\frac{1}{4 \det \kappa}\delta\hat A_a \ge^a \kappa^A.
\end{equation}
That this provides both a left and right inverse for $T_\kappa$ is easily checked using the identities $\kappa^\a_B\kappa^A_\a=-\det\kappa\, {\delta_B}^A$ and $\kappa^\a_A\kappa^A_\b=-\det\kappa\, {\delta^\a}_\b$. Furthermore, solutions of the linearized self-dual equations get mapped to solutions of the Dirac equation. This establishes that $T_\kappa$ is a family of isomorphisms between $\mathcal{S}$ and $T\fMM$, parameterized by those $\kappa \in S_{\rm smw}^+$ such that $\det\kappa\neq0$.

Furthermore those isomorphisms $T_\kappa$ also connect the quaternionic structures on both spaces.
As a first step in proving this, it is useful to define a canonical map
\begin{eqnarray} \label{so3map}
S^+_{\mathrm{smw}} & \rightarrow & SO(3) \cr
\kappa &\mapsto & R_\kappa\, ,
\end{eqnarray}
through the relation\footnote{This is the same as \eqref{Rotkappa}.  The existence of coefficients ${(R_{\kappa})^r}_s$ such that \eqref{so3def} is satisfied can be verified by direct calculation.  Note that a generic $\kappa \in S_{\rm smw}^+$ can be parameterized as $(\kappa)_{\alpha}^{\phantom{\alpha}A} = \left(\begin{array}{c c} z & w \\ \bar{w} & - \bar{z} \end{array}\right)$ for $z,w \in \mathbb{C}$; in particular $\det{\kappa} \leq 0$.}
\be
(\mathcal{J}^r)^A_{\phantom{A}B}\kappa^{B}=-(R_\kappa)^r_{\phantom{r}s}(-i\tau^s)\kappa^{A}\,.\label{so3def}
\ee
Using that both $\mathcal{J}^r$ and $-i\tau^r$ satisfy a quaternionic algebra it follows that $(R_\kappa)^r_{\phantom{r}t}(R_\kappa)^s_{\phantom{s}t}=\delta^{rs}$ and $\frac{1}{3!}\epsilon_{rst}\epsilon^{uvw}R^{r}_{\phantom{r}u}R^{s}_{\phantom{s}v}R^{t}_{\phantom{t}w}=1$\,, so we see that indeed $R_\kappa\in SO(3)$.  Secondly there is the identity
\be
i\tau^r\bar\tau_a=(-\bar{\eta}^r)_a^{\phantom{a}b}\bar\tau_b\qquad \mbox{for }r=1,2,3\,, \label{taudentity}
\ee
where the $\bar{\eta}^r$ are the anti-self-dual 't Hooft symbols \eqref{asdtHooft}.  Recall that these appear in the quaternionic structure on Euclidean four-space, \eqref{R4cs}, used to define the quaternionic structure on bosonic moduli space via \eqref{quatstructure}.  Using the map \eqref{so3map}, \eqref{so3def} and the relation \eqref{taudentity}, it is then a matter of algebra to verify that
\be \label{quatpres}
T_\kappa(\mathcal{J}^r(\rho))= \mathbb{J}^r\left(T_\kappa\left(\rho\right)\right)\,,
\ee
where $\mathbb{J}^r$ is the quaternionic structure on $T\fMM$.  (The simple form of this result is one of the reasons we chose to introduce $R_\kappa$ into the definition of the complex structures on $T\fMM$, \eqref{quatstructure}.)  In summary we see that the spaces of fermionic and bosonic zero modes are quaternion-isomorphic:
\begin{equation}
\mathcal{S} \overset{\mathbb{H}}{\cong} T \fMM \,.
\end{equation}

%%%%%%%%%%%%%%%%%%%%%
\subsection{Supersymmetric collective coordinate dynamics}\label{ssec:ccL}
%%%%%%%%%%%%%%%%%%%%%

We now have the necessary tools to determine the $O(1)$ part of the soliton sector Hamiltonian \eqref{solitonHam} corresponding to the dynamics of the collective coordinates.  Ideally, one would like to have the exact quantum Hamiltonian in the soliton sector, which one could then truncate to the collective coordinate degrees of freedom.  Unfortunately, as we discussed, the exact canonical transformation from perturbative sector field variables to soliton sector variables is generally not available for four-dimensional gauge theories.  Therefore a different approach is typically taken: one first truncates the classical theory to the collective coordinate degrees of freedom, resulting in a sigma model with target given by the soliton moduli space, and then one quantizes.\footnote{\label{fn:oporder}In general the operations of truncation and quantization do not commute.  From the sigma model point of view, there is a well-known operator ordering ambiguity that manifests itself in a potential energy term involving the Ricci scalar, whose coefficient cannot be determined starting from just the data of the sigma model \cite{DeWitt:1952js,DeWitt:1957at}.  Additionally there can be extrinsic curvature terms that encode how the soliton moduli space is embedded in the infinite-dimensional field configuration space.  See \cite{Fujii:1997pt,Moss:1998jf,Papageorgakis:2014jia} for discussions of this phenomenon in the field theory context.  Other operators in addition to the Hamiltonian can suffer such ambiguities.  In contrast, the field theory does not have these ambiguities, and the full canonical transformation to the soliton sector would determine the correct operator orderings for expressions built out of the collective coordinate variables.  For supersymmetric sigma models with four supercharges it is expected that the `quantum potential' corrections to the Hamiltonian vanish thanks to nonrenormalization theorems \cite{Howe:1987qv}.  This is obvious for the Ricci scalar in the situation considered here where the target manifold is hyperk\"ahler.  However for the Hamiltonian this discussion is somewhat moot in any event, since these potentials correspond to $O(g_{0}^2)$ terms in the semiclassical expansion \cite{Papageorgakis:2014jia}---the same order as other terms we neglect, due to \eqref{scassumption}.  For the remaining operators of interest we will also be able to bypass these issues.}  This is the approach we will be following here, where in this section we carry out the classical truncation and in the next we quantize.  Although the collective coordinate expansion of $\NN =2$ SYM is an old and well-understood subject in the vanilla case \cite{Gauntlett:1993sh,Gauntlett:1999vc,Gauntlett:2000ks}, we will find that a nonzero theta angle in the presence of \tHooft defects leads to a completely new set of terms in the collective coordinate Lagrangian.

% % % % % % % % % % % % % % %
\subsubsection{Collective coordinate ansatz}
% % % % % % % % % % % % % % % 

We begin by writing the field theory Lagrangian \eqref{action} in terms of variables that are adapted to the soliton analysis---the real scalars $X,Y$ and the symplectic Majorana-Weyl fermions $\rho,\lambda$.  See appendix \ref{N2conventions} for further details.  We also group together $\hat{A}_a = (A_i,X)$, and separate the terms involving these fields from those involving time derivatives.  The Lagrangian takes the form
\begin{align}\label{actionrealform}
L =&~ \frac{1}{g_{0}^2} \int_{\UU} \ed^3 x \Tr \bigg\{  (\hat{D}_a A_0 - \pd_0 \hat{A}_a)^2 - \half \hat{F}_{ab} \hat{F}^{ab} + (D_0 Y)^2 - \hat{D}_a Y \hat{D}^a Y  + \cr
&~ \quad - 2i \rho^A (D_0 \rho_A + [Y,\rho_A]) - 2i \lambda^A (D_0 \lambda_A - [Y,\lambda_A]) - 2 \lambda^A \bar{\tau}_a \hat{D}_a \rho_A + 2 \rho^A \tau^a \hat{D}_a \lambda_A \bigg\} + \cr
&~ + \frac{\theta_0}{4\pi^2} \int_{\UU} \ed^3 x \Tr \left\{ E^i B_i \right\} + L_{\rm def}~,
\end{align}
where the $\tau^a$ and $\bar{\tau}^a$ matrices were introduced in \eqref{taumat}.  We will sometimes write $\hat{E}_a := \hat{D}_a A_0 - \pd_0 \hat{A}_a$, viewing this as the electric field associated with the five-dimensional gauge field $(A_0,\hat{A}_a)$.

We will be expanding around a static magnetic solution, treating the collective coordinate velocities and secondary Higgs vev as small parameters, according to \eqref{scassumption}.  In addition to the bosonic collective coordinates $z^m(t)$ we introduce Grassmann-valued fermionic collective coordinates $\chi^m(t)$.  These are the coefficients of $\rho^A$ in an expansion along a basis of the space of fermionic zero modes, \eqref{fzmspace}, where we make use of the isomorphism $T_\kappa$ between this space and the space of bosonic zero modes.  The $z^m$ and $\chi^m$ will be related by supersymmetry and this dictates that we impose the scaling $\chi^m \sim O(g_{0}^{1/2})$ \cite{Gauntlett:2000ks}.  Thus we have
\begin{align}
\hat A_a(x) =&~ \hat A_a(\vec{x},z(t)) + O(g_{0}^2) \,,\nonumber\\
\rho^A(x) =&~ \frac{1}{2\sqrt{-\det\kappa}}\delta_m\hat A_a(-i\tau_a)\kappa^A\chi^m + O(g_{0}^{5/2})~, \label{modspacea0}
\end{align}
for the fields possessing zero mode fluctuations.  The condition that $\rho^A$ is symplectic Majorana--Weyl is equivalent to the condition that $\chi^m$ is real.  

The remaining fields $A_0,Y,\lambda^A$ are determined by solving their equations of motion in the presence of \eqref{modspacea0}.  These equations are
\begin{align}\label{constrainedeoms}
\hat{D}^a \hat{E}_a + [Y,[Y,A_0]] -i \left( [\rho^A, \rho_A] + [\lambda^A, \lambda_A]\right) =&~ 0~, \cr
\hat{D}^2 Y - D_{0}^2 Y + i \left([\rho^A, \rho_A] - [\lambda^A, \lambda_A]\right) =&~ 0 ~, \cr
 i \left( D_0 \rho_A + [Y,\rho_A] \right) - \tau^a \hat{D}_a \lambda_A =&~ 0 ~.
\end{align}
Under the scaling assumptions on $\dot{z}$ and $Y_\infty$, the solution for $A_0,Y$ should start at $O(g_0)$.  Recalling the defining property of the local gauge parameters $\varepsilon_m$ in \eqref{coordflows}, we also have that $\hat{D}^a (\pd_0 \hat{A}_a) = \dot{z}^m \hat{D}^a \pd_m \hat{A}_a = \dot{z}^m \hat{D}_a \varepsilon_m$.  Finally one can show
\begin{equation}
[\rho^A, \rho_A] = -\half [\delta_m \hat{A}^a, \delta_n \hat{A}_a] \chi^m \chi^n + O(g_{0}^3)~,
\end{equation}
evaluated on \eqref{modspacea0}.  Therefore \eqref{constrainedeoms} can be put in the form
\begin{align}\label{constrainteoms2}
\hat{D}^2 (A_0 - \dot{z}^m \varepsilon_m) + \frac{i}{2} [\delta_m \hat{A}^a, \delta_n \hat{A}_a] \chi^m \chi^n =&~ O(g_{0}^3)~, \cr
\hat{D}^2 Y  -  \frac{i}{2} [\delta_m \hat{A}^a, \delta_n \hat{A}_a] \chi^m \chi^n =&~ O(g_{0}^3)~, \cr
\tau^a \hat{D}_a \lambda_A =&~ O(g_{0}^{3/2})~.
\end{align}
Since the kernel of $L^\dag = -i \tau^a \hat{D}_a$ is trivial, the last equation implies $\lambda^A = O(g_{0}^3/2)$, which is consistent with our neglect of the $[\lambda^A,\lambda_A]$ terms in the first two equations.  In order to solve the first two equations one can make use of \eqref{unibundle} for the moduli space components of the curvature of the universal bundle.

Up to this point the analysis has proceeded in a formally identical fashion to the standard vanilla case.  Now, however, there is a new consideration that must be taken into account.  Namely, the harmonic parts of $A_0,Y$ are non-vanishing and must be chosen such that \emph{both} the defect and asymptotic boundary conditions are satisfied.  Since we take the \tHooft charges to be constant, $A_0$ must carry the poles that generate the singular $E$-field in \eqref{defectbcs}, and thus should have the same behavior as $Y$ in the vicinity of the defects.  Meanwhile we require $Y \to Y_\infty$ asymptotically while $A_0 \to 0$, in accordance with the global Gauss law constraint, \eqref{Gaussbc}.  There is a unique solution in both cases, a fact that follows from the arguments given around \eqref{curlyY}.  The harmonic part of $A_0$ is
\begin{equation}\label{A0har}
A_{0}^{\rm h} := - \tilde{\theta}_0 (X - \epsilon_{X_\infty})~.
\end{equation}
Recall that $\epsilon_H$ is defined as the unique solution in \eqref{globalgts} satisfying $\hat{D}^2 \epsilon_H = 0$ and having asymptotic limit $H \in \mathfrak{t}$.  Hence the Higgs field $X$, which also solves $\hat{D}^2 X = 0$, takes care of the required defect poles, while subtracting $\epsilon_X$ ensures that $A_{0}^{\rm h} \to 0$ asymptotically.  

It is convenient to write the harmonic part of $Y$ in terms of $A_{0}^{\rm h}$, namely $Y^{\rm h} = A_{0}^{\rm h} + \epsilon_{Y_\infty}$.  This is equivalent to the expression \eqref{curlyY} since $\YY = \epsilon_{\YY_{\infty}^{\rm cl}} = \frac{4\pi}{g_{0}^2} \epsilon_{Y_\infty} + \frac{\theta_0}{2\pi} \epsilon_{X_\infty}$.  Hence the solution to the first two of \eqref{constrainteoms2} is
\begin{align}\label{A0Ysolve}
A_0 =&~ \dot{z}^m \varepsilon_m  - \frac{i}{4} \phi_{mn} \chi^m \chi^n + A_{0}^{\rm h} + O(g_{0}^3)~,  \cr
Y  =&~ \epsilon_{Y_\infty} + \frac{i}{4} \phi_{mn} \chi^m \chi^n + A_{0}^{\rm h} + O(g_{0}^3)~.
\end{align}
In the absence of defects $X = \epsilon_{X_\infty}$, implying $A_{0}^{\rm h} = 0$, and these expressions reduce to their standard form.  Notice that $A_{0}^{\rm h} \sim O(g_{0}^2)$ and it is consistent to keep this term relative to the terms we neglected.  It plays a crucial role in generating the new terms of the collective coordinate dynamics we alluded to above.  The solutions \eqref{A0Ysolve} together with the observation $\lambda^A = O(g_{0}^{3/2})$ imply that the terms we neglected in the $\hat{A}_a$ and $\rho$ equations of motion when writing \eqref{modspacea0} are consistent with the indicated order of the corrections there.  The orders to which we have worked for each field are sufficient for capturing all terms through $O(g_0)$ in the collective coordinate Lagrangian.

The evaluation of the Lagrangian \eqref{actionrealform} on the configurations \eqref{modspacea0} and \eqref{A0Ysolve} is delicate and care must be taken regarding boundaries.  Terms from the expansion of the defect Lagrangian serve not only to cancel out divergences, but contribute to the collective coordinate dynamics as well.  The details of the computation are provided in appendix \ref{app:cc}, and the final result is
\begin{align}\label{Lcc}
L_{\rm c.c.} =&~  \frac{4\pi}{g_{0}^2} \left[ \half g_{mn} \bigg( \dot{z}^m \dot{z}^n + i \chi^m \DD_t \chi^n - {\rm G}(Y_\infty)^m {\rm G}(Y_\infty)^n \bigg) - \frac{i}{2} \chi^m \chi^n \nabla_m {\rm G}(Y_\infty)_n \right] + \cr
&~ - \frac{4\pi}{g_{0}^2} (\gm, X_\infty) + L_{\rm c.c.}^{\theta_0} + O(g_{0}^2)~, \raisetag{20pt}
\end{align}
where
\begin{equation}\label{Lcctheta0}
L_{\rm c.c.}^{\theta_0} :=  \frac{\theta_0}{2\pi} \bigg\{ (\gm, Y_\infty) +  g_{mn} (\dot{z}^m - {\rm G}(Y_\infty)^m) {\rm G}(X_\infty)^n -i \chi^m \chi^n \nabla_m {\rm G}(X_\infty)_n \bigg\}~.
\end{equation}
If we set $\theta_0 = 0$ then \eqref{Lcc} is the same, \emph{in form}, as the standard vanilla result \cite{Gauntlett:1999vc,Gauntlett:2000ks},\footnote{We remind that the metric here is related to the conventional  one via $g_{mn} = \frac{g_{0}^2}{4\pi} (g_{\rm phys})_{mn}$ and that it is $g$ that is being used to lower the index of the vector fields, ${\rm G}(H)_m = g_{mn} {\rm G}(H)^n$.} though the target of the sigma model is the moduli space of \emph{singular} monopoles, $\fMM$, as defined in \eqref{Mdef}.  $\nabla_m$ is the covariant derivative with respect to the Levi--Civita connection on $\fMM$, and $\DD_t \chi^n := \dot{\chi}^n + \dot{z}^p \Gamma^{n}_{\phantom{m}pq} \chi^q$ involves the pullback of the covariant derivative to the worldline $z^m(t)$.  Recall that for any $H \in \mathfrak{t}$, ${\rm G}(H)$ is a triholomorphic Killing vector defined by the map \eqref{Gdef}.  In particular ${\rm G}(Y_\infty)^m$ is denoted $G^m$ in references \cite{Gauntlett:1999vc,Gauntlett:2000ks}.  The constant term in the potential is the mass of the static monopole.  This quantity is $O(g_{0}^{-2})$ in units of the Higgs vev, while the collective coordinate sigma model is $O(1)$ under the assumptions \eqref{scassumption}, taking into account that $\chi^m \sim O(g_{0}^{1/2})$ and $g_{mn} \sim O(1)$.

The $\theta_0$ terms of \eqref{Lcctheta0} are new.  In the vanilla case ${\rm G}(X_\infty)$ is a covariantly constant vector field generating  the flat $\mathbb{R}_{X_\infty}$ direction in $\MM$, \eqref{Mfactor}.  Using that $g({\rm G}(X_\infty),{\rm G}(Y_\infty)) = (\gm,Y_\infty)$ in the vanilla case, we see that $L_{\rm c.c.}^{\theta_0}$ collapses to
\begin{equation}
L_{\rm c.c.}^{\theta_0} = \frac{\theta_0}{2\pi} g_{mn} \dot{z}^m {\rm G}(X_\infty)^n~, \qquad \textrm{(vanilla case)}~,
\end{equation}
which is a total time derivative.\footnote{Due to the factorization \eqref{Mfactor}, the only component of $g_{mn}$ that contributes to this expression is one with both legs along $\mathbb{R}_{X_\infty}$; this component is independent of the $z^m$ and hence constant in time.}  In contrast, when defects are present then ${\rm G}(X_\infty)$ is not covariantly constant and $L_{\rm c.c.}^{\theta_0}$ contributes to the dynamics.  $L_{\rm c.c.}^{\theta_0}$ is $O(g_0)$ in units of the Higgs vev $X_\infty$.  One might wonder if it is consistent to keep these terms given that there are $O(g_0)$ terms in the field theory interaction Hamiltonian, \eqref{solitonHam}, that were neglected.  However if we restrict ourselves to the `vacuum' of the soliton sector where perturbative particle states above the soliton are not excited, then these terms can only contribute through loops and will thus be $O(g_{0}^2)$ corrections to the collective coordinate dynamics.

% % % % % % % % % % % % % %
\subsubsection{Collective coordinate supersymmetry}
% % % % % % % % % % % % % % 

It is well known that the standard (\ie\ $\theta_0 = 0$) sigma model preserves four supersymmetries.  Let us define
\begin{equation}
\bbJ^a = ( \mathbb{J}^r, \mathbbm{1})~, \qquad  \tbbJ^a = (- \mathbb{J}^r, \mathbbm{1})~,
\end{equation}
where the $\mathbb{J}^r$ are the triplet of complex structures on $\fMM$, defined in \eqref{quatstructure}.  Their components with respect to the coordinate basis where given in \eqref{metJcomp}.  Then the transformations
\begin{align}\label{ccsusy}
\delta_\nu z^m =&~ -i \nu_a \chi^n (\tbbJ^a)_{n}^{\phantom{n}m} + O(g_{0}^{5/2})~, \cr
\delta_\nu \chi^m =&~ \nu_a \left[ (\dot{z}^n - {\rm G}(Y_\infty)^n) (\bbJ^a)_{n}^{\phantom{n}m} - i \chi^p (\bbJ^a)_{p}^{\phantom{p}n} \Gamma^{m}_{\phantom{m}nq} \chi^q \right] + O(g_{0}^3)~,
\end{align}
where $\nu_a$, $a=1,\ldots,4$ are Grassmann-real parameters, leave the standard sigma model action invariant \cite{Gauntlett:1999vc,Gauntlett:2000ks}.  To show this one needs to use that the complex structures are covariantly constant and that ${\rm G}(Y_\infty)$ is a triholomorphic Killing vector.  We have also indicated the order in $g_0$ at which corrections to these transformations can appear, consistent with the order of corrections to the Lagrangian \eqref{Lcc}.  

One can check that the $\theta_0$-Lagrangian, \eqref{Lcctheta0}, is separately invariant under the same transformations, provided ${\rm G}(Y_\infty),{\rm G}(X_\infty)$ are \emph{commuting} triholomorphic Killing vectors.  This is guaranteed: ${\rm G}$ is a Lie algebra homomorphism from the \emph{Cartan} subalgebra $\mathfrak{t} \in \mathfrak{g}$ to the space of triholomorphic Killing vectors.  Hence, the full action $\int \ed t L_{\rm c.c.}$ is invariant.  Taking the variation with time-dependent parameters $\nu_a(t)$ we find, up to total time derivatives,
\begin{equation}
\delta_{\nu} L_{\rm c.c.} = -i \dot{\nu}_a Q^a~,
\end{equation}
where the Noether charges are
\begin{equation}\label{Qcc}
Q^a   = \frac{4\pi}{g_{0}^2} \chi^m (\tbbJ^a)_{m}^{\phantom{m}n} (\dot{z}_n - {\rm G}(Y_\infty)_n) + O(g_{0}^{3/2})~.
\end{equation}

This supersymmetry descends from the preserved $\RR$-supersymmetries of the field theory.  Indeed, one can derive \eqref{ccsusy} directly from the $\RR$-supersymmetry transformations of $\hat{A}_a, \rho^A$ evaluated on \eqref{modspacea0}, where the embeddings of the supercharges and parameters into their field theory counterparts are\footnote{Defining $\delta_\nu \Phi := \nu_a [Q^a, \Phi ]$ and $\delta_\varepsilon \Phi := \varepsilon_A [\RR^A, \Phi]$, the precise relation is $\delta_\nu = \delta_\varepsilon + \delta_{\rm gauge}$, where the gauge transformation parameter is $\epsilon = \varepsilon_m \delta_\nu z^m$.  The gauge transformation restores the gauge-orthogonality condition for the field theory variation.}
\begin{align}\label{susyembedding}
\RR^A = \frac{Q^a}{\sqrt{-\det{\kappa}}}  (i \bar{\tau}_a) \kappa^A~, \qquad \varepsilon_A = \frac{\nu_a}{2 \sqrt{- \det{\kappa}}} (i \bar{\tau}^a) \kappa_A~.
\end{align}
The symplectic Majorana--Weyl condition satisfied by $\RR^A$ and $\varepsilon_A$ implies that $\nu_a$ and $Q^a$ are Grassmann-real.  Note if we identify the $\kappa$ appearing in these formulae with the one in the discussion of the protected spin character around \eqref{RRkappa} then $Q^4 \propto \RR_{\kappa} \equiv \kappa^{\alpha}_A \RR_{\alpha}^A$.  Later, when we study the algebra of the conserved charges of the collective coordinate theory, we will see that this identification is indeed appropriate.

Let us also comment further on the nature of the corrections to the supercharges \eqref{Qcc}.  The reason we have these corrections, even though we are at the moment working with the classical collective coordinate theory, is the following.  We are demanding that the collective coordinate ansatz for the fields, \eqref{modspacea0} and \eqref{A0Ysolve}, constitute an exact solution to the time-independent equations of motion when we take the $z^m,\chi^m$ to be constant.  However we are forced to solve these equations perturbatively in the quantities $\chi^m \chi^n$ and $\langle \alpha, Y_\infty\rangle/ \langle \alpha, X_\infty \rangle$, which are assumed to be $O(g_0)$.  (When we consider the dynamics of the moduli $z^m$ we also assume that $\dot{z}$ is small and of the same order for the reasons discussed around \eqref{manton}.)  The reason we demand an exact solution to the time-independent equations of motion is that the definition of the soliton state in the quantum theory is based on an expansion around a true local minimum of the Hamiltonian functional.  

% % % % % % % % % % % % %
\subsubsection{Phase space structure and Hamiltonian}
% % % % % % % % % % % % % 

Suppose we take $z^m,\chi^n$ as the basic coordinates.  Then the conjugate fermion momentum determined from \eqref{Lcc} is $(p^\chi)_m = \frac{2\pi i}{g_{0}^2} g_{mn} \chi^n$.  After constructing the $\mathbb{Z}_2$-graded Dirac bracket $\{~,~\}_{\rm \pm}$ for this second-class constraint, we find $\{ \chi^m, \chi^n \}_+ = - \frac{4\pi i}{g_{0}^2} g^{mn}$.  Due to the $z$-dependence of the metric, it is inconsistent to assume that the brackets of $\chi^m$ with both $z^m$ and its conjugate momentum $p_m$ vanish.  If we assume $\{z^m,\chi^n \}_- = 0$, then the Jacobi identity will imply that $\{\chi^m, p_n \}_-$ must be nonzero.  In other words, $\chi^m$ must depend on $z$: $\chi^m = \chi^m(z(t),t)$.

We can extract the $z$ dependence of $\chi^m$ by introducing a frame\label{framestuff} $\EE_{\um} = \EE_{\um}^{\phantom{\um} m} \pd_m$ on the tangent bundle and co-frame $e^{\um} = e^{\um}_{\phantom{\um} m} \ed z^m$ on the co-tangent bundle, with $e^{\um}_{\phantom{\um} m} = \delta^{\um\un} \EE_{\un}^{\phantom{\un}n} g_{nm}$, and defining 
\begin{equation}
\chi^{\um} := e^{\um}_{\phantom{\um}m} \chi^m~,
\end{equation}
as the fundamental fermionic coordinate.  Here we use underlined indices $\um,\un,\ldots$, for the local frame.  Noting that $e^{\um}_{\phantom{\um}n} \nabla_p \EE_{\un}^{\phantom{\un}n} = \omega_{p,\phantom{\um}\un}^{\phantom{p,}\um}$, the spin connection, the fermion kinetic terms of $L_{\rm c.c.}$ become
\begin{equation}
\frac{2\pi i}{g_{0}^2} g_{mn} \chi^m \DD_t \chi^n = \frac{2\pi i}{g_{0}^2} \delta_{\um\un} \chi^{\um} \left( \dot{\chi}^{\un} + \dot{z}^p \omega_{p,\phantom{\un}\uq}^{\phantom{p,}\un} \chi^{\uq} \right)~,
\end{equation}
and the conjugate momenta are
\begin{equation}
p_m = \frac{4\pi}{g_{0}^2} \left[ g_{mn} ( \dot{z}^n + \tilde{\theta}_0 {\rm G}(X_\infty)^n ) + \frac{i}{2} \omega_{m,\up\uq} \chi^{\up} \chi^{\uq} \right]~, \qquad (p^\chi)_{\um} = \frac{2\pi i}{g_{0}^2} \chi_{\um}~.
\end{equation}
These expressions are expected to receive corrections from the higher order terms in the $g_0$ expansion of the Lagrangian, but we suppress them here.  We construct the Dirac bracket, taking $(z^m,\chi^{\um}; p_n, (p^\chi)_{\un})$ as the basic phase space coordinates, and we find
\begin{equation}\label{Dbs}
\{ z^m, p_n \}_- = \delta^{m}_{\phantom{m}n}~, \quad \{\chi^{\um},\chi^{\un} \}_+ = - \frac{i g_{0}^2}{4\pi} \delta^{\um\un}~, \quad \{z^m, \chi^{\un} \}_- = 0 = \{p_m, \chi^{\un} \}_- ~. 
\end{equation}

Following \cite{Gauntlett:1999vc} it is useful to introduce the `super-covariant' momentum
\begin{equation}
\pi_m := p_m - \frac{2\pi i}{g_{0}^2} \omega_{m,\up\uq} \chi^{\up} \chi^{\uq}~.\label{scovmom}
\end{equation}
Notice that 
\begin{equation}\label{pizdotrel}
\pi_m = g_{mn} \left(\frac{4\pi}{g_{0}^2} \dot{z}^n + \frac{\theta_0}{2\pi} {\rm G}(X_\infty)^n\right) ~,
\end{equation}
and thus the supercharges \eqref{Qcc} are expressed simply in terms of $\pi_n$:
\begin{align}\label{Qccphase}
Q^a =&~ \chi^m (\tbbJ^a)_{m}^{\phantom{m}n} \left( \pi_n - \frac{4\pi}{g_{0}^2} {\rm G}(Y_\infty)_n - \frac{\theta_0}{2\pi} {\rm G}(X_\infty)_n \right) + O(g_{0}^{3/2}) \cr
=&~ \chi^m (\tbbJ^a)_{m}^{\phantom{m}n} \left( \pi_n - {\rm G}(\mathcal{Y}_{\infty}^{\rm cl})_n \right) + O(g_{0}^{3/2})~.
\end{align}

Having an expression for the classical collective coordinate Hamiltonian will be useful below, where we will need to disentangle the Hamiltonian from the central charge when writing the supersymmetry algebra.  We Legendre transform $L_{\rm c.c.}$, and find it convenient to write the result in the following form:
\begin{align}\label{Hcc}
H_{\rm c.c.} =&~ p_m \dot{z}^m + (p^\chi)_{\um} \dot{\chi}^{\um} - L_{\rm c.c.} \cr
=&~ M_{\gm}^{\rm cl} + \frac{g_{0}^2}{8\pi} \pi_m g^{mn} \pi_n + \frac{2\pi}{g_{0}^2} g_{mn} \left[ {\rm G}(Y_\infty)^m {\rm G}(Y_\infty)^n + 2 \tilde{\theta}_0  {\rm G}(Y_\infty)^m {\rm G}(X_\infty)^n \right] + \cr
&~  + \frac{i}{2} \chi^m \chi^n \nabla_m \left( \frac{4\pi}{g_{0}^2} {\rm G}(Y_\infty)_n + \frac{\theta_0}{2\pi} {\rm G}(X_\infty)_n \right) + \cr
&~  - \tilde{\theta}_0 \left( {\rm G}(X_\infty)^m \pi_m - \frac{2\pi i}{g_{0}^2} \chi^m \chi^n \nabla_m {\rm G}(X_\infty)_n \right) + O(g_{0}^2)~.
\end{align}
Here the constant term is the classical mass of the magnetic background, including the contribution from the Witten effect:
\begin{equation}\label{Mgmcl}
M_{\gm}^{\rm cl} = \frac{4\pi}{g_{0}^2} (\gm, X_\infty) - \frac{\theta_0}{2\pi} (\gm, Y_\infty)~.
\end{equation}
We note that \eqref{Hcc} can equivalently be written as
\begin{align}\label{Hcc2}
H_{\rm c.c.} =&~ M_{\gm}^{\rm cl} + \frac{g_{0}^2}{8\pi} \bigg\{ \pi_m g^{mn} \pi_n + g_{mn} {\rm G}(\YY_{\infty}^{\rm cl})^m {\rm G}(\YY_{\infty}^{\rm cl})^n + \frac{4\pi i}{g_{0}^2} \chi^m \chi^n \nabla_m {\rm G}(\YY_{\infty}^{\rm cl})_n \bigg\} + \cr
&~ +i \tilde{\theta}_0 \left( i {\rm G}(X_\infty)^m \pi_m + \frac{2\pi}{g_{0}^2} \chi^m \chi^n \nabla_m {\rm G}(X_\infty)_n \right) + O(g_{0}^2)~.
\end{align}

Passing to the Hamiltonian of the quantum theory requires two steps.  First, the classical mass \eqref{Mgmcl} should be replaced by its one-loop counterpart, \eqref{Qmonomass}, which accounts for the leading quantum effects of integrating out the fluctuation fields around the soliton.  Second, the collective coordinate dynamics must be quantized; this is the subject of subsection \ref{sec:quantize} below.

% % % % % % % % % % % % %
\subsubsection{Collective coordinate $SU(2)_R$ symmetry}
% % % % % % % % % % % % % 

The parent field theory possesses an $SU(2)_R$ symmetry under which the bosons are inert while the fermions $\rho^A,\lambda^A$ transform as doublets.  The generators are in fact precisely the quaternionic structure \eqref{quat2} on the space of symplectic Majorana--Weyl spinors.  We will normalize them so that
\begin{equation}\label{su2Rfzm}
\delta^{r}_{(I)} \rho^A := \half (-i \sigma^r)^{A}_{\phantom{A}B} \rho^B = \half (\mathcal{J}^r)^{A}_{\phantom{A}B} \rho^B~.
\end{equation} 
The subscript $(I)$ is a reminder that these transformations are associated with the generators $I^r$ of $\mathfrak{su}(2)_R$.  They satisfy the algebra $[\delta^{r}_{(I)},\delta^{s}_{(I)}] = \epsilon^{rs}_{\phantom{rs}t} \delta^{t}_{(I)}$.

We can use the collective coordinate ansatz \eqref{modspacea0} to infer the $SU(2)_R$ transformation properties of $z^m,\chi^n$.  In order to simplify the presentation we will work with the leading order quantities in the $g_0$ expansion and comment on corrections at the end.  Hence all statements until that point should be understood as leading order statements.  Then since $\delta_{(I)}^r \hat{A}_a = 0$ we should take $z^m$ to be invariant.  Making use of the quaternion-linear isomorphism $T_\kappa$, the expression for the collective coordinate expansion of $\rho^A$ is equivalent to
\begin{equation}
T_\kappa(\rho) = 2i \sqrt{-\det{\kappa}} \ \delta_m \hat{A}_a \chi^m~.
\end{equation}
Therefore, on the one hand,
\begin{equation}
\delta^{r}_{(I)} (T_\kappa(\rho)) = 2i \sqrt{-\det{\kappa}} \ \delta_m \hat{A}_a (\delta^{r}_{(I)} \chi^m)~.
\end{equation}
On the other hand, using the linearity of $T_\kappa$ and \eqref{quatpres},
\begin{align}
\delta^{r}_{(I)} (T_\kappa(\rho)) =&~ T_\kappa (\delta^{r}_{(I)} \rho) =  \half T_\kappa\left( \mathcal{J}^r(\rho) \right) = \half \mathbb{J}^r\left( T_\kappa(\rho)\right) =  \half  \left[ 2i \sqrt{-\det{\kappa}} (\mathbb{J}^r)_{m}^{\phantom{m}n} \delta_n \hat{A}_a \chi^m \right] \cr
=&~ 2i \sqrt{-\det{\kappa}} \ \delta_m \hat{A}_a \left[ \half \chi^n (\mathbb{J}^r)_{n}^{\phantom{n}m} \right]~. \raisetag{20pt}
\end{align} 
Hence,
\begin{equation}\label{su2Rcc}
\delta^{r}_{(I)} z^m = 0~, \qquad \delta^{r}_{(I)} \chi^m = \half \chi^n (\mathbb{J}^r)_{n}^{\phantom{n}m}~.
\end{equation}
We see that $\mathfrak{su}(2)_R$ acts not on the moduli space itself but rather on its tangent bundle via the endomorphisms $\mathbb{J}^r$.  The fact that the $\mathbb{J}^r$ satisfy the quaternion algebra ensures that the transformations \eqref{su2Rcc} satisfy the $\mathfrak{su}(2)_R$ algebra.

One can check that \eqref{su2Rcc} is indeed a symmetry of the collective coordinate action, $\int \ed t L_{\rm c.c}$.  We note that this computation relies on ${\rm G}(X_\infty), {\rm G}(Y_\infty)$ being triholomorphic Killing vectors.  Let us introduce a triplet of generating parameters $\vartheta_r$ and set $\delta_{\vartheta} := \vartheta_r \delta^{r}_{(I)}$.  Then, carrying out the transformation with time dependent $\vartheta_r$, we find $\delta_{\vartheta} L_{\rm c.c.} = - \dot{\vartheta}_r I^r$, with Noether charges
\begin{equation}\label{su2Rnoether}
I^r := \frac{4\pi}{g_{0}^2} \left(\frac{i}{4} (\sw^r)_{mn} \chi^m \chi^n\right) = \frac{i \pi}{g_{0}^2} (\sw^r)_{\um\un} \chi^{\um} \chi^{\un} ~,
\end{equation}
where we recall that $(\sw^r)_{mn}$ is the triplet of K\"ahler forms given in \eqref{metJcomp}.  With the aid of the Dirac brackets \eqref{Dbs} one can demonstrate
\begin{equation}\label{SSDalg}
\{ I^r, I^s \}_- = \epsilon^{rs}_{\phantom{rs}t} I^t ~.
\end{equation}

The exact expressions for the $SU(2)_R$ charges of the parent field theory take the form \eqref{su2Rnoether} to leading order in the $g_0$ expansion in the (magnetic charge $\gm$) soliton sector.  We expect corrections to \eqref{su2Rnoether} from the $O(g_{0}^2)$ corrections to the collective coordinate Lagrangian, \eqref{Lcc}.  They will be suppressed relative to the leading order result by $g_{0}^2$ and  should be such that the algebra \eqref{SSDalg} is maintained.

% % % % % % % %% % % % %
\subsubsection{Target space isometries: electric charge and angular momentum}\label{sec:angmom}
% % % % % % % % % % % % %

The moduli space isometries discussed in subsection \ref{sec:modisom} give rise to additional symmetries of the collective coordinate theory.  Recall that we denoted the general collection of Killing vectors as $\{ K^E \}$.  In the vanilla case this collection consists of $\{K^i, K^r, K^A\}$ originating from translational symmetry, rotational symmetry, and asymptotically nontrivial gauge transformations, respectively.  Defects break translational symmetry, but if we have only a single defect then rotational symmetry is preserved.  The triholomorphic isometries associated with effectively-acting asymptotically nontrivial gauge transformations are always present (except of course in the case $\fMM \cong \{ {\rm pt} \}$).

It is well-known from early work on supersymmetric sigma models that target space isometries induce symmetries of the sigma model.  In the case at hand the infinitesimal symmetry transformation associated with the Killing vector $K^E$ is
\begin{equation}\label{ccEvar}
\delta^{E}_{\rm c.c.} z^m = (K^E)^m ~, \qquad \delta^{E}_{\rm c.c.} \chi^m = \chi^n \pd_n (K^E)^m ~.
\end{equation}
Notice that there is no sign included in the variation of $z^m$ like there was in \eqref{symEonA}.  In order to study the symmetry of the sigma model one changes from an active to passive point of view, regarding the isometries of the target space as changes of coordinates.  This is why we have used a different notation, $\delta_{\rm c.c.}$, for the induced variations of the collective coordinate theory.  The relationship is $\delta^{E}_{\rm c.c.} = - \delta^E$.  Indeed one can check that the signs in \eqref{ccEvar} must be exactly as given in order for the variation to satisfy the symmetry algebra, 
\begin{equation}
\delta^{E}_{\rm c.c.} \delta^{F}_{\rm c.c.} - \delta^{F}_{\rm c.c.} \delta^{E}_{\rm c.c.} = f^{EF}_{\phantom{EF}G} \delta^{G}_{\rm c.c.}~,
\end{equation}
using \eqref{Kvecalg}.

Let $\delta_{\rm c.c.} := s_E \delta_{\rm c.c.}^E$ denote a general variation with time-dependent parameters $s_E(t)$.  Then straightforward computation leads to
\begin{equation}
\delta_{\rm c.c.} L_{\rm c.c.} = - \dot{s}_E N^E~,
\end{equation}
with Noether charges
\begin{align}\label{NoetherK}
N^E :=&~ -  \frac{4\pi}{g_{0}^2} \left( (K^E)^m g_{mn} (\dot{z}^n + \tilde{\theta}_0 {\rm G}(X_\infty)^n) - \frac{i}{2} (\nabla_m (K^E)_n) \chi^m \chi^n \right) + O(g_0) \cr
=&~ - \left( (K^E)^m \pi_m - \frac{2\pi i}{g_{0}^2} (\nabla_m (K^E)_n) \chi^m \chi^n \right) + O(g_0)~,
\end{align}
where in the second step we used the relation \eqref{pizdotrel}.  The corrections are $O(g_{0}^2)$ suppressed relative to the leading terms, and originate from the $O(g_{0}^2)$ corrections to $L_{\rm c.c.}$.  One may verify that the Noether charges represent the algebra with respect to the Dirac bracket,
\begin{equation}
\{ N^E, N^F \}_{-} = f^{EF}_{\phantom{EF}G} N^G~.
\end{equation}
We will denote the charges corresponding to translational isometries, rotational isometries, and effectively-acting asymptotically nontrivial gauge transformations as
\begin{equation}
\{ N^i, N^r, N^A \} \equiv \{ P^i, \II^r, N_{\rm e}^A \}~.
\end{equation}

Since the $N_{\rm e}^A$ are generated by gauge transformations that asymptote to the fundamental magnetic weights $h^{I_A}$, they should be interpreted as the coefficients of the electric charge along the simple roots:
\begin{equation}\label{ccecharge}
\gamma_{\rm e} = \sum_{A} N_{\rm e}^A \alpha_{I_A}~.
\end{equation}
As a check of this statement, we can compute $\langle \gamma_{\rm e}, h^{I_A} \rangle = (\gamma_{\rm e}^\ast, h^{I_A})$ as the flux of the combination $-\frac{2}{g_{0}^2} (\vec{E} + \tilde{\theta}_0 \vec{B})$ through the asymptotic two-sphere and traced against $h^{I_A}$, according to the definitions \eqref{flux}, evaluated on the collective coordinate ansatz.  First we evaluate this combination of electric and magnetic field on the collective coordinate ansatz using \eqref{A0har}, \eqref{A0Ysolve}:
\begin{align}
E_i + \tilde{\theta}_0 B_i =&~  \dot{z}^m (D_i \varepsilon_m - \pd_m A_i) - \frac{i}{4} (D_i\phi_{mn}) \chi^m \chi^n + D_i A_{0}^{\rm h} + \tilde{\theta}_0 D_i X + O(g_{0}^3) \cr
=&~ - \dot{z}^m \delta_m A_i -  \frac{i}{4} (D_i\phi_{mn}) \chi^m \chi^n + D_i \epsilon_{X_\infty} + O(g_{0}^3) ~.
\end{align}
Then, recalling that $\hat{D}_a \epsilon_H = - {\rm G}(H)^m \delta_m \hat{A}_a$, we can write
\begin{align}\label{ccechargecheck}
\langle \gamma_{\rm e}, h^{I_A} \rangle =&~ \frac{2}{g_{0}^2} \int_{S_{\infty}^2} \ed^2 S^i \Tr \left\{ \left( \dot{z}^m \delta_m A_i - D_i \epsilon_{X_\infty} + \frac{i}{4} (D_i \phi_{mn}) \chi^m \chi^n \right) h^{I_A} \right\} +O(g_0)\cr
=&~ \frac{2}{g_{0}^2} \int_{\UU} \ed^3 x \Tr \left\{ \left( \dot{z}^m \delta_m \hat{A}_a - \hat{D}_a \epsilon_{X_\infty} \right) \hat{D}^a \epsilon_{h^{I_A}} + \frac{i}{4} \chi^m \chi^n (\hat{D}^2 \phi_{mn}) \epsilon_{h^{I_A}} \right\} + O(g_0) \cr
=&~ - \frac{4\pi}{g_{0}^2} g_{mn} (\dot{z}^m + {\rm G}(X_\infty)^m) {\rm G}(h^{I_A})^n + \cr
&~ + \frac{i}{g_{0}^2} \chi^m \chi^n \int \ed^3 x \Tr \left\{ [\delta_m \hat{A}^a, \delta_n \hat{A}_a] \epsilon_{h^{I_A}} \right\} + O(g_0)~. \raisetag{20pt}
\end{align}
The second line is equal to the first using integration by parts, $\hat{D}^a \delta_m \hat{A}_a = 0$, and fact that $\int \ed^3 x \Tr \{ \hat{D}^a \phi_{mn} \hat{D}_a \epsilon_H \} = 0$ for any $H \in \mathfrak{t}$.  The last identity holds via integration by parts, using $\hat{D}^2 \epsilon_H = 0$ and asymptotic analysis of $\phi_{mn}$ and $\hat{D}_a \epsilon_H$ to show that there are no boundary terms.  The final integral in \eqref{ccechargecheck} using \eqref{covKilling}.  Thus we find
\begin{equation}\label{ccechargecheck2}
\langle \gamma_{\rm e}, h^{I_A} \rangle = - (K^A)^m \pi_m + \frac{2\pi i}{g_{0}^2} \chi^m \chi^n \nabla_m (K^A)_n + O(g_0) = N_{\rm e}^A~,
\end{equation}
in agreement with \eqref{ccecharge}.

Let us compare this expression for the electric charge with the one we obtained previously by studying solutions to the secondary BPS equation, \eqref{gecl}.  In that analysis we set the fermion fields to zero, so to compare we should set $\chi^m = 0$.  The solutions constructed there were static in the generalized temporal (gt) gauge where $A_{0}^{\rm (gt)} = Y$.  However in order to compare with expressions obtained in this section we need to be in a gauge where $A_0 \to 0$ as $|\vec{x}| \to \infty$.  This corresponds to making the gauge transformation discussed under \eqref{temporalgauge}.  The new fields, $\hat{A}$, are related to the old ones, $\hat{A}^{\rm (gt)}$, by $\hat{A}^{\rm (gt)} = \cg^{-1} (\hat{A} + \ed ) \cg$, where $\cg$ asymptotes to $\cg_\infty = \exp(-Y_\infty t)$.  We choose $\cg$ by requiring that the corresponding infinitesimal transformation be orthogonal to local gauge transformations.  This implies $\cg = \exp(- \epsilon_{Y_\infty} t)$, and therefore on the one hand,
\begin{equation}
(\pd_t \hat{A}_a)^{\perp} \bigg|_{t=0} = \hat{D}_a \epsilon_{Y_\infty} \bigg|_{t=0} = {\rm G}(Y_\infty)^m \delta_m \hat{A}_a \bigg|_{t=0}~.
\end{equation}
On the other hand, via the collective coordinate ansatz \eqref{modspacea0}, we have $\pd_t \hat{A}_a = \dot{z}^m \pd_m \hat{A}_a$, so that the gauge orthogonal piece is
\begin{equation}
(\pd_t \hat{A}_a)^\perp \bigg|_{t=0} = \dot{z}^m (\pd_m \hat{A}_a - \hat{D}_a \varepsilon_m) \bigg|_{t=0} = \dot{z}^m \delta_m \hat{A}_a \bigg|_{t=0}~.
\end{equation}
Comparing the two gives $\dot{z}^m = {\rm G}(Y_\infty)^{m}$ at $t=0$.  However one can show that 
\begin{equation}\label{dyonccsol}
\dot{z}^m = {\rm G}(Y_\infty)^m~, \qquad \textrm{and} \qquad  \chi^m = 0~,
\end{equation}
is a solution to the equations of motion following from the collective coordinate Lagrangian \eqref{Lcc}.  Hence it is these solutions that correspond to the classical dyon solutions of section \ref{sec:cldyon}.  Then on these solutions we have from \eqref{pizdotrel} that $\pi_m = g_{mn} {\rm G}(\YY_{\infty}^{\rm cl})^n$, and thus \eqref{ccechargecheck} reduces to \eqref{econstraints}:
\begin{equation}
N_{\rm e}^A ~ \xrightarrow{\chi^m = 0~,~ \dot{z}^n  = {\rm G}(Y_\infty)^n} ~  - g_{mn} {\rm G}(\YY_{\infty}^{\rm cl})^m (K^A)^n =  n_{\rm e}^{I_A}~.
\end{equation}

Next consider the charges $N^r \equiv \II^r$.  It is tempting to identify them with the conserved charges associated with angular momentum, since the Killing vectors $K^r$ implement infinitesimal angular momentum transformations on $\hat{A}_a(\vec{x},z^m)$ according to \eqref{symEonA}.  However a short computation shows that the Dirac bracket of $\II^r$ with the $\mathfrak{su}(2)_R$ charge $I^s$, \eqref{su2Rnoether}, does not vanish!  This makes the interpretation of $\II^r$ as the charge associated with angular momentum problematic since it is clear that $SO(3)$ and $SU(2)_R$ are commuting symmetries of the $\NN = 2$ field theory.  The form of $\{ \II^r, I^s \}_{-}$ suggests that the correct definition of angular momentum involves a shift by the $\mathfrak{su}(2)_R$ symmetry charge.  Note that $z^m$ does not transform under $\mathfrak{su}(2)_R$, so this would still be consistent with expectations from \eqref{symEonA}.

The observation that $\II^r$ cannot be the generator of angular momentum has been made in \cite{deVries:2008ic,deVries:2010vc}, where the correct shift by a term involving the complex structures was also determined.  The connection between the action of the complex structures and $\mathfrak{su}(2)_R$ was not discussed.  Understanding this shift will be important for the correct interpretation of quantum labels that can be ascribed to BPS states.\footnote{Note that an analogous phenomenon has been observed in the low energy core-halo approach of \cite{Kim:2011sc}.}  We therefore provide an independent derivation of the required shift by starting with the angular momentum transformation of the field theory fermion $\rho^A$ and utilizing the collective coordinate ansatz \eqref{modspacea0}.

On the one hand, the angular momentum transformation of the Weyl spinor $\rho^A$ is
\begin{align}
\delta^{r}_{(J)} \rho_{\alpha}^A =&~ - \epsilon^{rjk} x_j \pd_k \rho_{\alpha}^A - \epsilon^{rjk} (\sigma_{jk})_{\alpha}^{\phantom{\alpha}\beta} \rho_{\beta}^A \cr
=&~ -  \epsilon^{rjk} x_j \pd_k \rho_{\alpha}^A - \frac{i}{2} (\tau^r)_{\a}^{\phantom{\a}\b} \rho_{\b}^A~.
\end{align}
Let us evaluate this on the collective coordinate expression $\rho_{\alpha}^A(x) = \rho_{\alpha}^A(\vec{x}; z,\chi)$ given in \eqref{modspacea0}.  The only dependence on $x^k$ is through the bosonic zero mode so
\begin{align}\label{angrhovar1}
\delta_{(J)}^r \rho_{\alpha}^A =&~ \frac{-i}{2\sqrt{-\det{\kappa}}} \bigg\{ \left( - \epsilon^{rjk} x_j \pd_k \delta_m \hat{A}_a\right) (\tau^a)_{\alpha}^{\phantom{\alpha}\b} \kappa_{\b}^A \chi^m  -\half (\delta_m \hat{A}_b) (i \tau^r \tau^b)_{\a}^{\phantom{\a}\b} \kappa_{\b}^A \chi^m \bigg\}~.
\end{align}
One may verify that the following identities hold amongst the $\tau^a$, the anti-self dual 't Hooft symbols $\bar{\eta}^r$, and the $\mathfrak{so}(3)$ representation matrices $\ell^r$ of \eqref{elldef}:
\begin{align}
-\frac{i}{2} (\tau^r \tau^b)_{\alpha}^{\phantom{\alpha}\beta} =&~  (\tau^a)_{\a}^{\phantom{\a}\b} (\ell^r)_{a}^{\phantom{a}b} - \frac{i}{2} ( \tau^b \tau^r )_{\a}^{\phantom{\a}\b} \cr
=&~ (\tau^a)_{\a}^{\phantom{\a}\b} (\ell^r)_{a}^{\phantom{a}b} + \half (\tau^a)_{\a}^{\phantom{\a}\b} (\bar{\eta}^r)_{a}^{\phantom{a}b}~.
\end{align}
The first of these terms combines with the first term in \eqref{angrhovar1} to give the angular momentum variation of the bosonic zero mode, $\delta^{r}_{(J)} (\delta_m \hat{A}_a)$, which is of the same form as \eqref{angAvar}.  For the second term we use the relations \eqref{R4cs} and \eqref{quatstructure} to convert the 't Hooft symbol to a complex structure.  This brings us to
\begin{align}\label{angrhovar2}
\delta_{(J)}^r \rho_{\alpha}^A =&~ \frac{-i}{2\sqrt{-\det{\kappa}}} \bigg\{ \left( \delta^{r}_{(J)} \delta_m \hat{A}_a\right) (\tau^a)_{\alpha}^{\phantom{\alpha}\b} \kappa_{\b}^A \chi^m  + \half (R_{\kappa}^{-1})^{r}_{\phantom{r}s}(\mathbb{J}^ss)_{m}^{\phantom{m}n} \delta_n \hat{A}_a (\tau^a)_{\a}^{\phantom{\a}\b} \kappa_{\b}^A \chi^m \bigg\}~.
\end{align}
Finally we convert the variation of the bosonic zero mode to a Lie derivative via \eqref{Liezm}, and then write $(\Lie_{K^r} \delta_m \hat{A}_a) \chi^m = \Lie_{K^r} (\delta_m \hat{A}_a \chi^m) - \delta_m \hat{A}_a (\Lie_{K^r} \chi^m)$.  Since the bosonic zero mode and $\chi^m$ are the only $z$-dependent quantities we have
\begin{align}\label{angrhovar3}
\delta_{(J)}^r \rho_{\alpha}^A =&~ - \Lie_{K^r} \rho_{\alpha}^A -  \frac{i}{2\sqrt{-\det{\kappa}}} \delta_m \hat{A}_a (\tau^a)_{\a}^{\phantom{\a}\b} \kappa_{\b}^A \left\{ \Lie_{K_r} \chi^m + \half (R_{\kappa}^{-1})^{r}_{\phantom{r}s} \chi^n (\mathbb{J}^s)_{n}^{\phantom{n}m} \right\}~.
\end{align}

Now on the other hand we demand that the transformation $\delta^{r}_{(J)}$ be representable by a transformation on the collective coordinates $z^m, \chi^n$.  Since $\rho^A$, like $\hat{A}_a$, is a collective coordinate scalar, the contribution to the variation of $\rho^A$ due to the variation of $z^m$ should be as in \eqref{symEonA}, namely $- (K^r)^m \pd_m \rho^A  = - \Lie_{K^r} \rho^A$.  Meanwhile since $\rho$ is a linear function of $\chi$, the contribution to the variation of $\rho^A$ from the variation of $\chi$ will simply be $\rho(\vec{x};z,\delta_{(J)}^r \chi)$.  Hence
\begin{equation}\label{angrhovar4}
\delta_{(J)}^r \rho_{\alpha}^A = - \Lie_{K^r} \rho_{\alpha}^A - \frac{i}{2\sqrt{-\det{\kappa}}} \delta_m \hat{A}_a (\tau^a)_{\a}^{\phantom{\a}\b} \kappa_{\b}^A (\delta^{r}_{(J)} \chi^m)~.
\end{equation}
By comparing \eqref{angrhovar3} with \eqref{angrhovar4} we infer the required variation of $\chi^m$ under an angular momentum transformation:
\begin{equation}\label{activechivar}
\delta_{(J)}^r \chi^m = \Lie_{K^r} \chi^m + \half (R_{\kappa}^{-1})^{r}_{\phantom{r}s} \chi^n (\mathbb{J}^s)_{n}^{\phantom{n}m} ~.
\end{equation}

Note however that this is the active transformation; it goes hand in hand with $\delta^r z^m = - (K^r)^m$.  As we discussed below \eqref{ccEvar}, this variation is related to $\delta_{\rm c.c.}$ for the case of the bosonic $z^m$.  There is an additional subtlety in the case of the fermionic collective coordinate $\chi^m(z(t),t)$, which is that we wish to consider the variation $\delta^{r}_{\rm c.c.} \chi^m$ not at the new point ${z'}^m = z^m + \delta_{\rm c.c.}^r z^m$ but at the original point $z^m$.  Pulling back to the original point  cancels out the $(K^r)^n \pd_n \chi^m$ term of the Lie derivative since this term generates the translation to the new point.  Taking both of these into account brings us from \eqref{activechivar} to the collective coordinate angular momentum transformation
\begin{align}
\delta_{(J){\rm c.c.}}^r \chi^m :=&~ - \left( \delta^r \chi^m - (K^r)^n \pd_n \chi^m \right) \cr
=&~  \chi^n \pd_n (K^r)^m - \half (R_{\kappa}^{-1})^{r}_{\phantom{r}s} \chi^n (\mathbb{J}^s)_{n}^{\phantom{n}m}~.
\end{align}
The first term reproduces the transformation \eqref{ccEvar} for $K^r$, while the second term is a shift by an $\mathfrak{su}(2)_R$ transformation.  Comparing with \eqref{su2Rcc} we have
\begin{equation}
\delta_{(J){\rm c.c.}}^r = \delta^{r}_{\rm c.c.} - (R_{\kappa}^{-1})^{r}_{\phantom{r}s} \delta^{s}_{(I)} ~.
\end{equation}

This transformation will be a symmetry of the collective coordinate action since both $\delta^{r}_{\rm c.c.}$ and $\delta^{s}_{(I)}$ are.  The corresponding Noether charge is
\begin{align}
J^r :=&~ \II^r - (R_{\kappa}^{-1})^{r}_{\phantom{r}s} I^s \cr
=&~ -  \left((K^r)^m \pi_m - \frac{2 \pi i }{g_{0}^2} (\nabla_m (K^r)_n) \chi^m \chi^n \right) - \frac{i \pi}{g_{0}^2} (R_{\kappa}^{-1})^{r}_{\phantom{r}s} (\sw^s)_{mn} \chi^m \chi^n ~.
\end{align}
One can show that the $J^r$, unlike the $\II^r$, commute with the $\mathfrak{su}(2)_R$ symmetry charges $I^r$.    We will present the full algebra of Noether charges, $I^r, J^r, Q^a$ and discuss physical implications after passing to the quantum theory.  

The algebra in particular confirms the identification of $\{\II^r,J^r, I^r\}$ with the corresponding generators in \eqref{twisteddiag} and the identification of $\kappa$, and hence $R_{\kappa}$, between the two formulae.  From the point of view of the collective coordinate theory, the shift of $\II^r$ by the $\mathfrak{su}(2)_R$ generator can be understood as a compensating transformation that is needed because the collective coordinate ansatz forces a choice of isomorphism between bosonic and fermionic zero modes.  This choice is specified by $\kappa$ which equivalently determines a choice of diagonal subalgebra of $\mathfrak{so}(3) \oplus \mathfrak{su}(2)_R$.  Hence to make an angular momentum $SO(3)$ rotation, we are first making a rotation in the diagonal subgroup specified by $\kappa$, and then undoing that rotation in the $SU(2)_R$ factor.

%%%%%%%%%%%%%%%%%%%%%
\subsection{Quantization}\label{sec:quantize}
%%%%%%%%%%%%%%%%%%%%%

We canonically quantize the collective coordinate theory by promoting Dirac brackets to commutators, $[~,~]_{\pm} = i \{~,~\}_{\pm}$, whence
\begin{equation}\label{canonicalccs}
[\hat{z}^m, \hat{p}_n]_- = i \delta^{m}_{\phantom{m}n}~, \qquad [\hat{\chi}^{\um}, \hat{\chi}^{\un}]_+ = \frac{g_{0}^2}{4\pi} \delta^{\um\un}~, \qquad [\hat{z}^m, \hat{\chi}^{\un}]_- = 0 = [\hat{p}_m, \hat{\chi}^{\un}]_- ~.
\end{equation}
The Hermitian operators $\hat{\chi}^{\um}$ satisfy a Clifford algebra, so it is natural to take the wavefunctions to be sections of the Dirac spinor bundle over $\fMM$ and represent the $\hat{\chi}^{\um}$ by gamma matrices,
\begin{equation}\label{chigamma}
\hat{\chi}^{\um} = \frac{g_{0}}{2\sqrt{2\pi}} \gamma^{\um}~,
\end{equation}
so that $[\gamma^{\um}, \gamma^{\un}]_+ = 2\delta^{\um\un}$ \cite{Gauntlett:1999vc,Gauntlett:2000ks}.  If $\dim{\fMM} = 4N$ then the Dirac spinor represention is $2^{2N}$-dimensional.  In the case with defects the Hilbert space consists of $\Lsq$ sections; the innerproduct is 
\begin{equation}
\langle \Psi_1 | \Psi_2 \rangle := \int_{\fMM} e \overline{\Psi_1} \Psi_2~,\label{modspacespin}
\end{equation}
where $e := \det{ e^{\um}_{\phantom{\um}m}} = (\det{g_{mn}})^{1/2}$ and with the bar denoting the transpose conjugate.  In the vanilla case, however, we must remember that the BPS states are asymptotic particle states.  We require $\Lsq$ normalizability on the strongly centered moduli space, $\MM_0$, which represents the internal degrees of freedom of the particle state.

On a K\"ahler manifold there is an isomorphism between the Dirac spinor bundle and the space of $(0,\ast)$-forms tensored with a certain line bundle---a square root of the canonical bundle \cite{MR0358873}.  On a Ricci flat manifold, in particular a hyperk\"ahler manifold, the canonical bundle is trivial.  Hence, by choosing a distinguished complex structure on $\fMM$, one may also represent states as formal sums of $(0,q)$-forms for $0 \leq q \leq 2N$.  We will make some use of this construction later and have provided the relevant details in appendix \ref{app:holforms}.

Returning to the discussion of quantization via spinors, the momentum operator $\hat{p}_m$ acts via the ordinary coordinate derivative, twisted by the half-density $e^{1/2}$:
\begin{equation}
\hat{p}_m := -i e^{-1/2} \pd_m e^{1/2} = -i \left(\pd_m + \frac{1}{4} g^{np} \pd_m g_{np} \right) = -i \left( \pd_m + \half \Gamma^{n}_{\phantom{n}nm} \right)~.
\end{equation}
The twist is necessary in order for $\hat{p}_m$ to be Hermitian with respect to the innerproduct, \cite{DeWitt:1952js}.  See also the discussion in \cite{Davis:1984gh,Macfarlane:1984gr}.  It follows from the identification \eqref{chigamma} that the super-covariant momentum operator, $\hat{\pi}_m$, is  represented by the $e^{1/2}$-twisted covariant derivative on spinors:
\begin{equation}\label{spincovD}
\hat{\pi}_m = \hat{p}_m -\frac{2\pi i}{g_{0}^2} \omega_{m,\up\uq} \hat{\chi}^{\up} \hat{\chi}^{\uq} = -i \left( e^{-1/2} \pd_m e^{1/2} + \frac{1}{4} \omega_{m,\up\uq} \gamma^{\up\uq} \right) =: -i e^{-1/2} \DD_m e^{1/2} ~,
\end{equation} 
where $\gamma^{\um\un} := \half \gamma^{[\um} \gamma^{\un]}$.  Some useful commutators that follow directly from \eqref{canonicalccs} and the expression for $\hat{\pi}_m$ in terms of $\hat{p}_m$ and $\hat{\chi}^{\um}$ are
\begin{equation}\label{pihatcomrels}
[\hat{\pi}_m, \hat{\chi}^n]_- = i \Gamma^{n}_{\phantom{n}mp} \hat{\chi}^p~, \qquad [\hat{\pi}_m, \hat{\pi}_n]_- = - \frac{2\pi}{g_{0}^2} R_{mnpq} \hat{\chi}^p \hat{\chi}^q ~.
\end{equation}
Note here that we express these results in terms of $\hat{\chi}^m = \hat{\chi}^{\um} \EE_{\um}^{\phantom{\um}m}$.

In attempting to promote the various Noether charges constructed in the classical theory to operators in the quantum theory we encounter ordering ambiguities.  The supercharges \eqref{Qccphase} and isometry-induced Noether charges \eqref{NoetherK} both contain a term involving a product of $\hat{\pi}_m$ with a $\hat{z}$-dependent quantity.  (The $SU(2)_R$ symmetry charges \eqref{su2Rnoether} in contrast do not suffer ordering ambiguities.)  These ambiguities exist only because of our `truncation then quantization' approach; see footnote \ref{fn:oporder}.  However in the cases at hand it is easy to resolve them by demanding that the corresponding operator be Hermitian \cite{Davis:1984gh,Macfarlane:1984gr}.  For a classical quantity of the form $f(z,\chi) \pi_m$ this means a symmetrized prescription, $\widehat{f \pi_m} := \half [\hat{f},\hat{\pi}_m]_+$.  This will be more useful in the form
\begin{equation}\label{sympre}
\half [\hat{f}(\hat{z},\hat{\chi}),\hat{\pi}_m]_+ = \hat{f} \hat{\pi}_m - \half [\hat{f}, \hat{\pi}_m]_- = -i \hat{f} \left( \DD_m + \frac{1}{2} \Gamma^{n}_{\phantom{n}nm} \right) - \half [\hat{f}, \hat{\pi}_m]_- ~.
\end{equation}

Now in the case of the supercharges the relevant quantity is $\hat{f} = \hat{\chi}^n (\tbbJ^a)_{n}^{\phantom{n}m}$ and we have
\begin{equation}
[ \hat{\chi}^n (\tbbJ^a)_{n}^{\phantom{n}m}, \hat{\pi}_m]_- = i \hat{\chi}^n \pd_m ( \tbbJ^a)_{n}^{\phantom{n}m} - i \Gamma^{n}_{\phantom{n}mp} \hat{\chi}^p (\tbbJ^a)_{n}^{\phantom{n}m} = -i \Gamma^{m}_{\phantom{m}mp} (\tbbJ^a)_{n}^{\phantom{n}p} \hat{\chi}^n~,
\end{equation}
where we used that $\tbbJ^a$ is covariantly constant.  This serves to cancel the term in \eqref{sympre} involving the Christoffel symbol.  Thus we find
\begin{align}\label{Qhatcc}
\hat{Q}^a :=&~ \left\{ \half [ \hat{\chi}^n (\tbbJ^a)_{n}^{\phantom{n}m}, \hat{\pi}_m ]_+  - \hat{\chi}^n (\tbbJ^a)_{n}^{\phantom{n}m} {\rm G}(\YY_{\infty}^{\rm cl})_m \right\} \times \left(1 + O(g_{0}^2) \right) \cr
=&~  -\frac{i g_0}{2\sqrt{2\pi}} \gamma^n (\tbbJ^a)_{n}^{\phantom{n}m} \left( \DD_m - i {\rm G}(\YY_{\infty}^{\rm cl})_m \right) \times \left(1 + O(g_{0}^2) \right)~.
\end{align}
Meanwhile in the case of $\hat{N}^E$ the relevant quantity is $f = (K^E)^m(z)$ and we have that $[\hat{f}, \hat{\pi}_m] = i \pd_m (K^E)^m$.  This again cancels against the Christoffel term in \eqref{sympre} upon using the Killing equation in the form $\nabla_m (K^E)^m = 0$.  Hence
\begin{align}\label{Nhatcc}
\hat{N}^E :=&~ \left\{ -\half [ (K^E)^m, \hat{\pi}_m]_+  + \frac{2\pi i}{g_{0}^2} \left(\nabla_m (K^E)_n \right) \hat{\chi}^m \hat{\chi}^n \right\} \times \left( 1 + O(g_{0}^2) \right) \cr
=&~ i \left( (K^E)^m \DD_m + \frac{1}{4} \left(\nabla_m (K^E)_n \right) \gamma^{mn} \right) \times \left(1 + O(g_{0}^2) \right)  \cr
 =: &~ i \Lie_{K^E} \times \left(1+ O(g_{0}^2) \right) ~.
\end{align}
In the last step we observed that the quantity in parentheses is the Lie derivative, with respect to the Killing vector $K^E$, acting on sections of the Dirac spinor bundle \cite{MR0156292,Gauntlett:1997pk}.  Specializing to the cases $E = (i,r,A)$ we have the operators $\hat{N}^E = (\hat{P}^i, \hat{\II}^r, \hat{N}_{\rm e}^A)$ corresponding to translations (present in the vanilla case only), the diagonal of $\mathfrak{so}(3) \oplus \mathfrak{su}(2)_R$, and asymptotically nontrivial gauge transformations, respectively.  

Finally the generators of $\mathfrak{su}(2)_R$ are
\begin{equation}
\hat{I}^r := \frac{i \pi}{g_{0}^2} (\sw^{r})_{\underline{mn}} \hat{\chi}^{\um} \hat{\chi}^{\un} \left( 1 + O(g_{0}^2) \right) = \frac{i}{8} (\sw^{r})_{\underline{mn}} \gamma^{\underline{mn}} \left(1 + O(g_{0}^2) \right)~,
\end{equation}
and hence the generators of the angular momentum $\mathfrak{so}(3)$ are
\begin{equation}\label{Jhat}
\hat{J}^r := \hat{\II}^r - (R_{\kappa}^{-1})^{r}_{\phantom{r}s} \hat{I}^s = \left\{ i \Lie_{K^r} - \frac{i}{8} (R_{\kappa}^{-1})^{r}_{\phantom{r}s} (\sw^{s})_{\underline{mn}} \gamma^{\underline{mn}} \right\} \times \left( 1 + O(g_{0}^2) \right)~.
\end{equation}

The $\hat{N}_{\rm e}^A$ are the generators of gauge transformations that asymptote to the fundamental magnetic weights $h^{I_A}$.  Hence they are the components of the electric charge operator of the theory along the simple roots $\alpha_{I_A}$:
\begin{equation}\label{gammaehatcc}
\hat{\gamma}_{\rm e} := \sum_{A} \alpha_{I_A} \hat{N}_{\rm e}^A = i \sum_A \alpha_{I_A} \Lie_{K^A} \left(1+ O(g_{0}^2) \right) ~.
\end{equation}
This is simply the quantum version of \eqref{ccecharge}, and we showed there that the classical limit of this quantity indeed corresponds to the asymptotic flux of the appropriate linear combination of electric and magnetic fields.

We have emphasized that the various charges in \eqref{Qhatcc} through \eqref{gammaehatcc} are expected to receive corrections that are $O(g_{0}^2)$ suppressed relative to the leading terms.  These originate both from the corrections to the classical collective coordinate Lagrangian associated with expanding around an approximate classical solution---\ie\ the small velocity and weak potential energy approximations---and from integrating out the field fluctuations around the collective coordinate ansatz.

Now let us discuss the algebra of these charges.  We begin with the anticommutator of the supercharges.  On the one hand, a lengthy calculation starting directly from \eqref{Qhatcc} leads to
\begin{align}\label{QQalg}
[ \hat{Q}^a, \hat{Q}^b]_+ =&~ \frac{g_{0}^2}{4\pi} \delta^{ab} \bigg\{ -\frac{1}{\sqrt{g}} \DD_m \sqrt{g} g^{mn} \DD_n + g_{mn} {\rm G}(\YY_{\infty}^{\rm cl})^m {\rm G}(\YY_{\infty}^{\rm cl})^n + \frac{i}{2} \gamma^{mn} \nabla_m {\rm G}(\YY_{\infty}^{\rm cl})_n + \cr
&~ \qquad  \quad +2i \bigg({\rm G}(\YY_{\infty}^{\rm cl})^m \DD_m  + \frac{1}{4} \gamma^{mn} \nabla_m {\rm G}(\YY_{\infty}^{\rm cl})_n \bigg) \bigg\} \times\left( 1 + O(g_{0}^2) \right)~,  \raisetag{20pt}
\end{align}
which is consistent in form with \cite{Gauntlett:1999vc}.  Note the appearance of the spinorial Lie derivative with respect to ${\rm G}(\YY_{\infty}^{\rm cl})$ in the last line.  Recall that $\YY_{\infty}^{\rm cl} = \frac{4\pi}{g_{0}^2} Y_\infty + \frac{\theta_0}{2\pi} X_\infty$ and that in the weak potential energy approximation, $|\langle \alpha, Y_\infty \rangle |/ \langle \alpha, X_\infty \rangle \sim O(g_0)$.  Hence there are terms in the curly brackets of \eqref{QQalg} that are $O(g_{0}^2)$ suppressed relative to the leading terms.  Strictly speaking, they should be neglected in \eqref{QQalg} for consistency.  

On the other hand, we can invert the relation between $\hat{Q}^a$ and the supercharges $\hat{\RR}^A$ in \eqref{susyembedding} to obtain
\begin{equation}
\hat{Q}^a = \frac{1}{2\sqrt{-\det{\kappa}}} \kappa_{A} (-i\tau^a) \hat{\RR}^A~.
\end{equation}
Then from the algebra of the $\hat{\RR}^A$, \eqref{RTalgzeta}, we derive
\begin{equation}\label{QQalg2}
[\hat{Q}^a, \hat{Q}^b]_+ = 2 \delta^{ab} \left( \hat{H} + \Re(\zeta^{-1} \hat{Z}) \right)~,
\end{equation}
where $\hat{H}, \hat{Z}$ are the Hamiltonian and central charge operators.  

We want to compare \eqref{QQalg} and \eqref{QQalg2} to extract both $\hat{H}$ and $\Re (\zeta^{-1} \hat{Z})$.  In order to disentangle these two quantities we use that $\hat{H}$ must be consistent with \eqref{Hcc2} in the classical limit, where in particular $\DD_m \to i e^{1/2} \pi_m e^{-1/2} = i\pi_m$.  This allows us to determine the quantum Hamiltonian
\begin{align}\label{QHam}
\hat{H} =&~ \bigg\{ M_{\gm}^{\textrm{1-lp}} + \frac{g_{0}^2}{8\pi} \bigg[\frac{-1}{\sqrt{g}} \DD_m \sqrt{g} g^{mn} \DD_n + g_{mn} {\rm G}(\YY_{\infty}^{\rm cl})^m {\rm G}(\YY_{\infty}^{\rm cl})^n + \frac{i}{2} \gamma^{mn} \nabla_m {\rm G}(\YY_{\infty}^{\rm cl})_n \bigg] + \cr
&~  \qquad +i \tilde{\theta}_0 \Lie_{{\rm G}(X_\infty)} \bigg\} \times \left( 1 + O(g_{0}^2) \right)~, \raisetag{20pt}
\end{align}
Then, given \eqref{QQalg2} versus \eqref{QQalg}, we infer the central charge term:
\begin{align}\label{QZsc}
-\Re(\zeta^{-1} \hat{Z}) =&~ \left\{ M_{\gm}^{\textrm{1-lp}} - i \Lie_{{\rm G}(Y_\infty)} \right\} \left(1 +O(g_{0}^2) \right) \cr
=&~ \left\{ M_{\gm}^{\textrm{1-lp}} - \langle \hat{\gamma}_{\rm e}, Y_\infty \rangle \right\} \left(1 +  O(g_{0}^2) \right)~.
\end{align}

We will give a complete definition of framed and vanilla BPS states in sections \ref{ssec:fBPSspace} and \ref{ssec:vBPSspace} below, but let us already note here that the algebra \eqref{QQalg2} implies a bound on the spectrum: $\hat{H} \geq - \Re(\zeta^{-1} \hat{Z})$.  This bound is saturated by states---\ie\ $\Lsq$ sections of the Dirac spinor bundle---that are annihilated by any one, and hence all, of the supercharges.

Now let us consider some of the remaining commutation relations.  The $\hat{I}^r$ and $\hat{J}^r$ generate commuting $\mathfrak{su}(2)$ algebras:
\begin{equation}\label{commutingsu2s}
[ \hat{I}^r, \hat{I}^s ]_- = i\epsilon^{rs}_{\phantom{rs}t} \hat{I}^t ~, \qquad [ \hat{J}^r, \hat{J}^s ]_- = i\epsilon^{rs}_{\phantom{rs}t} \hat{J}^t~, \qquad [\hat{I}^r, \hat{J}^s]_- = 0~.
\end{equation}
The shift of $\hat{\II}^r$ by the $\mathfrak{su}(2)_R$ charge in $\hat{J}^r$ is crucial for this.  The algebra of the $\hat{Q}^a$ with the $\mathfrak{su}(2)_R$ charges is found to be
\begin{equation}\label{QIalg}
[ \hat{I}^r, \hat{Q}^s]_- = \frac{i}{2} \left( \delta^{rs} \hat{Q}^4 + \epsilon^{rs}_{\phantom{rs}t} \hat{Q}^t \right)~, \qquad  [\hat{I}^r, \hat{Q}^4]_- = -\frac{i}{2} \hat{Q}^r ~.
\end{equation}

Finally we compute the commutator of the supercharges with the generic isometry-induced Noether charges \eqref{Nhatcc}, finding
\begin{align}
[\hat{N}^E, \hat{Q}^r]_- =&~ \frac{g_0}{2\sqrt{2\pi}} \gamma^m \bigg\{ - \left( \Lie_{K^E} (\bbJ^r)_{m}^{\phantom{m}n}\right) ( \DD_n - i {\rm G}(\YY_{\infty}^{\rm cl})_n ) + i (\bbJ^r)_{m}^{\phantom{m}n} [K^E, {\rm G}(\YY_{\infty}^{\rm cl})]_n \bigg\}~, \cr
[\hat{N}^E, \hat{Q}^4]_- =&~ \frac{i g_0}{2\sqrt{2\pi}}  \gamma^m [ K^E, {\rm G}(\YY_{\infty}^{\rm cl})]_m~, \raisetag{20pt}
\end{align}
where on the right-hand sides $[~,~]$ denotes the commutator of vector fields.  In fact all of the Killing fields $K^E$ on $\fMM$ commute with ${\rm G}(\YY_{\infty}^{\rm cl})$.  In the case of $K^A = {\rm G}(h^{I_A})$ this is obvious since ${\rm G}$ is a Lie algebra homomorphism from the Cartan subalgebra into the space of vector fields.  It is also true for the $K^r$ corresponding to $\hat{\II}^r$, as well as the $K^i$ corresponding to translations in the vanilla case.  The reason is that the asymptotically nontrivial gauge transformation  corresponding to ${\rm G}(\YY_{\infty}^{\rm cl})$ is gauge equivalent to a spatially constant gauge transformation by $\epsilon(\vec{x}) = \YY_{\infty}^{\rm cl}$, which clearly commutes with rotations and translations.  Hence the vector field commutators can be dropped.  

Furthermore, in the case of the triholomorphic vector fields, $K^i$ and $K^A$, the Lie derivatives of the complex structures vanish as well.  The former means (in the vanilla case) that $[\hat{P}^i, \hat{Q}^a]_- = 0$, which is of course part of the Poincar\'e supersymmetry algebra, while the latter is summarized by the statement
\begin{equation}
[ \hat{\gamma}_{\rm e}, \hat{Q}^a]_- = 0~.
\end{equation}
In particular this means that the kernel of $\hat{Q}^a$ can be decomposed into eigenspaces of definite electric charge, a fact that will be important for the semiclassical construction of framed BPS states to follow.  

In the case of $\hat{\II}^r$ the Lie derivative term survives, but we can use \eqref{angJaction} to simplify it.  This leads to
\begin{equation}\label{QcurlyIalg}
[\hat{\II}^r, \hat{Q}^s]_- = i (R_{\kappa}^{-1})^{r}_{\phantom{r}u} \epsilon^{us}_{\phantom{us}t} \hat{Q}^t ~, \qquad [\hat{\II}^r, \hat{Q}^4]_- = 0~.
\end{equation}
As we have mentioned before, $\hat{\II}^r = \hat{J}^r + (R_{\kappa}^{-1})^{r}_{\phantom{r}s} \hat{I}^s$ generates the diagonal subgalgebra $\mathfrak{su}(2)_{\mathrm{d}}^{(\kappa)} \subset \mathfrak{so}(3) \oplus \mathfrak{su}(2)_R$ described around \eqref{twisteddiag}, while $\hat{Q}^4 \propto \RR_\kappa$.  Thus we see here the manifestation of the statement that $\RR_\kappa$ is a singlet of this diagonal subalgebra.  Finally we can combine \eqref{QcurlyIalg} with \eqref{QIalg} to infer the commutators of the supercharges with angular momentum,
\begin{equation}\label{JQcom}
 [ \hat{J}^r, \hat{Q}^s]_- = \frac{i}{2} (R_{\kappa}^{-1})^{r}_{\phantom{r}u} \left( -\delta^{us} \hat{Q}^4 + \epsilon^{us}_{\phantom{us}t} \hat{Q}^t \right)~, \qquad  [\hat{J}^r, \hat{Q}^4]_- = \frac{i}{2} (R_{\kappa}^{-1})^{r}_{\phantom{r}u} \hat{Q}^u ~.
\end{equation}

We have obtained the results \eqref{commutingsu2s} through \eqref{JQcom} by working with the leading order expressions for the Noether charges.  The corrections should be such that these commutation relations are preserved.

%%%%%%%%%%%%%%%%%%%%
\subsection{Comparing semiclassical and low energy analyses}\label{ssec:validity}
%%%%%%%%%%%%%%%%%%%%

In this paper we have discussed $\NN = 2$ SYM probed by line defects in two limits: the low energy quantum-exact one described in sections \ref{ssec:SWreview} through \ref{sec:wce}, and the semiclassical one reviewed and developed in this section.  Our main goal is to use results obtained in one description to learn about the other.  Hence it is imperative to understand the regime of overlapping validity of the two  limits.  It would perhaps make more pedagogical sense to have this discussion at the end of this section, after completing our semiclassical description of the remaining quantities of interest---namely the Hilbert spaces of (framed) BPS states and their protected spin characters.  However, we will use the comparison here to motivate a conjectural extension of the semiclassical results we have obtained thus far, and it will be convenient to have this in place before continuing further.

The Seiberg--Witten action is the leading set of terms in a spacetime derivative expansion.  The massless fields appearing in the effective action are assumed to be slowly varying functions of the spacetime coordinates, and supersymmetry ties the derivative expansion to an expansion in fermi fields.  The semiclassical approximation is a priori a weak coupling expansion; one assumes $g_0$ is small and computes the leading quantum corrections to classical quantities.  However in practice we have imposed additional approximations, which are of two types.  

The first is the Manton approximation, $\dot{z}^m \sim O(g_0)$; supersymmetry ties this to an expansion in the fermi collective coordinates, $\chi^m\chi^n \sim O(g_0)$, just as in field theory.  Taking the fields to be slowly varying in time is thus an approximation in both schemes.  Note however that in the semiclassical case, we do not require the fields to be slowly varying in space.  We are able to describe the wavefunctions of massive BPS states in a completely smooth and controlled fashion, something that is beyond the reach of the low energy analysis.  See \eg\ \cite{Chalmers:1996ya}.  In particular, our description of a framed BPS state is not limited to regions of the Coulomb branch near the walls of marginal stability, where the characteristic length scales of the state are much larger than those controlling the derivative expansion.

The second additional approximation we made on the semiclassical side is the weak potential energy approximation, $\langle \alpha, Y_\infty \rangle \sim O(g_0) \cdot \langle \alpha, X_\infty \rangle$.  We could in principal relax this approximation by working with a collective coordinate quantum mechanics based on the dyon moduli spaces $\fSigma$, \eqref{fBPSfc}, rather than the monopole moduli spaces.  For this purpose one might consider carrying out a reduction of the quantum mechanics on the ambient space $\fMM$ to one of the spaces $\fSigma$.  We will provide a natural ansatz below for incorporating these corrections, based on the comparison to the Seiberg--Witten description.

Hence in order to compare a generic observable in the two descriptions, we should restrict the Seiberg--Witten low energy effective description to the weak coupling regime, as described in \ref{sec:wce}, while the semiclassical description should be restricted to the regime of small spatial gradients.  However the latter can be relaxed if we consider quantities that are unaffected by the higher derivative terms of the low energy effective description.  

Indeed our main focus in this paper is the spectrum of BPS states as a function of the Coulomb branch parameters $\{ u \} \in \BB^\ast$.  This is controlled by the central charge operator: the real part of $\zeta^{-1} \hat{Z}$ determines the masses of framed BPS states, if they exist, while the imaginary part determines the walls of marginal stability.  (In the vanilla case it is the magnitude of $\hat{Z}$ that determines the masses and the phase difference between the central charges of constituents that determines the walls.)  In any case, this observable is believed to be protected from the higher derivative ``$D$-terms,'' such that \eqref{ZSW} is the exact expression in the full theory \cite{Seiberg:1994rs}.  We therefore expect our semiclassical analysis to reproduce the one-loop perturbative approximation to $\hat{Z}$, as determined by the one-loop dual coordinate, \eqref{pertaD}.  For future reference let us write this as
\begin{equation}\label{ZSWoneloop}
\hat{Z}^{\textrm{1-lp}} = (\hat{\gamma}_{\rm m} , a_{\mathrm{D}}^{\textrm{1-lp}} ) + \langle \hat{\gamma}_{\rm e}, a \rangle~,
\end{equation}
where $a_{\mathrm{D}}^{\textrm{1-lp}}$ is the one-loop approximation to the $\mathfrak{t}$-valued dual coordinate, \eqref{tvaluedaD}:
\begin{equation}\label{aDoneloop}
a_{\mathrm{D}}^{\textrm{1-lp}} := a_{\mathrm{D},I}^{\textrm{1-lp}} (\lambda^I)^\ast = \frac{i}{2\pi} \sum_{\alpha \in \Delta^+} \alpha^\ast \langle \alpha, a \rangle \left\{ \ln \left( \frac{ \langle \alpha, a \rangle^2}{2 \Lambda^2} \right) + 1 \right\}~.
\end{equation}
In charge sector $\{\hat{\gamma}_{\rm m}, \hat{\gamma}_{\rm e}\} = \{\gm,\gamma_{\rm e}\}$, \eqref{ZSWoneloop} is a scalar operator given by multiplication with $Z^{\textrm{1-lp}}_{\gamma} := (\gm, a_{\mathrm{D}}^{\textrm{1-lp}}) + \langle \gamma_{\rm e}, a \rangle$.

In the previous subsection we obtained the semiclassical formula \eqref{QZsc} for $\Re(\zeta^{-1} \hat{Z})$, in a fixed magnetic charge sector $\gm$.  It is a sum of magnetic and electric contributions, where the electric piece is an operator on the Hilbert space of $\Lsq$ sections of the Dirac spinor bundle over $\fMM$.  We have already considered the agreement of the magnetic piece with low energy formula, \eqref{ZSWoneloop}, in subsection \ref{sec:scphilo} around \eqref{Zgm1loop}, where we found
\begin{equation}\label{weakpotcor}
- \Re (\zeta^{-1} Z_{\gm}^{\textrm{1-lp}}) = - \Re \left\{ (\gm, \zeta^{-1} a_{\mathrm{D}}^{\textrm{1-lp}}) \right\} = M_{\gm}^{\textrm{1-lp}} \cdot \left( 1 + O\left( \frac{\langle \alpha, Y_\infty\rangle^2}{\langle \alpha, X_\infty \rangle^2}\right) \right)~.
\end{equation}
Meanwhile the electric piece of \eqref{QZsc}, acting on an eigenspace $\gamma_{\rm e}$ of $\hat{\gamma}_{\rm e}$, agrees exactly with the electric charge contribution to the real part of \eqref{ZSWoneloop}, using $\Re\left( \langle \gamma_{\rm e}, \zeta^{-1} a \rangle \right) = \langle \gamma_{\rm e}, Y_\infty \rangle$.  Hence the semiclassical and low energy results for $\Re(\zeta^{-1} \hat{Z})$ agree where they are supposed to.  

Furthermore we conjecture that all perturbatuve corrections to the semiclassical result, \eqref{QZsc}, come entirely from the weak potential energy approximation, and that the complete set of these corrections is captured by \eqref{ZSWoneloop} with \eqref{aDoneloop}.  This was written out explicitly in terms of $\langle \alpha, Y_\infty \rangle$ and $\langle \alpha, X_\infty\rangle$ in \eqref{Zgm1loop}.  

Note that $\Re(\zeta^{-1} \hat{Z})$ was determined semiclassically in \eqref{QZsc} via an indirect method.  We used the supersymmetry algebra together with knowledge of the one-loop corrected Hamiltonian to infer it.  One could have attempted a direct computation, and this would be necessary for an independent semiclassical computation of $\hat{Z}$ itself.  Schematically, the idea is as follows.  First recall that, classically, we have the formula \eqref{Zcl2},
\begin{equation}\label{Zclagain}
Z^{\rm cl} = \frac{2}{g_{0}^2} \int_{S_{\infty}^2} \ed^2 S^i \Tr \left\{ (i B_i - E_i)  \varphi \right\} =  (\gm, a_{\mathrm{D}}^{\rm cl}) + \langle \gamma_{\rm e}, a\rangle~.
\end{equation}
In the full quantum theory, we grade the Hilbert space by eigenvalues of the electromagnetic charge operator $\hat{\gamma} = \hat{\gamma}_{\rm m} \oplus \hat{\gamma}_{\rm e}$ and assume that, at least when restricted to the BPS sector of the Hilbert space, $\hat{Z}$ is a scalar operator in each charge sector.  Then the exact expression for $Z_\gamma$ is,
\begin{equation}\label{formalZ}
Z_\gamma = \langle s_\gamma | \hat{Z} | s_\gamma \rangle = \frac{2}{g_{0}^2} \int_{S_{\infty}^2} \ed^2 S^i \Tr \left\{ \langle s_\gamma | (i \hat{B}_i - \hat{E}_i) \hat{\varphi} | s_\gamma \rangle \right\}~,
\end{equation}
where $\hat{E}_i$ \etc.\ are the corresponding field operators and $|s_\gamma\rangle$ is a (soliton) state in charge sector $\gamma$.  To compute this matrix element semiclassically from first principles, we would make the canonical transformation to the appropriate soliton sector as described around \eqref{solitoncov}, where the soliton state can be defined in terms of creation operators of the fluctuation fields acting on a `vacuum state' in that sector.  The vacuum states are defined as those states annihilated by all of the annihilation operators in the fluctuation field, and are associated with the collective coordinate degrees of freedom.   Semiclassically they are in one to one correspondence with the Hilbert space of the collective coordinate quantum mechanics.  Since we are interested in the BPS sector we can take $|s_\gamma\rangle$ to be such a vacuum state.  Then the fluctuation fields, \eg\ denoted $\hat{a}$ in \eqref{solitoncov}, only contribute to \eqref{formalZ} through loops.

At tree level with respect to these fluctuation fields, the electric field term of \eqref{formalZ} can be computed in the collective coordinate quantum mechanics using the form of the electric charge operator, \eqref{gammaehatcc}.  The result is $Z_{\gamma_{\rm e}} = \langle \hat{\gamma}_{\rm e}, a \rangle$, where we used that $\hat{\varphi}$ restricted to the asymptotic two-sphere is the scalar multiplication operator $a(u)$ in vacuum $u \in \BB^\ast$.  For the magnetic term, the one-loop correction to the classical part \eqref{Zclagain} must be included, since it is the same order in the $g_0$ expansion as the leading electric term.  This computation has been considered in \cite{Rebhan:2004vn,Rebhan:2006fg}, at least in the limit $\langle \alpha, Y_\infty\rangle \to 0$, where agreement with the Seiberg--Witten expression was found.  We fully expect that this result can be extended to the class of $\NN = 2$ theories probed by line defects considered here.   

Consider now the framed case, where the walls of marginal stability, \eqref{fmsw}, are determined by the vanishing of the imaginary part of $\zeta^{-1} Z_\gamma$, (together with the condition that the real part is negative).  In the one-loop approximation, $\Im(\zeta^{-1} Z_\gamma)$ depends on the Coulomb branch parameters through $\Im(\zeta^{-1} a(u))$ and $\Im (\zeta^{-1} a_{\mathrm{D}}^{\textrm{1-lp}}(u))$.  How do we see walls of marginal stability in the semiclassical picture?  As we mentioned above, framed BPS states will be $\Lsq$ sections of the spinor bundle that sit the kernel of any one of (and hence all of) the supercharges, \eqref{Qhatcc}.  The leading semiclassical expression for this operator depends on $X_\infty = \Im(\zeta^{-1} a)$ through the metric on $\fMM$, and on $\YY_{\infty}^{\rm cl} = \Im(\zeta^{-1} a_{\mathrm{D}}^{\rm cl})$ through the twisting of the covariant derivative by ${\rm G}(\YY_\infty)_m$.  As we vary these quantities the kernel of the supercharges can jump: $\Lsq$ solutions can enter or leave as the continuum of the spectrum comes down to zero and the gap disappears.  These are co-dimension one walls (in $\{X_\infty, \YY_{\infty}^{\rm cl}\}$ space) where the supercharges fail to be Fredholm operators.  This is the semiclassical picture of wall crossing; we will see it very explicitly in examples later.

However in order for the semiclassical walls---that is the walls where the supercharge operators \eqref{Qhatcc} fail to be Fredholm---to have any chance of agreeing with the low energy ones, as we argued above they must, then it is clear that we must take into account the leading corrections to the supercharge operators.\footnote{Alternatively one could restrict attention on the low energy side to the \emph{classical} part of the central charge but, as we have argued, it is then generally inconsistent to consider the effects of electric charge.  These effects are responsible for much of the interesting wall crossing phenomena we will discuss, hence we must insist on treating the full perturbative central charge.}  There is an obvious and natural candidate that accounts for the required subset of perturbative corrections in $g_0$.  Namely, we should replace $\YY_{\infty}^{\rm cl} = \Im(\zeta^{-1} a_{\mathrm{D}}^{\rm cl})$ appearing in the argument of ${\rm G}$ with $\YY_{\infty}^{\textrm{1-lp}} = \Im(\zeta^{-1} a_{\mathrm{D}}^{\textrm{1-lp}})$.  For physical observables that are protected from higher derivative corrections, like those built from the central charge operator, perturbative quantum corrections truncate at one loop and it makes sense to inquire about nonperturbative corrections in the coupling, even though these are exponentially small in the weak coupling regime.  Indeed, the weak coupling expansion of the prepotential, \eqref{prepotexp}, is a convergent series in a neighborhood of $|\Lambda|/|\langle \alpha, a \rangle | \to 0, \forall \alpha \in \Delta$.  Hence it seems natural to conjecture that these effects can be included as well by simply taking the argument of ${\rm G}$ to be 
\begin{equation}\label{YYexact}
\YY_\infty := \Im(\zeta^{-1} a_{\mathrm{D}})~,
\end{equation}
where $a_{\mathrm{D}} = \sum_I a_{\mathrm{D},I} (\lambda^I)^\ast$ is the $\mathfrak{t}$-valued dual coordinate constructed from the exact prepotential in the appropriate weak coupling duality frame.  We will review how this duality frame is determined in the next section, when we describe the map between physics data and math data precisely.

We hope, however, that we have already motivated the following definition of a quartet of semiclassical, geometric supercharge operators.  Let line defect data $L = \{ \vec{x}_n,P_n \}_{n=1}^{N_t}$ and asymptotic data $\{\gm;X_\infty\}$ be given such that $\left(\fMM(L;\gm;X_\infty),g,\bbJ^r\right)$, defined via \eqref{Mdef}, \eqref{metC}, and \eqref{quatstructure}, is a hyperk\"ahler manifold.  (We assume $X_\infty \in \mathfrak{g}$ is regular so that it defines a unique Cartan subalgebra, $\mathfrak{t}$.)  Then for $\YY_\infty \in \mathfrak{t}$ we set
\begin{equation}\label{Qsc}
\hat{Q}_{\rm (sc)}^a := - \frac{i g_0}{2\sqrt{2\pi}} \gamma^m (\tbbJ^a)_{m}^{\phantom{m}n} \left( \DD_n -i  {\rm G}(\YY_\infty)_n \right)~,
\end{equation}
acting on sections of the Dirac spinor bundle over $\fMM$.  Recall that $\tbbJ^a = (\mathbbm{1},-\bbJ^r)$ and observe that $\hat{Q}_{\rm (sc)}^4$ is a Dirac operator, twisted by ${\rm G}(\YY_\infty)$---\ie\ we view ${\rm G}(\YY_\infty)_n$ as the gauge field of a connection on a trivial $U(1)$ bundle over $\fMM$.

What subset of the corrections in \eqref{Qhatcc} do we claim to capture with \eqref{YYexact} and \eqref{Qsc}?  Surely not everything.  The various corrections expected in \eqref{Qhatcc} are
\begin{itemize}
\item higher (time) derivative corrections---\ie\ higher $\hat{\pi}_m$ corrections---together with higher fermi collective coordinate corrections coupled to these by supersymmetry;  together we refer to these as ``$D$-term'' corrections;
\item corrections to the weak potential energy approximation---\ie\ corrections in the ratios $\langle \alpha, Y_\infty \rangle/\langle \alpha, X_\infty\rangle$;
\item quantum corrections from integrating out the fluctuation fields around the collective coordinate ansatz; these are expected to correct each order in the time-derivative expansion.
\end{itemize}
Based on the reasoning we have described in this section, we conjecture that \eqref{Qsc} captures all weak potential energy and quantum corrections to $\hat{Q}^a$ at \emph{leading order} in the time-derivative expansion, up to a renormalization of the overall coefficient such that the kernel is unaffected.  Notice that the kernel of $\hat{Q}^{a}_{\rm (sc)}$ only depends on the bare data $(g_0,\theta_0,\mu_0)$ through the dynamical scale $\Lambda$, \eqref{dyscale}, appearing in $a_{\mathrm{D}}$.  We anticipate that $\hat{Q}_{\rm (sc)}^a$ will receive higher derivative corrections, which will correspond to higher order differential operators, but we conjecture that the kernels of these higher-derivative corrected operators coincide with the kernel of $\hat{Q}_{\rm (sc)}^a$.  More precisely, the dimension of the kernel, its decomposition into eigenspaces of the electric charge operator, and the decomposition into eigenspaces of $\hat{\II}_3$ in the case of $\ker{\hat{Q}_{\rm(sc)}^4}$, should be unaffected by the higher derivative corrections.   

We note that the leading order forms of $\hat{\gamma}_{\rm e}, \hat{\II}^r$ are given in terms of a Lie derivative with respect to canonically defined Killing vectors, and are completely independent of $g_0$.  Furthermore their overall normalization is constrained by periodicity conditions and the $\mathfrak{su}(2)$ algebra respectively.  Hence we conjecture that their leading form given in \eqref{Nhatcc} is exact at leading order in the derivative expansion.  We will refer to these leading order quantities by the same notation, $\hat{\gamma}_{\rm e}$, $\hat{\II}^r$.

The motivation behind these conjectures is again the following.  On the one hand, the kernel of $\hat{Q}^a$ determines the semiclassical spectrum of framed BPS states as a piece-wise constant function on the Coulomb branch (through the map relating $X_\infty, \YY_\infty$ to $u$; see \eqref{mathxy}).  On the other hand, the Seiberg--Witten analysis determines the same data for the (possible) spectrum in terms of the central charge.  The central charge is protected from higher order $D$-terms and hence quantities determined from it should be reproduced by the semiclassical analysis, where the quantum corrections on each side can be matched order by order.  In other words, if we use the one-loop approximation to the central charge to determine the BPS spectrum and walls of marginal stability, we should be able to recover those results on the semiclassical side via the one-loop corrected semiclassical supercharges.  

The ansatz \eqref{Qsc} is a simple and natural generalization of the leading order result, \eqref{Qhatcc}, that contains both perturbative and nonperturbative quantum corrections.  These corrections appear through the same quantities, $X_\infty = \Im(\zeta^{-1} a(u))$ and $\YY_\infty = \Im(\zeta^{-1} a_{\mathrm{D}}(u))$, that play a role in the definition of the marginal stability walls \eqref{fmsw}.  Finally in sections \ref{Section:VanillaEx}, \ref{Section:FramedEx} we will see that the conjecture is confirmed in some nontrivial examples: the walls where the kernel of the Dirac operator $\hat{Q}_{\rm (sc)}^4$ jumps can be determined directly and agree exactly with  \eqref{vanillawalls}, \eqref{fmsw}.  These arguments certainly do not constitute a proof and we believe that further semiclassical computations, either providing evidence for or refuting \eqref{Qsc}, should be carried out.  It should at least be possible to determine the one-loop correction to the supercharges at leading order in the weak-potential approximation.

We have been specifically discussing the framed case in regards to \eqref{Qsc}.  However the leading order form of the supercharges $\hat{Q}^a$ is identical in the vanilla case, and the general reasoning of the last paragraph still holds.  Therefore we conjecture that \eqref{Qsc} is also the correct form of the semiclassical supercharges, relevant for determining the BPS spectrum, in the vanilla case.  The precise construction will be described in \ref{ssec:vBPSspace}.

%%%%%%%%%%%%%%% 
\subsection{Semiclassical identification of the space of framed BPS states}\label{ssec:fBPSspace}
%%%%%%%%%%%%%%%

The algebra \eqref{QQalg2} implies the framed BPS bound $M \geq -\Re(\zeta^{-1} Z)$, which is saturated on states $\Psi$ with $\hat{Q}^a \Psi = 0$ for any---and hence all---$a =1,\ldots,4$.  Thus states preserve either all four supercharges or none, consistent with the field theory interpretation.  We will work with the conjectural semiclassical approximation $\hat{Q}^a \to \hat{Q}_{\rm (sc)}^a$ discussed in the previous section.  Specifically, it is convenient to use $\hat{Q}_{\rm (sc)}^4$ in defining the space of framed BPS states.

Let us denote the Dirac-like operator constructed from the data $(L,\gm,\sx,\sy)$ as
\begin{equation}\label{fMdiracop}
\slashed{\DD}_{\fMM(L;\gm;\sx)}^{\rG(\sy)} := \gamma^m \left( \DD_m - i \rG(\sy)_m \right)~.
\end{equation}
We will sometimes refer to this as $\slashed{\DD}^{\rm G}$ for shorthand.  It is related to $\hat{Q}_{\rm (sc)}^4$ of \eqref{Qsc} via a simple rescaling.  Therefore the kernel of this operator coincides with that of the $\hat{Q}_{\rm (sc)}^a$ and hence, conjecturally, that of the exact collective coordinate supercharges $\hat{Q}^a$.  The Dirac operator $\slashed{\DD}^{\rG}$ commutes with the semiclassical electric charge operator $\hat{\gamma}_{\rm e} = i \sum_A \alpha_{I_A} \Lie_{K_A}$ and therefore its $\Lsq$ kernel decomposes into a direct sum of $\hat{\gamma}_{\rm e}$ eigenspaces:
\begin{equation}\label{framedDker}
\ker_{\Lsq} \left( \slashed{\DD}_{\fmMM(L;\gm;\sx)}^{\rG(\sy)} \right) = \bigoplus_{\gamma_{\rm e} \in \Lambda_{\rm rt}} \ker_{\Lsq}^{\gamma_{\rm e}} \left( \slashed{\DD}_{\fmMM(L;\gm;\sx)}^{\rG(\sy)} \right)~.
\end{equation}

We would like to relate these spaces to the physical BPS spaces $\HH_{L_\zeta,u,\gamma}^{\rm BPS}$, \eqref{Hframed}.  The first task is to describe precisely how the quantities $\{L,X_\infty,\YY_\infty,\gm,\gamma_{\rm e}\}$ in the mathematical construction are related to the quantities $\{L_\zeta,u,\gamma\}$ that define the physical space of BPS states.  Henceforth we will refer to $\{L,X_\infty,\YY_\infty,\gm, \gamma_{\rm e}\}$ as \emph{math data} and $\{L_\zeta,u,\gamma\}$ as \emph{physics data}.  The relationship is most easily described in a specific weak coupling duality frame; we summarize the procedure as follows.
\begin{enumerate}
\item  Recall that the special K\"ahler manifold $\BB^\ast$ admits an atlas of distinguished charts consisting of special coordinates $a^I(u)$, where there is a Lagrangian splitting $\Gamma_{L,u} \cong \Gamma_{L,u}^{\rm m} \oplus \Gamma_{L,u}^{\rm e}$ for all $u$ in the patch.  Such a coordinate patch and splitting is referred to as a duality frame.  There is an infinite set of weak coupling duality frames in the weak coupling regime of $\BB^\ast$.  Fix arbitrarily an integral basis of simple co-roots $\{ H_I \}$ for $\Lambda_{\rm cr}$.  Then an element of this set is specified by the following two choices:
\begin{enumerate}
\item\label{choice1}  Choose any $\langle \varphi \rangle = \varphi_\infty(u) \in \mathfrak{t}_{\mathbb{C}}$  that corresponds to the given $\{ u \}$, and set $a \equiv a^I H_I = \varphi_\infty(u)$.  There are $|W|$ such choices for $\varphi_\infty(u)$, leading to $|W|$ different sets of $a^I(u)$, where $W$ is the Weyl group of $\mathfrak{g}$.  They are permuted into each other\footnote{$a = \varphi_\infty$ is necessarily a regular element of $\mathfrak{g}_{\mathbb{C}}$ because $u \notin \BB^{\rm sing}$.} by the action of $W$ on $\mathfrak{t}$.  
\item\label{choice2} For each choice of solution in \ref{choice1}, choose a set of branches of the $\half (\dim{\mathfrak{g}} - r)$ logarithms in
\begin{equation}
a_{\rm D}^{\textrm{1-lp}} = \frac{i}{2\pi} \sum_{\alpha \in \Delta^+} \alpha^\ast \langle \alpha, a \rangle \left\{ \ln \left( \frac{\langle \alpha, a \rangle^2}{2\Lambda^2} \right) + 1 \right\}~.
\end{equation}
\end{enumerate}
These two choices determine a duality frame such that $u \mapsto (a_{\rm D}(u),a(u)) \in \mathfrak{t}_{\mathbb{C}} \times \mathfrak{t}_{\mathbb{C}}$.  Alternatively, we can eliminate choice \ref{choice2} by working on the universal cover $\widehat{\BB}$ of $\BB^\ast$.  Then there are only $|W|$ weak coupling duality frames, corresponding to choice 1.  We will work on $\widehat{\BB}$ and comment on projecting identifications to $\BB^\ast$ at the end.  
\item\label{wcRegime}  Given such a frame, the Seiberg--Witten prepotential will have an expansion of the form \eqref{prepotexp}.  Since the expansion is Weyl invariant, it will in fact take the same form in any of the $|W|$ weak coupling frames on $\widehat{\BB}$.  We define the \emph{weak coupling regime} of the Coulomb branch, $\widehat{\BB}_{\rm wc} \subset \widehat{\BB}$ to consist of those $u \in \widehat{\BB}$ such that the weak coupling expansion of the prepotential, or equivalently $a_{\mathrm{D}}(a)$, converges.  Note that the definition depends on the dynamical scale: $\widehat{\BB}_{\rm wc} = \widehat{\BB}_{\rm wc}(\Lambda)$.  We let $\BB_{\rm wc}^\ast(\Lambda) \subset \BB^\ast$ denote the projection of $\widehat{\BB}_{\rm wc}$.  We would like to make a conjecture concerning this definition.  Recall that the Coulomb branch $\widehat{\BB}$ can be partitioned into chambers in which the BPS spectrum is constant.  (Here we refer to the framed or vanilla spectrum depending on whether line defects are present or not.)  We call a chamber $c$ a \emph{weak coupling chamber} if its closure in $\widehat{\BB}$ is noncompact---\ie\ it exends to infinity in some direction---\emph{and} it does not contain any of the loci $\langle \alpha_I, a(u) \rangle = 0$ for $\alpha_I$ a simple root.  (The latter condition is imposed to remove the complex co-dimension one loci that exist for $r>1$, extend to infinity, and correspond classically to partial restoration of the non-Abelian gauge group.)  We conjecture that $\widehat{\BB}_{\rm wc}$ is contained in the union of the closures of all weak coupling chambers.  This is indeed true in the case of $SU(2)$, as illustrated in figure \ref{fig0}.
%
%%%%%%%%%%%%%%%%%%%%%%%%%
\begin{figure}
\begin{center}
\includegraphics[scale=0.5]{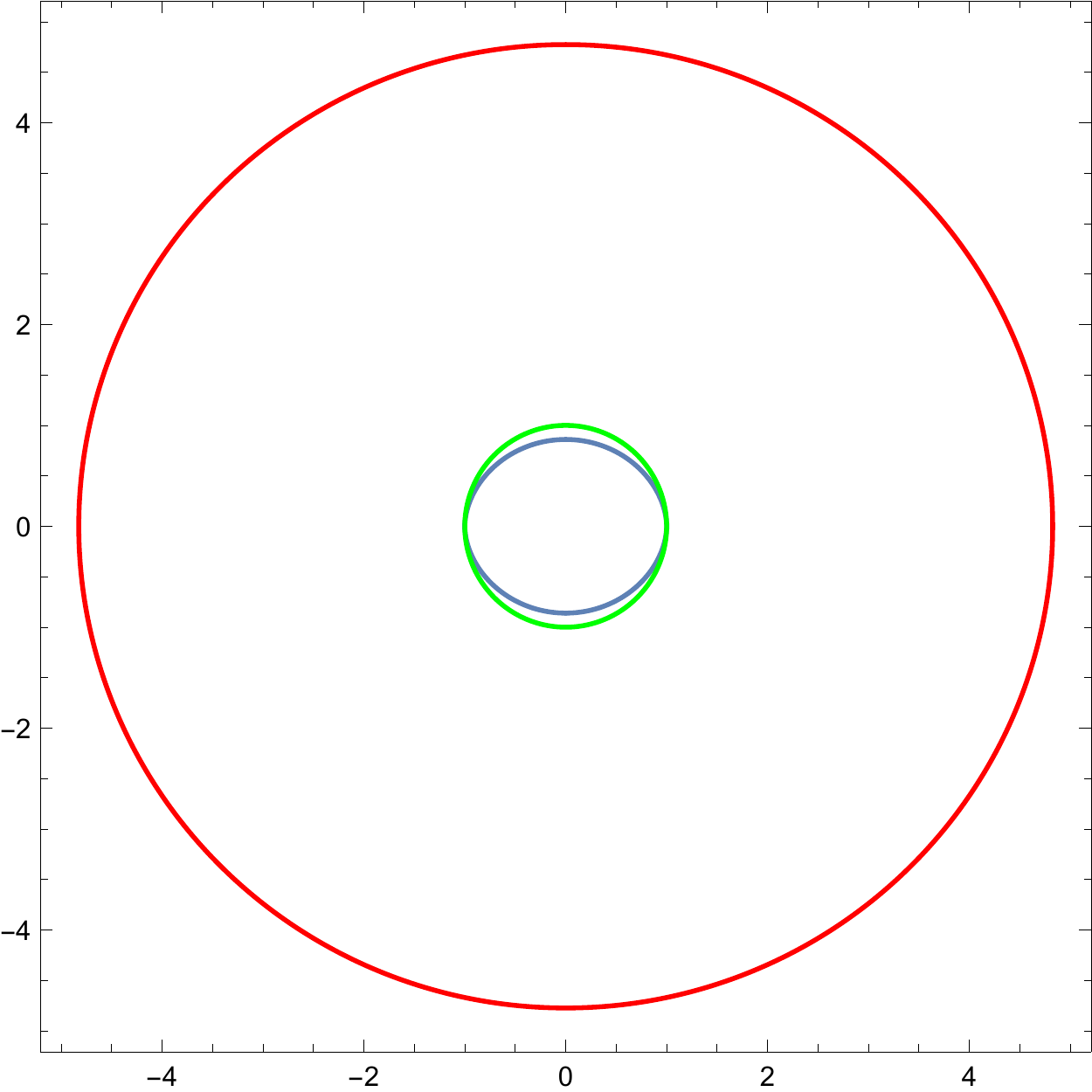}
\end{center}\caption{Here the $u$-plane is depicted and we have chosen $\Lambda=1$. In blue one recognizes the wall of marginal stability separating the weak coupling chamber (outside) and strong coupling chamber (inside). The red line is the boundary of the weak coupling regime $\BB_{\rm wc}$ (outside) which is defined in the main text as the region of convergence of $a_{\rm D}(a)$. We made this plot via an estimate of the radius of convergence using an exact computation of the first 400 coefficients of the prepotential. We would also like to point out that there are other similar but subtly different definitions of the weak coupling regime which one could consider. If one would rather use the convergence of  $a(u)$ or $a_{\rm D}(u)$ as the criterion this would lead to a region bounded by the curve in green. In all cases we see that the weak coupling regime is contained in the weak coupling chamber.}
\label{fig0}
\end{figure}
%%%%%%%%%%%%%%%%%%%%%%%%%
%  
\item\label{specialframe} Of the $|W|$ weak coupling duality frames covering $\widehat{\BB}_{\rm wc}$, we wish to work in a particular one.  In the framed case the choice is dictated by $\zeta$:  specifically, we require that $\Im(\zeta^{-1} a(u)) \in \mathfrak{t}$ be in the closure of the fundamental Weyl chamber with respect to our chosen basis of simple co-roots.  This fixes the duality frame uniquely for those $u$ such that this quantity is in the fundamental Weyl chamber, \ie\ $\langle \alpha, \Im(\zeta^{-1} a(u)) \rangle > 0, \forall \alpha \in \Delta^+$.  For now we stay away from the real co-dimension one loci where $\langle \alpha, \Im(\zeta^{-1} a(u)) \rangle$ = 0, and comment on crossing them at the end.  
\item We can now give the map between math data and physics data.  First, in this frame we identify $\Gamma_{L,u}^{\rm m} \cong \Lambda_{\rm cr} + \sum_n P_n$ and $\Gamma_{L,u}^{\rm e} \cong \Lambda_{\rm rt}$.  Let ${\rm m,e}$, denote the corresponding magnetic and electric trivialization maps, ${\rm m} : \Gamma_{L,u} \to \Lambda_{\rm cr} + \sum_n P_n$, and  ${\rm e}: \Gamma_{L,u} \to \Lambda_{\rm rt}$.  Then for a given section $\gamma \in \Gamma_{L}$ of the electromagnetic charge lattice we identify
\begin{equation}\label{mathgmap}
\gm = {\rm m}(\gamma)~, \qquad \gamma_{\rm e} = {\rm e}(\gamma)~,
\end{equation}
where $\gm$ is the asymptotic magnetic charge required to construct $\fMM$, and $\gamma_{\rm e}$ is an eigenvalue of the semiclassical electric charge operator, $\hat{\gamma}_{\rm e} = i \sum_A \alpha_{I_A} \Lie_{K^A}$.
\item Secondly we set
\begin{align}\label{mathxy}
X_\infty =&~ X_{\infty}(u,\zeta) := \Im(\zeta^{-1} a(u))~,  \cr
\YY_{\infty} =&~ \YY_\infty(u,\zeta;\Lambda) := \Im(\zeta^{-1} a_{\mathrm{D}}(u;\Lambda))~,
\end{align}
where we have noted the dependence of $\YY_\infty$ on the dynamical scale $\Lambda$.  This will usually be suppressed.  By construction $X_\infty$ satisfies $\langle \alpha, X_\infty(u,\zeta)\rangle \geq 0, \forall \alpha \in \Delta^+$, $u \in \widehat{\BB}_{\rm wc}$, and for now we consider only those $u$'s such that $\langle \alpha, X_\infty(u,\zeta) \rangle > 0, \forall \alpha \in \Delta^+$.
\end{enumerate}

Using this map we can now identify $\HH_{L_\zeta,u,\gamma}^{\rm BPS}$ with spaces appearing in \eqref{framedDker}.  For all $u \in \widehat{\BB}_{\rm wc}$ such that $\langle \alpha, \sx(u,\zeta)\rangle > 0, \forall \alpha \in \Delta^+$, we have
\begin{equation}\label{mainres}
\HH_{L_\zeta,u,\gamma}^{\rm BPS} \cong \ker_{\Lsq}^{{\rm e}(\gamma)} \left( \slashed{\DD}_{\fmMM(L; {\rm m}(\gamma) ;\sx(u,\zeta))}^{\rG(\sy(u,\zeta))} \right)~.
\end{equation}
This result extends previous work on semiclassical $\NN = 2$ SYM \cite{Gauntlett:1993sh,Sethi:1995zm,Gauntlett:1995fu,Lee:1996kz,Gauntlett:1999vc,Gauntlett:2000ks} in two directions.  First, we have identified the space of \emph{framed} BPS states for $\NN = 2$ SYM in the presence of supersymmetric 't Hooft defects specified by the line defect data $L_\zeta(\{\vec{x}_n,P_n\})$.  The data $\{\vec{x}_n,P_n\}$ enters through the determination of the singular monopole moduli space $\fmMM$, while the phase $\zeta$ enters through the specification of the duality frame and the map \eqref{mathxy}.  Second, we have extended the usual semiclassical constructs to capture, conjecturally, all perturbative and nonperturbative corrections that are relevant for the Hilbert space of BPS states, $\HH_{L_\zeta,u}^{\rm BPS}$.  We will describe an analogous result for the vanilla BPS spaces in the next section.

The identification \eqref{mainres} leads to several interesting mathematical applications.  Wall crossing properties of $\HH_{L_\zeta,u,\gamma}^{\rm BPS}$, as encoded by the protected spin characters $\fOmega$, imply a remarkable set of wall crossing formulae for certain index characters of the family of Dirac operators.  The absence of exotics---BPS states transforming in nontrivial $SU(2)_R$ representations---imposes strong constraints on the kernel of the Dirac operator.  Going in the other direction, we can make use of a Lichnerowicz--Weitzenb\"ock formula to find a special locus in $\widehat{\BB}_{\rm wc}$ where there are no framed BPS states other than the `vacuum', consisting of the pure 't Hooft defect.  Each of these will be discussed in section \ref{Section:Applications}, after describing the analog of \eqref{mainres} for vanilla BPS states.

Before turning to that we make a few observations.
\begin{itemize}
\item It follows from the construction of appendix \ref{app:embed} that for given $\{L,\gm\}$ such that the relative magnetic charge $\tilde{\gamma}_{\rm m} = \sum_{A=1}^d \tilde{n}_{\rm m}^{I_A} H_{I_A}$, the moduli space metric depends only on the components $\langle \alpha_{I_A}, X_\infty \rangle$ of $X_\infty$ and not the components $\langle \alpha_{I_M}, X_\infty \rangle$.  Furthermore ${\rm G}(\YY_\infty) = \sum_A \langle \alpha_{I_A}, \YY_\infty \rangle K^A$.  Hence the family of Dirac operators \eqref{fMdiracop} for such $\{L,\gm\}$ only depends on $2d$ real parameters, where $d = \rnk{\mathfrak{g}^{\rm ef}}$, rather than the $2r$ that are naively implied by the dependence on $\{X_\infty, \YY_\infty\}$.  Combined with \eqref{mainres} we learn the following.  For all $u$ such that the map \eqref{mathxy} applies and all $\{L_\zeta,\gamma\}$ such that $\tilde{\gamma}_{\rm m} = {\rm m}(\gamma) - \sum_{n} P_{n}^- = \sum_{A=1}^{d} \tilde{n}_{\rm m}^{I_A}$, the BPS Hilbert spaces $\HH_{L_\zeta,u,\gamma}$ are invariant along the $2(r-d)$ real dimensional surfaces in $\widehat{\BB}_{\rm wc}$ parameterized by $\langle \alpha_{I_M}, X_\infty(u,\zeta)\rangle$ and $\langle \alpha_{I_M},\YY_\infty(u,\zeta)\rangle$ for $M=1,\ldots,r-d$, with $\langle \alpha_{I_A}, X_\infty(u,\zeta)\rangle$ and $\langle \alpha_{I_A},\YY_\infty(u,\zeta)\rangle$ held fixed.
\item  As we mentioned, \eqref{mainres} is expected to be valid throughout $\widehat{\BB}_{\rm wc}$, except  where $\langle \alpha, \sx \rangle = 0$ for some nonzero root $\alpha$.  A semiclassical description valid on these real co-dimension one loci---if it can be given at all---would require a discussion of moduli spaces for massless monopoles with potentials---see \eg\ \cite{Lee:1999iia,Houghton:1999cm}---generalized to the case with defects.  We have a valid description on either side of such a ``wall'' and will content ourselves with understanding how the description changes when we cross it.\footnote{Note these walls do not, a priori, have anything to do with marginal stability walls; rather they are walls where our duality frame, \ie\ special coordinate patch, is breaking down and when we cross them we need to switch frames.  It might be that they also correspond to marginal stability walls for some BPS states.}  We start with $\sx$ in the fundamental Weyl chamber.  After crossing a wall corresponding to root $\alpha$, the choice in \ref{choice1} such that $\sx$ is in the fundamental Weyl chamber changes.  Indeed the new $a(u)$ is related to the old one by a Weyl reflection about the root $\alpha$.  This Weyl transformation acts as an $Sp(2r,\mathbb{Z})$ duality transformation on the doublets $\{a_{\mathrm{D}}^{\textrm{1-lp}},a\}$, hence $\{\YY_\infty,X_\infty\}$, and $\{\gm,\gamma_{\rm e}\}$.  (This transformation will be block diagonal with respect to the magnetic-electric splitting.)  If the wall is not a marginal stability wall, then \eqref{mainres} implies that the kernels of the Dirac operators corresponding to the old and new math data must be equivalent.  If it is a marginal stability wall, then the kernels will have to be related by the appropriate wall crossing transformation.  Note that this can even lead to relations among kernels of Dirac operators built on monopole moduli spaces of different dimensions!  An example of this is studied in section \ref{ssec:tropical}.
\item Projecting the map \eqref{mathgmap}, \eqref{mathxy} from $\widehat{\BB}_{\rm wc}$ to $\BB_{\rm wc}^\ast$ implies further relations among the kernels.  Suppose $u_{1,2} \in \widehat{\BB}_{\rm wc}$, with $u_1 \neq u_2$, project to the same $u \in \BB_{\rm wc}^\ast$.  The doublets $\{a_{\mathrm{D}}( u_{1,2}),a(u_{1,2})\}$ obtained by the above procedure for $u_1$ and $u_2$ will be related by an $Sp(2r,\mathbb{Z})$ duality transformation.  That same transformation will act on $\{\gm,\gamma_{\rm e}\}$, such that the physical quantities, $Z_\gamma(u)$, $\HH_{L_\zeta,u,\gamma}$, $\fOmega(L_\zeta,u,\gamma;y)$ remain invariant.  (This transformation will be of the $T$-type and corresponds to the usual Witten effect as described around \eqref{chargefibration}.)  The kernels of the corresponding Dirac operators must again be equivalent.
\item One might wonder if there is a role for the $S$-type duality transformations.  While one can certainly apply such a transformation to the math doublets $\{\YY_\infty,X_\infty\}$ and $\{\gm,\gamma_{\rm e}\}$ and consider the corresponding Dirac operator, there is no reason for its kernel to be physically meaningful.  The transformations discussed in the previous two items map weak coupling descriptions to weak coupling descriptions: the coordinates $a^I(u)$ are interpretable as the Cartan components of the vev of a non-Abelian Higgs field.   Furthermore the quantum corrections to the prepotential that defines $a_{\mathrm{D}}$ arise from integrating out massive fundamental degrees of freedom in a UV complete theory that are again associated with a non-Abelian gauge theory.  This connection to a non-Abelian gauge theory where the BPS states are soliton states is essential for the semiclassical construction of the Dirac operator.  The low energy degrees of freedom in an $S$-dual frame do not have such an interpretation.
\end{itemize}

%%%%%%%%%%%%%%% 
\subsection{Semiclassical identification of the space of vanilla BPS states}\label{ssec:vBPSspace}
%%%%%%%%%%%%%%%

Let us now turn to the vanilla case which, as usual, requires some additional steps due to the reducible structure of the moduli space.  

We first describe the modifications to the map between math and physics data.  In the vanilla case the particular duality frame we work in is dictated by the electromagnetic charge $\gamma \in \Gamma$ under consideration.  For a given $\gamma$ and duality frame choice $p$, we have $\{a_p(u),a_{\mathrm{D},p}(u)\} \in \mathfrak{t}_{\mathbb{C}} \times \mathfrak{t}_{\mathbb{C}}$ and a trivialization $\gamma \to \gamma_{\rm m}^p \oplus \gamma_{\rm e}^p \in \Lambda_{\rm cr} \oplus \Lambda_{\rm rt}$.  The central charge $Z_{\gamma}(u) = (\gamma_{\rm m}^p, a_{D,p}(u)) + \langle \gamma_{\rm e}^p, a_p(u)\rangle \in \mathbb{C}$ is a duality frame invariant; although we chose a frame to compute it we would get the same answer in any frame.  Define $\zeta_{\rm van}(u,\gamma) := - Z_{\gamma}(u)/ |Z_{\gamma}(u)|$, which is therefore also a duality invariant quantity.  

Then the duality frame of interest, \ie\ the analog of item \ref{specialframe} above, is the frame such that $\Im(\zeta_{\rm van}^{-1} a(u))$ is in the closure of the fundamental Weyl chamber.  This choice is unique for a given charge provided we are inside the chamber: $\langle \alpha, \Im(\zeta_{\rm van}^{-1} a(u)) \rangle > 0, \forall \alpha \in \Delta^+$.  We will restrict to $u \in \widehat{\BB}_{\rm wc}$ such that this is the case.  As in the framed case the analysis can be extended to those $u\in \widehat{\BB}_{\rm wc}$ such that $\langle \alpha, \zeta_{\rm van}^{-1} a(u) \rangle = 0$ for some $\alpha$ by appealing to the piecewise constancy of the spectrum and certain weak coupling electromagnetic duality transformations.  

We can now give the map between math data and physics data.  We identify $\gm$ and $\gamma_{\rm e}$ as the magnetic and electric components of the trivialization of the charge with respect to this duality frame, analogously to \eqref{mathgmap}.  Then we set
\begin{align}\label{mathxymapvan}
\sx =&~  \sx(u,\gamma) := \Im( \zeta_{\rm van}^{-1}(u,\gamma) a(u))~, \cr
\sy =&~   \sy(u,\gamma) := \Im( \zeta_{\rm van}^{-1}(u,\gamma) a_{\mathrm{D}}(u))~.
\end{align}
Note that in electromagnetic charge eigenspace $\gm \oplus \gamma_{\rm e}$, the identity $\Im(\zeta_{\rm van}^{-1} Z_\gamma) = 0$ implies the quantum generalization of the classical constraint \eqref{vconstraint2}:
\begin{equation}\label{Qvconstraint}
\langle \gamma_{\rm e}, X_\infty \rangle + (\gm, \sy) = 0~.
\end{equation}
Classically, this relation was a consequence of having a solution to the BPS equations; see \eqref{vanconstraint}.  Semiclassically, we will see below that this is a consequence of our assignments for the supercharge and electric charge operators.

The supercharge $\hat{Q}_{\rm (sc)}^4$ is proportional to a Dirac operator, $\slashed{\DD}_{\MM(\gm;X_\infty)}^{\rG(\sy)}$, which is the same in form as \eqref{fMdiracop}.  The space of vanilla BPS states is given in terms of the kernel of $\slashed{\DD}^{\rG}$, but this is not an $\Lsq$ kernel.  On the physics side the reason is that BPS states are asymptotic one-particle states; on the math side this is related to the reducible structure of $\MM$.  In order to clarify this we will decompose the Dirac operator into a sum of two terms---one that acts nontrivially on the center-of-mass $\mathbb{R}^4 = \mathbb{R}_{\rm cm}^3 \times \mathbb{R}_{X_\infty}$ factor of the simply-connected cover, $\widetilde{\MM}$, and one that acts nontrivially on the strongly centered moduli space $\MM_0$.  

We begin by recalling the direct product metric \eqref{productmetric} and introducing some notation for the center-of-mass and strongly centered coordinates:
\begin{align}\label{productmetric2}
\ed s_{\MM}^2 \equiv &~ g_{mn} \ed z^m \ed z^n = g_{ab} \ed z^a \ed z^b + g_{\tm\tn} \ed z^{\tm} \ed z^{\tn} \cr
=&~ (\gm, X_\infty) \left( \ed \vec{x}_{\rm cm} \cdot \ed \vec{x}_{\rm cm} + \frac{\ed \chi^2}{(\gm,X_\infty)^2} \right) + \ed s_{\MM_0}^2~.
\end{align}
Here $z^a = \{\vec{x}_{\rm cm},\chi\}$ are the (globally well-defined) center-of-mass coordinates and $z^{\tm}$, $\tm = 1,\ldots 4(|\gm|-1)$, denote local coordinates on $\MM_0$.  Correspondingly, let $\gamma^m = \{ \gamma^a,\gamma^{\tm} \}$ be an adapted basis of gamma matrices associated with the coordinate frame.  We take
\begin{equation}\label{cliffordfactor}
\gamma^a = \Gamma^a \otimes \mathbbm{1}~, \qquad \gamma^{\tm} = \overline{\Gamma} \otimes \gamma_{0}^{\tm}~,
\end{equation}
where the $\Gamma^a$ are four-dimensional matrices satisfying $[\Gamma^a,\Gamma^b]_+ = 2 g^{ab}$, the $\gamma_{0}^{\tm}$ are $4(|\gm|-1)$-dimensional gamma matrices satisfying $[ \gamma_{0}^{\tm}, \gamma_{0}^{\tn} ]_+ = 2 g^{\tm\tn}$, and $\overbar{\Gamma}$ is the chirality operator on the $\mathbb{R}^4$ center of mass factor, $\overline{\Gamma} := \Gamma^{\underline{1}} \cdots \Gamma^{\underline{4}} = (\det{g_{ab}})^{-1/2} \Gamma^{1} \cdots \Gamma^4$.  

We must also decompose ${\rm G}(\sy)$.  From \eqref{Cartandecomp} and \eqref{vGdecomp} we have that $\rG(\sy) = (\gm,\sy) \pd_\chi + \rG_0(\sy)$, where $\rG_0(\sy)$ is a triholomorphic Killing vector on $\MM_0$.  Then the full Dirac operator is
\begin{equation}\label{Diracdecomp}
\slashed{\DD}_{\mMM(\gm;\sx)}^{{\rm G}(\sy)} = \left[ \Gamma^i \pd_i + \Gamma^\chi \left( \pd_\chi - i \frac{(\gm,\sy)}{(\gm,\sx)} \right) \right] \otimes \mathbbm{1} + \bar{\Gamma} \otimes \slashed{\DD}_{\mMM_0(\gm;\sx)}^{{\rm G}_0(\YY_{\infty})} ~,
\end{equation}
where we have introduced the $\rG_0(\YY_{\infty})$-twisted Dirac operator on the strongly centered moduli space
\begin{equation}\label{M0diracop}
\slashed{\DD}_{\mMM_0(\gm;\sx)}^{{\rm G}_0(\YY_{\infty})} := \gamma_{0}^{\tm} \left( \DD_{\tm} -i {\rm G}_0(\YY_{\infty})_{\tm} \right)~.
\end{equation}
We will sometimes refer to this Dirac operator simply as $\slashed{\DD}_{0}^{\rG_0}$.  It follows from \eqref{Diracdecomp} that BPS states $\Psi$ must satisfy
\begin{align}\label{vanBPSconditions}
&\pd_i \Psi = 0~, \qquad  \left(\pd_\chi - i \frac{(\gm,\sy)}{(\gm,\sx)} \right) \Psi = 0~, \qquad \textrm{and} \quad  (\overline{\Gamma} \otimes \slashed{\DD}_{0}^{\rG_0} ) \Psi = 0~.
\end{align}
We introduce a factorization of $\Psi$ corresponding to the factorization \eqref{cliffordfactor}, $\Psi = \Psi_{\rm cm}(z^a) \otimes \Psi_0(z^{\tm})$, and deduce 
\begin{equation}\label{Psidecomp0}
\Psi = \psi_{\rm cm} e^{-i q_{\rm cm} \chi} \otimes \Psi_0(z^{\tm})~,
\end{equation}
where $\psi_{\rm cm}$ is a \emph{constant} four-component spinor on $\mathbb{R}^4$, and $q_{\rm cm}$ and $\Psi_0$ satisfy
\begin{equation}
q_{\rm cm} =  - \frac{(\gm,\sy)}{(\gm,\sx)}~, \qquad \slashed{\DD}_{0}^{\rG_0} \Psi_0 = 0 ~.
\end{equation}

Next we introduce the electric charge operator and describe how the decomposition of the kernel of $\slashed{\DD}^{\rm G}$ into electric charge eigenspaces is implemented with respect to the factorization.  Recall that the electric charge operator can be expressed in terms of spinorial Lie derivatives: $\hat{\gamma}_{\rm e} = \sum_A \alpha_{I_A} (i \Lie_{K^A})$.  Via the G-map, the change of basis $\{ h^{A} \} \mapsto \{h_{0}^{A}, h_{\rm cm} \}$ on $\mathfrak{t}^{\rm ef}$ gives a corresponding change of basis $\{ K^A \} \mapsto \{K_{0}^A, \pd_\chi \}$ on $\mathfrak{isom}_{\mathbb{H}}(\MM)$.  Linearity of the Lie derivative $\Lie_V$ with respect to $V$ leads to a decomposition of $\hat{\gamma}_{\rm e}$ completely analogous to \eqref{gebetaexp}:
\begin{equation}
\hat{\gamma}_{\rm e} = \gm^\ast (i\Lie_{\pd_\chi}) + \sum_{A=1}^{d-1} i_\ast(\beta_A) (i \Lie_{K_{0}^A}) =: \gm^\ast (i \Lie_{\pd_\chi}) + \hat{\gamma}_{{\rm e},0}~,
\end{equation}
where the last step defines a \emph{relative electric charge operator}, $\hat{\gamma}_{{\rm e},0}$, on $\MM_0$.  Then if $\Psi^{\gamma_{\rm e}}$ denotes an eigenspinor of $\hat{\gamma}_{\rm e}$ corresponding to electric charge eigenvalue $\gamma_{\rm e}$, we have
\begin{align}\label{echargeeigen}
i \Lie_{\pd_\chi} \Psi^{\gamma_{\rm e}} =&~  \frac{\langle \gamma_{\rm e}, X_\infty \rangle}{(\gm,X_\infty)} \Psi^{\gamma_{\rm e}} \equiv q_{\rm cm} \Psi^{\gamma_{\rm e}}  \qquad \textrm{and} \cr
\hat{\gamma}_{{\rm e},0} \Psi^{\gamma_{\rm e}} =&~  i\sum_{A=1}^{d-1} \left( \Lie_{K_{0}^A} \Psi^{\gamma_{\rm e}}\right) i_\ast(\beta_A) = \sum_{A=1}^{d-1} \left(N_{{\rm e},0}^A \, i_\ast(\beta_A)\right) \Psi^{\gamma_{\rm e}} = \gamma_{{\rm e},0} \Psi^{\gamma_{\rm e}}~,
\end{align}
where the $N_{{\rm e},0}^A$ are the integers defined in \eqref{echarge0def}.

Now let us impose both \eqref{vanBPSconditions} and \eqref{echargeeigen}.  Since $\Lie_{\pd_\chi} = \pd_\chi$, we see that the first condition of \eqref{echargeeigen} is compatible with the second condition of \eqref{vanBPSconditions} precisely because of \eqref{Qvconstraint}.  Meanwhile the operators $i \Lie_{K_{0}^A}$ act only on the $\Psi_0$ factor of $\Psi$ and furthermore commute with the Dirac operator $\slashed{\DD}_{0}^{\rG_0}$.  Thus we write
\begin{equation}\label{Psidecomp}
\Psi^{\gamma_{\rm e}} = \psi_{\rm cm} e^{-iq_{\rm cm} \chi} \otimes \Psi_{0}^{\gamma_{{\rm e},0}}~.
\end{equation}
For fixed data $\{\gm,X_\infty,\sy\}$, the $\Lsq$ kernel of $\slashed{\DD}_{0}^{\rG}$ decomposes into eigenspaces labeled by $\gerel$:
\begin{equation}\label{vanDker}
\ker_{\Lsq}\left( \slashed{\DD}_{0}^{\rG_0}\right) = \bigoplus_{\gerel} \ker_{\Lsq}^{\gerel} \left( \slashed{\DD}_{0}^{\rG_0}\right)~,
\end{equation}
where an $\Lsq$ spinor $\Psi_{0}^{\gerel}$ satisfies
\begin{equation}\label{Psi0conditions}
\Psi_{0}^{\gerel} \in \ker_{\Lsq}^{\gerel} \left( \slashed{\DD}_{0}^{\rG_0}\right) \quad \iff \quad \left\{ \begin{array}{l} \slashed{\DD}_{0}^{\rG_0} \Psi_{0}^{\gerel} = 0~,~~\textrm{and}  \\[1ex]  i \Lie_{K_{0}^A} \Psi_{0}^{\gerel} = N_{{\rm e},0}^A \Psi_{0}^{\gerel}~. \end{array} \right.
\end{equation}

Suppose we are given such a spinor $\Psi_{0}^{\gerel} \in \ker_{\Lsq}^{\gerel}(\slashed{\DD}_{0}^{\rG_0})$.  Then \eqref{Psidecomp} clearly defines a section of the Dirac spinor bundle over the universal cover $\widetilde{\MM}$.  However it does \emph{not} necessarily give a well-defined spinor on $\MM$ and hence does \emph{not} necessarily represent a BPS state.  In fact, we claim that it does descend to a well-defined spinor on $\widetilde{\MM}/\mathbb{D}_\gi$, but an additional equivariance condition on $\Psi_{0}^{\gerel}$ must be imposed in order to have the remaining $\mathbb{Z}_L$ quotient action of \eqref{Dhkfactor} be well-defined.

In order to explain these statements, we first construct a lift of the isometry, $\phi_\gi : \widetilde{\MM} \to \widetilde{\MM}$, that generates the subgroup of gauge-induced deck transformations, $\mathbb{D}_\gi$, to the space of sections of the spinor bundle, $\Gamma(\widetilde{\MM},\SS_{\rm D})$.  Recall that $\phi_\gi$ can be represented as $\exp(2\pi \rG(i_\ast(h_\gi)))$ for a particular $h_\gi \in \Lambda_{\rm mw}^{\rm ef}$ such that $\mu(h_\gi)$ generates the subgroup $\im(\mu)  \cong L \cdot \mathbb{Z} \cong \mathbb{D}_\gi $ in $\mathbb{Z} \cong \mathbb{D} $.  Furthermore we showed in \eqref{Nhatcc} that the Noether charge operator associated with a symmetry generated by Killing field $K$ is given in terms of the spinorial Lie derivative, $\hat{N}_K = i \Lie_{K}$.\footnote{We also argued later in section \ref{ssec:validity} that the corrections indicated in \eqref{Nhatcc} vanish.}  Hence the lift to $\Gamma(\widetilde{\MM},\SS_{\rm D})$ is
\begin{equation}
\widetilde{\phi}_\gi := \exp(2\pi i \hat{N}_{\rG(i_\ast(h_\gi))}) = \exp\left( -2\pi \Lie_{\rG(i_\ast(h_\gi))} \right) : \Gamma(\widetilde{\MM},\SS_{\rm D}) \to \Gamma(\widetilde{\MM},\SS_{\rm D})~.\label{liftedphi}
\end{equation}

Now, on the one hand, $\rG(i_\ast(h_\gi)) = \sum_A \langle \alpha_A, h_\gi \rangle K^A$ and $i \Lie_{K^A} \Psi^{\gamma_{\rm e}} = n_{\rm e}^{I_A} \Psi^{\gamma_{\rm e}}$.  Thus, since $i_\ast(h_\gi)$ is an element of the magnetic weight lattice and $\gamma_{\rm e} = \sum_{A} n_{\rm e}^{I_A} \alpha_{I_A}$ is an element of the (integral dual) root lattice, we have
\begin{equation}\label{Dhkinvariance}
\widetilde{\phi}_\gi(\Psi^{\gamma_{\rm e}}) = \exp\left(-2\pi i \sum_{A=1}^{d} \langle \alpha_A, h_\gi \rangle n_{\rm e}^{I_A} \right) \Psi^{\gamma_{\rm e}} = e^{-2\pi i\langle \gamma_{\rm e}, i_\ast(h_\gi)\rangle} \Psi^{\gamma_{\rm e}} = \Psi^{\gamma_{\rm e}}~.
\end{equation}
Hence $\Psi^{\gamma_{\rm e}}$ is invariant under the action of $\widetilde{\phi}_\gi$ and this implies that $\Psi^{\gamma_{\rm e}}$ is a well-defined section of the spinor bundle over the quotient $\widetilde{\MM}/\mathbb{D}_\gi$.

On the other hand, the action of $\phi_\gi$ factorizes according to \eqref{hkdeckgen} and hence so does the action of $\widetilde{\phi}_\gi$:
\begin{align}\label{liftedphith}
\widetilde{\phi}_\gi(\Psi) =&~ \left( \exp(-2\pi L \Lie_{\pd_\chi}) \cdot \Psi_{\rm cm}\right) \otimes \widetilde{\phi}_{\gi,0}(\Psi_0)~, \quad \textrm{where} \cr
\widetilde{\phi}_{\gi,0} :=&~ \exp\left(-2\pi  \sum_{A=1}^{d-1} \langle \beta_A, h_\gi\rangle \Lie_{K_{0}^A} \right) : \Gamma(\MM_0,\SS_{\rm D}) \to \Gamma(\MM_0,\SS_{\rm D})~.
\end{align}
Therefore the action on $\Psi^{\gamma_{\rm e}}$ takes the form
\begin{equation}
\widetilde{\phi}_\gi(\Psi^{\gamma_{\rm e}}) = \psi_{\rm cm} e^{-i q_{\rm cm} (\chi - 2\pi L)} \otimes \widetilde{\phi}_{\gi,0}\left(\Psi_{0}^{\gerel}\right)~.
\end{equation}
Given \eqref{Dhkinvariance} we conclude that $\widetilde{\phi}_{\gi,0}$ acts by a phase to cancel the phase induced by the action on the center of mass factor:
\begin{equation}\label{Dhk0action}
\widetilde{\phi}_{\gi,0}\left(\Psi_{0}^{\gerel}\right) = e^{-2\pi i q_{\rm cm} L} \Psi_{0}^{\gerel}~.
\end{equation}
Note that $q_{\rm cm}$ is generically an irrational real number, so no power of $\widetilde{\phi}_{\gi,0}$ will be trivial. 

The action \eqref{Dhk0action} is indeed compatible with the definition of $\widetilde{\phi}_{\gi,0}$ in \eqref{liftedphith}.  To observe this, recall the expansion of the electric charge eigenvalue \eqref{gebetaexp} and take the pairing of both sides with with $h_\gi$ to arrive at 
\begin{equation}\label{gehth}
\langle \gamma_{\rm e}, h_\gi\rangle = q_{\rm cm} L + \sum_{A=1}^{d-1} N_{{\rm e},0}^A \langle i_\ast(\beta_A), h_\gi \rangle~.
\end{equation}
The sum on the right-hand side (multiplied by $2\pi i$) is exactly what we would get by acting with $\widetilde{\phi}_{\gi,0}$, \eqref{liftedphith}, on $\Psi_{0}^{\gerel}$ using \eqref{Psi0conditions}.  Meanwhile $\langle \gamma_{\rm e}, h_\gi \rangle$ is an integer so $2\pi i$ times this quantity in the exponential gives one.  Hence we recover \eqref{Dhk0action} using the definition of $\widetilde{\phi}_{\gi,0}$ and \eqref{gehth}.

The result \eqref{Dhk0action} holds for any $\Psi_{0}^{\gerel} \in \ker_{\Lsq}^{\gerel}(\slashed{\DD}_{0}^{\rG_0})$ and ensures that the corresponding $\Psi^{\gamma_{\rm e}} \in \ker(\slashed{\DD}^{\rG})$, \eqref{Psidecomp}, is a well-defined spinor on $\widetilde{\MM}/\mathbb{D}_\gi$.  However it is not sufficient to ensure that $\Psi^{\gamma_{\rm e}}$ is well-defined $\MM$, which involves an additional $\mathbb{Z}_L$ quotient.  Recall that for $L > 1$ the $\mathbb{Z}_L$ action is generated by the isometry $\phi$, whose action factorizes into a uniform translation of $\mathbb{R}_{X_\infty}$ by $\chi \to \chi + 2\pi$, and an isometry $\phi_0$ of $\MM_0$ such that $\phi_{0}^L = \phi_{\gi,0}$; (see \eqref{deckgen}).  

Let us first analyze the situation from the point of view of $\MM_0$.  We know that the action of $\widetilde{\phi}_{\gi,0}$ is \eqref{liftedphith}, and on $\Psi_{0}^{\gerel}$ this is given by
\begin{equation}\label{liftedphithaction}
\widetilde{\phi}_{\gi,0}( \Psi_{0}^{\gerel}) = e^{-2\pi i L c}  \Psi_{0}^{\gerel}~,
\end{equation}
where for convenience we have introduced the notation $c =  c(\gm,X_\infty,\gerel)$ for the quantity
\begin{equation}
c(\gm,X_\infty,\gerel) := -\frac{1}{L} \sum_{A=1}^{d-1} \langle \beta_A, h_\gi\rangle N_{{\rm e},0}^A~.
\end{equation}
Here we are emphasizing that $c$ depends only on data intrinsic to $\MM_0$---the magnetic charge, the Higgs vev (which goes into the determination of the $\beta_A$ because of the condition $\langle i_\ast(\beta_A), X_\infty \rangle = 0$), and the eigenvalues of $i \LL_{K_{0}^A}$ that determine the relative electric charge $\gerel$.  Now let $\widetilde{\phi}_0$ denote the lift of $\phi_0$.  We require that $(\widetilde{\phi}_{0})^L = \widetilde{\phi}_{\gi,0}$, but this only determines its action on $\Psi_{0}^{\gerel}$ up to an $L^{\rm th}$ root of unity: $\Psi_{0}^{\gerel} \mapsto e^{-2\pi i (c + k/L)} \Psi_{0}^{\gerel}$ will satisfy this property for any $k \in \mathbb{Z}_L$.  Indeed, since \eqref{liftedphithaction} holds for any $\Psi_{0}^{\gerel}$ in the kernel, we have a decomposition
\begin{equation}\label{ZLgrading}
\ker_{\Lsq}^{\gerel}\left(\slashed{\DD}_{0}^{\rG_0}\right) = \bigoplus_{k=0}^{L-1} \ker_{\Lsq}^{\gerel}\left(\slashed{\DD}_{0}^{\rG_0}\right)^{(k)} ~, 
\end{equation}
where
\begin{equation}\label{ZLgrading2}
\ker_{\Lsq}^{\gerel}\left(\slashed{\DD}_{0}^{\rG_0}\right)^{(k)} := \left\{ \Psi_{0} \in \ker_{\Lsq}^{\gerel}\left(\slashed{\DD}_{0}^{\rG_0}\right) ~ \bigg| ~ \widetilde{\phi}_0(\Psi_{0}) = e^{-2\pi i (c + k/L)}\Psi_{0} \right\}~. 
\end{equation}
How do we choose a $k$?

The point is that we know that $\phi$ acts by a uniform translation $\chi \to \chi + 2\pi$ simultaneously with the action $\phi_{0}$ on $\MM_0$.  Let $\widetilde{\phi}$ denote the lift of $\phi$.  Then we fix $k$ by the requirement that $\widetilde{\phi}$ leave the total wavefunction $\Psi^{\gamma_{\rm e}}$ invariant.  We have $\widetilde{\phi} = \exp(-2\pi \Lie_{\pd_\chi}) \otimes \widetilde{\phi}_0$ and thus
\begin{equation}
\widetilde{\phi}(\Psi^{\gamma_{\rm e}}) = \Psi^{\gamma_{\rm e}} \quad \iff \quad \widetilde{\phi}_{0}(\Psi_{0}^{\gerel}) = e^{-2\pi i q_{\rm cm}} \Psi_{0}^{\gerel}~.
\end{equation}
Thus we require that $q_{\rm cm} = c + \frac{k}{L}$ modulo integers.  But now we can identify $k$ by comparing this condition with the equation \eqref{gehth}, which can be written as
\begin{equation}\label{qcmpieces}
q_{\rm cm} = \frac{1}{L} \langle \gamma_{\rm e}, h_\gi \rangle + c~.
\end{equation}
Hence we deduce $k = k_{\gamma_{\rm e}}$, where
\begin{equation}\label{kgedef}
k_{\gamma_{\rm e}} := \langle \gamma_{\rm e}, h_\gi \rangle ~{\rm mod}~L ~,
\end{equation}
and we have that $\Psi^{\gamma_{\rm e}}$ is a well-defined spinor on $\MM = \widetilde{\MM}/\mathbb{D}$:
\begin{equation}\label{ZLgrading3}
\Psi_{0}^{\gerel} \in \ker_{\Lsq}^{\gerel}\left(\slashed{\DD}_{0}^{\rG_0}\right)^{(k_{\gamma_{\rm e}})}  \qquad \Rightarrow \qquad \widetilde{\phi}(\Psi^{\gamma_{\rm e}}) = \Psi^{\gamma_{\rm e}}~.
\end{equation}
Note that neither $c$ nor $k_{\gamma_{\rm e}}$ is invariant under a shift of $h_\gi$ by an element in $\ker{\mu} = \Span_{\mathbb{Z}} \{h_{0}^A \}$.  However the combination $c + k/L$ is, modulo integers, and therefore the equivariance condition defining the $k_{\gamma_{\rm e}}$ subspace of the kernel is.

Now we are ready to identify the space of vanilla BPS states.  For a $\Psi^{\gamma_{\rm e}}$ of the form \eqref{Psidecomp} such that $\Psi_{0}^{\gerel}$ satisfies \eqref{ZLgrading3}, we observe that the degrees of freedom in $\psi_{\rm cm}$ correspond to the half-hypermultiplet factor in \eqref{hhfactor}.  One can, for example, decompose the angular momentum generators into center-of-mass and relative pieces: $\hat{J}^r = \hat{J}^{r}_{\rm cm} \otimes \mathbbm{1} + \mathbbm{1} \otimes \hat{J}^{r}_{0}$.  Only the $\frac{i}{8} (\mathbb{J}^r)_{mn} \gamma^{mn}$ term of \eqref{Jhat} contributes to $\hat{J}^{r}_{\rm cm}$ when acting on \eqref{Psidecomp}, and using this action it is easy to see that $\psi_{\rm cm}$ decomposes into two scalars and a spin $1/2$ representation.  This is the spin content of the half-hypermultiplet.  One can similarly check that the $\mathfrak{su}(2)_R$ content comes out correctly.  Thus the space of $\Psi_{0}^{\gerel}$'s identified in \eqref{ZLgrading3} is in one-to-one correspondence with $\HH_{0}^{\rm BPS}$ in \eqref{hhfactor}, where we use the map \eqref{mathxymapvan} to relate the parameters:
\begin{equation}\label{mainres2}
(\HH_0)_{u,\gamma}^{\rm BPS} \cong \ker_{\Lsq}^{{\rm e}(\gamma)_0} \left( \slashed{\DD}_{\MM_0({\rm m}(\gamma);\sx(u,\gamma))}^{ \rG_0(\YY_{\infty}(u,\gamma))} \right)^{(k_{{\rm e}(\gamma)})}~.
\end{equation}

This is the vanilla analog of \eqref{mainres} and our second main result of the paper.  The basic identification of the space of vanilla BPS states with the kernel of the $\rG(\YY_{\infty})$-twisted Dirac operator is not new; it was first obtained in \cite{Gauntlett:1999vc}.  However we have extended this identification in two directions.  First, we have stated and motivated a conjectural formula, \eqref{mathxymapvan}, that accounts for the full set of perturbative and nonperturbative quantum corrections to the Higgs vevs $\{X_\infty,\YY_\infty\}$ that determine the Dirac operator and its kernel in the weak coupling regime of the Coulomb branch.  

Second, we have clarified the equivariance condition that must be imposed on wavefunctions in the kernel of the Dirac operator on the strongly centered moduli space.  The electric charge eigenspaces of $\ker(\slashed{\DD}_{0}^{\rG_0})$ are $\mathbb{Z}_L$-graded, where $L$ is the greatest common divisor of the components of dual of the magnetic charge along the simple roots.  A wavefunction $\Psi_{0}^{\gerel} \in \ker(\slashed{\DD}_{0}^{\rG_0})$ is in the subspace corresponding to $k \in \mathbb{Z}_L$ if is satisfies the equivariance condition in \eqref{ZLgrading2}.  The space of BPS states corresponds to the $k=k_{\gamma_{\rm e}}$ subspace, \eqref{kgedef}.  Of course if $L=1$ then then $(\HH_{0}^{\rm BPS})_{u,\gamma}$ corresponds to the full (electric charge eigenspace of the) kernel.

The same comments as in the framed case concerning weak coupling duality transformations and independence of the kernel with respect to components $\langle \alpha_{I_M}, X_\infty\rangle, \langle \alpha_{I_M}, \YY_{\infty}\rangle$ of the Higgs vevs apply.  (Recall that these components are present when the magnetic charge is non-generic).  Further observations relevant to the vanilla case are as follows.
\begin{itemize}
\item $\rG_0$ does not depend on the component of $\YY_\infty$ parallel to $X_\infty$.  In terms of the change of basis $\{\alpha_{I_A} \} \mapsto \{i_\ast(\beta_A),\gm^\ast\}$, for the dual of $i_\ast(\mathfrak{t}^{\rm ef})$, $\rG_0$ depends only on the components $\langle i_\ast(\beta_A), \YY_\infty \rangle$ and not on $(\gm, \YY_\infty)$.  This is nicely consistent with the constraint \eqref{Qvconstraint}.  If the kernel of $\slashed{D}_{0}^{\rG_0}$ had depended on both $(\gm, \YY_\infty)$ and $\langle \gamma_{\rm e}, X_\infty \rangle$, then demanding that \eqref{Qvconstraint} hold would have imposed a rather artificial looking restriction on the data $\{\sx,\sy\}$ from a mathematical point of view.  The constraint can be viewed as fixing the component $(\gm, \YY_\infty)$ of $\YY_\infty$ that is not used in the construction of the Dirac operator.

In fact, from the mathematical point of view, it is far more natural to think of of the ``data'' as being $\{\gm, X_\infty, \YY_0 \}$ where, by definition,
\begin{equation}\label{mathY0def}
\YY_0 \in \mathfrak{t}_{\gm}^{\perp} := \{ H \in \mathfrak{t}~|~ (\gm,H) = 0 \}~,
\end{equation}
since $\rG_0$ only depends on the part of $\YY_{\infty}$ that is Killing-orthogonal to $\gm$.  Then we view \eqref{Qvconstraint} as a prescription for reconstructing an element $\YY_{\infty} \in \mathfrak{t}$, for each electric charge eigenvalue $\gamma_{\rm e}$, from the given data:
\begin{equation}\label{YinftyfromY0}
\YY_{\infty} = - \frac{\langle \gamma_{\rm e}, X_\infty \rangle}{(\gm,X_\infty)} X_\infty + \YY_0~.
\end{equation}
This is the point of view we will take in the next section. 

\item Related to this, recall that the preimage of $\{\sx,\sy\}$ under \eqref{mathxymapvan} in $\widehat{\BB}_{\rm wc}$ is a real one-dimensional curve.  In the asymptotic regime these curves are approximately circles parameterized by the overall phase of the $a^I(u)$; however these circles are deformed by $|\Lambda|/|\langle \alpha, a\rangle$ effects.  We claim there is a second family of real one-dimensional curves, naturally conjugate to these, along which $\HH_{u,\gamma}^{\rm BPS}$ is invariant.  These are the curves in $\widehat{\BB}_{\rm wc}$ that correspond to equal overall rescalings of $\sx,\sy$.  The constraint is invariant under such rescalings.  Furthermore by dimensional analysis the metric on $\MM_0$ and the $\rG(\YY_\infty)$ term must scale in the same way such that $\ker(\slashed{\DD}_{0}^{\rG_0})$ remains invariant.  Finally, $q_{\rm cm}$ in the equivariance condition defining the $k= k_{\gamma_{\rm e}}$ subspace of the kernel remains invariant.  Hence a nontrivial prediction of the identification \eqref{mainres2} is that the BPS spectrum must be invariant along these curves.  Asymptotically, such a rescaling of $\sx,\sy$ can be achieved by a uniform rescling $a^I(u)$.  However, like the circles, this is corrected by $|\Lambda|/|\langle \alpha, a\rangle|$ effects.
\item The previous item implies that $\widehat{\BB}_{\rm wc}$ is foliated by \emph{complex} dimension $r-d + 1$ curves along which the BPS spectrum is invariant.  The leaves of this foliation are labeled by the  values of $\langle \alpha_{I_A}, \sx \rangle$, $\langle i_\ast(\beta_A), \sy \rangle$, and the overall scale of $\sx,\sy$.  They are paramterized by holding these values fixed and varying $\langle \alpha_{I_M},\sx\rangle$, $\langle \alpha_{I_M}, \sy\rangle$, the overall scale, and the real parameter of the preimage of $\sx,\sy$ under \eqref{mathxymapvan}.  This has dramatic consequences in the case $d=1$, corresponding to a magnetic charge that has only one nonzero component.  For such cases we learn that $\HH_{u,\gamma}^{\rm BPS}$ is completely invariant over all of $\widehat{\BB}_{\rm wc}$.  In section \ref{ssec:vanlocus} we will use this observation, together with a vanishing theorem, to determine the complete spectrum of such states on $\widehat{\BB}_{\rm wc}$.  It consists of towers of Julia--Zee dyons, one for each simple root, and that's it.
\item Another consequence of the identification \eqref{mainres2} is that the existence of a $\Psi_{0}^{(\gamma_{\rm e})} \in \ker_{\Lsq}^{\gerel}(\slashed{\DD}_{0}^{\rG_0})^{(k_{\gamma_{\rm e}})}$ automatically guarantees the existence of a family of BPS states with the same magnetic charge and electric charges that differ from $\gamma_{\rm e}$ by \emph{integer} multiples of $\gm^\ast$.  Such shifts of $\gamma_{\rm e}$ lead to shifts of $q_{\rm cm}$ by integers and this does not affect the equivariance condition imposed by $\widetilde{\phi}_0$.\footnote{They also lead, via the map \eqref{mathxymapvan}, to a shift of $\YY_\infty$ by and integer amount along $X_\infty$ such that the constraint \eqref{Qvconstraint} is maintained, but again the Dirac operator does not depend on this component of $\YY_\infty$.}  Thus for each BPS state with charge ${\rm m}(\gamma) \oplus {\rm e}(\gamma) = \gm \oplus \gamma_{\rm e}$ we have a full \emph{Julia--Zee tower} of BPS states with charges $\{ \gm \oplus (\gamma_{\rm e} + n \gm^\ast), n\in \mathbb{Z} \}$.  This is completely consistent with the semiclassical monodromy---\ie\ Witten effect---discussed around \eqref{chargefibration}.

Observe however that we could shift $\gamma_{\rm e}$ by $L^{-1} \gm^\ast$ and still remain in the root lattice, since $L$ is the greatest common divisor of the components of $\gm^\ast$ along the simple roots.  This corresponds to shifting $q_{\rm cm}$ by $1/L$, which from \eqref{ZLgrading2} is equivalent to sending $k \to  k+1$.  Since the subspaces $\ker_{\Lsq}^{\gerel}(\slashed{\DD}_{0}^{\rG_0})^{(k)}$ need not be isomorphic for different values of $k$, the existence of a BPS state with electric charge $\gamma_{\rm e}$ does \emph{not} necessarily imply the existence of BPS states with electric charge $\gamma_{\rm e} + \frac{k}{L} \gamma_{\rm m}^\ast$ for $k = 1,\ldots,L-1$.

Let us introduce some notation to emphasize this point, and which will also be useful below.  We define equivalence classes of electric charges modulo integer shifts along the dual of the magnetic charge:
\begin{equation}\label{geZequiv}
[\gamma_{\rm e}]_{\rm JZ} = [\gamma_{\rm e}']_{\rm JZ} \quad \iff \quad \gamma_{\rm e}, \gamma_{\rm e}' \in \Lambda_{\rm rt} ~ \& ~ \gamma_{\rm e} - \gamma_{\rm e}' \in \Span_{\mathbb{Z}} \{ \gm^\ast \}~.
\end{equation}
The equivalence class $[\gamma_{\rm e}]_{\rm JZ}$ is the Julia--Zee tower for electric charge $\gamma_{\rm e}$, and the spaces $(\HH_{0}^{\rm BPS})_{u,\gamma}$ are isomorphic for each member of a given tower, for $u \in \widehat{\BB}_{\rm wc}$.  In order to specify a given $[\gamma_{\rm e}]_{\rm JZ}$ we must specify a $k_{\gamma_{\rm e}} \in \mathbb{Z}_L$ and a relative charge $\gerel$.  This is the data we see entering the right-hand side of \eqref{mainres2}.  Thus we have a one-to-one correspondence
\begin{equation}
[\gamma_{\rm e}]_{\rm JZ} \mapsto \{ k_{\gamma_{\rm e}}, \gerel \}~.
\end{equation}
The inverse map follows from \eqref{gebetaexp}, \eqref{qcmpieces}, and \eqref{kgedef}:
\begin{equation}\label{gerecon}
\gamma_{\rm e} = \left( n + \frac{k_{\gamma_{\rm e}}}{L} + c(\gm,X_\infty,\gerel) \right) \gm^\ast + \gerel~.
\end{equation}
where $n$ is an integer that drops out of $[\gamma_{\rm e}]_{\rm JZ}$.
\item Note the curious feature that the Dirac operator itself, at fixed $u \in \widehat{B}_{\rm wc}$, depends on the electromagnetic charge of interest through its dependence on $\sx,\sy$; in particular it depends on the equivalence class $[\gamma_{\rm e}]_{\rm JZ}$ of the electric charge.  Hence the full kernel of $\slashed{\DD}_{0}^{\rG_0}$ on a \emph{fixed} space $\MM_0$ is not physically relevant for the purposes of determining the BPS spectrum.  The dependence of $\MM_0$ and $\rG_0$ on the charge comes through the phase $\zeta_{\rm van}$.  In the classical approximation one ignores the electric charge dependence of this quantity, which is $O(g_{0}^2)$ suppressed relative to the magnetic charge contribution.  Only in that limit should one consider the full set of eigenspaces of the electric charge operator on a fixed $\MM_0$.  This might explain why this feature of \eqref{mainres2} seems not to have been noticed before.
\item Finally, throughout sections \ref{sec:clfBPS} and \ref{sec:sc} we have always restricted the vanilla discussion to BPS field configurations and BPS states that preserve the $\RR$-type supersymmetries.  This was convenient because some aspects of the analysis are then formally equivalent to the case of framed BPS field configurations and BPS states with $\zeta \to \zeta_{\rm van}$.  But there are also of course the vanilla BPS field configurations and BPS states that preserve the $\TT$-type supersymmetries.  According to the analysis around \eqref{RTalg} we can exchange the roles of $\RR$ and $\TT$ merely by sending $\zeta_{\rm van} \to - \zeta_{\rm van}$.  Via the change of field variables \eqref{realvars} this has the effect of sending $X \to - X$, $Y \to - Y$, $\rho^A \to \lambda^A$ and $\lambda^A \to - \rho^A$, while leaving the gauge field $A_\mu$ invariant.  Indeed we can see that the $\RR$-fixed point equations, \eqref{epsvar}-\eqref{fermvar2}, are mapped precisely to the $\TT$-fixed point equations, \eqref{etavar1}-\eqref{etavarlast}, under this transformation.  In this way $\RR$-type BPS solutions are mapped to $\TT$-type BPS solutions, as well as the collective coordinate expansions around them.  

Hence we find that in order to describe $\TT$-type BPS states we need only adjust  the map \eqref{mathxymapvan}:
\begin{equation}\label{Tvanxymap}
\begin{array}{r l} X_\infty(u,\gamma) =& - \Im(\zeta^{-1}(u,\gamma) a(u))~, \\ \YY_\infty(u,\gamma) =& - \Im(\zeta^{-1}(u,\gamma)a_{\mathrm{D}}(u))~, \end{array} \qquad (\textrm{for $\TT$-type susy}).
\end{equation}
Note we should still work in the same duality frame on $\widehat{\BB}_{\rm wc}$---defined by taking $\Im(\zeta^{-1}(u,\gamma)a(u))$ in the closure of the fundamental Weyl chamber---in order to meaningfully compare the semiclassical spectra of $\RR$-type and $\TT$-type BPS states.  Hence, for $\TT$-type supersymmetries, $X_\infty$ is restricted to the closure of the anti-fundamental Weyl chamber.  This means, for example, that the possible magnetic charges of $\TT$-type BPS states for $u \in \widehat{\BB}_{\rm wc}$ will be \emph{negative} combinations of simple co-roots: $n_{\rm m}^I \leq 0, \forall I$.  Indeed the Dirac operator and its decomposition into electric charge eigenspaces is completely invariant if we send the data $\gm \to -\gm$ and $\gamma_{\rm e} \to -\gamma_{\rm e}$ at the same time as sending $X_\infty \to - X_\infty$ and $\YY_\infty \to -\YY_\infty$.  Hence, for every $\RR$-type BPS state of charge $\gamma$ at $u \in \widehat{\BB}_{\rm wc}$, we have a $\TT$-type BPS state of charge $-\gamma$ \emph{at the same} $u \in \widehat{\BB}_{\rm wc}$.  From the point of view of the low energy effective Abelian gauge theory these are simply the charge conjugate states.
\end{itemize}

%%%%%%%%%%%%%%%%%%%%%%
%%%%%%%%%%%%%%%%%%%%%%
\section{Applications}\label{Section:Applications}
%%%%%%%%%%%%%%%%%%%%%%
%%%%%%%%%%%%%%%%%%%%%%

In this section we present some applications of our semiclassical identifications for the spaces of (framed) BPS states.

%%%%%%%%%%%%%%%%%%%%%%
\subsection{The complete framed BPS spectrum on a special locus}\label{ssec:vanlocus}
%%%%%%%%%%%%%%%%%%%%%%

Here we will make use of a simple Lichnerowicz formula to determine the complete framed BPS spectrum on a special locus in the weak coupling regime.  The results will be somewhat weaker in the vanilla case due to the charge dependence of the vanilla math-physics parameter map, \eqref{mathxymapvan}.  The application of the formula is valid when $\fmMM$ or $\mMM_0$ has positive dimension.  The case of zero dimension requires special treatment and it is useful to consider it first.

Let us in fact begin with the vanilla case.  We are considering pure $\NN = 2$ SYM with gauge algebra $\mathfrak{g}$ and dynamical scale $\Lambda$.  For a given electromagnetic charge $\gamma \in \Gamma$, we work in the distinguished weak coupling duality frame described at the beginning of \ref{ssec:vBPSspace}.  We employ the math-physics map of parameters, \eqref{mathxymapvan}, and restrict to those $u$ in the weak coupling regime $\widehat{\BB}_{\rm wc}$ such that $X_\infty(u,\gamma)$ is in the fundamental Weyl chamber.  By allowing $\YY_\infty = \YY_\infty(u,\gamma)$ to vary over all of $\mathfrak{t}$ we cover $\widehat{\BB}_{\rm wc}$ except for the real co-dimension one loci where $\langle \alpha_I, X_\infty \rangle = 0$ for some simple root $\alpha_I$.  In the distinguished weak coupling duality frame we have the magnetic and electric trivializations of the charge: ${\rm m}(\gamma) = \gamma_{\rm m} \in \Lambda_{\rm cr}$ and ${\rm e}(\gamma) = \gamma_{\rm e} \in \Lambda_{\rm rt}$.

The dimension of the strongly centered moduli space $\MM_0$ is given by $4(|\gm|  -1)$.  It is connected and therefore a point when the height of the magnetic charge is one.  This corresponds to $\gm = H_{I_1}$ for some fixed $I_1 \equiv I_{A=1}$.  The full moduli space is then $\MM = \mathbb{R}_{\rm cm}^3 \times S_{X_\infty}^1 \times \{ {\rm pt} \}$ where the metric in coordinates $(\vec{x}_{\rm cm},\chi)$ is given by \eqref{productmetric}, and $\chi \sim \chi + 2\pi \sp^1$, where $\sp^1 = 2/\alpha_{I_1}^2 \in \{1,2,3\}$.  The unique normalized wavefunction on $\MM_0$ is $\Psi_0 = 1$.  The electric charge operator acts via Lie derivative along $\pd_\chi$.  Let the possible electric charge eigenvalues be denoted $\gamma_{\rm e} = n_{\rm e} \alpha_{I_1}$, $n_{\rm e} \in \mathbb{Z}$.  Then the full wavefunction, $\Psi^{(n_{\rm e})}$ is
\begin{equation}\label{dyonwf}
\Psi^{(n_{\rm e})} =  \psi_{\rm cm} e^{-i n_{\rm e} \chi/\sp^1}~,
\end{equation}
where $\psi_{\rm cm}$ is a constant four-component spinor on $\mathbb{R}_{\rm cm}^3 \times S_{X_\infty}^1$.  The degrees of freedom of $\psi_{\rm cm}$ comprise the half-hypermultiplet and thus we have a half-hypermultiplet for every electric charge $n_{\rm e} \in \mathbb{Z}$.  Note the magnetic charge $\gm = H_{I_1}$ corresponds to a special case of $d \equiv \rnk{\mathfrak{g}^{\rm ef}} = 1$.  We argued on general grounds that the spectrum of BPS states associated with such charges is invariant throughout $\widehat{\BB}_{\rm wc}$.  This is completely obvious here as the wavefunction \eqref{dyonwf} is manifestly independent of the Coulomb branch parameters $u$.

We have focused on vanilla BPS states preserving the $\RR$-type supercharges.  However we can also consider BPS states that preserve the $\TT$-type supercharges, following the procedure outlined around \eqref{Tvanxymap}.  This leads to the tower of charge conjugate states $(\gm,\gamma_{\rm e}) = (-H_{I_1}, n_{\rm e} \alpha_{I_1})$, for $n_{\rm e} \in \mathbb{Z}$, which also exist throughout $\widehat{\BB}_{\rm wc}$.  Together, we will refer to this collection of vanilla BPS states as a \emph{dyon cohort}:
\begin{equation}\label{dyoncohort}
{\rm Coh}(\alpha_I) := \{ (\pm H_I,n \alpha_I) ~|~ n \in \mathbb{Z} \}~.
\end{equation}

For completeness we should also mention the massive vanilla BPS states in the perturbative vacuum sector, $\gm = 0$.  They consist of the $W$-bosons with electric charges $\gamma_{\rm e} = \alpha$ for each positive root $\alpha \in \Delta^+$ and their charge conjugates $\gamma_{\rm e} = -\alpha$.  The former preserve the $\RR$-type supersymmetries while the latter preserve the $\TT$-type supersymmetries.  Note we have $W$ bosons with charges $\gamma_{\rm e} = \pm \alpha$ for each \emph{positive root} $\alpha$ while there is a dyon cohort for each \emph{simple} root $\alpha_{I}$.  The $W$-bosons, like the dyon cohorts, exist throughout $\widehat{\BB}_{\rm wc}$.  Together they comprise all vanilla BPS states that carry a magnetic charge with height $|\gm| \leq 1$.  At height two and greater the strongly centered moduli space $\MM_0$ is nontrivial and the existence of BPS states at weak coupling is tied to the existence of $\Lsq$ sections in the kernel of the Dirac operator \eqref{M0diracop}.

Before turning to that, let us consider the framed case with $\dim{\fmMM} = 0$.  In addition to the gauge algebra $\mathfrak{g}$ and dynamical scale $\Lambda$, we specify a set of supersymmetric \tHooft defect data $L_\zeta( \{\vec{x}_n,P_n\})$.  The math-physics map of parameters is now given by \eqref{mathxy} and we restrict to those $u \in \widehat{\BB}_{\rm wc}$ such that $X_\infty(u,\zeta)$ is in the fundamental Weyl chamber.  The magnetic and electric trivializations of the charge are identified with $\gm,\gamma_{\rm e}$ as before, but now the magnetic charge takes values in the torsor, $\Lambda_{\rm cr} + \sum_n P_n$, of the co-root lattice.  

It is then the height of the relative magnetic charge, $\tilde{\gamma}_{\rm m} = \gm - \sum_n P_{n}^-$, that controls the dimension of $\fmMM$.  The case $\dim{\fmMM} = 0$ corresponds to $\gm = \sum_n P_{n}^-$.    As there are no collective coordinate degrees of freedom, the ``collective coordinate Hamiltonian,'' \eqref{QHam}, is simply a constant, $\hat{H} = M_{\gm}^{\textrm{1-lp}} + O(g_{0}^2)$, acting on a one-dimensional Hilbert space with a unique normalizable state.  This state is supersymmetric; the supercharge operators are represented as zero and annihilate it trivially.  Since there are no collective coordinate momenta, our discussion in section \ref{ssec:validity} implies that the corrections to the Hamiltonian are entirely from the weak-potential approximation and quantum corrections, and should serve to reconstruct the quantum-exact central charge, $H = - \Re(\zeta^{-1} Z_{\gm})$, as dictated by the framed BPS bound.  

Remember in the semiclassical quantization of section \ref{sec:sc} we are mostly discussing the truncated Hilbert space of a given soliton (\ie\ magnetic charge) sector, where we do not consider the continuum of perturbative particle excitations above the collective coordinate states.  In this somewhat degenerate case of $\dim{\fmMM} = 0$, it is especially important to keep this in mind.  The unique normalizable state of the ``collective coordinate theory'' we are referring to in this case is simply the unique vacuum state in this soliton sector, and we can of course consider the usual Fock space of perturbative particle excitations above it.  These, however, will not be supersymmetric, while the vacuum itself is.  This vacuum is in fact the analog of the perturbative vacuum when defects are present.  Indeed, since $\fmMM$ is only nonempty when the relative magnetic charge sits in the closure of the fundamental Weyl chamber, the ground state energy in any other soliton sector will be higher.

It is interesting to consider the question of framed wall crossing for these ``vacuum'' states, taking the halo particle to be a simple $W$-boson.  This leads quickly to connections with the notion of \emph{tropical labels} for the IR charges of line defects \cite{Gaiotto:2010be,Cordova:2013bza} and a semiclassical analog of the process of \emph{monopole extraction} described in \cite{MRVdimP2}.  It will be easier to investigate these issues once we have some other examples of framed BPS states under our belts, so we postpone this discussion until section \ref{ssec:tropical}.

In \cite{MRVdimP1,MRVdimP2} we argued that classical singular monopole field configurations exist if and only if the relative magnetic charge has the form $\tilde{\gamma}_{\rm m} = \sum_I \tilde{n}_{\rm m}^I H_I$, where the $H_I$ are the simple co-roots determined by $\sx$ and the $\tilde{n}_{\rm m}^I$ are all non-negative integers, in which case $\dim{\fmMM} = 4\sum_I \tilde{n}_{\rm m}^I$.  The conditions $\tilde{n}_{\rm m}^I \geq 0$ define a cone in the magnetic IR charge lattice $\Gamma_{L,u}^{\rm m} \cong \Lambda_{\rm cr} + (\sum_n P_n)$.  See \eg\ figure 1 of \cite{MRVdimP1}.  A consequence is that framed BPS states with magnetic charges outside of this cone do not exist, at least for $u \in \widehat{\BB}_{\rm wc}$.  The case where all $\tilde{n}_{\rm m}^I = 0$ is the tip of the cone, which we just discussed.  A single framed BPS state carrying this magnetic charge and zero electric charge exists for all $u$ in the weak coupling regime.  For magnetic charges with at least one $\tilde{n}_{\rm m}^I > 0$, one must analyze the $\Lsq$ kernel of the Dirac operator \eqref{fMdiracop}.

Determining the kernel of \eqref{fMdiracop} or \eqref{M0diracop} and its decomposition into electric charge eigenspaces for generic $u \in \widehat{\BB}_{\rm wc}$ is a difficult problem since it requires knowledge of the hyperk\"ahler metric on $\fmMM$ or $\mMM_0$.  However it might be possible to make progress given knowledge of the asymptotics of $\fmMM, \mMM_0$, as deduced in the vanilla case by \cite{Gibbons:1995yw,MR1361789,Lee:1996kz,MR1666843}.  Indeed there is an infinite family of $\mMM_0$'s known exactly \cite{Lee:1996kz}, and a complete analysis of the kernel of the Dirac operator was carried out for them in \cite{Stern:2000ie}.

However there is a special locus in $\widehat{\BB}_{\rm wc}$ where we can say precisely what the full framed BPS spectrum is:  the locus is defined by ${\rm G}(\sy(u,\zeta)) = 0$, and on it there are no framed BPS states (for \tHooft line defects) other than the pure 't Hooft defect vacuum state.  Since we are interested in ${\rm G}(\sy) = 0$ for generic magnetic charges, we assume $\ker{\rG} = \{0\}$ and so $\sy(u,\zeta) = 0$.  We cannot make such a strong statement for vanilla BPS states.  Rather, for each charge $\gamma \in \Gamma$ we can find a locus in $\widehat{\BB}_{\rm wc}$ where BPS states carrying that charge do not exist.  The locus is given by those $u$ such that ${\rm G}_0(\sy(u,\gamma)) = 0$.\footnote{However in the extreme weak coupling limit these loci coalesce since the electric charge dependence of $\YY_\infty(u,\gamma)$ is $O(g_{0}^2)$ suppressed relative to the leading behavior.}

The argument goes as follows.  When ${\rm G}(\sy)$ or ${\rm G}_0(\sy)$ vanish the operators \eqref{fMdiracop} or \eqref{M0diracop}, respectively, become ordinary Dirac operators.  Now recall the Lichnerowicz formula for the square of the Dirac operator, $\slashed{\DD} = \gamma^m \DD_m = \gamma^m (\pd_m + \frac{1}{4} \omega_{m,\underline{np}} \gamma^{\underline{np}})$, on a spin manifold:
\begin{equation}
(i \slashed{\DD})^2 = - \DD^{\dag m} \DD_m + \frac{1}{4} R~,
\end{equation}
where $R$ is the Ricci scalar and $\DD_{m}^\dag$ is the Hilbert space adjoint of $\DD_m$ acting on square-integrable sections of the Dirac spinor bundle.  

The Ricci scalar vanishes on a hyperk\"ahler manifold; in particular it vanishes for $\fmMM$ and $\mMM_0$.  Now suppose we have a nontrivial $\Psi \in \ker_{\Lsq}{\slashed{\DD}}$.  Then on these spaces we have
\begin{align}\label{vanishing}
\Psi \in \ker_{\Lsq}{\slashed{\DD}} \quad & \Rightarrow \quad \Psi \in \ker_{\Lsq}(\DD^{\dag m} \DD_m) \quad \Rightarrow \quad 0 = \int_{\fmMM} \overline{\Psi} \DD^{\dag m} \DD_m \Psi \cr
& \Rightarrow \quad 0 = \int_{\fmMM} \overline{\DD^m \Psi} \DD_m \Psi = \int_{\fmMM} || \DD_m \Psi ||^2 \cr
& \Rightarrow \quad 0 = \DD_m \Psi~.
\end{align}
In other words $\Psi$ (or $\Psi_0$ in \eqref{Psidecomp}) must be a covariantly constant spinor.  However non-compactness of $\fmMM$ or $\mMM_0$, and in particular the leading asymptotic form of the metric, implies that a covariantly constant spinor cannot be $\Lsq$---a contradiction.  Although the asymptotic form of the metric has not been worked out in generality for singular monopole moduli spaces as it has for the vanilla ones in \cite{Lee:1996kz}, it is intuitively clear that its leading behavior will be the metric on a locally flat Euclidean space just as in the vanilla case: the asymptotic region corresponds to moving the constituent fundamental monopoles far away from each other and far away from the defects.

Note that for any magnetic charge $\gm$ such that $\mathfrak{g}^{\rm ef} =\mathfrak{su}(2)$, we necessarily have $\rG_0(\YY_\infty)= 0$.  Thus, if $|\gm| \geq 2$ such that $\dim{\MM_0} > 0$, the $\Lsq$ kernel of the Dirac operator on $\MM_0$ is trivial.  Hence the only vanilla BPS states for $u \in \widehat{\BB}_{\rm wc}$ with magnetic charge along a single co-root are the dyon cohorts.  In particular, for pure $\mathfrak{su}(2)$ $\NN = 2$ SYM we recover the well known fact that the full weak coupling spectrum consists of the $W$-bosons and the dyon cohorts only.

%%%%%%%%%%%%%%%%%%%%%%
\subsection{The no-exotics theorem and the kernel of the Dirac operator}\label{ssec:exotics}
%%%%%%%%%%%%%%%%%%%%%%

One of the most remarkable properties of the BPS spectrum of $\NN = 2$ theories on the Coulomb branch is that it transforms trivially under the $SU(2)_R$ internal symmetry of the theory.  More precisely, we have
\begin{Description}
\item[No-exotics:] The Hilbert spaces of framed BPS states, $\HH_{L_\zeta,u,\gamma}^{\rm BPS}$, and the internal Hilbert spaces of vanilla BPS states, $(\HH_0)_{u,\gamma}^{\rm BPS}$, are $SU(2)_R$ singlets for all $u \in \BB^\ast$ and all electromagnetic charges.
\end{Description}
See section \ref{sec:wcf} for further discussion.  Here we work out some equally fascinating properties for the family of Dirac operators that follow from no-exotics and the map spelled out in sections \ref{ssec:fBPSspace} and \ref{ssec:vBPSspace}.  

A key result of the collective coordinate analysis is the semiclassical identification of the $\mathfrak{su}(2)_R$ generators $\hat{I}^r$.  In the representation of the Hilbert space via $\Lsq$ sections of the Dirac spinor bundle, we have
\begin{equation}\label{su2RDirac}
\hat{I}^r = \frac{i}{8} (\sw^r)_{\um\un} \gamma^{\um\un}~.
\end{equation}
We first give a geometric characterization of $SU(2)_R$ and state the no-exotics theorem in the language of Dirac spinors on $\fmMM$ or $\MM_0$.  We then translate this statement to the alternative picture in terms of $(0,\ast)$-forms, where it takes a rather striking form.

% % % % % % % % % % % % %
\subsubsection{Via spinors}
% % % % % % % % % % % % % 

Let the quaternionic dimension of our hyperk\"ahler manifold be $N$, so that $N = |\tilde{\gamma}_{\rm m}|$ or $|\gm|-1$ for $\fmMM(L;\gm;\sx)$ or $\mMM_0(\gm;\sx)$ respectively.  A useful characterization of the $SU(2)$ generated by the triplet \eqref{su2RDirac} is that it is a lift to $Spin(4N)$ of an embedding of the \emph{commutant} of the holonomy group $USp(2N,\mathbb{C}) \subset SO(4N)$, as we now explain.  

Recall that the geometric  origin of $\mathfrak{su}(2)_R$ from the moduli space point of view is that it acts naturally on the space fermion zero modes, \eqref{fzmspace}, via \eqref{su2Rfzm}.  This action is then mapped to one on the tangent space via the quaternion-linear isomorphism \eqref{Tmap}, \eqref{quatpres} that relates the spaces of fermionic and bosonic zero modes.  The resulting $SU(2)_R$ action on the tangent bundle is generated by the triplet of endomorphisms
\begin{equation}\label{Irtan}
I^{r}_{(T\fMM)} := \half \bbJ^r~,
\end{equation}
where we remind that the $\bbJ^r$ give a quaternionic structure on $\fmMM$,  the components of which with respect to a coordinate frame can be computed by \eqref{metJcomp}.  (In the vanilla case we take $I^{r}_{(T\MM_0)}$ to be one-half times the restriction of $\mathbb{J}^r$ to ${\rm End}(T\MM_0) \subset {\rm End}(T\MM)$.)  

Since parallel transport preserves the complex structures, the action of $SU(2)_R$ commutes with the action of the holonomy group, $USp(2N)$, on the $4N$ real-dimensional tangent space at any point.  In terms of the frame bundle, we have a reduction of structure group $USp(2N) \to SO(4N)$, and it is useful to describe $SU(2)_R$ via an explicit embedding $SU(2)_R \times USp(2N) \hookrightarrow SO(4N)$ as follows.  Working over a fixed point $[\hat{A}] \in \fmMM$, we identify the tangent space $T_{[\hat{A}]} \fmMM$ with $\mathbb{R}^{4N} \cong \mathbb{H}^N$.  Let 
\begin{equation}
q_{\alpha} = {\bf i} \EE_{1\alpha} + {\bf j} \EE_{2\alpha} + {\bf k} \EE_{3\alpha} + \EE_{4\alpha}
\end{equation}
be an orthonormal frame, where $\alpha =1,\ldots, N$.  We can choose the frame so that the action of $2 I^{1,2,3}_{(T\fMM)}$, \eqref{Irtan}, corresponds to right-multiplication of all $q_\alpha$ by ${\bf i},{\bf j},{\bf k}$ respectively.  The action on the $\EE_{a\alpha}$, $a = 1,\ldots,4$, is represented via the negative of the self-dual 't Hooft symbols:
\begin{equation}
2( I^{r}_{(T\fMM)})_{a\alpha,b\beta} = - (\eta^r)_{ab} \otimes \delta_{\alpha\beta}~.
\end{equation}
Meanwhile the action of $USp(2N,\mathbb{C}) \cong U(N,\mathbb{H})$ is represented by left-multiplication of the column vector $(q_\alpha)$ by $N\times N$ quaternionic matrices $U$ such that $U^{-1} = U^\dag$.  We note that left multiplication of a given $q_\alpha$ by ${\bf i},{\bf j},{\bf k}$ corresponds to the action of the anti-self-dual 't Hooft matrices, $(-\bar{\eta}^1,\bar{\eta}^2,\bar{\eta}^3)_{ab}$, on $\EE_{a\alpha}$.  The fact that left and right multiplication by quaternions commutes is expressed by the fact that the self-dual and anti-self-dual 't Hooft matrices commute.  A Cartan subalgebra for $\mathfrak{su}(2)_R \oplus \mathfrak{usp}(2N)$ consists of $2 I^3$ whose image in $\mathfrak{so}(4N)$ is $2 I^{3}_{(T\fMM)}$, together with $L_\gamma({\bf k})$, $\gamma = 1,\ldots,N$, for the $\mathfrak{usp}(2N)$ factor whose image in $\mathfrak{so}(4N)$ is represented by
\begin{equation}
(L_\gamma({\bf k}))_{a\alpha,b\beta} := (\bar{\eta}^3)_{ab} \otimes \delta_{\alpha\gamma}\delta_{\beta \gamma}~, \qquad \gamma = 1,\ldots, N~.
\end{equation}

Lie algebra homomorphisms are specified by their action on a Cartan subalgebra.  Thus we have just defined a homomorphism $\mathfrak{su}(2)_R \oplus \mathfrak{usp}(2N) \to \mathfrak{so}(4N)$.  To make this explicit, let
\begin{equation}
(T_{1\gamma2\gamma})_{a\alpha,b\beta} = \left( \begin{array}{c c c c} 0 & 1 & 0 & 0\\ -1 & 0 & 0 & 0\\ 0 & 0 & 0 & 0\\ 0 & 0 & 0 & 0\end{array}\right)_{ab} \otimes \delta_{\alpha\gamma} \delta_{\beta\gamma}~, \quad (T_{3\gamma4\gamma})_{a\alpha,b\beta} = \left( \begin{array}{c c c c} 0 & 0 & 0 & 0\\ 0 & 0 & 0 & 0 \\ 0 & 0 & 0 & 1 \\ 0 & 0 & -1 & 0 \end{array}\right)_{ab} \otimes \delta_{\alpha\gamma} \delta_{\beta\gamma}~. \quad
\end{equation}
so that $\{T_{1\alpha2\alpha},T_{3\alpha4\alpha} \}_{\alpha=1}^N$ generates a Cartan subalgebra of $\mathfrak{so}(4N)$.  Then the group homomorphism $\rho : SU(2)_R \times USp(2N) \hookrightarrow SO(4N)$ is defined via the action of its derivative, $\ed\rho$, on the Cartan subalgebra:
\begin{equation}
\ed\rho\left(2 I^{3}\right) = - \sum_{\alpha = 1}^N (T_{1\alpha2\alpha} + T_{3\alpha4\alpha})~, \qquad \ed\rho\left(L_\alpha({\bf k})\right) = T_{1\alpha2\alpha} - T_{3\alpha4\alpha}~.
\end{equation}
Letting $\psi \sim \psi + 2\pi$ parameterize the Cartan torus of $SU(2)_R$ and $\psi_\alpha \sim \psi_\alpha + 2\pi$ the Cartan torus of $USp(2N)$, we have
\begin{equation}\label{rhohomo}
\rho : \left\{ \begingroup \renewcommand{\arraystretch}{1.2} \begin{array}{l} \exp\left(2\psi I^{3}\right) \mapsto \prod_\alpha \exp( -\psi (T_{1\alpha2\alpha} + T_{3\alpha4\alpha}))~, \\  \exp\left(\psi_\alpha L_{\alpha}({\bf k})\right) \mapsto \exp(\psi_\alpha (T_{1\alpha2\alpha} - T_{3\alpha4\alpha}) )~. \end{array} \endgroup \right.
\end{equation}

Now we construct a lift $\tilde{\rho} : SU(2)_R \times USp(2N) \hookrightarrow Spin(4N)$ as follows.  Promote the frame to a Clifford algebra, $\EE_{a\alpha} \mapsto \tilde{\EE}_{a\alpha}$, with $[ \tilde{\EE}_{a\alpha}, \tilde{\EE}_{b\beta} ]_+ = 2\delta_{ab} \delta_{\alpha\beta}$, and define the generators of $Spin(4N)$ via
\begin{equation}
\tilde{T}_{a\alpha,b\beta} := \frac{1}{4}[ \tilde{\EE}_{a\alpha}, \tilde{\EE}_{b\beta}]_- = \half \tilde{\EE}_{a\alpha} \tilde{\EE}_{b\beta}~.
\end{equation}
Note that $\{ \tilde{T}_{1\alpha2\alpha}, \tilde{T}_{3\alpha4\alpha} \}_{\alpha=1}^N$ generates a Cartan subalgebra.  We then define $\tilde{\rho}$ through its action on the Cartan torus, just as in \eqref{rhohomo} but with $T \to \tilde{T}$.  On a generic element of the Cartan torus of $SU(2) \times USp(2N)$ we therefore have
\begin{align}\label{rhothomo}
& \tilde{\rho} \left( \exp(2\psi I^3), \prod_\alpha \exp(\psi_\alpha L_\alpha({\bf k})) \right) = \cr
& \qquad = \exp \left\{ \sum_\alpha \psi_\alpha (\tilde{T}_{1\alpha2\alpha} - \tilde{T}_{3\alpha4\alpha}) - \psi \sum_\alpha (\tilde{T}_{1\alpha2\alpha} + \tilde{T}_{3\alpha4\alpha}) \right\} \cr
& \qquad  = \prod_{\alpha =1}^N \left( \cos(\frac{_{\psi_\alpha - \psi}}{^2}) + \tilde{\EE}_{1\alpha} \tilde{\EE}_{2\alpha} \sin(\frac{_{\psi_\alpha - \psi}}{^2}) \right) \left(  \cos(\frac{_{\psi_\alpha + \psi}}{^2}) - \tilde{\EE}_{3\alpha} \tilde{\EE}_{4\alpha} \sin(\frac{_{\psi_\alpha + \psi}}{^2}) \right)~. \qquad
\end{align}
Nontrivially, this lift is well-defined as one sees by noting that when we set $\psi = 2\pi$ or any one $\psi_\alpha = 2\pi$, with all remaining $\psi$'s to zero, we get the identity.  The remaining central elements of $SU(2)_R \times USp(2N)$ are $(-1,1),(1,-1)$, and $(-1,-1)$, where $-1 \in USp(2N)$ is defined by $\psi_\alpha = \pi, \forall \alpha$.  With \eqref{rhothomo} we find
\begin{equation}\label{centermap}
\tilde{\rho}(-1,-1) = (-1)^N~, \qquad \tilde{\rho}(1,-1) = (-1)^N \omega~, \qquad \tilde{\rho}(-1,1) = \omega~,
\end{equation}
where $\omega := \prod_\alpha \tilde{\EE}_{1\alpha} \tilde{\EE}_{2\alpha} \tilde{\EE}_{3\alpha} \tilde{\EE}_{4\alpha}$ is the volume form on $Spin(4N)$.  The center of $Spin(4N)$ is $\mathbb{Z}_2 \times \mathbb{Z}_2$, consisting of $\{1,-1,\omega,-\omega\}$.  Hence $\tilde{\rho}$ maps onto the center when $N$ is odd and onto the $\mathbb{Z}_2$ subgroup $\{1,\omega\}$ when $N$ is even.

Furthermore, when acting via the Dirac spinor representation, $S_{\rm D} : Spin(4N) \to GL(\mathbb{C}^{2^{2N}})$, the image of $SU(2)_R$ under $\tilde{\rho}$ is indeed generated by the operators $\hat{I}^r$, \eqref{su2RDirac}.  In the Dirac representation the $\tilde{\EE}_{a\alpha}$ are represented by gamma matrices $\gamma_{a\alpha}$.  Then we have\footnote{The factor of $i$ appears because of our Lie algebra conventions.  We take generators of a Lie algebra such as $I^3$ to be represented by anti-Hermitian matrices, while conserved Noether charges in the collective coordinate quantum mechanics are Hermitian operators.}
\begin{align}\label{I3Ihat3}
S_{\rm D}(\ed \tilde{\rho}(I^3)) =&~ -\half \sum_\alpha S_{\rm D}( \tilde{T}_{1\alpha2\alpha} + \tilde{T}_{3\alpha4\alpha}) = -\frac{1}{4} \sum_\alpha (\gamma_{1\alpha} \gamma_{2\alpha} + \gamma_{3\alpha} \gamma_{4\alpha}) = \cr
=&~ - \frac{1}{8} (\eta_3)_{ab} \sum_\alpha \gamma_{a\alpha} \gamma_{b\beta} =  \frac{1}{8} (\sw^3)_{mn} \gamma^{mn} = -i \hat{I}^3~.
\end{align}

Let us consider the pullback, $\tilde{\rho}^\ast(S_{\rm D})$, of the Dirac spinor representation and determine how it decomposes into irreducible representations of $SU(2)_R$.  First note that $\omega$ is represented by the chirality operator $\bar{\gamma}$ in $S_{\rm D}$, and we have the decomposition $S_{\rm D} = S^+ \oplus S^-$ into $\pm 1$ eigenspaces of $\bar{\gamma}$.  Since $\bar{\gamma}$ acts as the identity on $S^+$,  we see from the last of \eqref{centermap} that $S^+$ must be pulled back to an $SU(2)_R \times USp(2N)$ representation on which $-1 \in SU(2)_R$ acts as the identity.  This means that $\tilde{\rho}^\ast(S^+)$ can contain only integer $SU(2)_R$ spin representations in its decomposition.  Similarly, $S^-$ is pulled back to a direct sum of half-integer spin representations.  In follows that the sign factor $(-1)^{2 \hat{I}^3}$ appearing in the definition of the protected spin characters \eqref{fPSC}, \eqref{PSC} is represented by $\bar{\gamma}$.  This will be useful in the next section.

A simple consequence of the no-exotics theorem is that the kernels of the Dirac operators \eqref{fMdiracop}, \eqref{M0diracop} sit in the positive chirality spinor bundle.  In particular their indices give the actual dimensions of the spaces of (framed) BPS states $\HH_{L_\zeta,u,\gamma}^{\rm BPS}$, $(\HH_{0})_{u,\gamma}^{\rm BPS}$.

However with the map \eqref{rhothomo} in hand we can make a much more precise statement by determining how the Dirac spinor representation pulls back as a direct sum of $SU(2)_R$ representations.  Consider the character
\begin{align}
\chi_{\tilde{\rho}^\ast(S_{\rm D})}(\psi;\psi_\alpha) :=&~ \Tr_{S_{\rm D}} \bigg\{ \prod_{\alpha =1}^N \left( \cos(\frac{_{\psi_\alpha - \psi}}{^2}) + \gamma_{1\alpha}\gamma_{2\alpha} \sin(\frac{_{\psi_\alpha - \psi}}{^2}) \right) \times \cr
&~ \qquad \qquad \qquad \qquad  \times \left(  \cos(\frac{_{\psi_\alpha + \psi}}{^2}) - \gamma_{3\alpha}\gamma_{4\alpha} \sin(\frac{_{\psi_\alpha + \psi}}{^2}) \right) \bigg\} \cr
=&~ \prod_{\alpha =1}^N \left( e^{i (\psi_\alpha - \psi)/2} + e^{-i (\psi_\alpha-\psi)/2} \right) \left( e^{i(\psi_\alpha+\psi)/2} + e^{-i(\psi_\alpha+\psi)/2} \right)~. \qquad
\end{align}
Specializing to $\psi_\alpha = 0$, we get the $SU(2)$ character
\begin{equation}
\chi_{\tilde{\rho}^\ast(S_{\rm D})}(\psi; 0) = \left( e^{i\psi/2} + e^{-i\psi/2} \right)^{2N}~.
\end{equation}
We can determine the decomposition of $\tilde{\rho}^\ast(S_{\rm D})$ into $n$-dimensional spin $j = \frac{n-1}{2}$ irreps of $SU(2)$, denoted $V_n$, by using completeness and orthonormality of the corresponding characters 
\begin{equation}\label{ndimcharacter}
\chi_n(\psi) = \frac{\sin{(n\psi)}}{\sin{\psi}}~,
\end{equation}
with respect to the measure $\pi^{-1} \sin^2(\psi) \, \ed \psi$ on $[0,2\pi]$.  The result is
\begin{equation}
\tilde{\rho}^\ast(S_{\rm D}) = \bigoplus_{n=1}^{N+1} V_n \otimes R_n~,
\end{equation}
where $R_n$ an a $USp(2N)$ representation with
\begin{equation}\label{dimRn}
\dim{R_n} = \frac{n}{N+1} \left( \begin{array}{c} 2N+2 \\ N + n + 1 \end{array}\right)~.
\end{equation}
As a simple check one finds $\sum_{n=1}^{N+1} n \dim R_n = 2^{2N}$.

This exercise in group theory has been carried out in the fiber over a given point in $\fmMM$.  The implication for the vector bundle of Dirac spinors is that it decomposes into a direct sum of subbundles that are invariant distributions for the Levi--Civita connection:
\begin{equation}\label{bundledecomp}
\SS_{\rm D} = \bigoplus_{n=1}^{N+1} ~~ \bigoplus_{m= - \frac{n-1}{2}}^{\frac{n-1}{2}} \SS_{n,m}~,
\end{equation}
In other words the covariant derivative maps sections of $\SS_{n,m}$ to sections of $\SS_{n,m}$.  The subbundles have $\rnk \SS_{n,m} = \dim R_n$ and are characterized by $(\hat{I}^r)^2 \Psi = \frac{1}{4} (n^2-1) \Psi$, $\hat{I}^3 \Psi = m \Psi$ for any $\Psi$ a section of $\SS_{n,m}$.  For example, observe that when $n = N+1$, we have $\rnk{\SS_{n,m}} = 1$.  Hence we have $N+1$ invariant line bundles; the constant sections of these are a basis for the space of covariantly constant spinors on the hyperk\"ahler manifold $\fmMM$.  The positive (negative) chirality spinor bundle corresponds to restricting the outer summand to odd (even) values of $n$ only.

The kernels of the Dirac operators \eqref{fMdiracop}, \eqref{M0diracop} are subspaces of the space of $\Lsq$ sections of $\SS_{\rm D}$.  Generically they would decompose into a direct sum of subspaces of $\Lsq$ sections of each $\SS_{n,m}$.  The no-exotics theorem implies that the kernel lies entirely within the space of $\Lsq$ sections of $\SS_{1,0}$ (\ie\ $\hat{I}^r \Psi = 0$):
\begin{equation}
\ker_{\Lsq} \left( \slashed{\DD}_{\fmMM(L;\gm;\sx)}^{\rG(\sy)} \right) \subset \Lsq(\fmMM, \SS_{1,0})~,
\end{equation}
and similarly for $\mMM_0$.

% % % % % % % % % % % % % %
\subsubsection{Via $(0,\ast)$-forms}\label{sssec:holforms}
% % % % % % % % % % % % % %

We now give an alternative characterization of the above statements via the description of BPS states in terms of $(0,\ast)$-forms.  We choose a distinguished complex structure, say $\bbJ^{3}$, and we introduce an adapted complex coordinate system on $\fmMM$.  Let
\begin{equation}\label{complexcoords}
\begingroup \renewcommand{\arraystretch}{1.2} \begin{array}{l} Z^{\bfn} = z^{\bfn} + i z^{2N + \bfn} \\ \Zbar^{\bfnbar}, = z^{\bfnbar} - i z^{2N+\bfnbar} \end{array} \endgroup~, \qquad \begingroup \renewcommand{\arraystretch}{1.2} \begin{array}{l} \XX^{\bfn} = \frac{1}{\sqrt{2}} (\chi^{\bfn} + i \chi^{2N+\bfn}) \\ \XXbar^{\bfnbar} = \frac{1}{\sqrt{2}}(\chi^{\bfnbar} - i \chi^{2N + \bfnbar}) \end{array} \endgroup~, \qquad \bfn = 1,\ldots,2N~,
\end{equation}
for the bosons and fermions, so that locally $\bbJ^{3} = i \pd_{\bfn} \otimes \ed Z^{\bfn} - i \pdbar_{\bfnbar} \otimes \ed \Zbar^{\bfnbar}$.  We remind the reader that the coordinates $z^n$, $n = 1,\ldots,4N$, are \emph{real}.  The space $(\fmMM,g,\bbJ^{3})$ is a K\"ahler manifold.  The K\"ahler form, defined by $\sw^3(U,V) = g(U,\bbJ^{3}(V))$ for all vector fields $U,V$, has the local form
\begin{equation}
\sw^3 = \frac{i}{2} g_{\bfm\bfnbar} \ed Z^{\bfm} \wedge \ed \Zbar^{\bfnbar}~,
\end{equation}
in terms of the Hermitian metric.  Further details on our K\"ahler conventions can be found in appendix \ref{app:holforms}.

The remaining two K\"ahler forms, $\sw^{1,2}(U,V) = g(U,\bbJ^{1,2}(V))$, can be combined into a closed holomorphic-symplectic form; we define
\begin{equation}\label{holsymform}
\sw_{\pm} := \sw^1 \pm i \sw^2~.
\end{equation}
It follows from the quaternionic algebra of the $\bbJ^{r}$ that $\sw_+ \in \Lambda^{(0,2)}$ while $\sw_- \in \Lambda^{(2,0)}$, where $\Lambda^{(p,q)}$ denotes the space of smooth $(p,q)$-forms.  Furthermore covariant-constancy of the $\bbJ^{r}$ imply $\ed \sw_{\pm} = 0$.  In particular, $\pdbar \sw_+ = 0$ and $\pd \sw_- = 0$, where $\ed = \pd + \pdbar$.

Let $\{ e^{\ubfn} = e^{\ubfn}_{\phantom{n}\bfn} \ed Z^{\bfn} \}$ be a coframe on the holomorphic cotangent bundle and $\{ \ebar^{\ubfnbar} \}$ its antiholomorphic counterpart, such that $g_{\bfm\bfnbar} = e^{\ubfm}_{\phantom{m}\bfm} \ebar^{\ubfnbar}_{\phantom{n}\bfnbar} \delta^{\ubfm\ubfnbar}$, and denote by $\{ \EE_{\ubfn}, \EEbar_{\ubfnbar} \}$ the inverse frame on the tangent bundle.  We take $\XX^{\ubfn} := e^{\ubfn}_{\phantom{n}\bfn} \XX^{\bfn}$, $\XXbar^{\ubfnbar} := \ebar^{\ubfnbar}_{\phantom{n}\bfnbar} \XXbar^{\bfnbar}$ as the basic fermionic coordinates.  After quantization they obey the standard anticommutation relations of fermionic creation and annihilation operators, with $(\hat{\XX}^{\ubfn})^\dag = \hat{\XXbar}{}^{\ubfnbar}$.  This can be represented as the exterior algebra of $(0,\ast)$-forms.  The Hilbert space is taken as $\HH \cong \Lsq(\fmMM,\Lambda^{(0,\ast)})$.  For
\begin{equation}
\HH \ni \lambda =  \sum_{q=0}^{2N} \frac{1}{q!} \lambda^{(q)}_{\ubfnbar_1 \cdots \ubfnbar_q} \ebar^{\ubfnbar_1} \wedge \cdots \wedge \ebar^{\ubfnbar_q}~,
\end{equation}
we have
\begin{equation}
\hat{\XXbar}{}^{\ubfnbar} \lambda := \ebar^{\ubfnbar} \wedge \lambda~, \qquad \hat{\XX}{}^{\ubfm} \lambda := \contract_{\EE_{\ubfm}} \lambda \equiv \contract^{\ubfm} \lambda~,
\end{equation}
where $\contract_V$ denotes the interior product, or contraction, with respect to the vector field $V$ and we introduce the notation $\contract^{\ubfm}$ for the contraction when $V$ is one of the tangent frame fields.  $\contract^{\ubfm}$ is indeed the adjoint of $\ebar^{\ubfmbar} \wedge$ with respect to the innerproduct
\begin{equation}\label{forminner}
\langle \lambda_1 | \lambda_2\rangle = \int_{\fmMM} \lambda_{1}^\ast \wedge \star \lambda_2~,
\end{equation}
where $\star$ is the standard Poincare dual with respect to the Riemannian volume form.

The relationship between this quantization scheme and the one involving spinors is well known.  We could have represented the fermions in terms of gamma matrices via $\hat{\XX}{}^{\ubfn} = \gamma^{\ubfn,-}$ and $\hat{\XXbar}{}^{\ubfnbar} = \gamma^{\ubfnbar,+}$, where $\gamma^{\ubfn,\pm} =  \half (\gamma^{\ubfn} \pm i \gamma^{2N + \ubfn})$.  Defining a ``ground state'' $\sqrt{\Omega}$ such that $\gamma^{\ubfn,-} \sqrt{\Omega} = 0, \forall \ubfn$, the generic state is 
\begin{equation}\label{spinors2forms}
\Psi = \sum_q \frac{1}{q!} \Psi^{(q)}_{\ubfnbar_1 \cdots \ubfnbar_q} \gamma^{+,\ubfnbar_1} \cdots \gamma^{+,\ubfnbar_q} \sqrt{\Omega}~.
\end{equation}
Locally we may identify $\Psi^{(q)}$ with $\lambda^{(q)}$.  On a general K\"ahler manifold this local isomorphism will only patch together to give a well-defined global isomorphism if one takes $\sqrt{\Omega}$ to transform as a section of a square root of the canonical bundle; see \eg\ \cite{MR0286136,MR0358873}.  (The lowest component of the spinor caries charge $-N$ under the central $U(1) \subset U(2N)$ of the holonomy group of a $2N$ complex-dimensional K\"ahler manifold, while a scalar has charge $0$.)  However on a hyperk\"ahler manifold the canonical bundle can be trivialized.  Furthermore if $\fmMM$ is simply-connected there is a unique square root line bundle, and we can take $\sqrt{\Omega}$ to be a covariantly constant section of it.\footnote{Otherwise we must simply make a choice  of square root, corresponding to a choice of spin structure.  Determining $\pi_1(\fMM)$ will require a better understanding of the singularity structure of $\fMM$, and we will not address that issue here.}

After describing how the basic operators are represented, one can go on to construct the analogs, for $\Lambda^{(0,\ast)}$, of the supercharges, Noether charges, \etc, that were given in section \ref{sec:quantize} for $\SS_{\rm D}$.  Details can be found in appendix \ref{app:holforms}; see also \cite{deVries:2008ic}.  Here we discuss only the ones relevant for the no-exotics theorem, starting with the following linear combination of supercharges \eqref{Qsc}:
\begin{equation}
\QQ^{ {\rm G}(\sy)}_{\fmMM(L;\gm;\sx)} := \frac{i \sqrt{\pi}}{g_0}\left( \hat{Q}_{\rm (sc)}^3 + i \hat{Q}_{\rm (sc)}^4 \right) =  \pdbar - i {\rm G}(\sy)^{(0,1)} \wedge ~,
\end{equation}
where ${\rm G}(\sy)^{(0,1)}$ is the antiholomorphic part of the one-form dual to the the vector field ${\rm G}(\sy)$.  We can view the one-form dual to the vector field ${\rm G}(\sy)$ as a connection on a trivial line bundle over $\fmMM$.  Since ${\rm G}(\sy)$ is triholomorphic, the curvature of this connection is type $(1,1)$ is all complex structures.  Hence this is an example of a \emph{hyperholomorphic bundle} with connection, as defined in \cite{MR1486984}, and $\QQ^{ {\rm G}(\sy)}_{\fmMM(L;\gm;\sx)}$ is the corresponding ${\rm G}(\sy)$-twisted Dolbeault operator.\footnote{This hyperholomorphic line bundle should not be confused with the hyperholomorphic line bundle constructed more recently in \cite{MR2394039}, which has also seen interesting applications in the areas of monopole moduli spaces \cite{MR3116317} and $\NN = 2$ theories \cite{Neitzke:2011za}.  We comment briefly on the connection to \cite{MR3116317} in appendix \ref{app:angmom}, after describing the action on $\lambda$ of the $SO(3)$ generators associated with spatial rotations.}  The notation is meant to remind that this operator is defined on a family of hyperk\"ahler manifolds with triholomorphic vector field, but for simplicity we will sometimes refer to it simply as $\QQ$.  The Dolbeault--Dirac operator, $\QQ + \QQ^\dag$, is the direct analog of \eqref{fMdiracop}.

$\QQ$ defines an elliptic complex when acting on square-integrable $(0,\ast)$-forms:
\begin{equation}
\Lsq(\fmMM,\Lambda^{(0,0)}) \xrightarrow{~~\QQ_0~~} \Lsq(\fmMM,\Lambda^{(0,1)}) \xrightarrow{~~\QQ_1~~}~ \cdots ~\xrightarrow{\QQ_{2N-1}} \Lsq(\fmMM,\Lambda^{(0,2N)}) ~,
\end{equation}
where $\QQ_{q}$ is the restriction of $\QQ$ to square-integrable $(0,q)$ forms.  BPS states are annihilated by both $\QQ$ and $\QQ^\dag$ and thus can be identified with the $\Lsq$-cohomology\footnote{Here we assume $\QQ$ is such that a Hodge theorem holds for the $\Lsq$ cohomology.  This is known when $(\fmMM,g)$ is complete and $\ker{\QQ}$ is finite-dimensional; see \eg\ part 1 of \cite{MR1924513}.} of $\QQ$.  We denote the cohomology spaces
\begin{equation}
H_{\Lsq}^q(\QQ) := \ker{\QQ_q} \bigg/ \Im{\,\QQ_{q-1}}~.
\end{equation}
There are analogs of the operators $\{ i \Lie_{K^r}, i \Lie_{K^A} \}$ that generate the action of the isometry group $SO(3) \times U(1)^d$ of $\fmMM$; see appendix \ref{app:holforms}.  They commute with $\QQ,\QQ^\dag$ and thus the cohomology, $\oplus_q H_{\Lsq}^q(\QQ)$, is a representation space.  We could, for example, consider the subspace corresponding to a fixed electric charge $\gamma_{\rm e}$ and construct its $SO(3)$ character.

Our interest here is rather in how the $SU(2)_R$ generators $\hat{I}^r$ are represented.  Letting $\hat{I}_{\pm} = \hat{I}^1 \pm i \hat{I}^2$, we find that
\begin{equation}\label{Lefschetz}
\hat{I}^3 \lambda^{(q)} = \half (q - N) \lambda^{(q)}~, \qquad \hat{I}_+ \lambda = i \, \sw_{+} \wedge \lambda~, \qquad \hat{I}_- \lambda = -i \, \contract_{\sw_{-}} \lambda~,
\end{equation}
where $\sw_{(+)-}$ is the (anti-)holomorphic-symplectic form \eqref{holsymform}, and $\contract_{\sw_-}$ is understood as the adjoint of $\sw_- \wedge$ with respect to \eqref{forminner}.  The commutation relations \eqref{QIalg} guarantee that this $\mathfrak{su}(2)$ action restricts to $\QQ$-cohomology.  The shift by $-N$ from the naive eigenvalue of $q$ for the action of $2\hat{I}^3$ is related to the twist by the square root of the canonical bundle in going from forms to spinors.  \eqref{Lefschetz} is reminiscent of the Lefschetz $\mathfrak{sl}(2)$ action on the ordinary de Rham cohomology of a K\"ahler manifold.  In fact, \eqref{Lefschetz} is precisely the construction described in \cite{MR1486984,MR1958088}, where it is  shown that a hyperk\"ahler manifold equipped with a hyperholomorphic bundle carries an $\mathfrak{sl}(2)$ Lefschetz action on its (bundle-valued) $\pdbar$-cohomology.  There, the Lefschetz triple is combined with four odd generators to form a supersymmetry algebra on $\Lambda^{(0,\ast)}$; this is isomorphic to the superalgebra generated by our $\hat{Q}^a, \hat{I}^r$.

Thus the $\mathfrak{su}(2)_R$ triple, $\hat{I}^r$, leads to a Lefschetz decomposition of the $\QQ$-cohomology.  The no-exotics theorem states that BPS states should be $\mathfrak{su}(2)_R$ singlets.  Hence from $\hat{I}^3 \lambda = 0$ we learn that they must be in the middle cohomology, $q = N$, and from $\hat{I}_{\pm} \lambda = 0$ we learn that they must be primitive---in the language of spin states $|j,m\rangle$, they cannot be $m = 0$ states in a $j > 0$ representation.  As a very basic consistency check with the spinor computation in the previous subsection, we can compute the rank of the subbundle of $\Lambda^{(0,N)}$ corresponding to the Lefschetz primitive states.  It should agree with $\rnk{\SS_{1,0}} = \dim{R_1}$, where $\dim{R_n}$ was given in \eqref{dimRn}.  Since $N$ is the middle degree, it follows from the basic structure of $\mathfrak{sl}(2)$ representations that $\hat{I}_+$ must map all of $\Lambda^{(0,N-2)}$ onto the subspace of non-primitives in $\Lambda^{(0,N)}$ in a one-to-one fashion.  Therefore denoting by $\Lambda_{\rm prim}^{(0,N)}$ the subbundle of degree $N$ primitive states, we indeed find\footnote{ABR thanks Daniel Robbins for this observation, which served as a helpful clue to unraveling the results described in this subsection.}
\begin{align}
\rnk{\Lambda_{\rm prim}^{(0,N)}} =&~  \rnk{\Lambda^{(0,N)}} - \rnk{\Lambda^{(0,N-2)}} = \left(\begin{array}{c} 2N \\ N \end{array} \right) - \left( \begin{array}{c} 2N \\ N-2 \end{array}\right) = \dim{R_1} ~.
\end{align}

To summarize, the no-exotics theorem for framed BPS states implies that all nontrivial $\Lsq$-cohomology of $\QQ_{\fmMM(L;\gm;\sx)}^{{\rm G}(\sy)}$ is in the middle degree:
\begin{equation}\label{nocoho}
H_{\Lsq}^{q}(\QQ) = \{0\}~,~~ \forall q \neq N~.
\end{equation}
Under the map spelled out in section \ref{ssec:fBPSspace}, we identify the middle cohomology with the space of framed BPS states for fixed magnetic charge:
\begin{equation}
\bigoplus_{{\rm e}(\gamma) \in \Lambda_{\rm rt}} \HH_{L_\zeta,u,\gamma}^{\rm BPS} \cong H_{\Lsq}^N\left(\QQ_{\fmMM(L;{\rm m}(\gamma);\sx(u))}^{{\rm G}(\sy(u))}\right)~,
\end{equation}
for all $u \in \widehat{\BB}_{\rm wc}$.  Analogous remarks apply regarding the no-exotics theorem for vanilla BPS states and the $\Lsq$-cohomology of $\QQ_{\mMM_0(\gm;\sx)}^{{\rm G}_0(\sy)}$.

The statement that all nontrivial cohomology is concentrated in the middle degree is reminiscent of Sen's famous conjecture \cite{Sen:1994yi,Segal:1996eb} concerning the existence of certain $\Lsq$ harmonic forms on the strongly centered vanilla space $\MM_0$.  Sen was considering the semiclassical description of soliton states in four-dimensional $\NN = 4$ SYM theory.  This is given by a supersymmetric quantum mechanics with the same monopole moduli spaces, $\MM(\gm;X_\infty)$, as target but with eight supercharges.  In this quantum mechanics states can be represented by general $\Lsq$ multi-forms and BPS states preserving all eight supercharges are represented by harmonic forms---$\Lsq$ normalizable zero modes of the Laplace--de Rham operator on $\MM_0$, \cite{Witten:1982df,Gauntlett:1993sh}.  Sen showed how $SL(2,\mathbb{Z})$ symmetry of the quantum SYM theory based on Lie algebra $\mathfrak{su}(2)$ implies that $\MM_0(\gm = L H_\alpha)$, with $\dim{\MM_0} = 4(L-1)$, must carry $\Lsq$ harmonic forms transforming with appropriate signs under the isometry generating the $\mathbb{D}/\mathbb{D}_\gi \cong \mathbb{Z}_L$ quotient in \eqref{su2Mfactor}, and furthermore that these forms must be (anti-)self-dual and hence exist \emph{only} in the middle degree cohomology, $q = 2(L-1)$.

While our result---which holds for a broad class of $\NN = 2$ SYM theories, given the validity of the no-exotics theorem---is the same in spirit, we would like to emphasize some key differences.  First, since we have less supersymmetry, the collective coordinate quantum mechanics is different.  The Hamiltonian has a potential energy term that is the norm of a distinguished triholomorphic vector field ${\rm G}(\YY_\infty)$.  Generic states are represented by $\Lsq$ normalizable $(0,\ast)$-forms (or equivalently Dirac spinors), while BPS states correspond to ${\rm G}(\YY_\infty)$-\emph{twisted} Dolbeault cohomology.  BPS states are in the middle antiholomorphic degree but have vanishing holomorphic degree, so their total form degree is one quarter the dimension of $\fMM$ (or $\MM_0$).  The modification of the $\pdbar$ operator by ${\rm G}(\YY_\infty)^{(0,1)}$ is essential for the existence of BPS states.  Indeed it was noted in \cite{Sen:1994yi} that an anti-self-dual form cannot be either purely holomorhpic or antiholomorphic in any complex structure.  If it could then this would have been in direct contradiction to our simple vanishing argument, given in Dirac spinor language, around \eqref{vanishing}.  

Of course there \emph{is} a direct analogy between $1/2$-BPS states in $\NN = 2$ SYM and $1/4$-BPS states in $\NN = 4$ SYM.  The latter context is in fact where the notion of ``moduli spaces with potentials'' was first uncovered \cite{Lee:1998nv}.  However in order to realize this embedding one must consider $\NN = 2$ theories with matter---particularly the $\NN = 2^\ast$ theory containing a single matter hypermultiplet in the adjoint representation.  An extension of the results in this paper to $\NN = 2$ theories with matter hypermultiplets is being considered \cite{BrennanMoore}.

The vanishing statement \eqref{nocoho} also has some striking similarities to results obtained by Verbitsky \cite{MR2371383}, however the assumptions appear different.  In that work the hyperk\"ahler manifold is compact and the line bundle is required to be topologically nontrivial.  Nevertheless we feel that further investigation into the possible connection with this work is warranted.

% % % % % % % % % % % % % % % % %
\subsubsection{$\QQ$ as a conjugated $\pdbar$ operator}
% % % % % % % % % % % % % % % % %

We note in passing that the $\rG(\sy)$-twisted Dolbeault operator can be expressed as a conjugated $\pdbar$-operator, along the lines of \cite{MR792703}.  Take $\mathbb{J}^3$ as the complex structure and $\sw = \sw^3$ as the associated K\"ahler form on $\fMM$ or $\MM_0$ respectively.  Let $V^A  = (i K^A)^{\bfm} \pd_{\bfm}$ denote the holomorphic part of the vector field $i K^A$ for each $A = 1,\ldots, d$ in the framed case, and similarly $V_{0}^A = (i K_{0}^A)^{\bfm} \pd_{\bfm}$ for $A = 1,\ldots, d-1$ in the vanilla case.\footnote{The factor of $i$ is inserted so that, around a fixed point in local holomorphic coordinates at $Z^{\bfm} = 0$, we have $V = \sum_{\bfm} \lambda^{\bfm} Z^{\bfm} \pd_{\bfm}$ for some collection of integers $\lambda^{\bfm}$ as in \cite{MR792703}.}  They are holomorphic Killing vectors generating $U(1)$ isometries of $\fMM$ and $\MM_0$ respectively.  The contractions $\contract_{V^A} \sw$ and $\contract_{V_{0}^A} \sw$ are thus $\pdbar$-closed.  Furthermore, simple-connectedness of $\fMM$ and $\MM_0$ ensures that they are $\pdbar$-exact.  Hence there exist functions $\{x^A\}_{A=1}^d$ and $\{x_{0}^A\}_{A=1}^{d-1}$ on $\fMM$ and $\MM_0$ respectively such that, \eg\
\begin{equation}\label{mommap}
\contract_{V^A} \sw =  \pdbar x^A~, \qquad  \contract_{V_{0}^A} \sw = \pdbar x_{0}^A~.
\end{equation}
Indeed, these functions are merely the third components, \eg\ $x^A = x^{3A}$ or $x_{0}^A = x_{0}^{3A}$, in a triplet $\vec{x}^A$ or $\vec{x}_{0}^A$ that is the hyperk\"ahler moment map for the triholomorphic isometry generated by $K^A$ or $K_{0}^A$ respectively.

Our notation for these functions is inspired by the asymptotic form of the moduli space metric, in which (we expect that) $\vec{x}^A$ can literally be identified with the position of the center of mass of monopoles of type $A$ relative to any fixed line defect.  (The position relative to a different defect will differ from $\vec{x}^A$ by a constant shift, and \eqref{mommap} only defines $\vec{x}^A$ up to constant shifts.)  Asymptotically, one also identifies a coordinate $\xi^A \sim \xi^A + 2\pi$ via $K^A = \pd_{\xi^A}$, where $\xi^A$ is the sum of the phases of all constituent monopoles of type $A$.  In the vanilla case, one can explicitly check these statements using the results of \cite{Lee:1996kz}:  $K_{0}^A$ can be identified with coordinates $\uppsi^A \sim \uppsi^A + 2\pi$ on the asymptotic strongly centered moduli space via $K_{0}^A = \pd_{\uppsi^A}$ which, roughly speaking, measure the relative total phases of the different types of fundamental monopoles.\footnote{In the case where the magnetic charges are $n_{\rm m}^{I_A} = 1, \forall A$, the one can literally take $\uppsi^A = \xi^A - \xi^{A+1}$.  The general formula could be deduced from the results in appendix \ref{app:Dquotient2}.}  Similarly, $\vec{x}_{0}^A$ measures the relative displacements of the centers of mass of the fundamental monopoles of each type.  Although these identifications make use of the asymptotic form of the moduli space, $K^A$ and $K_{0}^A$ are globally well defined, and hence so are their hyperk\"ahler moment maps $\vec{x}^A$, $\vec{x}_{0}^A$.

Using these we define the functions
\begin{equation}\label{morsefns}
\upmu := - \sum_{A=1}^d \langle \alpha_{I_A}, \YY_\infty\rangle x^A : \fMM \to \mathbb{R}~, \qquad \upmu_0 := - \sum_{A=1}^{d-1} \langle i_\ast(\beta_A), \YY_\infty\rangle x_{0}^A : \MM_0 \to \mathbb{R}~.
\end{equation}
It follows from \eqref{morsefns} and \eqref{mommap} that
\begin{equation}
-i {\rG}(\YY_{\infty})^{(0,1)} = \pdbar \upmu~, \qquad -i \rG_{0}(\YY_\infty)^{(0,1)} = \pdbar \upmu_0~,
\end{equation}
on $\fMM$ and $\MM_0$ respectively.  Hence the twisted Dolbeault operators, $\QQ$ and $\QQ_0$ on $\fMM$ and $\MM_0$ respectively, can be written as conjugations of the $\pdbar$ operator:
\begin{equation}\label{conjugatedQ}
\QQ = e^{-\upmu} \pdbar  e^{\upmu}~, \qquad \QQ_0 = e^{-\upmu_0} \pdbar e^{\upmu_0} ~.
\end{equation}
In particular they are members of one-parameter families of operators, $\QQ_{(s)} := e^{-s \upmu} \pdbar e^{s \upmu}$, $s \in [0,\infty)$, and similarly for $\QQ_{0(s)}$.

In \cite{MR792703} this construction was used advantageously to extract detailed information about the Dolbeault cohomology of compact K\"ahler manifolds admitting a holomorphic isometry generated by $V$.  In such a setting, the multiplication operator $e^{s \upmu}$ is invertible and maps the Dolbeault cohomology of $\QQ_{(0)} = \pdbar$ to $\QQ_{(s)}$ in a one-to-one fashion for any $s$.  It was shown that in the limit $s \to \infty$ the computation of the cohomology localizes to one around the fixed points of the isometry generated by the holomorphic vector field $V$, and that index characters of the cohomology can be expressed in terms of the Morse indices of $\upmu$, viewing $\upmu$ as a Morse function.

These techniques do not appear to be directly applicable in the noncompact setting here, where the Morse function $\upmu$ is unbounded and the multiplication operator $e^{s \upmu}$ is not generally invertible when acting on the $\Lsq$ cohomology.  A good illustration is the Taub--NUT manifold, which plays roles as both $\MM_0$ and $\fMM$ in the examples of sections \ref{Section:VanillaEx} and \ref{Section:FramedEx}.  Viewing Taub--NUT as a circle fibration over an $\mathbb{R}^3$ base, the triholomorphic vector field generates translation of the fiber and the corresponding Morse function is (say) the third Euclidean coordinate on the $\mathbb{R}^3$ base.  Nevertheless \eqref{conjugatedQ} with \eqref{morsefns} is a valid representation of the supercharge operator and might prove useful in future investigations.

%%%%%%%%%%%%%%%%%%%%%%
\subsection{Protected spin characters vs.~index characters for Dirac operators}\label{ssec:mathphys}
%%%%%%%%%%%%%%%%%%%%%%

The semiclassical construction of section \ref{sec:sc} relates  physical quantities---spaces of (framed) BPS states---to mathematical ones: kernels of a continuous family of Dirac operators on hyperk\"ahler manifolds, twisted by  triholomorphic vector fields.  Motivic wall crossing formulae for (framed) BPS spin characters imply a set of detailed predictions for the codimension-one loci where certain index characters of these Dirac operators jump, and for how they jump.  Of course it has long been recognized that spaces of BPS states can be identified semiclassically with kernels of such Dirac operators, and furthermore that such kernels can have jumping phenomena.  (See especially \cite{Stern:2000ie}.)  However the idea of turning this around and seeing what recent developments in wall crossing formulae for BPS states have to say about these kernels is, we believe, novel, and particularly worthwhile in light of our new understanding of the relation between physical and mathematical parameters, \eg\ \eqref{mathxy}.  It is especially interesting from a mathematical viewpoint as there are relatively few results available for $\Lsq$ index theorems on noncompact spaces.\footnote{One exception however is \cite{MR1936585}.  This work seems quite relevant and it would be interesting to make contact with it.}

Let us first summarize the relevant data and constructions on the mathematical side, introducing the index characters that will ultimately be equated to the protected spin characters.  In the case with defects the data consists of a simple compact Lie group $G$ with Lie algebra $\mathfrak{g}$ and a five-tuple $\{L,\gm,\gamma_{\rm e},\sx,\sy\}$.   The Lie algebra is equipped with a Killing form $(~,~)$.  $\sx$ is a regular element of $\mathfrak{g}$ which we use to define a Cartan subalgebra, $\sx \in \mathfrak{t} \subset \mathfrak{g}$; $\sy$ sits in the same Cartan subalgebra.  $L = \{\{\vec{x}_n, P_n\} \in \mathbb{R}^3 \times \Lambda_{G} \}$ is a set of 't Hooft line defect data consisting of a finite number of points in $\mathbb{R}^3$ with corresponding charges $P_n \in \Lambda_G \subset \mathfrak{t}$, in the co-character lattice of $G$.  The asymptotic magnetic charge $\gm$ sits in a torsor for the co-root lattice, $\gm \in (\Lambda_{\rm cr} + \sum_n P_n) \subset \mathfrak{t}$, while the asymptotic electric charge sits in the root lattice $\Lambda_{\rm rt} \subset \mathfrak{t}^\ast$.  We use $\sx$ to define a polarization of the root system and hence a basis of simple roots $\{ \alpha_I  \}$ and co-roots $\{ H_I \}$.  

In the case without defects we have a simple compact Lie algebra $\mathfrak{g}$ equipped with Killing form $(~,~)$, and a four-tuple $\{ \gm,[\gamma_{\rm e}]_{\rm JZ},\sx,\YY_0 \}$.  Again $X_\infty$ is assumed regular so that it defines a Cartan subalgebra $\mathfrak{t} \subset \mathfrak{g}$ and a set of simple roots.  The magnetic charge satisfies $\gm \in \Lambda_{\rm cr}$ while $[\gamma_{\rm e}]_{\rm JZ}$ denotes an equivalence class of elements of $\Lambda_{\rm rt}$, where the equivalence relation is \eqref{geZequiv}, identifying electric charges that differ by integer multiples of the dual of the magnetic charge.  The dual is defined with respect to the Killing form such that $(\alpha, \gm^\ast) = \langle \alpha, \gm \rangle$ $\forall \alpha\in \mathfrak{t}^\ast$.  Finally, $\YY_0 \in \mathfrak{t}_{\gm}^{\perp}$ is by definition Killing-orthogonal to the magnetic charge: $(\gm,\YY_0) = 0$; see \eqref{mathY0def}.  

The output of this data will be a set of $SU(2)$ characters $\fC(y),C(y)$, in the cases of one defect and no defects respectively, and a set of numbers $\fC$ in the case of multiple defects.  Here $y$ is a phase parameterizing a Cartan circle of $SU(2)$ in the usual way, so that the character of the $n$-dimensional representation, \eqref{ndimcharacter}, is $\chi_n(y) = (y^n - y^{-n})/(y-y^{-1})$.  It can be analytically continued to a parameter $y \in \mathbb{C}^\ast$.  The characters $\fC(y),C(y)$ are determined according to the following steps (a ``v'' indicates extra details that are specific to the vanilla case):
\begin{enumerate}
\item[1.] Construct the monopole moduli space $\fmMM(L;\gm;\sx)$ defined in \eqref{Mdef}, carrying the hyperk\"ahler metric \eqref{metC}.  Note the definition of the metric makes use of the Killing form $(~,~)$.  When no defects are present, this space is denoted $\mMM(\gm;\sx)$ and the strongly centered factor, $\mMM_0(\gm;\sx)$, is defined through \eqref{Mfactor}.  
\item[1v.]  In the latter case the group $\mathbb{D} \cong \mathbb{Z}$ of deck transformations acts on the universal cover $\widetilde{\MM} = \mathbb{R}_{\rm cm}^3 \times \mathbb{R}_{X_\infty} \times \MM_0$ via isometries.  Let the generator be denoted $\phi : \widetilde{\MM} \to \widetilde{\MM}$.  In the metric, \eqref{productmetric}, where $\chi$ parameterizes the $\mathbb{R}_{X_\infty}$ factor, $\phi$ acts via a simultaneous uniform translation $\chi \to \chi + 2\pi$ and an isometry $\phi_0 : \MM_0 \to \MM_0$. 
\item[2]  Let $P_{n}^- \in [P_n]$ denote the representative of the Weyl orbit of $P_n$ in the closure of the antifundamental Weyl chamber, and define the relative magnetic charge $\tilde{\gamma}_{\rm m} := \gm - \sum_n P_{n}^- \in \Lambda_{\rm cr}$.  $\fmMM$ is nonempty iff $\tilde{\gamma}_{\rm m} = \sum_I \tilde{n}_{\rm m}^I H_I$ with all $\tilde{n}_{\rm m}^I \geq 0$, in which case its real dimension is $4 |\tilde{\gamma}_{\rm m}| \equiv 4 \sum_I \tilde{n}_{\rm m}^{I}$.  $\mMM_0$ is nonempty iff $\gm = \sum_I n_{\rm m}^I H_I$ with all $n_{\rm m}^I \geq 0$ and $|\gm| > 0$, in which case its real dimension is $4(|\gm| -1)$.  Let $A = 1,\ldots,d$ label those components such that $\tilde{n}_{\rm m}^{I_A}$ or $n_{\rm m}^{I_A} > 0$.  Let $I_M$, $M = 1,\ldots \rnk(\mathfrak{g}) - d$, label the remaining simple (co-)roots such that $\{ \alpha_{I_A} \} \cup \{ \alpha_{I_M} \}$ is a partition of the simple roots of $\mathfrak{g}$.  
\item[3.]\label{gefcomment} Construct a semisimple Lie algebra $\mathfrak{g}^{\rm ef}$ as follows.  In the Dynkin diagram of $\mathfrak{g}$, keep those nodes corresponding to the simple roots $\alpha_{I_A}$ and the lines connecting them, while deleting the nodes corresponding to the $\alpha_{I_M}$ and any lines emanating from them.  $\mathfrak{g}^{\rm ef}$ is the Lie algebra of the resulting Dynkin diagram.  Denote by $\alpha_A$ the simple roots of $\mathfrak{g}^{\rm ef}$ and by $i_\ast$ the natural Lie algebra embedding with $i_\ast(H_A) = H_{I_A}$, $i_\ast(E_{\pm \alpha_A}) = E_{\pm \alpha_{I_A}}$.  $\mathfrak{t}^{\rm ef} := \Span\{H_A\}$ is a Cartan subalgebra for $\mathfrak{g}^{\rm ef}$.  For any element $H \in \mathfrak{t}$ we define the decomposition $H = i_\ast(H^{\rm ef}) + H^\perp$, where $\mathfrak{t}^{\rm ef} \ni H^{\rm ef} := \sum_A \langle \alpha_{I_A}, H \rangle h^A$ and $H^{\perp}$ is in the orthogonal complement of $i_\ast(\mathfrak{t}^{\rm ef})$ in $\mathfrak{t}$ with respect to the Killing form.  Here the $h^A$ are fundamental magnetic weights for $\mathfrak{t}^{\rm ef}$ and we note that a basis for the orthogonal complement is $\{ h^{I_M} \}$, so that $H^\perp = \sum_M \langle \alpha_{I_M}, H \rangle h^{I_M}$.  The Riemannian metric and quaternionic structure of $\fMM$ and $\MM$ depend only on $i_\ast(X_{\infty}^{\rm ef})$ and not on $X_{\infty}^{\perp}$.
\item[4] The spaces $\fmMM$ and $\mMM_0$ have $SO(3) \times U(1)^d$ and $SO(3) \times U(1)^{d-1}$ isometry groups respectively, where we are restricting to the case of a single defect for $\fmMM$ until further notice.  The $SO(3)$ factor is generated by a triplet of Killing vectors $\{K^r\}$, induced from the action of spatial rotations on solutions to the Bogomolny equations.  The torus factors are triholomorphic isometries generated by triholomorphic Killing vectors denoted $\{ K^A \}$ and $\{ K^{A}_0\}$, respectively, where in the latter case $A$ runs from $1$ to $d-1$ only.  These can be defined through the action of the ${\rm G}$-map (composed with $i_\ast$), \eqref{Gdef}, on the fundamental magnetic weights $h^{A}$, and on a certain basis $h_{0}^{A}$ of the $(d-1)$-dimensional space orthogonal to $\gm^{\rm ef} \in \mathfrak{t}^{\rm ef}$.  The $h^{A}$ generate curves of $2\pi$ periodicity in the Cartan torus of the adjoint form of the group, $T_{\rm ad}^{\rm ef}$, and the $h^{A}_0$ can be defined to do so as well.  (They generate the subtorus of $T_{\rm ad}^{\rm ef}$ orthogonal to the curve generated by $\gamma_{\rm m}^{\rm ef}$.)  Thus, since ${\rm G}$ is a Lie algebra homomorphism, the $K^A$ and $K^{A}_0$ generate isometries of $2\pi$ periodicity in $\fMM$ and $\MM_0$ respectively.  
\item[4v.] In the vanilla case there is also a triholomorphic isometry of the universal cover, $\phi_\gi : \widetilde{\MM} \to \widetilde{\MM}$, that generates a subgroup $\mathbb{D}_\gi \subset \mathbb{D}$ of the group of deck transformations with $\mathbb{D}/\mathbb{D}_\gi \cong \mathbb{Z}_L$, the cyclic group of order $L$.  These are the deck transformations associated with the action of gauge transformations on $\MM$.  Here, $L \equiv \gcd(\sp^1 n_{\rm m}^{I_1}, \ldots, \sp^d n_{\rm m}^{I_d})$ is the greatest common divisor of the coefficients of the dual of the magnetic charge with respect to the Killing form $(~,~)$ along the basis of simple roots; in particular $\sp^A \in \{1,2,3\}$ is the ratio of the length-squared of a long root to the length-squared of $\alpha_{I_A}$.  $\phi_\gi$ acts via a simultaneous uniform translation $\chi \to \chi + 2\pi L$ and a triholomorphic isometry $\phi_{\gi,0} : \MM_0 \to \MM_0$ such that $\phi_{\gi,0} = \phi_{0}^L$.  The action of $\phi_{\gi,0}$ is generated by a specific and computable linear combination of the $K_{0}^A$: $\phi_{\gi,0} = \exp(2\pi \sum_A c_A K_{0}^A)$, where the $c_A$ depend on $\gm,X_\infty$ and are irrational real numbers for generic $X_\infty$ such that no power of $\phi_{\gi,0}$ gives the identity.  Their form is fixed up to $GL(d-1,\mathbb{Z})$ transformations, the inverse of which acts on $K_{0}^A$ such that $\phi_{\gi,0}$ is uniquely defined.  See appendix \ref{app:Dquotient} for formulae with respect to an explicit choice of basis; the detailed form of the $c_A$ will not be needed in the following.
\item[5.] Let the bundle of Dirac spinors over $\fmMM$ or $\mMM_0$ be denoted $\SS_{\rm D} = \SS^+ + \SS^-$, where $\SS^{\pm}$ are the bundles of positive and negative chirality Weyl spinors.  We take the chirality operator $\bar{\gamma}$ to be the natural one induced by the quaternionic structure; see below \eqref{I3Ihat3}.  Then for the given $\sy$ or $\YY_0$ we have the self-adjoint Dirac operator \eqref{fMdiracop} or \eqref{M0diracop} respectively, mapping $\Lsq$ sections of $\SS^{\pm}$ to $\Lsq$ sections of $\SS^{\mp}$.  These operators commute with $\{ i \Lie_{K^r}, i \Lie_{K^A} \}$ and $\{ i \Lie_{K^r}, i \Lie_{K^{A}_0} \}$, which generate the action of the isometry groups on $\Lsq(\fmMM,\SS^{\pm})$ and $\Lsq(\mMM_0,\SS^{\pm})$ respectively.  In particular the action preserves chirality.  Hence the positive and negative chirality kernels of the Dirac operators, $\ker_{\Lsq}^{\pm} \slashed{\DD}$, furnish representations of the isometry groups.  The kernel of the ${\rm G}$-map is precisely the orthogonal complement of $i_\ast(\mathfrak{t}^{\rm ef})$ in $\mathfrak{t}$.  Therefore, given our previous comment in item \ref{gefcomment}, the Dirac operators are independent of the components $X_{\infty}^\perp$, $\YY_{\infty}^{\perp}$ (or $\YY_0^\perp$) of $X_\infty$, $\YY_\infty$ (or $\YY_0$).
\item[6.] Let $y^{2 \hat{\II}_3}$ denote an element of the Cartan torus of $SU(2)$, the simply-connected cover of the $SO(3)$ factor of the isometry group, where $\hat{\II}_3$ generates a Cartan subalgebra.  Let $t_A$ or $t_{0,A}$ denote a collection of phases such that $\fssy = (y^{2\hat{\II}^3} , t_A)$ or $\ssy = (y^{2\hat{\II}^3}, t_{0,A})$ is a generic element of the Cartan of the isometry group.  We use the same notation for the representative of this element acting on the kernel of $\slashed{\DD}^{\rm G}$ or $\slashed{\DD}_{0}^{\rG_0}$ respectively.  $\{N^A\} \in \mathbb{Z}^d$ and $\{N_{0}^A\} \in \mathbb{Z}^{d-1}$ will denote the weights determining a $U(1)^d$ or $U(1)^{d-1}$ representation.  
\item[6v.] In the case of $\MM_0$, let $\widetilde{\phi}_{\gi,0}$ and $\widetilde{\phi}_0$ be lifts of $\phi_{\gi,0}$ and $\phi_{0}$ to sections $\Gamma(\MM_0,\SS_{\rm D})$ such that $(\widetilde{\phi}_{0})^L = \widetilde{\phi}_{\gi,0}$.  For any spinor $\Psi_0$ carrying $U(1)^{d-1}$ weights $N_{0}^A$, the action of $\widetilde{\phi}_{\gi,0}$ on $\Psi_0$ is given by $\widetilde{\phi}_{\gi,0}(\Psi_0) = e^{-2\pi i L c} \Psi_0$, where $c := - \frac{1}{L} \sum_{A=1}^{d-1} c_A N_{0}^A$ depends only on $\gm$, $X_\infty$, and the $N_{0}^A$, and is an irrational real number for generic Higgs vevs $X_\infty$.  The action of $\widetilde{\phi}_0$ will therefore be of the form $\widetilde{\phi}_0(\Psi_0) = e^{-2\pi i (c + k/L)}$ for some $k \in \{0,\ldots,L-1\}$.  This determines a $\mathbb{Z}_L$ grading on the space of sections for each set of weights $N_{0}^A$ which furthermore descends to a $\mathbb{Z}_L$ grading on the kernel of $\slashed{\DD}_{0}^{\rG_0}$.  These subspaces are denoted $\ker_{\Lsq}(\slashed{\DD}_{0}^{\rG_0})^{(k)}$, such that $\ker_{L^2}(\slashed{\DD}_{0}^{\rG_0}) = \oplus_k \ker_{L^2}(\slashed{\DD}_{0}^{\rG_0})^{(k)}$.
\item[7.] Then we can define the index characters
\begin{align}\label{fDiracbigchar}
{\rm ind}\left( \fssy; \slashed{\DD}_{\fmMM(L;\gm;\sx)}^{{\rm G}(\sy)} \right) :=&~ \tr_{\ker_{\Lsq}( \slashed{\DD}^{\rG})} \left( \bar{\gamma}\, \fssy \right) \cr
\equiv &~ \sum_{\{N^A \in \mathbb{Z}\}} t_{1}^{N^1} \cdots t_{d}^{N^d} \, \fC^{\{N^A\}}_{L,\gm,\sx,\sy}(y)~, \qquad 
\end{align}
and
\begin{align}\label{vDiracbigchar}
{\rm ind}\left( \ssy; \slashed{\DD}_{\mMM_0(\gm;\sx)}^{{\rm G}_0(\YY_0)} \right)^{(k)} :=&~ \tr_{\ker_{\Lsq}(\slashed{\DD}_{0}^{\rG_0})^{(k)}} \left( \bar{\gamma}\, \ssy \right) \cr
\equiv &~ \sum_{\{N_{0}^{A} \in \mathbb{Z}\}} t_{0,1}^{N_{0}^1} \cdots t_{0,d-1}^{N_{0}^{d-1}} \, C^{\{k;N_{0}^A\}}_{\gm,\sx,\YY_0}(y)~. \qquad
\end{align} 
In the second lines we summed over all possible $U(1)$ weights associated with the triholomorphic isometries, thereby introducing collections $\fC{}^{\{N^A\}}$ and $C^{\{k; N_{0}^{A}\}}$ of $SU(2)$ characters.  A priori, these characters are virtual characters due to the presence of $\bar{\gamma}$ in the trace over the kernel.  However the no-exotics theorem implies that the negative chirality kernel is trivial, so
\begin{equation}
\tr_{\ker_{\Lsq}(\slashed{\DD})} \left( \bar{\gamma}\, \ssy \right) = \tr_{\ker_{\Lsq}(\slashed{\DD})} \left(\ssy \right) = \tr_{\ker_{\Lsq}^+(\slashed{\DD})} \left(\ssy \right)~,
\end{equation}
and hence the $\fC{}^{\{N^A\}}(y)$ and $C^{\{k;N_{0}^A\}}(y)$ are true characters.
\item[8.] Lastly, we encode the data $\gamma_{\rm e} \in \Lambda_{\rm rt}$, or $[\gamma_{\rm e}]_{\rm JZ} \subset \Lambda_{\rm rt}$, in these sets of $SU(2)$ characters by defining $\fC_{L,\gm,X_\infty,\YY_\infty}^{\gamma_{\rm e}}(y)$ and $C_{\gm,X_\infty,\YY_0}^{[\gamma_{\rm e}]_{\rm JZ}}(y)$ as follows.  First, expand the given charge along the basis of simple roots: $\gamma_{\rm e} =\sum_{A} n_{\rm e}^{I_A} \alpha_{I_A} + \sum_M n_{\rm e}^{I_M} \alpha_{I_M}$.  If any of the $n_{\rm e}^{I_M} \neq 0$, then we declare the corresponding characters to vanish:
\begin{equation}
\fC_{L,\gm,X_\infty,\YY_\infty}^{\gamma_{\rm e}}(y) := 0~, \qquad C_{\gm,X_\infty,\YY_0}^{[\gamma_{\rm e}]_{\rm JZ}}(y) := 0~, \qquad \textrm{when $n^{I_M}_{\rm e} \neq 0$ for any $M$}~.
\end{equation}
(Since $\gm^\ast$ does not have components along the $\alpha_{I_M}$, the $n_{\rm e}^{I_M}$ are uniquely defined for the equivalence class $[\gamma_{\rm e}]_{\rm JZ}$ in the vanilla case.)  Similarly, if the data $\{L;\gm;\sx\}$ is such that $\fmMM$ or $\MM_0$ is empty we declare the corresponding characters to vanish for all $\YY_\infty,\gamma_{\rm e}$ or all $\YY_0, [\gamma_{\rm e}]_{\rm JZ}$.  Now assume that the moduli spaces are nonempty and that only the $n_{\rm e}^{I_A}$ are nonzero.  In the case with defects we set $N^A = n_{\rm e}^{I_A} = \langle \gamma_{\rm e}, i_\ast(h^A)\rangle$.  If there is one defect then we identify
\begin{equation}\label{fDiracchar}
\fC_{L,\gm,\sx,\sy}^{\gamma_{\rm e}}(y) := \fC_{L,\gm,\sx,\sy}^{\{N^A = n_{\rm e}^{I_A} \} }(y)~.
\end{equation}
If there are multiple defects then $SO(3)$ is not part of the isometry group.  In this case the index characters $\fC(y)$ simply become numbers $\fC$.  In the vanilla case, the equivalence class $[\gamma_{\rm e}]_{\rm JZ}$ is in one-to-one correspondence with a set of integers $N_{0}^A = N_{{\rm e},0}^A$ and an element $k_{\gamma_{\rm e}} \in \mathbb{Z}_L$ via the relation
\begin{equation}
\gamma_{\rm e} = \left(n + \frac{k_{\gamma_{\rm e}}}{L} + c \right) \gm^\ast + \sum_{A=1}^{d-1} N_{{\rm e},0}^A \, i_\ast(\beta_A)~,
\end{equation}
where the $\beta_A$ are integral-dual to the $h_{0}^A$ used to define the $K_{0}^A$.  Here $n \in \mathbb{Z}$ is arbitrary and labels the representatives of the equivalence class $[\gamma_{\rm e}]_{\rm JZ}$.  In this way we identify
\begin{equation}\label{vDiracchar}
C_{\gm,\sx,\YY_0}^{[\gamma_{\rm e}]_{\rm JZ}}(y) := C_{\gm,\sx,\YY_0}^{\{ k = k_{\gamma_{\rm e}};N_{0}^A = N_{{\rm e},0}^A\} }(y)~.
\end{equation}
\end{enumerate}

On the physics side the simple compact Lie group $G$ together with a dynamical scale $\Lambda \in \mathbb{C}^\ast$ define a unique, UV-complete quantum $\NN =2$ SYM theory without hypermultiplets.  The 't Hooft defect data, augmented by a phase, $L \to L_\zeta( \{\vec{x}_n,P_n\} )$, specifies a set of supersymmetric 't Hooft defects.  As reviewed in section \ref{ssec:SWreview}, there is a Coulomb branch $\BB$ of vacua parameterized by local complex coordinates $\{u^s\}$, $s = 1,\ldots, r = \rnk{\mathfrak{g}}$, that correspond to the vacuum expectation value of gauge-invariant observables.  For example in the case of $\mathfrak{g} = \mathfrak{su}(N)$ we may take $u^s = \langle \tr_{\bf N} \varphi^{s+1} \rangle$.  In general the $\{u\}$ are an algebraically complete set of $r$ Weyl-invariant polynomials of the eigenvalues of $\langle \varphi \rangle$.  $\BB^\ast = \BB \setminus \BB^{\rm sing}$ denotes the Coulomb branch with complex co-dimension one singular loci removed; these are loci where the low energy effective description in terms of Abelian vector multiplets breaks down.  There is a local system of electromagnetic charges $\Gamma_L \to \BB^\ast$ with fiber $\Gamma_{L,u}$ a $\mathbb{Z}^{2r}$-torsor, equipped with an integral-valued symplectic pairing $\llangle~,~\rrangle$, and undergoing monodromy around the singular loci given by $Sp(2r,\mathbb{Z})$ duality transformations.  

At each $u \in \BB^\ast$ the Hilbert spaces of framed and vanilla BPS states are graded by electromagnetic charges $\gamma \in \Gamma_{L,u}$ and $\gamma \in \Gamma_u$ respectively: \eqref{Hframed} and \eqref{vanillagrading}.  The framed and vanilla protected spin characters, $\fOmega(L_\zeta,u,\gamma;y)$ and $\Omega(u,\gamma;y)$, are certain weighted traces over these spaces as defined in section \ref{sec:wcf}.  They obey wall crossing formulae.

In sections \ref{ssec:fBPSspace} and \ref{ssec:vBPSspace} we defined a map between the math data $\{L,\sx,\sy,\gm,\gamma_{\rm e}\}$ and physics data $\{L_\zeta,u,\gamma\}$ motivated by semiclassical analysis.  As described there, it is valid for those $u$ in the weak coupling regime, $\BB_{\rm wc}^\ast \subset \BB^\ast$, and is given in a preferred weak coupling duality frame.  The duality frame is dictated by $\zeta$ in the case with defects and by the particular electromagnetic charge $\gamma$ under consideration in the vanilla case.  The map takes the form \eqref{mathgmap} together with \eqref{mathxy} or \eqref{mathxymapvan} respectively.  In the vanilla case $\YY_\infty$ should be constructed from the given math data via \eqref{YinftyfromY0}.  Hence it can be viewed as a map from the independent math data $X_\infty, \YY_0$ to the Coulomb branch.  In \eqref{mainres} and \eqref{mainres2} we used these maps to identify the Hilbert spaces of framed and vanilla BPS states with appropriate electric charge eigenspaces of the kernels of the Dirac operators on $\fMM$ and $\MM_0$ respectively.  This identification, together with the above construction of the index characters $\fC$, $C$, leads to the corresponding result for the protected spin characters:
\begin{align}\label{inCequalsPSC}
\fOmega(L_\zeta,u,\gamma;y) =&~ \fC_{L,{\rm m}(\gamma),\sx(u,\zeta),\sy(u,\zeta)}^{{\rm e}(\gamma)}(y)~,  \nonumber \\[1ex]
\Omega(u,\gamma;y) =&~ C_{{\rm m}(\gamma),\sx(u,\gamma),\YY_0(u,\gamma)}^{[{\rm e}(\gamma)]_{\rm JZ}}(y)~.
\end{align}
In particular the vanilla identification makes it manifest that the protected spin characters are invariant under shifts of ${\rm e}(\gamma)$ by integer multiples of ${\rm m}(\gamma)^\ast$, for those $u$ in the weak coupling regime.  Note that the vanilla index characters are not defined when $\gm = 0$ since there is no moduli space and hence no Dirac operator to talk about.  In this case we take the identification \eqref{inCequalsPSC} to define them.  

A special case of the identification \eqref{inCequalsPSC} is given by setting $y=-1$ where we have an identification of the (framed) BPS indices with Dirac operator indices---or more precisely, indices for the restriction of the Dirac operator to the corresponding electric charge eigenspace.  No-exotics implies that the indices are in fact the dimensions of (electric charge eigen-subspaces of) the kernels.

We can now use these identifications to make predictions concerning the behavior of the kernels as the continuous parameters $\sx,\sy$ are varied.  Since $\fOmega,\Omega$ are piecewise constant functions of $u$ (and $\zeta$ in the framed case), the index characters $\fC,C$ are piecewise constant functions of $\sx,\sy$.  At certain co-dimension one walls in $\{X_\infty,\YY_\infty\}$ space the index characters will jump.  We begin with the vanilla case.

The vanilla PSC, $\Omega(u,\gamma;y)$, jumps at the walls \eqref{vanillawalls}.  These require a pair of charges $\gamma_{1,2}$ with $\gamma = \gamma_1 + \gamma_2$, $\llangle \gamma_1, \gamma_2\rrangle \neq 0$, $\Omega(u,\gamma_{1,2};y) \neq 0$, and such that $Z_{\gamma_1}(u)\overline{Z_{\gamma_2}(u)} \in \mathbb{R}_+$.  The marginal stability condition consists of two real conditions, $\Im(Z_{\gamma_1} \overline{Z_{\gamma_2}}) = 0$ and $\Re(Z_{\gamma_1} \overline{Z_{\gamma_2}}) > 0$.  The former is easily expressed in terms of $\sx,\sy$ as follows.  First, using linearity of $Z$, 
\begin{equation}\label{vanwallmanip}
\Im(Z_{\gamma_1} \overline{Z_{\gamma_2}}) = 0 \quad \iff \quad \Im(Z_{\gamma_1} \overline{Z_{\gamma}}) = 0~.
\end{equation}
Now set $\overline{Z_\gamma} = -\zeta_{\rm van}^{-1} | Z_\gamma|$, where $\zeta_{\rm van} = \zeta_{\rm van}(u,\gamma)$ is the phase of $-Z_\gamma$.  Then, working in the duality frame discussed above,
\begin{align}\label{vanImcondition1}
\Im(Z_{\gamma_1} \overline{Z_{\gamma_2}}) = 0 \quad & \iff \quad (\gamma_{1,{\rm m}}, \sy) + \langle \gamma_{1,{\rm e}}, \sx \rangle = 0~.
\end{align}

Since $\Im(Z_{\gamma_1} \overline{Z_{\gamma_2}}) = 0$ if and only if $\Im(Z_{\gamma_2} \overline{Z_{\gamma_1}}) = 0$ we can obviously derive the same result in terms of $\gamma_2$:
\begin{align}\label{vanImcondition2}
\Im(Z_{\gamma_1} \overline{Z_{\gamma_2}}) = 0 \quad & \iff \quad (\gamma_{2,{\rm m}}, \sy) + \langle \gamma_{2,{\rm e}}, \sx \rangle = 0~.
\end{align}
It is not independent, however.  If we are employing the math-physics map then one follows from the other using the vanilla constraint $(\gm, \sy) + \langle \gamma_{\rm e}, \sx \rangle = 0$.  Indeed we can use this constraint to write these conditions in terms of quantities $\YY_0$ and $\gerel$ that appear naturally in the mathematical construction.  Namely, \eqref{vanImcondition1} and \eqref{vanImcondition2} are equivalent to
\begin{equation}\label{vanImcondition}
(\gamma_{1,2,{\rm m}}, \YY_0) + \left\langle \gamma_{1,2,{\rm e}} - \frac{\langle \gamma_{\rm e}, X_\infty \rangle}{(\gm,X_\infty)} \gamma_{1,2,{\rm m}}^\ast~,~ X_\infty \right\rangle = 0~.
\end{equation}
The quantity that $X_\infty$ is paired with in the second term is precisely the part of the electric charge depending on the data $\gerel$.  In this form we also see that the two equations are equivalent: their sum vanishes identically, using $(\gm,\YY_0) = 0$.  Hence the solution space consists of a real co-dimension one wall in the space spanned by $X_\infty,\YY_0$.

The condition $\Re(Z_{\gamma_1} \overline{Z_{\gamma_2}}) > 0$ is not so easy to express directly in terms of $\sx,\YY_0$ as it would involve inverting the map to determine $a$ as a function of $\sx,\YY_0$.  Fortunately this is not necessary and one can assume that this condition is always satisfied when $\Im(Z_{\gamma_1} \overline{Z_{\gamma_2}}) = 0$, for any given $X_\infty, \YY_0$, provided $X_\infty$ is in the fundamental Weyl chamber and, in particular, nonzero.  The key point is that we actually have a full $\mathbb{C}^\ast$ family of parameter maps corresponding to the choice of dynamical scale $\Lambda$.  Any choice of $\Lambda$ such that the preimage of $X_\infty,\YY_0$ lies in $\widehat{\BB}_{\rm wc}$ can be used.  Let $W_{\mathfrak{t}}^+$ denote the fundamental Weyl chamber.  Then we claim:  
\begin{description}\label{wclemma}
\item[Lemma:]  Let $\{ \gm,\sx,\YY_0 \} \in \Lambda_{\rm cr} \times W_{\mathfrak{t}}^+ \times \mathfrak{t}_{\gm}^{\perp}$ be given such that $\MM_0(\gm;X_\infty)$ is nonempty.  Then there exists a $\mu \in \mathbb{R}_+$ such that for any $\Lambda \in \mathbb{C}^\ast$ with $|\Lambda| < \mu$ the following statements hold:
\begin{enumerate}
\item the preimage of $\{\sx,\YY_0\}$ with respect to the map \eqref{mathxymapvan} is in $\widehat{\BB}_{\rm wc}(\Lambda)$ and
\item for all pairs of charges, $\gamma_{1,2}$ satisfying ${\rm m}(\gamma_1 + \gamma_2) = \gm$, $\llangle \gamma_1, \gamma_2 \rrangle \neq 0$, and $\Omega(u,\gamma_{1,2};y) \neq 0$ for any $u$ in the preimage, we have $\Re( Z_{\gamma_1}(u) \overline{Z_{\gamma_2}(u)}) > 0$, whenever $\Im(Z_{\gamma_1}(u) \overline{Z_{\gamma_2}(u)}) = 0$.
\end{enumerate}
\end{description}
Hence for any finite path in the space $W_{\mathfrak{t}}^+ \times \mathfrak{t}_{\gm}^{\perp}$, there will be a minimum such $\mu$ along that path, and by taking the dynamical scale to be smaller in magnitude we ensure that the preimage of the path is in the weak coupling regime and that $\Re(Z_{\gamma_1} \overline{Z_{\gamma_2}}) > 0$ is satisfied whenever $\Im(Z_{\gamma_1} \overline{Z_{\gamma_2}}) = 0$ along the path for all constituent charges.  

The lemma is proven in appendix \ref{app:lemma}.  It involves the construction of a one-parameter family of physics data $u_t, \Lambda_t$, $0 \leq t < \infty$ that solves the math physics map for the given math data and is such that the conditions of the lemma hold for $t$ large enough.  In fact for large $t$ a second real parameter $\vartheta$ emerges; it parameterizes the one-dimensional preimage of $\sx,\YY_0$ under the vanilla math-physics map for the given $t$.  At large $t$ we have
\begin{align}\label{wcfam}
& a(u_t) = i \left(1 - \frac{i \vartheta}{t}\right) X_\infty + \frac{1}{t} \YY_0 + O(1/t^2)~, \qquad \Lambda_t = e^{-\pi t/h^\vee} \Lambda_0~,
\end{align}
where $\Lambda_0$ is some given and fixed dynamical scale.  For a fixed $\Lambda_t$ (\ie\ fixed $t$), this is a one-parameter family of inverses to \eqref{mathxymapvan} labeled by $\vartheta$.  The weak coupling duality frame that determines the function $a = a(u)$ appearing on the left is the frame such that $\Im(a(u))$ is in the fundamental Weyl chamber.

Now, as we noted previously, the BPS spectrum is invariant under a renormalization of scale, provided we scale all dimensionful quantities, including the dynamical scale.  Hence the physics data \eqref{wcfam} is equivalent to the data of a \emph{fixed} dynamical scale $\Lambda_0$, together with Coulomb branch parameters $u = u_t(\sx,\YY_0)$ determined via
\begin{equation}\label{inverseMPmap}
a(u_{t}(\sx,\sy)) := e^{\pi t/h^\vee} \left\{ \left(1 - \frac{i \vartheta}{t}\right) i X_\infty + \frac{1}{t} \YY_0 + O(1/t^2) \right\}~,
\end{equation}
as far as the BPS spectrum is concerned.  One can see from \eqref{mathxymapvan} that $\{a(u'),\Lambda'\} = e^{\pi t/h^\vee} \{a(u), \Lambda\}$ amounts to a rescaling of the math data, $\{X_{\infty}', \YY_0' \} = e^{\pi t/h^\vee} \{ \sx,\YY_0 \}$.  This leads to an overall rescaling of the Dirac operator and therefore the kernel is unaffected.  We can think of $\{ t,\vartheta \}$ in \eqref{inverseMPmap} as parameterizing the two-dimensional surfaces at fixed $\{ \sx,\YY_0 \}$ in the asymptotic regime of the Coulomb branch (\ie\ large $t$), along which the BPS spectrum is invariant.  We can use the large $t$ limit of \eqref{inverseMPmap} to give a convenient way of expressing the vanilla index characters in terms of the protected spin characters:
\begin{equation}\label{vanCOmega}
C_{{\rm m}(\gamma),\sx,\YY_0}^{[{\rm e}(\gamma)]_{{\rm m}(\gamma)}^{\mathbb{Z}}}(y) = \lim_{t \to \infty} \Omega\left(\gamma,u_t(\sx,\YY_0);y\right)~.
\end{equation}
In this limit we are ensured that the conditions of the lemma are satisfied and hence \eqref{vanImcondition}  gives the walls where the kernel will jump.  Hence we are led to the
\begin{description}
\item[Conjecture:]  Consider data $\{ \gm,\sx, \YY_0\} \in \Lambda_{\rm cr} \times W_{\mathfrak{t}}^+ \times \mathfrak{t}_{\gm}^\perp$ such that $\mMM_0(\gm;\sx)$ is non-empty.  The Dirac operators $\slashed{\DD}_{\mMM_0(\gm;\sx)}^{{\rm G}_0(\YY_0)}$ are Fredholm except at at the real co-dimension one walls in $W_{\mathfrak{t}}^+ \times \mathfrak{t}_{\gm}^\perp$ defined by
\begin{equation}\label{vanmathwalls}
(\gamma_{1,{\rm m}},\YY_0) + \left\langle \gamma_{1,{\rm e}} - \frac{\langle \gamma_{1,{\rm e}} + \gamma_{2,{\rm e}}, X_\infty \rangle}{(\gamma_{1,{\rm m}} + \gamma_{2,{\rm m}}, X_\infty)} \gamma_{1,{\rm m}}^\ast ~,~ \sx \right\rangle = 0~,
\end{equation}
where $\gamma_{1,2} = \gamma_{1,2,{\rm m}} \oplus \gamma_{1,2,{\rm e}} \in \Lambda_{\rm cr} \oplus \Lambda_{\rm rt}$ are any pair of electromagnetic charges satisfying
\begin{equation}
\gm = \gamma_{1,{\rm m}} + \gamma_{2,{\rm m}}~, \quad \llangle \gamma_1,\gamma_2\rrangle \neq 0~, \quad \textrm{and} ~~ C_{\gamma_{1,{\rm m}},\sx,\YY_0}^{[\gamma_{1,{\rm e}}]_{\rm JZ}}(y) \neq 0 \neq C_{\gamma_{2,{\rm m}},\sx,\YY_0}^{[\gamma_{1,{\rm e}}]_{\rm JZ}}(y)~.
\end{equation}
Across these walls the index characters $C_{\gm,\sx,\YY_0}^{[\gamma_{\rm e}]_{\rm JZ}}(y)$ of the Dirac operator, with $\gamma_{\rm e} = \gamma_{1,{\rm e}} + \gamma_{2,{\rm e}}$,~ jump in a way determined by the identification \eqref{vanCOmega} and the (motivic) Kontsevich--Soibelman wall crossing formulae for $\Omega(u,\gamma;y)$, \cite{2008arXiv0811.2435K}.
\end{description}

Observe that in order to determine how the index characters of the Dirac operator jump, one requires knowledge of the index characters for the constituent charges.  Fortunately there are three facts that, taken together, provide a bottom rung to this induction ladder.  They are
\begin{enumerate}
\item The moduli space $\MM_0(\gm;X_\infty)$ is empty unless $\gm = \sum_A n_{\rm m}^{I_A}$ with all $n_{\rm m}^{I_A} > 0$;
\item Walls in $W_{\mathfrak{t}}^+ \times \mathfrak{t}_{\gm}^\perp$, determined by \eqref{vanmathwalls}, only exist when both constituents have nonzero magnetic charge, $\gamma_{1,2,{\rm m}} \neq 0$; and
\item The index characters for the case where $\MM_0$ is a point, corresponding to $\gm = H_{I}$, a simple co-root, are equal to one, as the Dirac operator is trivial and the kernel is spanned by the constant wavefunction $\Psi_{0}^{\gerel} =1$.  ($L = 1$ and $\gamma_{{\rm e},0} = 0$ here, so there is only one equivalence class: $[\gamma_{\rm e}]_{\rm JZ} = \gm^\ast \cdot \mathbb{Z}$.)
\end{enumerate}
The last point is consistent with the fact that every member of the dyon cohort \eqref{dyoncohort} is a simple half-hypermultiplet with $\Omega = 1$.  The second point can be seen as follows.  Suppose one of the constituent magnetic charges vanishes and without loss of generality take it to be $\gamma_{1,{\rm m}} = 0$.  Then the condition for the wall gives us that $\langle \gamma_{1,{\rm e}}, X_\infty \rangle = 0$.  However we must also have that $\gamma_1$ corresponds to a populated state---this is the condition that the index character for the constituents be nonvanishing.  The only populated electric charges are those of the $W$-bosons with $\gamma_{\rm e} = \alpha \in \Delta$, a non-zero root.  However $\langle \alpha, X_\infty \rangle = 0$ is in contradiction with $X_\infty \in W_{\mathfrak{t}}^+$.

The dyon cohorts therefore give the basic building blocks for which we know the index characters.  We also know from the vanishing argument \eqref{vanishing} that, for $\dim(\MM_0(\gm;X_\infty)) > 0$, the $\Lsq$ kernel of the Dirac operator is always trivial on the locus $\YY_0 = 0$.  Hence these facts, together with the known wall crossing formulae for the $\Omega$, allow us in principle to determine the index characters for the whole family of Dirac operators labeled by $\{\gm,\sx,\YY_0\} \in \Lambda_{\rm cr} \times W_{\mathfrak{t}}^+ \times \mathfrak{t}_{\gm}^{\perp}$.

Turning to the case with defects, the framed PSC, $\fOmega(L_\zeta,u,\gamma;y)$, jumps at the walls \eqref{fmsw}.  These are defined by the marginal stability conditions $\zeta^{-1} Z_{\gamma_{\rm h}}(u) \in \mathbb{R}_-$, for halo charges in the vanilla lattice, $\gamma_{\rm h} \in \Gamma_u$, such that $\Omega(u,\gamma_{\rm h};y) \neq 0$.  The last condition ensures that the vanilla particle forming the supposed halo actually exists in the spectrum.  The condition $\Im(\zeta^{-1} Z_{\gamma_{\rm h}}) = 0$ is equivalent to $(\gamma_{\rm h,m},\sy) + \langle \gamma_{\rm h,e},\sx\rangle = 0$ via the map of parameters \eqref{mathxy}.  Meanwhile, a similar argument as above can be made to deal with the condition $\Re(\zeta^{-1} Z_{\gamma_{\rm h}}) < 0$.  Namely, in appendix \ref{app:lemma} we prove the following
\begin{description}
\item[Lemma:]  Let $\{\sx,\sy\} \in W_{\mathfrak{t}}^+ \times \mathfrak{t}$ be given.  Then there exists a $\mu \in \mathbb{R}_+$ such that for any $\Lambda \in \mathbb{C}^\ast$ with $|\Lambda| < \mu$,
\begin{enumerate}
\item the preimage of $\sx,\sy$ with respect to the map \eqref{mathxy} is in $\widehat{\BB}_{\rm wc}(\Lambda) \times \widehat{\mathbb{C}^\ast}$, and
\item For all populated vanilla charges $\gamma_{\rm h} \in \Gamma_u$ we have $\Re(\zeta^{-1} Z_{\gamma_{\rm h}}(u)) < 0$ whenever $\Im(\zeta^{-1} Z_{\gamma_{\rm h}}(u)) = 0$.
\end{enumerate}
\end{description}

The lemma is proven by again constructing a family of physics data $u_t,\Lambda_t$, for any given $\zeta \in \widehat{\mathbb{C}^\ast}$, such that $(u_t,\zeta) \in \widehat{\BB}_{\rm wc} \times \widehat{\mathbb{C}^\ast}$ is in the inverse image of $\sx,\sy$ with respect to \eqref{mathxy} and the conditions of the lemma are satisfied for $t$ large enough.  Importantly, the minimum bound on $t$ such that the conditions hold does not depend on $\zeta$.  The explicit family takes the form
\begin{align}\label{wcfamf}
a(u_t) =&~ \zeta \bigg\{ \left[ 1 - \frac{i}{t} \left( \frac{h^\vee (\pi +2\arg(\zeta))}{2\pi} -\frac{\theta_0}{2\pi}\right) \right] i X_\infty + \frac{1}{t} \YY_\infty + O(1/t^2) \bigg\}~, \cr
\Lambda_t =&~ e^{-\pi t/h^\vee} |\zeta| \Lambda_0~, \raisetag{20pt}
\end{align}
at large $t$, where $\Lambda_0$ is some fixed and given dynamical scale.  The function $a = a(u_t)$ is again specified by the requirement that $\Im(a(u)) \in W_{\mathfrak{t}}^+$.  By an overall renormalization of scale, the physics data \eqref{inverseMPmapf} is equivalent to the data of a fixed dynamical scale $\Lambda_0$, together with Coulomb branch parameters, $u_t = u_t(\sx,\sy;\zeta)$, determined via
\begin{align}\label{inverseMPmapf}
a(u_t(\sx,\sy;\zeta)) :=&~ \frac{\zeta}{|\zeta|} e^{\pi t/h^\vee} \bigg\{ \left[ 1 - \frac{i}{t} \left( \frac{h^\vee (\pi +2\arg(\zeta))}{2\pi} -\frac{\theta_0}{2\pi}\right) \right] i X_\infty + \cr
&~ \qquad \qquad \qquad \qquad \qquad \qquad \qquad \quad+  \frac{1}{t} \YY_\infty + O(1/t^2) \bigg\}~, \qquad
\end{align}
as far as the BPS spectrum is concerned.  Taking the $t \to \infty$ limit gives an expression analogous to \eqref{vanCOmega} for the framed index characters:
\begin{equation}\label{fCOmega}
\fC_{L,{\rm m}(\gamma),\sx,\YY_0}^{\gamma_{\rm e}}(y) = \lim_{t \to \infty} \fOmega\left(L_\zeta,\gamma,u_t(\sx,\sy;\zeta);y\right)~.
\end{equation}
In this limit we are ensured that the conditions of the lemma are satisfied and hence $(\gamma_{\rm h,m}, \sy) + \langle \gamma_{\rm h,e}, \sx \rangle = 0$ determines the walls where the kernel will jump.  We thus have the
\begin{description}
\item[Conjecture:]  Consider data $\{L,\gm,\sx,\sy\}$ such that $\fmMM(L;\gm;\sx)$ is nonempty.  The Dirac operators $\slashed{\DD}_{\fmMM(L;\gm;\sx)}^{{\rm G}(\sy)}$ are Fredholm except at the real co-dimension one walls  defined by
\begin{equation}\label{framedmathwalls}
(\gamma_{\rm h,m}, \sy) + \langle \gamma_{\rm h,e}, \sx \rangle = 0~,
\end{equation}
for any $\gamma_{\rm h} = \gamma_{\rm h,m} \oplus \gamma_{\rm h,e} \in \Lambda_{\rm cr} \oplus \Lambda_{\rm rt}$ such that the vanilla index character is nonvanishing, $C_{\gamma_{\rm h,m},\sx,\YY_0}^{[\gamma_{\rm h,e}]_{\gamma_{\rm h,m}}^{\mathbb{Z}}}(y) \neq 0$.  Across these walls the index characters $\fC_{L,\gm,\sx,\sy}^{\gamma_{\rm e}}(y)$ for all $\gamma_{\rm e} \in \Lambda_{\rm rt}$ jump in a way determined by the identification \eqref{fCOmega} and the (motivic) framed wall crossing formulae \cite{Gaiotto:2010be}.
\end{description}
We note that in our formulation of the walls, \eqref{fmsw}, it is possible that the charges $\gamma = \gm \oplus \gamma_{\rm e}$ of all framed BPS states associated with a given line defect will have zero symplectic pairing with a given halo charge.  In this case the wall crossing formula of \cite{Gaiotto:2010be} implies that the kernel will not jump.  These walls are referred to as invisible walls; see footnote \ref{fn:invisible}.

These statements are conjectures because they rely on our conjectural relation between math and physics data described in sections \ref{ssec:fBPSspace} and \ref{ssec:vBPSspace}.  In the next two sections we will verify these conjectures by direct computation in cases where the kernel of the Dirac operator can be determined explicitly.  We expect, however, that it should be possible to prove that the operators are Fredholm except at the stated walls, \eqref{vanmathwalls} and \eqref{framedmathwalls}, by making use of the asymptotic form of the moduli space metric and appropriately generalizing the arguments in \cite{Stern:2000ie}.

%%%%%%%%%%%%%%%%%%%%%%
%%%%%%%%%%%%%%%%%%%%%%
\section{Reviewing a vanilla example: bound states for $|\gm| = 2$}\label{Section:VanillaEx}
%%%%%%%%%%%%%%%%%%%%%%
%%%%%%%%%%%%%%%%%%%%%%

In this section and the next we consider the case where the dimension of $\MM_0$ and $\fMM$ is four.  In this situation the kernel of the Dirac operator can be determined explicitly, and we can compare with predictions from Sieberg--Witten theory.  We begin with the vanilla case, which is mostly a review of work carried out in \cite{Lee:1996kz,Gauntlett:1999vc}. 

We are considering pure $\NN = 2$ gauge theory with Lie algebra $\mathfrak{g}$.  Our discussion in \ref{ssec:vanlocus} showed that there are no BPS states in the weak coupling regime beyond the $W$-bosons and dyon cohorts for $\mathfrak{g} = \mathfrak{su}(2)$, so we assume $\rnk{\mathfrak{g}} > 1$.  Then, when the height of the magnetic charge is two, there are three possibilities:
\begin{enumerate}
\item  $\mathfrak{g}^{\rm ef} = \mathfrak{su}(2) \oplus \mathfrak{su}(2)$.  The magnetic charge $\gm = \sum_A n_{\rm m}^{I_A} H_{I_A}$ has two nonzero components with $n_{\rm m}^{I_1} = 1 = n_{\rm m}^{I_2}$.  Furthermore the corresponding nodes in the Dynkin diagram are not connected.
\item $\mathfrak{g}^{\rm ef}$ is a rank two simple Lie algebra, either $\mathfrak{su}(3)$, $\mathfrak{so}(5) \cong \mathfrak{sp}(2)$, or $\mathfrak{g}_2$.  Again there are two nonzero components of the magnetic charge,  $n_{\rm m}^{I_1} = 1 = n_{\rm m}^{I_2}$.  Now the corresponding nodes are connected, such that the sugdiagram is the Dynkin diagram for $\mathfrak{su}(3)$, $\mathfrak{so}(5) \cong \mathfrak{sp}(2)$, or $\mathfrak{g}_2$.
\item $\mathfrak{g}^{\rm ef} = \mathfrak{su}(2)$.  The magnetic charge has only one nonzero component, with $n_{\rm m}^{I_1} = 2$.
\end{enumerate}
Only the second case admits BPS states.  The absence of BPS states in the last case, for which $\MM_0$ is the double cover of the Atiyah--Hitchin manifold, follows from the vanishing theorem as explained under \eqref{vanishing}.  In the first case we are embedding two charge one $\mathfrak{su}(2)$ solutions along disjoint simple roots.    The moduli space will be a product of two single monopole moduli spaces, $\mMM \cong (\mathbb{R}^3 \times S^1)^2$.  Without going into precise details it is clear that the centered moduli space will be metrically $\mMM_0 \cong \mathbb{R}^3 \times S_{\uppsi}^1$ and ${\rm G}_0(\sy) \propto \pd_\uppsi$, where $\uppsi$ parameterizes the circle in $\mMM_0$.  The corresponding Dirac operator, \eqref{M0diracop}, does not admit $\Lsq$ zero modes.

In the second case $\mMM_0$ and ${\rm G}_0(\sy)$ are such that the Dirac operator can have a nontrivial kernel with intricate jumping phenomena.  We recount the essential details below and compare with expectations from Seiberg--Witten theory.  Let us immediately summarize the results here: we find perfect agreement for the locations of the walls, the jumping of the spectrum, and for the bound-state radii.  This should be contrasted with \cite{Fraser:1996pw}, where all of these walls were missed because the electric charge contribution to the Seiberg--Witten prepotential was neglected.  As we emphasized in section \ref{sec:scphilo}, the electric terms of the prepotential, along with the one-loop corrections to the magnetic terms, are of the same order as the collective coordinate dynamics in the Manton approximation and must be included.

%%%%%%%%%%%%%%%%%%%%%%
\subsection{The monopole moduli space}
%%%%%%%%%%%%%%%%%%%%%%

Lee, Weinberg and Yi (LWY) \cite{Lee:1996kz}, extending the approach of Gibbons and Manton \cite{Gibbons:1995yw}, derived the asymptotic form of the vanilla moduli space $\mMM(\gm;\sx)$ and its metric, corresponding to widely separated monopoles, for general gauge algebra $\mathfrak{g}$ and magnetic charge $\gm$. In certain situations, namely  when all $n_{\rm m}^{I_A}= 1$, the space is free of singularities and it is expected that their construction gives the exact metric; see \cite{Murray:1996hi,Chalmers:1996jd}.  The case we are considering here, with two $n_{\rm m}^{I_A} = 1$, was investigated previously in \cite{Gauntlett:1996cw,Lee:1996if}, and in this case it can be proven that the asymptotic metric is the exact metric.

In this case the LWY metric is
\begin{align}
\ed s^2 =&~ \ed s_{\rm cm}^2 + \ed s_{0}^2~, \qquad \textrm{where} \label{metricLWY} \\
\ed s_{\rm cm}^2 =&~ M \ed \vec{x}_{\rm cm} \cdot \ed \vec{x}_{\rm cm} + M^{-1} \ed \chi^2~, \\
\ed s_{0}^2 =&~ \mu \bigg[ H(r) \ed \vec{r} \cdot \ed \vec{r} + \left( \frac{\sp}{2\mu} \right)^2 H(r)^{-1} (2 \ed \uppsi + \vec{w}(\vec{r}) \cdot \ed \vec{r} )^2 \bigg]~, \label{su3TN}
\end{align}  
where
\begin{equation}
H(r) = 1 + \frac{\sp}{2\mu r}~, \qquad \vec{\nabla} \times \vec{w} = - \frac{\vec{r}}{r^3}~,
\end{equation}
and
\begin{equation}
\mu := \frac{m_1 m_2}{m_1 + m_2}~, \qquad M = m_1 + m_2~, \qquad \textrm{with} \quad  m_A = (H_{I_A},\sx)~.
\end{equation}
Without loss of generality we take $\alpha_{I_1}$ to be a long root.  Then $\sp = \sp^2$ is the integer, first appearing in \eqref{dualgm}, for $\alpha_{I_2}$; $\sp = 1,2,3$ for $\mathfrak{g}_{\rm ef} = \mathfrak{su}(3),\mathfrak{so}(5),\mathfrak{g}_2$ respectively.  

The coordinates $(\chi,\uppsi)$ give an explicit parameterization of the triholomorphic $U(1)^2$ isometry induced from asymptotically nontrivial gauge transformations.  In order to make the connection with our general discussion, let $\xi^A$ be coordinates such that
\begin{equation}\label{xicoords}
K^A \equiv \rG(i_\ast(h^A)) = \frac{\pd}{\pd \xi^A}~, \qquad \xi^A \sim \xi^A + 2\pi~,
\end{equation}
where the $h^A$ are the fundamental magnetic weights of $\mathfrak{su}(3)$.  Recall that $\pd_\chi = \rG(i_\ast(h_{\rm cm}))$.  If we identify $\pd_\uppsi$ with the $2\pi$-periodic triholomorphic Killing vector $K_{0}^1 = \rG(i_\ast(h_{0}^1))$ on $\MM_0$, then the change of basis formula for $\{h^1,h^2\} \mapsto \{h_{0},h_{\rm cm} \}$ described in appendix \ref{app:Dquotient2}, with $(\ell^1,\ell^2) =(1,\sp)$, gives\footnote{Since $d=2$ here and there is only a single $h_{0}^A$ we simply write $h_{0} \equiv h_{0}^1$ and similarly for its dual element $\beta \equiv \beta_1$ and the component of the relative electric charge $N_{{\rm e},0} \equiv N_{{\rm e},0}^1$ below.}
\begin{align}\label{chipsixi}
\frac{\pd}{\pd\chi} =&~ \rG(i_\ast(h_{\rm cm})) = \frac{m_1}{M} \cdot \frac{\pd}{\pd \xi^1} + \frac{m_2}{\sp M}  \cdot\frac{\pd}{\pd \xi^2}~, \cr
\frac{\pd}{\pd\uppsi} =&~ \rG(i_\ast(h_{0})) = \sp \frac{\pd}{\pd \xi^1} - \frac{\pd}{\pd \xi^2}~.
\end{align}
These imply the relations
\begin{equation}\label{chipsi2xi}
\chi = \xi^1 + \sp \xi^2~, \qquad \uppsi = \frac{m_2 \xi^1 - m_1 \sp \xi^2}{\sp M}~.
\end{equation}

As a check of these identifications, let $\vec{x}_{1,2}$ be constituent position vectors defined through the relations
\begin{equation}
\vec{x}_{\rm cm} = \frac{m_1 \vec{x}_1 + m_2 \vec{x}_2}{M}~, \qquad \vec{r} = \vec{x}_1 - \vec{x}_2~.
\end{equation}
Then, in the limit of infinite separation, $r \to \infty$, the metric takes the form
\begin{align}
\lim_{r\to \infty} \ed s^2 =&~ M \ed \vec{x}_{\rm cm} \cdot  \ed \vec{x}_{\rm cm} + \mu \ed \vec{r} \cdot \ed \vec{r} + M^{-1} \ed\chi^2 + \mu^{-1} \sp^2 \ed\uppsi^2 \cr
=&~ m_1 \ed \vec{x}_1 \cdot \ed \vec{x}_1 + \frac{(\ed \xi^1)^2}{m_1} + m_2 \ed \vec{x}_2 \cdot \vec{x}_2 + \frac{(\sp \ed \xi^2)^2}{m_2}~,
\end{align}
a sum of correctly normalized flat metrics for the constituents, $\lim_{r\to \infty} \MM = (\mathbb{R}^3 \times S^1)^2$, with correct periodicities for the constituent phases.

Hence $\pd_\uppsi$ is correctly identified with the $2\pi$-periodic Killing vector $K_{0}^1$, and we see that $\uppsi \sim \uppsi + 2\pi$ implies that $(\MM_0, \ed s_{0}^2)$ is the smooth, single-centered Taub--NUT manifold.  The periodicities \eqref{xicoords} imply a further identification on $\widetilde{\MM}$, which is associated with the generator of deck transformations.  Note that $L = \gcd(\ell^1,\ell^2) = 1$ in this case so the group of deck transformations is generated by the gauge-induced isometry $\phi_\gi = \exp(2\pi \rG(i_\ast(h_\gi)))$.  In this simple example we could of course easily obtain the identification in terms of $(\chi,\uppsi)$ from \eqref{chipsi2xi}, but let us instead take the opportunity to illustrate the general formulae of appendix \ref{app:Dquotient2}.  The element $h_\gi \in \Lambda_{\rm mw}^{\rm ef}$ can be expanded in the basis $\{h_{0},h_{\rm cm}\}$, with the component along $h_{\rm cm}$ being $L = 1$.  The component along $h_{0}$, from \eqref{hthbetacomp}, is
\begin{equation}
\langle \beta, h_\gi \rangle = \frac{x_{1,\sp} m_2}{\sp M} - \frac{y_{1,\sp} m_1}{M}~,
\end{equation}
where $x_{1,\sp},y_{1,\sp} \in \mathbb{Z}$ are any solution to $x_{1,\sp} + \sp y_{1,\sp} = 1$.  The general solution in this case is $\{x_{1,\sp},y_{1,\sp}\} = \{1,0\} + n \{\sp,-1\}$ for any $n \in \mathbb{Z}$, which would give $\langle \beta, h_\gi \rangle = \frac{m_2}{\sp M} + n$.  Now $\exp(2\pi n \rG(i_\ast(h_{0}))) = \exp(2\pi n \pd_{\uppsi})$ is trivial for any $n \in \mathbb{Z}$, so we might as well choose $n=0$, in which case
\begin{equation}
h_\gi = h_{\rm cm} + \frac{m_2}{\sp M} h_{0}~.
\end{equation}
Thus we have that the generator of deck transformations is
\begin{equation}
\phi = \phi_\gi = \exp\left\{ 2\pi \pd_\chi + \frac{2\pi m_2}{\sp M} \pd_\uppsi \right\}~,
\end{equation}
which imposes on $\MM$ the identification
\begin{equation}\label{Dquotientsu3}
(\chi, \uppsi) \sim \left( \chi + 2\pi, \uppsi + \frac{2\pi m_2}{\sp M}\right)~.
\end{equation}
For generic $X_\infty$, $m_2/M$ is irrational and no power of $\phi$ acts trivially on $\MM_0$.

Let us also consider the decomposition of the triholomorphic Killing vector $\rG(H)$ with respect to $\pd_\chi, \pd_\uppsi$, for generic $H \in \mathfrak{t}$.  One can use ${\rm G}(H) = \langle \alpha_{I_1}, H \rangle \pd_{\xi^1} + \langle \alpha_{I_2}, H \rangle \pd_{\xi^2}$ and then (the inverse of) \eqref{chipsixi}.  Equivalently, one can use $\rG(H) = (\gm,H) \pd_\chi + \langle \beta, H\rangle \pd_\uppsi$ and \eqref{betares1}, \eqref{betares2}.    Either way, the result is
\begin{equation}
{\rm G}(H) = (\gm, H) \frac{\pd}{\pd \chi} + \frac{1}{\sp} \left[ \frac{m_2}{M} \langle \alpha_{I_1}, H \rangle - \frac{m_1}{M} \sp \langle \alpha_{I_2}, H \rangle \right] \frac{\pd}{\pd \uppsi}~.
\end{equation}
In particular, the projection along $T\MM_0$ relevant for the Dirac operator $\slashed{\DD}_{0}^{\rG_0}$ can be written as
\begin{equation}
{\rm G}_0(\YY_0) = \frac{1}{\sp} \left[ \frac{ (H_{I_2},\sx) (H_{I_1},\YY_0) - (H_{I_1},\sx) (H_{I_2},\YY_0) }{(\gm, \sx)} \right] \frac{\pd}{\pd \uppsi} ~.
\end{equation}
Recall that $\rG_0(\sy) = \rG_0(\YY_0)$ only depends on the component of $\YY_\infty$ that is Killing orthogonal to $\gm$, which is what we have denoted $\YY_0$.  In this case $(\gm,\YY_0) = 0$ implies that $(H_{I_1},\YY_0) = - (H_{I_2},\YY_0)$.  Using this we simply have
\begin{equation}\label{G0Y}
{\rm G}_0(\YY_0) = \frac{1}{\sp} (H_{I_1},\YY_0) \pd_{\uppsi} ~.
\end{equation}

Finally, we will also need the decomposition of the generic electric charge eigenvalue,
\begin{equation}
\gamma_{\rm e} = n_{\rm e}^{I_1} \alpha_{I_1} +  n_{\rm e}^{I_2} \alpha_{I_2}~,
\end{equation}
with respect to the basis $\{ i_\ast(\beta), \gm^\ast\}$, dual to $\{ i_\ast(h_{0}), i_\ast(h_{\rm cm}) \}$.  We have
\begin{equation}
\gamma_{\rm e} = q_{\rm cm} \gm^\ast + N_{{\rm e},0} \, i_\ast(\beta)~,
\end{equation}
with
\begin{align}
N_{{\rm e},0} =&~ \langle \gamma_{\rm e}, i_\ast(h_{0}) \rangle = \sp n_{\rm e}^{I_1} - n_{\rm e}^{I_2}~, \cr
q_{\rm cm} =&~ \langle \gamma_{\rm e}, i_\ast(h_{\rm cm}) \rangle = \frac{n_{\rm e}^{I_1} \langle \alpha_{I_1},X_\infty \rangle + n_{\rm e}^{I_2} \langle \alpha_{I_2}, X_\infty \rangle }{(\gm, X_\infty)} =  \frac{n_{\rm e}^{I_1} \sp m_1 + n_{\rm e}^{I_2} m_2}{\sp M}~.
\end{align}
Note that $q_{\rm cm}$ can also be written as
\begin{equation}\label{qcmsu3}
q_{\rm cm} = - N_{{\rm e},0} \frac{m_2}{\sp M} + n_{\rm e}^{I_1} = -N_{{\rm e},0} \langle \gamma_{\rm e}, h_\gi \rangle + n_{\rm e}^{I_1}~,
\end{equation}
in agreement with the form \eqref{qcmpieces} for $k_{\gamma_{\rm e}} = 0$ (as required for $L=1$) and with $n = n_{\rm e}^{I_1}$ labeling the Julia--Zee tower.

%%%%%%%%%%%%%%%%%%%%%%%
\subsection{Spectrum and wall crossing from the Dirac kernel}
%%%%%%%%%%%%%%%%%%%%%%%

As we reviewed in section \ref{ssec:vBPSspace} the semiclassical BPS spectrum can be identified with zero modes of the $\rG(\sy)$-twisted Dirac operator on $\MM$.  Furthermore the Dirac operator commutes with the electric charge operator so we can consider the Dirac operator within each electric charge eigenspace.  Finally, in the vanilla case there is always the flat factor in moduli space which gives rise to the center of mass half-hypermultiplet. The nontrivial part of the spectrum is contained in the $\rG_0(\YY_0)$-twisted Dirac operator, \eqref{M0diracop}, on the strongly centered moduli space.  In the case at hand, BPS states with electric charge $\gamma_{\rm e}$ are represented by spinors on $\MM$ of the form
\begin{equation}\label{Psigesu3}
\Psi^{\gamma_{\rm e}} = \psi_{\rm cm} e^{-i q_{\rm cm} \chi} \otimes \Psi_{0}^{(N_{{\rm e},0})}~,
\end{equation}
where $\psi_{\rm cm}$ is a constant four-component spinor on $\mathbb{R}^4$, $q_{\rm cm}$ is given by \eqref{qcmsu3}, and $\Psi_{0}^{(N_{{\rm e},0})}$ is an $\Lsq$-normalizable spinor on $\MM_0$ satisfying
\begin{equation}
\slashed{\DD}_{\MM_0}^{\rG_0(\YY_0)} \Psi_{0}^{(N_{{\rm e},0})} = 0 \qquad \textrm{and} \qquad i \Lie_{\pd_\uppsi} \Psi_{0}^{(N_{{\rm e},0})} = N_{{\rm e},0} \Psi_{0}^{(N_{{\rm e},0})}~.
\end{equation}
The four real degrees of freedom in $\psi_{\rm cm}$ correspond to the half-hypermultiplet factor in \eqref{hhfactor}.

The explicit zero modes of precisely this Dirac operator were obtained originally in \cite{Pope:1978zx}.  (See also \cite{Jante:2013kha} for a recent analysis.)  Due to the importance of this result and for completeness we provide yet a different derivation in appendix \ref{appendix:TN}.  The relations between the parameters of the problem at hand and those used in the appendix are as follows:
\begin{equation}\label{TNparams}
m=\m\,,\quad \ell=\frac{\sp}{2\mu}\,,\quad x^4 = 2\uppsi\,,\quad k = 1\,,\quad C  = \frac{\sp}{2} (H_{I_1}, \YY_0)  \frac{(\gm, X_\infty)}{(H_{I_1},X_\infty) (H_{I_2},X_\infty)} \,.
\end{equation}
In the appendix $m$ is the overall mass scale of the Taub--NUT metric, $\ell$ is the Taub--NUT radius, and $C$ is the coefficient of the canonical one-form $H(r)^{-1} (2\ed\uppsi + \vec{w}(\vec{r}) \cdot \ed\vec{r})$ that we twist the Dirac operator by.  This one-form is, up to a constant factor, the metric dual of $\pd_\uppsi$.  More precisely, if $\rG_0(\YY_0) = \tilde{C} \pd_{\uppsi}$ then $C =  2 m \ell^2 \tilde{C}$.  One can use this to see that \eqref{G0Y} leads to \eqref{TNparams}.  We solve the problem on a $\mathbb{Z}_k$ quotient of Taub-NUT---equivalently, $k$-centered Taub--NUT with coincident centers.  (This will be useful in the framed example of the next section but here we are interested in $k=1$.)  The coordinate $x^4$ parameterizes the circle fiber with $4\pi/k$ periodicity.

In the appendix we also use a parameter $\nu$, which is the eigenvalue of $-\frac{i}{2} \Lie_{\pd_{\uppsi}} = -\frac{i}{2} \pd_{\uppsi} = -i \pd_{x^4}$, and can take integer or half-integer values.  Thus the relation between $\nu$ and the electric charge parameters is  
\begin{equation}
2\nu = - N_{{\rm e},0} = n_{\rm e}^{I_2} - \sp n_{\rm e}^{I_1}~.
\end{equation}

The kernel of the Dirac operator exhibits wall crossing, i.e. a jump in the spectrum, when the value of $C$ passes integer and half-integer values---the possible values of $\nu$.  When $|C| \leq \half$ the kernel is empty.  When $|C|$ crosses $1/2$, an $\mathfrak{su}(2)$ multiplet of spin $j = 0$ is added to the spectrum.  This continues each time $|C|$ crosses a positive half integer or integer, $|\nu| \in \half \mathbb{N}$, where now a new multiplet of spin $j = |\nu| - \half$ is created.  The sign of $C$ is also correlated with that of $\nu$, so that only one spin $j = |\nu| - \half$ multiplet is created when $|C|$ increases past a new value of $|\nu|$.  Remember, though, that due to the center of mass factor, each of these spin $j$ multiplets corresponds to a full Julia--Zee tower of spin $j$ half-hypermultiplets, where $n_{\rm e}^{I_1} \in \mathbb{Z}$ in $q_{\rm cm}$, \eqref{qcmsu3}, runs over the elements of the tower.  Thus, in the chamber $|\nu|<|C|< |\nu| + \half$ we have a Julia--Zee tower of spin $j$ hypermultiplets for each $j \in \{0,\half,1,\frac{3}{2},\ldots, |\nu|-\half \}$.  The corresponding relative electric charges of these hypermultiplets are $N_{{\rm e},0} = - \sgn(C) \times \{1,2,3,4,\ldots, 2|\nu| \}$.

\begin{figure}
\begin{center}
\includegraphics{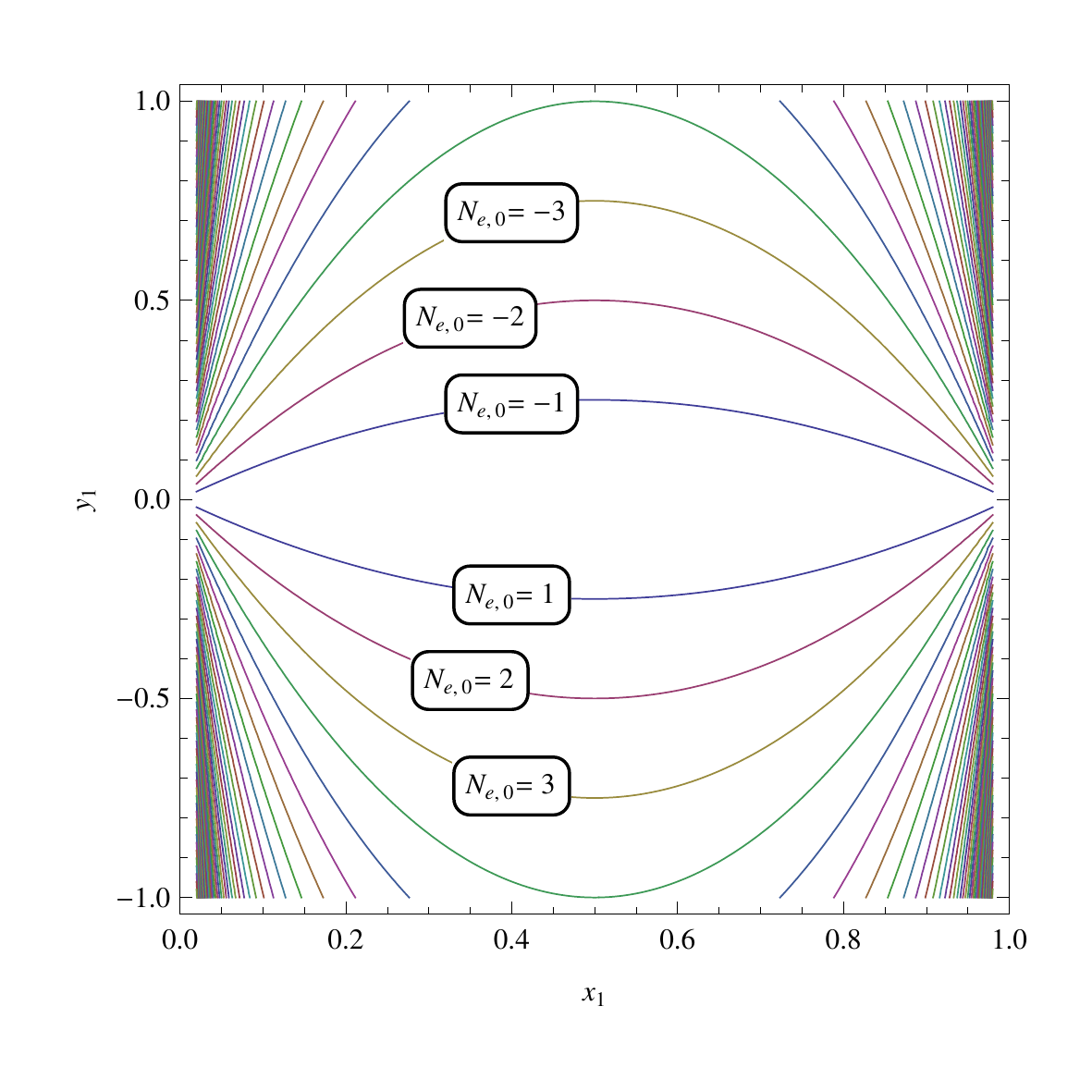}
\caption{The chamber structure for vanilla BPS states with fixed magnetic charge of the form $\gm = H_{I_1} + H_{I_2}$ and effective gauge group $\mathfrak{g}^{\rm ef} = \mathfrak{su}(3),\mathfrak{so}(5)$, or $\mathfrak{g}_2$ for $\sp = 1,2$, or $3$ respectively.  The axes are $x_1 := m_1/M = (H_1,X_\infty)/(\gm,X_\infty)$ and $y_1 := \sp(H_1, \YY_0)/(\gm,X_\infty)$.  The walls are labeled by nonzero integer values of the relative electric charge $N_{{\rm e},0}$.  Starting from the chamber in the middle where there are no BPS states (with this magnetic charge), when we cross the line labeled by $N_{{\rm e},0}$, a new tower of BPS states enters the spectrum with electric charges $\gamma_{\rm e} = n \alpha_{I_1} + (\sp n - N_{{\rm e},0}) \alpha_{I_2}$ for each $n \in \mathbb{Z}$.  Note that as $x_1$ approaches $0$, $1$, the first constituent monopole or second constituent monopole is becoming massless, respectively.  We must have $0 < x_1 < 1$ in order for $X_\infty$ to remain in the fundamental Weyl chamber.}
\label{fig1}
\end{center}
\end{figure}

To say things in a different way, if we \emph{fix} a nonzero relative electric charge $N_{{\rm e},0}$ then there is a single wall at $C = -\half N_{{\rm e},0}$.  If $N_{{\rm e},0} > 0$ then a state with this relative charge exists when $C < -\half N_{{\rm e},0}$ and does not exist when $C > -\half N_{{\rm e},0}$.  If $N_{{\rm e},0} < 0$, then the state with this charge exists when $C > -\half N_{{\rm e},0}$ and does not exists when $C < -\half N_{{\rm e},0}$.  There are no walls for $N_{{\rm e},0} = 0$ and states carrying this value of the relative charge do not exist for any value of $C$.  Figure \ref{fig1} displays the chamber structure as a function of the dimensionless ratios $x_1 := (H_{I_1},X_\infty)/(\gm,X_\infty)$ and $y_1 := \sp (\YY_0, H_1)/(\gm,X_\infty)$.

We can also summarize the content of the kernel by making use of the index characters introduced in section \ref{ssec:mathphys}.  The quantum numbers $j = \half (|N_{{\rm e},0}| - 1)$ and $m_j$ are associated with the $SU(2)$ isometry of Taub-NUT and are the relevant ones for determining the $SU(2)$ characters.  Let the $SU(2)$ character for the $n = 2j+1$ dimensional representation be $\chi_n(y) = (y^n - y^{-n})/(y - y^{-1})$.  Then for magnetic and electric charge
\begin{align}
\gm =&~ H_{I_1} + H_{I_2} ~,  \cr
\gamma_{\rm e} =&~ n \alpha_{I_1} + (\sp n - N_{{\rm e},0}) \alpha_{I_2} \quad \Rightarrow \quad [\gamma_{\rm e}]_{\rm JZ} = [ -N_{{\rm e},0} \alpha_{I_2}]_{\rm JZ}~,
\end{align}
we have the characters
\begin{equation}\label{CDiracsu3}
C_{\gm,\sx,\YY_0}^{[\gamma_{\rm e}]_{\rm JZ}}(y) = \left\{ \begin{array}{l l} \chi_{|N_{{\rm e},0}|}(y)~,~~ & \begin{array}{l} C(X_\infty,\YY_0) < -\half N_{{\rm e},0} < 0~, ~ \textrm{or} \\ C(X_\infty,\YY_0) > -\half N_{{\rm e},0} > 0~, \end{array} \\[2ex] 0~, & \textrm{otherwise}~, \end{array} \right.
\end{equation}
where $C = C(X_\infty, \YY_0)$ is the function given in \eqref{TNparams}.

%%%%%%%%%
%%%%%%%%%

The $\Lsq$ wavefunctions, $\Psi_{0}^{(N_{{\rm e},0})}$ on $\MM_0$ can be given explicitly and it is instructive to do so.  Let $\vec{r} = (r\sin{\theta}\cos{\phi},r\sin{\theta}\sin{\phi},r\cos{\theta})$ parameterize the $\mathbb{R}^3$ base with the nut at $r=0$, and let $\uppsi \sim \uppsi+2\pi$ parameterize the circle fiber.  We take corresponding gamma matrices $\gamma_{0}^{\underline{i}}$, $i=1,2,3$, and $\gamma_{0}^{\underline{4}}$, respectively, with
\begin{equation}
\gamma_{0}^{\underline{\tilde{m}}} = \left( \begin{array}{c c} 0 & \bar{\tau}^{\underline{\tilde{m}}} \\ \tau^{\underline{\tilde{m}}} & 0 \end{array} \right)~, \qquad \tau^{\underline{\tilde{m}}} = (\vec{\sigma}, -i \mathbbm{1})~, \qquad \bar{\tau}^{\underline{\tilde{m}}} = (\vec{\sigma}, i \mathbbm{1})~.
\end{equation}
In this basis the chirality operator is $\bar{\gamma}_{0} = \gamma_{0}^{\underline{1}} \gamma_{0}^{\underline{2}} \gamma_{0}^{\underline{3}} \gamma_{0}^{\underline{4}} = {\rm diag}(\mathbbm{1},-\mathbbm{1})$.  One finds that all zero modes are of positive chirality, $\Psi_0 = (\psi_+,0)^T$, with the two-component spinors $\psi_+$ given by
\begin{equation}\label{psiexpsu3}
\psi_{+,m_j}^{(N_{{\rm e},0})} = N^{(N_{{\rm e},0})}  e^{-i N_{{\rm e},0} \uppsi} \cdot \frac{ (r/\ell)^{(|N_{{\rm e},0}|-1)/2} }{\sqrt{1 + (r/\ell)}} e^{-|2C + N_{{\rm e},0}| r/(2\ell)} \cdot \breve{\psi}_{+,m_j}^{(N_{{\rm e},0})}(\theta,\phi)~,
\end{equation}
where $\breve{\psi}_{+,m_j}^{(N_{{\rm e},0})}(\theta,\phi)$ is a unit normalized spinor on $S^2$ whose explicit form can be found in the appendix.  Here $m_j$ runs from $-j$ to $j$ in integer steps, filling out a spin $j$ representation, where $j = (|N_{{\rm e},0}| - 1)/2$.  Finally, the prefactor
\begin{equation}
N^{(N_{{\rm e},0})} = \frac{ (|2C + N_{{\rm e},0}|)^{(|N_{{\rm e},0}|+1)/2} }{\ell^{3/2} \sqrt{4\pi (|N_{{\rm e},0}|!)} }~,
\end{equation}
is such that the wavefunctions are unit-normalized:
\begin{equation}
1 = \int_{\mMM_0} \ed^4 x \sqrt{g_{\MM_0}} (\Psi_{0,m_j}^{(N_{{\rm e},0})})^\dag \Psi_{0,m_j}^{(N_{{\rm e},0})} = 2\pi \int_{\mathbb{R}^3} \ed^3r H(r) (\psi_{+,m_j}^{(N_{{\rm e},0})})^\dag \psi_{+,m_j}^{(N_{{\rm e},0})}~.
\end{equation}

 In the case of a four-dimensional hyperk\"ahler manifold such as Taub--NUT, the positive chirality spinor bundle coincides with the subbundle $\SS_{1,0}$ of the Dirac spinor bundle, \eqref{bundledecomp}, on which $SU(2)_{R}$ symmetry acts trivially.  It follows that all of these BPS states are $SU(2)_R$ singlets, in line with the no-exotics theorem.  Furthermore the quantum numbers $\{j,m_j\}$ associated with the diagonal $SU(2)$ of $R$-symmetry and angular momentum can in fact be identified as angular momentum quantum numbers.
 
 As we pointed out under equation \eqref{Dhkinvariance}, wavefunctions $\Psi^{\gamma_{\rm e}}$ corresponding to properly quantized electric charge eigenvalues will automatically be well-defined on $\widetilde{\MM}/\mathbb{D}_\gi$ without the need to impose any equivariance condition.  In this case, since $\mathbb{D}_\gi = \mathbb{D}$, the wavefunctions \eqref{Psigesu3} should be well-defined on $\MM$.  Using the explicit form of $\Psi_{0}^{(N_{{\rm e},0})}$ in \eqref{psiexpsu3} together with the form of $q_{\rm cm}$ in \eqref{qcmsu3} we can see that $\Psi^{\gamma_{\rm e}}$ indeed respects the identification \eqref{Dquotientsu3}. 

\begin{figure}
\begin{center}
\includegraphics{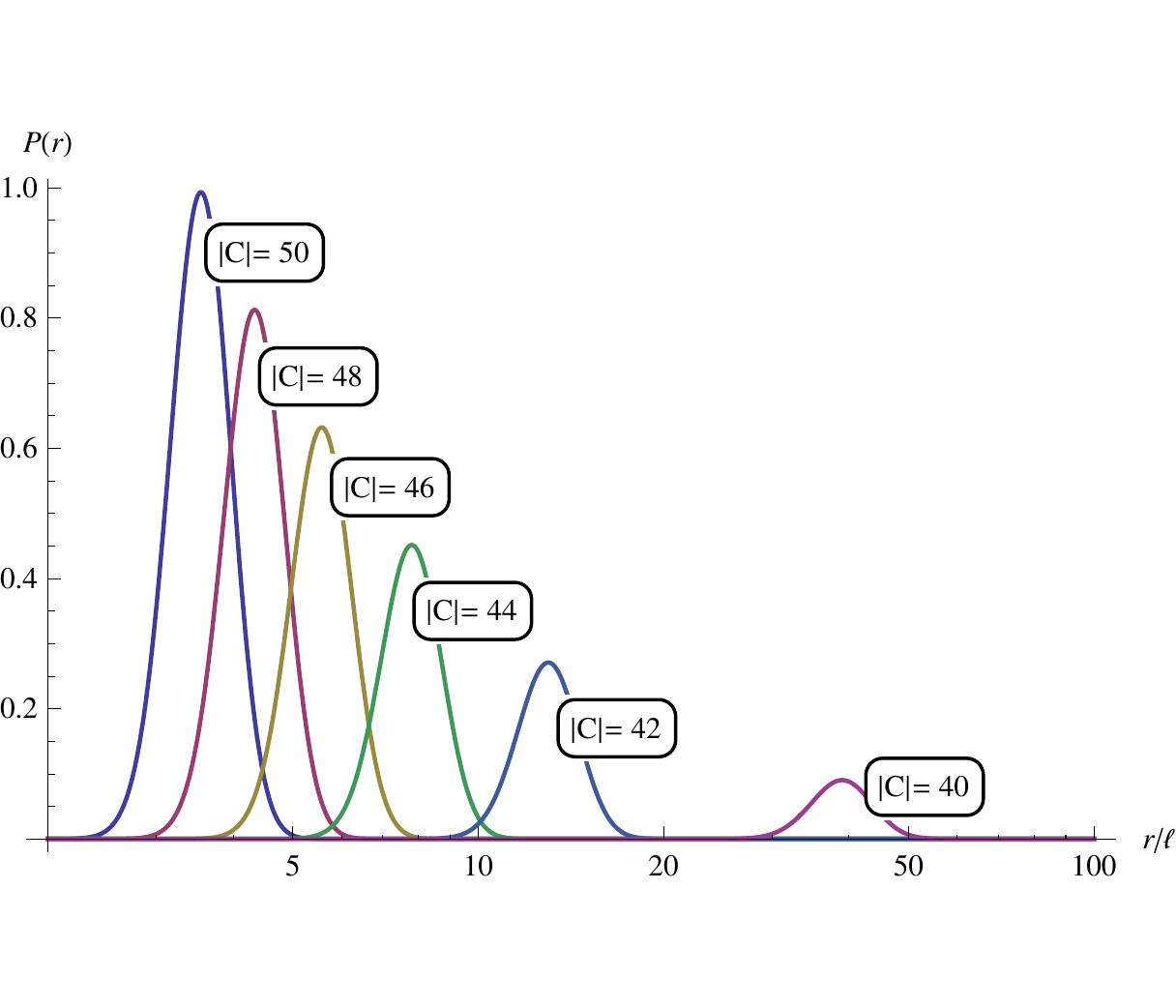}
\caption{Bound state radial probability functions on Taub--NUT, for relative electric charge $|N_{{\rm e},0}| = 78$.  We consider different values of $|C|$ starting at $50$ and approaching $\half |N_{{\rm e},0}| = 39$.  As $|C|$ approaches the wall at $\half |N_{{\rm e},0}|$ we can see the corresponding probability function move out to large values of $r/\ell$ and spread out.  We plotted the radial distance on a log scale in order to aid visualization; the spreading is quite severe as the wall is approached.}
\label{fig2}
\end{center}
\end{figure}

The exponential damping factor, $e^{-|2C + N_{{\rm e},0}|r/(2\ell)}$, controls normalizability.  As $C \to -\half N_{{\rm e},0}$, \ie\ as we approach the wall for the states with relative charge $N_{{\rm e},0}$ from the side where they exist, we can literally see these bound states leave the spectrum and merge with the continuum.  One way to quantify this is to define the radial probability density
\begin{equation}
P^{(N_{{\rm e},0})}(r) := 2\pi r^2 H(r) \int_{S^2} \ed\Omega_2 (\psi_{+,m_j}^{(N_{{\rm e},0})})^\dag \psi_{+,m_j}^{(N_{{\rm e},0})}~,
\end{equation}
such that $\int_{0}^{\infty} \ed r P(r) = 1$ and compute its extremum, $P'(r_{\rm ext}) = 0$.  This is given by 
\begin{equation}\label{rext}
r_{\rm ext} = \frac{|N_{{\rm e},0}|}{|2C + N_{{\rm e},0}|} \ell = - \frac{N_{{\rm e},0}}{2C + N_{{\rm e},0}} \ell~,
\end{equation}
where we used that either $C < -\half N_{{\rm e},0} < 0$ or $C > -\half N_{{\rm e},0} > 0$ in order for bound states to exist.  We see that $r_{\rm ext} \to \infty$ as $C \to -\half N_{{\rm e},0}$.  We can do better, however, and compute the radial expectation value exactly:
\begin{equation}\label{rexp}
r_{\rm bnd} := \langle \Psi_{0,m_j}^{(N_{{\rm e},0})} | r |  \Psi_{0,m_j}^{(N_{{\rm e},0})} \rangle = \int_{\mMM_0} \ed^4 x \sqrt{g_{\MM_0}} \, ( \Psi_{0,m_j}^{(N_{{\rm e},0})})^\dag r \,  \Psi_{0,m_j}^{(N_{{\rm e},0})} =  \frac{(|N_{{\rm e},0}| + 1)}{|2C + N_{{\rm e},0}|} \ell~.
\end{equation}
This agrees with $r_{\rm ext}$ for large relative charge, $|N_{{\rm e},0}|$.  We plot $P(r)$ for several values of $|C|$ approaching $\half |N_{{\rm e},0}|$ from above in figure \ref{fig2}.

%%%%%%%%%
%%%%%%%%%

Next we compare these expressions for the walls of marginal stability and bound state radii with the corresponding expressions from the low energy Seiberg--Witten analysis.

%%%%%%%%%%%%%%%%%%%%
\subsection{Comparison with low energy effective theory}
%%%%%%%%%%%%%%%%%%%%

We work in the weak coupling regime of the Coulomb branch and apply the math-physics map of parameters.  The map depends on the electromagnetic charge we wish to consider and is given by
\begin{equation}
X_\infty = \Im(\zeta_{\rm van}^{-1}(u,\gamma) a(u))~, \qquad \YY_{\infty} = \Im(\zeta_{\rm van}^{-1}(u,\gamma) a_{\rm D}(u))~,
\end{equation}
where the duality frame which we use to evaluate $a(u)$ is such that $X_\infty \in W_{\mathfrak{t}}^+$ and the corresponding charge trivialization maps, ${\rm m} : \Gamma_u \to \Lambda_{\rm cr}$ and ${\rm e} : \Gamma_\mu \to \Lambda_{\rm wt}$, are such that ${\rm m}(\gamma) = \gm$ and ${\rm e}(\gamma) = \gamma_{\rm e}$.  The map implies the relation $(\gm, \sy) + \langle \gamma_{\rm e}, \sx \rangle = 0$.  We use this to fix the component of $\sy$ along $X_\infty$, writing $\YY_\infty = -\frac{\langle \gamma_{\rm e}, X_\infty \rangle}{(\gm,X_\infty)} X_\infty + \YY_0$, where $\YY_0$ satisfies $(\gm, \YY_0) = 0$.  

Given these identifications, we argued that the walls of marginal stability, \eqref{vanillawalls}, are given by \eqref{vanmathwalls}:
\begin{equation}\label{vmw2}
(\gamma_{1,{\rm m}},\YY_0) + \left\langle \gamma_{1,{\rm e}} - \frac{\langle \gamma_{1,{\rm e}} + \gamma_{2,{\rm e}}, X_\infty \rangle}{(\gamma_{1,{\rm m}} + \gamma_{2,{\rm m}}, X_\infty)} \gamma_{1,{\rm m}}^\ast ~,~ \sx \right\rangle = 0~,
\end{equation}
where $\gamma_{1},\gamma_{2}$ is any pair of charges such that $\gamma_1 + \gamma_2 = \gamma$, $\llangle \gamma_1,\gamma_2\rrangle \neq 0$, and BPS states exist in the spectrum for these charges.  As we pointed out under \eqref{vanmathwalls}, walls for which $X_\infty \in W_{\mathfrak{t}}^+$ require that both constituents have non-zero magnetic charge.  Hence the full set of constituents corresponding to $\gm = H_{I_1} + H_{I_2}$ for which all of these conditions hold is
\begin{equation}
\gamma_1 = \gamma_{1,{\rm m}} \oplus \gamma_{1,{\rm e}} = H_{I_1} \oplus n_{\rm e}^{I_1} \alpha_{I_1} \qquad \& \qquad \gamma_2 = \gamma_{2,{\rm m}} \oplus \gamma_{2,{\rm e}}  = H_{I_2} \oplus n_{\rm e}^{I_2} \alpha_{I_2}~,
\end{equation}
where we require
\begin{equation}\label{conpairing}
\llangle \gamma_1, \gamma_2\rrangle = n_{\rm e}^{I_1} \langle \alpha_{I_1}, H_{I_2} \rangle - n_{\rm e}^{I_2} \langle \alpha_{I_2}, H_{I_1} \rangle = n_{\rm e}^{I_2} - \sp n_{\rm e}^{I_1} = - N_{{\rm e},0} \neq 0~.
\end{equation}
The constituents are members of dyon cohorts for any $n_{\rm e}^{I_{1,2}} \in \mathbb{Z}$ and are present in the spectrum throughout the weak coupling regime.

Now let us compute the second term of \eqref{vmw2}.  We find
\begin{align}\label{ge1perpX}
& \left\langle n_{\rm e}^{I_1} \alpha_{I_1} - \frac{\langle n_{\rm e}^{I_1} \alpha_{I_1} + n_{\rm e}^{I_2} \alpha_{I_2}, X_\infty \rangle}{(H_{I_1} + H_{I_2}, X_\infty)} H_{I_1}^\ast ~,~ \sx \right\rangle \cr
&  = \frac{1}{(\gm,X_\infty)} \bigg\{ n_{\rm e}^{I_1} \left[ \langle \alpha_{I_1}, X_\infty\rangle (\gm, X_\infty) - \langle \alpha_{I_1}, X_\infty\rangle (H_{I_1}, X_\infty) \right] - n_{\rm e}^{I_2} \langle \alpha_{I_2}, X_\infty \rangle (H_{I_1}, X_\infty) \bigg\} \cr
&  =  \frac{1}{(\gm,X_\infty)} \bigg\{ n_{\rm e}^{I_1} \langle \alpha_{I_1}, X_\infty \rangle (H_{I_2},X_\infty) - n_{\rm e}^{I_2} \langle \alpha_{I_2}, X_\infty \rangle (H_{I_1}, X_\infty) \bigg\} \cr
& = \frac{ (H_{I_1}, X_\infty) (H_{I_2}, X_\infty)}{\sp (\gm, X_\infty)} (\sp n_{\rm e}^{I_1} - n_{\rm e}^{I_2})~.
\end{align}
Thus observe that the walls of marginal stability, \eqref{vmw2}, agree perfectly with the walls where the Dirac kernel jumps:
\begin{align}
& (\gamma_{1,{\rm m}},\YY_0) + \left\langle \gamma_{1,{\rm e}} - \frac{\langle \gamma_{1,{\rm e}} + \gamma_{2,{\rm e}}, X_\infty \rangle}{(\gamma_{1,{\rm m}} + \gamma_{2,{\rm m}}, X_\infty)} \gamma_{1,{\rm m}}^\ast ~,~ \sx \right\rangle = 0  \cr
\iff \qquad & (H_{I_1}, \YY_0) + \frac{ (H_{I_1}, X_\infty) (H_{I_2}, X_\infty)}{\sp (\gm, X_\infty)} (\sp n_{\rm e}^{I_1} - n_{\rm e}^{I_2}) = 0 \cr
\iff \qquad & \sp (H_{I_1}, \YY_0) \frac{(\gm, X_\infty)}{(H_{I_1},X_\infty) (H_{I_2}, X_\infty)} + (\sp n_{\rm e}^{I_1} - n_{\rm e}^{I_2}) = 0 \cr
\iff \qquad & 2C + N_{{\rm e},0} = 0~,
\end{align}
including the fact that there is no wall when $N_{{\rm e},0} = 0$.

In section \ref{ssec:vBPSspace} we emphasized that the identification \eqref{mainres2} implies that the weak coupling regime of the $2r$ real dimensional Coulomb branch is foliated by surfaces of real dimension $2(r-d+1)$, $d \equiv \rnk{\mathfrak{g}^{\rm ef}}$, along which the spectrum is invariant.  In the case at hand this leaves $2(d-1) = 2$ directions on which the BPS spectrum can depend; these directions can be taken as $x_1 = x_1(u,\gamma)$ and $y_1 = y_1(u,\gamma)$ used in figure \ref{fig1}, for fixed overall scale of $\sx,\YY_0$.  Hence this figure, pulled back to the Coulomb branch via the math-physics map, fully captures the chamber structure of the weak coupling regime for all BPS states with $\gm = H_{I_1} + H_{I_2}$.

The kernel of the Dirac operator is also consistent with the predictions of vanilla wall crossing formulae.  The constituent charges $\gamma_{1,2}$ are primitive, so we can apply the primitive wall crossing formula \eqref{primwc}.  Being members of dyon cohorts, they have protected spin characters $\Omega(u,\gamma_{1,2};y) = 1$ throughout the weak coupling regime.  Hence after crossing the wall $\widehat{W}(\gamma_1,\gamma_2)$, the protected spin character of the newly created BPS space $(\HH_{0}^{\rm BPS})_{u,\gamma}$ of charge $\gamma = \gamma_1 + \gamma_2$ should be
\begin{equation}
\Omega(u,\gamma;y) = \chi_{|\llangle \gamma_1, \gamma_2\rrangle|}(y) = \chi_{|N_{{\rm e},0}|}(y)~,
\end{equation}
using \eqref{conpairing}.  This agrees precisely with the corresponding index character of the Dirac operator, \eqref{CDiracsu3}, as it should according to \eqref{inCequalsPSC}.

Finally, let us consider the Denef bound state radius, \eqref{rDenef}.  As a first step we compute
\begin{align}
\Im\left(\zeta_{\rm van}^{-1}(u,\gamma) Z_{\gamma_1}(u) \right) =&~ \left( \gamma_{1,{\rm m}}, \Im(\zeta^{-1}_{\rm van} a_{\rm D}) \right) + \langle \gamma_{1,{\rm e}}, \Im(\zeta_{\rm van}^{-1} a) \rangle \cr
=&~ (\gamma_{1,{\rm m}}, \YY_\infty) + \langle \gamma_{1,{\rm e}}, X_\infty \rangle \cr
=&~ (H_1, \YY_0) - \bigg\langle H_{1}^\ast, \frac{\langle \gamma_{\rm e}, X_\infty \rangle}{(\gm,X_\infty)} X_\infty \bigg\rangle + n_{\rm e}^{I_1} \langle \alpha_{I_1}, X_\infty \rangle \cr
=&~ (H_1, \YY_0) + \frac{ (H_{I_1}, X_\infty) (H_{I_2}, X_\infty)}{\sp (\gm, X_\infty)} (\sp n_{\rm e}^{I_1} - n_{\rm e}^{I_2})~,
\end{align}
where in the last step we plugged in \eqref{ge1perpX}.  Now, using \eqref{TNparams} and setting $\sp n_{\rm e}^{I_1} - n_{\rm e}^{I_2} = N_{{\rm e},0}$, we find that this can be expressed as
\begin{equation}
\Im\left(\zeta_{\rm van}^{-1}(u,\gamma) Z_{\gamma_1}(u) \right) = \frac{2 C \mu}{\sp} + \frac{\mu}{\sp} N_{{\rm e},0} = \frac{\mu}{\sp} (2C + N_{{\rm e},0})~.
\end{equation}
Employing also \eqref{conpairing}, we then see that
\begin{align}
r_{\rm Denef} =&~ \frac{ \langle \gamma_1, \gamma_2 \rangle }{2 \Im(\zeta_{\rm van}^{-1} Z_{\gamma_1})} = - \frac{N_{{\rm e},0}}{(2C + N_{{\rm e},0})} \left( \frac{\sp}{2\mu} \right)  = - \frac{N_{{\rm e},0}}{(2C + N_{{\rm e},0})} \ell \cr
=&~ r_{\rm ext} ~,
\end{align}
a perfect agreement with the extremal radius \eqref{rext}!  Note $r_{\rm Denef}$ only agrees with the expectation value, \eqref{rexp}, in the limit of large $|N_{{\rm e},0}|$.  This is expected since the Denef formula is a classical result (in the low energy effective theory) that only holds in the limit of large charges (see also \cite{Denef:2002ru}).

%%%%%%%%%%%%%%%%%%%%%%
%%%%%%%%%%%%%%%%%%%%%%
\section{Framed example: singular monopoles in $\mathfrak{su}(2)$ SYM}\label{Section:FramedEx}
%%%%%%%%%%%%%%%%%%%%%%
%%%%%%%%%%%%%%%%%%%%%%

Now we apply the formalism discussed in this paper to a concrete \emph{framed} example where the explicit semiclassical description is known, namely the case of a single smooth monopole bound to a pure 't Hooft defect in the $\mathfrak{g} = \mathfrak{su}(2)$ theory. First we briefly review the spectrum of framed BPS states in the low energy $\mathfrak{su}(2)$ theory obtained in \cite{Gaiotto:2010be}.  We extend that analysis to obtain the complete line defect generating functional \eqref{Fgenfun} (with full $y$ dependence) for the case of pure 't Hooft defects in the weak coupling regime. We then discuss the explicit classical solutions of \cite{Cherkis:2007jm,Cherkis:2007qa}, (see also \cite{Blair:2010vh}), describing a single smooth $\mathfrak{su}(2)$ monopole bound to a pure 't Hooft defect, and their moduli space. The Dirac operator and its kernel are constructed and we show how the semiclassical framed BPS states defined this way match perfectly with the prediction from \cite{Gaiotto:2010be}. The result for the line defect generating funcational also contains predictions for the kernels of twisted Dirac operators on an infinite family of hyperk\"ahler manifolds.  We describe these predictions in detail for a class of eight-dimensional $\fMM$'s, members of which have been considered in \cite{Cherkis:1997aa,Dancer:1992kn,Dancer:1992km,Houghton:1997ei,Houghton:1999qu}.  We close with a discussion of how the concept of `tropical labels' can be understood semiclassically.

%%%%%%%%%%%%%%%%%%%%%%
\subsection{Framed BPS states in the $\mathfrak{su}(2)$ theory}\label{sec:fbps}
%%%%%%%%%%%%%%%%%%%%%%

When the gauge algebra is $\mathfrak{su}(2)$ the line defect lattice $\LL$ \eqref{lineoplattice} will be a sublattice of $(\Lambda_{\mathrm{mw}}\oplus\Lambda_{\mathrm{wt}})/\mathbb{Z}_2$, where the magnetic weight lattice is generated by $\frac{1}{2}H_\a$ and the weight lattice by $\frac{1}{2}\alpha$. This means we can represent the charges of any line defect by a pair of integers $(p,q)$ such that:
\begin{equation}\label{su2opcharges}
P \oplus Q=   \frac{p}{2}H_\a \oplus \frac{q}{2}\a ~.
\end{equation}
In this way we establish an isomorphism $\Lambda_{\rm mw} \oplus \Lambda_{\rm wt} \cong \mathbb{Z} \times \mathbb{Z}$, with $P\oplus Q \mapsto \{p,q\}$.  The action of the $\mathbb{Z}_2$ Weyl group is simply $\{p,q\}\rightarrow \{-p,-q\}$ and so we should physically identify such charges.  Mutual locality of line defects determines the possible sublattices by requiring the symplectic pairing to be even \cite{Kapustin:2005py, Gaiotto:2010be, Aharony:2013hda}: $q_1p_2-q_2p_1\in 2\mathbb{Z}$.  There are three inequivalent possibilities:
\begin{align}
& \LL_1 \cong (2\mathbb{Z}\times \mathbb{Z})/\mathbb{Z}_2 \,,\\
& \LL_2 \cong  (\mathbb{Z}\times 2\mathbb{Z})/\mathbb{Z}_2 \,,\\
& \LL_3 \cong \left[ \left(2\mathbb{Z}\times 2\mathbb{Z}\right)\cup\left( (2\mathbb{Z}+1)\times (2\mathbb{Z}+1)\right) \right]/ \mathbb{Z}_2 \,.
\end{align}
The first corresponds to the $G = SU(2)$ theory.  The latter two correspond to $G = SO(3)$ theories (denoted $SO(3)_{\pm}$ in \cite{Aharony:2013hda}); they are mapped into each other via the Witten effect for line defects under $\theta_0 \to \theta_0 + 2\pi$.  In particular, $\theta_0$ should be taken $4\pi$ periodic in $SO(3)$ gauge theory, where instanton winding numbers are multiples of $1/2$.  In the following we will restrict ourselves to pure 't Hooft defects for which $q = 0$ and the lattices $\LL_1$ and $\LL_3$ are indistinguishable.  

We will work in the weak coupling duality frame defined by $\langle \alpha, X_\infty \rangle > 0$, where $X_\infty =  \Im(\zeta^{-1} a(u))$.  We identify the vanilla lattice according to $\Gamma_u \cong \Lambda_{\rm cr} \oplus \Lambda_{\rm wt}$, and we note that electric charges will be confined to the sublattice $\Lambda_{\rm rt} \subset \Lambda_{\rm wt}$.  Then the IR charges of framed BPS states can be labeled by integers $\{\tilde{n}_{\rm m}, n_{\rm e}\}$ such that
\begin{equation}\label{genIRcharge}
\gamma_{\mathrm{m}} \oplus \gamma_{\mathrm{e}} =  (\tilde{n}_{\rm m} - \frac{{}_{|p|}}{{}^{2}} )H_\a \oplus n_{\rm e} \alpha  \in (\Lambda_{\rm cr} + P) \oplus \Lambda_{\rm rt} ~.
\end{equation}
Note that the torsor $\Lambda_{\rm cr} + P$ only depends on the Weyl orbit of $P$, and that $\tilde{n}_{\rm m}$ gives the relative magnetic charge:
\begin{equation}
\tilde{\gamma}_{\rm m} \equiv \gamma_{\rm m} - P^- = \gm + \frac{|p|}{2} H_\alpha = \tilde{n}_{\rm m} H_\alpha~.
\end{equation}

Now consider the line defect generating functional, \eqref{Fgenfun}.  For a generic IR charge $\gamma = \gm \oplus \gamma_{\rm e}$ of the form \eqref{genIRcharge}, we have
\begin{equation}
X_{\gamma} = y^{n_{\rm e}(2\tilde{n}_{\rm m}-|p|)}  X_{\gm} X_{\gamma_{\rm e}} = y^{n_{\rm e}(2\tilde{n}_{\rm m}-|p|)}X_{1}^{\tilde{n}_{\rm m} - |p|/2} X_{2}^{n_{\rm e}}~, \qquad \textrm{where} \quad \left\{ \begin{array}{l} X_1 := X_{H_\alpha} ~, \\ X_2 := X_\alpha ~. \end{array} \right.
\end{equation}
Hence the generating functional takes the form
\begin{equation}\label{generalF}
F(L_{\zeta}(p,0),u,\{X_1,X_2\};y) = X_{1}^{-|p|/2} \sum_{\mathclap{\tilde{n}_{\rm m},n_{\rm e} \in \mathbb{Z}}} ~ \fOmega\left(L_\zeta(p,0),u,(\gm,\gamma_{\rm e});y \right)y^{n_{\rm e}(2\tilde{n}_{\rm m}-|p|)} X_{1}^{\tilde{n}_{\rm m}} X_{2}^{n_{\rm e}} ~.
\end{equation}

The dependence on the Coulomb branch parameter $u$ is piecewise constant and set by wall crossing. One of the main observations of \cite{Gaiotto:2010be}, reviewed in section \ref{sec:corehalo}, is that the wall crossing of the framed BPS spectrum can be completely characterized in terms of the creation or annihilation of halos of particles that are themselves elements of the vanilla spectrum. In this case the vanilla theory is the pure $\NN=2$, $\mathfrak{su}(2)$ Yang--Mills theory,\footnote{The pure glue vanilla theory does not depend on the global form of the Lie group.} and the spectrum is well known.  In the weak coupling regime of interest, it contains two electrically charged vector multiplets (the massive $W$-bosons) and a dyon cohort of hypermultiplets:
\begin{equation}\label{vanBPSspectrum}
\mathbf{\mathrm{vm}}:\ \gamma=\pm\alpha\,,\quad \mathbf{\mathrm{hm}}:\  \gamma=\pm H_\a \oplus n\a\,, ~ n\in\mathbb{Z} \,.
\end{equation}
These states exist throughout the weak coupling region of the Coulomb branch.  

We can consider the framed marginal stability walls \eqref{fmsw} associated with each, taking each of these charges to be a halo charge.  We postpone discussion of the vector multiplet walls until section \ref{ssec:tropical}, and focus here on the hypermultiplets.  Consider the halo charge $\gamma_{\rm h} =\gamma_n \equiv H_\a \oplus n\a$.  (There is an analogous story for the charge conjugate states $\gamma_{\rm h} = (- H_\alpha) \oplus (- n\alpha)$ in the duality frame where $\langle \alpha, \sx \rangle < 0$.)  The walls are determined by the marginal stability condition $\zeta^{-1} Z_{\gamma_{\rm h}}(u) \in \mathbb{R}_-$.  Making use of the math-physics map, \eqref{mathgmap}, \eqref{mathxy}, the vanishing of the imaginary part of $\zeta^{-1} Z_{\gamma_{\rm h}}$ is equivalent to $(H_\alpha, \sy) + n \langle \alpha, \sx \rangle = 0$.  Meanwhile, negativity of the real part is automatic if we take the dynamical scale small enough, according to the lemma in \ref{wclemma}.  Thus in the weak coupling regime, we have the walls $\widehat{W}_n \equiv \widehat{W}(\gamma_n)$ for a framed BPS state of charge $\gamma$ given by
\begin{equation}\label{walls}
\widehat{W}_n = \left\{(u,\zeta) \in \widehat{B} \times \widehat{\mathbb{C}}^\ast ~|~  (H_\alpha, \sy(u,\zeta)) + n \langle \alpha, \sx(u,\zeta) \rangle = 0 \right\}~.
\end{equation}
Note however that if the core is purely magnetic, as is the case here, then the $n=0$ case is an invisible wall in the language of footnote \ref{fn:invisible}.  Therefore we do not include $\widehat{W}_0$ in our set of physical walls in the following.

\begin{figure}
\begin{center}
\includegraphics{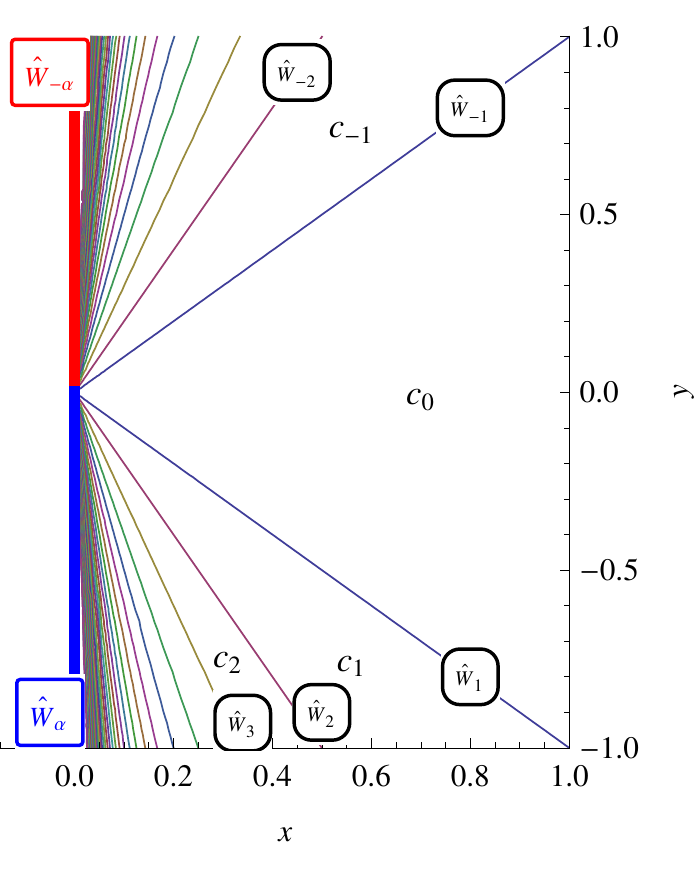}
\caption{BPS walls in $X_\infty$-$\YY_\infty$ space, with $x := \langle \alpha, X_{\infty}\rangle$ and $y := (H_\alpha, \YY_{\infty})$.  We restrict $X_\infty$ to the fundamental Weyl chamber so that $x > 0$ and only the walls for dyon cohorts with $+H_{\alpha}$ magnetic charge are visible.  A mirror reflection about $x = 0$ will describe the walls and chamber structure for cohorts with magnetic charge $-H_{\alpha}$.  The walls for the vector multiplets will be discussed further in section \ref{ssec:tropical}.}\label{wallfig}
\end{center}
\end{figure}

These walls are easily visualized in the $\sx$-$\sy$ plane; see figure \ref{wallfig}.  They can be mapped to $(u,\zeta)$ space via \eqref{mathxy}.  Note that this map is defined in a specific weak coupling duality frame which does not extend into the strong coupling region of the Coulomb branch; see discussion around item \ref{wcRegime} in section \ref{ssec:fBPSspace}.  Smaller $\sx,\sy$ can be accommodated by choosing smaller values of the dynamical scale in accord with the lemma in \ref{wclemma}.  Since $\widehat{W}_0$ is not a physical wall, let us denote the \emph{single} chamber between $\widehat{W}_{\pm1}$ as $c_0$.  (We use the same name for the chambers in $X_\infty$-$\YY_\infty$ space and their premiages in $\widehat{\BB}_{\rm wc} \times \widehat{\mathbb{C}^\ast}$.)  We define the chamber $c_n$, $n\neq 0$, to be the chamber between $\widehat{W}_{n}$ and $\widehat{W}_{n+1}$ when $n>0$ and the chamber between $\widehat{W}_n$ and $\widehat{W}_{n-1}$ when $n$ is negative.  As the framed BPS spectrum is constant in such chambers we can define
\begin{equation}
F_n(p,\{X_1,X_2\};y) := F(L_\zeta(p,0),u,\{X_1,X_2\};y) \big|_{(u,\zeta) \in c_n} .
\end{equation}
%  

%%%%%%%%%%%%%%%%%%
%%%%%%%%%%%%%%%%%%

A closed-form expression for precisely this quantity was obtained in \cite{Gaiotto:2010be} in the special case $y=1$, where $X_{1,2}$ become commuting variables, by making use of an alternative formulation of framed BPS states inspired by connections to the six-dimensional $(2,0)$ theory compactified on a Riemann surface. Adapted to the notations here\footnote{To be precise, $X_{\mathrm{GMN}}=(X_1X_2^n)^{-1},\ Y_{\mathrm{GMN}}=X_1X_2^{n+1},\ n_{\mathrm{GMN}}=n+1,\ \alpha_{\mathrm{GMN}}=f_n$ and we used the Chebyshev relation $T_{n+1}(x)=xU_n(x)-U_{n-1}(x)$\,.}, the result is
\begin{equation}\label{mfgenfun}
F_n(p,\{X_1,X_2\};y=1)=\left[X_1^{-1/2}X_2^{-n/2}\left( U_{n}(f_{n})-X_2^{-1/2}U_{n-1}(f_{n})\right)\right]^{|p|}\,,
\end{equation}
where
\begin{equation}
f_n:=\frac{X_2^{1/2}+X_2^{-1/2}\left(1+X_1X_2^{n+1}\right)}{2}
\end{equation}
and we used the Chebyshev polynomials defined as
\begin{equation}
U_{-1}(x):= 0\,,\qquad U_0(x):= 1\,,\qquad U_{n+1}(x):=2x U_n(x)-U_{n-1}(x)\,.
\end{equation}
The formula \eqref{mfgenfun} is valid for the chambers $c_n$ with $n\geq 0$, and we will focus on this case for the time being.  

This generating function \eqref{mfgenfun} contains all the framed BPS `indices'\footnote{The actual indices correspond to $y = -1$.  However, by the no-exotics theorem, the $\fOmega$ at $y=1$ give the actual dimensions of framed BPS Hilbert spaces.} for any pure 't Hooft defect in the $\mathfrak{su}(2)$ theories, in the weak coupling chambers $c_n$, $n\geq 0$. As we show in appendix \ref{wcapp}, the dependence of the generating function on the chamber number $n$ satisfies the wall crossing formula \eqref{genWCtransfo} which in this case takes the particular form
\begin{equation}
F_{n}(p,\{X_1,X_2\};y=1) = F_{n-1}(p,\{(1+X_1X_2^n)^{-2n}X_1,(1+X_1X_2^n)^{2}X_2\};y=1)\,.\label{relationwalls}
\end{equation}
The generating function \eqref{mfgenfun} provides a wide number of predictions for the realization of the framed BPS states as zero modes of a Dirac operator on moduli space. There are a few conclusions that easily follow---the simplest, obtained from inspection of \eqref{mfgenfun}, being that
\begin{equation}
F_0(p,\{X_1,X_2\};y=1) = X_1^{-|p|/2}\,.\label{Fnzero}
\end{equation} 
Then observe that, when $X_2=0$, this extends via the identity \eqref{relationwalls} to all positive chambers:
\begin{equation}\label{Fx2zero}
F_n(p,\{X_1,0\};y=1) = X_1^{-|p|/2}\, , \qquad (n \geq 0)~.
\end{equation}
Let us compare this result with the general expression \eqref{generalF}.  The fact that $F_n$ is finite when $X_2 \to 0$ implies that there are no framed BPS states carrying negative electric charge, \ie\ $\fOmega(n_{\rm e} < 0;1) = 0$, in the chambers $c_n$, $n\geq 0$.  Furthermore there is a single framed BPS state carrying zero electric charge---it is the pure 't Hooft defect with zero relative magnetic charge, $\tilde{n}_{\rm m} = 0$, and it is present in all weak coupling chambers.  This is consistent with our discussion in section \ref{ssec:vanlocus}.  

Similarly, in the limit $X_1\rightarrow 0$ one can easily extend \eqref{Fnzero} to $n\geq 0$ via \eqref{relationwalls}, leading to
\begin{equation}\label{Fx1zero}
 \lim_{X_1\rightarrow 0}  X_{1}^{|p|/2}  F_n(p,\{X_1,X_2\};1)  =\lim_{X_1\rightarrow 0}  X_{1}^{|p|/2}  F_0(p,\{X_1,X_2\};y=1) = 1\,.
\end{equation}
This implies that there are no framed BPS states with $\tilde{n}_{\rm m} <0$.  This is in perfect agreement with our conjecture from \cite{MRVdimP1} that the moduli space of singular monopoles should be empty in this case. 

As a last observation one can inspect the large $X_1$ limit:
\begin{equation}\label{largeX1}
F_n(p,\{X_1,X_2\};y=1) = X_1^{|p|(n-1/2)}X_2^{|p|n^2} \left( 1 + O(X_{1}^{-1}) \right)~.
\end{equation}
This means that in $c_n$ there is a unique state of maximal relative magnetic charge, $\tilde n_{\mathrm{m}}=|p|n$, that furthermore has electric charge $n_\mathrm{e}=|p|n^2$.  This implies that for any $\sx,\sy$ in $c_n$ there exists a moduli space $\fmMM$ of maximal dimension $4|p|n$ on which the Dirac operator \eqref{fMdiracop} has a nontrivial kernel, and that in this maximal case there is a unique zero mode!

It is beyond the scope of this work to review the techniques of \cite{Gaiotto:2010be} that led to the result \eqref{mfgenfun}.  Since the result is important, however, we give an alternative derivation via the core-halo picture of wall crossing.  This will also be helpful for extracting the remaining BPS degeneracies of interest and furthermore will allow us to obtain the answer for arbitrary $y$.  In order to apply this method we need to know the exact spectrum in a starting chamber.  Fortunately we know this on the vanishing locus $\sy = 0$ discussed in section \ref{ssec:vanlocus}, where the exact spectrum consists of the pure 't Hooft defect only.  This locus sits in the middle of $c_0$ and hence we know that the pure 't Hooft defect is the only framed BPS state in this entire chamber.  (This is, by the way, consistent with the $n=0$ case of the formula \eqref{mfgenfun}, but we did not need to use that formula to infer it.)

We can now infer the spectrum in the chambers $c_n$, $n > 0$ by creating new halos of dyons with halo charges $\gamma_n = H_\alpha \oplus n \alpha$ when going from $c_{n-1}$ to $c_{n}$. What are all bound states that have total relative magnetic charge $\tilde{n}_{\rm m}$? As each vanilla particle that can bind has exactly magnetic charge $H_\alpha$, we are considering bound states of $\tilde{n}_{\rm m}$ particles to the core. Each particle has an electric charge, which can run from $\alpha$ to $n \alpha$ in the $n$'th chamber. Due to the order of the chambers the particles with lower electric charge bind first.  So the possible bound states of $\tilde{n}_{\rm m}$ particles in chamber $c_n$ will be of the form 
\begin{equation}
\left\{r_{n}\gamma_{n},\left\{r_{n-1}\gamma_{n-1},\left\{\ldots ,\left\{r_1 \gamma_{1},\gamma_{\rm c} \right\} \right\} \right\} \right\}~,
\end{equation}
where $\vec{r} = (r_i)$ is a collection of nonnegative integers and we impose the constraint
\begin{equation}\label{magnorm}
\lVert \vec{r}\, \rVert_{\rm m} := \sum_i r_i = \tilde{n}_{\rm m}~.
\end{equation}
Here $\gamma_{\rm c}$ is the (IR) charge of the initial core particle in chamber $c_0$, corresponding to the pure 't Hooft defect: $\gamma_{\rm c} = -\frac{|p|}{2} H_\alpha$.  The notation $\{a,\{b,c\}\}$ means that $a$ binds as a halo to a core which itself is a halo of $b$ bound to $c$, \etc. Note that some of the $r_i$ can be zero.

The degeneracy of such bound configurations can be easily derived using the principles of \cite{Denef:2002ru,Denef:2007vg}. For two bound pointlike dyons of charges $\gamma_{1,2}$ the number of states is simply an angular momentum multiplet: $2J_{12}+1$, where the angular momentum vector is directed along the line connecting the dyons and has magnitude
\begin{equation}
J_{12}=\frac{|\llangle \gamma_1,\gamma_2\rrangle|-1}{2} ~.  \label{angmomtdyons}
\end{equation}
A halo then consists of $r_i$ fermionic particles distributed among these states, giving a binomial degeneracy. Finally we should take into account that the core itself carries internal degrees of freedom (it is itself a `sub-halo') and multiply by those. It then follows that, for the configuration of halos described above and specified by the vector $\vec{r} = (r_i)$, the number of states gained\footnote{Note that if this number is negative this should be interpreted as the number of states {\it lost}. This can happen if the top entry in a binomial in $N_n(p,\vec{r};y=1)$ becomes negative at non-zero $r_i$, which requires $\tilde n_{\rm m}\geq 2$. In this case halos that were not created at the previous wall, dissapear. This indicates a transformation of the bound state structure of the states inside the chamber. It would be interesting to understand the physics of this phenomenon in some more detail.} in crossing the wall from $c_{n-1}$ to $c_n$, $n > 0$, is
\begin{equation}\label{numofstates}
N_n(p,\vec{r};y=1\,) :=\prod_{i=1}^{n}\begin{pmatrix}
\left\llangle \gamma_c+\sum_{j=1}^{i-1}r_j\gamma_j~, \gamma_i \right\rrangle \\ r_i
\end{pmatrix}=\prod_{i=1}^{n}\begin{pmatrix}
|p| i-2\sum_{k=1}^{i-1}r_k(i-k)\\r_i
\end{pmatrix} ~.
\end{equation}

In order to obtain the total number of framed BPS states in chamber $c_n$ carrying relative magnetic charge $\tilde{n}_{\rm m}$, we must sum over these degeneracies for all possible $\vec{r}\,$'s subject to the constraint \eqref{magnorm}.  However we must also keep track of the electric charges of these states.  Since the core is purely magnetic the electric charge of a given configuration, specified by $\vec{r}$, is simply the sum of the electric charges of the halo particles:
\begin{equation}
n_{\rm e} = \sum_{l} l r_l =: \lVert \vec{r} \, \rVert_{\rm e}~.
\end{equation}
Let us therefore denote the set of all possible $\vec{r}\,$'s in chamber $n$ subject to these two constraints as follows:
\begin{equation}
S_{n}^{\tilde{n}_{\rm m},n_{\rm e}} :=  \left\{ \vec{r} = (r_1,\ldots,r_{n}) \in \mathbb{Z}_{\geq 0}^{n} ~\bigg|~ \lVert \vec{r}\, \rVert_{\rm m} = \tilde{n}_{\rm m} ~~ \& ~~ \lVert \vec{r}\, \rVert_{\rm e} = n_{\rm e} \right\}~.
\end{equation}
The number of states in $c_n$ with charges $\{\tilde{n}_{\rm m},n_{\rm e}\}$ is the sum of degeneracies \eqref{numofstates} over this set.  Hence this argument shows that the generating function \eqref{mfgenfun} can equivalently be written as\footnote{Again, the identification between the number of states and the protected spin character at $y=1$ only holds when all states have trivial $R$-charge. Here, where the states can all be identified as halos of $R$-neutral particles around an $R$-neutral core this is indeed the case.}
\begin{equation}\label{mfgenfun2pos}
F_n(p,\{X_1,X_2\};y=1)  =  X_1^{- |p|/2} \sum_{\tilde{n}_{\rm m} = 0}^{\infty} \sum_{n_{\rm e} = 0}^{\infty} \left\{ ~~~\sum_{\mathclap{\vec{r} \in S_{n}^{\tilde{n}_{\rm m},n_{\rm e}}}} ~N_n(p,\vec{r};y=1) \right\} X_{1}^{\tilde{n}_{\rm m}} X_{2}^{n_{\rm e}}~, \quad (n \geq 0)~.
\end{equation}
The equality of \eqref{mfgenfun} and \eqref{mfgenfun2pos} is established by observing that both agree at $n=0$ and that they both satisfy the same recursion relation \eqref{relationwalls}, as we show in appendix \ref{wcapp}.

Now let us consider the chambers $c_n$ for $n < 0$.  Relative to the previous analysis there are two additional signs to consider.  The first comes from the fact that in going from chamber $c_{n+1}$ to $c_n$ we are binding halos with electric charge $-|n|$ instead of $|n|$.  This leads to an overall relative sign in the top factor of the binomial in \eqref{numofstates}.  However we get a second sign there because we are now crossing walls in the opposite direction---instead of moving counterclockwise in figure \ref{wallfig} we are moving clockwise.  These signs cancel.  We summarize this by saying $N_{n}(\vec{r}\,) = N_{|n|}(\vec{r}\,)$.  Then, in order to construct the generating function we merely need to sum over $\vec{r}\,$'s in $S_{|n|}^{\tilde{n}_{\rm m},|n_{\rm e}|}$.  Thus we can write the generating function for arbitrary $n \in \mathbb{Z}$ as
\begin{equation}\label{mfgenfun2}
F_n(p,\{X_1,X_2\};y=1)  =  X_1^{- |p|/2} \sum_{\tilde{n}_{\rm m} = 0}^{\infty} \sum_{s = 0}^{\infty} \left\{ ~~~\sum_{\mathclap{\vec{r} \in S_{|n|}^{\tilde{n}_{\rm m},s}}} ~N_{|n|}(p,\vec{r};y=1) \right\} X_{1}^{\tilde{n}_{\rm m}} X_{2}^{\sgn(n)\cdot s}~, 
\end{equation}
where comparison with the general form \eqref{generalF} identifies the electric charges with $n_{\rm e} = \sgn(n) \cdot s$.  Hence we can read off the framed protected spin characters in chamber $c_n$ at $y=1$:
\begin{equation}
\fOmega\left(L_{\zeta_n}(p,0),u_n,(\tilde{n}_{\rm m} - \frac{{}_{|p|}}{{}^2})H_\alpha \oplus n_{\rm e} \alpha; y=1\right) = ~~~~~\sum_{\mathclap{\vec{r} \in S_{|n|}^{\tilde{n}_{\rm m},\sgn(n)\cdot n_{\rm e}}}} ~N_{|n|}(p,\vec{r};y=1)~.
\end{equation}
Note if $\tilde{n}_{\rm m}$ is negative, or if the signs of $n$ and $n_{\rm e}$ disagree, then the set $S_{|n|}^{\tilde{n}_{\rm m},\sgn{n}\cdot n_{\rm e}}$ is empty and we get $\fOmega = 0$.

The discussion and results above, given at $y=1$, strongly suggest what the generalization to arbitrary $y$ should be. The number of states $N_n(p,\vec{r};y=1)$ is composed of binomial coefficents $({}^{a}_{b})$  as our halos are associated to $b$ fermions distributed over $a$ possible angular momentum states. To keep track not only of the total number of states but also the quantum number $2 \II^3 = 2 J^3$ of each state, one can introduce the character of the corresponding angular momentum representation. Here, as we are dealing with the $b^{\rm th}$ antisymmetric power of the $a$-dimensional representation of $\mathfrak{so}(3)$, this character is given by the $q$-binomial coefficient:
\begin{equation}
\begin{bmatrix}
a\\ b
\end{bmatrix}_{\! y}:=\frac{\prod_{i=1}^a(y^{i}-y^{-i})}{\prod_{i=1}^b(y^{i}-y^{-i})\prod_{i=1}^{a-b}(y^{i}-y^{-i})} ~.
\end{equation}
It is natural to propose that the proper generalization to arbitrary $y$ of \eqref{numofstates} is\footnote{As pointed out before, it can happen that the top entry in the binomial coefficient becomes negative. Just like a regular binomial coefficient, the $q$-binomial coefficient with negative top entry is defined as \begin{equation}
\begin{bmatrix}
a\\ b
\end{bmatrix}_{\! y}:=(-1)^b\begin{bmatrix}
b-a-1\\ b
\end{bmatrix}_{\! y}\qquad (a<0)\,.
\end{equation}}
\begin{equation}\label{yN}
N_n(p,\vec{r};y\,) :=\prod_{i=1}^{n}\begin{bmatrix}
|p| i-2\sum_{k=1}^{i-1}r_k(i-k)\\r_i
\end{bmatrix}_{\! y} ~,
\end{equation}
and that the generating function in chamber $c_n$ is then
\begin{equation}\label{mfgenfuny}
F_n(p,\{X_1,X_2\};y)  =  X_1^{- |p|/2} \sum_{\tilde{n}_{\rm m} = 0}^{\infty} \sum_{s = 0}^{\infty} \left\{ ~~~\sum_{\mathclap{\vec{r} \in S_{|n|}^{\tilde{n}_{\rm m},s}}} ~N_{|n|}(p,\vec{r};y) \right\} y^{n_{\rm e}(2\tilde{n}_{\rm m}-|p|)}X_{1}^{\tilde{n}_{\rm m}} X_{2}^{\sgn(n)\cdot s}~.
\end{equation}
This clearly gives the correct answer in $c_0$ and in appendix \ref{wcapp} we check that \eqref{mfgenfuny} satisfies the required wall crossing formulae.  Hence this proves that indeed it is correct for all $n$. 

In addition, one can find the generalization of the form \eqref{mfgenfun}, which has the advantage of manifestly showing the spectrum is finite\footnote{For the extension of \eqref{mfgenfunuber} to $n<0$ one can use the relation \eqref{ntominusn}.}.  Remarkably, the form in terms of Chebyshev functions remains unchanged while the quantity $f_n$ picks up a simple $y$-dependence: 
\begin{equation}\label{mfgenfunuber}
F_n(p,\{X_1,X_2\};y)=\left[X_1^{-1/2}X_2^{-n/2}\left( U_{n}(f_{n})-X_2^{-1/2}U_{n-1}(f_{n})\right)\right]^{|p|}~,\quad (n\geq 0)~,
\end{equation}
where now
\begin{equation}
f_n:=\frac{X_{2}^{1/2}+ X_2^{-1/2}\left(1+y^{2n+3}X_1X_2^{n+1}\right)}{2} ~.
\end{equation}
Again we refer to appendix \ref{wcapp} for the derivation of this equality.

The protected spin characters are easily read off by comparing \eqref{mfgenfuny} with \eqref{generalF}:
\begin{align}\label{haloPSC}
\fOmega_n(p,\tilde{n}_{\rm m},n_{\rm e};y) \equiv \fOmega\left(L_{\zeta_n}(p,0),u_n, (\tilde{n}_{\rm m} - \frac{{}_{|p|}}{{}^2})H_\alpha \oplus n_{\rm e} \alpha ;y\right) =&~ ~~~~~ \sum_{\mathclap{\vec{r} \in S_{|n|}^{\tilde{n}_{\rm m},\sgn(n)\cdot n_{\rm e}}}} ~N_{|n|}(p,\vec{r};y)  ~. \raisetag{10pt}
\end{align}
Via \eqref{inCequalsPSC} this leads directly to a set of predictions for index characters of Dirac operators on various hyperk\"ahler manifolds, $\fmMM\left(P = \frac{p}{2}H_\alpha;\gm = (\tilde{n}_{\rm m} - \frac{|p|}{2})H_\alpha;\sx \right)$, of dimension $4 \tilde{n}_{\rm m}$.  We will verify these predictions---both the location of the walls and the spectrum in each chamber---below via an explicit semiclassical analysis for $\tilde{n}_{\rm m}=1$.  In this case it is easy to evaluate the $N_{|n|}$ more explicitly and we find
\begin{equation}\label{haloIndex}
\fOmega_n(p,\tilde{n}_{\rm m} =1,n_{\rm e};y) = \left\{ \begin{array}{c l} \displaystyle \chi_{|p n_{\rm e}|}(y) ~, & ~~1 \leq |n_{\rm e}| \leq \sgn(n_{\rm e}) \cdot n~, \\ [1ex] 0~, & ~~\textrm{otherwise}~. \end{array} \right.
\end{equation}

In section \ref{ssec:nm2} we work out the details for $\tilde{n}_{\rm m} = 2$, giving a set of predictions for the Dirac kernel on a family of eight-dimensional manifolds, some of which have been studied in the literature previously.

%%%%%%%%%%%%%%%%%%%%%
\subsection{The Blair--Cherkis--Durcan monopole and its moduli space}\label{CDmonopole}
%%%%%%%%%%%%%%%%%%%%%

In this subsection we briefly review the Blair--Cherkis--Durcan (BCD) solution \cite{Cherkis:2007jm, Cherkis:2007qa, Blair:2010vh} that describes a single non-Abelian magnetic monopole in the presence of a magnetic singularity.  Those readers mainly interested in the comparison of \eqref{haloIndex} with the index of the semiclassical Dirac operator on moduli space can safely skip the discussion here.  The only result of relevance for the next subsection, where this comparison is made, is that the moduli space of this singular monopole configuration is a discrete quotient of Taub--NUT space.  The purpose of the current subsection is to provide an explicit example of a singular monopole configuration that allows us to illustrate some of the general properties and definitions of \cite{MRVdimP1}, reviewed in section \ref{sec:clfBPS}, in a nontrivial setting.  

In this way we will be taking---or at least summarizing---a rather pedagogical approach to obtaining the moduli space of interest, more or less directly from the definitions, as done in \cite{Shah}.  The same result was obtained much earlier via other methods: brane constructions relating this space to the Coulomb branch of certain three-dimensional supersymmetric field theories \cite{Hanany:1996ie,Seiberg:1996nz}; the Nahm transform for singular monopoles \cite{Cherkis:1997aa}; a twistor formulation of singular monopoles \cite{Cherkis:1998xca,Cherkis:1998hi}; and, more recently, the bow formalism \cite{Cherkis:2008ip,Cherkis:2010bn,Blair:2010vh}.  These approaches are undoubtedly more powerful and yield the result with much greater ease.  However we find the direct approach instructive and appealing, especially if one does not wish to assume familiarity with the more sophisticated machinery employed by other methods.\footnote{Although, in truth, the bosonic zero modes were obtained in \cite{Shah} via the singular Nahm transform rather than directly from the background solution.  The reason is that it is difficult to obtain the gauge parameters $\varepsilon_m$, in \eqref{coordflows}, necessary for gauge orthogonality of the zero modes, by directly solving the Poisson equation that defines them.  Of course the background solution itself was also obtained via the singular Nahm transform, and later the bow construction.}

%%%%%%%%%%%%%%%%%%%%%
\begin{figure}
\begin{center}
\qquad\qquad\qquad\qquad\qquad\includegraphics[scale=1]{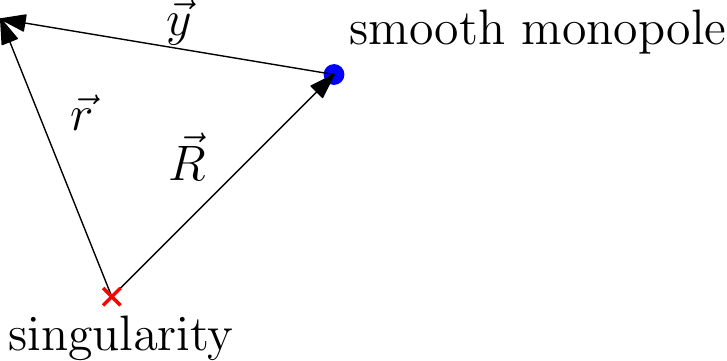}
\end{center}
\caption{Coordinates}\label{CDfig}
\end{figure}
%%%%%%%%%%%%%%%%%%%%%  

Let a singularity of charge $k$ be placed at the origin and denote the general position vector with respect to this origin by $\vec{r}$.  The precise relation between the parameter $k$ and the 't Hooft charge will be explained below.  We introduce a fixed displacement vector $\vec{R}$ that can be thought of as the position of the non-Abelian monopole relative to the singularity, and we define the relative displacement $\vec{y}=\vec{r}-\vec{R}$.  See see figure \ref{CDfig}.  The corresponding Euclidean distances will be denoted $r,R$, and $y$, and the corresponding unit vectors $\hat{r},\hat{R}$ and $\hat{y}$.  Finally, let the mass of the $W$-boson be $m_W$,\footnote{We will see below that this corresponds to the physical mass of the $W$-boson in the Yang--Mills--Higgs theory with a single adjoint-valued Higgs field $X$.  In the $\NN = 2$ context it would only correspond to the mass on the locus of $\widehat{\BB}_{\rm wc}$ where $\YY_{\infty} = 0$.} and let $T^i$ be an (anti-Hermitian) basis for $\mathfrak{su}(2)$ satisfying $[T^i,T^j] = \epsilon^{ij}_{\phantom{ij}k} T^k$, along which the Higgs and gauge field are expanded: $X = X^i T^i$, \etc. Then the $\mathfrak{su}(2)$ solution from \cite{Blair:2010vh} reads:
\begin{align}\label{singmon}
X^i =&~ \left(\frac{1}{y}-\left(m_W+\frac{k}{2r}\right)\coth(m_Wy+v)\right)\frac{y^i}{y}+\frac{k(y^2\delta^{ij}-y^iy^j)}{yr \sinh(m_Wy+v)} \cdot \frac{R^j}{(R+r)^2-y^2}~, \cr
A^i =&~ \frac{1}{y}\left(\frac{1}{\sinh(m_Wy+v)}\left[m_W + \frac{k(R+r)}{(R+r)^2-y^2}\right]-\frac{1}{y}\right)\epsilon^{ijk}y^j\ed r^k + \cr
&~ +\frac{k}{2} \coth(m_Wy+v)\,\frac{y^i}{y} \cdot \frac{\epsilon^{jkl}R^j r^k \ed r^l}{r(rR+R^ir^i)}-\frac{k(y^2\delta^{ij}-y^iy^j)}{yr\sinh(m_Wy+v)} \cdot \frac{\epsilon^{jkl}r^k \ed r^l}{(R+r)^2-y^2}~, \qquad  \raisetag{20pt}
\end{align}
where
\begin{equation}
v := \frac{k}{2}\log \left( \frac{r+R+y}{r+R-y}\right) \,.
\end{equation}
Here we see a typical hedehog-like construction, where directions in physical space are correlated with directions in the Lie algebra.  Upon setting $k=0$, which implies that $v \to 0$ and all $\vec{R}$ dependence drops out, it is easy to see that one recovers the classic Prasad--Sommerfield solution \cite{Prasad:1975kr} for the smooth monopole.

In order to clarify the relation between the parameters of the solution and its physical properties, consider the asymptotics.  As $r \to 0$ the fields are singular and one finds\footnote{The appearance of $|k|$ in this expression might seem surprising but comes about as follows.  When $r \to 0$ we also have $R \to y$.  Hence $v$ is blowing up, but its sign depends on the sign of $k$.  The hyperbolic trigonometric functions must be expanded accordingly.  One has, for example,
$$
\sinh(m_W y + v) = |k| \left( \frac{2 R}{r (1 + \hat{R} \cdot \hat{r})} \right)^{|k|/2} e^{\frac{k}{|k|} m_W R} \left\{ 1 + O(r/R) \right\}~,
$$
as $r \to 0$, where we used $r+R - y = r(1 + \hat{R} \cdot \hat{r}) + O(r^2/R)$.}
\begin{equation}\label{Xsing}
X^i  = \frac{|k|}{2r}\hat{R}^i+ \left\{  \frac{e^{-\frac{k}{|k|}m_W R}}{2R} \left(\frac{1+\hat{r}^j\hat{R}^j}{2R}\right)^{\frac{|k|}{2}-1} \left(\delta^{ij}-\hat{R}^i\hat{R}^j\right)\hat r^j \right\} r^{\frac{|k|}{2}-1} (1 + O(r/R))~.
\end{equation}
There are two things to note about this expression.  First, the leading singularity is consistent with a 't Hooft defect of charge $P = -|k| \hat{R}^i T^i$.  Now remember that $\vec{R}$ is simply a triplet of fixed parameters.  Thus for convenience we are free to choose our Cartan subalgebra to be spanned by the combination $\hat{R}^i T^i$, and a properly normalized co-root is
\begin{equation}\label{corootchoice}
H_\alpha = 2 \hat{R}^i T^i~,
\end{equation}
whence the 't Hooft charge
\begin{equation}\label{CDtHooftcharge}
P = -\frac{|k|}{2} H_\alpha~.
\end{equation}
Comparing with \eqref{su2opcharges}, we can then identify $p = -|k|$.  Second, notice that the subleading term has a $1/\sqrt{r}$ divergence if $|k| = 1$.  (The coefficient of this divergence even depends on the direction from which one approaches the origin!)  This gives a very concrete motivation for the generalized boundary conditions introduced in \cite{MRVdimP1} and discussed in appendix \ref{app:bcs}.  This subleading singular behavior is only possible for $G = SO(3)$ gauge group.

Now consider the large $r$ asymptotics:
\begin{equation}
X = - \left[ m_W - \left(1 - \frac{k}{2} \right) \frac{1}{r} \right] \hat{r}^i T^i + o(1/r)~.
\end{equation}
This is not in the standard gauge that we assumed in \eqref{Mdef}, but is easily brought to it by making an asymptotically nontrivial patchwise gauge transformation that conjugates $\hat{r}^i T^i$ into $-\hat{R}^i T^i$.  After doing so the gauge transformed field is
\begin{equation}\label{gtedX}
X' = \left[ m_W - \left(1 - \frac{k}{2} \right) \frac{1}{r} \right] \hat{R}^i T^i + o(1/r)~,
\end{equation}
from which, using \eqref{corootchoice} and comparing with \eqref{Mdef}, we read off both the vev and asymptotic magnetic charge:
\begin{equation}
\sx \equiv X_\infty = \frac{m_W}{2} H_\alpha~, \qquad \gm = \left(1 - \frac{k}{2}\right) H_\alpha~.
\end{equation}
As advertised, $m_W = \langle \alpha, \sx \rangle$ is identified with the $W$-boson mass.  Given the 't Hooft charge \eqref{CDtHooftcharge}, we see that the asymptotic magnetic charge indeed sits in the torsor $\Lambda_{\rm cr} + P$.  One has the regularized BPS mass \eqref{MBPScl}:
\begin{equation}\label{BCDBPSmass}
M^{\rm cl}_{\gm} =\frac{4\pi}{g_{0}^2} (\gm,\sx) = \frac{4\pi m_W}{g_0^2}\left(1-\frac{k}{2}\right)~.
\end{equation}

To study the quantum BPS states related to this solution we need to understand its moduli space $\fmMM_k$.  The dimension can be computed via the framed dimension formula \eqref{dim1}:
\begin{equation}
\tilde \gamma_{\mathrm{m}}=\gamma_{\mathrm{m}}-P^{-}=\frac{2+|k|-k}{2}H_\alpha\quad\Rightarrow\quad \dim \fmMM_{k}=\begin{cases}4\quad\qquad\qquad\mbox{when}\ k\geq 0~,\\
4(1+|k|)\quad\ \mbox{when}\ k\leq 0~.\end{cases}
\end{equation}
Note that this implies that when $k<0$ the BCD solution actually describes $1-k$  non-Abelian monopoles where, by construction, $-k$ of the monopoles are constrained to lie atop the singularity. It would be interesting to see if one can deform the BCD solution in this case to one where some of the additional monopoles move off the singularity.  When $k >0$ the BCD solution describes a single smooth monopole in the presence of the defect.  As $k$ increases the BPS mass \eqref{BCDBPSmass} decreases indicating a stronger and stronger binding energy. 

For the purposes of this paper we will restrict ourselves to the case $k>0$, so that the moduli space is four-dimensional.  As we mentioned, there are several paths that lead to the hyperk\"ahler metric on this space.  We give here a very brief summary of the approach taken in \cite{Shah} which is conceptually, though not computationally, the most straightforward.  This involves computing the metric directly from the definition \eqref{metC} in terms of the bosonic zero modes.  The zero modes, $\delta_a \hat{A} = (\delta_a A, \delta_a X)$, $a = i,4$, can in principal be obtained from the background, but in practice are more easily obtained via the Nahm transform for singular monopoles.  They are quite nontrivial and here we merely quote the bare minimum of data required to compute the metric.  Namely, since both the $g_{44}$ and $g_{4i}$ components can be reduced to a boundary integral over the asymptotic two-sphere at spatial infinity, one requires only the large $r$ (or equivalently, large $y$) asymptotics:  
\begin{align}\label{asympL}
\hat y^i\delta_i \hat A_j =&~ \hat y^kT^k\frac{\Omega_j}{y^2H}\left(1-(1+2m_W R)\hat y\cdot \hat R\right)+O(y^{-3})~,  \cr
\delta_4\hat A_a =&~ \hat D_a\Lambda_4\,,\quad \Lambda_4=2\hat y^iT^i\left(1- \frac{1}{m_W y H}\right)+ O(y^{-2})~, 
\end{align}
where
\begin{equation}
H=1+\frac{k}{2m_W R}\,,\quad\mbox{and}\quad \vec{\nabla}_R\times\vec{\Omega}=\vec{\nabla}_RH\,.
\end{equation}
Hyperkahlarity then fixes the $g_{ij}$ components\footnote{Up to a constant that can be fixed by demanding that when $R\rightarrow\infty$ and the smooth monopole gets infinitely separated from the singularity, the moduli space reduces to that of the free smooth non-Abelian monopole.} and leads to the metric:
\begin{equation}\label{BCDmetric}
\ed s^2= m_W \left( H \ed R^i \ed R^i+H^{-1}\left(2m_W^{-1}\ed\uppsi+\Omega^i \ed R^i\right)^2\right)\,.
\end{equation}

Locally this metric is that of (single-centered) Taub--NUT space (TN), but the precise global structure depends on the periodicity of $\uppsi$.  The coordinate $\uppsi$ is, per definition, the one associated to the global gauge mode, generated by $\Lambda_4$ given above.  Rather, if we make the asymptotically nontrivial gauge transformation to put the fields their standard form as discussed around \eqref{gtedX}, then the relevant gauge parameter is
\begin{equation}
\Lambda_{4}' = -2 \hat{R}^i T^i \left(1- \frac{1}{m_W y H}\right)+ O(y^{-2})~,
\end{equation}
which asymptotes to $-2 \hat{R}^i T^i = - H_\alpha$.  This acts on the solution $\hat{A}'(\vec{r};\vec{R})$ to exhibit dependence on all four moduli $z^a = (\vec{R},\uppsi)$ via
\begin{equation}\label{adac}
\hat{A}''(\vec{r};z^a) := U^{-1}\hat{A}' U -U^{-1} \ed U\,,\quad U= \exp({\Lambda_{4}' \uppsi}) ~.
\end{equation} 
Since the gauge transformation is an adjoint action the relevant exponential is the one for the adjoint form of the group, $G_{\rm ad} = SO(3)$, for which we have $\exp(\pi H_\alpha) = \mathbbm{1}$.  Hence we should take $\uppsi \sim \uppsi+\pi$. 

This periodicity of $\pi$ allows us to identify the metric \eqref{BCDmetric} with our canonically normalized metric for  the $\mathbb{Z}_k$ quotient of Taub-NUT in \eqref{TN}, upon taking the periodic coordinate to be $x^4 = \frac{4}{k} \uppsi$.  Thus we have
\begin{equation}\label{singmonmod}
\fmMM\left(P = \frac{{}_{p}}{{}^2} H_\alpha~; \gm = (1- \frac{{}_{|p|}}{{}^2}) H_\alpha~; \sx\right) \, \cong  \,\mathrm{TN}(m,\ell)/\mathbb{Z}_{|p|}\,.
\end{equation}
with the identification of parameters
\begin{equation}\label{singexid1}
m = m_W = \langle \alpha,\sx\rangle  \qquad \ell=\frac{|p|}{2\langle \alpha, \sx\rangle} \,.  
\end{equation}
%

%%%%%%%%%%%%%%%%%%%%%%%
\subsection{Spectrum and wall crossing from the Dirac kernel}
%%%%%%%%%%%%%%%%%%%%%%%

In this subsection we discuss the kernel of the ${\rm G}$-twisted Dirac operator on the singular monopole moduli space \eqref{singmonmod}, and compare the corresponding indices with those predicted by \eqref{haloPSC} for $\tilde{n}_{\rm m} = 1$.  For the Dirac operator we require the triholomorphic Killing vector ${\rm G}(\sy) = \langle \alpha, \sy \rangle K$, where $K = \frac{2}{|p|} \pd_{x^4}$ generates a $U(1)$ triholomorphic isometry with periodicity $2\pi$.  This should be compared with \eqref{G} to extract the constant $C$:
\begin{equation}\label{Chiggs}
\frac{2\langle \alpha, \sy \rangle}{|p|} \pd_{x^4} = {\rm G}(\sy) \equiv \frac{C}{m\ell^2} \pd_{x^4} \quad \Rightarrow \quad C = \frac{2m\ell^2}{|p|} \langle \alpha, \sy \rangle = \frac{|p|}{2} \frac{\langle \alpha,\sy \rangle}{\langle \alpha,\sx \rangle}~.
\end{equation}
As explained in appendix \ref{appendix:TN}, the kernel of the Dirac operator on $\mathrm{TN}/\mathbb{Z}_{|p|}$ and its wall crossing are completely encoded in $C$.  

The explicit wavefunctions $\Psi = (\psi_+,0)^T$ can be found in \eqref{zmodesTN}.  The kernel sits inside the positive chirality spinor bundle, which coincides with $\SS_{1,0}$ in \eqref{bundledecomp} for four-dimensional hyperk\"ahler manifolds.  Hence all states are $SU(2)_R$ singlets, consistent with no-exotics, and the quantum numbers associated with the $SU(2)$ isometry of Taub--NUT correspond to ordinary angular momentum.  

Whenever $C$ crosses a number $\nu$ with $|\nu| \in \frac{|p|}{2} \mathbb{N}$, an angular momentum multiplet with spin $j = |\nu| - \half$ is created.  When $|C|<\frac{|p|}{2}$ the spectrum is empty and when $|C| > \frac{|p|}{2}$ the spectrum is made up of angular momentum multiplets for each spin 
\begin{equation}
j \in \left\{ \frac{|p|}{2} - \half~, |p| - \half~, \frac{3|p|}{2} - \half~, \ldots, \frac{|p|}{2} \left\lfloor \frac{2|C|}{|p|} \right\rfloor - \half \right\}~,
\end{equation}
where $\lfloor x \rfloor = \max \{ s \in \mathbb{N} ~|~ s < x \}$ is the left-continuous floor function.  If we identify
\begin{equation}
\nu = - \frac{|p| n}{2}~, ~~ n \in \mathbb{Z} \setminus \{ 0 \}~,
\end{equation}
then we see that the equation for the walls is equivalent to \eqref{walls}:
\begin{equation}
C = \nu \qquad \Rightarrow \qquad  \frac{\langle \alpha, \sy \rangle}{\langle \alpha,\sx\rangle} = -n~,
\end{equation}
upon recalling that $\langle \alpha, \sy\rangle = (H_\alpha,\sy)$ for the simple $\mathfrak{su}(2)$ root.  Hence we identify $-\frac{|p|}{2} < C < \frac{|p|}{2}$ with the chamber $c_0$, $\frac{|p| n}{2} < (-C) < \frac{|p|(n+1)}{2}$ with the chamber $c_n$ for $n > 0$, and $\frac{|p|(n-1)}{2} < (-C) < \frac{|p| n}{2}$ with the chamber $c_n$ for $n < 0$.

Not only the location of the walls, but also the BPS spectrum inside each chamber matches in the two pictures.  To see this we must first account for the electric charge.   One observes from the explicit form of the wavefunction that $-\nu$ is the eigenvalue of the spinorial Lie derivative $i \Lie_{\pd_{x^4}}$.  Thus taking into account the $4\pi/|p|$ periodicity of $x^4$, the electric charge operator acts as
\begin{equation}
\hat{\gamma}_{\rm e} \Psi \equiv i \alpha \Lie_{K} \Psi = \left(-\frac{|p|\nu}{2} \alpha \right) \Psi~.
\end{equation}
Identifying the eigenvalue with $\gamma_{\rm e} = n_{\rm e} \alpha$, we deduce
\begin{equation}
\nu = -\frac{|p| n_{\rm e}}{2}~.
\end{equation}
Thus the magnitude of the electric charge is correlated with the spin such that $j = \frac{|p n_{\rm e}|}{2} - \half$, and the sign is opposite to the sign of $C$.  

Let us denote an angular momentum representation of spin $j$ and electric charge $n_{\rm e}$ by $[ 2j+1]_{n_{\rm e}}$.  Then if $(\sx_n,\sy_n)$ denotes any $(\sx,\sy) \in c_n$, and $\fmMM(L(p,0);\tilde{n}_{\rm m} = 1; \sx_n) \equiv \fmMM(\sx_n)$, we have
\begin{align} 
\ker_{\Lsq}^+ \left( \slashed{\DD}_{\fmMM(\sx_n)}^{\rm G(\sy_n)} \right) \cong&~ \left\{ \begin{array}{c c} \displaystyle\bigoplus_{s=1}^{|n|} \, [s |p|]_{\sgn(n) \cdot s}~, & ~~n \neq 0 \label{kerp} \\[2ex]
\{ 0 \}~, & ~~n = 0 \end{array} \right. ~, \\[1ex]
\ker_{\Lsq}^- \left( \slashed{\DD}_{\fmMM(\sx_n)}^{\rm G(\sy_n)} \right) =&~~ \{ 0 \}~,~\forall n~. \label{kerm}
\end{align}
This space of states is in complete agreement with the results described in section \ref{sec:fbps}.  The index characters following from \eqref{kerp}, \eqref{kerm} are
\begin{equation}
\fC_{L(p,0),(1-\frac{|p|}{2})H_\alpha,\sx_n,\sy_n}^{n_{\rm e} \alpha}(y) =  \left\{ \begin{array}{c l} \chi_{|pn_{\rm e}|}(y)~, & ~~1 \leq |n_{\rm e}| \leq \sgn(n_{\rm e}) \cdot n~, \\ 0~, & ~~\textrm{otherwise}~. \end{array} \right.
\end{equation}
These coincide perfectly with the protected spin characters $\fOmega$. 

%%%%%%%%%%%%%%%%%% 
\subsection{Predictions for the Dirac kernel on manifolds with $|\tilde{\gamma}_{\rm m}| = 2$}\label{ssec:nm2}
%%%%%%%%%%%%%%%%%%

Let us illustrate \eqref{haloPSC} further in the case $\tilde n_\mathrm{m}=2$, which starts to show how nontrivial this formula really is.  We assume that the signs of $n$ and $n_{\rm e}$ agree, otherwise the set $S_{|n|}^{\tilde{n}_{\rm e}, \sgn(n) n_{\rm e}}$ is empty.  Now consider the form of $\vec{r}\in S_{|n|}^{2,|n_{\rm e}|}$.  We solve the condition $\lVert \vec{r}\, \rVert_{\rm m} = 2$ by writing $r_i = \delta_{i,j} + \delta_{i,k}$ where without loss of generality we can assume $1 \leq k \leq j \leq n$.  The condition $\lVert \vec{r}\, \rVert_{\rm e} = |n_{\rm e}|$ gives $j + k = |n_{\rm e}|$ which we can use to eliminate $j$.  Thus the set of $\vec{r}^{(k)}$ we must sum over is parameterized by an integer $k$ such that
\begin{equation}\label{vecrparam}
r_{i}^{(k)} = \delta_{i,k} + \delta_{i,|n_{\rm e}|-k}~, \qquad \textrm{with} \quad \max\{1,|n_{\rm e}|-|n| \} \leq k \leq \left\lfloor \frac{|n_{\rm e}|}{2} \right\rfloor~.
\end{equation}
The lower and upper bounds can be understood as follows.  First we observe that if $|n| < |n_{\rm e}| - 1$, then it is possible for there to be no solution for $j$ if $k$ is too small as it would require $j > n$.  This leads to the lower bound.  Similarly since $k \leq j$, we must have $k \leq |n_{\rm e}|/2$ to find a solution.  In \eqref{vecrparam} $\left\lfloor x \right\rfloor=\max \left\{m\in \mathbb{Z}\,|\, m \leq x  \right\}$ indicates the left-continuous floor function.  

Observe that if $|n| < |n_{\rm e}|/2$ then the range of $k$ is empty, and there is no solution to the constraints in this case.  Thus we already learn that states with electric charge $n_{\rm e}$ exist only in chambers $c_n$ with $\sgn(n) = \sgn(n_{\rm e})$ and $|n| \geq |n_{\rm e}|/2$.  We also see that if $|n_{\rm e}| =0$ or $1$ the range is empty: there are no framed BPS states with $\tilde{n}_{\rm m} = 2$ carrying  electric charge $|n_{\rm e}| = 0,1$.  This is obvious from the halo picture since we need two halo particles to account for the relative magnetic charge and they both necessarily carry nonzero electric charge of the same sign.  Therefore we have the protected spin characters
\begin{align}\label{fPSCnm2}
\fOmega_n(p, 2, n_{\rm e};y) =&~ \left\{ \begin{array}{c l} \displaystyle\sum_{k = \max\{1,|n_{\rm e}|-|n|\}}^{\lfloor \frac{|n_{\rm e}|}{2} \rfloor} N_{|n|}(p ,\vec{r}^{(k)};y)~, &  2 \leq |n_{\rm e}| \leq \sgn(n_{\rm e}) 2n~, \\[4ex]
 0~, &  \textrm{otherwise} ~. \end{array} \right. 
\end{align}

Now let us evaluate the $N_{|n|}$'s, assuming the range in \eqref{vecrparam} is non-empty.  There are two qualitatively different cases to consider depending on whether $k < |n_{\rm e}|/2$ or $k = |n_{\rm e}|/2$, corresponding to whether $\vec{r}^{(k)}$ has two distinct entries with value $1$, or a single entry with value $2$.  Note the latter case is only possible when $|n_{\rm e}|$ is a positive even integer.  In the former case $N_{|n|}(p,\vec{r};y)$ has the structure $[{}^{a}_1]_{y} \cdot [{}^{b}_1]_{y}$ and in the latter it has the structure $[{}^{a}_{2}]_{y}$.  The first case corresponds to an onion-like structure where we have a halo around another halo, while the second case corresponds to a single halo formed by two identical dyons.  We find from \eqref{yN} that
\begin{align}\label{Nypgen}
N_{|n|}(p,\vec{r}^{(k)};y) =&~ \left[ \begin{array}{c} |p| k \\ 1 \end{array} \right]_{y} \cdot \left[ \begin{array}{c} (|p|-2) (|n_{\rm e}|-k) +2k \\ 1 \end{array} \right]_{y} ~, \qquad k < |n_{\rm e}|/2~, \cr
N_{|n|}(p,\vec{r}^{(k)};y) =&~ \left[ \begin{array}{c} |p n_{\rm e}|/2 \\ 2 \end{array} \right]_y ~, \qquad k = |n_{\rm e}|/2 ~.
\end{align}
Notice that the second factor of the top line has an upper entry that can be negative when $|p| = 1$ but is otherwise positive.  For $a > 0$, we have that $[{}^{a}_1]_y = \chi_{a}(y)$ and $[{}^{a}_2]_y = \chi_{a}(y) \chi_{a-1}(y)/\chi_{2}(y)$.  (The last case actually requires $a \geq 2$, and vanishes when $a = 1$.)  If $a < 0$ then $[{}^{a}_{1}]_{y} = -\chi_{|a|}(y)$.  We examine the case $|p|=1$ in some detail and briefly comment on $|p|=2$.

In the case $|p| = 1$ we have
\begin{align}\label{Nyp1}
N_{|n|}(\pm 1,\vec{r}^{(k)};y) =&~ \sgn(3k - |n_{\rm e}|) \chi_{k}(y) \chi_{|3k - |n_{\rm e}||}(y) ~, \quad \max\{1,|n_{\rm e}|-|n|\} \leq k < |n_{\rm e}|/2~, \cr
N_{|n|}(\pm 1,\vec{r}^{(k)};y) =&~ \frac{\chi_{\frac{|n_{\rm e}|}{2}}(y) \chi_{\frac{|n_{\rm e}|}{2} -1}(y)}{\chi_{2}(y)} ~, \quad k = \frac{|n_{\rm e}|}{2} > 1~~ \& ~~ |n| \geq \frac{|n_{\rm e}|}{2}~, \cr
N_{|n|}(\pm 1,\vec{r}^{(k)};y) =&~0 ~, \quad k = \frac{|n_{\rm e}|}{2} = 1~.  \raisetag{20pt}
\end{align}
Note that $\sgn(0)$ is to be understood as $0$ here.  First we observe from the last line that there are no BPS states with $|n_{\rm e}| = 2$.  This can be understood as follows: when $|p| = 1$, halo particles in the first chamber, $c_{\pm 1}$, with $\gamma_{\rm h} = H_{\alpha} \oplus (\pm \alpha)$ bind in an angular momentum singlet state.  Being fermions, the Pauli Exclusion Principle does not allow for two identical particles in such a configuration.  Next consider $|n_{\rm e}| = 3$.  The only possible value of $k$ is $k=1$ but we see that the $\sgn$ factor vanishes.  Hence there are no states with $|n_{\rm e}| = 3$ either!  We can also understand this physically.  Such a configuration would correspond to a halo of a $\gamma_{\rm h} = H_{\alpha} \oplus (\pm 2\alpha)$ particle bound to an inner core-halo system of a $\gamma_{\rm h} = H_\alpha \oplus (\pm \alpha)$ particle bound to the core.  However this inner system has a net charge $(1-\half)H_{\alpha} \oplus (\pm \alpha)$ that is parallel to the would-be outer halo particle.  Therefore they do not bind.

Next consider $|n_{\rm e}| = 4$.  For $k = 1$ we are in the case of the first line \eqref{Nyp1}, which gives $- \chi_1(y)^2 = -1$, but this is only possible for chambers $c_{|n|}$ with $|n| \geq 3$.  For $k =2$ we are instead in the case of the second line, which gives $1$, starting in chamber $c_{|n|}$ with $|n| \geq 2$.  Hence, with $\sigma = \pm$ the sign of the electric charge we have
\begin{equation}\label{p1ne4}
\fOmega_{n}(\pm 1, 2,4\sigma ;y) = \left\{ \begin{array}{c l} 1~,  & n = 2\sigma~, \\ 0~, & \textrm{otherwise}~. \end{array} \right.  
\end{equation}
As we start out from the middle chamber $c_{0}$, heading in the direction of $c_{\sigma |n|}$, a single framed BPS state with $\{ \tilde{n}_{\rm m}, n_{\rm e} \} = \{2, 4\sigma \}$---an angular momentum singlet---enters the spectrum in the second chamber, $c_{2\sigma}$, and then leaves after passing to $c_{3\sigma}$.  Notice that $\fOmega(y=1)$ always remains nonnegative, consistently with no-exotics.  Our results thus far are also nontrivially consistent with the statements under \eqref{largeX1}.  In this example where $|p|=1$ and $\tilde{n}_{\rm e} = 2$, the large $X_1$ asymptotics of the generating function implies that the first nonempty chamber $c_n$ for $n > 0$ should be $c_2$, where we should find a single state with electric charge $n_{\rm e} = 2^2 = 4$.

For $|n_{\rm e}| = 5$ we again only have contributions from $k = 1,2$, and this time they are both given by the first line of \eqref{Nyp1}.  The $k=1$ case is only possible starting in chamber $c_{4\sigma}$ due to the lower bound, while the $k = 2$ case starts in chamber $c_{3\sigma}$.  The $k=2$ case is a positive contribution of $\chi_2(y)$, while the $k = 1$ case is a negative contribution of $\chi_{2}(y)$.  Hence we get an angular momentum doublet state that appears in the third chamber (in the direction of the sign of the electric charge) and then promptly vanishes in the fourth:
\begin{equation}
\fOmega_{n}(\pm 1, 2,5\sigma ;y) = \left\{ \begin{array}{c l} \chi_2(y) = y + y^{-1}~,  & n = 3\sigma~, \\ 0~, & \textrm{otherwise}~. \end{array} \right.  
\end{equation}

Let's look at one more case, $|n_{\rm e}| = 6$.  $k$ can run over $1,2,3$, but $k = 2$ gives a vanishing contribution due to $\sgn(0) = 0$.  The $k=1$ case is only possible starting in chamber $c_{5\sigma}$ and gives a negative contribution of $-\chi_3(y)$, the character of the triplet or adjoint representation.  Meanwhile the $k=3$ contribution comes from the second line of \eqref{Nyp1}, is present starting in chamber $c_{3\sigma}$, and is given by $\chi_3(y)$.  Hence
\begin{equation}
\fOmega_{n}(\pm 1, 2,6\sigma ;y) = \left\{ \begin{array}{c l} \chi_{3}(y) = y + 1 + y^{-1}~,  & n =  3\sigma,4\sigma~, \\ 0~, & \textrm{otherwise}~. \end{array} \right.  
\end{equation}
Notice how the $n_{\rm e} = 4\sigma$ state remains the unique state in chamber $c_{2\sigma}$, in accord with the comments under \eqref{largeX1}.  The pattern we see is that the negative contributions always start in later chambers relative to the positive ones (moving outward from $c_0$) so that $\fOmega(y = 1)$ remains nonnegative in all chambers.  No-exotics demands that this be true for all $n_{\rm e}$, since $\fOmega(y=1)$ gives the dimension of the Hilbert space of framed BPS states carrying the corresponding electromagnetic charge.

The case of $|p| = 2$ is also interesting, as there is a simplification in the second factor of the top line of \eqref{Nypgen} such that
\begin{align}
N_{|n|}(\pm 2, \vec{r}^{(k)}; y) =&~ \left\{ \begin{array}{c l} \chi_{2k}(y)^2 ~, &  \max\{1, |n_{\rm e}| -|n| \} \leq k < |n_{\rm e}|/2~,  \\[1ex] \chi_{|n_{\rm e}|}(y) \chi_{|n_{\rm e}| -1}(y) /\chi_{2}(y)~, & k = |n_{\rm e}|/2~, ~~ \& ~~ |n| \geq |n_{\rm e}|/2~, \end{array} \right. 
\end{align}
where again the second case is only possible when $n_{\rm e}$ is even and nonzero.  The sum over $k$ in \eqref{fPSCnm2} can then be computed straightfowardly.  In this case and in general for $|p| > 1$ we only gain, not lose, halos as we move outward from the chamber $c_0$.

The corresponding moduli spaces $\fMM\left(P = \frac{p}{2} H_\alpha; \gm = (2-\frac{|p|}{2})H_\alpha; X_\infty\right)$ of singular monopoles, denoted henceforth by $\fMM(p,\tilde{n}_{\rm m} = 2)$, describe two smooth 't Hooft--Polyakov monopoles in the presence of the charge $p$ defect, and are fascinating eight-dimensional hyperk\"ahler manifolds.  The $|p|=1$ case has been studied in considerable detail by Dancer \cite{Dancer:1992kn,Dancer:1992km} and can be identified with the strongly centered moduli space $\MM_0$ for $SU(3)$ monopoles with charge $\gm = 2 H_1 + H_2$.  See \cite{Houghton:1997ei,Houghton:1999qu}.  

The hypkerk\"ahler quotient of $\fMM(1,2)$ by the triholomorphic $U(1)$ corresponding to the conserved electric charge gives the four-dimensional $D_1$ ALF space.  This is a one-parameter generalization of the Atiyah--Hitchin manifold.  Briefly, the hyperk\"ahler moment map is a three-vector that specifies the position of the center of mass of the two smooth monopoles, relative to the position of the defect.  The parameter of the one-parameter deformation is the magnitude of this displacement---\ie\ the distance of the center of mass of the two monopole system from the defect.  The four-dimensional $D_1$ ALF space (also known as the Dancer manifold), then describes the motion of the two smooth monopoles, relative to their fixed center of mass.  In the limit that the distance from the center of mass to the defect tends to infinity, the Dancer manifold approaches the Atiyah--Hitchin manifold describing two smooth monopoles in the absence of defects.

This picture generalizes.  The hyperk\"ahler quotient of $\fMM(p,2)$ by its triholomorphic $U(1)$ yields a partially resolved $D_{|p|}$ ALF space with a $\mathbb{Z}_{|p|}$ orbifold singularity.  The case $|p| = 2$ was described in some detail in \cite{Houghton:1997ei} as a certain infinite mass limit of an $\gm = H_1 + 2 H_2 + H_3$ monopole in $\mathfrak{su}(4)$ gauge theory.  See \cite{Cherkis:1998xca,Cherkis:1998hi} for a construction of $D_k$ ALF spaces from the point of view of singular monopoles and Nahm data, and \cite{Cherkis:2003wk} for an explicit construction of their metrics.  These spaces also arise in other physical contexts; see section 5 of \cite{MRVdimP1} for a discussion.  There we exhibited an explicit two-parameter family of spherically symmetric field configurations, corresponding to the locus of $\fMM(2,2)$ describing configurations in which both smooth monopoles are atop the defect.\footnote{In fact the monopoles do not have a point-particle interpretation on this locus.  Rather the energy density of the fields is spread on a spherical shell around the defect.}

The results of this section give explicit predictions for the kernel of the $\rG(\YY_\infty)$-twisted Dirac operator on all of these spaces.  It would be fascinating to try to reproduce some of these results by direct analysis, especially for the $|p| = 1$ case of the Dancer manifold, where the kernel exhibits some rather remarkable structure.

%%%%%%%%%%%%%%%%%%%%%%
\subsection{Electric walls, tropical labels, and magnetic anti-walls}\label{ssec:tropical}
%%%%%%%%%%%%%%%%%%%%%%

In this section we return to the list of vanilla particles \eqref{vanBPSspectrum} and consider the case of the vector multiplets, $\gamma_{\rm h} = \pm \alpha$.  These have nonzero pairing with the pure 't Hooft defect, $\gamma_{\rm c} = -\frac{|p|}{2} H_\alpha$, and so the walls are defined by $\zeta^{-1} Z_{\pm \alpha} \in \mathbb{R}_-$, leading to\footnote{We restrict ourselves to the regime described around \eqref{inverseMPmapf} for large $t$, where we see that the sign of $\Re(\zeta^{-1} Z_{\alpha}(u)) = \langle \alpha, \Re(\zeta^{-1} a(u)) \rangle$ is the same as the sign of $\langle \alpha, \YY_\infty \rangle$ when $X_\infty = 0$.}
\begin{align}
\widehat{W}(+\alpha) =&~ \left\{ (u,\zeta)~|~ \langle \alpha, \sx \rangle = 0~,~ \langle \alpha, \sy \rangle < 0 \right\}~, \cr
\widehat{W}(-\alpha) =&~ \left\{ (u,\zeta)~|~ \langle \alpha, \sx \rangle = 0~,~ \langle \alpha, \sy \rangle > 0  \right\}~.
\end{align}

We will confine our attention here to the fate of the pure 't Hooft defect as we approach $\widehat{W}(+\alpha)$ from the side $\langle \alpha, \sx \rangle > 0$.  We know that the pure 't Hooft defect does not bind to a purely electric particle, and hence we are crossing from an unstable region to a stable one.  But what is the new bound state on the other side of the wall?

In order to answer this question it is important to first realize that the new state will not carry the magnetic charge of a pure't Hooft defect---indeed it cannot: the semiclassical analysis demonstrates that purely electric particles do not bind to the pure 't Hooft defect in any (weak coupling) chamber.  The asymptotic magnetic charge of the state does not change, as the binding particle does not  carry magnetic charge.  Rather, it is our designation of which asymptotic magnetic charge corresponds to the `pure 't Hooft defect' state that changes when $\langle \alpha,\sx \rangle$ goes from positive to negative:  our definition of positive root changes and hence the \emph{relative} magnetic charge, which depends on choosing the representative of $P$ in the anti-fundamental Weyl chamber, changes.  If the relative magnetic charge vanished when $\langle \alpha, \sx \rangle > 0$, then it will be $\tilde{\gamma}_{\rm m} = |p| H_\alpha$ when $\langle \alpha, \sx \rangle < 0$, after crossing the wall.

The IR charge label that we associate to the pure 't Hooft defect is an example of a \emph{tropical label}, as defined in \cite{Gaiotto:2010be} and elaborated on in \cite{Cordova:2013bza}.  The pure 't Hooft defect state is the vacuum state of the theory in the presence of the UV line defect, and so dominates the asymptotics of the line defect vevs discussed in \cite{Gaiotto:2010be}.  It was argued there that the label should change along BPS \emph{anti-walls}, defined by
\begin{equation}
\widecheck{W}(\gamma) := \left\{ (u,\zeta) ~|~ \zeta^{-1} Z_{\gamma}(u) \in -i \mathbb{R}_+ \right\}~,
\end{equation}
where $\gamma = \gamma_2 - \gamma_1$, with $\gamma_1(\gamma_2)$ the label before(after) crossing the anti-wall.\footnote{We drop the condition in \cite{Gaiotto:2010be} that $\gamma$ be a populated charge in the vanilla spectrum; this is appropriate when the line defect under consideration is `simple', which here corresponds to the case $|p|=1$.}  Now we observe that the electric wall $\widehat{W}(+\alpha)$ coincides with magnetic anti-walls, $\widecheck{W}(s H_\alpha)$, for any $s \in \mathbb{N}$.  Meanwhile the asymptotic charge label goes from $\gamma_1 = -\frac{|p|}{2} H_\alpha$ to $\gamma_2 = + \frac{|p|}{2} H_\alpha$.  This is consistent with crossing the magnetic anti-wall corresponding to $s = |p|$.

Thus let us return to the fate of the pure 't Hooft defect as we cross the wall.  Since the asymptotic magnetic charge does not change, it is not a pure 't Hooft defect on the other side.  Rather, since the relative magnetic charge is $|p| H_\alpha$, we are dealing with a bound state of $|p|$ smooth monopoles and one $W$-boson to the pure 't Hooft defect.  In the special case $|p| = 1$ we can identify the bound state precisely.  It is the zero mode of the ${\rm G}$-twisted Dirac operator on Taub-NUT space with electric charge $\alpha$ (corresponding to $\nu = -1/2$ in the language of appendix \ref{appendix:TN}) and hence angular momentum $j=0$.  The fact that the new bound state is an angular momentum singlet is consistent with \eg\ the primitive wall crossing formula.  The halo particle ($W$-boson) had no internal degeneracy and its charge pairing with the pure 't Hooft defect gives $\langle \alpha, -\half H_\alpha \rangle = -1$.  Hence the pure 't Hooft defect becomes a bound state of a dyon with the pure 't Hooft defect upon crossing the wall $\widehat{W}(\alpha)$.

What state, then, becomes the pure 't Hooft defect?  Clearly it is the reverse process.  Start at $\langle \alpha, \sx \rangle > 0$ with the $\gamma_{\rm e} = \alpha$ angular momentum singlet state on Taub--NUT, corresponding to $|p| = 1$ and asymptotic magnetic charge $\gm = +\half H_\alpha$.  Now the pairing of $\gamma_{\rm h} = \alpha$ with the charge of this state has the same magnitude but opposite sign.  Hence we are losing a halo.  When we cross the wall two things happen: the bound state radius of the $W$-boson goes to infinity and it unbinds, and the magnetic charge $+\half H_\alpha$ becomes that of the pure 't Hooft defect.  The smooth monopole that accounted for the initially nonzero relative magnetic charge is swallowed by the defect.

The process of creating or destroying smooth monopoles when the wall $\langle \alpha, \sx \rangle$ is crossed was described in detail from a D-brane perspective in \cite{MRVdimP2}, where we dubbed it \emph{monopole extraction}.  The wall crossing of the framed BPS states just described is the semiclassical analog of that classical process.  This makes it clear that monopole extraction is a physically distinct process from monopole bubbling.  Monopole extraction corresponds to wall crossing on the Coulomb branch of the supersymmetric gauge theory in the presence of a fixed UV line defect, while monopole bubbling corresponds to a process in which the UV defect itself is changed.

%%%%%%%%%%%%%%%%%%%%%%
%%%%%%%%%%%%%%%%%%%%%%
\section{Further Directions}\label{Section:Further}
%%%%%%%%%%%%%%%%%%%%%%
%%%%%%%%%%%%%%%%%%%%%%

In conclusion, we mention here some potentially interesting
future directions.

We have made a number of statements about the $\Lsq$-kernel
of Dirac-like operators on monopole moduli spaces. Some
of these such as the chiral nature of the kernel, and the
wall crossing behavior are rather general. It would be gratifying
if these properties could be confirmed using rigorous
mathematical arguments. For example, it  is heuristically
clear that on the walls of stability the Dirac-like operators
fail to be Fredholm because a gap between the boundstate and the
continuum closes. It would be nice to verify this in more detail.
We have also made a number of specific claims regarding the
dimensions of the $\Lsq$-kernels on certain moduli spaces of
arbitrarily high dimension. It seems challenging to verify
these statements using standard mathematical techniques.

As far as we are aware, standard index theorems do not apply
to the Dirac-like operators we have been discussing, although special cases have been
discussed in \cite{Stern:2000ie}. Nevertheless, a physical argument
suggests that the dimensions of the kernels of Dirac-like operators should
be expressible as integrals of characteristic classes over the monopole moduli
spaces for the following reason. We consider the path integral of the
$\NN=2$ supersymmetric Yang-Mills theory on $\mathbb{R}^3\times S^1$
with a supersymmetric Wilson-'t Hooft operator wrapping the circle.
On the one hand, given its relation to topological field theory \cite{Witten:1988ze}
one expects that the path integral will localize
to an integral of characteristic classes on the moduli space of
periodic instantons. On the other hand, the ``Darboux expansion''
described in \cite{Gaiotto:2010be} together with the results of the present
paper show that these integrals of characteristic classes should
give the dimension (and spin character) of the kernel of the
Dirac operator.

Another future direction is to use the results of \cite{Sethi:1995zm,Gauntlett:1995fu,Gauntlett:2000ks,Tong:2014yla}  to
generalize our equations \eqref{mainres}, \eqref{mainres2} and their
wall crossing properties to general supersymmetric Yang-Mills theory
coupled to hypermultiplet matter in the presence of general
't Hooft-Wilson lines. We expect the generalization to be straightforward
but there are certainly many nontrivial details.  These are currently being worked out in \cite{BrennanMoore}.

%%%%%%%%%%%%%%%%%
%%%%%%%%%%%%%%%%%
\section*{Acknowledgements}
%%%%%%%%%%%%%%%%%
%%%%%%%%%%%%%%%%%%

It is a pleasure to thank Sergey Cherkis, Sebastian Guttenberg, Nigel Hitchin, Claude LeBrun, Andy Neitzke, Daniel Robbins, Erick Weinberg and Edward Witten for helpful discussions.  GM and ABR thank the Aspen Center for Physics  and the NSF grant \#1066293 for hospitality during the completion of this work.  DVdB is partially supported by TUBITAK grant 113F164 and by the Bo\u{g}azi\c{c}i University Research Fund under grant number 13B03SUP7, GM is supported by the U.S. Department of Energy under grant DE-FG02-96ER40959, and ABR is supported by the Mitchell Family Foundation.  GM also thanks the KITP for hospitality during the final phases of the preparation of this paper (National Science Foundation under Grant No. NSF PHY11-25915); ABR thanks the NHETC at Rutgers University.

%%%%%%%%%%%%%%%%%%%%%%%%%%%%%%%%%%%%%%%%%%%%%%%%%%%
\appendix

%%%%%%%%%%%%%%%%%%%%
%%%%%%%%%%%%%%%%%%%%
\section{$\NN = 2$ supersymmetry: notation and conventions}\label{N2conventions}
%%%%%%%%%%%%%%%%%%%%
%%%%%%%%%%%%%%%%%%%%

We study $\NN = 2$, $d=4$ SYM with gauge group $G$ and no hypermultiplets.  The fermionic completion of the bosonic action (\ref{Scl}) is
\begin{align}\label{action}
S  =&~ -\frac{1}{g_0^2}\int d^4 x \Tr \left\{ \frac{1}{2} F_{\m\n} F^{\m\n} +  D_\m \varphi D^\m \varphibar -\left(i \psibar^A \overbar{\sigma}^\m D_\m \psi_A + i \psi_A \sigma^\mu D_\mu \psibar^A\right) +  \right. \cr
&~ \left. + i\epsilon^{AB} \psi_A [ \psi_B, \varphibar ]  +i \epsilon_{AB} \psibar^A [ \psibar^B , \varphi ]  - \frac{1}{4} [ \varphi, \varphibar ]^2  \right\} + \frac{\theta_0}{8\pi^2} \int \Tr F \wedge F + S_{\rm def} \cr
\equiv &~ S_{\rm van} + S_{\rm def}~.
\end{align}
Here $S_{\rm van}$ is the ordinary ``vanilla'' $\NN = 2$ action in the absence of defects while $S_{\rm def}$ contains the additional terms to be studied in appendix \ref{app:bcs}.  The two fermi kinetic terms are the same up to a boundary term, however on a manifold with boundary this makes a difference and it is \eqref{action} that is real.  Our conventions follow Bagger and Wess \cite{Wess:1992cp} with regards to the Lorentz structure of the anti-commuting spinors, the $\sigma$-matrices, and canonical index placement. We do however use a different orientation, $\epsilon_{0123}=1$ , so that for us $\sigma^{\m\n}=-\frac{i}{2}\epsilon^{\m\n\r\s}\sigma_{\r\s}$ .

Due to the presence of an $SU(2)_R$ symmetry the Weyl spinors appearing in (\ref{action}, \ref{susyvarGMN}) are furthermore valued in the fundamental $SU(2)$ representation: $\psi(x), \xi(x) \in S^+\otimes \mathbb{C}^2 \otimes \mathfrak{g}$, with $S^+$ the space of positive chirality Weyl fermions. As the $SU(2)_R$ will play an important role in this paper we will almost always make it manifest by explicitly writing its representation index, for which we use capital Latin script, \eg\ $\psi^A, \xi^A$. We take the $SU(2)_R$ to be generated by ($-\frac{i}{2}$ times) the three Pauli matrices $(\sR^{r})^{A}_{\phantom{A}B}$. One can consistently extend the complex conjugation rules of \cite{Wess:1992cp} to the spinor doublets as follows
\begin{equation}\label{conjdef}
(\psi_{\alpha A})^\ast = - \psibar_{\dot{\alpha}}^A~, \quad (\psi^{\alpha}_{A})^\ast = - \psibar^{\dot{\alpha} A}~, \quad (\psi_{\alpha}^A)^\ast = \psibar_{\dot{\alpha} A}~, \quad (\psi^{\alpha A})^\ast = \psibar^{\dot{\alpha}}_A~.
\end{equation}

An antisymmetric $\epsilon$ tensor is introduced for each type of index and can be used to raise or lower it. Our conventions on all $\epsilon$ tensors will be $\epsilon^{12} = -\epsilon_{12} = 1$; explicitly we have
\begin{equation}\label{greekeps}
\psi_{\alpha} = \epsilon_{\alpha\beta} \psi^\beta~, \quad \psibar_{\dot{\alpha}} = \epsilon_{\dot{\alpha}\dot{\beta}} \psibar^{\dot{\beta}}~,\quad \psi_A=\epsilon_{AB}\psi^B~,\quad \psibar_A=\epsilon_{AB}\psibar^B~ .
\end{equation}

The  vanilla action $S_{\rm van}$ of \eqref{action} is invariant under the supersymmetry transformations
\begin{align}\label{susyvarGMN}
& \delta_{\xi} \varphi  = -2 \xi_A \psi^A ~, \qquad \delta_{\xi} \varphibar  = 2 \xibar^A \psibar_A~, \cr
& \delta_{\xi} A_\m =   -\xi_A \sigma_\m \psibar^A + \xibar^A \overbar{\sigma}_\m \psi_A ~, \cr
& \delta_{\xi} \psi_{A} = -i \sigma^{\m\n}\xi_{A} F_{\m\n} +i  \sigma^\m \xibar_{A} D_\m \varphi  + \frac{i}{2} \xi_{ A} [\varphi, \varphibar] ~, \cr
& \delta_{\xi} \psibar^{A} = i  \overbar{\sigma}^{\m\n}\xibar^{A}F_{\m\n} - i  \overbar{\sigma}^\m\xi^{ A} D_\m \varphibar  + \frac{i}{2} \xibar^{A} [\varphi, \varphibar] ~,
\end{align}
up to boundary terms.  Specifically, allowing for a spatially varying susy parameter as well, a tedious but straightforward computation leads to
\begin{equation}\label{Sbulksusyvar}
\delta_{\xi} S_{\rm van} = \int \ed^4 x \left\{ (\pd_\mu \xi_A) S^{\mu A} - (\pd_\mu \xibar^A) \overbar{S}_{A}^\mu + \pd_\mu \left( \xi_A \BB^{\mu A} - \xibar^A \overbar{\BB}_{A}^\mu \right) \right\}~,
\end{equation}
where the supercurrent $S_{\alpha}^{\mu A}$ is given by
\begin{align}
S_{\alpha}^{\mu A} :=&~ \frac{1}{g_{0}^2} \Tr \bigg\{ 2 \left(F^{\mu\nu} - \frac{i}{2} \epsilon^{\mu\nu\rho\lambda} F_{\rho\lambda} \right) (\s_\nu)_{\alpha\dot{\alpha}} \psibar^{\dot{\alpha} A} +  \cr
&~ \qquad \qquad + 2 \left( \eta^{\mu\nu} {\delta_\alpha}^\beta + 2{(\s^{\mu\nu})_\alpha}^\beta \right) \psi_{\beta}^A D_\nu \varphibar + (\s^\mu)_{\alpha\dot{\alpha}} \psibar^{\dot{\alpha} A} [\varphi,\varphibar] \bigg\}~,
\end{align}
up to possible improvement terms of the form $\pd_\nu \mathcal{X}^{[\nu\mu]}$ that will not concern us.  Meanwhile the boundary current is given by
\begin{align}\label{bsusycurrent}
\BB_{\alpha}^{\mu A} :=&~ \frac{1}{g_{0}^2} \Tr \bigg\{ \left(F^{\mu\nu} - \frac{\theta_0 g_{0}^2}{8\pi^2} \epsilon^{\mu\nu\rho\lambda} F_{\rho\lambda} + \frac{i}{2} \epsilon^{\mu\nu\rho\lambda} F_{\rho\lambda} \right) (\s_\nu)_{\alpha\dot{\alpha}} \psibar^{\dot{\alpha} A}  + \cr
& \qquad \qquad - {(\s^\mu \sb^\nu)_{\alpha}}^\beta \psi_{\beta}^A D_\nu \varphibar  -  \half (\s^\mu)_{\alpha\dot{\alpha}} \psibar^{\dot{\alpha} A} [\varphi,\varphibar] \bigg\}~.
\end{align}

In the absence of defects, the standard asymptotic falloff conditions ensure that the boundary term of \eqref{Sbulksusyvar}, associated with the two-sphere at spatial infinity, vanishes.  Thus one finds that the action is invariant under rigid $\NN = 2$ supersymmetry.  The spatial integrals of the time components of the supercurrents give the field representation of the supercharges,
\begin{align}\label{Qfield}
Q_{\alpha}^A =&~ \frac{2}{g_{0}^2} \int \ed^3 x \Tr \bigg\{ (E_i + i B_i) (\s^i)_{\a \adot} \psibar^{\adot A} - \psi_{\a}^A D_0 \varphibar + {(\s^0 \sb^i)_{\a}}^{\b} \psi_{\b}^A D_i \varphibar + \cr
&~ \qquad \qquad \qquad + \half (\s^0)_{\a\adot} \psibar^{\adot A} [\varphi,\varphibar] \bigg\}~,
\end{align}
This expression can be used to compute the canonical bracket of two supercharges as in \cite{Witten:1978mh}.  Setting $\{ Q_{\alpha}^A, Q_{\beta}^B \}_+ = 2 \epsilon_{\alpha\beta} \epsilon^{AB} \Zbar$, one obtains the classical central charge
\begin{equation}
Z^{\rm cl} = \frac{2}{g_{0}^2} \int \ed^3 x \pd^i \Tr \left\{ (-E_i + i B_i) \varphi \right\}~,
\end{equation}
which reduces to a boundary contribution from the asymptotic two-sphere.

%%%%%%%%%%%%%%%%%%%%
\subsection{Real variables and $\RR$ supersymmetries}\label{app:real}
%%%%%%%%%%%%%%%%%%%%

We will call a Dirac spinor $\chi=(\alpha_\a,\overbar{\beta}^{\dot\a})^T$ a Majorana spinor if it satisfies the reality condition
\be
\overbar\beta=\overbar\sigma^0\alpha\,. \label{Majorana}
\ee
Recall that in four dimensions there are no nontrivial Majorana-Weyl spinors, as the condition $\psibar=\overbar{\sigma}^0\psi$ implies that $\psi=(\overbar{\sigma}^0\psi)^*=-\sigma^0\psibar=-\psi$. However, as the fermions appearing in the $\NN=2$ theory all come in $SU(2)_R$ doublets, we can use this to define a symplectic-Majorana-Weyl condition:
\begin{equation}
\overbar\lambda^A=\overbar{\sigma}^0\lambda^A ~.
\end{equation}
Note that this has non-zero solutions as $\overbar{\lambda}^{A}=\epsilon^{AB}(\lambda^B)^*$. We will denote the space of symplectic-Majorana-Weyl spinors as
\begin{equation}\label{smwspace}
S_{\mathrm{smw}}^{+}=\left\{\lambda\in S^{+}\otimes \mathbb{C}^2\,|\, \overbar{\lambda}^A=\overbar{\sigma}^0\lambda^A\right\}\,.
\end{equation}

The following change of variables is often employed:
\begin{equation}\label{realvars}
\varphi = \zeta (Y+iX)~, \qquad \psi^A = \zeta^{\half} (\rho^A + i \lambda^A)~, \qquad \psibar^A =  \zeta^{-\half} \sb^0 (\rho^A - i \lambda^A)~,
\end{equation}
where $\rho^A,\lambda^A \in S_{\rm smw}^+$ are symplectic--Majorana--Weyl spinors.  In terms of these variables the vanilla action takes the form
\begin{align}\label{action2ndform}
S_{\rm van} =&~ -\frac{1}{g_{0}^2} \int \ed^4 x \Tr \bigg\{ \half F_{\mu\nu} F^{\mu\nu} + D_\mu X D^\mu X + D_\mu Y D^\mu Y + [X, Y]^2 +  \cr
& \qquad \qquad \qquad \quad + 2i \rho^A D_0 \rho_A + 2i \lambda^A D_0 \lambda_A  + 2 \lambda^A \s^{0i} D_i \rho_A - 2 \rho^A \s^{0i} D_i \lambda_A + \cr
& \qquad \qquad \qquad \quad  -2i \left( \lambda^A [Y, \lambda_A] - \rho^A [Y,\rho_A] \right)  + 2i \left( \rho^A [ X, \lambda_A] + \lambda^A [ X, \rho_A] \right) \bigg\} \cr
&~ + \frac{\theta_0}{8\pi^2} \int \Tr F \wedge F~.
\end{align}

We define the supersymmetry generators, corresponding to the transformations (\ref{susyvarGMN}), through their action on the fields as
\begin{equation}\label{susyvargen}
\delta_{\xi} \Phi := [ Q_A \xi^{A} + \overbar{Q}^A\xibar_A, \Phi ]~,
\end{equation}
Line defects of type $\zeta$ should preserve the $\RR_{\alpha}^A$ supersymmetries, \eqref{Rsusys}, while the orthogonal linear combinations, $\TT_{\alpha}^A$, \eqref{Tsusys}, are broken.

We split the complex supersymmetry parameter $\xi$ and into two symplectic-Majorana-Weyl components, $\varepsilon, \eta$:
\bea\label{susysplits}
\xi^A &=& \zeta^{\frac{1}{2}} ( \varepsilon^A + i\eta^A)~.
\eea
From this and \eqref{QRT} it follows that \eqref{susyvargen} can be written as
\be\label{RTvar2}
\delta_{\xi} \Phi = - \left[ \eta_A \TT^A,\Phi\right] + \left[ \varepsilon_A \RR^A, \Phi \right] \equiv \delta_\varepsilon\Phi+\delta_\eta\Phi \,.
\ee
Hence $\eta_A$ generates the $\TT^A$ supersymmetries while $\varepsilon_A$ generates the $\RR$-supersymmetries.  More explicitly, using \eqref{susysplits} and the field redefinitions \eqref{realvars} in the variations \eqref{susyvarGMN}, one finds that the action on the fields is
\bea\label{epsvar}
\delta_\ve X&=&2i\ve_A\r^A\\
\delta_\ve Y&=&-2i\ve_A\l^A\\
\delta_\ve A_0&=&-2i\ve_A\l^A\\
\delta_\ve A_i&=&2\ve_A\s^0\bar\s_i\r^A\\
\delta_\ve \r^A&=&\left[-(D_0X-[Y,X])+i(E_i-D_iY)\s^0\bar\s^i\right]\ve^A \label{fermvar1}\\
\delta_\ve \l^A&=&\left[D_0Y+i(B_i-D_iX)\s^0\bar\s^i\right]\ve^A \label{fermvar2}
\eea
\bea
\delta_\eta X&=&-2i\eta_A\l^A \label{etavar1} \\
\delta_\eta Y&=&-2i\eta_A\r^A\\
\delta_\eta A_0&=&2i\eta_A\r^A \label{etavar2}\\
\delta_\eta A_i&=&2\eta_A\s^0\bar\s_i\l^A \\
\delta_\eta \r^A&=&\left[D_0Y-i(B_i+D_iX)\s^0\bar\s^i\right]\eta^A\\
\delta_\eta \l^A&=&\left[D_0X +[Y,X]+i(E_i+D_iY)\s^0\bar\s^i\right]\eta^A \label{etavarlast}
\eea

The conditions for the invariance of a field configuration under $\RR$-supersymmetry, \ie\ the $\RR$-fixed point equations, are simply $\delta_{\varepsilon} \Phi = 0$.  In the case of a purely bosonic field configuration only the variations \eqref{fermvar1}, \eqref{fermvar2} are nontrivial. The condition for them to vanish is equivalent to the BPS equations (\ref{BPSreal}).

%%%%%%%%%%%%%%%%%%%%%
%%%%%%%%%%%%%%%%%%%%%
\section{Defects and boundaries}\label{app:bcs}
%%%%%%%%%%%%%%%%%%%%%
%%%%%%%%%%%%%%%%%%%%%

%%%%%%%%%%%%%%%%%%%%%
\subsection{UV defects and boundary terms}
%%%%%%%%%%%%%%%%%%%%%

The variation of the vanilla action $S_{\rm van}$ leads to terms proportional to the equations of motion plus a set of boundary terms.  The equations of motion are
\begin{align}\label{cmplxeoms}
0=&~ D^\mu F_{\mu\nu} - \half [\varphi, D_\nu \varphibar] - \half [\varphibar, D_\nu \varphi] - i [\psibar^A, \overbar{\sigma}_\nu \psi_A] ~, \cr
0 =&~ D_\mu D^\mu \varphi - i [\psi_A, \psi^A] - \half [\varphi,[\varphi,\varphibar] ]~, \cr
0 =&~  \overbar{\sigma}^\mu D_\mu \psi_A + [\varphi, \psibar_A]~.
\end{align}
When they are satisfied the variation reduces to
\begin{align}\label{Svaronshell}
\delta S_{\rm van} =&~  - \frac{1}{g_{0}^2} \int \ed^4 x \pd_i \Tr \displaystyle\biggl\{ \delta A_\nu \left( 2 F^{i \nu} - \frac{g_{0}^2 \theta_0}{8\pi^2} \epsilon^{i \nu \rho\sigma} F_{\rho\sigma} \right) + \cr
& \qquad \qquad \qquad \qquad  + \delta \varphi D^i \varphibar + \delta \varphibar D^i \varphi -   i \psibar^A \overbar{\sigma}^i \delta \psi_A  + i (\delta \psibar^A) \overbar{\sigma}^i \psi_A \displaystyle\biggr\}  \cr
=&~ \frac{1}{g_{0}^2} \int \ed^4 x \pd^i \Tr \bigg\{ \delta A_0 \left( 2 E_i + \frac{g_{0}^2 \theta_0}{4\pi^2} B_i \right) + \delta A^j \left( -2 F_{ij} + \frac{g_{0}^2 \theta_0}{4\pi^2} \epsilon_{ijk} E^k \right) + \cr
& \qquad \qquad \qquad \quad - 2 \delta X D_i X - 2 \delta Y D_i Y + i \psibar^A \sb_i \delta \psi_A - i (\delta \psibar^A) \sb_i \psi_A \bigg\}~. \quad
\end{align}
Here we have assumed that the fields satisfy Dirichlet conditions on the spacelike boundaries at $t =\pm \infty$ and therefore have restricted attention to the timelike boundaries of the form $\mathbb{R}_t \times \pd \UU$, where $\UU = \mathbb{R}^3 \setminus \{ \vec{x}_n \}_{n=1}^{N_t}$.  $\pd \UU$ consists of the asymptotic two-sphere $S_{\infty}^2$ at spatial infinity and infinitesimal two-spheres $S_{\varepsilon_n}^2$ of radius $\varepsilon_n$ around each $\vec{x}_n$.  Let us focus on the infinitesimal two-spheres.

We want to impose boundary conditions consistent with the insertion of an 't Hooft defect.  These should include $1/r_{n}^2$ singularities in $B_i$ and $D_i X$.  The $\delta A_0$ term in \eqref{Svaronshell} suggests a corresponding singularity in the $E$-field when $\theta_0 \neq 0$, and this is consistent with expectations from the Witten effect for line defects \cite{Kapustin:2005py,Henningson:2006hp,Gomis:2011pf}.  Then, having boundary conditions that are consistent with some supersymmetry will require a singularity in $Y$ as well.  Hence our defect boundary conditions are taken to be
\begin{align}\label{defectbcstrial}
& B^i = \frac{P}{2r_{n}^2} \hat{r}_{n}^i + O(r_{n}^{-2+\updelta})~, \qquad  & X = -\frac{P}{2r_n} + O(r_{n}^{-1+\updelta})~, \cr
& E_i = -\frac{\theta_0 g_{0}^2}{8\pi^2} \cdot \frac{P}{2r_{n}^2} \hat{r}_{n}^i + O(r_{n}^{-2+\updelta})~, \qquad & Y = \frac{\theta_0 g_{0}^2}{8\pi^2} \cdot \frac{P}{2r_n} + O(r_{n}^{-1+\updelta})~,
\end{align}
as $r_n \equiv |\vec{x} - \vec{x}_n| \to 0$, where $\updelta > 0$ is to be determined.  Since $P$ is assumed time independent, these magnetic and electric fields correspond to a gauge field of the form
\begin{equation}
A = \frac{P}{2} (\pm 1 - \cos{\theta}_n) \ed \phi_n + \frac{g_{0}^2 \theta_0}{8\pi^2} \cdot \frac{P}{2r_n} \ed t + O(r^{-1+\updelta})~.
\end{equation}
We also assume that the fermions $\psi_A = O(r_{n}^{-1+\updelta})$, which we will show is consistent with the equations of motion.  $\updelta$ controls the subleading behavior of the fields in the vicinity of the defect and the leading behavior of their variations, $\delta A_\mu, \delta X,\delta Y$.  Hence we see there are terms in \eqref{Svaronshell} that behave as $O(\varepsilon_{n}^{-3+\updelta})$ on the infinitesimal two-sphere $S_{\varepsilon_n}^2$, diverging as $\varepsilon_n \to 0$ even for $\updelta = 1$.  

These divergences must be canceled by the variation of the defect action in order for the boundary conditions \eqref{defectbcstrial} to be consistent with the variational principle.  In \cite{MRVdimP1}, where we studied the truncation of the above theory to Yang--Mills--Higgs theory with a real scalar $X$ and $\theta_0 = 0$, we showed that there is a natural choice for the defect action such that its variation cancels these divergences.  Furthermore this action leads to an energy functional that is finite in the presence of defects and satisfies the standard Bogomolny bound.  The generalization to the full $\NN = 2$ theory is
\begin{equation}\label{Sbndryapp}
S_{\rm def} = -\frac{2}{g_{0}^2}\int \ed t \sum_n \int_{S_{\varepsilon_n}^2} \ed^2 \Omega_n r_{n}^2 \hat{r}_{n}^i \Tr \left\{ E_i Y + B_i X \right\}~,
\end{equation}
the variation of which takes the form
\begin{align}\label{bndrybosvar}
\delta S_{\rm def} =&~ -\frac{2}{g_{0}^2}\int \ed t \sum_n \int_{S_{\varepsilon_n}^2} \ed^2 \Omega_n r_{n}^2 \hat{r}_{n}^i \Tr \left\{ Y \delta E_i + E_i \delta Y + \delta A^j \epsilon_{ijk} D^k X + B_i \delta X \right\}~.
\end{align}
Here we have used that $\delta B_i = \epsilon_{ijk} D^j \delta A^k$ involves only tangential derivatives and thus can be integrated by parts.  However $\delta E_i = D_i \delta A_0 - D_0 \delta A_i$ involves a normal derivative and must be treated as an independent variation on the boundary.

We assume standard falloff conditions for the fields as $|\vec{x}| \to \infty$, such that the asymptotic two-sphere does not contribute to the variation.  Then, combining \eqref{Svaronshell} with \eqref{bndrybosvar}, we have
\begin{align}\label{Svar}
\delta S =&~  \frac{2}{g_{0}^2} \int \ed t \sum_n \int_{S_{\varepsilon_n}^2} \ed^2 \Omega_n r_{n}^2 \hat{r}_{n}^i \Tr \bigg\{ - \delta A_0 \left( E_i + \frac{g_{0}^2 \theta_0}{8\pi^2} B_i \right) - \delta E_i Y - \frac{g_{0}^2 \theta_0}{8\pi^2} \delta A^j \epsilon_{ijk} E^k + \cr
&~  \qquad \qquad + \delta A^j ( F_{ij} - \epsilon_{ijk} D^k X) + \delta X (D_i X - B_i) + \delta Y (D_i Y - E_i) + \cr
&~ \qquad \qquad - \frac{i}{2} \left( \psibar^A \sb_i \delta \psi_A - (\delta \psibar^A) \sb_i \psi_A\right)\bigg\}~. \raisetag{20pt}
\end{align}
For the last term in the first line we use that $E^k = D^k Y + O(\varepsilon_{n}^{-2+\updelta})$, and integrate by parts to write
\begin{equation}\label{deltaEB}
- \delta E_i Y - \frac{g_{0}^2 \theta_0}{8\pi^2} \delta A^j \epsilon_{ijk} E^k = -Y \left( \delta E_i + \frac{g_{0}^2 \theta_0}{8\pi^2} \delta B_i \right) + O(\varepsilon_{n}^{-3 + 2\updelta})~.
\end{equation}
We then make this replacement in \eqref{Svar}.  Now, given \eqref{defectbcstrial}, we have that all the other terms of $\delta S$ \emph{naively} go as $O(\varepsilon_{n}^{-3+2\updelta},\varepsilon_{n}^{-5/2 + \updelta})$.  Hence we would have concluded that any $\updelta > \half$ will lead to a consistent variational principle; \ie\ $\delta S = 0$ on a solution to the equations of motion.  However the right-hand side of \eqref{deltaEB} is still \emph{naively} $O(\varepsilon^{-3 + \updelta})$, which would seem to require $\updelta > 1$.

As we discussed in \cite{MRVdimP1}, there are explicit constructions of singular monopole solutions where the subleading behavior of the gauge and Higgs field is $O(r_{n}^{-1/2})$, corresponding to the value $\updelta = \half$.  It is important for the consistency of these solutions that $\updelta = \half$ be an admissible value.  In \cite{MRVdimP1} we demonstrated that any solution to the \emph{second order} equations of motion, satisfying the defect boundary conditions, also solves the first order BPS equations at the first subleading order.  In other words, although \eqref{defectbcstrial} only implies $B_i - D_i X = O(\varepsilon_{n}^{-2+\updelta})$, any solution to the equations of motion will in fact have $B_i - D_i X = O(\varepsilon_{n}^{-2+\updelta + \updelta'})$ for some $\updelta' > 0$.  Clearly we need a generalization of this argument here, which also reduces the naive degree of divergence from \eqref{deltaEB} sufficiently.

For completeness we provide this argument in the subsection below; it relies on some details that were worked out in \cite{MRVdimP1}.  The main result can be summarized as follows.  The equations of motion \eqref{cmplxeoms} together with the defect boundary conditions \eqref{defectbcstrial} imply that the strongest possible subleading behavior of the fields corresponds to $\updelta = \half$ and furthermore that
\begin{equation}\label{defectbcsaux}
B_i - D_i X = O(\varepsilon_{n}^{-1/2})~, \quad E_i - D_i Y = O(\varepsilon_{n}^{-1/2})~, \quad E_i + \frac{\theta_0 g_{0}^2}{8\pi^2} B_i = O(\varepsilon_{n}^{-1/2})~,
\end{equation}
rather than the naive $O(\varepsilon_{n}^{-3/2})$ as follows from \eqref{defectbcstrial} alone.  These conditions are sufficient to ensure consistency of the variational principle when $\updelta = \half$.

%%%%%%%%%%%%%%%%%%%%%
\subsection{Admissibility of defects with $r^{-1/2}$ subleading behavior}\label{app:deltahalf}
%%%%%%%%%%%%%%%%%%%%%

The idea is to analyze the equations of motion perturbatively in the distance from the defect to determine if the first subleading terms near the defect automatically satisfy the same boundary condition that the leading terms do.  For this analysis it is convenient to use the variables \eqref{realvars}.  Now we introduce the following notation.  Define $A_M \equiv (A_\mu, X,Y)$, thinking of $X,Y$ as the fourth and fifth components of a 6D gauge field.  All fields will be independent of these extra coordinates, so $D_4 = \ad(X)$ and $D_5 = \ad(Y)$.  We also denote the first four spatial directions by an index $a,b,\ldots$, so $A_M = (A_0, A_a,Y)$.  We  define Euclidean sigma matrices $\tau^a, \bar{\tau}^a$
\begin{equation}
{(\bar{\tau}^a)_{\a}}^{\b} = \left( ( {\sigma^{0i})_{\a}}^{\b} , i {\delta_\a}^{\b} \right)~, \qquad {(\tau^a)_{\a}}^{\b} = \left( ( {\sigma^{0i})_{\a}}^{\b} , -i {\delta_\a}^{\b} \right)~,
\end{equation}
so that $\tau^a = (\vec{\sigma}, - i \mathbbm{1} )$, $\bar{\tau}^a = (\vec{\sigma}, i \mathbbm{1})$.  Then one may show that the action \eqref{action2ndform} takes the form
\begin{align}
S_{\rm van} =&~ - \frac{1}{g_{0}^2} \int \ed^4 x \Tr \bigg\{ \half F_{MN} F^{MN} + 2i \rho^A \left( D_0 \rho_A + [Y, \rho_A] \right) + \cr
& \qquad \qquad \qquad \quad + 2i \lambda^A \left( D_0 \lambda_A - [Y, \lambda_A] \right) + 2\lambda^A \bar{\tau}^a D_a \rho_A -2 \rho^A \tau^a D_a \lambda_A \bigg\} + \cr
&~ + \frac{\theta_0}{8\pi^2} \int \Tr F \wedge F~,
\end{align}
which is more suitable for the linearized analysis to follow.  The $\theta_0$ term can be ignored in this subsection as it plays no role in the analysis of the equations of motion.

We analyze the equations of motion around a particular defect and choose our coordinate system so that this defect is located at $\vec{x} = 0$.  Thus we expand around a bosonic background $\tilde{A}_M$, where we take
\begin{align}\label{Cartanbkgrnd}
& \tilde{A}_0 = \frac{g_{0}^2 \theta_0}{8\pi^2} \cdot \frac{P}{2r} ~, \qquad & \tilde{A}_i = \frac{P}{2} (\pm 1 - \cos{\theta}) \ed\phi_i ~, \cr
& \tilde{X} = -\frac{P}{2r}~, \qquad  & \tilde{Y} = \frac{g_{0}^2 \theta_0}{8\pi^2} \cdot \frac{P}{2r}~,
\end{align}
which is a (Cartan-valued) exact solution to the equations of motion, satisfying the defect boundary conditions.  We write
\begin{align}
& A_M = \tilde{A}_M + \delta A_M~,  \qquad \rho^A = \delta \rho^A~, \qquad \lambda^A = \delta \lambda^A~.
\end{align} 
where we assume that $\delta A_M, \delta \rho, \delta \lambda = O(r^{-1+\updelta})$.  By analyzing the equations below we will see if this is a consistent assumption and if so, determine $\updelta$.
  
We want to determine what the equations of motion imply for the leading order behavior of the fluctuations.  For this it is sufficient to obtain the equations of motion linearized in the fluctuation fields.  Choosing the background Lorentz gauge,
\begin{equation}
\tilde{D}^M \delta A_M = 0~,
\end{equation}
one gets the standard form of the linearized gauge equation together with the Fermi equations:
\begin{align}\label{lineoms}
& ( \eta_{MN} \tilde{D}^P \tilde{D}_P + 2 \ad(\tilde{F}_{MN}) ) \delta A^N = O(\textrm{fermi}^2)~, \cr
& i (\tilde{D}_0 \lambda_A - [\tilde{Y}, \lambda_A]) + \bar{\tau}^a \tilde{D}_a \rho_A = 0~, \cr
& i (\tilde{D}_0 \rho_A + [\tilde{Y},\rho_A]) - \tau^a \tilde{D}_a \lambda_A = 0~,
\end{align}
where we neglected the fermi-squared terms in the bosonic equation since they are second order in fluctuations (and hence we will see from analyzing the fermi equations that they are suppressed relative to the other terms in the bosonic equation).

Let us begin with the $\rho$ equation.  Noting that $\tilde{A}_0 = \tilde{Y}$ we have
\begin{equation}
i \pd_0 \lambda_A + \bar{\tau}^a \tilde{D}_a \rho_A = 0~.
\end{equation}
Under our assumption $\delta \rho, \delta \lambda = O(r^{-1+\updelta})$, we see that the $\pd_0 \lambda_A$ term is suppressed relative to the remaining terms at small $r$, since the background $\tilde{A}_a$ goes as $O(1/r)$.  Hence, in order to obtain the leading order behavior of $\delta \rho_A$ we can drop this term so that
\begin{equation}
 \bar{\tau}^a \tilde{D}_a \rho_A = O(r^{-1+\updelta})~,
 \end{equation}
where the terms on the left are $O(r^{-2+\updelta})$.  Thus, at leading order, nontrivial $\delta \rho_A$ sit in the kernel of the Dirac operator $\bar{\tau}^a \tilde{D}_a$.  This is exactly the operator we analyzed in the section 4 of \cite{MRVdimP1}.  If we make a root expansion, $\rho_A = \rho_{A}^{(\alpha)} E_\alpha$ in the Lie algebra, then the leading behavior of the components are $r^{-1 + |p_\alpha|/2}$ where $p_\alpha = \langle \alpha, P \rangle$.  Hence for any $\alpha$ such that $\langle \alpha, P \rangle = 1$ we can have $r^{-1/2}$ behavior, and this is the strongest possible.  Thus we learn $\updelta = 1/2$.

Now we can plug the leading order $\rho$ solution into the $\lambda$ equation.  In the $\lambda$ equation $\tilde{A}_0$ and $\tilde{Y}$ add so that
\begin{equation}
\tau^a \tilde{D}_a \lambda_A = i[ \tilde{A}_0 + \tilde{Y}, \rho_A] + O(r^{-1+\delta})~,
\end{equation}
where we dropped the subleading $\pd_0 \lambda_A$.  Given a specific $\rho_A$ we can solve this inhomogeneous equation.  We also know that the kernel of $\tau^a \tilde{D}_a$ is trivial, so there is a unique solution for a given $\rho_A$.  The leading-$r$ behaviors of both sides match up if we take $\lambda = O(r^{-1/2})$, which is consistent with our assumption.

This means that the $O(\textrm{fermi}^2)$ terms in the bosonic equation are $O(r^{-1})$ and this is indeed subleading to the terms we have kept.  Writing out the bosonic equations, we first note that
\begin{align}
\tilde{D}^P \tilde{D}_P =&~ - \tilde{D}_{0}^2 + \tilde{D}^a \tilde{D}_a + \ad(\tilde{Y})^2 \cr
=&~ - \pd_{0}^2 - \pd_0 \cdot \ad(\tilde{A}_0) - \ad(\tilde{A}_0) \cdot \pd_0 + \tilde{D}^a \tilde{D}_a \cr
=&~ \tilde{D}^a \tilde{D}_a + O(r^{-1})~,
\end{align}
where the $\ad(\tilde{A}_0)^2$ term canceled against the $\ad(\tilde{Y})^2$ term, and the terms involving time derivatives are subleading to $\tilde{D}^a \tilde{D}_a$.  We also have that
\begin{align}
& \tilde{F}_{04} = \tilde{D}_0 \tilde{X}  = 0~, \qquad \tilde{F}_{05} = \tilde{D}_0 \tilde{Y} = 0~, \qquad \tilde{F}_{45} = [\tilde{X}, \tilde{Y}] = 0~, \qquad \tilde{F}_{i5} = \tilde{D}_i \tilde{Y} = \tilde{E}_i ~.
\end{align}
Then the $\delta A_0$ equation is
\begin{equation}
\tilde{D}^a \tilde{D}_a \delta A_0 + 2 \ad(\tilde{F}_{0i}) \delta A^i = O(r^{-2+\updelta}) \quad \Rightarrow \quad \tilde{D}^a \tilde{D}_a \delta A_0 = 2 [ \tilde{E}_i, \delta A^i] + O(r^{-2+\updelta})~,
\end{equation}
while the $\delta Y$ equation is
\begin{align}
& \tilde{D}^a \tilde{D}_a \delta Y  -2 \ad(\tilde{E}_i) \delta A^i = O(r^{-2+\updelta}) \quad \Rightarrow \quad  \tilde{D}^a \tilde{D}_a \delta Y = 2 [\tilde{E}_i, \delta A^i] + O(r^{-2+\updelta})~.
\end{align}
We also get $O(r^{-2+2\updelta})$ terms from the fermi terms but this is subleading to the $O(r^{-2+\updelta})$ that we have displayed.  Notice that these two equations are identical to the order displayed.  Taking the difference,
\begin{equation}
\tilde{D}^a \tilde{D}_a (\delta A_0 - \delta Y) = O(r^{-2 + \updelta})~.
\end{equation}

Thus either $\delta A_0 - \delta Y$ is annihilated by $\tilde{D}^a \tilde{D}_a$ or else is $O(r^{\updelta})$.  (The kernel of $\tilde{D}^a \tilde{D}_a$ is trivial when acting on $\Lsq$ sections, but we are doing an infinitesimal analysis around $r = 0$ and cannot make any assumptions about the large $r$ asymptotics of $\delta A_M$.)  Using the explicit background \eqref{Cartanbkgrnd}, it is straightforward to find the general set of solutions to 
\begin{equation}
\tilde{D}^a \tilde{D}_a f = 0~,
\end{equation}
for an adjoint-valued section $f$.  For the Cartan components of $f$ this equation reduces to $\pd^i \pd_i f = 0$ and the leading admissible behavior is $O(1)$.  For the components along root directions $E_\alpha$, we use $[P, E_\alpha] = -i \langle \alpha, P \rangle E_\alpha \equiv -i p_\alpha E_\alpha$ and find that the general form of the solution is
\begin{equation}
f^\alpha = e^{\pm i p_\alpha \phi/2} \sum_{j,m} a_{j,m}^\alpha r^j d_{p_\alpha/2,m}^j(\theta) e^{im\phi}~,
\end{equation}
where the $a_{j,m}^\alpha$ are constants, $j$ starts at $|p_\alpha|/2$ and increases in integer steps, $m$ runs from $-j$ to $j$ in integer steps, and $d_{m',m}^j$ is a Wigner little $d$ function.  The $\pm$ corresponds to the northern/southern patch of the two-sphere and is correlated with the sign in the background gauge field $\tilde{A}_i$.  This solution applies equally well for the Cartan components of $f$, provided we set $p_\alpha =0$, and this confirms that the leading behavior of $f$ in this case is constant.  Meanwhile, the leading behavior of the root components is $O(r^{|p_\alpha|/2})$, which is at least $O(r^{1/2})$.  Applying this result to $f = \delta A_0 - Y$ we have that
\begin{equation}\label{A0Ysubleading}
\delta A_0 - \delta Y = (\textrm{Cartan-valued}) \cdot O(1) + O(r^{\updelta},r^{1/2})~.
\end{equation}
Note that the background covariant derivative of the Cartan-valued constant vanishes, and thus taking $\tilde{D}_i$ of both sides we conclude
\begin{equation}\label{BPSbonus1}
E_i - D_i Y = O(r^{-1 + \updelta},r^{-1/2})~.
\end{equation}

Then consider the $\delta A_b$ equations:
\begin{align}
& ( \tilde{D}^c \tilde{D}_c + O(r^{-1})) \delta A_i + 2 \ad(\tilde{F}_{i0}) \delta A^0 + 2 \ad(\tilde{F}_{ib}) \delta A^b + 2 \ad(\tilde{F}_{i5}) \delta Y = O(r^{-2+2\updelta})  \cr
\Rightarrow \quad &  \tilde{D}^c \tilde{D}_c \delta A_i + 2 \ad(\tilde{F}_{ib}) \delta A^b + 2 \ad(\tilde{E}_i) ( \delta Y - \delta A_0) = O(r^{-2+\updelta}) \cr
\Rightarrow \quad & \tilde{D}^c \tilde{D}_c \delta A_i + 2 \ad(\tilde{F}_{ib}) \delta A^b = O(r^{-2+\updelta})~,
\end{align}
and
\begin{align}
&  ( \tilde{D}^c \tilde{D}_c + O(r^{-1})) \delta X + 2 \ad(\tilde{F}_{4b}) \delta A^b = O(r^{-2+2\updelta}) \cr
\Rightarrow \quad &   \tilde{D}^c \tilde{D}_c  \delta X + 2 \ad(\tilde{F}_{4b}) \delta A^b = O(r^{-2+\updelta})~,
\end{align}
or combining,
\begin{equation}\label{deltaAbeqn}
( \delta_{ab} \tilde{D}^c \tilde{D}_c + 2 \ad(\tilde{F}_{ab}) ) \delta A^b = O(r^{-2+\updelta})~.
\end{equation}
This leading order equation is identical to the leading order equation we found in our analysis of Yang--Mills--Higgs theory \cite{MRVdimP1}---notice the crucial cancellation between the $\delta Y$ and $\delta A_0$ terms in the $\delta A_i$ equation.  As we showed there, the kernel of the operator $ \delta_{ab} \tilde{D}^c \tilde{D}_c + 2 \ad(\tilde{F}_{ab})$ agrees with the kernel of the operator obtained by combining the Bogomolny equation with the gauge-orthogonality constraint, $\tilde{D}^a \delta A_a = 0$.  Fortunately this gauge orthogonality constraint is consistent with our Lorentz gauge condition at leading order, using that $\tilde{A}_0 = \tilde{Y}$ and \eqref{BPSbonus1}:
\begin{align}
& \tilde{D}^M \delta A_M = -\pd_0 \delta A_0 - [\tilde{A}_0, \delta A_0] + \tilde{D}^a \delta A_a + [\tilde{Y}, \delta Y] = 0 \cr
\Rightarrow \quad & \tilde{D}^a \delta A_a = O(r^{-1+\updelta})~.
\end{align}
Hence the leading order behavior of $\delta A^b$ as determined from \eqref{deltaAbeqn} must be the same as the leading order behavior of the tangent vectors to monopole moduli space, and we know that the strongest possible behavior of these is $O(r^{-1/2})$ which occurs for Lie algebra components along $E_\alpha$ such that $|\langle \alpha, P\rangle| = 1$.  Hence we have found that $\updelta = 1/2$ for the bosons as well.

Since $r = 0$ is a regular singular point of the differential equation determining $\delta A_a$, \eqref{deltaAbeqn}, subleading corrections go as $O(r^{-1/2 +n})$ for integer $n$.  Furthermore, given that the unknown terms in \eqref{deltaAbeqn} are suppressed by one power of $r$, it follows that $\delta A_b$ will differ from the BPS quantity by terms of $O(r^\updelta)$.  Hence,
\begin{equation}\label{BPSbonus2}
B_i - D_i X = O(r^{-1/2})~,
\end{equation}
on any solution to the second-order equations satisfying the defect boundary conditions, instead of the naive $O(r^{-3/2})$.

Finally observe the following.  The $\delta X$ equation can be rewritten as
\begin{align}
& \tilde{D}^c \tilde{D}_c \delta X - 2 \ad(\tilde{D}_i \tilde{X}) \delta A^i = O(r^{-3/2}) \cr
& \Rightarrow  \quad \tilde{D}^c \tilde{D}_c \delta X = 2 [\tilde{B}_i, \delta A^i] + O(r^{-3/2})~,
\end{align}
while the $\delta Y$ equation is
\begin{equation}
\tilde{D}^c \tilde{D}_c \delta Y = 2 [\tilde{E}_i, \delta A^i] + O(r^{-3/2}) = - \frac{\theta_0 g_{0}^2}{8\pi^2} \cdot 2 [\tilde{B}_i, \delta A^i] + O(r^{-3/2})~.
\end{equation}
Hence,
\begin{equation}
\tilde{D}^a \tilde{D}_a \left( \frac{\theta_0 g_{0}^2}{8\pi^2} \delta X + \delta Y \right) = O(r^{-3/2})~.
\end{equation}
By the same arguments that led to \eqref{A0Ysubleading} we conclude that
\begin{equation}\label{XYsubleading}
\delta Y + \frac{\theta_0 g_{0}^2}{8\pi^2} \delta X  = (\textrm{Cartan-valued}) \cdot O(1) + O(r^{1/2})~.
\end{equation}
Taking $\tilde{D}_i$ derivatives of both sides, and making use of \eqref{BPSbonus1} and \eqref{BPSbonus2}, we learn that
\begin{equation}\label{BPSbonus3}
E_i = - \frac{\theta_0 g_{0}^2}{8\pi^2} B_i + O(r^{-1/2})~,
\end{equation}
on any solution to the equations of motion satisfying defect boundary conditions.  \eqref{BPSbonus1}, \eqref{BPSbonus2}, and \eqref{BPSbonus3} establish \eqref{defectbcsaux}.

Finally, let us note that we could have started with the general eigenvalue equations in \eqref{lineoms} rather than specializing to the zero eigenvalue case, and the analysis resulting in \eqref{defectbcsaux} would go through just the same.  The reason is that the eigenvalue term is never more divergent than the terms we kept track of.  This shows that the enhanced defect boundary conditions \eqref{defectbcsaux} hold for any field fluctuations around a background satisfying the line defect boundary conditions \eqref{defectbcs}.  In other words, \eqref{defectbcsaux} hold off shell.  This will be important for the demonstration of supersymmetry invariance in the following subsection. 

%%%%%%%%%%%%%%%%%%%%%
\subsection{Invariance of the vanilla plus defect action under $\RR$-supersymmetry}\label{app:defsusy}
%%%%%%%%%%%%%%%%%%%%%

In the presence of defects the boundary terms of \eqref{Sbulksusyvar} are divergent on the infinitesimal two-spheres, $S_{\varepsilon_{n}}^2$, and therefore the vanilla action $S_{\rm van}$ is not invariant under any supersymmetry.  It is a nontrivial check of our proposed defect action, \eqref{Sbndryapp}, that once included, the total action is invariant under a subset of the original supersymmetry.  This subset should be the supersymmetries generated by the $\RR$-supercharges, \eqref{Rsusys}.  In order to check this it is useful to describe the supersymmetry transformations in terms of the $\RR$ and $\TT$ supersymmetries directly.

Consider the variation of the full vanilla plus defect action, $S = S_{\rm van} + S_{\rm def}$, with $S_{\rm def}$ given by \eqref{Sbndryapp}.  Under rigid supersymmetry we have
\begin{equation}
\delta_{\xi} S = - \int \ed t \sum_n \int_{S_{\varepsilon_n}^2} \ed^2 \Omega_n r_{n}^2 \hat{r}_{n}^i \left\{ \xi_A \BB_{i}^A - \xibar^A \bar{\BB}_{i A} + \frac{2}{g_{0}^2} \delta_\xi \Tr(E_i Y + B_i X) \right\}~.
\end{equation}
First we express the terms coming from $S_{\rm van}$ in terms of the supersymmetry parameters $\varepsilon,\eta$, and the field variables \eqref{realvars}.  Plugging these relations into \eqref{bsusycurrent} we find, after some algebra,
\begin{align}\label{Sbulksusyvar2}
& \frac{g_{0}^2}{2} \left( \xi_A \BB_{i}^A - \xibar^A \bar{\BB}_{i A} \right) = \cr
& = \varepsilon_A \Tr \bigg\{ \left[ -i \left( E_i + \frac{\theta_0 g_{0}^2}{4\pi^2} B_i - D_i Y\right) + \left(\epsilon_{ijk} (E^k - D^k Y) - \delta_{ij} (D_0X - [Y,X]) \right) \s^{j0} \right] \lambda^A  \cr
& \qquad \quad + \left[ -i (B_i + D_i X) + \left( \epsilon_{ijk} \left( D^k X + B^k - \frac{\theta_0 g_{0}^2}{4\pi^2} E^k \right) + \delta_{ij} D_0Y \right) \s^{j0} \right] \rho^A \bigg\} + \cr
& \, + \eta_A \Tr \bigg\{ \left[ i(-B_i + D_i X) + \left( \epsilon_{ijk} \left( B^k - \frac{\theta_0 g_{0}^2}{4\pi^2} E^k - D^k X \right) - i \delta_{ij} D_0 Y \right) \s^{j0} \right] \lambda^A + \cr
&  \qquad \quad + \left[ i \left( E_i + \frac{\theta_0 g_{0}^2}{4\pi^2} B_i + D_i Y\right) - \left( \epsilon_{ijk} (E^k + D^k Y) + \delta_{ij} (D_0 X + [Y,X]) \right) \s^{j0} \right] \rho^A \bigg\}~.  \cr
\end{align}

This is the exact bulk variation.  Some terms can be canceled provided we make use of both \eqref{defectbcs} \emph{and} \eqref{defectbcsaux}.  Above we argued that \eqref{defectbcsaux} hold for a complete basis of field fluctuations around any background satisfying \eqref{defectbcs}.  Therefore they can be used in checking supersymmetry, which should hold off shell.  Then there are some cancelations and we find
\begin{align}\label{Sbulksusyvar3}
& \xi_A \BB_{i}^A - \xibar^A \bar{\BB}_{i A} = \cr
& = \frac{2}{g_{0}^2} \varepsilon_A \Tr \bigg\{-i \frac{\theta_0 g_{0}^2}{4\pi^2} B_i \lambda^A + \left[ -2 i B_i + \epsilon_{ijk} \left( 2 D^k X - \frac{\theta_0 g_{0}^2}{4\pi^2} E^k \right) \s^{j0} \right] \rho^A \bigg\}+  \cr
& \, +  \frac{2}{g_{0}^2} \eta_A \Tr \bigg\{ - \frac{\theta_0 g_{0}^2}{4\pi^2} \epsilon_{ijk} E^k \s^{j0} \lambda^A - \left( 2 \epsilon_{ijk} D^k Y+ 2\delta_{ij} [Y,X] \right) \s^{j0} \rho^A \bigg\} + O(r^{-3/2})~. \qquad 
\end{align}

Next we consider the variation of the defect action \eqref{Sbndryapp}.  In order to take the variation of the $E$-field, which involves a normal derivative, we use the fact that \eqref{defectbcsaux} implies that the first subleading behavior of $E_i$ is related to that of $B_i$.  This means that the variations agree at leading order:
\begin{equation}
\delta_\xi E_i = - \frac{\theta_0 g_{0}^2}{8\pi^2} \delta_\xi B_i + O(r^{-1/2})~.
\end{equation}
Since the leading behavior of $Y$ is $O(r^{-1})$, the corrections to this relation can be neglected.  Thus we have
\begin{equation}
\delta_\xi \Tr \left( E_i Y + B_i X \right) = \Tr \left\{ \delta_\xi B_i \left( X - \frac{\theta_0 g_{0}^2}{8\pi^2}Y \right) + E_i \delta_\xi Y + B_i \delta_\xi X \right\} + O(r^{-3/2})
\end{equation}

To compute the variations we work directly with $\varepsilon$ and $\eta$.  These have been defined (see discussion around \eqref{RTvar2}) so that $\delta_\xi = \delta_\varepsilon + \delta_\eta$, and the $\varepsilon,\eta$ variations of all of the fields have been obtained in \eqref{epsvar}-\eqref{etavarlast}.  Using these we find
\begin{align}
\delta_\xi \Tr \left( E_i Y + B_i X \right) =&~ \varepsilon_A \Tr \left\{ - \epsilon_{ijk} \left( 2 D^k X - \frac{\theta_0 g_{0}^2}{4\pi^2} D^k Y \right) \s^{j0} \rho^A - 2i E_i \lambda^A + 2i B_i \rho^A \right\} + \cr
& + \eta_A \Tr \left\{ - \epsilon_{ijk} \left( 2D^k X - \frac{\theta_0 g_{0}^2}{4\pi^2} D^k Y \right) \s^{j0} \lambda^A - 2i E_i \rho^A - 2i B_i \lambda^A \right\} + \cr
&~ + O(r^{-3/2})~. \raisetag{20pt}
\end{align}
All of the displayed terms can be either finite or divergent and must be kept.  Adding $2/g_{0}^2$ times this result to \eqref{Sbulksusyvar3} we find
\begin{align}
& \xi_A \BB_{i}^A - \xibar^A \bar{\BB}_{i A} + \frac{2}{g_{0}^2} \delta_\xi \Tr \left( E_i Y + B_i X \right) = \cr
& \qquad = -\frac{4}{g_{0}^2} \eta_A \Tr \bigg\{ \left[ B_i - \epsilon_{ijk} D^k X \s^{j0} \right] \lambda^A + \left[ i E_i + (\epsilon_{ijk} D^k Y + \delta_{ij}[Y,X] ) \s^{j0} \right] \rho^A \bigg\} + \cr
& \qquad \qquad + O(r^{-3/2})~,
\end{align}
where we again used the enhanced boundary conditions \eqref{defectbcsaux}.  All terms proportional to $\varepsilon_A$ have canceled out in a very nontrivial fashion.  However, the remaining terms proportional to $\eta_A$ are clearly divergent and/or finite.  Hence the total vanilla plus defect action is invariant under the $\RR$-supersymmetries, but not the $\TT$-supersymmetries,
\begin{equation}
\delta_{\varepsilon} S = 0~, \qquad \delta_{\eta} S \neq 0~,
\end{equation}
provided we impose the boundary conditions \eqref{defectbcs} and \eqref{defectbcsaux}.

%%%%%%%%%%%%%%%%%%%%%
\subsection{IR defects and boundary terms}\label{app:IRbndry}
%%%%%%%%%%%%%%%%%%%%%

In this subsection we show that the low energy effective action, $S_{\rm SW} = \int \ed^4 x \LL_{\rm SW}$, \eqref{lel}, should be supplemented by the boundary terms \eqref{SIRbndry} in the presence of IR line defects.  Given the previous analysis we do not expect the fermions to play a role in the discussion, so we restrict ourselves to the bosonic degrees of freedom.  Then, on a solution to the equations of motion, the variation of $S_{\rm SW}$ restricts to the following set of boundary terms:
\begin{align}
\delta S_{\rm SW}^{\rm bos} \bigg|_{\textrm{on-shell}} =&~ - \frac{1}{4\pi} \int \ed^4 x \pd^\mu \bigg\{ \Im(\tau_{IJ}) \left( \pd_\mu a^I \delta \bar{a}^J + \pd_\mu \bar{a}^I \delta a^J \right) +  \cr
&~ \qquad \qquad \qquad  \quad + \left( 2 \Im(\tau_{IJ}) F_{\mu\nu}^I - \Re(\tau_{IJ}) \epsilon_{\rho\sigma\mu\nu} F^{\rho\sigma I}\right) \delta A^{\nu J} \bigg\}~. \qquad
\end{align}
For static configurations the contributions from $t = \pm \infty$ cancel and we only need to worry about the timelike boundaries $\mathbb{R}_t \times S_{\infty}^2$ and $\mathbb{R}_t \times S_{\varepsilon_{n}}^2$.  Thus we have
\begin{align}\label{varSbulk}
\delta S_{\rm SW}^{\rm bos} \bigg|_{\textrm{on-shell}} =&~ \frac{1}{4\pi} \int \ed t \left( \lim_{r \to \infty}  \int_{S_{\infty}^2} \ed\Omega r^2 \hat{r}^i - \sum_n \int_{S_{\varepsilon_n}^2} \ed\Omega_n \varepsilon_{n}^2 \hat{r}_{n}^i \right) \times \cr
&~ \times \bigg\{ - \Im(\tau_{IJ}) (\pd_i a^I \delta \bar{a}^J + \pd_i \bar{a}^I \delta a^J) - 2(\Im(\tau_{IJ}) E_{i}^I + \Re(\tau_{IJ}) B_{i}^I) \delta A^{0 J} + \cr
&~ \qquad + 2\epsilon_{ijk} \left( - \Im(\tau_{IJ})  B^{k I}  + \Re(\tau_{IJ}) E^{k I} \right) \delta A^{j J} \bigg\}~. 
\end{align}
Note in the last line that $ \Re(\tau_{IJ}) E^{k I} - \Im(\tau_{IJ})  B^{k I} = \Re\left[ \tau_{IJ} (E^{k I} + i B^{k I}) \right]$.

The boundary term from the asymptotic two-sphere will vanish provided that the variations of the fields go to zero as $r = |\vec{x}| \to \infty$, and this is consistent with the asymptotic boundary conditions we wish to impose on the fields.  The boundary terms from the infinitesimal two-spheres, however, need not vanish when we plug in the behavior of the fields on the dyon solution.  In particular, $\pd_i a, E_i, B_i$ all go as $\varepsilon_{n}^{-2}$, so in order for these terms to vanish one would have to impose that the field variations go to zero as $|\vec{x} - \vec{x}_n| \to 0$.  This is an unreasonable restriction on the space of field configurations.  Hence we require a boundary action whose variation will cancel these terms.

The ansatz for our IR defect action is \eqref{SIRbndry}:
\begin{equation}
S_{\rm def}^{\rm IR} =  \frac{1}{2\pi} \sum_n \int \ed t  \int_{S_{\varepsilon_n}^2} \left\{ \Re \left[ \zeta^{-1} (F^I a_{\mathrm{D},I} + G_I a^I) \right] - \half q_{I}^{(n)} A_{0}^I \ed \Omega_n \right\} ~.
\end{equation}
Let $S^{\rm IR} = S_{\rm SW} + S_{\rm def}^{\rm IR}$.  Then we immediately note that the $A_0$ term in the boundary action is precisely what is required to kill the $\delta A_{0}$ term in the variation:
\begin{equation}\label{dSdA0}
\delta S^{\rm IR} = \sum_n \frac{1}{2\pi} \int \ed t \int_{S_{\varepsilon_n}^2} \ed \Omega_n \varepsilon_{n}^2 \hat{r}_{n}^i \left\{ \Im(\tau_{IJ}) E_{i}^J + \Re(\tau_{IJ}) B_{i}^J + q_{I}^{(n)} \frac{\hat{r}_{(n)i}}{2 \varepsilon_{n}^2} \right\} \delta A^{0I} + \cdots
\end{equation}
The leading behavior at the singularity cancels among the three terms, using \eqref{IRdefectbcs}.\footnote{Here we are using the fact that the variation of the other boundary terms cannot contain $\delta A_0$.  The reason is that these boundary terms depend on the fieldstrengths, and $\delta E_i$ in an \emph{independent} variation from $\delta A_0$ as $E_i$ and $A_0$ are related by a \emph{normal} derivative to the boundary.}

Actually, in order to conclude that this term vanishes we need to know that the subleading behavior of the quantities in curly brackets times the leading behavior of the variation $\delta A^0$ is less divergent than $\varepsilon_{n}^{-2}$ as $\varepsilon_{n} \to 0$.  In the non-Abelian case we had to work hard for this because we allowed for the possibility that the subleading behavior of $E_i, B_i$ is $O(\varepsilon_{n}^{-3/2})$ and that of $\delta A_0$ is $O(\varepsilon_{n}^{-1/2})$.  The reason this was necessary can be traced to the behavior of the zero mode fluctuations around the line defect background fields which can be $O(\varepsilon_{n}^{-1/2})$.  Here, the fields are Abelian and there are no zero modes.  This means that any solution to the equations of motion will agree with the form of the dyon solution in the vicinity of the defect, up to regular, \eg\ $O(1)$, corrections.  Hence it is consistent to take the variations of the fields to be $O(1)$ in the vicinity of the defect and then it follows, in particular, that \eqref{dSdA0} vanishes.

In fact we will be slightly more general and allow for field variations that diverge logarithmically, $\delta A_\mu, \delta a, \delta \bar{a} \sim \log(\varepsilon_n)$.  The reason for this is the following.  From the dyon solution, we see that the $E$-field, and hence the real part of $\pd_i (\zeta^{-1}a)$, involves $\tau_{IJ}$.  Now, as $\vec{x} \to \vec{x}_n$, $a^I$ is diverging and this corresponds to going far out on the Coulomb branch to the weak coupling region.  There we know that the one-loop part of $\tau_{IJ}$ gives the leading behavior, and this is logarithmic in $a^I$, and hence logarithmic in $\varepsilon_n = |\vec{x} - \vec{x}_n|$.  (The instanton corrections go to zero with power-law behavior).  It follows that the leading behavior of the fields $A_\mu, a$ may be either $O(\varepsilon_{n}^{-1})$ or $O(\varepsilon_{n}^{-1} \cdot \log{\varepsilon_n})$, and hence we should allow for variations which are $O(1)$ or $O(\log{\varepsilon_n})$.  This does not affect our conclusion that \eqref{dSdA0} vanishes as $\varepsilon_n \to 0$.

Now let us consider the variation of the remaining terms of $S_{\rm def}^{\rm IR}$.  We have
\begin{align}
\int_{S_{\varepsilon_n}^2} \delta (F^I a_{\mathrm{D},I} + G_{I} a^I) =&~ \int_{S_{\varepsilon_n}^2} \ed \Omega_n \varepsilon_{n}^2 \hat{r}_{n}^i \cdot \delta \left\{ B_{i}^I a_{\mathrm{D},I} - ( \Im(\tau_{IJ}) E_{i}^I + \Re(\tau_{IJ}) B_{i}^I) a^J \right\} \cr
=&~  \int_{S_{\varepsilon_n}^2} \ed \Omega_n \varepsilon_{n}^2 \hat{r}_{n}^i \bigg\{ \delta B_{i}^I a_{\mathrm{D},I}  + B_{i}^I (\tau_{IJ} - \Re(\tau_{IJ})) \delta a^J  + \cr
&~   \qquad \qquad \qquad \quad  + \Im(\tau_{IJ}) E_{i}^I \delta a^J - a^J \delta \tilde{G}_{i I} \bigg\}~. \raisetag{24pt}
\end{align}
Here we used that $\delta a_{\mathrm{D},I} = (\pd_J a_{\mathrm{D},I}) \delta a^J = \tau_{IJ} \delta a^J$ and introduced the notation $\tilde{G}_{i I} = - (\Im(\tau_{IJ}) E_{i}^J + \Re(\tau_{IJ}) B_{i}^J) \equiv \half \epsilon_{ijk} G^{jk I}$.  Now, $\delta B_{i}^I = \epsilon_{ijk} \pd^j \delta A^{k I}$, and we can integrate by parts because this involves only tangential derivatives.  This brings us to
\begin{align}
\int_{S_{\varepsilon_n}^2} \delta (F^I a_{\mathrm{D},I} + G_{I} a^I) =&~ \int_{S_{\varepsilon_n}^2} \ed \Omega_n \varepsilon_{n}^2 \hat{r}_{n}^i \bigg\{ \epsilon_{ijk} (\tau_{IJ}\pd^k a^J)\delta A^{j I} + \Im(\tau_{IJ}) (i B_{i}^I - E_{i}^I) \delta a^J + \cr
&~  \qquad \qquad \qquad \quad + a^I \delta \tilde{G}_{i I} \bigg\}~. \raisetag{24pt}
\end{align}
from which we can easily infer $\delta S_{\rm bndry}^{\rm IR}$.

Combining with \eqref{varSbulk}, and taking into account the vanishing of \eqref{dSdA0}, we have that
\begin{align}
\delta S^{\rm IR}  =&~ \frac{1}{4\pi} \int_{S_{\varepsilon_{n}^2}} \ed \Omega_n \varepsilon_{n}^2 \hat{r}_{n}^i \bigg\{ \Im(\tau_{IJ}) \left[ \pd_i a^I - \zeta (E_{i}^I + i B_{i}^I) \right] \delta \bar{a}^J + \cr
&~ \qquad \qquad \qquad \qquad +  \Im(\tau_{IJ}) \left[ \pd_i \bar{a}^I - \zeta^{-1} (E_{i}^I - i B_{i}^I) \right] \delta a^J + \cr
&~ \qquad \qquad \qquad \qquad+ 2 \epsilon_{ijk} \Re \left[ \tau_{IJ} \left( \pd^k (\zeta^{-1} a^I) - E^{kI} - i B^{kI} \right) \right] \delta A^{j J} + \cr
&~ \qquad \qquad \qquad \qquad + 2 \Re(\zeta^{-1} a^I) \delta \tilde{G}_{i I} \bigg\}~.
\end{align}
Observe that the $\delta A^{j}, \delta a, \delta \bar{a}$ terms all involve the quantity $\pd_i (\zeta^{-1} a^I) - (E_{i}^I + i B_{i}^I)$ or its conjugate, or its real and imaginary parts.  The vanishing of this quantity is the BPS equation, and the line defect boundary conditions solve the BPS equation at leading order.  This is sufficient to ensure that all of these terms vanish: we have that $\pd_i (\zeta^{-1} a^I) - (E_{i}^I + i B_{i}^I)$ is no more divergent than $O( \varepsilon_{n}^{-1} \cdot \log{\varepsilon_n})$ and the variations $\delta A^j, \delta a, \delta \bar{a}$ are no more divergent than $O(\log{\varepsilon_n})$.   

This leaves only the $\delta \tilde{G}_{i I}$ term.  Naively, we have that
\begin{equation}
\tilde{G}_{i I} = q_{I}^{(n)} \frac{\hat{r}_i}{2 \varepsilon_{n}^2} + O(\varepsilon_{n}^{-1} \cdot \log{\varepsilon_n}) \quad \Rightarrow \quad \delta \tilde{G}_{i I} = O(\varepsilon_{n}^{-1} \cdot \log{\varepsilon_n})~,
\end{equation}
and hence there could be trouble since $\Re(\zeta^{-1} a^I) = O(\varepsilon_{n}^{-1} \cdot \log{\varepsilon_n})$.  Thus what we would like to show is that on any solution to the equations of motion we in fact have the stronger result $\delta \tilde{G}_{i I} = O(\log{\varepsilon_n})$.  Recall that the gauge field equations of motion are precisely $\ed G_I = 0$, or equivalently $\epsilon^{\mu\nu\rho\sigma} \pd_\nu G_{\rho\sigma} = 0$.  In particular, the $\mu = 0$ component is equivalent to $\pd^i \tilde{G}_{i I} = 0$.  However any divergent terms in $\tilde{G}_i$ should also be due to the fields created by the defect and thus should be spherically symmetric.  The leading $\hat{r}_i/r^2$ term in $\tilde{G}_i$ is the only such divergent term that is consistent with the equations of motion.  (There of course may be contributions to $\tilde{G}_i$ that are regular at $\vec{x}_n$ and satisfy the equation of motion.)  Hence we in fact must have that
\begin{equation}
\tilde{G}_{i I} = q_{I}^{(n)} \frac{\hat{r}_i}{2 \varepsilon_{n}^2} + O(1)~,
\end{equation}
and thus $\delta \tilde{G}_i = O(1)$.  Therefore all terms in $\delta S^{\rm IR}$ vanish and we conclude that the addition of the boundary action leads to a consistent variational principle.

%%%%%%%%%%%%%%%%%%%%%
%%%%%%%%%%%%%%%%%%%%%
\section{Supplementary material for monopole moduli spaces}\label{app:monmod}
%%%%%%%%%%%%%%%%%%%%%
%%%%%%%%%%%%%%%%%%%%%

%%%%%%%%%%%%%%%%%%%%%
\subsection{An embedding of monopole moduli spaces}\label{app:embed}
%%%%%%%%%%%%%%%%%%%%%

In this appendix we fill in some details of the construction described in subsection \ref{sssec:isoG}.  We construct a family of dimension-preserving embeddings of singular monopole moduli spaces, where a moduli space associated with a gauge group of lower rank is embedded into one associated with a gauge group of larger rank.  The fact that we have a family of embeddings rather than a single canonical one is related to the fact that we allow the magnetic charge of the moduli space associated with the higher-rank gauge group to be arbitrary.  We then describe how this construction can be used to determine the kernel of the ${\rm G}$ map introduced in \ref{sssec:isoG}.

In order to determine those $H \in \mathfrak{t}$ such that the gauge transformation $\epsilon_H$ gives a nontrivial action, $\delta_{\epsilon_H} \hat{A} \neq 0$, we must characterize precisely, in some convenient gauge, which root directions the field is excited along.  We start with the relative magnetic charge, $\tilde{\gamma}_{\rm m} = \gm - \sum_n P_{n}^- = \sum_I \tilde{n}_{\rm m}^I H_I$ and recall the partition \eqref{corootpart}:
\begin{equation}
\{ I_A ~|~ \tilde{n}_{\rm m}^{I_A} > 0~,~ A = 1,\ldots,d \} ~\cup~ \{ I_M ~|~ \tilde{n}_{\rm m}^{I_M} = 0~,~ M = 1,\ldots,r-d \}~,
\end{equation}
where $r = \rnk{\mathfrak{g}}$ and $d$ satisfies $0 < d \leq r$, as $d=0$ implies $\dim{\fMM} = 0$.  The set $\{ H_{I_A}, h^{I_M} \}$ is a basis for $\mathfrak{t}$, where $h^{I_M}$ are the fundamental magnetic weights integral-dual to the $\alpha_{I_M}$.  In fact, defining
\begin{equation}\label{tpartperp}
\mathfrak{t}^{\parallel} = \Span \{ H_{I_A} \}~, \qquad \mathfrak{t}^\perp = \Span \{ h^{I_M} \}~,
\end{equation}
we have that $\mathfrak{t} = \mathfrak{t}^{\parallel} \oplus \mathfrak{t}^\perp$ as vector spaces and that $\mathfrak{t}^\perp$ is the orthogonal complement of $\mathfrak{t}^{\parallel}$ with respect to the Killing form $(~,~)$:
\begin{equation}
(H_{I_A}, h^{I_M}) = \frac{2}{\alpha_{I_A}^2} (\alpha_{I_A}^\ast , h^{I_M}) = \frac{2}{\alpha_{I_A}^2} \langle \alpha_{I_A} , h^{I_M} \rangle = \frac{2}{\alpha_{I_A}^2} {\delta_{I_A}}^{I_M} = 0~.
\end{equation}
Hence any element $H \in  \mathfrak{t}$ can be uniquely decomposed, $H = H^{\parallel} + H^{\perp}$, with $H^{\parallel} \in \mathfrak{t}^{\parallel}$ and $H^\perp \in \mathfrak{t}^\perp$.  In particular $\tilde{\gamma}_{\rm m} = \tilde{\gamma}_{\rm m}^{\parallel}$; \ie\ $\tilde{\gamma}_{\rm m}^{\perp} = 0$.

The subspace $\mathfrak{t}^\parallel$ may be identified with the Cartan subalgebra of a semisimple Lie algebra as follows.  Each simple root $\alpha_I$ corresponds to a node in the Dynkin diagram, $\mathfrak{D}_{\mathfrak{g}}$, of $\mathfrak{g}$.  Let $\mathfrak{D}^{\rm ef}$ denote the Dynkin diagram obtained from $\mathfrak{D}_{\mathfrak{g}}$ by deleting those nodes corresponding to the $\alpha_{I_M}$.  This gives the Dynkin  diagram of a semisimple Lie algebra $\mathfrak{g}^{\rm ef}$.  (This diagram may have disconnected components; these correspond to the simple factors of the semi-simple Lie algebra.)  Let $\alpha_A$ denote the simple roots of $\mathfrak{g}^{\rm ef}$.  Then there is a natural embedding
\begin{equation}\label{liealgembed}
i_\ast : \mathfrak{g}^{\rm ef} \hookrightarrow \mathfrak{g}~,
\end{equation}
where
$i_\ast( E_{\alpha_A}) = E_{\alpha_{I_A}}$.  We can identify the corresponding Cartan subalgebra:
\begin{equation}
i_\ast(\mathfrak{t}^{\rm ef}) = \mathfrak{t}^{\parallel}~.
\end{equation}
Indeed if $H_A = H_{\alpha_A}$ are the simple co-roots of $\mathfrak{g}^{\rm ef}$ then $i_\ast(H_A) = H_{I_A}$.  Note the crucially important fact that if $T \in \mathfrak{g}^{\rm ef}$ and $H \in \mathfrak{t}$, then
\begin{equation}\label{decouple}
\ad(i_\ast(T))(H^\perp) = 0~.
\end{equation}
In general if $H \in \mathfrak{t}$ we define $H^{\rm ef} \in \mathfrak{t}^{\rm ef}$ such that $i_\ast(H^{\rm ef}) = H^\parallel$.

Now consider the expansion of $X_\infty$ in the above basis,
\begin{equation}
X_\infty = x^{I_A} H_{I_A} + x_{I_M} h^{I_M}~.
\end{equation}
Then $\langle \alpha_{I_A}, X_\infty \rangle = C_{I_A I_B} x^{I_B} = (C^{\rm ef})_{AB} x^{I_B}$, where $C_{IJ}$ are the components of the Cartan matrix of $\mathfrak{g}$ and $(C^{\rm ef})_{AB}$ are the components of the Cartan matrix of $\mathfrak{g}^{\rm ef}$.  Hence if $(C^{\rm ef})^{AB}$ denotes the inverse of $(C^{\rm ef})_{AB}$, then
\begin{equation}
X_{\infty}^{\rm ef} = (C^{\rm ef})^{AB} \langle \alpha_{I_B}, X_\infty \rangle H_A = \langle \alpha_{I_B}, X_\infty \rangle h^B~,
\end{equation}
where $h^B$ are the fundamental magnetic weights of $\mathfrak{g}^{\rm ef}$.  It follows that if $X_\infty$ is in the fundamental Weyl chamber of $\mathfrak{t}$, then $X_{\infty}^{\rm ef}$ is in the fundamental Weyl chamber of $\mathfrak{t}^{\rm ef}$.  Similarly, with
\begin{equation}
(P_{n}^-)^{\rm ef} = \langle \alpha_{I_A}, P_{n}^- \rangle h^A~,
\end{equation}
we see that $P_{n}^- \in \Lambda_{G} \Rightarrow \langle \alpha_{I_A} , P_{n}^- \rangle \in \mathbb{Z} \Rightarrow (P_{n}^-)^{\rm ef} \in \Lambda_{\rm mw}^{\rm ef} = \Lambda_{G_{\rm ad}^{\rm ef}}$, where $G_{\rm ad}^{\rm ef}$ is the semisimple Lie group with Lie algebra $\mathfrak{g}^{\rm ef}$ and trivial center, while $P_{n}^-$ in the antifundamental Weyl chamber of $\mathfrak{t}$ implies $(P_{n}^-)^{\rm ef}$ in the antifundamental Weyl chamber of $\mathfrak{t}^{\rm ef}$.  It follows that $((\vec{x}_m, (P_{n}^-)^{\rm ef}); \gm^{\rm ef}; X_{\infty}^{\rm ef})$ comprise data for a singular $G_{\rm ad}^{\rm ef}$-monopole moduli space\footnote{Monopole moduli spaces, $\fMM^{\rm ef}$, for semisimple gauge groups are direct products of the moduli spaces for each simple factor.}
\begin{equation}
\fMM^{\rm ef} := \fMM\left((\vec{x}_n, (P_{n}^-)^{\rm ef}); \gm^{\rm ef}; X_{\infty}^{\rm ef} \right)~.
\end{equation}

The embedding \eqref{liealgembed} induces an embedding of singular monopole moduli spaces
\begin{equation}
\hat{\imath} : \fMM^{\rm ef} \hookrightarrow \fMM~,
\end{equation}
as follows.  First let $\cg \notin \GG_{\{ P_n \}}^0$ be a gauge transformation that goes to the identity at infninty and conjugates each $P_{n}^-$ to $P_n$.  Such a transformation is unique modulo $\GG_{\{ P_n \}}^0$.  Then if $[\hat{A}^{\rm ef}] \in \fMM^{\rm ef}$ we set
\begin{align}\label{nicegauge}
& A = \Ad(\cg) \left(  i_\ast(A^{\rm ef}) + \sum_n \frac{(P_{n}^-)^\perp}{2} (\pm 1 - \cos{\theta_n}) \ed\phi_n \right) \equiv \Ad(\cg)(A^-)~, \cr
& X = \Ad(\cg) \left(  i_\ast(X^{\rm ef}) + X_{\infty}^{\perp} - \sum_n \frac{(P_{n}^-)^\perp}{2 r_n} \right) \equiv \Ad(\cg)(X^-) ~,
\end{align}
and we claim that $[\hat{A}] \in \fMM((\vec{x}_n,P_n);\gm;X_\infty)$.  Consider first the arguments, $\hat{A}^- = \{A^-,X^-\}$ of $\Ad(\cg)$.  They comprise a sum of two solutions to the Bogomolny equation which in this case is also a solution.  First $i_\ast(\hat{A}^{\rm ef})$ is a solution since $\hat{A}^{\rm ef}$ is, while the remaining terms comprise a Cartan-valued solution.  Furthermore the sum is also a solution because the cross-terms from the nonlinear terms in the Bogomolny equation vanish due to \eqref{decouple}.  The asymptotics are such that this configuration represents a point in $\fMM((\vec{x}_n,P_{n}^-);\gm;X_\infty)$, using in particular that $\gm^\perp = \sum_n (P_{n}^-)^\perp$.  Finally the gauge transformation maps this to a representative in $\fMM((\vec{x}_n,P_n);\gm;X_\infty)$.  The fact that $\hat{\imath} : [\hat{A}^{\rm ef}] \mapsto [\hat{A}]$ is an embedding follows from $i_\ast$ having this property together with the linearity of the construction.  In particular the derivative map acts as $(\eD\hat{\imath})_{[\hat{A}^{\rm ef}]} : \delta \hat{A}^{\rm ef} \mapsto i_\ast(\delta \hat{A}^{\rm ef})$ and is clearly injective.  We also see from the form of the derivative map that the embedding respects the Riemannian metrics and quaternionic structures.  Note that injectivity of the derivative map on the zero modes means, in particular, that the metric and quaternionic structure of $\fMM$ depend only on the components $\langle \alpha_{I_A}, X_\infty \rangle$ of the Higgs vev and not on the components $\langle \alpha_{I_M},X_\infty \rangle$.

The key point of this construction is that the dimension of $\fMM^{\rm ef}$ and $\fMM$ are the same.  One observes that the relative magnetic charge of $\hat{A}^{\rm ef}$, $\widetilde{\gm^{\rm ef}} := \gm^{\rm ef} - \sum_n (P_{n}^-)^{\rm ef}$, satisfies 
\begin{equation}
i_{\ast}(\widetilde{\gm^{\rm ef}}) = \tilde{\gamma}_{\rm m}^{\parallel} =  \tilde{\gamma}_{\rm m}~,
\end{equation}
and thus $\widetilde{\gm^{\rm ef}} = \sum_A \tilde{n}_{\rm m}^{I_A} H_A$.  We expect that $\hat{\imath}$ is in fact an isomorphism of hyperk\"ahler manifolds, though we are unable to rule out the possibility of a discrete cover.  If this is the case then we are guaranteed that any $[\hat{A}] \in \fMM$ can be represented in a gauge of the form \eqref{nicegauge}.  Even if there is a discrete cover we will still assume that any $[\hat{A}] \in \fMM$ can be represented in a gauge of this form.

Given \eqref{nicegauge} we can now understand the kernel of ${\rm G}$.  First we note that if $\hat{A} = \Ad(\cg)(\hat{A}^-)$, then $\epsilon_H \in {\rm Lie}(\TT_{\{ P_n\}}^0)$ satisfies $\hat{D}^2 \epsilon_H = 0$ with $\lim_{|\vec{x}| \to \infty} \epsilon_H = H$ if and only if $\epsilon_H = \Ad(\cg)(\epsilon_{H}^-)$ where $\epsilon_{H}^-$ satisfies $(\hat{D}^-)^2 \epsilon_{H}^- = 0$ and $\lim_{|\vec{x}| \to \infty} \epsilon_{H}^- = H$.  Thus the bosonic zero mode $\delta_H \hat{A}$ corresponding to ${\rm G}(H) = \delta_H$ is non-zero at $[\hat{A}] \in \fMM$ if and only if $[\hat{A}^-,H] \neq 0$.  Given \eqref{decouple}, it is clear that $[\hat{A}^-, H^\perp] = 0$ and hence $\mathfrak{t}^\perp \subseteq \ker{\rm G}$.

We claim that the opposite containment holds as well and $\mathfrak{t}^\perp = \ker{\rm G}$.  To prove this it would be sufficient to show that there exist points $[\hat{A}] \in \fMM$ such that, in the gauge \eqref{nicegauge}, $\hat{A}^{\rm ef}$ has non-zero components along every simple root direction $E_{\alpha_A}$ of $\mathfrak{g}^{\rm ef}$.  This is intuitively clear given the physical interpretation of having $\tilde{n}_{\rm m}^{I_A} > 0$ mobile fundamental monopoles of type $A$ for each $A$.  There should be asymptotic regions of moduli space where an exact solution $\hat{A}^{\rm ef}$ is approximated exponentially well in some spatial region surrounding each constituent by a superposition of basic building block solutions.\footnote{The building blocks in this case would be the solutions \cite{Cherkis:2007jm, Cherkis:2007qa, Blair:2010vh} for one smooth monopole in the presence of an arbitrary number of \tHooft defects.}  The building block solution for a fundamental monopole of type $A$ would have nonzero components along $E_{\pm \alpha_A}$.  In the case of smooth monopoles a result along these lines is available \cite{Taubes:1981gw}, and we will assume that the construction can be generalized to the case with defects.

Noting that $\mathfrak{t} /\ker{\rm G} \cong \mathfrak{t}^{\rm ef}$, we can construct an injective homomorphism, ${\rm G} \circ i_\ast : \mathfrak{t}^{\rm ef} \to \mathfrak{isom}_{\mathbb{H}}(\fMM)$ mapping elements of $\mathfrak{t}^{\rm ef}$ to triholomorphic Killing vectors.  Exponentiating both sides, we obtain a torus action of triholomorphic isometries on $\fMM$.  In order to have an effective torus action we should use the exponential map of the adjoint form of the group.  ${\rm G}$ acts by infinitesimal gauge transformations, which act through the adjoint representation.  This representation is only faithful for the adjoint form of the group.  This gives a $T_{\rm ad}^{\rm ef}$ action on $\fMM$ via triholomorphic isometries.  A natural basis of generators for $T_{\rm ad}^{\rm ef}$ consists of the fundamental magnetic weights $h^A$ which generate $2\pi$-periodic cycles.  Now,
\begin{align}
i_\ast(h^A) =&~ (C^{\rm ef})^{AB} i_\ast(H_B) = (C^{\rm ef})^{AB} H_{I_B} = (C^{\rm ef})^{AB} \left(C_{I_BI_C} h^{I_C} + C_{I_B I_M} h^{I_M}\right)  \cr
=&~ h^{I_A} + (C^{\rm ef})^{AB} C_{I_B I_M} h^{I_M}~,
\end{align}
so the difference between $h^{I_A}$ and $i_\ast(h^A)$ is in the kernel of ${\rm G}$ and this implies the second equality of \eqref{KAdef}.

%%%%%%%%%%%%%%%%%%%%%
\subsection{Gauge-induced deck transformations and the homomorphism $\mu$}\label{app:Dquotient}
%%%%%%%%%%%%%%%%%%%%%

We give an argument via the rational map construction that the homomorphism $\mu$ determining the subgroup $\mathbb{D}_\gi \subset \mathbb{D}$ is given by $\mu(\lambda) = (\gm,\lambda)$.

Consider the action of the Cartan torus of the adjoint group $T_{\rm ad} \subset G_{\rm ad}$ on $\MM$, by asymptotically nontrivial gauge transformations, through any point $m_0 \in \MM$. The induced homomorphism
\be\label{mupi1map}
\pi_1(T_{\rm ad}, 1) \rightarrow \pi_1(\MM,m_0)
\ee
can be considered as a group homomorphism
\be
\mu: \Lambda_{\rm mw} \to \mathbb{Z}\,.
\ee
We claim this is given by
\be\label{eq:mu-hom}
\mu(\lambda) = (\gm,\lambda)~,
\ee
where on the right hand side we   are using
the Killing form normalized so that the long roots
have  square-length equal to two.  To compare with what we claimed in \eqref{muclaim}, set $\lambda = i_\ast(h) + \lambda^\perp$, where $h \in \Lambda_{\rm mw}^{\rm ef}$ and $\lambda^\perp \in \mathfrak{t}^\perp$, \eqref{tpartperp}, and note that on the one hand any such $\lambda^\perp$ generates a gauge transformation that acts trivially on $\MM$ and on the other $(\gm,\lambda^\perp) = 0$.

To prove \eqref{eq:mu-hom} we note first that since the domain and
codomain of $\mu$ are torsion-free it suffices to prove
the result for $\lambda$ given by the simple co-roots:
\be\label{eq:muHI}
\mu(H_I)= \sum_J n_{\rm m}^J \sp^J C_{JI} ~,
\ee
where $\sp^J = 2/(\alpha_J,\alpha_J) \in \{1,2,3\}$.

Now, to prove this it is useful to view the moduli space $\MM$ as
the space of basepoint-preserving rational maps to the flag variety:
\be
\mathbb{C}\mathbb{P}^1 \rightarrow \FF ~,
\ee
where  the flag variety, $\FF$, can be written
as $\tilde G^c/B$ where $\tilde G^c$ is the complexification
of $\tilde G$ and $B$ is a Borel subgroup.  ($\tilde{G}$ is the simply-connected cover of $G$).  This is a standard result in monopole theory.
See \cite{Atiyah:1988jp,Donaldson:1985id,MR804459,Hurtubise:1989wh,Hurtubise:1989qy,MR1625475,MR1625471}. In these terms,
the $T$-action is just given by conjugating the map by $T$.

Recall that for a single $SU(2)$ monopole the map is
\be
[z:1] \rightarrow \left[ s(z;a,y) \right]~,
\ee
where
\be
s(z;a,y) = \begin{pmatrix} z-a & y^{-1} \\ - y & 0 \\  \end{pmatrix}~,
\ee
and $[s]$ denotes the equivalence class of $s \in SU(2)^c$ modulo the Borel.  Here $(a,y) \in \mathbb{C} \times \mathbb{C}^*$ and we used stereographic projection so that $z$ is in
the plane. The position of the monopole in $\mathbb{R}^3 \cong \mathbb{C} \times \mathbb{R}$ is
encoded in  $(a, \log \vert y \vert)$ while the argument of $y$ encodes phase of the monopole.

Now, let $\varphi_I: SU(2) \rightarrow \tilde G$ be the canonical embedding into the simply-connected cover of $G$ along a co-root $H_I$. That is, $\varphi_I(H) = H_I$
and $\varphi(E^\pm) = E^\pm_I$. For a point $m_0\in \MM$ corresponding to widely separated
monopoles the rational map is given by
\be
[z:1] \rightarrow [S(z)]~,
\ee
where $S(z)$ can be approximated by
\be\label{eq:Smtx}
S(z;\{ a_{k_I}, y_{k_I} \} )  :=  \prod_{I} \prod_{k_I=1}^{n_m^I}   \varphi_I( s(z;a_{k_I}, y_{k_I}))~.
\ee
The approximation becomes arbitrarily good as the positions encoded in $(a_{k_I}, y_{k_I})$
are taken arbitrarily far from each other. (The factors in the product do not commute,
not even approximately, and a specific ordering is appropriate for a specific configuration
of centers.) For our homotopy-theoretic considerations
we can replace $S(z)$ by $S(z;\{ a_{k_I}, y_{k_I} \} )$. Therefore, we want to compute the element
of
\be
\pi_1({\rm Map}^{\rm Rat}(\mathbb{C}\mathbb{P}^1, \FF), m_0)
\ee
given by $S^1 \to \MM \cong {\rm Map}^{\rm Rat}(\mathbb{C}\mathbb{P}^1, \FF)$ with
\be
e^{i \theta} \mapsto \Biggl\{ [z:1] \mapsto \left[ e^{\theta H_I} S(z;\{ a_{k_I}, y_{k_I} \} ) e^{-\theta H_I} \right] \Biggr\}~,
\ee 
as this is what the map $(e^{i\theta} \mapsto e^{\theta H_I})$ maps to under \eqref{mupi1map}.  The winding numbers of the factors in \eqref{eq:Smtx} add since, as we will soon explain,
\be
 \pi_1({\rm Map}^{\rm Rat}(\mathbb{CP}^1, \FF), m_0) \cong \mathbb{Z}~.
\ee

Now, according to \cite{Taubes:1984je,MR1294670} there is a stable homotopy
equivalence allowing us to replace
${\rm Map}^{\rm Rat}(\mathbb{CP}^1, \FF)$ with ${\rm Map}^{\rm Cont.}(\mathbb{CP}^1, \FF)$,
at least when computing $\pi_1$.
A loop in the latter space is just a continuous map
\be
f: S^1 \times S^2 \to \FF ~.
\ee
If we fix a point in $S^2$ the map is homotopically trivial since
$\pi_1(\FF)=0$. If we fix a point on $S^1$ the map defines an
element of $\pi_2(\FF) \cong \Lambda_{\rm cr}$. This homotopy class
just corresponds to the magnetic charge, and is fixed by the choice of
basepoint $m_0$.  The only other homotopy invariant is
the element of $\pi_3(\FF)$ given by $[f\circ p]$ where
$p: S^3 \to S^1 \times S^2$ is any degree one map. On the
other hand, by the exact sequence for a fibration we have a
natural isomorphism (given by the homotopy lifting property)
\be
\pi_1({\rm Map}^{\rm Cont.}(\mathbb{C}\mathbb{P}^1,\FF),m_0) \cong \pi_3(\FF) = \pi_3(\tilde G^c/B) \cong \pi_3(\tilde G^c) \cong \pi_3(\tilde G) \cong \mathbb{Z} ~.
\ee

To compute $\mu(H_I)$ we need to add the contributions to $\pi_3(\FF) \cong \mathbb{Z}$
from the different factors in \eqref{eq:Smtx}.
The first step is to observe that
\be\label{eq:IJ-comm}
\begin{split}
e^{\theta H_I } \varphi_J (s(z;a,y) ) e^{-\theta H_I} &   = \varphi_J( s(z;a, y e^{i \theta C_{JI} } )) \\
\end{split} ~,
\ee
where we used $[i H_I , E^\pm_J] = \pm C_{JI} E^\pm_J$.

Next, for fixed $(a,y) \in \mathbb{C} \times \mathbb{C}^*$ let $f_{I,a,y}$ denote the map
\be
f_{I,a,y}: S^1 \times S^2 \to \FF ~,
\ee
defined by
\be
f_{I,a,y}(e^{i \theta}, [z:1]) = \left[ \varphi_I(s(z; a, e^{i \theta} y)) \right]~.
\ee
We claim that under the isomorphism $\pi_3(\FF)\cong \mathbb{Z}$ defined above,
\be\label{eq:pI-homotop}
[f_{I,a,y}\circ p] = \sp^I ~.
\ee

One way to understand \eqref{eq:pI-homotop} is to note that $f_{I,a,y}$ corresponds
to the loop around the center of mass for a single Prasad-Sommerfield monopole
embedded by $\varphi_I$. That is, for a monopole of magnetic charge $\gm = H_I$.
By careful normalization of the Lee--Weinberg--Yi metric one can check that this is indeed $\sp^I$
times the generator of $\pi_1(\MM(\gamma_m=H_I;X)) = \pi_1( \mathbb{R}^3 \times S^1) \cong \mathbb{Z}$.

Another way to prove \eqref{eq:pI-homotop} is the following: By mutliplying on
the right by a suitable element of the Borel we can bring $s(z;a,y)$ to the form
\be
\hat s(z;a,y)   =
\frac{1 }{\sqrt{\vert y \vert^2  +  \vert z-a \vert^2}}
\begin{pmatrix}   (z-a) &  \bar y \\
- y &  (\bar z - \bar a) \\
\end{pmatrix}
\ee
and then define $m_{a,y}: S^1 \times S^2 \to SU(2)$ by
\be
m_{a,y}(e^{i \theta}, [z:1]) := \hat s(z;a, e^{i \theta} y)
\ee
Now consider the diagram:
\be
\xymatrix{
          &  SL(2,\mathbb{C}) \ar[r]^{\varphi_I} & \tilde G^c \ar[d]^\pi \\
SU(2) \ar[ru]^{\rm Id} \ar[r]_p & S^1 \times S^2 \ar[u]_{m_{a,y}}\ar[r]_{f_{I,a,y}} & \FF \\
}
\ee
The square commutes, but the
 triangle on the left only commutes up to homotopy. The bottom row defines an element of
$\pi_3(\FF)$ which is the homotopy class we seek. But this is given by the homotopy
class of $\varphi_I: SU(2) \to \tilde{G}^c$, corresponding to composing the
upper arrows of the diagram. We next prove that the upper arrows define a homotopy
class given by $\sp^I$.

 We know that $\pi_3(\tilde G) \cong \mathbb{Z}$ and moreover
a generator is given by \underline{some} embedded $SU(2)$.
Therefore, if we choose a basis of simple co-roots it must be
that one of the $\varphi_I$ provide a generator. Now the multiple
of the generator can be measured by a suitable multiple of
$(g^{-1} dg, [g^{-1}dg, g^{-1} dg])$. Therefore the short roots
should provide generators and in general the homotopy class of
the  map $\varphi_I$, thought of as a map $S^3 \to G$ is
$\sp^I$ times the generator.\footnote{Incidentally, this  provides a rather nice homotopy-theoretic
interpretation of $\sp^I$. Note there is no need to choose a normalization
of the Killing form in this characterization of $\sp^I$. }
This finally establishes \eqref{eq:pI-homotop}.

We now have all the pieces of the puzzle, and we simply add the contributions
from \eqref{eq:Smtx} using equations \eqref{eq:IJ-comm} and \eqref{eq:pI-homotop}:
\be
\mu(H_I) = \sum_J  \sum_{i=1}^{n_m^J}  \sp^J C_{JI} = \sum_J n_m^J \sp^J C_{JI} ~,
\ee
as we wanted to show. $\spadesuit$

%%%%%%%%%%%%%%%%%%%%%
\subsection{Bases and dual bases for $\ker{\mu}$}\label{app:Dquotient2}
%%%%%%%%%%%%%%%%%%%%%

In this appendix we describe an explicit basis of generators for the kernel and image of the group homomorphism $\mu : \Lambda_{\rm mw}^{\rm ef} \to \pi_1(\MM) \cong \mathbb{Z}$.  Let $\{ h^A \}_{A=1}^d$ be the integral basis of fundamental magnetic weights for $\Lambda_{\rm mw}^{\rm ef}$.  With respect to this basis (and the standard basis $\{1\}$ for $\mathbb{Z}$) the matrix representation of $\mu$ takes the form
\begin{equation}
\mu = (\ell^1, \ldots, \ell^d)~,
\end{equation}
where the integers $\ell^A$ are strictly positive.  Given that $\ker(\mu)$ is a $(d-1)$-dimensional sublattice of $\Lambda_{\rm mw}^{\rm ef}$ and that $\im(\mu) \cong L \mathbb{Z}$, with $L = \gcd(\ell^1,\ldots,\ell^d)$, there exists a change of basis matrix $V \in GL(d,\mathbb{Z})$ such that
\begin{equation}
\mu V = \left( 0,\ldots,0, L \right)~.
\end{equation}

The transformation can be found by working inductively with respect to $d$.  Let's consider the $d=2$ case.  B\'ezout's identity guarantees the existence of integers $x_{\ell^1,\ell^2},y_{\ell^1,\ell^2}$ such that $x_{\ell^1\ell^2} \ell^1 + y_{\ell^1\ell^2} \ell^2 = \gcd(\ell^1,\ell^2)$.  Then
\begin{equation}
(\ell^1,\ell^2) \left(\begin{array}{c c} \frac{\ell^2}{\gcd(\ell^1,\ell^2)} & x_{\ell^1,\ell^2}  \\[1ex] -\frac{\ell^1}{\gcd(\ell^1,\ell^2)} & y_{\ell^1,\ell^2} \end{array} \right) = (0,\gcd(\ell^1,\ell^2))~.
\end{equation}
The two-by-two matrix has integer entries and determinant $\frac{1}{\gcd(\ell^1,\ell^2)} (x_{\ell^1,\ell^2} \ell^1 + y_{\ell^1,\ell^2} \ell^2) = 1$.  Now in $d$ dimensions we work iteratively as follows.  At the first step we embed the above two-by-two matrix into the upper left block of a $d$-by-$d$ matrix with $1$'s on the remaining diagonal and $0$'s elsewhere:
\begin{align}
\mu V_{(1)} =&~  (\ell^1,\ell^2,\ldots,\ell^d) \left( \begin{array}{c c c c c} \frac{\ell^2}{\gcd(\ell^1,\ell^2)} & x_{\ell^1,\ell^2} & 0 & \cdots & 0 \\[1ex] -\frac{\ell^1}{\gcd(\ell^1,\ell^2)} & y_{\ell^1,\ell^2} & 0 & \cdots & 0 \\ 0 & 0 & 1 & \cdots & 0 \\ \vdots & \vdots & \vdots & \ddots & \vdots \\ 0 & \cdots & \cdots & 0 & 1 \end{array} \right) = (0,\gcd(\ell^1,\ell^2),\ell^3,\cdots, \ell^d)~.
\end{align}
Note $V_{(1)} \in GL(d,\mathbb{Z})$.  At the second step we multiply this result on the right by a matrix $V_{(2)}$ whose $22$, $23$, $32$, and $33$ entries are of the same form as the two-by-two matrix above, but with $\ell^1 \to \gcd(\ell^1,\ell^2)$ and $\ell^2 \to \ell^3$.  Specifically, we take
\begin{align}
& (V_{(2)})_{2}^{\phantom{2}2} = \frac{\ell^3}{\gcd(\ell^1,\ell^2,\ell^3)} ~, \qquad (V_{(2)})_{2}^{\phantom{2}3} = x_{\gcd(\ell^1,\ell^2),\ell^3} ~,  \cr
& (V_{(2)})_{3}^{\phantom{3}2} = - \frac{\gcd(\ell^1,\ell^2)}{\gcd(\ell^1,\ell^2,\ell^3)} ~, \qquad (V_{(2)})_{3}^{\phantom{3}3} = y_{\gcd(\ell^1,\ell^2),\ell^3} ~, \cr
& (V_{(2)})_{A}^{\phantom{A}A} = 1~, \textrm{ when $A \neq 2,3$}~, \qquad (V_{(2)})_{A}^{\phantom{A}B} = 0~, \textrm{ otherwise}~,
\end{align}
where we used that $\gcd(\gcd(\ell^1,\ell^2),\ell^3) = \gcd(\ell^1,\ell^2,\ell^3)$.  Then we will find
\begin{equation}
\mu V_{(1)} V_{(2)} = \left(0,0,\gcd(\ell^1,\ell^2,\ell^3),\ell^4,\cdots,\ell^d\right)~.
\end{equation}

Continuing in this way, at the $k^{\rm th}$ step we define
\begin{align}
& (V_{(k)})_{k}^{\phantom{k}k} = \frac{\ell^{k+1}}{\gcd(\ell^1,\ldots,\ell^{k+1})} ~, \qquad (V_{(k)})_{k}^{\phantom{k}k+1} = x_{\gcd(\ell^1,\ldots, \ell^{k}),\ell^{k+1}} ~,  \cr
& (V_{(k)})_{k+1}^{\phantom{k+1}k} = - \frac{\gcd(\ell^1,\ldots,\ell^k)}{\gcd(\ell^1,\ldots,\ell^{k+1})} ~, \qquad (V_{(k)})_{k+1}^{\phantom{k+1}k+1} = y_{\gcd(\ell^1,\ldots, \ell^{k}),\ell^{k+1}} ~, \cr
& (V_{(k)})_{A}^{\phantom{A}A} = 1~, \textrm{ when $A \neq k,k+1$}~, \qquad (V_{(k)})_{A}^{\phantom{A}B} = 0~, \textrm{ otherwise}~.
\end{align}
The process terminates at the $(d-1)^{\rm th}$ step where we have
\begin{equation}
\mu V = \left(0,\ldots,0,\gcd(\ell^1,\ldots,\ell^d)\right)~, \qquad \textrm{with} \quad V := V_{(1)} \cdots V_{(d-1)}~.
\end{equation}
$V$ is a product of elements of $GL(d,\mathbb{Z})$ and therefore is an element of $GL(d,\mathbb{Z})$.

Let the new basis of $\Lambda_{\rm mw}^{\rm ef}$, with respect to which $\mu = (0,\ldots,0,L)$, be denoted $\{h_{0}^A\}_{A=1}^{d-1} \cup \{h_\gi\}$.  It is given in terms of the old basis by the transpose of $V$:
\begin{equation}
h_{0}^A = \sum_{B=1}^d (V^T)^{A}_{\phantom{A}B} h^B~, \quad A = 1,\ldots,d-1~, \qquad h_\gi = \sum_{B=1}^d (V^T)^{d}_{\phantom{d}B} h^B~.
\end{equation}
Now, $V^T = V_{(d-1)}^T \cdots V_{(1)}^T$ and the simplicity of the individual $V_{(k)}^T$ allows for a simple recursive formula for the new basis.  The key observation is that after the $k^{\rm th}$ factor, $V_{(k)}^T$, acts, $h_{0}^A$ is unaffected by the remaining factors for $A \leq k$.  In particular we have
\begin{align}
& \left( \begin{array}{c} h_{0}^1 \\ {h'}^2 \\ h^3 \\ \vdots \\ h^d \end{array} \right) = V_{(1)}^T \left( \begin{array}{c} h^1 \\ h^2 \\ h^3 \\ \vdots \\ h^d \end{array} \right)~, \qquad \left( \begin{array}{c} h_{0}^1 \\ h_{0}^2 \\ {h'}^3 \\\vdots \\ h^d \end{array} \right) = V_{(2)}^T \left( \begin{array}{c} h_{0}^1 \\ {h'}^2 \\ h^3 \\ \vdots \\ h^d \end{array} \right)~,
\end{align}
\etc.  At the first step we have
\begin{equation}
h_{0}^1 = \frac{1}{\gcd(\ell^1,\ell^2)} (\ell^2 h^1 - \ell^1 h^2)~, \qquad {h'}^2 = x_{\ell^1,\ell^2} h^1 + y_{\ell^1,\ell^2} h^2~.
\end{equation}
Now observe at the second step we can use $y_{\ell^1,\ell^2} = \frac{1}{\ell^2} (\gcd(\ell^1,\ell^2) -\ell^1 x_{\ell^1, \ell^2})$ to write
\begin{align}
h_{0}^2 =&~ \frac{\ell^3}{\gcd(\ell^1,\ell^2,\ell^3)} {h'}^2 - \frac{\gcd(\ell^1,\ell^2)}{\gcd(\ell^1,\ell^2,\ell^3)} h^3 \cr
=&~ \frac{\ell^3}{\gcd(\ell^1,\ell^2,\ell^3)} \left[ x_{\ell^1,\ell^2} \left( h^1 - \frac{\ell^2}{\ell^1} h^2\right) + \frac{\gcd(\ell^1,\ell^2)}{\ell^1} \right]  - \frac{\gcd(\ell^1,\ell^2)}{\gcd(\ell^1,\ell^2,\ell^3)} h^3 \cr
=&~ \frac{\gcd(\ell^1,\ell^2)}{\ell^2 \gcd(\ell^1,\ell^2,\ell^3)} \left( \ell^3 h^2 - \ell^2 h^3 + \ell^3 x_{\ell^1,\ell^2} h_{0}^1 \right)~.
\end{align}
More generally at the $A^{\rm th}$ step we have
\begin{align}
h_{0}^A =&~ \frac{\ell^{A+1}}{\gcd(\ell^1,\ldots,\ell^{A+1})} {h'}^A - \frac{\gcd(\ell^1,\ldots,\ell^A)}{\gcd(\ell^1,\ldots,\ell^{A+1})} h^{A+1}~, \quad \textrm{and} \cr
{h'}^{A+1} =&~ x_{\gcd(\ell^1,\ldots,\ell^{A}),\ell^{A+1}} \gcd(\ell^1,\ldots,\ell^A) +  y_{\gcd(\ell^1,\ldots,\ell^{A}),\ell^{A+1}} \ell^{A+1}~.
\end{align}
Eliminating $y$ in ${h'}^{A}$ in favor of $x$ using the corresponding B\'ezout identity, one can show that
\begin{equation}\label{h0Ares}
h_{0}^A = \frac{\gcd(\ell^1,\ldots,\ell^A)}{\ell^A \gcd(\ell^1,\ldots,\ell^{A+1})} \left( \ell^{A+1} h^A - \ell^A h^{A+1} + \ell^{A+1} x_{\gcd(\ell^1,\ldots,\ell^{A-1}),\ell^A} h_{0}^{A-1} \right)~,
\end{equation}
holds for all $A = 2,\ldots,d-1$.  It also holds for $A = 1$ if we define $h_{0}^0 := 0$.  In this form, the fact that $h_{0}^A$ is an \emph{integral} linear combination of the $h^A$ is not manifest, but this is guaranteed by the construction.  It is manifest from \eqref{h0Ares}, however, that $h_{0}^A \in \ker(\mu)$ since $\mu(h^A) = \ell^A$.

We can also obtain an equally compact expression for $h_\gi$.  At the $(d-1)^{\rm th}$ step we have
\begin{align}\label{hthres}
h_\gi =&~ x_{\gcd(\ell^1,\ldots,\ell^{d-1}),\ell^d} {h'}^{d-1} + y_{\gcd(\ell^1,\ldots,\ell^{d-1}),\ell^d} h^d \cr
=&~ x_{\gcd(\ell^1,\ldots,\ell^{d-1}),\ell^d} \left( {h'}^{d-1} - \frac{\gcd(\ell^1,\ldots,\ell^{d-1})}{\ell^d} h^d \right) + \frac{L}{\ell^d} h^d \cr
=&~ \frac{L}{\ell^d} \left( h^d + x_{\gcd(\ell^1,\ldots,\ell^{d-1}),\ell^d} h_{0}^{d-1} \right)~.
\end{align}
Again, the fact that $h_\gi$ is an integral combination of the $h^A$ is not manifest in this form, but it is guaranteed.  We do see quite directly from this expression that $\mu(h_\gi) = L$.

We are also interested in a slightly different change of basis for $\mathfrak{t}^{\rm ef}$ where we again take the $h_{0}^{A}$ as the first $d-1$ basis vectors, but now take the final basis vector to be
\begin{align}\label{hcmexp}
h_{\rm cm} =&~ \frac{1}{(\gm,X_\infty)} X_{\infty}^{\rm ef} = \frac{1}{(\gm,X_\infty)} \sum_{A=1}^d \langle \alpha_{I_A}, X_\infty \rangle h^A \equiv \frac{1}{M} \sum_{A=1}^d m_A \frac{h^A}{\ell^A}~,
\end{align}
where in the last step we defined
\begin{equation}
m_A := \ell^A \langle \alpha_{I_A}, X_\infty \rangle = n_{\rm m}^{I_A} (H_{I_A},X_\infty)~, \qquad M := \sum_{A=1}^d m_A = (\gm,X_\infty)~.
\end{equation}
These have a physical interpretation as the mass of all fundamental monopoles of type $A$, and the total mass.

The basis $\{ h_{0}^A \}_{A=1}^{d-1} \cup \{ h_{\rm cm}\}$ is adapted to the factorization of the vanilla moduli space, and thus it is important to be able to expand generic elements of $\mathfrak{t}^{\rm ef}$ in terms of this basis.  This requires knowledge of its the integral-dual basis on $(\mathfrak{t}^{\rm ef})^\ast$.  Recall that the simple roots $\{ \alpha_A \}_{A=1}^d$ furnish the integral-dual basis to the fundamental magnetic weights, $\{h^A \}$.  Let us denote the basis of $(\mathfrak{t}^{\rm ef})^\ast$ that is integral-dual to $\{ h_{0}^A \}_{A=1}^{d-1} \cup \{ h_{\rm cm}\}$ by $\{ \beta_A \}_{A = 1}^d$.  As discussed in the text, we already know that $\beta_d = (\gamma_{\rm m}^{\rm ef})^\ast = \sum_{A=1}^d \ell^A \alpha_A$, for this manifestly satisfies the required properties
\begin{equation}
\langle  (\gamma_{\rm m}^{\rm ef})^\ast, h_{0}^A \rangle = \mu(h_{0}^A) = 0~, \qquad \langle (\gamma_{\rm m}^{\rm ef})^\ast , h_{\rm cm} \rangle = 1~.
\end{equation}
However here we will derive this directly from the inverse of the transformation \eqref{h0Ares} and \eqref{hcmexp}, as well as obtain explicit expressions for the $\{\beta_A\}_{A=1}^{d-1}$ in terms of the $\alpha_A$.  These $\beta_A$ are required to satisfy
\begin{equation}
\langle \beta_A, h_{0}^{B} \rangle = \delta_{A}^{\phantom{A}B}~, \qquad \langle \beta_A, h_{\rm cm} \rangle = 0~.
\end{equation}

To obtain this transformation it is useful to first rewrite \eqref{h0Ares} and \eqref{hcmexp} in terms of certain rescaled quantities.  Let
\begin{align}
& \tilde{h}^A = \frac{h^A}{\ell^A}~,~ A = 1,\ldots,d~, \cr
& \tilde{h}_{0}^A := \frac{\gcd(\ell^1,\ldots,\ell^{A+1})}{\ell^{A+1} \gcd(\ell^1,\ldots,\ell^{A})} h_{0}^A~,~ A = 1,\ldots,d-1~, \qquad \tilde{h}_{0}^d := h_{\rm cm}~, \cr
& \tilde{x}_{A} := \frac{\gcd(\ell^1,\ldots,\ell^A)}{\gcd(\ell^1,\ldots,\ell^{A+1})} x_{\gcd(\ell^1,\ldots,\ell^A),\ell^{A+1}}~,~ A = 1,\ldots,d-2~.
\end{align}
Then one can check that \eqref{h0Ares} and \eqref{hcmexp} are equivalent to
\begin{equation}\label{U1U2}
\sum_{B=1}^d (U_{(1)})^{A}_{\phantom{A}B} \tilde{h}^B = \sum_{C=1}^d (U_{(2)})^{A}_{\phantom{A}C} \tilde{h}_{0}^C~, \qquad A = 1,\ldots,d~,
\end{equation}
where the matrix $U_{(1)}$ is given by
\begin{align}
& (U_{(1)})^{k}_{\phantom{k}k} = 1 = - (U_{(1)})^{k}_{\phantom{k}k+1} ~,~ k =1,\ldots,d-1~, \qquad (U_{(1)})^{d}_{\phantom{d}k} = \frac{m_k}{M}~,~ k=1,\ldots,d~, \cr
& (U_{(1)})^{A}_{\phantom{A}B} = 0~,~ \textrm{otherwise}~, \raisetag{18pt}
\end{align}
and the matrix $U_{(2)}$ is given by
\begin{align}
& (U_{(2)})^{k}_{\phantom{k}k} = 1~,~k = 1,\ldots, d~, \qquad (U_{(2)})^{k+1}_{\phantom{k+1}k} = - \tilde{x}_k~,~ k =1,\ldots, d-2~, \cr
& (U_{(2)})^{A}_{\phantom{A}B} = 0~,~ \textrm{otherwise}~.
\end{align}

Similarly, define the rescaled dual quantities
\begin{align}
& \tilde{\alpha}_A = \ell^A \alpha_A~, ~ A = 1,\ldots,d~, \cr
& \tilde{\beta}_A = \frac{\ell^{A+1} \gcd(\ell^1,\ldots,\ell^{A})}{\gcd(\ell^1,\ldots,\ell^{A+1})} \beta_A~,~ A = 1,\ldots,d-1~, \qquad \tilde{\beta}_d = \beta_d~,
\end{align}
such that $\langle \tilde{\alpha}_A, \tilde{h}^B \rangle = \delta_{A}^{\phantom{A}B} = \langle \tilde{\beta}_A, \tilde{h}_{0}^B \rangle$.  Then, starting from \eqref{U1U2}, one can show that
\begin{equation}\label{alpha2beta}
\tilde{\beta}_A = \sum_{B=1}^d (U_{(2)}^T (U_{(1)}^{-1})^T)_{A}^{\phantom{A}B} \tilde{\alpha}_B~.
\end{equation}
In particular we need the inverse of $U_{(1)}$.  This is a matrix with all entries nonzero; we find
\begin{align}
& (U_{(1)}^{-1})^{A}_{\phantom{A}B} = a_A~,~ \textrm{for}~ 1 \leq A \leq B < d~, \cr
& (U_{(1)}^{-1})^{A}_{\phantom{A}B} = b_A~,~ \textrm{for}~ 1 \leq B < A \leq d~, \cr
& (U_{(1)}^{-1})^{A}_{\phantom{A}d} = 1~, ~ \textrm{for}~ A = 1,\ldots,d~,
\end{align}
where
\begin{equation}
a_A :=  \frac{m_{A+1} + \cdots + m_d}{M}~, \qquad b_A = -\frac{m_1 + \cdots + m_A}{M}~.
\end{equation}
Then from \eqref{alpha2beta} one indeed recovers
\begin{equation}
\beta_d = \sum_{A=1}^d \ell^A \alpha_A = (\gamma_{\rm m}^{\rm ef})^\ast~,
\end{equation}
while for the other $\beta_A$ one finds
\begin{equation}\label{betares1}
\beta_A = \frac{\gcd(\ell^1,\ldots,\ell^{A+1})}{\ell^{A+1} \gcd(\ell^1,\ldots,\ell^{A})} \times \left\{ \begin{array}{l l} \beta_{A}' - \tilde{x}_A \beta_{A+1}' ~, & 1 \leq A \leq d-2 \\ \beta_{d-1}' ~, & A = d-1~, \end{array} \right.
\end{equation}
where the $\beta_{A}' := \sum_{B=1}^{d} ((U_{(1)}^{-1})^T)_{A}^{\phantom{A}B} \tilde{\alpha}_B$, are given by
\begin{equation}\label{betares2}
\beta_{A}' = a_A \sum_{B=1}^{A} \ell^B \alpha_B + b_A \sum_{B = A+1}^d \ell^B \alpha_B ~, \quad A = 1,\ldots,d-1~.
\end{equation}

Let us use these results to compute the decomposition of $h_\gi$, \eqref{hthres} with respect to the basis $\{h_{0}^A\}_{A=1}^{d-1} \cup \{h_{\rm cm}\}$.  First, the component along $h_{\rm cm}$ is
\begin{equation}
\langle \beta_d, h_\gi\rangle = \langle (\gamma_{\rm m}^{\rm ef})^\ast, h_\gi\rangle = \mu(h_\gi) = L~.
\end{equation}
Keeping in mind that $\langle \beta_A, h_{0}^B \rangle = \delta_{A}^{\phantom{A}B}$, we first have
\begin{align}
\langle \beta_A, h_\gi \rangle =&~ \frac{L}{\ell^d} \langle \beta_A, h^d \rangle~, ~A = 1,\ldots,d-2~, \cr
\langle \beta_{d-1}, h_\gi \rangle =&~ \frac{L}{\ell^d} \left( \langle \beta_{d-1}, h^d \rangle + x_{\gcd(\ell^1,\ldots,\ell^{d-1}),\ell^d} \right)~.
\end{align}
Now for all $\beta_{A}'$ we have that $\langle \beta_{A}' ,h^d \rangle = \ell^d b_A$.  Then, with $b_d = -1$, we find that both cases can be expressed in the same form:
\begin{equation}
\langle \beta_A, h_\gi \rangle = \frac{L \gcd(\ell^1,\ldots, \ell^{A+1})}{\ell^{A+1} \gcd(\ell^1,\ldots,\ell^A)} \left\{ b_A - \frac{\gcd(\ell^1,\ldots,\ell^A)}{\gcd(\ell^1,\ldots,\ell^{A+1})} x_{\gcd(\ell^1,\ldots,\ell^A),\ell^{A+1}} b_{A+1} \right\}~,
\end{equation}
for $A = 1,\ldots,d-1$.  Using B\'ezout's identity for $x_{\gcd(\ell^1,\ldots,\ell^A),\ell^{A+1}},y_{\gcd(\ell^1,\ldots,\ell^A),\ell^{A+1}}$ and that $b_A - b_{A+1} = m_{A+1}/M$, we can also write this as
\begin{equation}\label{hthbetacomp}
\langle \beta_A, h_\gi \rangle = L \left\{ \frac{x_{\gcd(\ell^1,\ldots,\ell^A),\ell^{A+1}}}{\ell^{A+1}} \cdot \frac{m_{A+1}}{M} - \frac{y_{\gcd(\ell^1,\ldots,\ell^A),\ell^{A+1}}}{\gcd(\ell^1,\ldots,\ell^A)} \cdot \frac{m_1 + \cdots + m_A}{M} \right\}~.
\end{equation}
%

%%%%%%%%%%%%%%%%%%%%%
%%%%%%%%%%%%%%%%%%%%%
\section{Collective coordinate expansion in the presence of defects}\label{app:cc}
%%%%%%%%%%%%%%%%%%%%%
%%%%%%%%%%%%%%%%%%%%%

In this appendix we provide the details of the collective coordinate expansion leading to \eqref{Lcc}.  The various terms in the vanilla part of the Lagrangian \eqref{actionrealform} contribute to the integrand as follows when evaluated on the field configurations \eqref{modspacea0}, \eqref{A0Ysolve}:
\begin{align}
\Tr\left\{ \hat{E}_{a} \hat{E}^a \right\} =&~ \Tr\left\{ \dot{z}^m \dot{z}^n \delta_{m} \hat{A}^a \delta_n \hat{A}_a + \hat{D}_a A_{0}^{\rm h} \hat{A}^a A_{0}^{\rm h} - f_1 - f_2 \right\}  + \cr
&~ -2 \dot{z}^m \pd^i \Tr \left\{ A_{0}^{\rm h} \delta_i A_i \right\} + O(g_{0}^4)~, \label{magkin} \\
\Tr \left\{-\half \hat{F}^{ab} \hat{F}_{ab} \right\} =&~ -2 \pd^i \Tr \left\{ X B_i + \hat{F}_{ib} \delta^{(2)} \hat{A}^b\right\} + O(g_{0}^4)~, \label{magpot} \\
\Tr \left\{(D_0 Y)^2 \right\} =&~ O(g_{0}^4)~, \label{Ykin} \\
\Tr \left\{ - \hat{D}^a Y \hat{D}_a Y \right\} =&~ \Tr \left\{ - \hat{D}^a \epsilon_{Y_\infty} \hat{D}_a \epsilon_{Y_\infty} - \hat{D}^a A_{0}^{\rm h} \hat{D}_a A_{0}^{\rm h} - f_1 + f_2 \right\} + \cr
&~ - 2 \pd^i \Tr \left\{ A_{0}^{\rm h} D_i \epsilon_{Y_\infty} \right\} + O(g_{0}^4)~, \label{Ypot} \\
\Tr \left\{ \frac{g_{0}^2 \theta_0}{4\pi^2} B^i E_i \right\} =&~  \pd^i \Tr \left\{ \frac{g_{0}^2 \theta_0}{4\pi^2} \left( - \dot{z}^m X \delta_m A_i  + B_i A_{0}^{\rm h} \right) + f_1 \right\} + O(g_{0}^4)~, \label{thetaccterm}
\end{align}
and
\begin{align}
& \Tr \left\{-2i \rho^A (D_0 \rho_A + [Y,\rho_A]) \right\} = \cr
& \qquad \qquad = i \Tr \bigg\{ \delta_m \hat{A}^a \chi^m \left( \delta_n \hat{A}_a \dot{\chi}^n + \dot{p}^p D_p \delta_n \hat{A}_a \chi^n \right)  - \epsilon_{Y_\infty} [\delta_m \hat{A}^a, \delta_n \hat{A}_a] \chi^m \chi^n  + 2f_1 \bigg\} + \cr
&  \qquad \qquad \quad - i \chi^m \chi^n \pd^i \Tr \left\{ A_{0}^{\rm h} D_i \phi_{mn} \right\} + O(g_{0}^4)~, \label{rhokin} \\
& \Tr \left\{ -2i \lambda^A (D_0 \lambda_A - [Y,\lambda_A]) \right\} = O(g_{0}^4)~, \label{lambdakin} \\
& \Tr \left\{ -2 \lambda^A \bar{\tau}^a \hat{D}_a \rho_A + 2 \rho^A \tau^a \hat{D}_a \lambda_A \right\} = O(g_{0}^4)~. \label{fermpot}
\end{align}
In these expressions $f_{1,2}$ are shorthand for the quantities
\begin{align}
f_1 =&~ \frac{i}{2} \chi^p \chi^q \Tr \left\{ \hat{D}^a A_{0}^{\rm h} \hat{D}_a \phi_{pq} \right\}  \cr
=&~ -i \chi^p \chi^q \Tr \left\{ A_{0}^{\rm h} [\delta_p \hat{A}^a, \delta_q \hat{A}_a] \right\} + \frac{i}{2} \chi^p \chi^q \pd^i \Tr \left\{ A_{0}^{\rm h} D_i \phi_{pq} \right\} \cr
=&~  \frac{i}{2} \chi^p\chi^q \pd^i \Tr \left\{ \phi_{pq} D_i A_{0}^{\rm h} \right\}~, \cr
f_2 =&~ \frac{1}{16} \chi^m \chi^n \chi^p \chi^q \Tr \left\{ \hat{D}^a \phi_{mn} \hat{D}_a \phi_{pq} \right\}~.
\end{align}
In several places we used $\hat{D}^2 A_{0}^{\rm h} = 0$ and $\hat{D}^a \delta_m \hat{A}_a = 0$ to write terms as total derivatives.  The quantity $\delta^{(2)} \hat{A}_b$ in \eqref{magkin} denotes the $O(g_{0}^2)$ correction to $\hat{A}_b$ in \eqref{modspacea0} which we have not computed.  This boundary term can receive a finite contribution from the infinitesimal two-spheres, where $\hat{F}_{ab} \sim O(\varepsilon_{n}^{-2})$, but we will see that it is canceled by the boundary action.  Also we have kept the terms $(\hat{D}_a A_{0}^{\rm h})^2$ and $g_{0}^2 B^i D_i \hat{A}_{0}^{\rm h}$ even though they are $O(g_{0}^4)$ because they lead to divergent contributions (which must be canceled by the boundary action).

The vanilla Lagrangian is the sum total of \eqref{magkin} through \eqref{fermpot} integrated with measure $\frac{1}{g_{0}^2} \int_{\UU} \ed^3 x$.  In the sum there are simplifications; for example the $f_2$ terms cancel out.  Several of the boundary terms evaluate to zero as well:
\begin{equation}
\int_{\pd U} \ed^2 S^i \Tr \left\{ A_{0}^{\rm h} \delta_m A_i \right\} = \int_{\pd U} \ed^2 S^i \Tr \left\{ A_{0}^{\rm h} D_i \epsilon_{Y_\infty} \right\} = \int_{\pd U} \ed^2 S^i \Tr \left\{ A_{0}^{\rm h} D_i \phi_{mn} \right\} = 0~.
\end{equation}
The first vanishes because $\delta_m A_i$ is $O(\varepsilon_{n}^{-1/2})$ on the infinitesimal two-spheres and $O(r^{-2})$ asymptotically while $A_0$ is $O(\varepsilon_{n}^{-1})$ on the infinitesimal two-spheres and $O(r^{-1})$ asymptotically.  The second and third vanish for the same reason, after using that $D_i \epsilon_{Y_\infty} = - {\rm G}(Y_\infty)^m \delta_m A_i$ and $D_i \phi_{mn} = -2 D_{[m} \delta_{n]} A_i$.  Recall, in general, $\hat{D}_a \epsilon_H = - {\rm G}(H)^m \delta_{m} \hat{A}_a$, where ${\rm G}(H)$ is a triholomorphic Killing vector on $\fMM$.  Taking these into account we find
\begin{align}\label{Lvancc1}
L_{\rm van} \bigg|_{\textrm{c.c.ans.}} =&~ \frac{4\pi}{g_{0}^2} \left[ \half g_{mn} \left( \dot{z}^m \dot{z}^n + i \chi^m \DD_t \chi^n - {\rm G}(Y_\infty)^m {\rm G}(Y_\infty)^n \right) - \frac{i}{2} \chi^m \chi^n \nabla_m {\rm G}(Y_\infty)_n \right] + \cr
&~ + \frac{\theta_0}{2\pi} g_{mn} \dot{z}^m {\rm G}(X_\infty)^n + \frac{1}{g_{0}^2} \int_{\UU} \ed^3 x f_1 + \cr
&~ - \frac{2}{g_{0}^2} \int_{\pd\UU} \ed^2 S^i \Tr \left\{ X B_i + \hat{F}_{ib} \delta^{(2)} \hat{A}^b - \frac{g_{0}^2 \theta_0}{8\pi^2} B_i A_{0}^{\rm h} \right\} + O(g_{0}^2)~. \quad \raisetag{20pt}
\end{align}
In this expression $\nabla_m$ is the covariant derivative with respect to the Levi--Civita connection while $\DD_t \chi^n := \dot{\chi}^n + \dot{z}^p \Gamma^{n}_{\phantom{n}pq} \chi^q$ is the pullback of the covariant derivative to the worldline $z^m(t)$.  In obtaining \eqref{Lvancc1} we used \eqref{metC} and \eqref{ccChristoffel} for the metric and Christoffel symbols.

The $\theta_0$ term in the second line of \eqref{Lvancc1} originates from
\begin{align}
\int_{\pd\UU} \ed^2 S^i \Tr \left\{ X \delta_{m} A_i \right\} =&~ \lim_{r \to \infty} \int_{S_{\infty}^2} \ed^2 \Omega r^2 \hat{r}^i \left\{ X_\infty \delta_{m} A_i \right\} \cr
=&~ \int_{\UU} \ed^3 x \Tr \left\{ \hat{D}^a \epsilon_{X_\infty} \delta_{m} \hat{A}_a \right\} \cr
=&~ - 2\pi g_{mn} {\rm G}(X_\infty)^n ~. 
\end{align}
Of the remaining terms in the second and third line of \eqref{Lvancc1}, only the $X B_i$ term gives a contribution on the asymptotic two-sphere, which evaluates to $-\frac{4\pi}{g_{0}^2} (\gm, X_\infty)$.  In contrast all of them give contributions on the infinitesimal two-spheres.  In detail,
\begin{align}\label{Lvanccboundary}
&  \frac{1}{g_{0}^2} \int_{\UU} \ed^3 x f_1 - \frac{2}{g_{0}^2} \int_{\pd\UU} \ed^2 S^i \Tr \left\{ X B_i + \hat{F}_{ib} \delta^{(2)} \hat{A}^b - \frac{g_{0}^2 \theta_0}{8\pi^2} B_i A_{0}^{\rm h} \right\} = \cr
&  = \frac{2}{g_{0}^2}  \sum_n \int_{S_{\varepsilon_n}^2} \ed^2 \Omega_n \varepsilon_{n}^2 \hat{r}_{n}^i \bigg\{ X B_i + \epsilon_{ijk} B^k \delta^{(2)} A^j + B_i \delta^{(2)} X + A_{0}^{\rm h} D_i A_{0}^{\rm h}  -\frac{i}{4} \chi^p \chi^q \phi_{pq} D_i A_{0}^{\rm h} \bigg\} + \cr
& \quad \,  -\frac{4\pi}{g_{0}^2} (\gm,X_\infty) ~.
\end{align}

We have yet to consider, however, the expansion of the defect Lagrangian:
\begin{align}\label{Ldefcc}
L_{\rm def} =&~ -\frac{2}{g_{0}^2} \sum_n \int_{S_{\varepsilon_n}^2} \ed^2 \Omega_n \varepsilon_{n}^2 \hat{r}_{n}^i \left\{ X B_i + Y E_i \right\} \bigg|_{\textrm{c.c.~ansatz}} ~,
\end{align}
where
\begin{align}
X B_i \to &~  X B_i + B_i \delta^{(2)} X + X \epsilon_{ijk} D^j \delta^{(2)} A^k + O(g_{0}^4)~, \cr
Y E_i \to &~ \left( \epsilon_{Y_\infty} + \frac{i}{4} \phi_{pq} \chi^p \chi^q + A_{0}^{\rm h} \right)\left( -\dot{z}^m \delta_m A_i - \frac{i}{4} \chi^m \chi^n D_i \phi_{mn} + D_i A_{0}^{\rm h} \right) + O(g_{0}^4) \cr
 \to &~ \epsilon_{Y_\infty} D_i A_{0}^{\rm h} + \frac{i}{4} \chi^p \chi^q \phi_{pq} D_i A_{0}^{\rm h} + A_{0}^{\rm h} D_i A_{0}^{\rm h} + O(g_{0}^4)~,
\end{align}
on the collective coordinate ansatz.  In the second step of evaluating $Y E_i$, we kept only the terms that are finite or divergent in the limit $\varepsilon_n \to 0$.  Notice that all $X B_i$ terms cancel when \eqref{Ldefcc} is added to \eqref{Lvanccboundary}, as do the divergent $A_{0}^{\rm h} D_i A_{0}^{\rm h}$ terms.  The remaining terms are finite and can be evaluated explicitly:
\begin{align}\label{newint1}
& -\frac{2}{g_{0}^2} \sum_n \lim_{\varepsilon_n \to 0} \int_{S_{\varepsilon_n}^2} \ed^2 \Omega_n \varepsilon_{n}^2 \hat{r}_{n}^i \Tr \left\{ \epsilon_{Y_\infty} D_i A_{0}^{\rm h} \right\} = \frac{\theta_0}{2\pi} \sum_n (P_n, \epsilon_{Y_\infty}(\vec{x}_n) ) = \cr
& \qquad \qquad = \frac{\theta_0}{2\pi} \left[ (\gm, Y_\infty) - g_{mn} {\rm G}(X_\infty)^m {\rm G}(Y_\infty)^n \right]~,
\end{align}
and
\begin{align}\label{newint2}
& -\frac{i}{g_{0}^2} \chi^p \chi^q \sum_n \lim_{\varepsilon_n \to 0} \int_{S_{\varepsilon_n}^2} \ed^2 \Omega_n \varepsilon_{n}^2 \hat{r}_{n}^i \Tr \left\{ \phi_{pq} D_i A_{0}^{\rm h} \right\}  \cr
& \qquad \qquad = -\frac{i \theta_0}{8\pi^2} \chi^p \chi^q \int_{\UU} \ed^3 x \Tr \left\{ \hat{D}^a \phi_{pq} \hat{D}_a (X - \epsilon_{X_\infty}) \right\} \cr 
& \qquad \qquad = \frac{i \theta_0}{8\pi^2} \chi^p \chi^q \int_{\UU} \ed^3 x \Tr \left\{ (X - \epsilon_{X_\infty}) \hat{D}^2 \phi_{pq}\right\} \cr
& \qquad \qquad = -\frac{i \theta_0}{2\pi} \chi^p \chi^q \nabla_p {\rm G}(X_\infty)_q~.
\end{align}
Thus adding \eqref{Lvancc1} and \eqref{Ldefcc}, we arrive at the collective coordinate Lagrangian \eqref{Lcc}.

In order to obtain \eqref{newint1} we first combine our earlier observation \eqref{XYC}, \eqref{strangeid} with the moduli space expression for the electric charge, \eqref{gecl}, which leads to
\begin{align}\label{strangeid2}
\sum_n (P_n, \YY(\vec{x}_n)) =&~ (\gm, \YY_{\infty}^{\rm cl}) + \langle \gamma_{\rm e}, X_\infty \rangle \cr
=&~ \left(\gm, \frac{4\pi}{g_{0}^2} Y_\infty +\frac{\theta_0}{2\pi} X_\infty \right) - g_{mn} {\rm G}(\YY_\infty)^m {\rm G}(X_\infty)^n ~.
\end{align}
Then we use $\YY = \frac{4\pi}{g_{0}^2} \epsilon_{Y_\infty} + \frac{\theta_0}{2\pi} \epsilon_{X_\infty}$.  Since $\epsilon_{X_\infty}$ and $\epsilon_{Y_\infty}$ are well defined at $\vec{x} = \vec{x}_n$ and $\theta_0$-independent, while \eqref{strangeid2} must hold for all $\theta_0$, if follows that
\begin{align}
\sum_n (P_n, \epsilon_{Y_\infty}(\vec{x}_n)) =&~ (\gm, Y_\infty) - g_{mn} {\rm G}(Y_\infty)^m {\rm G}(X_\infty)^n~, \cr
\sum_n (P_n, \epsilon_{X_\infty}(\vec{x}_n)) =&~ (\gm, X_\infty) - g_{mn} {\rm G}(X_\infty)^m {\rm G}(X_\infty)^n~.
\end{align}
The first of these was used in \eqref{newint1}.  Note that in the absence of defects, where $\epsilon_{Y_\infty} \to Y$ and $\epsilon_{X_\infty} \to X$, the right-hand sides can be shown to vanish as required.

Meanwhile for \eqref{newint2}, one can use $\hat{D}_a \phi_{mn} = -2 D_{[m} \delta_{n]} \hat{A}_a$ and the asymptotics of the bosonic zero modes to argue that there are no boundary terms in going from the second to third line.  Then for the final step we first plug in $\hat{D}^2 \phi_{pq} = 2 [\delta_p \hat{A}_a, \delta_q \hat{A}^a]$, and then use \eqref{covKilling} and \eqref{intid1} to evaluate each term.

%%%%%%%%%%%%%%%%%%%%%
%%%%%%%%%%%%%%%%%%%%%
\section{Details on quantization via $(0,\ast)$-forms}\label{app:holforms}
%%%%%%%%%%%%%%%%%%%%%
%%%%%%%%%%%%%%%%%%%%%

In the main text, section \ref{sec:quantize}, we discussed how the Hilbert space of the monopole quantum mechanics has a natural representation in terms of square integrable sections of the Dirac spinor bundle over $\fmMM$ in the framed case, or $\mMM_0$ in the vanilla case.  In this appendix we provide some details on an equivalent representation in terms of complex differential forms (valued in the square root of the canonical bundle).  See also \cite{Gauntlett:1993sh,deVries:2008ic}.  The conclusion is that framed BPS states also correspond to elements of (${\rm G}(\sy)$-twisted) Dolbeault cohomology.  This could be anticipated from the classic result of Hitchin \cite{MR0358873} that the space of harmonic spinors on a K\"ahler manifold is in one-to-one correspondence with Dolbeault cohomology, valued in a square root of the canonical bundle.

On a K\"ahler manifold of complex dimension $D$, the Dirac spinor bundle $\SS$ is in one-to-one correspondence with the tensor product of the bundle of $(0,\ast)$-forms, $\Lambda^{(0,*)} = \oplus_{q=0}^D \Lambda^{(0,q)}$, with a square root of the canonical bundle $\Lambda^{(D,0)}$ which we denote by $\sqrt{\Lambda^{(D,0)}}$.  The K\"ahler manifold is spin if and only if such a square root exists, and the different possible square roots are in one-to-one correspondence with $\pi_1$ and with the different possible spin structures.  As $\fmMM$ is hyperk\"ahler we have that an everywhere non-vanishing covariantly constant section of the canonical bundle exists---which we denote $\Omega$.  If $\fMM$ is simply-connected there is furthermore a unique square root bundle, or else we simply choose a square root.  The latter has a covariantly constant section that we denote $\sqrt{\Omega}$.  In this case, then, the space of standard $(0,q)$-forms is isomorphic to the space of $(0,q)$-forms valued in the square root of the canonical bundle.  

In summary, the Hilbert space $\HH$ of our quantum mechanics can be identified with any of these:
\begin{equation}\label{Hilbertoptions}
\HH \cong \Lsq(\fmMM,\SS)\cong \Lsq(\fmMM,\sqrt{\Lambda^{(D,0)}}\otimes\Lambda^{(0,*)})\cong \Lsq(\fmMM,\Lambda^{(0,*)}) 
\end{equation}
where the isomorphism for the last equivalence is simply given by tensoring with $\sqrt{\Omega}$:
\begin{equation}
f: \Lambda^{(0,*)}(\fmMM,\mathbb{C})\rightarrow \Lambda^{(0,*)}(\fmMM,\sqrt{\Lambda^{(D,0)}}): \lambda \mapsto \sqrt{\Omega}\,\lambda\,.
\end{equation}
An explicit construction of the middle isomorphism between spinors and $\sqrt{\Lambda^{(D,0)}}$-valued forms is well known and is summarized in the paragraph containing \eqref{spinors2forms}.  In this appendix we will work in terms of the more standard $\mathbb{C}$-valued forms\footnote{Only when discussing the $R$-symmetry we will see some trace of the connection to the $\sqrt{\Lambda^{(D,0)}}$-valued forms.}. All operators discussed below can be translated to $\sqrt{\Lambda^{(D,0)}}$-valued forms by simply replacing $\OO\mapsto \tilde\OO=\sqrt{\Omega}\OO\sqrt{\Omega}^{-1}$.

%%%%%%%%%%%%%%%%
\subsection{Basic definitions}
%%%%%%%%%%%%%%%%

To continue let us make a choice of complex coordinates on the monopole moduli space $\fmMM$ (or $\mMM_0$).  We will denote these coordinates $Z^{\bfm}$, $\bfm=1,\ldots,D$, their conjugates $\Zbar^{\bfmbar}$, and the Hermitian metric in these coordinates by $g_{\bfm\bfnbar}$. Besides the bosonic coordinates $z^m$ we also complexify the fermions $\chi^m$ appearing in the monopole mechanics accordingly, writing them as $\XX^\bfm$ and $\XXbar^{\bfmbar}$.  (See \eqref{complexcoords}, where $N= 2D$.)  We will assume that these coordinates are adapted to the third complex structure so that the corresponding K\"ahler form is
\begin{equation}
\sw := \sw^3 = \frac{i}{2}g_{\bfm\bfnbar}\ed Z^{\bfm} \wedge \ed\Zbar^{\bfnbar}\,.
\end{equation}
Our conventions for the normalization of the Hermitian metric are such that
\begin{equation}
\ed s^2 = \half \left( g_{\bfm \bfnbar} \ed Z^{\bfm} \otimes \ed \Zbar^{\bfnbar} + g_{\bfmbar \bfn} \ed \Zbar^{\bfmbar} \otimes \ed Z^{\bfn} \right)~,
\end{equation}
with $(g_{\bfm\bfnbar})^\ast = g_{\bfn\bfmbar}$.  Hence if $V^m$ is a vector field then $V_{\bfm} = \half g_{\bfm\bfnbar} V^{\bfnbar}$, $V^{\bfmbar} = 2 g^{\bfmbar\bfn} V_{\bfn}$, \etc, where $g^{\bfmbar\bfm} g_{\bfm\bfnbar} = \delta^{\bfmbar}_{\phantom{\bfm}\bfnbar}$.  Flat space corresponds to $g_{\bfm\bfnbar} = \delta_{\bfm\bfnbar}$.

The relation between the triplet of K\"ahler forms and complex structures is $g(U,\bbJ^{r} V) = \sw^r(U,V)$.  In components, $(\sw^r)_{mn} = g_{mp} (\bbJ^{r})_{n}^{\phantom{n}p}$, or equivalently 
\begin{equation}\label{omega2J}
g^{qm} (\sw^r)_{mn} = - (\sw^r)_{nm} g^{mq} = (\bbJ^{r})_{m}^{\phantom{m}q}~.
\end{equation}
Define the two-forms
\begin{equation}
\sw_{\pm} := \sw^1 \pm i \sw^2 \, .
\end{equation}
It follows from the quaternionic algebra that $\sw_+ \in\Lambda^{(0,2)}$ and $\sw_- \in\Lambda^{(2,0)}$. 

It will be useful to work with a unitary local frame: 
\begin{equation}
e^{\ubfm}:=e^{\ubfm}_{\phantom{m}\bfm} \ed Z^{\bfm} \,,
\end{equation} 
such that $g_{\bfm\bfnbar}=e^{\ubfm}_{\phantom{m}\bfm} \ebar^{\ubfnbar}_{\phantom{n}\bfnbar} \delta_{\ubfm\ubfnbar}$.  The absolute value of its determinant is the half-density, $e^{1/2}$, that plays a role in the construction of the momentum operators conjugate to the coordinates:
\begin{equation}
e^{1/2} = \sqrt{\det (e^{\ubfm}_{\phantom{m}\bfm})\det (\ebar^{\ubfnbar}_{\phantom{n}\bfnbar})}=g^{1/4}\,.
\end{equation}

As Hilbert space we will take the set of square integrable sections of the bundle $\Lambda^{(0,*)}$.  In the local complex coordinates introduced above we can write such a section as
\begin{equation}
\lambda=\sum_{q=0}^D\lambda^{(q)}=  \sum_{q=0}^D \frac{1}{q!} \lambda_{\bfnbar_1 \cdots \bfnbar_q} \ed\Zbar^{\bfnbar_1}\wedge\cdots\wedge \ed\Zbar^{\bfnbar_q}~.
\end{equation}
It will also be useful to express the same form with respect to the frame basis:
\begin{equation}
\lambda^{(q)}= \frac{1}{q!} \lambda_{\ubfnbar_1 \cdots \ubfnbar_q} \ebar^{\ubfnbar_1}\wedge\cdots\wedge \ebar^{\ubfnbar_q} ~.
\end{equation}
The inner product we will use is the standard one:
\begin{equation}\label{formsmetric}
\langle\lambda_1|\lambda_2\rangle := \int_{\fmMM}\lambda_{1}^\ast \wedge \star \lambda_2 = \sum_q \frac{1}{q!} \int_{\fmMM} \ed^DZ \ed^D\Zbar \,e\, \lambda_{1\, \ubfnbar_1 \cdots \ubfnbar_q}^\ast\lambda_{2}^{\, \ubfnbar_1 \cdots \ubfnbar_q} ~.
\end{equation}

Now that we have specified our Hilbert space, our starting point is the complexified version of the canonical commutation relations \eqref{canonicalccs}:
\begin{equation}
\left[ \hat{Z}^{\bfm}, \hat{P}_{\bfn} \right]_- = i \delta^{\bfm}_{\phantom{\bfm}\bfn}~, \qquad \left[ \hat{\Zbar}{}^{\bfmbar} , \hat{\Pbar}_{\bfnbar} \right]_- = i \delta^{\bfmbar}_{\phantom{\bfm}\bfnbar}~, \qquad \left[ \hat{\XX}^{\ubfm}, \hat{\XXbar}{}^{\ubfnbar} \right]_+ = \frac{g_{0}^2}{4\pi} \delta^{\ubfm\ubfnbar}~\,,\label{complcom}
\end{equation}
with all others vanishing.  Here the barred operators are the adjoints of the unbarred ones, $\hat{\Pbar}_{\bfnbar} = (\hat{P}_{\bfn})^\dag$, \etc, with respect to the innerproduct \eqref{formsmetric}.  One can then check that these commutation relations are represented on the Hilbert space $\Lsq(\fmMM,\Lambda^{(0,\ast)})$ by defining the operators as follows:
\begin{align}
\hat Z^{\bfm} \lambda :=&~ Z^{\bfm} \lambda~, \qquad \hat{\Zbar}{}^{\bfnbar} \lambda := \Zbar^{\bfnbar} \lambda~, \\
\hat P_{\bfm} \lambda :=&~  -i\sum_{q}\frac{1}{q!}\left( e^{-1/2} \partial_{\bfm} \left(e^{1/2} \lambda_{\ubfnbar_1\cdots\ubfnbar_q}\right)\right) \ebar^{\ubfnbar_1}\wedge\cdots\wedge \ebar^{\ubfnbar_q}~,  \\
\hat{\Pbar}_{\bfnbar}\lambda :=&~ i\sum_{q}\frac{1}{q!}\left(e^{-1/2} \pdbar_{\bfnbar} \left(e^{1/2} \lambda_{\ubfnbar_1\cdots \ubfnbar_q}\right)\right) \ebar^{\ubfnbar_1}\wedge\cdots\wedge \ebar^{\ubfnbar_q} ~, \\
\hat{\XX}{}^{\ubfm} \lambda :=&~ \frac{g_{0}}{\sqrt{2\pi}} \contract^{\ubfm} \lambda~, \qquad \hat{\XXbar}{}^{\ubfnbar} \lambda := \frac{g_{0}}{\sqrt{2\pi}} \ebar^{\ubfnbar} \wedge \lambda~.  \label{XXops}
\end{align}
Here we used a shorthand $\contract^{\ubfm} \equiv \contract_{\EEbar^{\ubfm}}$ for the interior product with the tangent frame vectors $\EEbar^{\ubfm} = \delta^{\ubfm\ubfnbar} \EEbar_{\ubfnbar}$, which is defined by
\begin{equation}
\contract^{\ubfm} e^{\ubfn} =0\,,\qquad \contract^{\ubfm} \ebar^{\ubfnbar}=\delta^{\ubfm\ubfnbar}\,,
\end{equation}
and satisfies the usual $\mathbb{Z}_2$-graded Leibniz rule.

Using this basic representation one can then find the action of other operators expressed in terms of these canonical ones.  For example the super-covariant momentum operator is
\begin{equation}
\hat{\pi}_{\bfp} := \hat{P}_{\bfp} - \frac{4\pi i}{g_{0}^2} \sw_{\bfp,\ubfnbar\ubfm}(  \hat{\XXbar}{}^{\ubfnbar} \hat{\XX}{}^{\ubfm} - \hat{\XX}{}^{\ubfm} \hat{\XXbar}{}^{\ubfnbar}  ) = \hat{P}_{\bfp} - \frac{8\pi i}{g_{0}^2} \sw_{\bfp,\ubfnbar\ubfm} \hat{\XXbar}{}^{\ubfnbar} \hat{\XX}{}^{\ubfm}~,
\end{equation}
where we are using that Hermiticity of the connection implies $\sw_{\bfp,\ubfm\ubfn} = \sw_{\bfp,\ubfmbar\ubfnbar} = 0$.  One can compute that
\begin{equation}\label{XbarXaction}
(\hat{\XXbar}{}^{\ubfnbar} \hat{\XX}{}^{\ubfm} \lambda^{(q)} )_{\ubfnbar_1 \cdots \ubfnbar_q} = \frac{g_{0}^2}{4\pi} \delta^{\ubfm\ubfmbar} \sum_{i=1}^q \delta^{\ubfnbar}_{\phantom{\bfn}\ubfnbar_i} \lambda^{(q)}_{\ubfnbar_1 \cdots \ubfnbar_{i-1} \ubfmbar \, \ubfnbar_{i+1} \cdots \ubfnbar_q} ~.
\end{equation}
Then, remembering that $\sw_{\bfp,\ubfnbar\ubfm} (2\delta^{\ubfm\ubfmbar}) = \sw_{\bfp,\ubfnbar}{}^{\ubfmbar}$, we find that $\hat{\pi}_{\bfp}$ acts each component of $\lambda$ as
\begin{align}\label{formcodiv}
\hat{\pi}_{\bfp} \lambda^{(q)} =&~ -i e^{-1/2} \left[ \partial_{\bfp} \left(e^{1/2} \lambda_{\ubfnbar_1\cdots\ubfnbar_q}\right) + e^{1/2} \left(\sw_{\bfp,\ubfnbar_1}^{\phantom{p,n_1}\ubfnbar} \lambda_{\ubfnbar\ubfnbar_2 \cdots \ubfnbar_{q}} + \cdots + \sw_{\bfp,\ubfnbar_q}^{\phantom{p,n_q}\ubfnbar} \lambda_{\ubfnbar_1 \cdots \ubfnbar_{q-1} \ubfnbar} \right)  \right] \times \cr
&~ \qquad \times \ebar^{\ubfnbar_1}\wedge\cdots\wedge \ebar^{\ubfnbar_q} \cr
=&~ -i \left[ e^{-1/2} \DD_{\bfp} \left( e^{1/2} \lambda_{\ubfnbar_1 \cdots \ubfnbar_q} \right) \right] \ebar^{\ubfnbar_1} \wedge \cdots \wedge \ebar^{\ubfnbar_q} \cr
=&~ -i e^{-1/2} \nabla_{\bfp} (e^{1/2} \lambda^{(q)} ) = -i \left( \nabla_{\bfp} + \half \Gamma^{\bfm}_{\phantom{m}\bfm\bfp} \right) \lambda^{(q)}~, \raisetag{24pt}
\end{align}
and so $\hat{\pi}_{\bfp} \lambda = -i e^{-1/2} \nabla_{\bfp} (e^{1/2} \lambda)$.  The conjugate $(\hat{\pi}_{\bfp})^\dag = \hat{\pibar}_{\bfpbar}$ acts as $\hat{\pibar}_{\bfpbar} \lambda = i e^{-1/2} \nabla_{\bfpbar} (e^{1/2} \lambda)$.  In these expressions $\nabla_{\bfp}$ is the (holomorphic part of the) Levi--Civita covariant derivative while $\DD_{\bfp}$ is the associated covariant derivative on the frame bundle.

We list a number of further results in the following subsections.

%%%%%%%%%%%%%%%%%%%
\subsection{$R$-symmetry operators}
%%%%%%%%%%%%%%%%%%%

The $SU(2)_R$ generators were defined in section \ref{sec:quantize} as
\begin{equation}
\hat{I}^r=\frac{i\pi}{g_{0}^2}(\sw^r)_{mn}\hat{\chi}^m\hat{\chi}^n~.
\end{equation}
Then in the complex coordinate representation we have
\begin{align}
\hat{I}^3 =&~ \frac{2\pi i}{g_{0}^2} \sw_{\bfm\bfnbar} \left( \hat{\XX}{}^{\bfm} \hat{\XXbar}{}^{\bfnbar} - \hat{\XXbar}{}^{\bfnbar} \hat{\XX}{}^{\bfm} \right) = \frac{\pi}{g_{0}^2} g_{\bfm\bfnbar} \left( \hat{\XXbar}{}^{\bfnbar} \hat{\XX}{}^{\bfm} - \hat{\XX}{}^{\bfm} \hat{\XXbar}{}^{\bfnbar} \right)  \cr
=&~  \frac{2\pi}{g_{0}^2} g_{\bfm\bfnbar}  \hat{\XXbar}{}^{\bfnbar} \hat{\XX}{}^{\bfm} - \frac{D}{4} ~, \cr
\hat{I}_{+} :=&~ \hat{I}^1 + i \hat{I}^2 = \frac{2\pi i}{g_{0}^2} (\sw_+)_{\bfmbar\bfnbar} \hat{\XXbar}{}^{\bfmbar} \hat{\XXbar}{}^{\bfnbar} ~, \cr
\hat{I}_- :=&~  \hat{I}^1 - i \hat{I}^2 = \frac{2\pi i}{g_{0}^2} (\sw_-)_{\bfm\bfn} \hat{\XX}{}^{\bfm} \hat{\XX}{}^{\bfn} ~.
\end{align}
Note from \eqref{XXops} we have that $\hat{\XX}{}^{\bfm} = \frac{g_{0}}{\sqrt{2\pi}} g^{\bfm\bfnbar} \contract_{\pdbar_{\bfnbar}}$ and $\hat{\XXbar}{}^{\bfnbar} = \frac{g_0}{\sqrt{2\pi}} \ed \Zbar^{\bfnbar}$.  Furthermore the coordinate frame version of \eqref{XbarXaction} is
\begin{equation}\label{XbarXaction2}
(\hat{\XXbar}{}^{\bfnbar} \hat{\XX}{}^{\bfm} \lambda^{(q)} )_{\bfnbar_1 \cdots \bfnbar_q} = \frac{g_{0}^2}{4\pi} g^{\bfm\bfmbar} \sum_{i=1}^q \delta^{\bfnbar}_{\phantom{\bfn}\bfnbar_i} \lambda_{\bfnbar_1 \cdots \bfnbar_{i-1} \bfmbar \, \bfnbar_{i+1} \cdots \bfnbar{q}}^{(q)}~.
\end{equation}
It then follows that
\begin{eqnarray}
\hat I^3\lambda^{(q)}&=& \half \left(q-\frac{D}{2}\right)\lambda^{(q)} ~, \label{formR3}\\
\hat I_+\lambda&=& i \, \sw_+ \wedge\lambda~,\\
\hat I_-\lambda&=&-i \, \contract_{\sw_-} \lambda~,
\end{eqnarray}
where we've defined the contraction with respect to a two-form:
\begin{equation}\label{2formcontract}
(\contract_{\sw_-} \lambda^{(q)})_{\bfnbar_1 \cdots \bfnbar_{q-2}} := \half (\sw_{-})^{\bfmbar\bfnbar} \lambda^{(q)}_{\bfmbar\bfnbar\bfnbar_1 \cdots \bfnbar_{q-2}}~.
\end{equation}
With this definition, $\contract_{\sw_-}$ is the adjoint of $\sw_+ \wedge$ with respect to the innerproduct \eqref{formsmetric}.  As we are on a hyperk\"ahler manifold, $D$ is even and therefore the eigenvalues of $\hat{I}^3$ are half-integer as required.

Note that the action \eqref{formR3} of $2 \hat I^3$ on a $(0,q)$-form is not exactly the natural geometric action of the corresponding complex structure on this form. There is a shift by $-D/2$ that appears.  This shift has exactly the interpretation as the action of the complex structure on a section of the square root of the canonical bundle $\sqrt{\Omega} \in \sqrt{\Lambda^{(D,0)}}$.  Hence, although this section is covariantly constant, its presence in the isomorphism between spinors and forms, \eqref{Hilbertoptions}, is crucial for the correct interpretation of $R$-symmetry.  Indeed this shows that $R$-symmetry singlets lie in the middle degree, $\Lambda^{(0,D/2)}$, not the zero degree.  Furthermore it allows $\{ 2\hat{I}^3, \hat{I}_{\pm}\}$ to be interpreted as a Lefschetz $\mathfrak{sl}(2)$ triple, acting on the space of antiholomorphic forms \cite{MR1958088}.  

%%%%%%%%%%%%%%%%%%%%%%%%%%%%
\subsection{Operators corresponding to Killing vectors}
%%%%%%%%%%%%%%%%%%%%%%%%%%%%

Given a Killing vector $K$, recall the associated Noether charge operator, \eqref{Nhatcc},\footnote{To ease notation we momentarily suppress the index $E$ that labels the $K$ and $\hat{N}$ corresponding to different types of symmetries.}
\begin{align}
\hat{N} :=&~ - \half [ K^m, \hat{\pi}_m]_+ + \frac{2\pi i}{g_{0}^2} (\nabla_m K_n) \hat{\chi}^m \hat{\chi}^n \cr
=&~ - K^m \hat{\pi}_m + \frac{i}{2}( \pd_m K^m)  + \frac{2\pi i}{g_{0}^2} (\pd_{[m} K_{n]}) \hat{\chi}^m \hat{\chi}^n~,
\end{align}
where the symmetrization in the first term gives the correct operator ordering prescription, and in going to the second line we used the Killing vector equation $\nabla_{(m} K_{n)} = 0$.  When one acts this on $\lambda$, using the representation above, one finds that the $(\pd_m K^m)$ term combines with the term involving the explicit Christoffel symbol in the last line of \eqref{formcodiv} to give $\nabla_m K^m = 0$.   Thus we can write
\begin{align}
\hat{N} \lambda^{(q)} =&~ i \bigg\{ K^{\bfm} \nabla_{\bfm} + K^{\bfnbar} \nabla_{\bfnbar} + \frac{4\pi}{g_{0}^2} (\nabla_{\bfnbar} K_{\bfm}) ( \hat{\XXbar}{}^{\ubfnbar} \hat{\XX}{}^{\ubfm} -  \hat{\XX}{}^{\ubfm} \hat{\XXbar}{}^{\ubfnbar} ) + \cr
&~ \qquad + \frac{2\pi}{g_{0}^2}(\ed K)_{\bfmbar\bfnbar} \hat{\XXbar}{}^{\bfmbar} \hat{\XXbar}{}^{\bfnbar} +  \frac{2\pi}{g_{0}^2}(\ed K)_{\bfm\bfn} \hat{\XX}{}^{\bfm} \hat{\XX}{}^{\bfn}  \bigg\} \lambda^{(q)} \cr
=&~ \frac{i}{q!} \bigg\{ \left( K^{\bfm} \nabla_{\bfm} + K^{\bfnbar}\nabla_{\bfnbar} + \half \nabla^{\bfm} K_{\bfm} \right) \lambda^{(q)}_{\bfnbar_1 \cdots  \bfnbar_q} +\cr
&~ \qquad + (\nabla_{\bfnbar} K_{\bfm}) (2g^{\bfm\bfmbar}) \sum_{i=1}^q \delta^{\bfnbar}_{\phantom{\bfn}\bfnbar_i} \lambda^{(q)}_{\bfnbar_1 \cdots \bfnbar_{i-1} \bfmbar \, \bfnbar_{i+1} \cdots \bfnbar_q} \bigg\} \ed\Zbar^{\bfnbar_1} \cdots \ed\Zbar^{\bfnbar_q} + \cr
&~ +  i\left(  \ed K^{(0,2)} \wedge \lambda^{(q)} - \contract_{\ed K^{(2,0)}} \lambda^{(q)} \right)~,
\end{align} 
where we made use of \eqref{XbarXaction2} and \eqref{2formcontract}.  

The terms in curly brackets are closely related to the Lie derivative.  However for a general Killing vector field the Lie derivative does not preserve the $(p,q)$ decomposition of forms.  Acting on a $(0,q)$-form we obtain both a $(0,q)$-form and a $(1,q-1)$ form, with components
\begin{align}
(\Lie_K \lambda^{(q)})_{\bfnbar_1 \cdots \bfnbar_q} =&~ (K^{\bfm} \nabla_{\bfm} + K^{\bfnbar} \nabla_{\bfnbar}) \lambda^{(q)}_{\bfnbar_1 \cdots \bfnbar_{q}} + \sum_{i=1}^q (\nabla_{\bfnbar_i} K^{\bfnbar}) \lambda_{\bfnbar_1 \cdots \bfnbar_{i-1} \bfnbar \, \bfnbar_{i+1} \bfnbar_q}~, \cr
(\Lie_K \lambda^{(q)})_{\bfm \bfnbar_1 \cdots \bfnbar_{q-1}} =&~ (\nabla_{\bfm} K^{\bfnbar}) \lambda_{\bfnbar \bfnbar_1 \cdots \bfnbar_{q-1}}^{(q)}~.
\end{align}
We define the projected Lie derivative on the multi-form $\lambda$ so that the $(1,q-1)$-forms are omitted:
\begin{equation}
\Lie_{K}^{(0,\ast)} \lambda := \sum_{q=1}^D\frac{1}{q!} (\Lie_K \lambda^{(q)})_{\bfnbar_1 \cdots \bfnbar_q} \ed \Zbar^{\bfnbar_1} \cdots \ed \Zbar^{\bfnbar}_q~.
\end{equation}
Hence we arrive at
\begin{equation}\label{genNformaction}
\hat{N} \lambda = i \left\{ \Lie_{K}^{(0,\ast)} + \half \nabla_{\bfm} K^{\bfm} + \ed K^{(0,2)} \wedge~  - \contract_{\ed K^{(2,0)}}  \right\} \lambda~.
\end{equation}

There are essentially two sets of Killing vectors that interest us here.  The first set is comprised of   triholomorphic Killing vectors, $K = K^A$, which satisfy $\Lie_{K^A} \bbJ^r = 0$, $\forall r$.  These vectors correspond to asymptotically nontrivial gauge transformations, and the associated Noether charges are the electric charges.  The second set is a triplet of Killing vectors, $K^r$, associated with rotational symmetry.  They are not all holomorphic with respect to any one complex structure, but rather satisfy $\Lie_{K^r} \bbJ^s = (R_{\kappa}^{-1})^{r}_{\phantom{r}u}\epsilon^{us}_{\phantom{us}t} \bbJ^{t}$.  The Noether charges are the symmetry generators for a \emph{diagonal subgroup} of $SU(2)_R$ and the $SO(3)$ of angular momentum.

% % % % % % % % % %  % % %
\subsubsection{Triholomorphic Killing vectors and the electric charge operator}
% % % % % % % % % % % % % 

First, if $K$ is a holomorphic vector field, $\Lie_{K} \bbJ^{3} = 0$, then in our adapted complex coordinate system we have $\nabla_{\bfm} K^{\bfnbar} = 0 = \nabla_{\bfnbar} K^{\bfm}$.  (This is equivalent to $\pd_{\bfm} K^{\bfnbar} = 0 = \pdbar_{\bfnbar} K^{\bfm}$ since the Levi--Civita connection is Hermitian.)  By lowering with the metric this also implies $\ed K^{(2,0)} = 0 = \ed K^{(0,2)}$.  Thus if $K$ is a holomorphic Killing field, then the last two terms of \eqref{genNformaction} vanish, while the projected Lie derivative becomes the ordinary Lie derivative.

Now suppose that $K$ is a triholomorphic Killing field.  Since $\Lie_K \bbJ^{r} = 0$ and $\Lie_K g = 0$, $K$ also preserves the K\"ahler forms.  In particular, consider the following manipulation.  On the one hand,
\begin{equation}\label{LKpinitial}
(\Lie_{K} \sw_+)_{\bfmbar\bfnbar} = (\nabla_{\bfmbar} K^{\bfpbar}) (\sw_+)_{\bfpbar\bfnbar} + (\nabla_{\bfnbar} K^{\bfpbar}) (\sw_+)_{\bfmbar\bfpbar}~.
\end{equation}
Now contract both sides with $g^{\bfmbar\bfm} g^{\bfnbar\bfn} (\sw_-)_{\bfm\bfn}$:
\begin{align}
(\Lie_{K} \sw_+)_{\bfmbar\bfnbar} g^{\bfmbar\bfm} g^{\bfnbar\bfn} (\sw_-)_{\bfm\bfn} =&~ (\nabla_{\bfmbar} K^{\bfpbar}) \left( (\sw_+)_{\bfpbar\bfnbar} g^{\bfnbar\bfn}\right) \left(g^{\bfmbar\bfm} (\sw_-)_{\bfm\bfn}\right) + \cr
& + (\nabla_{\bfnbar} K^{\bfpbar}) (\sw_+)_{\bfmbar\bfpbar} \left( - (\sw_+)_{\bfpbar\bfmbar} g^{\bfmbar\bfm} \right) \left( -g^{\bfnbar\bfn} (\sw_+)_{\bfn\bfm} \right)~. \cr
\end{align}
Then apply \eqref{omega2J} on the right-hand side:
\begin{align}\label{LKpprefinal}
(\Lie_{K} \sw_+)_{\bfmbar\bfnbar} g^{\bfmbar\bfm} g^{\bfnbar\bfn} (\sw_-)_{\bfm\bfn} =&  - \frac{1}{4} (\nabla_{\bfmbar} K^{\bfpbar}) (\bbJ_{+}')_{\bfpbar}^{\phantom{\bfp}\bfn} (\bbJ_{-}')_{\bfn}^{\phantom{\bfn}\bfmbar} - \frac{1}{4} (\nabla_{\bfnbar} K_{+}^{\bfpbar}) (\bbJ_{+}')_{\bfpbar}^{\phantom{\bfp}\bfm} (\bbJ_{-}')_{\bfm}^{\phantom{\bfm}\bfnbar} \cr
=&~ (\nabla_{\bfmbar} K^{\bfpbar}) \delta_{\bfpbar}^{\phantom{\bfp}\bfmbar} + (\nabla_{\bfnbar} K^{\bfpbar}) \delta_{\bfpbar}^{\phantom{\bfp}\bfnbar} \cr
=&~ 2 \nabla_{\bfnbar} K^{\bfnbar}~, \raisetag{20pt}
\end{align}
where in the second step we used the algebra $\bbJ_{+}' \bbJ_{-}' = -2 \mathbbm{1} - 2i \bbJ_{3}'$.  On the other hand $\Lie_K \sw_+ = 0$.  Thus
\begin{equation}\label{LKpfinal}
\Lie_K \sw_+ = 0 \quad \Rightarrow \quad \nabla_{\bfnbar} K^{\bfnbar} = 0 \quad \iff \quad \nabla_{\bfm} K^{\bfm} = 0~,
\end{equation}
where the last implication follows from the fact that $K$ is Killing and hence $\nabla_m K^m = 0$.

Hence if $K$ is a triholomorphic Killing field then \eqref{genNformaction} collapses to the ordinary Lie derivative.  In particular, the electric charge operator $\hat{\gamma}_{\rm e} = \sum_A \alpha_{I_A} \hat{N}^A$, is
\begin{equation}
\hat{\gamma}_{\rm e} = i \sum_A \alpha_{I_A} \Lie_{K_A}~,
\end{equation}
where the $K^A$ generate $2\pi$-periodic isometries of $\fmMM$.

% % % % % % % % % % % %
\subsubsection{Spatial rotation Killing vectors and angular momentum}\label{app:angmom}
% % % % % % % % % % % %

Suppose instead we have a triplet of Killing vectors satisfying $\Lie_{K^r} \bbJ^{s} = (R_{\kappa}^{-1})^{r}_{\phantom{r}u}\epsilon^{us}_{\phantom{us}t} \bbJ^{t}$.  Then the same relation holds for the K\"ahler forms, and we investigate the consequences of the latter.  So as not to carry around the $SO(3)$ rotation matrix $R_\kappa$, let us define a rotated triplet of Killing vectors such that
\begin{equation}
\tilde{K}^r = (R_{\kappa})^{r}_{\phantom{r}s} K^s~, \qquad \Lie_{\tilde{K}^r} \sw^s = \epsilon^{rs}_{\phantom{rs}t} \sw^t~.
\end{equation}
Furthermore it is convenient to work with $\tilde{K}_{\pm} = \tilde{K}^1 \pm i \tilde{K}^2$.  In terms of these, the relations are
\begin{align}
& \Lie_{\tilde{K}_{\pm}} \sw_{\pm} = 0~,  \qquad \Lie_{\tilde{K}_{\pm}} \sw_{\mp} = \mp 2i \, \sw~,  \qquad \Lie_{\tilde{K}_{\pm}} \sw = \pm i \, \sw_{\pm}~, \cr
& \Lie_{\tilde{K}^3} \sw_{\pm} = \mp i \, \sw_{\pm}~, \qquad \Lie_{\tilde{K}^3} \sw =  0~.
\end{align}

Let us focus on the equations involving $\tilde{K}_+$ first.  By applying the same manipulations that led from \eqref{LKpinitial} to \eqref{LKpfinal}, we deduce
\begin{equation}\label{omegapmcontract}
\Lie_{\tilde{K}_+} \sw_+ = 0 \quad \Rightarrow  \quad  \nabla_{\bfm} \tilde{K}_{+}^{\bfm} = 0~,  
\end{equation}
Meanwhile from the $\bfm\bfn$ and $\bfmbar\bfnbar$ components of the $\sw$ equation we find
\begin{align}
(\Lie_{\tilde{K}_+} \sw)_{\bfm\bfn} = 0 \quad & \Rightarrow \quad (\nabla_{[\bfm} \tilde{K}_{+}^{\bfpbar}) \sw_{\bfn] \bfpbar} = 0 \quad \Rightarrow \quad \nabla_{[\bfm} (\tilde{K}_+)_{\bfn]} = 0~, \cr
(\Lie_{\tilde{K}_+} \sw)_{\bfmbar \bfnbar} = i \, (\sw_+)_{\bfmbar\bfnbar} \quad & \Rightarrow \quad 2( \nabla_{[\bfmbar} \tilde{K}_{+}^{\bfp}) \sw_{|\bfp| \bfnbar]} = i (\sw_+)_{\bfmbar\bfnbar} \cr
& \Rightarrow \quad 2 \nabla_{[\bfmbar} (\tilde{K}_+)_{\bfnbar]} = (\sw_+)_{\bfmbar\bfnbar}~. 
\end{align}
The remaining components of the other equations give results equivalent to these.  The full set of conditions can be summarized as follows:
\begin{equation}\label{Kplusres}
\nabla_{\bfm} \tilde{K}_{+}^{\bfm} = 0~, \qquad \ed \tilde{K}_{+}^{(2,0)} = 0~, \qquad \ed \tilde{K}_{+}^{(0,2)} = \sw_+ ~.
\end{equation}
We similarly find the following for $\tilde{K}_-$:
\begin{equation}\label{Kminusres}
\nabla_{\bfm} \tilde{K}_{-}^{\bfm} = 0~, \qquad \ed \tilde{K}_{-}^{(2,0)} = \sw_- ~, \qquad \ed \tilde{K}_{-}^{(0,2)} = 0 ~.
\end{equation}
Note that $\ed \tilde{K}_{+}^{(2,0)} = 0$ (and the Killing spinor equation) imply $\nabla_{\bfm} \tilde{K}^{\bfnbar} = 0$, and hence $\Lie_{\tilde{K}_+}^{(0,\ast)} = \Lie_{\tilde{K}_+}$ when acting on the multi-form $\lambda \in \Gamma(\fmMM,\Lambda^{(0,\ast)})$.  However for $\tilde{K}_-$ this is not the case: $\Lie_{\tilde{K}_-}^{(0,\ast)} \neq \Lie_{\tilde{K}_-}$.

Next consider the equations for $\tilde{K}^3$.  Since $\Lie_{\tilde{K}^3} \bbJ^{3} = 0$, $\tilde{K}^3$ is a holomorphic Killing vector this immediately implies that the curvature two-form $\ed \tilde{K}^3$ is type $(1,1)$.  It also implies that the projected Lie derivative is the ordinary one when acting on $\lambda$.    Now from \eqref{LKpprefinal} with $K = \tilde{K}^3$ we have on the one hand
\begin{equation}
(\Lie_{\tilde{K}^3} \sw_+)_{\bfmbar\bfnbar} g^{\bfmbar\bfm} g^{\bfnbar\bfn} (\sw_-)_{\bfm\bfn} = 2 \nabla_{\bfnbar} (\tilde{K}^{3})^{\bfnbar}~.
\end{equation}
On the other hand, $\Lie_{\tilde{K}^3} \sw_+ = -i \sw_+$, and therefore the same quantity satisfies
\begin{align}
(\Lie_{\tilde{K}^3} \sw_+)_{\bfmbar\bfnbar} g^{\bfmbar\bfm} g^{\bfnbar\bfn} (\sw_-)_{\bfm\bfn} =&~ -i \left((\sw_+)_{\bfmbar\bfnbar} g^{\bfnbar\bfn}\right) \left( g^{\bfmbar\bfm} (\sw_-)_{\bfm\bfn} \right) \cr
=&~ \frac{i}{4} (\bbJ_{+}')_{\bfmbar}^{\phantom{\bfm}\bfn} (\bbJ_{-}')_{\bfn}^{\phantom{\bfm}\bfmbar} \cr
=&~ -i \delta_{\bfmbar}^{\phantom{\bfm}\bfmbar} = -i D~.
\end{align}
Comparing these two gives $\nabla_{\bfnbar} (\tilde{K}^{3})^{\bfnbar} = - \frac{i D}{2}$.  Thus, altogether,
\begin{equation}\label{K3res}
\nabla_{\bfm} (\tilde{K}^3)^{\bfm} =  \frac{i D}{2}~, \qquad (\ed \tilde{K}^{3})^{(2,0)} = 0 ~, \qquad (\ed \tilde{K}^{3})^{(0,2)} = 0 ~.
\end{equation}

The Noether charges associated these Killing vectors $K^r$ were denoted $\hat{N}^r \equiv \hat{\II}^r$ and correspond to the diagonal $\mathfrak{su}(2)_{\mathrm{d}}^{(\kappa)} \subset \mathfrak{su}(2)_R \oplus \mathfrak{so}(3)$ discussed around \eqref{twisteddiag}.  As they are linear in $K$, the rotated charges
\begin{equation}
\hat{\widetilde{\II}}{}^r := (R_\kappa)^{r}_{\phantom{r}s} \hat{\II}^s~,
\end{equation}
will correspond to $\hat{N}_K$ with $K = \tilde{K}^r$.  Making use of \eqref{Kplusres}, \eqref{Kminusres}, and \eqref{K3res} we then have
\begin{align}
\hat{\widetilde{\II}}_+ \lambda =&~ i \left\{ \Lie_{\tilde{K}_+} + \sw_+ \wedge \right\} \lambda~, \cr
\hat{\widetilde{\II}}_- \lambda =&~  i \left\{ \Lie_{\tilde{K}_-}^{(0,\ast)} - \contract_{\sw_-} \right\} \lambda~, \cr
\hat{\widetilde{\II}}{}^3 \lambda =&~ \left\{ i \Lie_{\tilde{K}^3} - \frac{D}{4} \right\} \lambda~.
\end{align}

Meanwhile the generators of angular momentum,
\begin{equation}
\hat{J}^r = \hat{\II}^r - (R_{\kappa}^{-1})^{r}_{\phantom{r}s} \hat{I}^s = (R_{\kappa}^{-1})^{r}_{\phantom{r}s} (\hat{\widetilde{\II}}{}^s - \hat{I}^s )~,
\end{equation}
are given by
\begin{align}
& (R_\kappa \hat{J})_+ \lambda = i \Lie_{K_+} \lambda~, \quad (R_\kappa \hat{J})_- \lambda = i \Lie_{K_-}^{(0,\ast)} \lambda~, \quad (R_\kappa \hat{J})^3 \lambda^{(q)} =  \left\{ i \Lie_{K^3} - \frac{q}{2} \right\} \lambda^{(q)}~.
\end{align}
Here as usual we have defined $(R_\kappa \hat{J})_{\pm} := (R_\kappa \hat{J})^1 \pm i (R_\kappa \hat{J})^2$, and furthermore have chosen to express the results in terms of the rotated angular momentum generators $(R_\kappa \hat{J})^r = (R_\kappa)^{r}_{\phantom{r}s} \hat{J}^s$.  Note that these are the generators that naturally appear in, \eg, the algebra with the supercharges, \eqref{JQcom}.  

One can check that this triplet satisfies the $\mathfrak{so}(3)$ algebra and commutes with the $R$-symmetry generators $\hat{I}^r$.  The role of the projected Lie derivative is crucial in verifying these relations.  For example, if $(R_\kappa \hat{J})_-$ was constructed with the ordinary Lie derivative then, due to the property $[\Lie_V, \Lie_W] = \Lie_{[V,W]}$, there would be no way to generate the shift by $-q/2$ that appears in $(R_\kappa \hat{J})^3$.  Indeed, one can verify the following modified commutator for the projected Lie derivative acting on the multi-form $\lambda \in \Gamma(\fmMM,\Lambda^{(0,\ast)})$:
\begin{align}
& \left( [\Lie_{V}^{(0,\ast)}, \Lie_{W}^{(0,\ast)}] \lambda \right)_{\bfnbar_1 \cdots \bfnbar_q} = \left( \Lie_{[V,W]}^{(0,\ast)} \lambda \right)_{\bfnbar_1 \cdots \bfnbar_q} + \cr
& \qquad \qquad \qquad - \sum_{i=1}^q \left( (\nabla_{\bfnbar_i} V^{\bfm}) \nabla_{\bfm} W^{\bfnbar} - (\nabla_{\bfnbar_i} W^{\bfm}) \nabla_{\bfm} V^{\bfnbar} \right) \lambda_{\bfnbar_1 \cdots \bfnbar_{i-1} \bfnbar \, \bfnbar_{i+1} \cdots \bfnbar_q} ~. \qquad \qquad
\end{align}
Then taking $V = \tilde{K}_+$ and $W = \tilde{K}_-$ and using \eqref{Kplusres}, \eqref{Kminusres}, we have
\begin{align}
& \left( [(R_\kappa \hat{J})_+, (R_\kappa \hat{J})_-]\lambda \right)_{\bfnbar_1 \cdots \bfnbar_q} =  -\left( [\Lie_{\tilde{K}_+}^{(0,\ast)}, \Lie_{\tilde{K}_-}^{(0,\ast)}] \lambda \right)_{\bfnbar_1 \cdots \bfnbar_q} \cr
& \qquad \qquad =  2i \left( \Lie_{\tilde{K}^3}^{(0,\ast)} \lambda \right)_{\bfnbar_1 \cdots \bfnbar_q} + \sum_{i=1}^q (g^{\bfm\bfmbar} (\sw_+)_{\bfnbar_{i}\bfmbar}) (g^{\bfnbar\bfn}(\sw_-)_{\bfm\bfn})  \lambda_{\bfnbar_1 \cdots \bfnbar_{i-1} \bfnbar \, \bfnbar_{i+1} \cdots \bfnbar_q} \cr
& \qquad \qquad = 2i \left( \Lie_{\tilde{K}^3} \lambda \right)_{\bfnbar_1 \cdots \bfnbar_q} - \sum_{i=1}^q \delta_{\bfnbar_i}^{\phantom{\bfn_i}\bfnbar}  \lambda_{\bfnbar_1 \cdots \bfnbar_{i-1} \bfnbar \, \bfnbar_{i+1} \cdots \bfnbar_q} \cr
& \qquad \qquad =  \left[ 2\left( i\Lie_{\tilde{K}^3} - \frac{q}{2} \right)\lambda \right]_{\bfnbar_1 \cdots \bfnbar_q} = \left( 2 (R_\kappa \hat{J})^3 \lambda \right)_{\bfnbar_1 \cdots \bfnbar_q}~,
\end{align}
so that $[\hat{J}_+,\hat{J}_-] \lambda = 2\hat{J}^3 \lambda$ as required.

The relationship between the Killing vectors $\tilde{K}^r$ and the K\"ahler forms, summarized by \eqref{Kplusres}, \eqref{Kminusres}, and \eqref{K3res}, can be more elegantly stated as follows:
\begin{equation}
\ed ({}^\ast \tilde{K}^r) = \sw^r + F^r~,
\end{equation}
where ${}^\ast \tilde{K}^r$ denotes the dual one-form and each $F^r$, $r = 1,2,3$, is a two-form of type $(1,1)$ \emph{in all complex structures}.  For a given $r$, $F^r$ can be viewed as the curvature two-form of a hyperholomorphic line bundle over $\fmMM$ that can be constructed naturally from the data of the circle action generated by the corresponding $\tilde{K}^r$.  The $SO(3)$ isometries of monopole moduli spaces were  discussed in precisely this context in section 3.6 of \cite{MR3116317}, where an explicit formula was additionally derived for $F^r$ in terms of the quadrupole moment of the Higgs field.

%%%%%%%%%%%%%%%
\subsection{Supercharges}
%%%%%%%%%%%%%%%

The semiclassical collective coordinate supercharges were defined in \eqref{Qsc} and are given by
\begin{equation}
\hat{Q}_{\rm (sc)}^a = \half [ \hat{\chi}^n (\tbbJ^a)_{n}^{\phantom{n}m}, \hat{\pi}_n]_+ - \hat{\chi}^n (\tbbJ^a)_{n}^{\phantom{n}m} {\rm G}(\sy)_m~,
\end{equation}
where the symmetrization in the first term gives the appropriate operator ordering prescription.  We found in section \ref{sec:quantize} that
\begin{equation}
\half [ \hat{\chi}^n (\tbbJ^a)_{n}^{\phantom{n}m}, \hat{\pi}_n]_+ = \hat{\chi}^n (\tbbJ^a)_{n}^{\phantom{n}m} \hat{\pi}_m + \frac{i}{2} \Gamma^{p}_{\phantom{p}pn} (\tbbJ^a)_{m}^{\phantom{m}n} \hat{\chi}^n~,
\end{equation}
and just as there, the second term in this expression serves to cancel out the connection term within $\hat{\pi}_m$.  (See last line of \eqref{formcodiv} above.)

In the context here it is natural to consider the complex combinations of (rescaled) supercharges, 
\begin{equation}
\hat{\QQ} := \left(\frac{i\sqrt{2\pi}}{g_0}\right)\frac{1}{\sqrt{2}} (\hat{Q}^3 + i \hat{Q}^4)~, \qquad \hat{\SS} := \left(\frac{i\sqrt{2\pi}}{g_0}\right)\frac{1}{\sqrt{2}} (\hat{Q}^1 + i \hat{Q}^2)~.
\end{equation}
Then using, \eg, 
\begin{equation}
\frac{1}{\sqrt{2}} \hat{\chi}^n (\tbbJ^3)_{n}^{\phantom{n}m} {\rm G}(\sy)_m = -\frac{1}{\sqrt{2}} \hat{\chi}^n (\bbJ^3)_{n}^{\phantom{n}m} {\rm G}(\sy)_m = - i (\hat{\XX}{}^{\bfm} {\rm G}(\sy)_{\bfm} - \hat{\XXbar}{}^{\bfnbar} {\rm G}(\sy)_{\bfnbar})~,
\end{equation}
we find
\begin{equation}
\hat{\QQ} \lambda = \left( \pdbar -i {\rm G}(\sy)^{(0,1)} \wedge \right) \lambda~,
\end{equation}
where $\pdbar = \ed \Zbar^{\bfnbar} \pdbar_{\bfnbar}$ is the Dolbeault operator and ${\rm G}(\sy)^{(0,1)} = {\rm G}(\sy)_{\bfnbar} \ed \Zbar^{\bfnbar}$.  We can promote the one-form that is the metric dual of the vector field ${\rm G}(\sy)$ to a connection on a trivial line bundle over $\fmMM$, and we refer to $\hat{\QQ}$ as the ${\rm G}(\sy)$-twisted Dolbeault operator.  Triholomorphicity of ${\rm G}(\sy)$ implies that the curvature of this connection is type $(1,1)$ in all complex structures.  In particular, $\pdbar ({\rm G}(\sy)^{(0,1)}) = 0$ and this shows that $\hat{\QQ}^2 = 0$.

An explicit expression for the remaining complex supercharge is most easily obtained by making use of the algebra with the $R$-symmetry generators, \eqref{QIalg}.  These relations imply
\begin{equation}
[ \hat{I}_-, \hat{\QQ}] = \SS^\dag~, \qquad [\hat{I}_+, \hat{\QQ}^\dag] = -\SS~,
\end{equation}
so \eg\,
\begin{equation}
\SS^\dag \lambda = -i \left[ \contract_{\sw_-}, \left( \pdbar - i {\rm G}(\sy)^{(0,1)} \wedge \right) \right] \lambda~.
\end{equation}
%

%%%%%%%%%%%%%%%%%%%%%
%%%%%%%%%%%%%%%%%%%%%
\section{A couple of weak coupling lemmata}\label{app:lemma}
%%%%%%%%%%%%%%%%%%%%%
%%%%%%%%%%%%%%%%%%%%%

In this appendix we prove the two lemmata given in section \ref{wclemma}.  In each case we start by studying a certain one-parameter family of physics data, which will eventually be related to the math data of the lemma but is, for the moment, independent of it.  

%%%%%%%%%%%%%%%%%%%%%
\subsection{Vanilla case}
%%%%%%%%%%%%%%%%%%%%%

Let some initial $u_0 \in \widehat{\BB}$ and $\Lambda_0 \in \mathbb{C}^\ast$ be given.  In the vanilla case we fix a weak coupling duality frame by requiring that $\Im(a(u_0))$ is in the fundamental Weyl chamber, $W_{\mathfrak{t}}^+$.  (We assume $u_0$ is generic such that the imaginary part of the corresponding $\varphi_\infty$ is a regular element of $\mathfrak{g}$.)  We denote the corresponding charge trivialization maps $\tilde{{\rm m}}: \Gamma_{u_0} \to \Lambda_{\rm cr}$ and $\tilde{{\rm e}} : \Gamma_{u_0} \to \Lambda_{\rm wt}$.  Then let $u_{t},\Lambda_{t}$, $0 \leq t < \infty$ be one-parameter families such that
\begin{equation}\label{wcfamily}
a(u_t) =  \frac{\Re(a(u_0))}{1+t} + i \Im(a(u_0))~, \qquad \Lambda_t = e^{-\pi t/h^\vee} \Lambda_0~.
\end{equation}
The main point is that $\Lambda_t$ should go to zero exponentially fast as $t \to \infty$; the precise definition given here will be convenient for the comparison with math data later.  Note that the imaginary part of $a(u_t)$ does not vary, so that duality frames defined by the condition that $\Im(a(u_t)) \in W_{\mathfrak{t}}^+$ are the same.

The definition of the weak coupling regime for a given dynamical scale $\Lambda$, $\widehat{\BB}_{\rm wc}(\Lambda) \in \widehat{\BB}$, is that it consists of those $u \in \widehat{\BB}$ such that the weak coupling expansion of the prepotential, (which takes the same form in any weak coupling duality frame), is convergent.  Therefore, on the one hand, there exists an $R_{\rm wc} \in \mathbb{R}_+$ such that
\begin{equation}
\min_{\alpha \in \Delta^+} \left\{ \frac{ | \langle \alpha, a(u) \rangle |}{|\Lambda|} \right\} > R_{\rm wc} \quad \Rightarrow \quad u \in \widehat{\BB}_{\rm wc}~.
\end{equation}
On the other hand, considering the $t\to \infty$ limit of our family, we have
\begin{equation}
\min_{\alpha \in \Delta^+} \left\{ \frac{ | \langle \alpha, a(u_{t}) \rangle |}{|\Lambda_{t}|} \right\} =  e^{\pi t/h^\vee} \left[ \min_{\alpha \in \Delta^+} \left\{ \frac{\langle \alpha, \Im(a(u_0)) \rangle }{|\Lambda_0|} \right\} + O(1/t) \right]~.
\end{equation}
Hence there exists some $t_{\rm wc} < \infty$ such that $u_t \in \widehat{\BB}_{\rm wc}(\Lambda_t)$, $\forall t > t_{\rm wc}$.

Let us then restrict to those $t > t_{\rm wc}$ such that $u_{t} \in \widehat{\BB}_{\rm wc}$ and consider the formula for the dual coordinate:
\begin{equation}
a_{\mathrm{D}}(u_{t}) = \frac{i}{2\pi} \sum_{\alpha \in \Delta_+} \alpha^\ast \langle \alpha, a(u_{t})\rangle \left[ \ln \left( \frac{ \langle \alpha, a(u_{t})\rangle^2}{2\Lambda_{t}^2} \right) + 1 \right] + a_{\mathrm{D}}^{\rm np}(u_{t})~,
\end{equation}
where $a_{\mathrm{D}}^{\rm np}$ is the convergent instanton sum.  Plugging in \eqref{wcfamily} we have that $a_{\rm D}^{\rm np}$ falls off exponentially fast as $t \to \infty$ and so
\begin{align}
a_{\mathrm{D}}(u_{t}) =&~ \frac{i}{2\pi} \sum_{\alpha \in \Delta^+} \alpha^\ast \left\langle \alpha, i \Im(a(u_0)) + \Re(a(u_0))/t + O(1/t^2) \right\rangle \times \cr
& \times \left[ \frac{2\pi t}{h^\vee} + \ln\left( \frac{\langle \alpha, \Im(a(u_0))\rangle^2}{2 |\Lambda_0|^2}\right) + 1 + i \left(\pi - \frac{\theta_0}{h^\vee}\right) + O(1/t)\right] + O(e^{-t})~. \qquad
\end{align}
Hence we have
\begin{align}\label{wcdualfamily}
\Im(a_{\mathrm{D}}(u_{t})) =&~\sum_{\alpha \in \Delta^+} \alpha^\ast \left[ \frac{1}{h^\vee} \langle \alpha, \Re(a(u_0))\rangle + \left( \frac{\theta_0}{2\pi h^\vee} - \frac{1}{2} \right) \langle \alpha, \Im(a(u_0)) \rangle \right] + O(1/t)~, \cr
\Re(a_{\mathrm{D}}(u_{t})) =&~ - \frac{t}{h^\vee} \sum_{\alpha \in \Delta^+} \alpha^\ast \langle \alpha, \Im(a(u_0)) \rangle \left\{ 1 + O(1/t)) \right\}~. \raisetag{24pt}
\end{align}
Notice the real part diverges like $t$ while the imaginary part is finite as $t \to \infty$.

Now introduce the math data $\{ \gm,X_\infty, \YY_0 \} \in \Lambda_{\rm cr} \times W_{\mathfrak{t}}^+ \times \mathfrak{t}_{\gm}^{\perp}$ of the vanilla lemma.  This is to be related to some set of physics data, $(u,\gamma)$, via the math-physics map described around \eqref{mathxymapvan}.  Can the physics data be taken as a member of the one-parameter family we have been discussing?  The necessary and sufficient conditions are that
\begin{enumerate}
\item the choice of duality frame for the family made above coincides with the preferred frame in the math-physics map, and that 
\item it is possible to adjust the initial $u_0$ such that the identification \eqref{mathxymapvan} can be achieved for the given $X_\infty,\YY_\infty$ and that member of the family.
\end{enumerate}

To address the first item, recall that the preferred duality frame in the math-physics map is the one such that the $X_\infty$ constructed from the physics data via $X_\infty(u,\gamma) \equiv \Im(\zeta_{\rm van}^{-1}(u,\gamma) a(u))$ is in the fundamental Weyl chamber.  Since this condition can only be true in a single weak coupling duality frame, we simply need to check that it holds if we assume the frames coincide.  Therefore let $\gamma \in \Gamma_{u_{t}}$ be such that $\tilde{{\rm m}}(\gamma) = {\rm m}(\gamma) = \gm$, while $\tilde{{\rm e}}(\gamma) = {\rm e}(\gamma) \equiv \gamma_{\rm e}$ is some element of the root lattice.  Then we can compute $\zeta_{\rm van}(\gamma,u_{t})$ as follows.  First, from \eqref{wcfamily} and \eqref{wcdualfamily} we have
\begin{align}\label{wcZvan}
Z_\gamma(u_{t}) =&~ (\gm,a_{\mathrm{D}}(u_{t})) + \langle \gamma_{\rm e}, a(u_{t}) \rangle \cr
=&~ -t \left( \gm, \Im(a(u_0)) \right) + i \left( \gm, \Re(a(u_0)) \right) + i \langle \gamma_{\rm e}, \Im(a(u_0)) \rangle + \cr
& + i \left( \gm, \Im(a(u_0)) \right) \left( \frac{\theta_0}{2\pi} - \frac{h^\vee}{2} \right) + \cr
& - \frac{1}{2\pi} \sum_{\alpha \in \Delta^+} \langle \alpha, \gm \rangle \langle \alpha, \Im(a(u_0))\rangle \left[ \ln\left( \frac{\langle \alpha, \Im(a(u_0))\rangle^2}{2|\Lambda_0|^2}\right) + 1 \right] + O(1/t)~, \qquad
\end{align}
where $(~,~)$ is the Killing form of the physical theory defined in \eqref{Trdef} and we used \eqref{adtr}.  But since ${\rm m}(\gamma) = \gm$ and $\MM(\gm;X_\infty)$ is nonempty by the assumption in the lemma, we know from semiclassical analysis that $\gm$ is necessarily a nonnegative integral combination of simple co-roots.  Hence, as $\tilde{{\rm m}}(\gamma) = \gm$ as well, $(\gm,\Im(a(u_0))) > 0$ and therefore
\begin{align}\label{wczetavan}
\zeta_{\rm van}(u_{t},\gamma) =&~ - \frac{Z_\gamma(u_{t})}{|Z_\gamma(u_{t})|} = 1 + \frac{i \alpha}{t} + O(1/t^2)~, \qquad \textrm{where} \cr
\alpha =&~ \frac{h^\vee}{2} - \frac{\theta_0}{2\pi} - \frac{ \left( \gm, \Re(a(u_0)) \right) + \langle \gamma_{\rm e}, \Im(a(u_0)) \rangle }{ \left( \gm, \Im(a(u_0)) \right) } ~.
\end{align}
Hence there exists a $t_{\rm wc} \leq t_\ast < \infty$ such that
\begin{align}
\Im\left( \zeta_{\rm van}^{-1}(u_{t},\gamma)a(u_{t})\right) =&~ \Im(a(u_0)) + O(1/t) \in W_{\mathfrak{t}}^+~, \quad \forall t > t_\ast~.
\end{align}
This establishes the first condition.

Now let some $t' > t_\ast$ be given.  The second condition is concerned with the existence of a $u_0 \in \widehat{\BB}$ such that
\begin{equation}\label{XYwcidentify}
X_\infty = \Im\left( \zeta_{\rm van}^{-1}(u_{t'},\gamma) a(u_{t'}) \right)~, \qquad \YY_\infty = \Im \left( \zeta_{\rm van}^{-1}(u_{t'},\gamma) a_{\mathrm{D}}(u_{t'}) \right)~,
\end{equation}
can be satisfied.  On the left-hand side of the second equation, $\YY_\infty$ is defined to be,
\begin{equation}\label{YtildeY}
\YY_\infty := - \frac{\langle \gamma_{\rm e}, X_\infty \rangle}{(\gm, X_\infty)} X_\infty + \YY_0~.
\end{equation}
in terms of the given math data and the electric charge $\gamma_{\rm e} = {\rm e}(\gamma)$.  This ensures that the constraint $(\gm, \YY_\infty) + \langle \gamma_{\rm e}, X_\infty \rangle = 0$, that follows from \eqref{XYwcidentify}, is automatic.  Using the expressions \eqref{wcfamily} and \eqref{wcdualfamily} we see that \eqref{XYwcidentify} can be solved perturbatively in $1/t'$ to obtain $a(u_{0})$ (and hence $u_{0}$ in principle) in terms of the given $X_\infty$, $\YY_0$.  We denote this solution $u_0 = u_{0}^{(t')}(X_\infty,\YY_0)$.  For example it follows from \eqref{wczetavan} and \eqref{wcfamily} that $\Im(a(u_{0}^{(t')})) = X_\infty$ at leading order and that the first correction actually vanishes:
\begin{equation}\label{Xwcsoln}
\Im(a(u_{0}^{(t')})) = X_\infty  + O(1/{t'}^2)~.
\end{equation}

Meanwhile \eqref{wczetavan} and \eqref{wcdualfamily} can be used to obtain an expression for the real part.  There are some cancelations and using \eqref{Xwcsoln} we we find
\begin{align}
\YY_\infty =&~ \frac{1}{h^\vee} \sum_{\alpha\in\Delta^+} \alpha^\ast \left\langle \alpha, \Re(a(u_{0}^{(t')}))\right\rangle + \cr
&~ - \frac{1}{h^\vee} \left[ \frac{ \left( \gm, \Re(a(u_{0}^{(t')})) \right)}{(\gm,X_\infty)} + \frac{\langle \gamma_{\rm e}, X_\infty\rangle}{(\gm,X_\infty)} \right] \sum_{\alpha \in \Delta^+} \alpha^\ast \langle \alpha, X_\infty \rangle + O(1/t')~. \qquad
\end{align} 
Pairing both sides with a generic simple root $\alpha_I$ and using \eqref{YtildeY}, we have
\begin{equation}
\langle \alpha_I, \YY_0 \rangle = \bigg\langle \alpha_I, \Re(a(u_{0}^{(t')})) - \frac{(\gm, \Re(a(u_{0}^{(t')})))}{(\gm,X_\infty)} X_\infty \bigg\rangle + O(1/t')~.
\end{equation}
In other words, the components of $\Re(a(u_{0}^{(t')}))$ and $\YY_\infty$ orthogonal to the magnetic charge with respect to the Killing form must agree at leading order.  The component of $\YY_\infty$ orthogonal to the magnetic charge is precisely the math data $\YY_0$.  The leading order component of $\Re(a(u_{0}^{(t')}))$ along $X_\infty$ is not fixed by \eqref{XYwcidentify}.  Let us denote this component
\begin{equation}
\vartheta := \lim_{t \to \infty} \frac{(\gm, \Re(a(u_{0}^{(t)})) )}{(\gm, \Im(a(u_{0}^{(t)})) )}~,
\end{equation}
so that our solution for $\Re(a(u_{0}^{(t')})$ can be written as
\begin{equation}\label{Ywcsoln}
\Re(a(u_{0}^{(t')})) = \YY_0 + \vartheta X_\infty + O(1/t')~.
\end{equation}
In the large $t$ limit $\vartheta$ is precisely the overall phase of $a$ as discussed under \eqref{vanconstraint}, parameterizing the one-dimensional preimage of $\sx,\sy$ under the math-physics map of parameters.

Then for a given $\gamma \in \Gamma_u$, the math-physics map is established for $X_\infty = X_\infty(u,\gamma)$ and $\YY_0 = \YY_0(u,\gamma)$ by taking $u = u_{t}^{(t)}(X_\infty,\YY_0)$ and $\Lambda = \Lambda_{t}$ for any $t > t_\ast$.  As $t\to \infty$ this solution takes the form
\begin{align}
a(u_{t}^{(t)}(\sx,\sy)) =&~ i \left(1 - \frac{i \vartheta}{t}\right)X_\infty + \frac{1}{t} \YY_0 + O(1/t^2)~, \cr
\Lambda_t =&~ e^{-\pi t/h^\vee} \Lambda_0~.
\end{align}
Furthermore, for such $t$'s we ensure that $u \in \widehat{\BB}_{\rm wc}(\Lambda_t)$.  Hence,  by setting $\mu = |\Lambda_{t_\ast}|$, the first part of the lemma is proven.\footnote{There exists a $t_{\ast}' < \infty$ such that the series solutions \eqref{Xwcsoln} and \eqref{Ywcsoln} converge for $t' > t_{\ast}'$.  It might then be necessary to redefine $t_{\ast}^{\rm old} \to t_{\ast}^{\rm new} = \max\{ t_{\ast}^{\rm old}, t_{\ast}' \}$.}

For the second part of the lemma on p\pageref{wclemma} we use \eqref{wcZvan}.  Set $u = u_{t}^{(t)}(\sx,\YY_0)$ for some $t > t_\ast$, and let $\gamma_{1,2}$ be given such that ${\rm m}(\gamma_1 + \gamma_2) \equiv \gamma_{1,{\rm m}} + \gamma_{2,{\rm m}} = \gm$, $\langle \gamma_1, \gamma_2 \rangle \neq 0$, and $\Omega(u,\gamma_{1,2};y) \neq 0$.  (The last condition simply means that the there exists vanilla BPS states carrying the given charges: $\HH_{\gamma_{1,2},u}^{\rm BPS} \neq \{0 \}$.)  Now since $\gamma_{1,{\rm m}}$ and $\gamma_{2,{\rm m}}$ have to sum $\gm$ such that $\MM(\gm;X_\infty)$ nontrivial, and they have to be magnetic charges of actual vanilla BPS states in the weak coupling regime, there are two possibilities.  Either 
\begin{enumerate}
\item both $\gamma_{1,{\rm m}},\gamma_{2,{\rm m}}$ are nonzero and nonnegative integral combinations of simple co-roots, or 
\item only one of them is---without loss of generality say $\gamma_{1,{\rm m}}$---while $\gamma_2$ corresponds to a $W$-boson: $\gamma_{2,{\rm m}} \oplus \gamma_{2,{\rm e}} = \alpha$ for some $\alpha \in \Delta^+$.
\end{enumerate}
However in the second case it is not possible to have simultaneously that $\Im(Z_{\gamma_1} \overline{Z_{\gamma_2}}) = 0$, for this would imply $(\gamma_{2,{\rm m}}, \YY_\infty) + \langle \gamma_{2,{\rm e}},X_\infty\rangle = 0$, hence $\langle \alpha, X_\infty \rangle = 0$, in contradiction to our assumption that $X_\infty \in W_{\mathfrak{t}}^+$.  Therefore we need only consider the first case.  But in that case we have
\begin{align}
Z_{\gamma_{1,2}}(u) =&~  -t \left(\gamma_{1,2,{\rm m}}, \Im(a(u_{0}^{(t)}(\sx,\YY_0)))\right) \times \left\{ 1 + O(1/t) \right\} \cr
=&~ - t (\gamma_{1,2,{\rm m}}, X_\infty) \times \left\{ 1 + O(1/t) \right\}~.
\end{align}
Hence the real part of $Z_{\gamma_{1,2}}(u)$ is negative and diverging like $-t$ while the imaginary part is $O(1)$.  It follows that there exists an $t_{\ast\ast}$, with $t_{\rm wc} \leq t_{\ast\ast} < \infty$, such that $\Re(Z_{\gamma_1}(u) \overline{Z_{\gamma_2}(u)}) > 0$ for $t > t_{\ast\ast}$.  By taking $\mu = \min\{ |\Lambda_{t_\ast}|, |\Lambda_{t_{\ast\ast}}| \}$, both parts of the vanilla lemma are proven.

%%%%%%%%%%%%%%%%%%%%%
\subsection{Framed case}
%%%%%%%%%%%%%%%%%%%%%

We will be briefer here since the ideas are the same.  Let $u_0 \in \widehat{\BB}$ and $\Lambda_0 \in \mathbb{C}^\ast$ be given, and suppose supersymmetric 't Hooft defects of type $\zeta$ are present.  We allow $\zeta \in \widehat{\mathbb{C}^\ast}$, the universal cover of $\mathbb{C}^\ast$, though physical defects require $|\zeta| = 1$.  Fix a duality frame by requiring that $\Im(\zeta^{-1} a(u_0)) \in W_{\mathfrak{t}}^+$ and denote the corresponding charge trivialization maps $\tilde{{\rm m}} : \Gamma_{L_{\zeta},u_0} \to \Lambda_{\rm cr} + \sum_n P_n$ and $\tilde{{\rm e}} : \Gamma_{L_{\zeta},u_0} \to \Lambda_{\rm wt}$.  Consider the one parameter family $u_t, \Lambda_t$ defined by
\begin{equation}\label{wcfamilyf}
\zeta^{-1} a(u_t) =  \frac{\Re(\zeta^{-1} a(u_0))}{1+t} + i \Im(\zeta^{-1} a(u_0))~, \qquad \Lambda_t = e^{-\pi t/h^\vee} |\zeta| \Lambda_0~.
\end{equation}

As before there exists some $t_{\rm wc} < \infty$ such that $u_{t} \in \widehat{\BB}_{\rm wc}(\Lambda_t)$, $\forall t > t_{\rm wc}$.  For $t > t_{\rm wc}$ we thus have
\begin{equation}
\zeta^{-1} a_{\mathrm{D}}(u_{t}) = \frac{i}{2\pi} \sum_{\alpha \in \Delta_+} \alpha^\ast \langle \alpha, \zeta^{-1} a(u_{t})\rangle \left[ \ln \left( \frac{ \langle \alpha, \zeta^{-1} a(u_{t})\rangle^2}{2\zeta^{-2} \Lambda_{t}^2} \right) + 1 \right] + \zeta^{-1} a_{\mathrm{D}}^{\rm np}(u_{t})~,
\end{equation}
where $a_{\rm D}^{\rm np}(u_t)$ falls off exponentially fast as $t \to \infty$.  Plugging in \eqref{wcfamilyf} we find
\begin{align}\label{wcdualfamilyf}
\Im(\zeta^{-1} a_{\mathrm{D}}(u_{t})) =&~\sum_{\alpha \in \Delta^+} \alpha^\ast \bigg[ \frac{1}{h^\vee} \langle \alpha, \Re(\zeta^{-1} a(u_0))\rangle + \cr
&~ \qquad \qquad + \left( \frac{\theta_0}{2\pi h^\vee} - \frac{1}{2} - \frac{\arg(\zeta)}{\pi} \right) \langle \alpha, \Im(\zeta^{-1} a(u_0)) \rangle \bigg] + O(1/t)~, \quad \cr
\Re(\zeta^{-1} a_{\mathrm{D}}(u_{t})) =&~ - \frac{t}{h^\vee} \sum_{\alpha \in \Delta^+} \alpha^\ast \langle \alpha, \Im(\zeta^{-1} a(u_0)) \rangle \left\{ 1 + O(1/t)) \right\}~. \raisetag{24pt}
\end{align}
We note for future reference that the higher order terms in the $1/t$ expansion of the imaginary part of $\zeta^{-1} a_{\rm D}$ will not depend on $\arg(\zeta)$; only the leading term does.

Now let the math data $\{\sx,\sy\} \in W_{\mathfrak{t}}^+ \times \mathfrak{t}$ of the lemma be given.  We first want to argue that for $t'$ large enough there exists a solution $u_0 = u_{0}^{(t')}(\sx,\sy)$ to the math-physics map
\begin{equation}
\sx = \Im(\zeta^{-1} a(u_{t'}))~, \qquad \sy = \Im(\zeta^{-1} a_{\rm D}(u_{t'}))~,
\end{equation}
such that the trivializations $\tilde{{\rm m}},\tilde{{\rm e}}$ agree with ${\rm m}, {\rm e}$.  The latter correspond to the duality frame determined by the condition $X_\infty \in W_{\mathfrak{t}}^{+}$.  This is more straightforward than the vanilla case since the phase $\zeta$ is fixed and independent of $u_{t'}$.  Indeed, using \eqref{wcfamilyf} and \eqref{wcdualfamilyf} the solution is specified by
\begin{align}\label{XYwsolf}
\Im(\zeta^{-1} a(u_{0}^{(t')})) =&~ X_\infty~, \cr
 \Re(\zeta^{-1} a(u_{0}^{(t')})) =&~ \YY_\infty + \left(\frac{h^\vee}{2} + \frac{h^\vee \arg(\zeta)}{\pi} - \frac{\theta_0}{2\pi} \right) X_\infty + O(1/t')~.
\end{align}
In particular the solution for the imaginary part is exact in $t'$ while there will be some $t_{\ast}' < \infty$ such that the solution for the real part converges for $t > t_{\ast}'$.  The higher order terms in the second of \eqref{XYwsolf} will not have any explicit dependence on $\zeta$ once expressed in terms of $\sx,\sy$ and therefore $t_{\ast}'$ is independent of $\zeta$.  Thus for a given $t > t_\ast := \max\{t_{\rm wc}, t_{\ast}'\}$, we have that $u = u_{t}^{(t)}(\sx,\sy)$ defined by
\begin{equation}
\zeta^{-1} a(u_{t}^{(t)}(\sx,\sy)) = i \left[ 1 - \frac{i}{t} \left(\frac{h^\vee}{2} + \frac{h^\vee \arg(\zeta)}{\pi} - \frac{\theta_0}{2\pi} \right)\right] X_\infty + \frac{1}{t} \YY_\infty + O(1/t^2)~,
\end{equation}
gives a family of inverses to the math-physics map in $\widehat{\BB}_{\rm wc}(\Lambda_t) \times \widehat{\mathbb{C}^\ast}$, parameterized by $\zeta$.  Furthermore since the solution \eqref{XYwsolf} for $X_\infty$ is exact we trivially see that the duality frames agree: $\tilde{{\rm m}} = {\rm m}$, $\tilde{{\rm e}} = {\rm e}$.  Therefore by taking $\mu_\ast = |\Lambda_{t_\ast}|$ the first part of the framed lemma is verified.

For the second part, suppose that $\gamma_{\rm h}$ with ${\rm m}(\gamma_{\rm h}) \in \Lambda_{\rm cr}$ and ${\rm e}(\gamma_{\rm h}) \in \Lambda_{\rm rt}$ is a halo charge associated with a vanilla BPS state in $\HH_{u_t,\gamma_{\rm h}}^{\rm BPS}$, $t > t_\ast$.  Then for such $t$ observe from \eqref{wcdualfamilyf} that
\begin{align}
\Re(\zeta^{-1} Z_{\gamma}(u)) =&~ \left( {\rm m}(\gamma_{\rm h}), \Re(\zeta^{-1} a_{\rm D}(u)) \right) + \langle {\rm e}(\gamma_{\rm h}), \Re(\zeta^{-1} a(u)) \rangle \cr
=&~ - \frac{t}{h^\vee} \sum_{\alpha \in \Delta^+} \langle \alpha, {\rm m}(\gamma_{\rm h}) \rangle \langle \alpha, X_\infty \rangle \left\{ 1 + O(1/t) \right\} \cr 
=&~ - t \, \left( {\rm m}(\gamma_{\rm h}), X_\infty\right) \left\{ 1 + O(1/t) \right\}~. 
\end{align}
By the same reasoning as in the vanilla case we can assume that ${\rm m}(\gamma_{\rm h}) \neq 0$, for otherwise the condition to be at a wall, $\Im(\zeta^{-1} Z_{\gamma_{\rm h}}(u)) = 0$, for such a charge would imply that $X_\infty \notin W_{\mathfrak{t}}^+$.  But now we know that the only occupied vanilla magnetic charges in the weak coupling regime are \emph{non-negative} integer linear combinations of simple co-roots.  Hence we necessarily have $\left( {\rm m}(\gamma_{\rm h}), X_\infty\right) > 0$.  Hence there exists a $t_{\ast\ast} < \infty$ such that $\Re(\zeta^{-1} Z_{\gamma}(u)) < 0$ for all $t > t_{\ast\ast}$.  By taking $\mu = \min\{ |\Lambda_{t_\ast}|, |\Lambda_{t_{\ast\ast}}| \}$ both parts of the lemma are verified.

%%%%%%%%%%%%%%%%%%%%%
%%%%%%%%%%%%%%%%%%%%%
\section{Zero modes of the $\rG$-twisted Dirac operator on Taub--NUT}\label{appendix:TN}
%%%%%%%%%%%%%%%%%%%%%
%%%%%%%%%%%%%%%%%%%%%

In this appendix we review the computation of the zero modes of a Dirac operator on Euclidean Taub--NUT, twisted with respect to an anti-self-dual $U(1)$ connection. This problem has been considered before \cite{Pope:1978zx}, where it was solved using the Newman--Penrose formalism. Here we follow a slightly different route, inspired by the observation that the $U(1)$ connection on Taub--NUT reduces to a monopole bundle on $\mathbb{R}^3$, a problem for which the Dirac zero mode problem was solved in \cite{HarishChandra:1948zz}. Recently we revisited this problem in \cite{MRVdimP1}, and in this appendix we will use some of the notation and results of that work.  

The connection between the Taub--NUT Dirac operator and the $\mathbb{R}^3$ monopole Dirac operator has also been explored in depth recently in \cite{Jante:2013kha}.  These authors focused on the (anti-)holomorphic description of the zero modes and on their transformation properties under the action of the $SU(2)$ double cover of the $SO(3)$ isometry group, which we have identified here as the diagonal subgroup of $SU(2)$ $R$-symmetry and angular momentum.  In this appendix we focus on the representation of the zero modes as spinors since this is the point of view we have taken throughout most of the paper.

%%%%%%%%%%%%%%%%%%%%%%%%
\subsection{Geometry of Taub--NUT}
%%%%%%%%%%%%%%%%%%%%%%%%

We use coordinates $x^a = (x^i, x^4)$, with $i = 1,2,3$, and $a = 1,2,3,4$ and write the metric for Taub-NUT as\footnote{Note that the parameter $m$ can in principle be absorbed into $\ell$ by using the scaling symmetry of the metric: $x^i\rightarrow \lambda x^i, \ell \rightarrow  \lambda \ell$ and $m\rightarrow \lambda^{-2} m$. We will keep $m$ explicit as this factor naturally appears in metric of the monopole moduli space, where it has the physical interpretation of a mass.}
\begin{align}\label{TN}
& \ed s^2 \equiv g_{ab} \ed x^a \ed x^b = m \left( H \ed x^i \ed x^i + H^{-1} (\ell \ed x^4 + \Omega)^2 \right)~, \qquad \textrm{where} \cr
& H = 1 + \frac{\ell}{r}~, \qquad \ed \Omega = \star_3 \ed H~, \qquad x^4 \sim x^4 + \frac{4\pi}{k}\,.
\end{align}
Here $\star_3$ is the Hodge dual on flat $\mathbb{R}^3$.  Standard Taub-NUT, which when equiped with the metric and coordinates above we will refer to as $\mathrm{TN}(m,\ell)$, has $k=1$ in the periodicity for $x_4$. We will also be interested in $\mathbb{Z}_k$ quotients of Taub--NUT. These correspond to the same metric but with the periodicity of $x^4$ generalized to any $k\in\mathbb{N}$ as above.  We will soon switch to standard spherical coordinates on the $\mathbb{R}^3$ base:
\begin{equation}\label{spherical}
x^1 = r \sin{\theta} \cos{\phi}~, \qquad x^2 = r \sin{\theta} \sin{\phi}~, \qquad x^3 = r \cos{\theta}~.
\end{equation}
Observe that $H$ is harmonic on $\mathbb{R}^3 \setminus \{0\}$ and $\Omega$ is only fixed up to addition of an exact piece.  We take
\begin{equation}\label{omega1}
\Omega = \ell (\cos{\theta} - \epsilon) \ed \phi~,
\end{equation}
where $\epsilon = \pm 1$.  The change of variables $r = \rho^2/4\ell$ reveals that the space is smooth as $r \to 0$ for $k=1$, but has a conical deficit when $k>1$.

Our conventions for orthonormal frames are that $\ua = \underline{1},\ldots,\underline{4}$, and $\underline{i} = \underline{1},\underline{2},\underline{3}$ are tangent space indices corresponding to the $a$ and $i$ coordinate indices.  Our coframe on the cotangent bundle is
\begin{equation}\label{cotframe}
e^{\underline{i}} = m^{1/2} H^{1/2} \ed x^i~, \qquad e^{\underline{4}} = m^{1/2} H^{-1/2} (\ell \ed x^4 + \Omega)~,
\end{equation}
and the dual frame on the tangent bundle is found to be
\begin{align}\label{tanframe}
& \EE_{\underline{1}} = m^{-1/2} H^{-1/2} \left( \pd_1 + \frac{\sin{\phi}}{r\sin{\theta}}(\cos{\theta}-\epsilon) \pd_4 \right) = m^{-1/2} H^{-1/2} \left( \pd_1 - \ell^{-1} \Omega_1 \pd_4 \right)~, \cr
& \EE_{\underline{2}} = m^{-1/2} H^{-1/2} \left( \pd_2 - \frac{\cos{\phi}}{r \sin{\theta}} (\cos{\theta}-\epsilon) \pd_4\right) = m^{-1/2} H^{-1/2} \left( \pd_2 - \ell^{-1} \Omega_2 \pd_4 \right)~, \cr
& \EE_{\underline{3}} = m^{-1/2} H^{-1/2} \pd_3 = m^{-1/2} H^{-1/2} \left( \pd_3 - \ell^{-1} \Omega_3 \pd_4 \right)~, \cr
&\EE_{\underline{4}} = m^{-1/2} H^{1/2} \ell^{-1} \pd_4~.
\end{align}
These satisfy the usual realtions: $e^{\ua} (\EE_{\ub}) = \delta^{\ua}_{\phantom{a}\ub}$ and $\ed s^2 = \delta_{\underline{ab}} e^{\ua} \otimes e^{\ub}$, where $\delta_{\underline{ab}}$ is the flat Euclidean metric on the tangent space.

One can compute the components of the spin connection referred to the coframe, $\omega_{\uc,\underline{ab}}$, with the formulae
\begin{align}\label{spincon}
& \ed e_{\ua} = \half \xi_{\ua,\underline{bc}} e^{\ub} \wedge e^{\uc}~, \qquad \xi_{\ua,\underline{bc}} = - \xi_{\ua,\underline{cb}}~, \cr
& \omega_{\uc,\underline{ab}} = \half \left( \xi_{\ua,\underline{bc}} + \xi_{\ub,\underline{ca}} - \xi_{\uc,\underline{ab}} \right)~.
\end{align}
We find
\begin{align}\label{TNspin}
& \omega_{\underline{i},\underline{jk}} = \frac{1}{2 \sqrt{m} H^{3/2}} \left( \delta_{ij} \pd_k H - \delta_{ik} \pd_j H \right)~, \qquad \omega_{\underline{i},\underline{j 4}} =  \frac{1}{2 \sqrt{m} H^{3/2}} \epsilon_{ijk} \pd_k H~, \cr
& \omega_{\underline{4},\underline{ij}} = - \frac{1}{2 \sqrt{m} H^{3/2}} \epsilon_{ijk} \pd_k H~, \qquad \omega_{\underline{4},\underline{4i}} = - \frac{1}{2 \sqrt{m} H^{3/2}} \pd_i H~,
\end{align}
where we have used $(\ed \Omega)_{ij} = -\epsilon_{ijk} \pd_k H$, with $\epsilon_{ijk}$ the Levi--Civita tensor on $\mathbb{R}^3$, $\epsilon_{123} = 1$.  Our index use is a little sloppy here.  On the left of these equations we have tangent space indices and on the right we have coordinate indices.  What we are really doing is employing a triad $(\tilde{e})^{\underline{i}}_{\phantom{\underline{i}}j}$, with $\tilde{e} = \mathbf{1}$, on the flat base space which converts coordinate space indices to tangent space indices.  This would lead to overly cluttered formulae, so we leave it implicit.

A particular Killing vector of interest for us will be
\begin{equation}\label{G}
\rG = \frac{C}{m\ell^{2}} \pd_4~,
\end{equation}
where $C$ is a constant.  Since the topology of Taub-NUT is trivial, the dual one-form,
\begin{equation}\label{Gform}
{}^\ast\rG = C H^{-1} (\ed x^4 + \ell^{-1} \Omega)~,
\end{equation}
can be promoted to a connection on a trivial $U(1)$ bundle, which we can then couple to a Dirac operator.  The connection $\rG$ is everywhere smooth and its curvature is anti-self-dual if we take the volume form to be $\sqrt{g} \ed^3x \ed x^4$.  

%%%%%%%%%%%%%%%%%%%
\subsection{Zero modes of the $\rG$-twisted Dirac operator}
%%%%%%%%%%%%%%%%%%%

Introduce Hermitian $\gamma$ matrices satisfying the Clifford algebra
\begin{equation}\label{gamma}
[ \gamma^{\ua}, \gamma^{\ub} ]_+ = 2 \delta^{\underline{ab}}~.
\end{equation}
We work in a Weyl basis where 
\begin{equation}\label{taus}
\gamma^a = \left( \begin{array}{c c} 0 & \bar{\tau}^a \\ \tau^a & 0 \end{array} \right)~, \qquad \tau^a = (\vec{\sigma}, -i \mathbf{1})~, \qquad \bar{\tau}^a = (\vec{\sigma}, i \mathbf{1})~,
\end{equation}
and define
\begin{equation}\label{gammabar}
\bar{\gamma} := \gamma^{\underline{1}} \gamma^{\underline{2}} \gamma^{\underline{3}} \gamma^{\underline{4}} = \left( \begin{array}{c c} \mathbbm{1} & 0 \\ 0 & -\mathbbm{1} \end{array}\right)~.
\end{equation}

We are interested in $\Lsq$-normalizable solutions to the Dirac equation,
\begin{equation}\label{Dirac}
\slashed{\DD}^{\rm G} \Psi \equiv \gamma^{\ua} \EE^{a}_{\phantom{a} \ua} \left( \pd_a + \frac{1}{4} \omega_{a, \underline{bc}} \gamma^{\underline{bc}} - i \rG_a \right) \Psi = 0~.
\end{equation}
Using the conventions and notation outlined above one computes that
\begin{align}\label{subresultsdirac}
& \gamma^{\ua} \omega_{\ua,\underline{bc}} \gamma^{\underline{bc}} = \frac{1}{\sqrt{m} H^{3/2}} \gamma^{\underline{k}} \pd_k H \left( \mathbbm{1} + \bar{\gamma} \right)~,\\
& \gamma^{\ua} \EE^{a}_{\phantom{a} \ua} \pd_\mu = \frac{1}{\sqrt{m H}} \gamma^{\underline{i}} \pd_i + \left[ \frac{ (\cos{\theta} - \epsilon)}{r \sin{\theta} \sqrt{m H}} \left( \sin{\phi} \gamma^{\underline{1}} - \cos{\phi} \gamma^{\underline{2}} \right) + \frac{\sqrt{H}}{\ell\sqrt{m}} \gamma^{\underline{4}} \right] \pd_4~,\\
& \gamma^{\ua} \EE^{a}_{\phantom{a}\ua} \rG_a = \gamma^{\ua} e_{\ua b} \rG^b = \frac{C}{\ell \sqrt{m H}} \gamma^{\underline{4}} ~.
\end{align} 

Hence \eqref{Dirac} is equivalent to
\begin{align}\label{Dirac2a}
& \displaystyle\biggl\{ \frac{1}{\sqrt{H}} \gamma^{\underline{i}} \pd_i + \frac{1}{4 H^{3/2}} \gamma^{\underline{k}} \pd_k H \left( \mathbbm{1} + \bar{\gamma} \right) - \frac{i C}{\ell \sqrt{H}} \gamma^{\underline{4}} +  \cr
&  \qquad \qquad + \left[ \frac{ (\cos{\theta} - \epsilon)}{r \sin{\theta} \sqrt{H}} \left( \sin{\phi} \gamma^{\underline{1}} - \cos{\phi} \gamma^{\underline{2}} \right) + \frac{\sqrt{H}}{\ell} \gamma^{\underline{4}} \right] \pd_4 \displaystyle\biggr\} \Psi = 0~.
\end{align}

We now introduce Weyl spinors $\psi_{\pm}$ such that $\Psi = (\psi_+, \psi_-)^T$.  The Dirac equation decomposes into two separate equations (signs correlated):
\begin{equation}\label{Weyl}
 \left\{ \vec{\sigma} \cdot \vec{\pd} \mp \frac{C}{\ell} +  \left[ \frac{ (\cos{\theta} - \epsilon)}{r \sin{\theta}} \left( \sin{\phi} \sigma^1 - \cos{\phi} \sigma^2 \right) \mp i \frac{H}{\ell} \right] \pd_4 \right\} \tilde{\psi}_{\pm} = 0~,
\end{equation}
where $\vec{\sigma}$ are the Pauli matrices, $\vec{\del}$ is the standard gradient operator on $\mathbb{R}^3$, and where we have introduced
\begin{equation}\label{psit}
\tilde{\psi}_+ := \sqrt{H} \psi_+~, \qquad \tilde{\psi}_- := \psi_- ~.
\end{equation}
The rescaling of $\psi_+$ accounts for the extra term in its equation relative to the $\psi_-$ equation, due to its coupling to the spin connection.

Up to the trivial appearance of $\pd_4$ this equation looks very similar to that of an electron coupled to a Dirac monopole. As in \cite{HarishChandra:1948zz} we make use of the following identities:
\begin{align}
& \vec{\sigma} \cdot \vec{\pd} =  U\left\{ \sigma^3\left(\pd_r+\frac{1}{r}\right)+\frac{\sigma^1}{r}\left[ \pd_\theta+\frac{i\sigma^3}{\sin\theta}\left( \pd_\phi-\frac{i\sigma^3}{2}\cos\theta \right)\right]\right\} U^{-1}\,,\\
&  \sin\phi\sigma^1-\cos\phi\sigma^2  =  -U\sigma^2U^{-1}\,,
\end{align}
where
\begin{equation}\label{Urot}
U=e^{-i\phi\sigma^3/2}e^{-i\theta\sigma^2/2}\,.
\end{equation}
With these we can now rewrite the differential operator appearing on the left-hand side of \eqref{Weyl} as
\begin{equation}
U\left\{ \sigma^3\left(\pd_r+\frac{1}{r}\right) \mp \frac{1}{\ell}(C+i H\pd_4)+\frac{\sigma^1}{r}\left[ \pd_\theta+\frac{i\sigma^3}{\sin\theta} \left(\pd_\phi + \epsilon\pd_4- (2\pd_4 + i\sigma^3)\frac{\cos\theta}{2} \right)\right]\right\} U^{-1}~.
\end{equation}
This suggests the separation of variables,
\begin{equation}\label{sepvar}
\tilde{\psi}_{\pm}=e^{i\nu x_4}e^{i(\epsilon\nu - m)\phi}U\hat\psi_{\pm}
(r,\theta)~,
\end{equation} 
for which the equation \eqref{Weyl} above simplifies to
\begin{equation}\label{reducedop}
\left[\sigma^3\left(\pd_r+\frac{1}{r}\right) \mp \frac{1}{\ell}(C-\nu H)+\frac{K}{r}\right]\hat\psi_{\pm}=0
\end{equation}
where the operator $K$ only concerns the $\theta$ dependence:
\begin{equation}
K=\sigma^1\left[\pd_\theta+\frac{\sigma^3}{\sin\theta}\left(m+\frac{1}{2}(2\nu+\sigma^3)\cos\theta\right)\right] ~.
\end{equation}

Now that the variables have been separated all that is left is to solve some known ODE's. We did this in detail in appendix A of \cite{MRVdimP1}, where the same equation appears. Indeed, under the identification of $2\nu\rightarrow -p_\mu$ and $\frac{C-\nu}{\ell}\rightarrow-x_\mu$, \eqref{reducedop} becomes identical to (C.11) in \cite{MRVdimP1}, where we discussed the general solution. Before we borrow the results from that discussion, let us point out that although the remaining equations are identical, there is a difference in the notion of normalizability in the two setups. In \cite{MRVdimP1} one looks for spinors that are square integrable on $\mathbb{R}^3$, while here we are interested in spinors on Taub-NUT where the scalar product is naturally defined as
\begin{equation}
\langle \Psi_1,\Psi_2\rangle := \frac{1}{m^2}\int \ed^3x \ed x^4 \sqrt{g}\, \overline{\Psi}_1\Psi_2 = \int \ed^3x\ed x^4 H\, \overline{\Psi}_1\Psi_2 ~.
\end{equation}
Using the redefinitions \eqref{psit} we find that
\begin{equation}
\langle \psi_+,\psi_+\rangle=\int \ed^3x \ed x^4 \, \overline{\tilde\psi}_+\tilde\psi_+ ~, \qquad \langle \psi_-,\psi_-\rangle=\int \ed^3x\ed x^4 \, H\overline{\tilde\psi}_-\tilde\psi_- ~.
\end{equation}
Note that by \eqref{sepvar} the integrands are actually $x^4$ independent. We hence see that for $\tilde\psi_+$ the normalization is identical to the standard normalization on $\mathbb{R}^3$ and the results from \cite{MRVdimP1} carry over immediatly. For $\tilde\psi_-$ the situation is slightly different.  Since $H\rightarrow 1$ at infinity and $H \sim r^{-1}$ as $r\rightarrow 0$, it is true however that if a solution $\tilde\psi_-$ is not normalizable on $\mathbb{R}^3$ it will also not be normalizable on Taub-NUT.  In \cite{MRVdimP1} we found that there are no normalizable $\tilde{\psi}_-$ solutions on $\mathbb{R}^3$, and hence by this logic there are no normalizable $\tilde\psi_-$ solutions on Taub-NUT.

Translating the discussion in \cite{MRVdimP1} to our setup here, the normalizable solutions for $\hat{\psi}_{+}(r,\theta)$ appearing in \eqref{sepvar} are
\begin{equation}
\hat{\psi}_{+}(r,\theta) = c \, r^{|\nu| -1} e^{- |C - \nu| r/\ell} \times \left\{ \begin{array}{l l} d^{j}_{m,j}(\theta) \left( \begin{array}{c} 1  \\ 0 \end{array} \right)~, & ~C < \nu < 0~, \\ \\  d^{j}_{m,-j}(\theta) \left( \begin{array}{c} 0 \\ 1\end{array} \right)~, & ~C > \nu > 0~, \end{array} \right.
\end{equation}
where $c$ is a normalization constant, $j = |\nu|-1/2$, and $d^{j}_{m,m'}(\theta)$ are the Wigner little $d$ matrices.\footnote{We follow the conventions of \cite{Sakurai} for Wigner $d$ functions and $SU(2)$ representation matrices.  The combination $e^{-i m \phi} d^{j}_{m,m'}(\theta)$ can also be expressed in terms of spin-weighted spherical harmonics, ${}_{m'}Y_{jm}$.}  If $|C| \leq |\nu|$ then there are no normalizable solutions.  We can now use \eqref{psit}, \eqref{Urot}, and \eqref{sepvar} to reconstruct the spinor $\psi_+$.  We fix the constant $c$ by demanding that $\langle \psi_+ , \psi_+ \rangle = 1$.  The result can be expressed in the form 
\begin{equation}\label{zmodesTN}
\psi_+ = \frac{\sqrt{k}}{\sqrt{4\pi (2|\nu|)!}} \left( \frac{2|C-\nu|}{\ell}\right)^{|\nu|+1/2} \frac{r^{|\nu|-1}}{\sqrt{H}} e^{-|C-\nu|r/\ell} \breve{\psi}_{+;\nu,m}(\theta,\phi)~,
\end{equation}
where
\begin{equation}
\breve{\psi}_{+;\nu,m}(\theta,\phi) = \sqrt{\frac{|\nu|}{2\pi}} e^{i(\nu \epsilon - m)\phi} \times \left\{ \begin{array}{l l} d^{|\nu|-1/2}_{m,|\nu|-1/2}(\theta) \left(\begin{array}{c}  e^{-i\phi/2}\cos\frac{\theta}{2}\\e^{i\phi/2}\sin\frac{\theta}{2} \end{array} \right)~, & ~C < \nu < 0~, \\ \\  d^{|\nu|-1/2}_{m,-(|\nu|-1/2)}(\theta) \left(  \begin{array}{c} -e^{-i\phi/2}\sin\frac{\theta}{2} \\ e^{i\phi/2}\cos\frac{\theta}{2} \end{array} \right)~, & ~0 < \nu < C~. \end{array} \right.
\end{equation}
We have normalized $\breve{\psi}_+$ so that $\int_{S^2} \sin{\theta} \ed\theta \ed\phi \, \breve{\psi}_{+}^\dag \breve{\psi}_+ = 1$.

Apart from the condition $|\nu| < |C|$ arising from normalizability, there are further regularity and periodicity constraints. As $x_4$ is a periodic coordinate with period $4\pi/k$, periodicity of $\psi_+$ requires the quantization $\nu\in \frac{k}{2}\mathbb{Z}$. Furthermore the Wigner small $d$-matrix $d^{j}_{m,m^\prime}(\theta)$ is only regular for integer or half-integer $j$, and with $m,m^\prime \in\{-j,-j+1,\ldots,j\}$. In our case $j=|\nu|-1/2$, so we see that the quantization of $j$ is consistent with that of $\nu$.

%%%%%%%%%%%%%%%%%%%%%
\subsection{Summary and physical interpretation of the zero mode spectrum}\label{physspec}
%%%%%%%%%%%%%%%%%%%%%

From the discussion above it follows that all zero modes of the Dirac equation are positive chirality Weyl spinors. Depending on the value of $C$ the precise number of zero modes changes.

From \eqref{zmodesTN} it follows that when $|C| \leq \frac{k}{2}$ then there are no zero modes.  In particular, there are no zero modes for the pure (\ie\ untwisted) Dirac operator on Taub--NUT.

When $|C| > \frac{k}{2}$, we have angular momentum multiplets of spin $j$ for each 
\begin{equation}
j \in  \left\{\frac{k}{2} -\half ,k - \half, \ldots, \frac{k}{2} \left\lfloor\frac{2|C|}{k}\right\rfloor - \half \right\}\,,
\end{equation}
where $\left\lfloor x \right\rfloor=\max \left\{m\in \mathbb{Z}\,|\, m \leq x  \right\}$ is the left-continuous floor function.  Each multiplet occurs with multiplicity one.  To count the total number of states we can introduce an integer $l=\frac{2|\nu|}{k}$, such that $0<l\leq\left\lfloor\frac{2|C|}{k}\right\rfloor$. Furthermore for each fixed value of the angular momentum $j=|\nu|-\half$ there are $2|\nu|=k l$ states so that
\begin{eqnarray}
\dim{\ker^{(l)}{\slashed{\mathcal{D}}_{+}^{\rm G}}}&=&\begin{cases}
k l &\quad\mbox{when}\ 1\leq l \leq \lfloor\frac{2|C|}{k}\rfloor\\
0&\quad\mbox{otherwise}
\end{cases}\label{gradeddim}\\
\dim{\ker{\slashed{\mathcal{D}}_{+}^{\rm G}}} &=& k \sum_{l=1}^{\lfloor\frac{2|C|}{k}\rfloor} l = \frac{k}{2}\left(\left\lfloor\frac{2|C|}{k}\right\rfloor^2+\left\lfloor\frac{2|C|}{k}\right\rfloor\right)~.\label{BPSdimension}
\end{eqnarray}
Since the kernel of $\slashed{\mathcal{D}}_{-}^{\rm G}$ is empty, the second quantity above also equals the index of the twisted Dirac operator, $\slashed{\mathcal{D}}^{\rm G}$. These results agree with the original calculation of \cite{Pope:1978zx} upon setting $k=1$.  We see that there is wall crossing for this index; whenever $|C|$ passes through an element of $|\nu| \in \frac{k}{2}\mathbb{N}$ a complete multiplet of angular momentum $|\nu|-\frac{1}{2}$ is gained or lost, depending on whether $|C|$ is increasing or decreasing, respectively.  

%%%%%%%%%%%%%%%%%%%%%%
%%%%%%%%%%%%%%%%%%%%%%
\section{Framed wall crossing identities for $\mathfrak{su}(2)$ SYM}\label{wcapp}
%%%%%%%%%%%%%%%%%%%%%%
%%%%%%%%%%%%%%%%%%%%%%

In this appendix we present some details on the framed wall crossing formulae in the case of the hypermultiplet walls in the weak coupling regime of $\mathfrak{su}(2)$ SYM. We then show how the generating function of framed BPS states discussed in section \ref{Section:FramedEx} indeed satisfies these formulae.

%%%%%%%%%%%%%%%%%%%%%%%
\subsection{Framed wall crossing through (half) hypermultiplet walls}
%%%%%%%%%%%%%%%%%%%%%%%

In the framed case the BPS walls are fully specified by a halo charge $\gamma_{\mathrm{h}}$. The halo particles themselves are elements of the single particle vanilla spectrum and can in principle sit in any short representation of the $\NN=2$ supersymmetry algebra. Here we will focus on the case where the halo particle sits in the simplest such representation possible, the so-called half hypermultiplet. It transforms in the $(0;\frac{1}{2})\oplus(\frac{1}{2};0)$ representation of the bosonic $\mathfrak{so}(3)_\mathrm{rot}\oplus \mathfrak{su}(2)_\mathrm{R}$ subalgebra. This case is simple because $\Omega(u,\gamma_{\mathrm{h}};y)=1$ throughout the weak coupling regime, which in practice means we can treat the halo particle as a fermion without internal degrees of freedom. The general framed wall crossing formula \eqref{genWCtransfo} of \cite{Gaiotto:2010be} then simplifies to
\begin{equation}\label{hyperWCtransfo}
F^-(\{X_\gamma\})=\left(\prod_{\gamma_\mathrm{h}}\Phi^{-1}(X_{\gamma_\mathrm{h}})\right)F^+(\{X_\gamma\})\left(\prod_{\gamma_\mathrm{h}}\Phi(X_{\gamma_\mathrm{h}})\right)~.
\end{equation}
Here we suppessed a few arguments, but $F^\pm(\{X_\gamma\})$ is the generating function \eqref{Fgenfun} of framed protected spin characters on the $\pm \Im \zeta^{-1}Z_{\gamma_\mathrm{h}}(u)>0$ side of the BPS wall $\widehat{W}(\gamma_{\rm h})$ defined in \eqref{fmsw}, and the product is over all $\gamma_{\mathrm{h}}$ associated to this wall. The function $\Phi$ is known as the quantum dilogarithm,
\begin{eqnarray}\label{dilogdef}
\Phi(\xi)&:=&\prod_{k=1}^{\infty}\left(1+y^{2k-1}\xi \right)^{-1} ~.
\end{eqnarray}
Finally one should remember that the formal variables $X_\gamma$ satisfy the typically non-commuting multiplication rule
\begin{equation}\label{XprodApp}
X_{\gamma}X_{\gamma'}=y^{\llangle\gamma,\gamma'\rrangle}X_{\gamma+\gamma'} ~.
\end{equation}
Noting that the generating function $F$ is actually linear in $X_\gamma$, we can equivalently think of \eqref{hyperWCtransfo} as originating from a change of variables when going from the $+$ to the $-$ side of the wall:
\begin{equation}\label{conjwcf}
X_\gamma\rightarrow \left(\prod_{\gamma_\mathrm{h}}\Phi^{-1}(X_{\gamma_{\rm h}})\right)X_\gamma\left(\prod_{\gamma_\mathrm{h}}\Phi(X_{\gamma_{\rm h}})\right)~.
\end{equation}

We will now present an equivalent form of \eqref{conjwcf} that will be of practical use in the remainder of this appendix. We proceed by a power series expansion of the quantum dilogarithm and its inverse:
\begin{eqnarray}\label{dilogexps}
\Phi(\xi)&=&\sum_{k=0}^\infty\frac{y^{-\frac{k(k-1)}{2}}\xi^k}{[k]_{y}!(y-y^{-1})^k} ~, \\
\Phi(\xi)^{-1}&=&\sum_{k=0}^\infty\frac{y^{\frac{k(k-1)}{2}}\xi^k}{[k]_{y}!(y^{-1}-y)^k}~, 
\end{eqnarray}
where
\begin{equation}
[k]_{y}! = \prod_{i=1}^k \chi_i(y) \equiv \prod_{i=1}^k\frac{y^i-y^{-i}}{y-y^{-1}} ~.  \,
\end{equation}
Using these and some well-known generating functions and identities for $q$-binomial coefficients one can derive that
\begin{equation}
\Phi^{-1}(X_{\gamma_\mathrm{h}})X_\gamma\Phi(X_{\gamma_\mathrm{h}})=X_\gamma\Psi_{\llangle\gamma,\gamma_{\mathrm{h}}\rrangle}(X_{\gamma_{\mathrm{h}}}) ~,
\end{equation}
where\footnote{Note that the function $\Psi_a$ we use here is related to the function $\Phi_a$ as defined in \cite{Gaiotto:2010be}, as follows: $\Psi_a(\xi)=\Phi_a^{\mathrm{sgn}(a)}(\xi)$}
\begin{equation}\label{psidef}
\Psi_a(\xi):=\sum_{k=0}^{\infty} \begin{bmatrix}a\\k\end{bmatrix}_{\! y} y^{-ak}\xi^k ~,
\end{equation}
and here we introduced the $q$-binomial coefficients, defined for $a\in \mathbb{Z}$, $b \in \mathbb{Z}_{\geq 0}$ by
\begin{equation}\label{qbinomialdef}
\begin{bmatrix}a\\b\end{bmatrix}_{\! y}:=\begin{cases}\frac{\prod_{i=1}^a(y^{i}-y^{-i})}{\prod_{j=1}^b(y^{j}-y^{-j})\prod_{k=1}^{a-b}(y^{k}-y^{-k})}&\quad\mbox{when } a\geq b\geq 0~, \\[2ex]
0&\quad\mbox{when } 0 \leq a < b~, \\
(-1)^b\begin{bmatrix}b-a-1\\b\end{bmatrix}_{\! y}&\quad\mbox{when } a<0, b\geq 0~.
\end{cases}
\end{equation}

Again using some properties of $q$-binomial coefficients one can find the following alternative product form:
\begin{equation}
\Psi_a(\xi)=\begin{cases}\prod_{k=1}^a(1+y^{-(2k-1)}\xi)&\quad\mbox{when }a>0~,\\
1&\quad\mbox{when }a=0~,\\
\prod_{k=1}^{|a|}(1+y^{2k-1}\xi)^{-1}&\quad\mbox{when }a<0~.
\end{cases}\label{psi2}
\end{equation}
This product form is easily inverted, and expanding that inverse in a series again one obtains the useful formula
\begin{equation}
\label{psiinverse}
\Psi_a^{-1}(\xi)=\sum_{k=0}^{\infty} \begin{bmatrix}-a\\k\end{bmatrix}_{\! y} y^{-ak}\xi^k ~.
\end{equation}

To conclude let us summarize in the case where the framed BPS wall is associated to a {\it single} half hypermultiplet halo particle. Crossing such a wall from the $+$ to the $-$ side amounts to the replacement
\begin{equation}\label{finalwcf}
X_\gamma\rightarrow X_\gamma\Psi_{\llangle\gamma,\gamma_{\mathrm{h}}\rrangle}(X_{\gamma_{\rm h}})=\sum_{k=0}^{\infty} \begin{bmatrix}\llangle\gamma,\gamma_{\mathrm{h}}\rrangle\\k\end{bmatrix}_{\! y}X_{\gamma+k\gamma_\mathrm{h}}~,
\end{equation}
while crossing from the $-$ to the $+$ side is equivalent to the inverse transformation
\begin{equation}\label{finalwcfinverse}
X_\gamma\rightarrow X_\gamma\Psi_{\llangle\gamma,\gamma_{\mathrm{h}}\rrangle}^{-1}(X_{\gamma_{\rm h}})=\sum_{k=0}^{\infty} \begin{bmatrix}-\llangle\gamma,\gamma_{\mathrm{h}}\rrangle\\k\end{bmatrix}_{\! y}X_{\gamma+k\gamma_\mathrm{h}} ~.
\end{equation}
The formulas written this way have the advantage over the form \eqref{conjwcf} that the limit $y\rightarrow\pm 1$ is explicitly non-singular and can be easily read off.  To be crystal clear, the analog of \eqref{hyperWCtransfo}, using \eqref{finalwcf}, is
\begin{equation}\label{hyperWCtransfo2}
F^- \left( \{ X_\gamma \} \right) = F^+ \left( \{ X_\gamma \Psi_{\llangle \gamma, \gamma_{\rm h}\rrangle}(X_{\gamma_{\rm h}}) \} \right)~.
\end{equation}

The physical interpretation is also clearer in this form.  Consider \eqref{finalwcf}, thinking of $\gamma$ as the electromagnetic charge of a core in the core-halo picture of framed BPS states, $\gamma = \gamma_{\rm c}$.  The formula for the Denef radius, \eqref{rbound}, shows that the stable side of the wall is where the sign of $\llangle \gamma_{\rm c},\gamma_{\rm h}\rrangle$ is opposite the sign of $\Im[\zeta^{-1} Z_{\gamma_{\rm h}}(u)]$, while the unstable side is where they have the same sign.  Therefore, in \eqref{finalwcf}, where we are going from the $\Im[\zeta^{-1} Z_{\gamma_{\rm h}}(u)] > 0$ side to the $\Im[\zeta^{-1} Z_{\gamma_{\rm h}}(u)] <0$ side, this corresponds to \emph{gaining} a halo when $\llangle \gamma, \gamma_{\rm h} \rrangle > 0$.  Now when $a \geq b \geq 0$ the $q$-binomial coefficients \eqref{qbinomialdef} are none other than
\begin{equation}
\begin{bmatrix}a\\b\end{bmatrix}_{\! y} = \frac{ [a]_{y}!}{[b]_{y}! [a-b]_{y}!} = \frac{ \prod_{i=1}^a \chi_{i}(y)}{\prod_{j=1}^b \chi_j(y) \prod_{k=1}^{a-b} \chi_k(y)} = \chi_{\Lambda^a \rho_b}(y)~,
\end{equation}
the character of the $a^{\rm th}$ antisymmetric product of the $b$-dimensional $\mathfrak{su}(2)$ irrep.  The halo particles obey fermi statistics and each particle can be thought of as carrying a spin $j = \half (\llangle \gamma_{\rm c}, \gamma_{\rm h}\rrangle - 1)$ $\mathfrak{su}(2)$ representation associated with the electromagnetic field.  The $k^{\rm th}$ term in the sum \eqref{finalwcf} thus corresponds to the configuration of $k$ halo particles surrounding the core.  Summing over all $k$ therefore accounts for all new states in the BPS Fock space associated with core charge $\gamma$ when $\widehat{W}(\gamma_{\rm h})$ is crossed.  If $\llangle \gamma, \gamma_{\rm h} \rrangle <0$ in \eqref{finalwcf} then we are crossing from the stable to the unstable side of the wall, and instead of gaining these states we are losing them.  In \eqref{finalwcfinverse}, where we are going from the $\Im[\zeta^{-1} Z_{\gamma_{\rm h}}(u)] < 0$ to $\Im[\zeta^{-1} Z_{\gamma_{\rm h}}(u)] > 0$ side, we simply interchange the role of the sign of $\llangle \gamma, \gamma_{\rm h} \rrangle$ in the above discussion.

%%%%%%%%%%%%%%%%%%%%%%%%%
\subsection{The $\mathfrak{su}(2)$ SYM case}
%%%%%%%%%%%%%%%%%%%%%%%%%

We now specialize to the case of framed BPS states in pure $\mathfrak{su}(2)$ SYM. A detailed discussion of the possible line defects, vanilla spectrum, and BPS walls can be found in section \ref{sec:fbps}. Let us quickly review the main points of relevance for convenience. Here we will consider line defects of 't Hooft charge $P=\frac{p}{2} H_{\alpha}$.  The charges of possible framed BPS states are labelled by two integers $\tilde n_{\mathrm{m}}, n_\mathrm{e}$:
\begin{equation}
\gamma=(\tilde{n}_{\mathrm{m}}-\frac{|p|}{2})H_\alpha \oplus n_\mathrm{e}\alpha\,.\label{genchargeapp}
\end{equation}
This implies we can rewrite the associated variables \eqref{XprodApp} as
\begin{equation}
X_\gamma=y^{(2\tilde n_\mathrm{m}-|p|)n_\mathrm{e}}X_1^{-\frac{|p|}{2}+\tilde n_\mathrm{m}}X_2^{n_\mathrm{e}}  ~,  \qquad \textrm{where} \quad X_1:=X_{H_\a}\,,\ X_2:=X_{\a} ~.
\end{equation}
The vanilla spectrum of $\mathfrak{su}(2)$ SYM contains dyons of charge $\gamma_{n}:=H_\a \oplus n\a$ which form a half\footnote{The other half is formed by the anti-particles $\bar \gamma_{n}:=(-H_\a) \oplus (-n\a)$, we will not explicitly consider them in the discussion, but the arguments presented here extend straightforwardly to their case as well.} hypermultiplet representation of the superalgebra. They can bind to other framed BPS states in halos when passing through the associated BPS wall $\widehat{W}_n$; see \eqref{walls} for a precise definition. With $c_n$ we denote the chamber between $\widehat{W}_n$ and $\widehat{W}_{n+1}$ when $n>0$, between $\widehat{W}_n$ and $\widehat{W}_{n-1}$ when $n<0$, and between $\widehat{W}_1$ and $\widehat{W}_{-1}$ when $n=0$. Inside such a chamber $c_n$ the generating function of protected spin characters is constant and we call that constant $F_n(p,\{X_1,X_2\};y)$.    

Given a wall $\widehat{W}_{n}$ let us identify the $\pm \Im \zeta^{-1}Z_{\gamma_\mathrm{h}}(u)>0$ sides. To do so, note that
\begin{equation}
I_n:=\Im\left[\zeta^{-1}Z_{\gamma_{n}}(u)\right]=(H_\a,\sy)+n\langle\alpha,\sx\rangle~.
\end{equation} 
This has the obvious properties
\begin{equation}
I_{n+1}=I_n+\langle\alpha,\sx\rangle~, \qquad \left.I_n\right|_{\widehat W_{n}}=0~.
\end{equation}
But then it immediately follows that\footnote{Remember that, by our definition of positive root, $\langle\alpha,\sx\rangle>0$.}
\begin{equation}
\left.I_{n}\right|_{\widehat W_{n-1}}=\left.\langle\alpha,\sx\rangle\right|_{\widehat W_{n-1}}>0~, \qquad \qquad\left.I_{n}\right|_{\widehat W_{n+1}}=-\left.\langle\alpha,\sx\rangle\right|_{\widehat W_{n-1}}<0 ~.
\end{equation} 
We thus see that for the wall  $\widehat{W}_{n}$ we have the identification
\begin{equation}
\begin{cases}
F^+=F_{n-1}\quad F^-=F_{n}&\qquad \mbox{when }n>0~,\\
F^+=F_{n}\quad F^-=F_{n+1}&\qquad \mbox{when }n<0~.
\end{cases}\label{chamberid}
\end{equation}

When $n>0$ the wall crossing formula \eqref{finalwcf} applies to moving from $c_{n-1}$ into $c_n$ through $\widehat{W}_n$ and amounts to the change of variables
\begin{align} \label{wcfus}
&y^{(2\tilde n_\mathrm{m}-|p|)n_\mathrm{e}}X_1^{-\frac{|p|}{2}+\tilde n_\mathrm{m}}X_2^{n_\mathrm{e}}\rightarrow  \cr
& \qquad \rightarrow \sum_{k=0}^{\infty} \begin{bmatrix}(|p|-2\tilde n_\mathrm{m})n+2n_\mathrm{e}\\k\end{bmatrix}_{\!y} y^{(2\tilde n_\mathrm{m}+2k-|p|)(n_\mathrm{e}+nk)}X_1^{-\frac{|p|}{2}+\tilde n_\mathrm{m}+k}X_2^{n_\mathrm{e}+nk}~. \qquad
\end{align}
When $n<0$ one can use the inverse wall crossing formula \eqref{finalwcfinverse} to move from $c_{n+1}$ into $c_{n}$ through $\widehat{W}_{n}$, it is equivalent to
\begin{align} \label{wcfusneg}
& y^{(2\tilde n_\mathrm{m}-|p|)n_\mathrm{e}}X_1^{-\frac{|p|}{2}+\tilde n_\mathrm{m}}X_2^{n_\mathrm{e}}\rightarrow \cr
& \qquad \rightarrow \sum_{k=0}^{\infty} \begin{bmatrix}-(|p|-2\tilde n_\mathrm{m})n-2n_\mathrm{e}\\k\end{bmatrix}_{\!y} y^{(2\tilde n_\mathrm{m}+2k-|p|)(n_\mathrm{e}+nk)}X_1^{-\frac{|p|}{2}+\tilde n_\mathrm{m}+k}X_2^{n_\mathrm{e}+nk} ~. \qquad
\end{align}
Note that the transformation \eqref{wcfusneg} is exactly equal to the transformation \eqref{wcfus}, after we map $\{n,n_\mathrm{e},X_2\} \rightarrow \{-n,-n_\mathrm{e},X_2^{-1}\}$.  Now suppose $F_0$ is given and is invariant under the transformation $\{n_{\rm e}, X_2\} \to \{-n_{\rm e}, X_{2}^{-1}\}$, as is the case in our $\mathfrak{su}(2)$ example where we have $F_0 = X_{1}^{-|p|/2}$.  Then the transformation properties \eqref{wcfus}, \eqref{wcfusneg} determine $F_n$ uniquely and thus the symmetry between them implies that
\begin{equation}
F_{n}(p,\{X_1,X_2\};y)=F_{-n}(p,\{X_1,X_2^{-1}\};y) ~. \label{ntominusn}
\end{equation}
Indeed, the explicit form of $F_n$ presented in \eqref{mfgenfuny} manifestly satisfies this relation.  In the remainder of this section we can thus restrict ourselves to $n>0$, and it will be sufficient to show that the various forms of $F_n$ presented in section \ref{sec:fbps} transform correctly under \eqref{wcfus}. First we will consider the simple case of $y=1$ and then at the end generalize to arbitrary $y$.

% % % % % % % % % % % % % % % % % % % 
\subsubsection{Wall crossing identities for $y=1$}
% % % % % % % % % % % % % % % % % % % 

When $y=1$ the protected spin characters are essentially counting the number of states and things simplify considerably as then $X_1$ and $X_2$ commute. Using the alternative form \eqref{psi2} we can rewrite \eqref{wcfus} simply as
\begin{eqnarray}
X_1&\rightarrow& \tilde X_1:=X_1(1+X_1X_2^n)^{-2n}~,  \label{Xts}\\
X_2&\rightarrow& \tilde X_2:=X_2(1+X_1X_2^n)^{2}~,
\end{eqnarray}
and wall crossing through $\widehat{W}_n$ becomes equivalent to the recursion relation
\begin{equation}
F_{n}(p,\{X_1,X_2\},y=1)=F_{n-1}(p,\{\tilde X_1,\tilde X_2\},y=1) ~.  \label{wcrecurs}
\end{equation}
In section \ref{sec:fbps} two forms of $F_n(y=1)$ are presented and here we check both solve the recursion relation above.

\paragraph{Chebyshev form.}
The first way to write $F_n(y=1)$ is
\begin{equation}\label{mfgenfunapp}
F_n(p,\{X_1,X_2\};y=1)=\left[X_1^{-1/2}X_2^{-n/2}\left( U_{n}(f_{n})-X_2^{-1/2}U_{n-1}(f_{n})\right)\right]^{|p|}\,,
\end{equation}
where $U$ are Chebyshev polynomials, defined through the recursion relation\footnote{One can also find the explicit expression:
\begin{equation*}
U_n(x)=\sum_{k=0}^{\lfloor\frac{n}{2}\rfloor}(-1)^k\frac{(n-k)!}{k!(n-2k)!}(2x)^{n-2k}~.
\end{equation*}}
\begin{eqnarray}
U_{-1}(x):= 0\,,\quad U_0(x):= 1\,,&\quad& U_{n+1}(x):=2x U_n(x)-U_{n-1}(x)~, \label{chebydefs}
\end{eqnarray}
and we introduced the following intermediate object:
\begin{equation}
f_n:=\frac{X_2^{1/2}+X_2^{-1/2}\left(1+X_1X_2^{n+1}\right)}{2}\,.
\end{equation}
Now observe that
\begin{equation}\label{fnrel}
\tilde f_{n-1}=f_n\,,
\end{equation}
where we used the obvious notation that a tilde denotes the corresponding object evaluated at the $\tilde X$, defined in \eqref{Xts}.

One can then explicitly check that \eqref{mfgenfunapp} satisfies the recursion relation \eqref{wcrecurs}:
\begin{align}
F_{n-1}(p,\{\tilde X_1,\tilde X_2\};y=1) =&~ \left( \tilde X_1^{-1/2}\tilde X_2^{-n/2}\left[ \tilde X_2^{1/2}U_{n-1}(\tilde f_{n-1})-U_{n-2}(\tilde f_{n-1})\right]\right)^{|p|} \cr
=&~ \left(X_1^{-1/2}X_2^{-n/2}\left[ X_2^{1/2}(1+X_1X_2^n)U_{n-1}( f_{n})-U_{n-2}(f_{n})\right]\right)^{|p|} \cr 
=&~ \left(X_1^{-1/2}X_2^{-n/2} \times \right. \cr
&~ \quad \times \left. \left[2f_nU_{n-1}(f_n)-U_{n-2}(f_{n}) -X_2^{-1/2}U_{n-1}(f_{n})\right]\right)^{|p|} \cr
=&~ \left(X_1^{-1/2}X_2^{-n/2}\left[U_{n}(f_{n}) -X_2^{-1/2}U_{n-1}( f_{n})\right]\right)^{|p|} \cr
=&~  F_n(p,\{X_1,X_2\};y=1)~.
\end{align}

\paragraph{Power series form.}
The second way to write $F_n(y=1)$ is
\begin{equation}\label{mfgenfun2app}
F_n(p,\{X_1,X_2\};y=1)  =  X_1^{- |p|/2} \sum_{\tilde{n}_{\mathrm m} = 0}^{\infty} \sum_{n_{\mathrm e} = 0}^{\infty} \left\{ ~~~\sum_{\mathclap{{\vec{r} \in S_{n}^{\tilde{n}_{\mathrm m},n_{\mathrm e}}}}} N_n(p,\vec{r};y=1) \right\} X_{1}^{\tilde{n}_{\mathrm m}} X_{2}^{n_{\mathrm e}}~,
\end{equation} 
where the set $S_{n}^{\tilde{n}_{\rm m},n_{\rm e}}$ is defined by
\begin{align}
S_{n}^{\tilde{n}_{\mathrm m},n_{\mathrm e}} :=&~  \left\{ \vec{r} = (r_1,\ldots,r_{n}) \in \mathbb{Z}_{\geq 0}^{n} ~\bigg|~ \lVert \vec{r}\, \rVert_{\mathrm m} = \tilde{n}_{\mathrm m} ~~ \& ~~ \lVert \vec{r}\, \rVert_{\mathrm e} = n_{\mathrm e} \right\}~,
\end{align}
with
\begin{equation}\label{menorms}
\lVert \vec{r}\, \rVert_{\mathrm m} := \sum_{k=1}^n r_k~,\qquad \lVert \vec{r} \, \rVert_{\mathrm e} := \sum_{k=1}^n k r_k~,
\end{equation}
and the quantities $N_{n}(p,\vec{r} ;y=1)$ are defined by
\begin{equation}
N_n(p,\vec{r};y=1\,)= \prod_{i=1}^{n}\begin{pmatrix}
|p| i-2\sum_{k=1}^{i-1}r_k(i-k)\\r_i
\end{pmatrix} ~.
\end{equation}

We check that also \eqref{mfgenfun2app} satisfies the recursion relation \eqref{wcrecurs}.  Here can make use of \eqref{wcfus} with $y  =1$, noting that in this limit the $q$-binomial becomes the ordinary binomial.  We also introduce the shorthand $\tilde{m}_{\rm m} := \tilde{n}_{\rm m} + k$, $m_{\rm e} := n_{\rm e} + n k$, and
\begin{equation}
a := 2n_\mathrm{e}+n(|p|-2\tilde{n}_{\rm m}) = 2m_\mathrm{e}+n(|p|-2\tilde{m}_{\rm m})~.
\end{equation}
Then with $F_{n-1}(p,\{\tilde{X}_1,\tilde{X}_2\};y=1) = F_{n-1}( \{ \tilde{X} \})$ we have
\begin{align}
F_{n-1}(\{\tilde{X} \}) =&~ \tilde X_1^{- |p|/2} \sum_{\tilde{n}_{\mathrm m} = 0}^{\infty} \sum_{n_{\mathrm e} = 0}^{\infty} \left\{ ~~~\sum_{\mathclap{{\vec{r} \in S_{n-1}^{\tilde{n}_{\mathrm m},n_{\mathrm e}}}}} N_{n-1}(p,\vec{r};y=1) \right\} \tilde X_{1}^{\tilde{n}_{\mathrm m}} \tilde X_{2}^{n_{\mathrm e}}~ \cr
=&~ X_1^{- |p|/2} \sum_{\tilde{n}_{\mathrm m} = 0}^{\infty} \sum_{n_{\mathrm e} = 0}^{\infty} \left\{ \sum_{k=0}^\infty \begin{pmatrix} a\\k  \end{pmatrix} ~~ \sum_{\mathclap{{\vec{r} \in S_{n-1}^{\tilde{n}_{\mathrm m},n_{\mathrm e}}}}} N_{n-1}(p,\vec{r};y=1) \right\}  X_{1}^{\tilde{n}_{\mathrm m}+k}  X_{2}^{n_{\mathrm e}+nk} \cr
=&~ X_1^{- |p|/2} \sum_{\tilde{m}_{\mathrm m} = 0}^{\infty} \sum_{m_{\mathrm e} = 0}^{\infty} \left\{ \sum_{k=0}^\infty~\begin{pmatrix} a\\k  \end{pmatrix} ~~~~~~ \sum_{\mathclap{\vec{r} \in S_{n-1}^{\tilde{m}_{\mathrm m}-k,m_{\mathrm e}-nk}}} ~N_{n-1}(p,\vec{r};y=1) \right\}  X_{1}^{\tilde{m}_{\mathrm m}}X_{2}^{m_{\mathrm e}} ~. \cr 
\end{align}
In the second step we used \eqref{wcfus} and in the last step we changed the summation indices $\{ \tilde{n}_{\rm m}, n_{\rm e}, k\} \to \{\tilde{m}_{\rm m}, m_{\rm e}, k\}$.  In general the latter would restrict the range of $k$, however here we are using the fact that $S_{n-1}^{\tilde{n}_{\rm m},n_{\rm e}}$ is empty if either $\tilde{n}_{\rm m}$ or $n_{\rm e}$ is negative.  Therefore these two sums can be trivially extended to sums over all of $\mathbb{Z}$ and that enables us to make the final equality.  

Comparing the last line with the form of $F_{n}(\{X\})$, what remains to be shown is
\begin{equation}\label{Nidentity}
\sum_{\mathclap{{\vec{r} \in S_{n}^{\tilde{n}_{\mathrm m},n_{\mathrm e}}}}} ~N_{n}(p,\vec{r};y=1) \stackrel{\textrm{?}}{=} \sum_{k\geq 0}~ \sum_{\vec{s} \in S_{n-1}^{\tilde{n}_{\mathrm m}-k,n_{\mathrm e}-nk}}\begin{pmatrix} 2n_\mathrm{e}+n(|p|-2\tilde n_m)\\k \end{pmatrix} ~N_{n-1}(p,\vec{s};y=1) ~.
\end{equation}
Let us start by decomposing the set over which we sum on the left-hand side of this expression. We do this by decomposing $\vec{r}\in \mathbb{Z}_{\geq 0}^{n}$ as $\vec{r}=(\vec{s},k)$ with $\vec{s}\in \mathbb{Z}_{\geq 0}^{n-1}$ and $k\in \mathbb{Z}_{\geq 0}$. It is then easy to analyze the constraints.  Using \eqref{menorms} (with the appropriate upper limit of $n$ or $n-1$ in the case of $\vec{r}$ or $\vec{s}$ respectively), we have
\begin{align}
& \tilde{n}_{\rm m} = \lVert \vec{r}\, \rVert_{\mathrm m} = \sum_{i=1}^n r_i = k + \sum_{i=1}^{n-1} s_i = k + \lVert \vec{s}\, \rVert_{\mathrm m}~, \cr
&  \tilde{n}_{\rm m} = \lVert \vec{r}\, \rVert_{\mathrm e} = \sum_{i=1}^n i r_i = n k + \sum_{i=1}^{n-1} i s_i = n k + \lVert \vec{s}\, \rVert_{\mathrm e}~,
\end{align}
and this implies
\begin{equation}\label{Ssum1}
S_{n}^{\tilde{n}_{\mathrm m},n_{\mathrm e}}=\bigcup_{k=0 }^\infty S_{n-1}^{\tilde{n}_{\mathrm m}-k,n_{\mathrm e}-nk} ~.
\end{equation}
The last piece of the puzzle then falls into place by the observation that
\begin{align}\label{Ssum2}
N_n(p,\vec{r};y=1\,)=&~ \prod_{i=1}^{n}\begin{pmatrix}  |p| i-2\sum_{l=1}^{i-1}r_l(i-l)\\r_i\end{pmatrix} \cr
=&~  \begin{pmatrix}  |p| n-2\sum_{l=1}^{n-1}s_l(n-l)\\k\end{pmatrix}N_{n-1}(p,\vec{s};y=1\,) \cr
=&~ \begin{pmatrix}  2n_\mathrm{e}+n(|p|-2\tilde n_m)\\k\end{pmatrix}N_{n-1}(p,\vec{s};y=1\,)~.
\end{align}
Results \eqref{Ssum1} and \eqref{Ssum2} establish the validity of \eqref{Nidentity}, whence $F_{n-1}(\{ \tilde{X} \}) = F_{n}( \{ X\})$.

% % % % % % % % % % % % % % % % % % % % % % 
\subsubsection{Wall crossing identities for arbitrary $y$}
% % % % % % % % % % % % % % % % % % % % % %

Here we will present the two forms of the generating function at arbitrary $y$ and show that they satisfy the correct ($y$-dependent) wall crossing. Because both forms of $F_n$ are manifestly equal at $n=0$ and furthermore solve the same wall crossing recursion relation this proves that they are equal for aribitrary $n$.

\paragraph{Chebyshev form.}
At arbitrary $y$, the expression \eqref{mfgenfunuber} for the generating function in terms of Chebyshev polynomials becomes
\begin{equation}\label{mfgenfunuberapp}
F_n(p,\{X_1,X_2\};y)=\left[ X_1^{-1/2}X_2^{-n/2}\left( U_{n}(f_{n})- X_2^{-1/2}U_{n-1}(f_{n})\right)\right]^{|p|}~, \quad (n\geq 0)~,
\end{equation}
where now
\begin{equation}\label{ydepfn}
f_n: = \frac{X_{2}^{1/2}+ X_2^{-1/2}\left(1+y^{2n+3} X_{1} X_{2}^n \right)}{2} = \frac{X_{2}^{1/2}+ X_2^{-1/2}\left(1+y X_{\gamma_{n+1}}\right)}{2} \,.
\end{equation}
In the second step we used \eqref{XprodApp} to note that
\begin{equation}\label{Xgammanpieces}
X_{\gamma_n} = X_{H_{\alpha} \oplus n\alpha} = y^{2n} X_{H_{\alpha}} X_{n\alpha} = y^{2n} X_1 X_{2}^n~.
\end{equation}
The latter form of $f_n$ in terms of $X_{\gamma_{n+1}}$ will be more convenient in the following.  Since $U_0(x) = 1$ and $U_{-1}(x) = 0$, we can observe that $F_0 = X_{1}^{-|p|/2}$ as required.  

Then, using the identification of the chambers \eqref{chamberid}, the wall crossing formula we wish to check is
\begin{equation}\label{yrecurs}
F_n(p,\{X_1,X_2\};y) =F_{n-1}(p,\{\tilde X_1,\tilde X_2\};y) ~, 
\end{equation}
where, with the aid of \eqref{psi2}, the relation between new and old variables is
\begin{align}\label{X1X2ytrans}
\tilde{X}_1 =&~ X_1 \Psi_{-2n}(X_{\gamma_n}) = X_1\prod_{k=1}^{2n}(1+y^{2k-1} X_{\gamma_n})^{-1} ~,  \cr
\tilde{X}_2 =&~ X_2 \Psi_{2}(X_{\gamma_n}) =  X_2(1+y^{-1} X_{\gamma_n})(1+y^{-3} X_{\gamma_n})~.
\end{align}

These transformations are quite complicated, especially since the $X$'s do not commute but rather satisfy the relation
\begin{equation}
X_2^{\alpha}X_1^{\beta}=y^{4\alpha\beta}X_1^\beta X_2^{\alpha} ~.
\end{equation}
The key step is to realize that $F_{n-1}$ can equivalently be thought of as a function in the variables $X_1X_2^n$ and $X_2^{1/2}$. These have much simpler transformation properties.  Indeed, by pushing all $X_2$ factors through to the right, one can show that
\begin{equation}
\left[  X_2(1+y^{-1} X_{\gamma_n})(1+y^{-3} X_{\gamma_n}) \right]^n = \left[ \prod_{k=1}^{2n} (1 + y^{2k-1} X_{\gamma_{n}}) \right] X_2~,
\end{equation}
whence it immediately follows that
\begin{equation}
\tilde{X}_1 \tilde{X}_{2}^n = X_1 X_{2}^n~.
\end{equation}
Now, although $X_{\gamma_n}$ is not equal to $X_{1} X_{2}^n$, it is proportional to it---see \eqref{Xgammanpieces}---and this is sufficient to guarantee that $\tilde{X}_{\gamma_{n}} = X_{\gamma_n}$.  Meanwhile, by $X_{2}^{1/2}$ we mean $X_{\gamma = \lambda}$, where $\lambda$ is the fundamental magnetic weight with $\alpha = 2\lambda$, such that $X_2 = X_{2\lambda} = X_{\lambda}^2$.  Then one finds
\begin{align} 
\tilde{X}_{2}^{1/2} =&~ X_{2}^{1/2} \Psi_1(X_{\gamma_n}) = X_{2}^{1/2} (1 +y^{-1} X_{\gamma_n})~, \cr
\tilde{X}_{2}^{-1/2} =&~ X_{2}^{-1/2} \Psi_{-1}(X_{\gamma_n}) = X_{2}^{-1/2} (1 + y X_{\gamma_n})^{-1}~.
\end{align}
By squaring both sides of the first equation one can recover the second of \eqref{X1X2ytrans}.  A final identity that will be useful is
\begin{align}\label{Xgammanid}
X_{\gamma_{n+1}} =&~ y^{2(n+1)} X_{1} X_{2}^{n+1} = y^{2(n+1)} y^{-4} X_2 X_1 X_{2}^{n+1} = y^{-2} X_2 (y^{2n} X_1 X_{2}^n)  \cr
=&~ y^{-2} X_2 X_{\gamma_n}~.
\end{align}

Using these, one first establishes the direct analog of \eqref{fnrel}:
\begin{align}
\tilde{f}_{n-1} =&~ \half \left\{ \tilde{X}_{2}^{1/2} + \tilde{X}_{2}^{-1/2} ( 1 + y \tilde{X}_{\gamma_n}) \right\} \cr
=&~ \half \left\{ X_{2}^{1/2} (1 + y^{-1} X_{\gamma_{n}}) + X_{2}^{-1/2} \right\} \cr
=&~ \half \left\{ X_{2}^{1/2} + y X_{2}^{-1/2} X_{\gamma_{n+1}} + X_{2}^{-1/2} \right\} \cr
=&~ f_n~.
\end{align}
Now note that for any $a$, $X_{1}^{a} X_{2}^{na} = y^{-2a} X_{a \gamma_n} = y^{-2a} \tilde{X}_{a \gamma_n} = \tilde{X}_{1}^a \tilde{X}_{2}^{na}$.  Therefore we have
\begin{align}
F_{n-1}( \{\tilde{X}\} ) =&~ \left\{ \tilde{X}_{1}^{-1/2} \tilde{X}_{2}^{-(n-1)/2} \left[ U_{n-1}(\tilde{f}_{n-1}) - \tilde{X}_{2}^{-1/2} U_{n-2}(\tilde{f}_{n-1}) \right] \right\}^{|p|} \cr
=&~  \left\{ X_{1}^{-1/2} X_{2}^{-n/2} \left[ \tilde{X}_{2}^{1/2} U_{n-1}(f_n) - U_{n-2}(f_n) \right] \right\}^{|p|} \cr
=&~ \left\{ X_{1}^{-1/2} X_{2}^{-n/2} \left[ U_{n}(f_n) + \left( X_{2}^{1/2} (1 + y^{-1} X_{\gamma_n}) - 2 f_n\right) U_{n-1}(f_n) \right] \right\}^{|p|} \cr
=&~  \left\{ X_{1}^{-1/2} X_{2}^{-n/2} \left[ U_{n}(f_n) - X_{2}^{-1/2} U_{n-1}(f_n) \right] \right\}^{|p|} \cr
=&~ F_n(\{X\})~.
\end{align}
In the last step we plugged in $f_n$, \eqref{ydepfn}, and used \eqref{Xgammanid} again.

\paragraph{Power series form.}
For generic $y$ the generating function $F_n$ can also be presented as a direct generalization of \eqref{mfgenfun2app}:
\begin{equation}\label{mfgenfunyapp}
F_n(p,\{X_1,X_2\};y)  =  X_1^{- |p|/2} \sum_{\tilde{n}_{\mathrm m} = 0}^{\infty} \sum_{n_{\mathrm e} = 0}^{\infty} \left\{ ~~~ \sum_{\mathclap{{\vec{r} \in S_{n}^{\tilde{n}_{\mathrm m},n_{\mathrm e}}}}} ~N_n(p,\vec{r};y) \right\}y^{(2\tilde n_\mathrm{m}-|p|)n_\mathrm{e}}X_{1}^{\tilde{n}_{\mathrm m}} X_{2}^{n_{\mathrm e}} ~,
\end{equation} 
where now
\begin{equation}
N_n(p,\vec{r};y\,) = \prod_{i=1}^{n}\begin{bmatrix}|p| i-2\sum_{k=1}^{i-1}r_k(i-k)\\r_i\end{bmatrix}_{\!y} ~.
\end{equation}
To check that \eqref{mfgenfunyapp} satisfies the correct recursion, we begin by applying the change of variables \eqref{wcfus} to $F_{n-1}(p,\{\tilde{X}_1,\tilde{X}_2\};y)$:
\begin{align}
F_{n-1}(\{\tilde{X} \}) =&~ \tilde X_1^{- |p|/2} \sum_{\tilde{n}_{\mathrm m} = 0}^{\infty} \sum_{n_{\mathrm e} = 0}^{\infty} \left\{ ~~~\sum_{\mathclap{{\vec{r} \in S_{n-1}^{\tilde{n}_{\mathrm m},n_{\mathrm e}}}}} N_{n-1}(p,\vec{r};y) \right\} y^{(2\tilde n_\mathrm{m}-|p|)n_\mathrm{e}}  \tilde{X}_{1}^{\tilde{n}_{\mathrm m}} \tilde{X}_{2}^{n_{\mathrm e}} \cr
=&~ X_1^{- |p|/2} \sum_{\tilde{n}_{\mathrm m} = 0}^{\infty} \sum_{n_{\mathrm e} = 0}^{\infty} \left\{ ~~~\sum_{\mathclap{{\vec{r} \in S_{n}^{\tilde{n}_{\mathrm m},n_{\mathrm e}}}}} N_{n-1}(p,\vec{r};y) \right\} \sum_{k=0}^{\infty}\begin{bmatrix}(|p|-2\tilde n_\mathrm{m})n+2n_\mathrm{e}\\k\end{bmatrix}_{\!y}\times \cr
&~ \times y^{(2\tilde n_\mathrm{m}+2k-|p|)(n_\mathrm{e}+nk)}X_1^{-\frac{|p|}{2}+\tilde n_\mathrm{m}+k}X_2^{n_{\rm e}+nk} \cr
=&~ X_1^{- |p|/2} \sum_{\tilde{m}_{\mathrm m} = 0}^{\infty} \sum_{m_{\mathrm e} = 0}^{\infty} \left\{\sum_{k=0}^{\infty} ~~~~~~~ \sum_{\mathclap{{\vec{r} \in S_{n}^{\tilde{m}_{\mathrm m}-k,m_{\mathrm e}-nk}}}} N_{n-1}(p,\vec{r};y) \begin{bmatrix}(|p|-2\tilde m_\mathrm{m})n+2m_\mathrm{e}\\k\end{bmatrix}_{\!y} \right\}\times \cr
&~ \times y^{(2\tilde{m}_\mathrm{m}-|p|)m_\mathrm{e}}X_1^{-\frac{|p|}{2}+\tilde{m}_\mathrm{m}}X_2^{m_{\rm e}} ~.
\end{align}
In the last step we made the same relabeling, $\tilde{m}_{\rm m} = \tilde{n}_{\rm m} + k$ and $m_{\rm e} = n_{\rm e} + nk$, as in the $y=1$ case.  We see that the last line will agree with \eqref{mfgenfunyapp} if
\begin{equation}
\sum_{\mathclap{\vec{r} \in S_{n}^{\tilde{n}_{\mathrm m},n_{\mathrm e}}}} ~N_n(p,\vec{r};y) \stackrel{\textrm{?}}{=} \sum_{k=0}^{\infty} ~~~~~~~  \sum_{\mathclap{{\vec{s} \in S_{n}^{\tilde{n}_{\mathrm m}-k,n_{\mathrm e}-nk}}}} N_{n-1}(p,\vec{s};y)\begin{bmatrix}(|p|-2\tilde n_\mathrm{m})n+2n_\mathrm{e}\\k\end{bmatrix}_{\!y} ~.
\end{equation}
The proof of this relation is completely analogous to that of \eqref{Nidentity} presented above; see \eqref{Ssum1} and \eqref{Ssum2}.

%%%%%%%%%%%%%%%%%%%%%%%%
%%%%%%%%%%%%%%%%%%%%%%%%
\section{List of symbols and notation}\label{app:notation}
%%%%%%%%%%%%%%%%%%%%%%%%
%%%%%%%%%%%%%%%%%%%%%%%%

In this appendix we have collected a list of symbols that appear frequently throughout the paper. Certain symbols that only appear locally, i.e. in a short discussion on a few pages or only in an appendix are not included.
{\setlength{\tabcolsep}{-5pt}\begin{longtable}{lcc}
Symbol& definition/introduction & corresponding object\\
\hline
$\hat{\cdot}$ &section \ref{sec:quantize}& operator obtained after quantization\\
$[\cdot]$&\dpageref{PQdef}& Weyl orbit\\
$(\, \cdot \,, \, \cdot \,)$&\eqref{Trdef}& positive definite quadratic form\\
$\langle\, \cdot \,, \, \cdot \,\rangle$&\dpageref{canpair}& canonical pairing\\
$\llangle\, \cdot \,, \, \cdot \,\rrangle$&\dpageref{dszpairing}& symplectic pairing on $\Gamma_u,\Gamma_{L,u}$\\
$a$&\dpageref{scalaradef}& Cartan valued part of Higgs field\\
$a^I$&\dpageref{spcodef}& special coordinates\\
$a_{\mathrm{D}}$&\eqref{tvaluedaD}& Cartan valued dual special coordinate\\
$a_{\mathrm{D},I}$&\eqref{pvector}& dual special coordinates\\[1ex]
$a_{\mathrm{D}}^{\textrm{1-lp}}$&\dpageref{aDoneloop}& one-loop approx.~to dual special coord.\\
$\mathfrak{a}_I$&\eqref{Darboux}& half Darboux basis\\
$A$&\dpageref{Adef}& UV gauge field\\
$\hat{A}$&\eqref{hatA}& Euclidean gauge field\\
$A^I$&\dpageref{AIdef}& IR gauge field\\
$\alpha$&\eqref{rootdecomp}& root\\
$\alpha_I$&\dpageref{defsimpleroots}& simple root\\
$\alpha_{I_A}$&\dpageref{GXKexpand}& subset of simple roots\\
$B_i$&\dpageref{EBdef}& UV magnetic field\\
$B_i^I$& \eqref{HSW}& IR magnetic field\\
$\mathfrak{b}^I$&\eqref{Darboux}& half Darboux basis\\
$\BB$&\dpageref{coulombbdef}& Coulomb branch\\
$\BB^*$&\dpageref{coulombast}& non-singular part of Coulomb branch\\
$\BB^\mathrm{sing}$&\dpageref{coulombast}& singular part of Coulomb branch\\
$\widehat{\BB}$&\dpageref{vanillawalls}& universal cover of non-sing C. branch\\
$\widehat{\BB}_\mathrm{wc}$&\dpageref{wcRegime}& weak coupling regime\\
$\chi^m$&\eqref{modspacea0}& fermionic collective coordinate\\
$\chi_n(y)$&\dpageref{ncharacter}& character of $n$-dim'l $SU(2)$ irrep \\
$\chi$&\dpageref{pdchi}& global coordinate on $\mathbb{R}_{X_\infty}$\\
$\XX^{\bfn}$&\eqref{complexcoords}& complexified fermi collective coord's\\
$d$&\dpageref{defd}& rank of the effective lie algebra\\
$D$&\dpageref{Ddef}& gauge covariant derivative\\
$\hat{D}$&\dpageref{hatA}& Euclidean covariant derivative\\
$\mathbb{D}$&\dpageref{Mfactor}& group of deck transformations\\
$\mathbb{D}_\gi$&\dpageref{Dhkfactor}& group of gauge-induced deck transfo's\\[1ex]
$\slashed{\DD}_{\mMM(\gm;\sx)}^{{\rm G}(\sy)}$&\eqref{Diracdecomp}& $\rG(\YY_\infty)$-twisted Dirac op.~on $\MM$\\[1ex]
$\slashed{\DD}_{\mMM_0(\gm;\sx)}^{{\rm G}_0(\YY_{\infty})}$&\eqref{M0diracop}& $\rG_0(\YY_{\infty})$-twisted Dirac op.~on $\MM_0$\\[1ex]
$\slashed{\DD}_{0}^{\rG_0}$&\dpageref{M0diracop}&shorthand for $\slashed{\DD}_{\mMM_0(\gm;\sx)}^{{\rm G}_0(\YY_{\infty})}$\\[1ex]
$\slashed{\DD}_{\fMM(L;\gm;\sx)}^{\rG(\sy)}$&\eqref{fMdiracop}& $\rG(\YY_\infty)$-twisted Dirac op.~on $\fMM$\\[1ex]
$\slashed{\DD}^{\rG}$&\dpageref{fMdiracop}& shorthand for $\slashed{\DD}_{\fMM(L;\gm;\sx)}^{\rG(\sy)}$\\[1ex]
$\delta\hat{A}_a$&\dpageref{linearsd}& bosonic zeromode\\
$\Delta$&\eqref{rootdecomp}& set of nonzero roots\\
$\Delta^+$&\dpageref{posrootdef}& set of positive roots\\
$e^{\um}$&\dpageref{framestuff}& tangent space frame\\
$e^{\ubfm}$&\dpageref{holsymform}& holomorphic tangent space frame\\
$\mathrm{e}$&\dpageref{mathgmap}& electric trivialization map\\
$E_i$&\dpageref{EBdef}& UV electric field\\
$E_i^I$&\eqref{HSW}& IR electric field\\
$E_\alpha$&\dpageref{rootdecomp}& raising/lowering operator\\
$\EE_{\um}$&\dpageref{framestuff}& inverse frame\\
$\epsilon_H$&\dpageref{Gdef}&gauge parameter asymptoting to $H$\\
$\varepsilon_m$&\eqref{coordflows}& compensating gauge parameter\\
$F$&\eqref{Fgenfun}& generating function\\
$\hat{F}$&\dpageref{hatA}& Euclidean field strength\\
$F^I$&\eqref{IRcharges}& IR field strength\\
$\mathbb{F}$&\eqref{sdF}& imaginary selfdual field strength\\
$\varphi$&\dpageref{phidef}& complex Higgs field\\
$\varphi_\infty$&\dpageref{phidef}& asymptotic Higgs field\\
$\phi_{mn}$&\eqref{modunicurve}& curvature on universal bundle\\
$\phi_\gi$&\eqref{phiiso}&generator of $\mathbb{D}_{\gi}$\\
$\phi_{\gi,0}$&\eqref{phihk0}& action of $\phi_{\gi}$ on $\MM_0$\\
$\widetilde{\phi}_\gi$&\eqref{liftedphi}&lift of $\phi_{\gi}$ to spinor bundle\\
$\widetilde{\phi}_{\gi,0}$&\eqref{liftedphith}& lift of $\phi_{\gi,0}$ to spinor bundle\\
$g$&\eqref{metC}&metric on moduli space\\
$g_\mathrm{phys}$&footnote \ref{fn:physmet}& metric on mod.~space, physics normaliz'n\\
$\mathfrak{g}$&\dpageref{liealg}& simple compact Lie algebra\\
$\mathfrak{g}_{\mathbb{C}}$&\dpageref{complexliealg}& complexified Lie algebra\\
$\mathfrak{g}^\ast$&\dpageref{liealgdual}& Lie algebra dual\\
$g_0$&\dpageref{tau0def}& classical Yang-Mills coupling\\
$G$&\dpageref{liegroup}& simple compact Lie group\\
$\tilde G$&\dpageref{crlatdef}& simply-connected Lie group\\
$\mathrm{G}$&\eqref{Gdef}& G-homomorphism\\
$\rG_0$&\dpageref{vGdecomp}& projected G-homomorphism\\
$G_\mathrm{ad}$&\dpageref{mwlatdef}& adjoint Lie group\\
$G_I$&\eqref{IRcharges}& dual IR field strengths\\
$\GG^0_{\{P_n\}}$&\eqref{localgts}& group of local gauge transformations\\
$\gamma$&\dpageref{lowecharge}& electromagnetic charge\\
$\gamma_{\rm c}$&\dpageref{sec:corehalo}& core charge\\
$\gamma_{\rm e}$&\dpageref{flux}& quantized electric charge\\
$\gamma_{\rm e}^\ast$&\dpageref{flux}& dual of the electric charge\\
$\gamma_{\rm e}^{\rm phys}$&\eqref{flux}& physical electric charge\\
$\gamma_{{\rm e},0}$&\eqref{geequiv}& relative electric charge\\
$\gamma_{\rm def}$&\dpageref{defcharge}& charge of an IR defect\\
$\gamma_{\rm m}$&\eqref{flux}& magnetic charge\\
$\gamma_{\rm m}^{\rm ef}$&\dpageref{mudef}& effective magnetic charge\\
$\tilde \gamma_{\rm m}$&\eqref{relcharge}&relative magnetic charge\\
$\gamma_{{\rm e},I}$&\dpageref{chargecomps}& electric charge comp's w.r.t.~Darboux basis\\
$\gamma_{{\rm e},I}^{\rm phys}$&\dpageref{physEflux}& phys. electric charge comp's w.r.t.~Darboux b.\\
$\gamma_{\rm h}$&\dpageref{Ehalo}& halo charge\\
$\gamma_{{\rm m}}^{I}$&\dpageref{chargecomps}& magnetic charge comp's w.r.t.~Darboux basis\\
$\gamma_L$&\eqref{torsor}& IR representative of UV defect charge\\
$[\gamma_{\rm e}]_{\rm JZ}$&\eqref{geZequiv}& Julia-Zee tower\\
$\Gamma_u$&\dpageref{lowecharge}& electromagnetic charge lattice\\
$\Gamma_u^\mathrm{e}$&\dpageref{emlatticesplit}& electric charge lattice\\
$\Gamma_u^\mathrm{m}$&\dpageref{emlatticesplit}& magnetic charge lattice\\
$\Gamma_{L,u}$&\eqref{torsor}& shifted charge lattice\\
$ \Gamma^{p}_{\phantom{p}mn}$&\eqref{ccChristoffel}& Christoffel symbols\\
$h^A$&\dpageref{KAdef}&fundamental magnetic weight (fmw) of $\mathfrak{t}^{\rm ef}$\\
$h_{0}^A$&\dpageref{K0Adef}& linear combo.~of fmw orthog'l to $\gamma_{\rm m}^{\rm ef}$\\
$h_{\rm cm}$&\dpageref{hcmdef}& ele.~of $\mathfrak{t}^{\rm ef}$ corresponding to $X_\infty$\\
$h_\gi$&\dpageref{K0Adef}& linear combo.~of fmw generating $\im{\mu}$\\
$h^{I_A}$&\dpageref{KAdef}&fundamental magnetic weight of $\mathfrak{t}$\\
$h^{I_M}$&\dpageref{KAdef}&fundamental magnetic weight of $\mathfrak{t}$\\
$H$&\dpageref{rootdecomp}&generic element of Cartan subalgebra\\
$H_\alpha$&\dpageref{rootdecomp}& co-root\\
$H_I$&\dpageref{defsimpleroots}& simple co-root of $\mathfrak{t}$\\
$H_A$&\dpageref{corootpart}& simple co-root of $\mathfrak{t}^{\rm ef}$\\
$\HH_L$&\dpageref{hilbdef}& Hilbert space of framed theory\\
$\HH_{L_\zeta}^{\mathrm{BPS}}$&\dpageref{hilbdef}& BPS Hilbert space of framed theory\\
$\HH_{u,\gamma}^{\rm BPS}$&\eqref{vanillagrading}& fixed charge Hilbert space\\
$(\bar\eta^r)_{ab}$&\eqref{asdtHooft}& anti-selfdual 't Hooft symbols\\
$\mathcal{I}^r$&\dpageref{twisteddiag}& twisted diagonal generator\\
$i_\ast$&\dpageref{iastdef}& embedding of Lie algebras $\mathfrak{g}^{\rm ef} \to \mathfrak{g}$\\
$I^r$&\dpageref{twisteddiag}& generator of $R$-symmetry\\
$J^r$&\dpageref{twisteddiag}& generator of spatial rotations\\
$\mathcal{J}^r$&\eqref{quat2}& quaternionic structure on $S^+_{\mathrm{smw}}$\\
$\mathbbm{j}^r$&\eqref{R4cs}& quaternionic structure on euclidean space\\
$\mathbb{J}^r$&\eqref{quatstructure}& quaternionic structure on moduli space\\
$\bbJ^a$&\dpageref{ccsusy}& extended quaternionic structure\\
$\tbbJ^a$&\dpageref{ccsusy}& extended quaternionic structure\\
$\ker^{\gamma_{\rm e}}$&\eqref{framedDker}& kernel component\\
$\ker^{\gerel}$&\eqref{vanDker}& kernel component\\
$K^A$&\dpageref{Killvecs}& Killing vector from global gauge symmetry\\
$K_{0}^A$&\eqref{K0Adef}& projected triholomorphic Killing vector\\
$K^E$&\dpageref{symEonA}& Killing vector of moduli space\\
$K^i$&\dpageref{Killvecs}& translational Killing vector\\
$K^r$&\dpageref{Killvecs}& rotational Killing vector\\
$\kappa_A$&\dpageref{twisteddiag}& symplectic-Majorana spinor\\
$L$&\dpageref{Ldef}& line defect\\
$L$&\dpageref{dualgm}& greatest common divisor of the $\ell^A$\\
$\ell^A$&\dpageref{dualgm}& components of $\gamma_{\rm m}^\ast$ along $\alpha_{I_A}$\\
$\LL$&\eqref{lineoplattice}& defect lattice\\
$\lambda^I$&\dpageref{defsimpleroots}& fundamental weight\\
$\lambda^A$&\dpageref{smwdef}& symplectic-Majorana-Weyl fermion\\
$\Lambda$&\eqref{dyscale}& dynamical scale\\
$\Lambda_{\mathrm{cr}}$&\dpageref{crlatdef}& coroot lattice\\
$\Lambda_{\mathrm{mw}}$&\dpageref{mwlatdef}& magnetic weight lattice\\
$\Lambda_{\rm mw}^{\rm ef}$&\dpageref{mudef}& effective magnetic weight lattice\\
$\Lambda_{\mathrm{rt}}$&\dpageref{crlatdef}& root lattice\\
$\Lambda_{\mathrm{wt}}$&\dpageref{repdef}& weight lattice\\
$\Lambda_G$&\eqref{cochar}& cocharacter lattice\\
$\Lambda_G^\vee$&\eqref{cochar}& character lattice\\
$\Lambda^{(p,q)}$&\dpageref{holsymform}& bundle of $(p,q)$ differential forms\\
$\mathrm{m}$&\dpageref{mathgmap}& magnetic trivialization map\\
$M$&\eqref{RTalg}& mass\\
$M_\gamma$&\dpageref{vanillagrading}& mass of charged particle\\
$M_\gamma^\mathrm{cl}$&\eqref{MBPScl}& classical BPS mass\\[1ex]
$M_{\gm}^{\textrm{1-lp}}$&\eqref{Qmonomass}& 1 loop corrected monopole mass\\
$\MM$&\dpageref{Mdef}& vanilla monopole moduli space\\
$\MM_0$&\dpageref{sec:modisom}& strongly centered monopole moduli space\\
$\widetilde{\MM}$&\dpageref{sec:modisom}&universal cover of moduli space\\
$\fMM$&\eqref{Mdef}& moduli space of singular monopoles\\
$\mu$&\dpageref{mudef}& group homomorphism\\
$\mu_0$&\dpageref{dyscale}& momentum scale\\
$\tilde n_\mathrm{m}^I$&\dpageref{dim3}& integer relative magnetic charge\\
$\tilde{n}_{\rm m}^{I_A}$&\eqref{corootpart}& non-zero integer relative magnetic charge\\
$n_\mathrm{e}^I$&\dpageref{econstraints}& integer electric charge\\
$n_{\rm e}^{I_A}$&\dpageref{neIA}& subset of integer electric charges\\
$N_{\mathrm{e}}^A$&\dpageref{ccecharge}&moduli space electric charge\\
$N_{\mathrm{e},0}^A$&\eqref{echarge0def}& integer components of rel.~electric charge\\
$\sw^r$&\eqref{comtripple}& Kahler form on moduli space\\
$\omega_{p,\phantom{\um}\un}^{\phantom{p,}\um}$&\dpageref{framestuff}&spin connection\\
$\Omega$&\eqref{PSC}&protected spin character\\
$\fOmega$&\eqref{fPSC}&framed protected spin character\\
$p^I$&\dpageref{defcharge}&magnetic components of defect charge\\
$p_m$&\dpageref{solitoncov}& momentum in moduli space mechanics\\
$\sp^A$&\dpageref{dualgm}& length of simple root of $\mathfrak{g}^{\rm ef}$\\
$\sp^I$&appendix \ref{app:Dquotient}& length of simple root of $\mathfrak{g}$\\
$P$&\dpageref{PQdef}& UV magnetic defect charge\\
$\pi_m$&\eqref{scovmom}& super-covariant momentum\\
$\psi_A$&\dpageref{fermdef}& Weyl fermion\\
$\psi_{\mathrm{cm}}$&\eqref{Psidecomp}&constant spinor on $\mathbb{R}^4$\\
$\Psi$&\dpageref{modspacespin}& spinor on moduli space\\
$q^I$&\dpageref{defcharge}&electric components of defect charge\\
$q_\mathrm{cm}$&\eqref{echarge0def}&component of electric charge along $\gm^\ast$\\
$Q$&\dpageref{PQdef}& UV electric defect charge\\
$Q^a$&\eqref{Qcc}& moduli space supersymmetry generator\\
$Q^A$&\dpageref{Qdef}& supersymmetry generator\\
$\hat{Q}_{\rm (sc)}^a$&\eqref{Qsc}& semi-classical supercharge operator\\
$R_\kappa$&\dpageref{twisteddiag}& rotation matrix\\
$\RR^A$&\eqref{Rsusys}& unbroken supersymmetry generator\\
$\mathbb{R}_{\rm cm}^3$&\eqref{ucoverfactor}& space of monopole c.o.m. positions\\
$\mathbb{R}_{X_\infty}$&\eqref{ucoverfactor}& space of monopole c.o.m. phase\\
$\rho$&\dpageref{repdef} &Lie algebra representation\\
$\rho^A$&\dpageref{smwdef}& symplectic-Majorana-Weyl fermion\\
$S_{\mathrm{van}}$& \eqref{action} & vanilla action\\
$S_{\mathrm{def}}$&\eqref{Sbndryapp} & defect action\\
$S_\mathrm{smw}^+$&\eqref{smwdef}& space of pos.~chirality sympl'c Majorana-Weyl spinors\\
$\mathfrak{so}(3)$&\dpageref{rotationalg}& spatial rotation algebra\\
$SU(2)_R$&\dpageref{SU2Rdef}& R-symmetry group\\
$\mathcal{S}_{[\hat{A}]}$&\eqref{fzmspace}& space of fermionic zeromodes\\
$\SS_{\rm D}$&\dpageref{liftedphi}& Dirac spinor bundle\\
$\mathfrak{su}(2)_R$&\dpageref{twisteddiag}& R-symmetry algebra\\
$\mathfrak{su}(2)_\mathrm{d}^{(\kappa)}$&\dpageref{twisteddiag}& twisted diagonal algebra\\
$\Sigma$&\eqref{BPSfc}& moduli space of BPS field configurations\\
$\Sigma_0$&\eqref{relBPSfc}& strongly centered mod~.space of BPS field config's\\
$\fSigma$&\eqref{fBPSfc}& moduli space of framed BPS field configurations\\
$\mathfrak{t}$&\dpageref{rootdecomp}& Cartan subalgebra\\
$\mathfrak{t}^\ast$&\dpageref{rootdecomp}& dual Cartan subalgebra\\
$\mathfrak{t}_\mathbb{C}$&\dpageref{rootdecomp}& complexified Cartan subalgebra\\
$\mathfrak{t}^\mathrm{ef}$&\dpageref{iastdef}& effictive Cartan subalgebra\\
$\mathfrak{t}_{\gm}^{\perp}$&\eqref{mathY0def}&subspace of $\mathfrak{t}$ Killing-orth.~to $\gm$\\
$T$&\eqref{cochar}& Cartan torus\\
$T_{\rm ad}^{\rm ef}$&\dpageref{Tefad}& effective adjoint Cartan torus\\
$T_{[\hat{A}]} \fMM$&\dpageref{linearsd}& tangent space of moduli space\\
$T_\kappa$&\eqref{Tmap}& quaternionic isomorphism\\
$\TT^A$&\eqref{Tsusys}& broken supersymmetry generator\\
$\TT_{ \{ P_n \} }$&\eqref{globalgts}& group of global gauge transformations\\
$\tau_0$&\dpageref{tau0def}& classical complexified coupling\\
$\tau_{IJ}$&\eqref{lel}& IR coupling matrix\\
$\tau^a$&\eqref{taumat}& Euclidean sigma matrices\\
$\theta_0$&\dpageref{tau0def}& classical theta-angle\\
$\tilde\theta_0$&\eqref{t0tilde}& rescaled theta-angle\\
$\tr$&\eqref{Trdef}& Cartan-Killing form\\
$\Tr$&\eqref{Trdef}& positive definite quadratic form\\
$u^s$&\eqref{coulombbdef}& Coulomb branch coordinate\\
$\mathcal{U}$&\dpageref{Ham1}& space with singularities removed\\
$V_u$&\eqref{sympvecspace}& symplectic vectorspace\\
$V_u^{\rm e}$&\eqref{sympvecspace}& electric part of the symplectic vectorspace\\
$V_u^{\rm m}$&\eqref{sympvecspace}& magnetic part of the symplectic vectorspace\\
$W$&\eqref{lineoplattice}& Weyl group\\
$\widehat{W}(\cdot,\cdot)$ &\eqref{vanillawalls}& vanilla wall\\
$\widehat{W}(\cdot)$ &\eqref{fmsw}& framed wall\\
$\vec{x}_n$&\dpageref{defposdef}& position of $n$'th defect\\
$\vec{x}_{\mathrm{cm}}$&\dpageref{productmetric}&center of mass coordinates\\
$X$&\eqref{XYdef}& imaginary part of Higgs field\\
$X_\infty$&\eqref{XYasymptotic}& asymptotic value of $X$\\
$Y$&\eqref{XYdef}& real part of Higgs field\\
$Y_\infty$&\eqref{XYasymptotic}& asymptotic value of $Y$\\
$\mathcal{Y}$&\eqref{XYC}& combination of $X$ and $Y$\\
$\mathcal{Y}_\infty^\mathrm{cl}$&\eqref{YYcl}& asymptotic classical value of $\mathcal{Y}$\\
$\mathcal{Y}_\infty$&\eqref{YYexact}& imaginary part of dual special coordinate\\
$\mathcal{Y}_0$&\eqref{mathY0def}&an element of $\mathfrak{t}_{\gm}^{\perp}$\\
$z^m$&\dpageref{coordflows}& real coordinate on moduli space\\
$\dot z^m$&\dpageref{manton}& coordinate velocities on moduli space\\
$Z$&\eqref{RTalg}& central charge\\
$Z_\gamma$&\dpageref{vanillagrading}& central charge as a function of charge\\
$Z^\mathrm{cl}$&\eqref{Zcl}& classical central charge\\
$Z_{\gm}^{\textrm{1-lp}}$&\dpageref{Zgm1loop}&1-loop corrected monopole central charge\\
$Z^{\bfn}$&\eqref{complexcoords}& complex coordinate on moduli space\\
$\zeta$&\dpageref{zetadef}& line defect phase\\
$\zeta_\mathrm{van}$&\dpageref{zetavangen}& phase of minus the central charge\\
\end{longtable}}
\vspace{2cm}\noindent
We use the following notation for indices and coordinates:
{\setlength{\tabcolsep}{2pt}\begin{longtable}{lcc}
index& range & space whose directions it parameterizes\\
\hline
$\mu,\nu,\ldots$& $0,\ldots,3$& 4d Lorentzian space\\
$i,j,\ldots$& $1,2,3$ &3d spatial directions\\
$a,b,\ldots$& $1,\ldots,4$&4d Euclidean lift\\
$m,n,\ldots$& $1,\ldots, 4N$& moduli space (real coordinates)\\
$\um,\un,\ldots$&$1,\ldots, 4N$&\mbox{moduli space (orhtonormal frame)}\\
$\bfm,\bfn,\ldots$&$1,\ldots,2N$&\mbox{moduli space (complex coordinates)}\\
$\ubfm,\ubfn,\ldots$&$1,\ldots,2N$&\mbox{moduli space (unitary frame)}\\
$r,s\ldots$&$1,2,3$&\mbox{$SU(2)$ adjoint and quaternionic structure}\\
$A,B\ldots$&$1,2$&\mbox{$SU(2)_R$ fundamental index on fermions}\\
$\a,\beta\ldots$& 1,2& Weyl indices\\
$\dot\a,\dot\beta\ldots$&1,2&Weyl indices\\
$s,\ldots$ &$1,\ldots r=\rnk\mathfrak{g}$&Coulomb branch\\
$I,J, \ldots $&$1,\ldots,r=\mathrm{rnk}\mathfrak{g}$& simple roots/coroots\\
$I_A,J_A,\ldots$& $1,\ldots, d=\mathrm{rnk}\,\mathfrak{g}^{\mathrm{ef}}$& image of effective co-roots (see \eqref{corootpart})\\
$I_M, J_M\ldots $&$1,\ldots, r-d$& complement of im.~of eff.~co-roots (see \eqref{corootpart})\\
$A,B\ldots$&$1,\ldots,d$& effective co-roots
\end{longtable}}

%%%%%%%%%%%%%%%%%%%%%%%%%%%%%%%%%%%%%%%%%%%%%%%%%%%

\bibliographystyle{utphys}
\bibliography{MRVbiblio}

\end{document}